# V1: A Visual Query Language for Property Graphs

**Lior Kogan**[1]

The *property graph* is an increasingly popular data model. Pattern construction and pattern matching are important tasks when dealing with property graphs. Given a *property graph schema S*, a *property graph G,* and a *query pattern P,* all expressed in language $L = (L_S, L_G, L_P, L_R)$, *pattern matching* is the process of finding, merging, and annotating subgraphs of $G$ that *match* $P$. The syntaxes of sublanguages $L_S$, $L_G$, $L_P$, and $L_R$ define what and how symbols may be combined to create well-formed schemas, graphs, patterns, and query results, respectively. A semantics of $L_P$ is a mapping $(S, G, P) \rightarrow R$: which subgraphs of $G$ match $P$ and how to merge and annotate them. Expressive pattern languages support topological constraints, property value constraints, negations, quantifications, aggregations, and path semantics. *Calculated properties* may be defined for vertices, edges, and subgraphs, and constraints may be imposed on their evaluation result.

Query posers (e.g., analysts, investigators, researchers) would like to construct patterns with minimal effort, minimal trial and error, and in a manner that is coherent with the way they think. The ability to express patterns in a way that is aligned with their mental processes is crucial to the flow of their work and to the quality of the insights they can draw. Many potential query posers will not use textual query languages (e.g., Gremlin [1], GSQL [2], Cypher [3], PGQL [4], G-CORE [5], and the proposed GQL [6]), as the learning curve may be too sharp for someone with little or no prior experience in programming. Moreover, even experienced query posers who do use textual query languages often spend much time on the technicalities and may be distracted from their line of inquiry.

Since the capabilities of the human visual system with respect to pattern perception are remarkable, it is a matter of course that query patterns were to be expressed visually. Indeed, five of the abovementioned languages use ASCII-art syntax for expressing topological constraints. Needless to say, this type of 'visualization' is quite limited. While the use of ASCII-art declined in the 1990s in favor of graphical images, query languages began to adopt ASCII-art only recently. Visual (graphical, diagrammatic) query languages have the potential to be much more 'user-friendly' than their textual counterparts in the sense that patterns may be constructed and understood much more quickly and with much less mental effort. Given a schema, interactive tools can allow query posers to construct valid patterns with minimal typing. A long-standing challenge is to design a visual query language that is generic, has rich expressive power, and is highly receptive and productive. V1 attempts to answer this challenge.

V1[2] is a declarative visual pattern query language for schema-based property graphs. V1 supports property graphs with mixed (both directed and undirected) edges and unary edges, with multivalued and composite properties, and with *null* property values. V1 supports temporal data types, operators, and functions and can be extended to support additional data types, operators, and functions (one spatiotemporal model is presented). V1 is generic, concise, has rich expressive power, and is highly receptive and productive.

[1] ✉ koganlior1@gmail.com

[2] V1 is named after the primary visual cortex in our brain, which is also known as visual area one (V1).

## 1. THE PROPERTY GRAPH DATA MODEL

A *graph* consists of a set of *vertices* (*nodes*) V and a set of *binary edges* (*edges*) E. A graph may be *directed* (i.e., *digraph*), where each edge consists of an ordered pair $(u, v) \in V^2$, *undirected*, where each edge consists of an unordered pair $\{u, v\} \in V^2$, or *mixed*, where both directed and undirected edges may exist. A *pseudograph* is a graph in which both *loops* (an edge between a vertex and itself) and *multiple edges* (two or more edges connecting the same pair of vertices) are allowed. Sometimes *unary edges* (*half-edges* – edges attached to only one vertex) are allowed as well.

An *attributed graph* is a graph where nodes and/or edges are annotated with attributes or sets of attributes (i.e., *multi-attributed graph*). *Attributes* can be nominal, ordinal, key-value pairs, and so on.[3]

A *property graph* (also called *labeled property graph*) is a multi-attributed pseudograph[4,5] where

- Each vertex has a nominal attribute called *label* (*vertex-labeled graph*), or, in some definitions, a set of labels (*vertex multi-labeled graph*)
- Each edge has a nominal attribute called *label* (*edge-labeled graph*)
- Each vertex, edge, and unary edge has a set of key-value pair attributes called *properties*

Labels are usually strings – for example, *Person* and *member of*. The set of vertex-labels and the set of edge-labels are usually disjoint.

Each property has a distinct nominal name (key), usually a string, and a value of a certain data type (e.g., *string*, *integer*, *float*). For example: *weight*: *integer* = 450.

*Multivalued properties* (ordered or unordered, with or without duplicates) and *composite properties* are sometimes supported as well. Composite properties are composed of a set of sub-properties, each has a name and a value of a certain data type. For example: *name* = (*first*: *string* = "Brandon", *last*: *string* = "Stark").

*null* is a valid value for any *nullable property* and sub-property regardless of its data type.

The following semantics is commonly applied:

- Graph vertices represent entities. An *entity* is a physical, conceptual, virtual, or fictional object or 'thing' (e.g., a certain person, a certain guild, or a certain dragon). Any two objects or 'things' are distinguishable. A vertex's label denotes the *entity's type* (e.g., *Person*, *Guild*, and *Dragon*).
- Graph edges represent relationships. A *relationship* is an *association* or an *interaction* between a pair of entities. Each relationship is either *directional* (*unidirectional*, *asymmetric*) (e.g., an *owns* relationship between a *Person* entity and a *Horse* entity; an *offspring* relationship between two *Person* entities) or *bidirectional* (*non-directional*, *symmetric*, *reciprocal*) (e.g., a *sibling* relationship between two *Person* entities). Directed graph edges represent directional relationships, while undirected edges represent bidirectional relationships. An edge's label denotes the *relationship's type* (e.g., *owns* and *member of*).
- Unary edges, if allowed, usually represent entities' *actions* (e.g., *sleeps* action for a *Dragon* entity). A unary edge's label denotes the *action's type* (e.g., *sleeps*).
- Properties and sub-properties represent *features* (*characteristics*) and sub-features of entities (e.g., *name* for a *Person* entity), relationships (e.g., *timeframe* for an *owns* interaction), and actions (e.g., *timeframe* for a *sleeps* action).

---

[3] The term '*attributes'* is sometimes used in a limited scope to refer only to key-value pairs, while the term '*labels'* is used to refer to nominal attributes.

[4] A property graph is usually defined as a directed graph. V1 supports directed, undirected, and mixed graphs.

[5] A property graph is usually defined as a multigraph. We use the term *pseudographs* to disambiguate the allowance of loops.

- *Null-valued* [sub]property indicates that a [sub]property value is absent – either missing (unknown, classified, etc.) or inapplicable. For a birth date property, a *null* value would likely represent an unknown birth date. For a death date property, a *null* value may represent either that the date on which the person died is unknown (missing) or that the person is still alive (inapplicable). What *null* value means for each *nullable property* is outside the scope of the data model.

The term *property graph* was popularized by Rodriguez and Neubauer [7, 8], though other terms were used for similar data models. Tsai and Fu's *attributed relational graph* [9] is a directed multigraph where both nodes and edges have labels, and each label defines a set of numerical or logical attributes. Shaw et al. [10] used the term *Heterogeneous graph* for the same construct. Gallagher used the term *data graph* [11] to refer to graphs where vertices and/or edges may be typed and/or attributed. Singh et al. [12] used the term *M*3* (multi-modal, multi-relational, multifeatured) *network* to refer to graphs with multiple entity-types, multiple relationship-types, and multiple descriptive features for nodes and edges. Krause et al. [13] used the term *typed graph* to refer to graphs with typed nodes, typed edges, and typed node properties.

Various extensions were proposed, including support of schema-level and data-level *metaproperties* (properties of properties – e.g., units of measure, accuracy, and reliability).

## 2. THE PROPERTY GRAPH SCHEMA

A *schema* is a model for describing the structure of information in a certain domain using a certain data model. A *property graph schema* defines the entity-types, the relationship-types, and the properties of thereof.

A property graph may be:

- *Schema-free* (*schemaless*, *schema-independent*). A schema-free property graph neither defines nor enforces entity-types or relationships-types; each vertex and edge, regardless of its label, may have properties with any name and of any data type.
- *Schema-based* (*schema-driven*, *schema-full*, *schema-dependent*). A schema-based property graph is a property graph that conforms to a given schema.
- *Schema-mixed* (*schema-hybrid*), where a schema is defined, but additional elements (e.g., additional properties) may be used.

It is much easier to define patterns when the information is presented consistently. For example, to match patterns such as *Any person who owns a white horse,* one would first:

- Define entity-types *Person* and *Horse*
- Define a relationship-type *owns* that holds from entities of type *Person* to entities of type *Horse*
- For the *Horse* entity-type, define a *color* property with a nominal data type
- Ensure that all the information is structured accordingly

Though suggested property graph and property graph schema definitions have much in common (See Angles [14], Wu [15], Hartig and Hidders [16], and Thakkar et al. [17]), to date, there is neither a de jure nor a de facto standard definition (and hence, no standard property graph schema definition language).

The following *property graph schema model* (i.e., *property graph metamodel*) is assumed in this paper:

A *property graph schema* is defined by:

- A finite set of user-defined data types (on top of the built-in data types)
  - *Categorical (nominal* and *ordinal*) *data types* (e.g., *gender*: nominal *{male, female}*)
  - *Multivalued property data types* (e.g., *nicknames*: set(*string*))
  - *Composite data types* (e.g., *name* {*first*: *string*, *last*: *string*})
  - *Interval data types* (e.g., *span*: interval(*date*))
- A finite set of *entity-types*. For each entity-type:
  - A unique name
  - A set of *properties*. For each property:
    - A unique name
    - A data type
    - For properties and sub-properties with numeric data types, intervals of numeric data types, or multivalued numeric data types: an optional schema-level *units* metaproperty representing units of measure (e.g., Kg, cm, seconds)
- A finite set of *relationship-types*. For each relationship-type:
  - The relationship-type's directionality: directional or bidirectional
  - A unique name
  - A set of pairs of entity-types for which the relationship-type is applicable (e.g., *owns*: {(*Person*, *Horse*), (*Person*, *Dragon*)}. When a pair is of the same type (e.g., (*Dragon*, *Dragon*)), loops can be allowed or disallowed
  - A set of *properties* – similar to entity-types' properties

A predefined property-less entity-type *Null* serves two purposes:

- Realizing unary edges (action-types) as edges (relationship-types): an action-type can be realized as a relationship-type that is applicable between some entity-type and the *Null* entity-type. For example, a *sleeps*: {(*Dragon*, *Null*)} relationship-type realizes a sleep action-type for the *Dragon* entity-type.
- Realizing relationships to unknown or unimportant entities: sometimes a real entity is unknown or unimportant, but the existence of a relationship and the values of the relationship's properties – are important. For example, we may know that some dragons were owned in given timeframes, but we do not know or do not care who/what owned them. Still – we want to be able to store and query such information. *owns*: {(*Person*, *Dragon*), (*Guild*, *Dragon*), (*Null*, *Dragon*)} allows us to realize this.

Each entity of type *Null* is connected by exactly one relationship. A relationship between two *Null* entities is not allowed.

Property graph schema definitions may vary in many aspects, including:

- Supported *schema constraints* (properties *uniqueness* and *nullableness*, *property value constraints*, *relationships cardinality constraints*, disallow loops for certain relationship-types, etc.)
- Supported ways to declare user-defined data types
- Properties may be either:
  - Defined globally and assigned to one or more entity/relationship-types, or
  - Defined per entity/relationship-type: different entity/relationship-types may have a property with the same name but with a different data type
- *Directional relationship-types naming*:
  - A unique name for each direction (e.g., *owns*, *owned by*; *parent of*, *offspring of*), or
  - A unique name for only one direction
- Support of *derivation* (*specialization*) of entity-types, relationship-types, and property types

V1 can be utilized with most definitions with minimal changes.

## 3. PATTERNS AND PATTERN LANGUAGES

A *pattern* defines a set of topological and property value constraints on property graphs. Each property (sub)graph either *matches* the pattern or not. For some patterns, a given (sub)graph may match a pattern in more than one way.

When patterns are described in a natural language, they may be ambiguous, and the description may miss some nuances expressed in formal languages. Nevertheless, all the patterns below are described in English so that the reader may gain an intuitive understanding of their meaning.

Here are two examples:

- ***P1:*** *Any person who owns at least five white horses* (see Q101)

  *P1* defines the set of (sub)graphs where

  - There is a vertex *p* with a label *Person*
  - There are at least five vertices $h1..h_n$, each with a label *Horse*
  - Each of $h1..h_n$ has a *color* property, and its value is *white*
  - There are relationships from *p* to $h1..h_n$, each with a label *owns*

- ***P2:*** *Any person whose date of birth is between January 1, 970 and January 1, 980, who owns a white Horse, who owns a dragon whose name starts with 'M' and over the last month froze at least three dragons belonging to members of the Masons Guild*

  *P2* defines the set of (sub)graphs where

  - There is a vertex *p* with a label *Person*
  - *p* has a *birthdate* property of type *date*, and its value is between January 1, 970 and January 1, 980
  - There is at least one vertex *h* with a label *Horse*
  - There is a relationship from *p* to *h* with a label *owns*
  - *h* has a *color* property, and its value is *white*
  - There is at least one vertex *d* with a label *Dragon*
  - There is a relationship from *p* to *d* with a label *owns*
  - *d* has a *name* property with a value that starts with 'M'
  - There are at least three vertices $d1..d_n$, each with a label *Dragon*
  - There are relationships from *d* to any of $d1..d_n$, each with a label *freezes*
  - Each of these relationships has a *tf* property (stands for "time frame") with a *since* sub-property whose value is in the range [*now*()-*months*(3) .. *now*()]
  - There is at least one vertex *g* with a label *Guild*
  - *g* has a *name* property, and its value is *Masons*
  - There are one or more vertices $p1..p_n$, each with a label *Person*
  - There are relationships from any of $p1..p_n$ to *g*, each with a label *member of*
  - There are relationships from each of $p1..p_n$ to one or more of $d1..d_n$, each with a label *owns*. Each of $d1..d_n$ is connected by at least one of these relationships

The terms *entity* and *relationship* denote both pattern elements and graph elements. When the context may be ambiguous, we use the terms *pattern-entity* and *pattern-relationship* to refer to pattern elements, and the terms *graph-entity* and *graph-relationship* to refer to graph elements.

Given a property graph schema *S*, a property graph *G,* and a query pattern *P,* all expressed in some language $L = \{L_S, L_G, L_P\}$, *pattern matching* is the process of finding, merging, and annotating subgraphs of *G* that *match P*. Any valid subgraph that matches the pattern is called *an assignment.* We use *assignment* *to X* where *X* is a pattern-entity, a pattern-relationship or a set of thereof, to denote the graph-entity, the graph-relationship, or the set of thereof that matches *X* as part of an assignment.

In the patterns given below, unless otherwise stated, each reported assignment should include the graph-entity assigned to each mentioned pattern-entity and the graph-relationship assigned to each mentioned pattern-relationship. Hence, any reported assignment to *P1* should be composed of:

- A *Person* graph-entity
- Five or more *Horse* graph-entities, each of which has a *color* property, and its value is *white*
- The *owns* graph-relationships between the *Person* graph-entity to those *Horse* graph-entities

Consider the following alternative patterns:

- ***P1':*** *Any person who owns at least five white horses. Report only the person*
- ***P1":*** *Any person who owns at least five white horses. Report only the horses*

An answer to a query pattern may be:

- The set of all assignments
- The set of all entity assignments, but for each assignment – the union of all relationship assignments
- The union of all assignments. This is sometimes preferred since it avoids exponential or factorial explosion for many queries (e.g., if a person owns ten white horses, any subset of five or more horses compose an assignment to *P1*). However, for some patterns, individual assignments cannot be derived from their union.

Implementations may support one or more of the above.

Pattern languages differ in many aspects, including:

- *Genericity* – general-purpose (e.g., schema-driven) vs. domain-specific
- *Pattern representation* – *textual* vs. *visual* (*graphical*, *diagrammatic*)
- *Declarative / Imperative* – *Declarative languages* describe patterns but do not necessarily specify how to match them. *Imperative languages* describe patterns in terms of the steps required to match them on a given computation model. Some languages provide both declarative and imperative constructs
- *Expressive power* – The breadth of patterns that can be expressed. Formal pattern languages are always limited in a Gödelian sense
- *Receptivity* and *Productivity* (i.e., *readability and writability*) – How simple and intuitive is it to understand existing patterns and to construct new ones
- *Conciseness* – The fewness of symbols and symbol-types required for expressing patterns
- *Aesthetics* – The quality of patterns being visually appealing

V1 introduces the concept of *calculated properties* – non-inherent properties of graph-entities, graph-relationships, and subgraphs, defined as part of a pattern. Each calculated property's evaluation result is part of the reported query results, extending V1 capabilities beyond 'simple' pattern matching. For example, *The average number of horse ownerships per person* – a calculated property of the set of all graph-entities of type *Person* can be defined as part of a pattern. See Q356).

## 4. A SONG OF ICE AND FIRE

The following scenario, loosely based on George R. R. Martin's *A Song of Ice and Fire* [18], will serve us to demonstrate the language's expressive power.

The subjects of Sarnor, Omber, and the other kingdoms of the known world like their horses. There is one thing they love even more – that is their dragons. They own dragons of ice and fire. Like all well-behaved dragons, their dragons love to play. Dragons always play in couples. When playing, dragons often get furious, fire at each other (fire breath), and freeze one another (cold breath). Dragons usually freeze one another for periods of several minutes, but on occasion, when they are furious, they can freeze one another for periods of several hours. The subjects enjoy watching their dragons play. Fascinated by these magnificent creatures, they wrote thousands of books documenting each fire breath and each cold breath over the last 900 years.

The kings of Sarnor and Omber regularly pose queries about their history. More than often, it takes the royal historians and royal analysts several days to come up with answers, during which the kings tend to get pretty impatient. Lately, the high king of Sarnor posed a very complex query, and after waiting for results for more than two moons, he ordered the chief analyst to be executed. He then summoned his chief mechanics and ordered them to develop an apparatus that he could use to pose queries and get results quickly.

The engineers started by collecting all queries posed by their master over the last few years. Then they constructed a property graph schema using which these queries can be expressed.

The schema was composed of the following entity-types (and their properties):

- ***Person:*** *name* {*first*: *string*, *last*: *string*}, *gender*: nominal *{male, female}*, *birthDate*: *date*, *deathDate*: *date*, *height*: *int* [cm]
- ***Dragon:*** *name*: *string*, *color*: nominal *{black, white, …}*
- ***Horse:*** *name*: *string*, *color*: nominal *{black, white, …}*, *weight*: *int* [Kg]
- ***Guild:*** *name*: *string*
- ***Kingdom:*** *name*: *string*

the following directional relationship-types (and their properties):

- ***owns***: {(*Person*, *Horse*), (*Person*, *Dragon*), (*Guild*, *Horse*), (*Guild*, *Dragon*)} – *df*: *dateframe*
- ***fires at***: {(*Dragon*, *Dragon*)} – *time*: *datetime*; no loops allowed
- ***freezes***: {(*Dragon*, *Dragon*)} – *tf*: *datetimeframe*; no loops allowed
- ***offspring of***: {(*Person*, *Person*)}; no loops allowed
- ***member of***: {(*Person*, *Guild*)} – *df*: *dateframe*
- ***subject of***: {(*Person*, *Kingdom*)}

and of the following bidirectional relationship-type (and its properties):

- ***friend of***: {(*Person*, *Person*)} – *since*: *date*; no loops allowed

*Person*'s *name* is a composite property. The *date*, *datetime*, *dateframe,* and *datetimeframe* data types are defined in section 7 – Data Types, Operators, and Functions.

The engineers then represented the whole known history using a property graph that conforms to this schema.

## 5. V1 BASICS

The following sections describe the syntax and the semantics of the V1 language[6]. We start with the basics, adding more language elements as we go along.

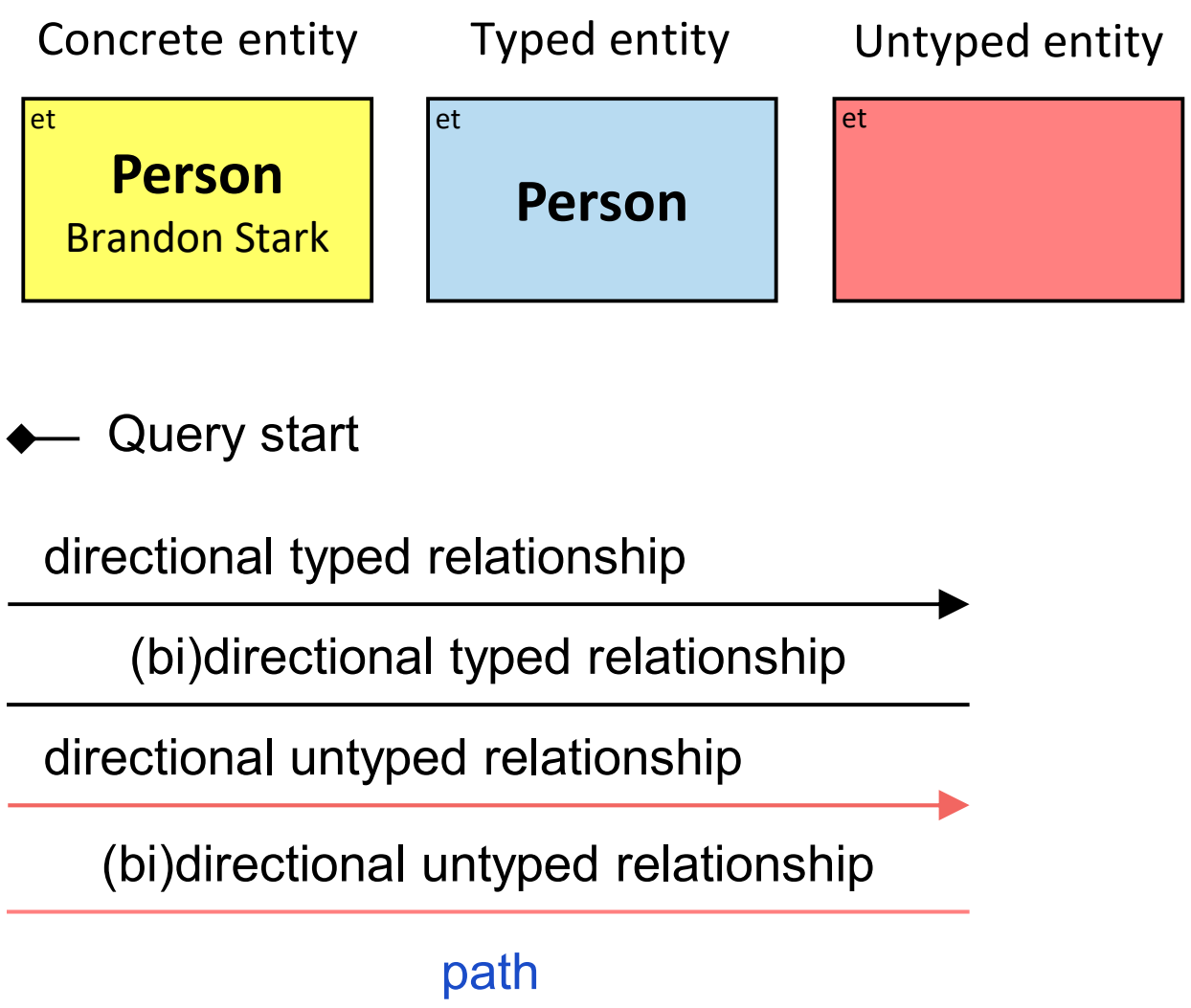

Patterns are generally read from left to right. Each pattern starts with **a small black diamond**, denoting the pattern start. The most straightforward patterns are structured as a sequence of rectangles, where consecutive rectangles are connected with an arrow or a line.

Yellow, blue, and red rectangles represent *concrete*, *typed* and *untyped* pattern-entities, respectively.

**A yellow rectangle** represents a *concrete entity*: a specific person, a specific horse, etc. A concrete entity has a single assignment – a specific graph-entity. The text inside the rectangle denotes the entity-type and the value of a *visualization expression* defined for this entity-type. For example, the visualization expression for the *Person* entity-type may be *name.first* || ' ' || *name.last* and its value, for a specific graph-entity, would be 'Brandon Stark.'

**A blue rectangle** represents a *typed entity*. The text inside the rectangle denotes an entity-type. Only graph-entities of this type may be assigned to the pattern-entity.

[6] V1 has two equivalent syntaxes for expressing patterns: A visual syntax – described here, and a textual syntax (JSON-based). There is a bijective mapping between patterns expressed in these two syntaxes. A description of the textual syntax, as well as sample textual patterns, are available at https://github.com/LiorKogan/V1.

**A red rectangle** represents an *untyped entity*. Graph-entities of different types may be assigned to an untyped pattern-entity (subject to *type constraints*. See section 16 – Untyped Entities).

Two consecutive rectangles can be connected with:

- A horizontal **black arrow**, representing a *directional typed relationship*,
- A horizontal **black line**, representing either a *bidirectional typed relationship* or a directional typed relationship where either direction is acceptable,
- A horizontal **red arrow**, representing an *untyped directional relationship* (See section 18 – Untyped Relationships),
- A horizontal **red line**, representing either an *untyped bidirectional relationship* or an *untyped directional relationship* where either direction is acceptable, or
- A horizontal **blue line**, representing a *path* (See section 21 – Paths)

Each black arrow/line has a label on top. The label denotes a relationship-type. For arrows – the label is aligned to the arrow's origin. For lines – the label is centered. Only graph-relationships of this type can match the pattern-relationship.

For every blue rectangle, red rectangle, black arrow, and black line, a pattern matching engine would look in the property graph for assignments. Graph entities are assigned to pattern entities. Graph relationships are assigned to pattern relationships. An *assignment* to the pattern is a set of graph-entities and graph-relationships that matches the whole pattern.

The relationship-type between any two entities must be valid with respect to the schema.

**Q1:** *Any dragon owned by Brandon Stark* (two versions)

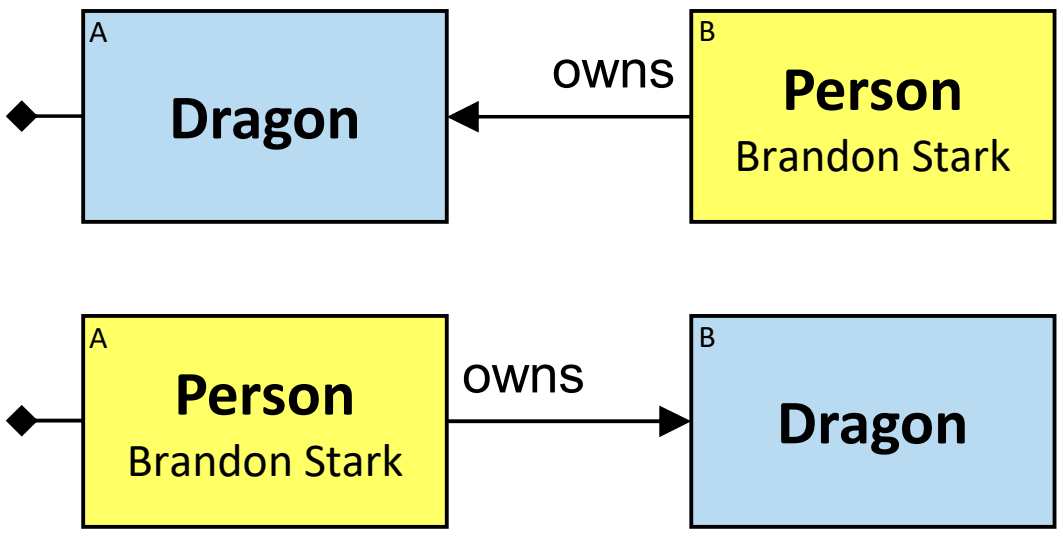

**Q2:** *Any dragon C that at least once had been frozen by a dragon owned by Brandon Stark*

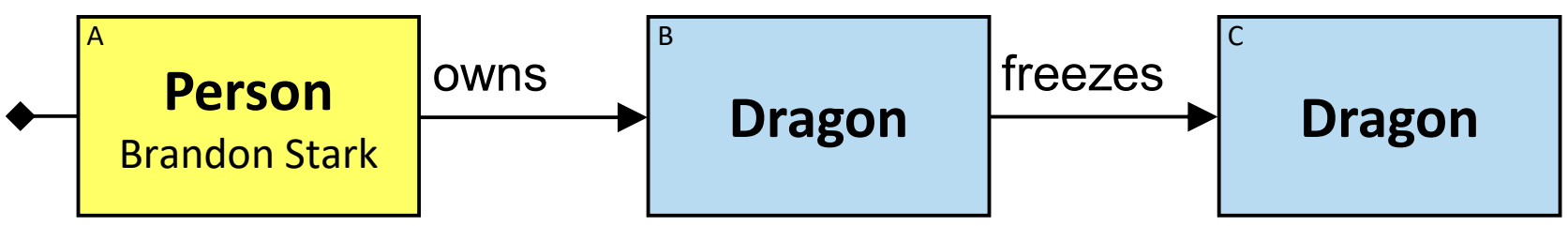

**Q184:** *Any dragon C that at least once froze a dragon owned by Brandon Stark or was frozen by a dragon owned by Brandon Stark*

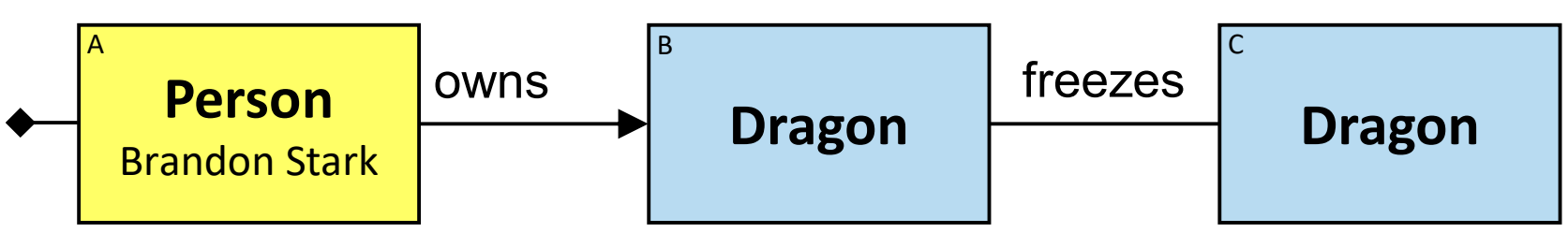

Both directions of the *freezes* relationship are acceptable. Therefore – a line (instead of an arrow) is used in the pattern.

## 6. EXPRESSIONS AND EXPRESSION CONSTRAINTS

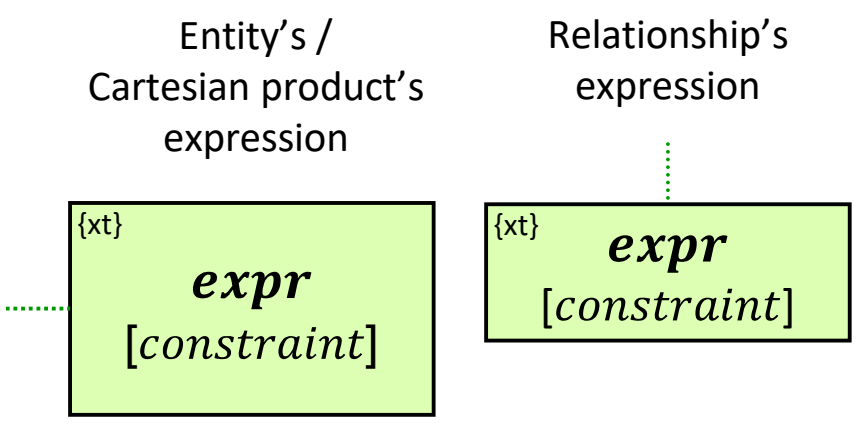

A **green rectangle** represents an *expression*. It contains:

- An *expression-tag* ('{xt}') (see section 24 – Referencing Expression-Tags)
- An *expression* ('*expr*')
- An optional *constraint on the result of the evaluation of the expression*, composed of:
    - A constraint operator
    - A constraint expression (except when the constraint operator is *is null* or *not null*)
- When units of measure are defined for the expression (based on the units of measures of the properties and the operators that compose the expression) – they are depicted as well (see Q117, Q304, Q265, Q95).

*expr* is an entity's expression, a relationship's expression, a Cartesian product's expression, or a global expression.

- An *entity's expression*

  The expression is composed of or depends on at least one property (inherent or calculated) of the connected entity and no properties of other entities/relationships.

  The green rectangle is connected to a pattern-entity (concrete, typed, or untyped) on its left.

  An expression-tag of an entity's expression is a *property of* each unique assignment to the pattern-entity.

- A *relationship's expression*

  The expression is composed of or depends on at least one property (inherent or calculated) of the connected relationship and no properties of other entities/relationships.

  The green rectangle is connected to a pattern-relationship on its top.

  An expression-tag of a relationship's expression is a *property of* each unique assignment to the pattern-relationship (Note that it is not a property of an assignment to the Cartesian product of the two related entities).

- A *Cartesian product's expression*

  The expression is composed of or depends on properties (inherent or calculated) of at least two entities (see {3} in Q207, {2}, {3} and {4} in Q340), at least two relationships (see {2} in Q267v2), or at least one entity and one relationship (see {2} and {4} in G3).

  The green rectangle is connected to one of the entities/relationships or appears on the same level as the leftmost entities (when there is no single leftmost entity) (see Q207).

An expression-tag of a Cartesian product's expression is a *property of* each unique assignment to the Cartesian product.

See also *extended Cartesian product's expression* in section 36 – Extended Aggregators.

- A *global expression*

  The expression is composed of and depends on no entity's expression, relationship's expression, nor Cartesian product's expressions.

  The green rectangle appears on the same level as the leftmost entities (see Q375).

  A global expression is a *global property.*

Any expression-tag of an expression that is not a property name, a sub-property name, or a constant is called a *calculated property*.

An *expression* is

- A literal – *string*, *integer*, or *float*
  Note: *date*, *datetime*, and *duration* literals are represented using the functions *date*(*string*), *datetime*(*string*) and *duration*(*string*), respectively. In visual syntax, these function names are omitted, and expressions are formatted according to the regional settings
- <*inherent property name*> (of a connected entity/relationship)
  (valid for a Cartesian product's expression only if it is connected to an entity/relationship),
- <*inherent property name*>.<*sub-property name*>[.<*sub-property name*> ...] (of a connected entity/relationship)
  (valid for a Cartesian product's expression only if it is connected to an entity/relationship),
- An *expression-tag* or an *aggregation-tag* (e.g., '{1}'),
- *op expr*, where *op* is a unary operator (e.g., '- {1}'),
- *expr op expr*, where *op* is a binary operator (e.g., '3 + {1}'),
- (*expr*),
- '*f*()' where *f* is a parameterless function (e.g., '*now*()'. See G11),
- '*f*(e1, e2, ...)' where *f* is a function and e1, e2, ... are expression (see Q353),
- 'e1.*f*' – equivalent to *f*(e1), where *f* is a function and e1 is an expression,
- 'e1.*f*(e2, e3, ...)' – equivalent to *f*(e1, e2, e3, ...), where *f* is a function and e1, e2, e3, ... are expressions,
- An *interval expression* (see Q327),
- A *set expression* (see Q318),
- A *bag expression* (see Q315), or
- A *list expression*

An *interval expression* can be explicitly constructed using the following syntaxes:

- (*expr1* .. *expr2*) – an open interval
- (*expr1* .. *expr2*] – a half-open interval
- [*expr1* .. *expr2*) – a half-open interval
- [*expr1* .. *expr2*] – a closed interval

Both *expr1* and *expr2* are of the same ordinal data type.
If *expr2* < *expr1* – *expr1* and *expr2* are swapped.
If either *expr1* or *expr2*, or both are evaluated to *null* – the interval is evaluated to *null*.

A *set* is an unordered collection of zero or more *non-null* values (called *elements*) of the same data type in which each element may occur only once.

A set can be explicitly constructed using the following syntax:

- {*expr*, *expr*, ...} – zero or more expressions of the same data type. *null* values are ignored.
  Duplicate values are merged into one distinct value.
  A comma after the last element is optional, except for a single-element set where it is mandatory.

A *bag* (*multiset*) is an unordered collection of zero or more *non-null* values (called *elements*) of the same data type in which elements may occur more than once.

A bag can be explicitly constructed using the following syntax:

- [*expr*, *expr*, ...] – zero or more expressions of the same data type. *null* values are ignored.
  A comma after the last element is optional.

A *list* is an ordered collection of values (called *elements*) of the same data type in which elements may occur more than once.

A list can be explicitly constructed using the following syntax:

- (*expr*, *expr*, ...) – zero or more expressions of the same data type.
  A comma after the last element is optional, except for a single-element list where it is mandatory.

Expressions must match the data types defined for each operator and function.

A constraint filters assignments; an assignment is valid only if the result of the expression's evaluation *satisfies* the constraint.

A constraint cannot be defined for a concrete entity's expression.

For untyped entities, expressions can be composed only of properties that are common to all valid entity-types. Valid entity-types for an untyped entity are defined implicitly (according to the types of the pattern-entities and pattern-relationships which are connected to the untyped entity) or explicitly (using entity-type constraints – see later) (see Q291).

A sub-property of a composite property is denoted as *property-name.sub-property-name* (e.g., *name.first*, *tf.since*).

**Q3:** *Any person who owns a dragon and whose first name is Brandon* (version 1)

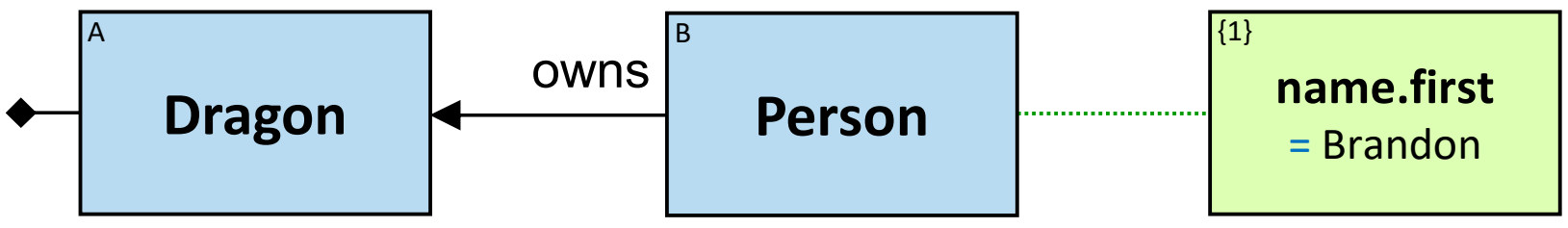

{1} is a property of each unique assignment to B.

**Q190:** *Any person who become a dragon owner at 1011 or later* (two versions)

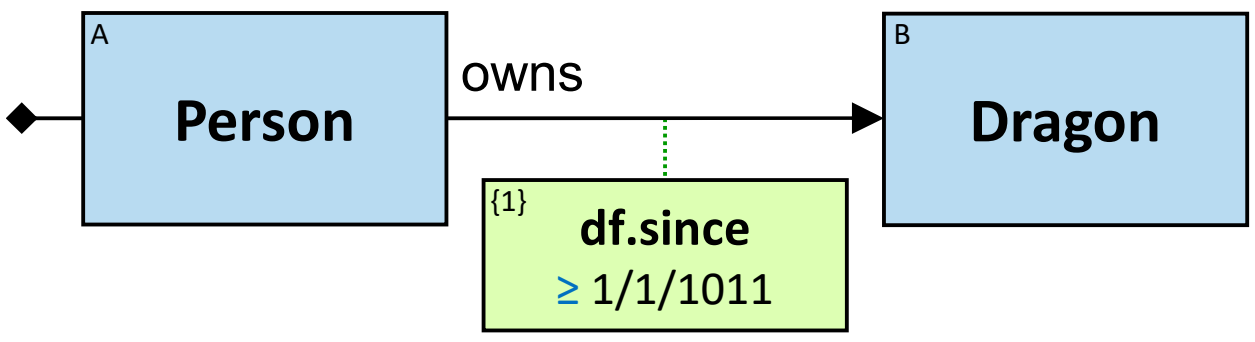

{1} is a property of each unique assignment to the *owns* relationship.

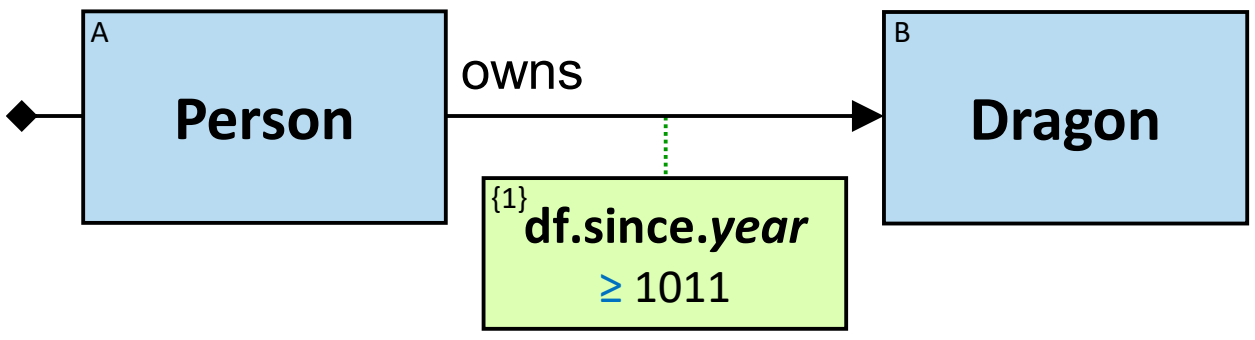

*year* is a function (see next section).

All V1 constraint operators are first evaluated using Kleene's three-valued logic [19] and then mapped to a two-valued logic: the constraint is either *satisfied* or *not satisfied*.

- *null* values are considered unknowns
- Each constraint is evaluated to *true*, *false*, or *unknown*
- For all constraint operators except *is null* and *not null*: if one or both operands are *null* – the result is *unknown*. The result of *5 = null is unknown.* The result of *null* = *null* is also *unknown*

Each constraint operator except *is null* and *not null* can be either blue or red.

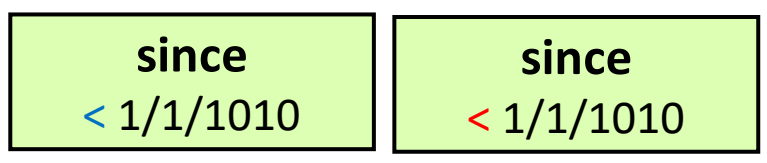

- **A blue constraint operator**: the constraint is satisfied if and only if it is evaluated to *true*
- **A red constraint operator**: the constraint is satisfied if and only if it is evaluated to *true* or to *unknown*

The following constraint operators can be only blue:

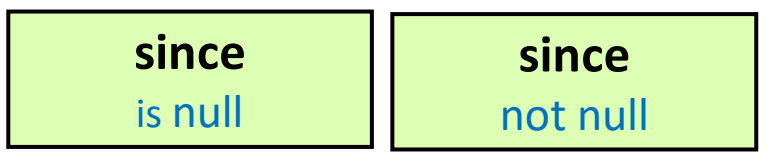

- A *is null* constraint is satisfied if and only if the expression is evaluated to *null*
- A *not null* constraint is satisfied if and only if the expression is not evaluated to *null*

See Q8, Q11, Q267

All V1's operators and all functions are well-defined when one or more of the operands or parameters are *null* or evaluated to *null* (e.g., 1 + *null* = *null*, *max*(*5*, *null*) = *null*).

## 7. DATA TYPES, OPERATORS, AND FUNCTIONS

One design goal of V1 is to make it applicable with many property graph database management systems. Therefore, we do not define a formal type system nor a formal set of operators and functions.

To present V1, we use the following loosely-defined type system:

**Built-in basic data types:**

| **Type** | **Notes** |
|---|---|
| *int* | Integer |
| *float* | Floating-point |
| *date* | Date. For simplicity, we will not consider time zones here. |
| *datetime* | Date and time. For simplicity, we will not consider time zones here. |
| *duration* | Can be negative |
| *string* | Unicode string |

**Built-in composite data types:**

| **Type** | **Notes** |
|---|---|
| *dateframe* | {*since*: *date*, *till*: *date*} |
| *datetimeframe* | {*since*: *datetime*, *till*: *datetime*} |

**Literals**:

| **Type** | **Examples** |
|---|---|
| *integer* | 12, -3 |
| *float* | 3., 3.12, -1.78e-6, NaN, -INF, +INF |
| *string* | "abc", 'abc' |

$S_t$ denotes a set of elements of type $t$<br>
$B_t$ denotes a bag of elements of type $t$<br>
$L_t$ denotes a list of elements of type $t$<br>
$I_t$ denotes an interval of ordinal type $t$

**Operators:**

| **Operator (*op*)** | **Operands and result type (result may be *null* as well)** |
|---|---|
| +, - (unary) | *op int* → *int*<br>*op* float → *float*<br>If the operand is *null* – the result is *null* |
| +, - (binary) | *int op int* → *int*<br>*float op float* → *float*<br>*duration op duration* → *duration*<br>*duration* + *date* → *date*<br>*date op duration* → *date*<br>*duration* + *datetime* → *datetime*<br>*datetime op duration* → *datetime*<br>If one or both operands are *null* – the result is *null* |
| * | *int* * *int* → *int*<br>*float* * *float* → *float*<br>*float* * *duration* → *duration*<br>*duration* * *float* → *duration*<br>If one or both operands are *null* – the result is *null* |
| / | *int* / *int* → *int* (truncated towards zero) |
| | *float* / *float* → *float*<br>*duration* / *float* → *duration*<br>If one or both operands are *null* – the result is *null* |
| % (modulo) | *int* % *int* → *int* (reminder has the same sign as the dividend)<br>If one or both operands are *null* – the result is *null* |
| ∪, ∩, -, △<br>(union, intersection, difference, symmetric difference) | $S_t$ *op* $S_t$ → $S_t$ (*t* is any type)<br>$B_t$ *op* $B_t$ → $B_t$ (*t* is any type)<br>If one or both operands are *null* – the result is *null* |
| ‖ (concatenation) | string ‖ string → string. *s* ‖ null = *null* ‖ *s* = *null*<br>$L_t$ ‖ $L_t$ → $L_t$ (*t* is any type). *null* ‖ *L* = *L* ‖ *null* = *null*<br>*t* ‖ $L_t$ → $L_t$ (*t* is any type). *null* ‖ *L* = (*null*, …). *t* ‖ *null* = *null*<br>$L_t$ ‖ *t* → $L_t$ (*t* is any type). *L* ‖ *null* = (…, *null*). *null* ‖ *t* = *null* |

**Constraint Operators:**

| **Operator (*op*)** | **Operands type (result is false / true / unknown)** |
|---|---|
| is null, not null (unary) | any_type *op*<br>(note that an empty set / bag / list is not a *null* value) |
| =, ≠ | both operands of the same type (any type) |
| <, >, ≤, ≥ | both operands of the same ordinal type:<br>*int* / *float* / *date* / *datetime* / *duration* / *string* / other ordinal |
| ∈, ∉ ([not] in) | left operand: any type *t*. right operand: $S_t$ / $B_t$ / $L_t$<br>left operand: any ordinal type *t*. right operand: $I_t$ |
| ∋, ∌ ([not] contains) | right operand: any type *t*. left operand: $S_t$ / $B_t$ / $L_t$<br>right operand: any ordinal type *t*. left operand: $I_t$ |
| ⊃, ⊇, ⊅, ⊉ ([not] [proper] super of) | both operands of the same type: string / $S_t$ / $B_t$ / $L_t$ (*t* is any type)<br>*t* is ordinal, and both operands of the same type: $I_t$ |
| ⊂, ⊆, ⊄, ⊈ ([not] [proper] sub of) | both operands of the same type: string / $S_t$ / $B_t$ / $L_t$ (*t* is any type)<br>*t* is ordinal, and both operands of the same type: $I_t$ |
| ⊳, ⋫ ([not] starts with) | both operands: *string*<br>left operand: $L_t$. right operand: *t* / $L_t$ (*t* is any type) |
| ⊲, ⋪ ([not] ends with) | both operands: *string*<br>left operand: $L_t$. right operand: *t* / $L_t$ (*t* is any type) |
| ≍, ≭ ([not] match) | both operands: *string* (right operand is a regex string) |

**Implicit type coercion:**

| **From type (t1)** | **To type (t2)** | **Examples** |
|---|---|---|
| *int* | *float* | 5 + 3. → 8. |
| *date* | *datetime* (00:00) | *date*("2018-04-05") = *datetime*("2018-04-05T00:00:00") → true |
| *dateframe* | interval(*date*) | *df*.*duration*, where *df* is a *dateframe* property |
| interval(*date*) | *Dateframe* | (*date*("2018-04-05") .. *date*("2018-05-05")).since → *date*("2018-04-05") |
| *datetimeframe* | interval(*datetime*) | *tf*.*duration*, where *tf* is a *datetimeframe* property |
| interval(*datetime*) | *Datetimeframe* | (*date*("2018-04-05") .. *datetime*("2018-05-05T00:00")).since → *datetime*("2018-04-05T00:00") |

Also, based on any of these coercion rules:

| **From type** | **To type** | **Examples** |
|---|---|---|
| {} (empty set) | set(t2) of any type t2 | {} ∪ {5} → {5} |
| [] (empty bag) | bag(t2) of any type t2 | [] ∪ [5] → [5], see Q349 |
| () (empty list) | list(t2) of any type t2 | () ‖ (5,) → (5,) |
| set(t1) | set(t2) | {3, 5, 8} ∪ {3., 8.} → {3., 5., 8.} |
| bag(t1) | bag(t2) | [3, 5, 8] ∪ [3., 8.] → [3., 3., 5., 8., 8.] |
| list(t1) | list(t2) | (3, 5, 8) ‖ (3., 8.) → (3., 5., 8., 3., 8.) |
| interval(t1) | interval(t2) | (3 .. 8) = (3. .. 8.) → true (3 .. 8) ⊃ (3. .. 5.) → true |

| {t1, t2, ...} | set(t2) * | {3, 5.} → {3., 5.} |
|---|---|---|
| [t1, t2, ...] | bag(t2) * | [3, 5.] → [3., 5.] |
| (t1, t2, ...) | list(t2) * | (3, 5.) → (3., 5.) |
| t1 .. t2 | interval(t2) * | [3 .. 5.) → [3. .. 5.) |

* where there is implicit type coercion from t1 to t2

### Functions

Functions over *float* expressions:

| Function | Notes |
|---|---|
| *trunc*(*float*) → *int* | truncates toward zero |
| *round*(*float*) → *int* | rounds to the nearest integer |
| *mRound*(*float*, *int*) → *int* | rounds to the nearest multiple of a given integer<br>(see G9, G10) |
| *seconds*(*float*) → *duration* | (e.g., seconds(6) is a duration of six seconds) |
| *minutes*(*float*) → *duration* | (see G10) |
| *hours*(*float*) → *duration* | |
| *days*(*float*) → *duration* | (see Q216, Q289) |
| *weeks*(*float*) → *duration* | One week = seven days |
| *months*(*float*) → *duration* | One month = 30.4367 days (see Q110) |
| *years*(*float*) → *duration* | One year = 365.24 days (see Q317) |

Functions over *string* expressions:

| Function | Notes |
|---|---|
| *length*(*string*) → *int* | (see Q255) |
| *toLower*(*string*) → *string* | (see Q308) |
| *date*(*string*) → *date* | String format: YYYY-MM-DD (e.g., “2018-04-23”)<br>In visual syntax, the function name is omitted, and the string is formatted according to the regional settings. |
| *datetime*(*string*) → *datetime* | String format: YYYY-MM-DD**T**HH:MM[:SS[.sss]]<br>(e.g., "2018-04-23T12:34:00")<br>In visual syntax, the function name is omitted, and the string is formatted according to the regional settings. |
| *duration*(*string*) → *duration* | String format adapted from Cypher:<br>**P**[n**Y**][n**M**][n**W**][n**D**][**T**[n**H**][n**M**][n**S**]]<br>(e.g., "P1Y2M10DT12H45M30.25S") Y: years, M: months, W: weeks, D: days, H: hours, M: minutes, S: seconds<br>In visual syntax, the function name is omitted, and the string is formatted. |

Functions over *datetime* expressions:

| Function | Notes |
|---|---|
| *date*(*datetime*) → *date* | (see Q158) |
| *year*(*datetime*) → *int* | (see Q185) |
| *month*(*datetime*) → *int* | The month of the year (1-12) |
| *day*(*datetime*) → *int* | The day of the month (1-31) |
| *hour*(*datetime*) → *int* | 0-23 (see G4) |
| *minute*(*datetime*) → *int* | 0-59 |
| sec(*datetime*) → *int* | 0-59 |
| *yearsSinceEpoch*(*datetime*) → *float* | |
| *monthsSinceEpoch*(*datetime*) → *float* | |
| *weeksSinceEpoch*(*datetime*) → *float* | |

| *daysSinceEpoch*(*datetime*) → *float* | |
|---|---|
| *hoursSinceEpoch*(*datetime*) → *float* | |
| *minsSinceEpoch*(*datetime*) → *float* | |
| *span*(*datetime*, *datetime*) → *duration* | Positive difference (see Q374) |

Functions over *date* expressions:

| **Function** | **Notes** |
|---|---|
| *year*(*date*) → *int* | |
| *month*(*date*) → *int* | The month of the year (1-12) |
| *day*(*date*) → *int* | The day of the month (1-31) |
| *yearsSinceEpoch*(*date*) → *float* | |
| *monthsSinceEpoch*(*date*) → *float* | |
| *weeksSinceEpoch*(*date*) → *float* | |
| *daysSinceEpoch*(*date*) → *float* | |

Functions over *dateframe* expressions:

| **Function** | **Notes** |
|---|---|
| *duration*(*dateframe*) → *duration* | |
| *overlap*(*dateframe*, *dateframe*) → *duration* | Always non-negative<br>(see Q267v2) |

Functions over *datetimeframe* expressions:

| **Function** | **Notes** |
|---|---|
| *duration*(*datetimeframe*) → *duration* | (see Q110) |
| *overlap*(*datetimeframe*, *datetimeframe*) → *duration* | Always non-negative |

Functions over *duration* expressions:

| **Function** | **Notes** |
|---|---|
| *years*(*duration*) → *float* | One year = 365.24 days |
| *months*(*duration*) → *float* | One month = 30.4367 days |
| *weeks*(*duration*) → *float* | One week = 7 days |
| *days*(*duration*) → *float* | (see Q328) |
| *hours*(*duration*) → *float* | |
| *minutes*(*duration*) → *float* | |
| *seconds*(*duration*) → *float* | |

Functions over *set* expressions:

| **Function** | **Notes** |
|---|---|
| *count*($S_t$) → *int* | Number of elements |
| *bag*($S_t$) → $B_t$ | Set to bag |
| *list*($S_t$) → $L_t$ | Set to list |
| *el*($S_t$) → *t* | (see section 44 – Multivalued Functions and Expressions) |
| *subset*($S_t$) → $S_t$ | (see section 44 – Multivalued functions and Expressions) |
| *min*($S_t$) → *t*<br>*max*($S_t$) → *t* | *t* is an ordinal type<br>*null* when $S_t$ is *null* or when it is empty |
| *avg*($S_t$) → *t* or *float* | *t* is an ordinal type<br>if *t* is *int* – the result is *float*<br>*null* when $S_t$ is *null* or when it is empty |
| *sum*($S_t$) → *t* | *t* is *int* / *float* / *duration* (not *date* / *time* / datetime)<br>*null* when $S_t$ is *null*; zero when it is empty |

| *min*($S_t$, *n*) → $S_t$<br>*max*($S_t$, *n*) → $S_t$ | *t* is an ordinal type<br>set of (up to) *n* smallest/largest elements<br>*null* when $S_t$ is *null*, {} when it is empty |
|---|---|
| *overlap*($S_{dateframe}$) → *duration*<br>*overlap*($S_{datetimeframe}$) → *duration* | The duration of the overlap between all members of *S*<br>Always non-negative (see Q371) |
| *union*($S_{S_t}$) → $S_t$<br>*union*($S_{B_t}$) → $B_t$ | The union of all members of a set of sets/bags (*t* is any type) |
| *intersection*($S_{S_t}$) → $S_t$<br>*intersection*($S_{B_t}$) → $B_t$ | The intersection of all members of a set of sets/bags (*t* is any type) |

Functions over *bag* expressions:

| **Function** | **Notes** |
|---|---|
| *count*($B_t$) → *int* | Number of elements |
| *distinct*($B_t$) → *int* | Number of distinct elements |
| *multiplicity*($B_t$, *t*) → *int* | Number of times *t* occurs in $B_t$ |
| *set*($B_t$) → $S_t$ | Bag to set |
| *list*($B_t$) → $L_t$ | Bag to list |
| *el*($B_t$) → *t* | (see section 44 – Multivalued Functions and Expressions) |
| *subbag*($B_t$) → $B_t$ | (see section 44 – Multivalued Functions and Expressions) |
| *min*($B_t$) → *t*<br>*max*($B_t$) → *t* | *t* is an ordinal type<br>*null* when $B_t$ is *null* or when it is empty |
| *avg*($B_t$) → *t* or *float* | *t* is an ordinal type<br>if *t* is *int* – the result is *float*<br>*null* when $B_t$ is *null* or when it is empty |
| *sum*($B_t$) → *t* | *t* is *int* / *float* / *duration* (not *date* / *time* / *datetime*)<br>*null* when $B_t$ is *null*; zero when it is empty |
| *min*($B_t$, *n*) → $B_t$<br>*max*($B_t$, *n*) → $B_t$ | *t* is an ordinal type<br>bag of (up to) *n* smallest/largest elements<br>*null* when $B_t$ is *null*, [] when it is empty |
| *overlap*($B_{dateframe}$) → *duration*<br>*overlap*($B_{datetimeframe}$) → *duration* | The duration of the overlap between all members of *B*<br>Always non-negative |
| *union*($B_{S_t}$) → $S_t$<br>*union*($B_{B_t}$) → $B_t$ | The union of all members of a bag of sets/bags (*t* is any type) |
| *intersection*($B_{S_t}$) → $S_t$<br>*intersection*($B_{B_t}$) → $B_t$ | The intersection of all members of a bag of sets/bags (*t* is any type) |

Functions over *list* expressions:

| **Function** | **Notes** |
|---|---|
| *count*($L_t$) → *int* | Number of elements |
| *distinct*($L_t$) → *int* | Number of distinct *non-null* elements |
| *multiplicity*($L_t$, *t*) → *int* | Number of times *t* occurs in $L_t$ |
| *at*($L_t$, *n*) → *t* | $n^{th}$ element (1-based)<br>*null* if *n* is out of range |
| *set*($L_t$) → $S_t$ | List to set |
| *bag*($L_t$) → $B_t$ | List to bag |
| *min*($L_t$) → *t*<br>*max*($L_t$) → *t* | *t* is an ordinal type<br>*null* values are ignored<br>*null* when $L_t$ is *null* or when it contains no *non-null* elements |
| *avg*($L_t$) → *t* or *float* | *t* is an ordinal type<br>if *t* is *int* – the result is *float*<br>*null* values are ignored<br>*null* when $L_t$ is *null* or when it contains no *non-null* elements |
| *sum*($L_t$) → *t* | t is *int* / *float* / *duration* (not *date* / *time* / *datetime*)<br>*null* values are ignored |
| | zero when $L_t$ is *null* or when it contains no *non-null* elements |
| *min*($L_t$, *n*)→ $L_t$<br>*max*($L_t$, *n*)→ $L_t$ | *t* is an ordinal type<br>*null* values are ignored<br>list of (up to) *n* smallest/largest elements<br>*null* when $L_t$ is *null*, () when it contains no *non-null* elements |
| *sort*($L_t$) → $L_t$ | *t* is an ordinal type<br>Sorted list |
| *invsort*($L_t$) → $L_t$ | *t* is an ordinal type<br>Inverse-sorted list |

Functions over *interval* expressions:

| **Function** | **Notes** |
|---|---|
| *low*($I_t$) → *t* | Lower bound<br>*null* when $I_t$ is *null* |
| *high*($I_t$) → *t* | Higher bound<br>*null* when $I_t$ is *null* |
| *set*($I_t$) → $S_t$ | Interval to set<br>*t* is discrete (*int*, *datetime*, or another ordinal type)<br>*null* when $I_t$ is *null* |
| *bag*($I_t$) → $B_t$ | Interval to bag<br>*t* is discrete (*int*, *datetime*, or another ordinal type)<br>*null* when $I_t$ is *null* |
| *list*($I_t$) → $L_t$ | Interval to list<br>*t* is discrete (*int*, *datetime*, or another ordinal type)<br>*null* when $I_t$ is *null* |

Other functions:

| **Function** | **Notes** |
|---|---|
| *now*() → *datetime* | |
| *date*(year, month, day) → *date* | construct date using three integers (see Q353) |
| *min*(*t*, *t*, …) → *t*<br>*max*(*t*, *t*, …) → *t* | one or more values of the same ordinal type<br>*null* when at least one value is *null*<br>*min*({*t*, *t*, …})and *max*({*t*, *t*, …}) ignore *null* values (see Q317) |

Implementations may support additional data types, operators, functions, and constraints. Implementations may support *opaque data types* – data types where the internal data representation is not exposed. For each opaque data type – a set of functions and operators may be defined (see *location* date type in Section 45 – Application: Spatiotemporality).

## 8. QUANTIFIERS

A **vertical purple rectangle** represents a *vertical quantifier*. It contains a quantifier type.

Vertical quantifiers (or simply 'quantifiers') can be used when there is a need to satisfy more than one constraint.

**Q3:** *Any person who owns a dragon and whose first name is Brandon* (version 2)

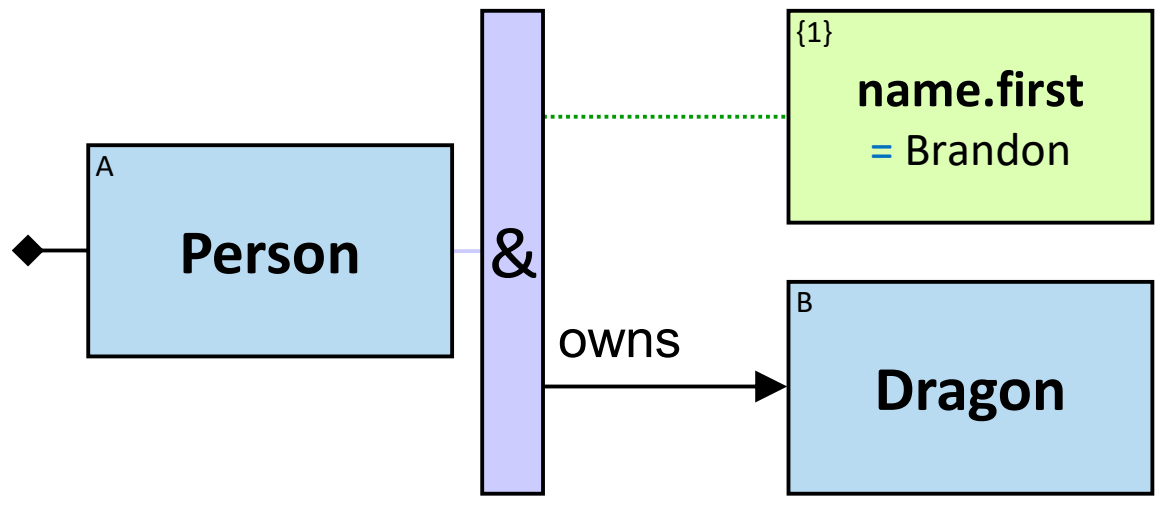

**Q219:** *Any person who owns a white horse weighing more than 200 Kg*

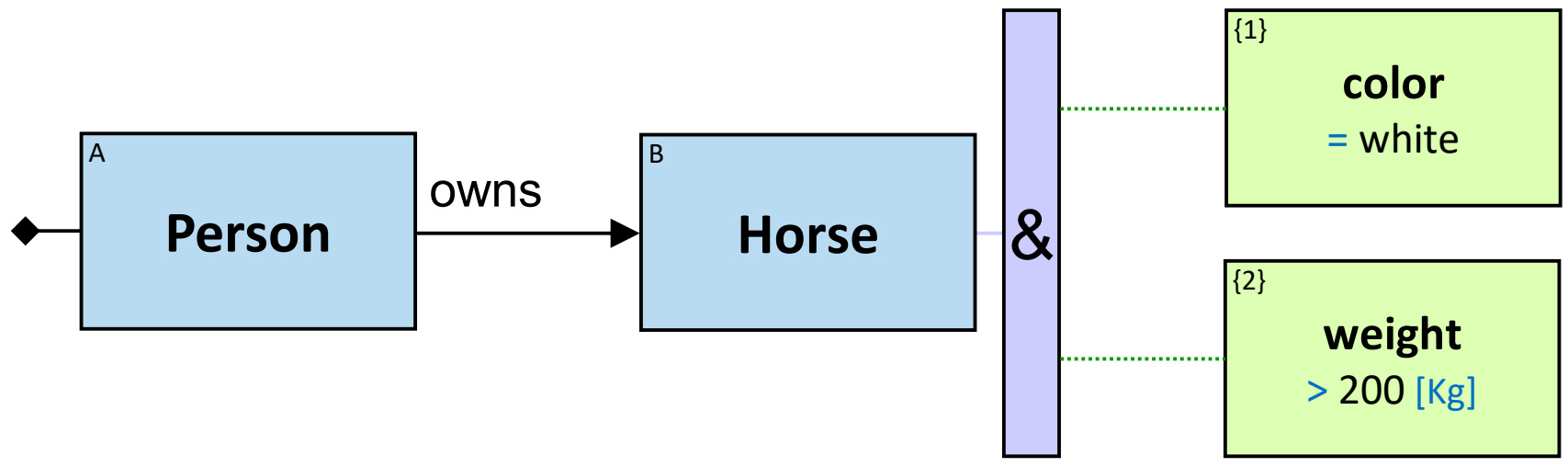

**Q304:** *Any person who owns a white horse and who owns a horse weighing more than 200 Kg*

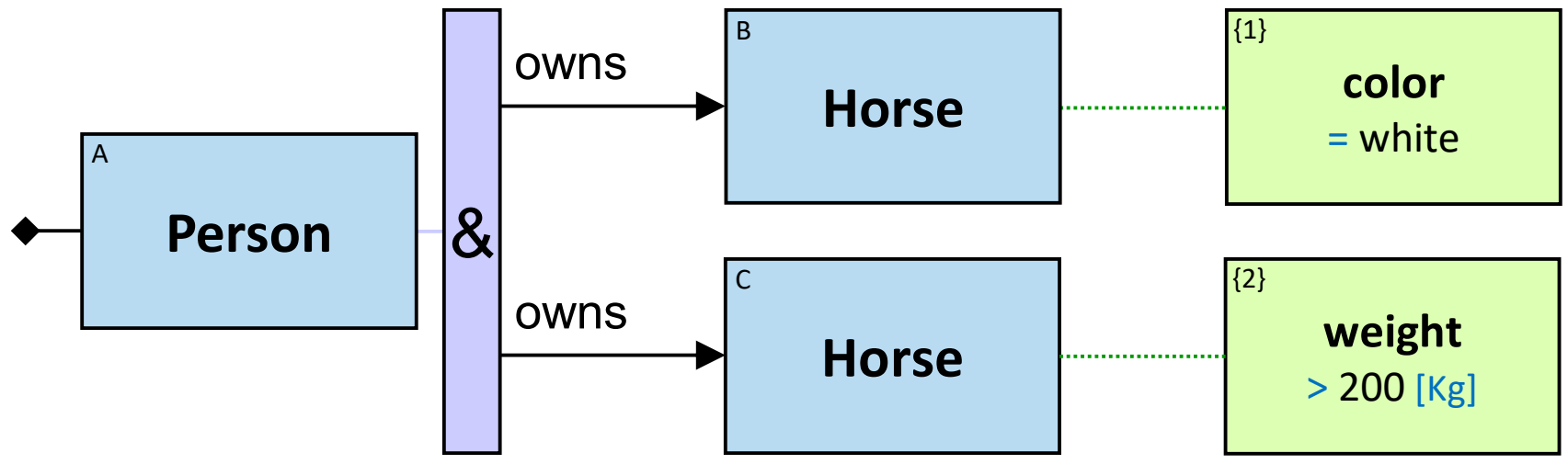

Note that the same graph-entity can match more than one pattern-entity. Either the same horse or different horses may be assigned to B and C (this can be avoided: see *identicality*, *nonidenticality*, and *order constraints* in section 9 – Entity-tags).

A vertical quantifier has one connection on its left side and zero or more *branches* on its right side. On its left side, there is an entity, a quantifier, or the pattern's start. Except at the pattern's start, a quantifier may be wrapped with an 'O' (see Q147, Q149).

When a quantifier is directly left of the pattern start, each branch may start with:

- An entity (see Q108),
- A Cartesian product's expression (see Q207), or
- A quantifier (see Q332v2)

When a quantifier is directly left of an entity, each branch may start with:

- A relationship/path, optionally with a relationship/path-negator, optionally wrapped with a negator or with an 'O' (see Q358, Q359),
- An entity's expression (see Q3),
- A Cartesian product's expression (see Q340), or
- A quantifier (see Q8)

A quantifier's branch composed of an entity's expression with no constraint does not affect the quantifier's evaluation (see Q109).

Let $b$ denote the number of branches, excluding branches composed of entity's expression with no constraint and excluding branches that start with an 'O'.

We will name the left side of the quantifier *the left component*, and anything that follows a branch, up to the branch's end, *a right component*.

12 quantifier types are defined:

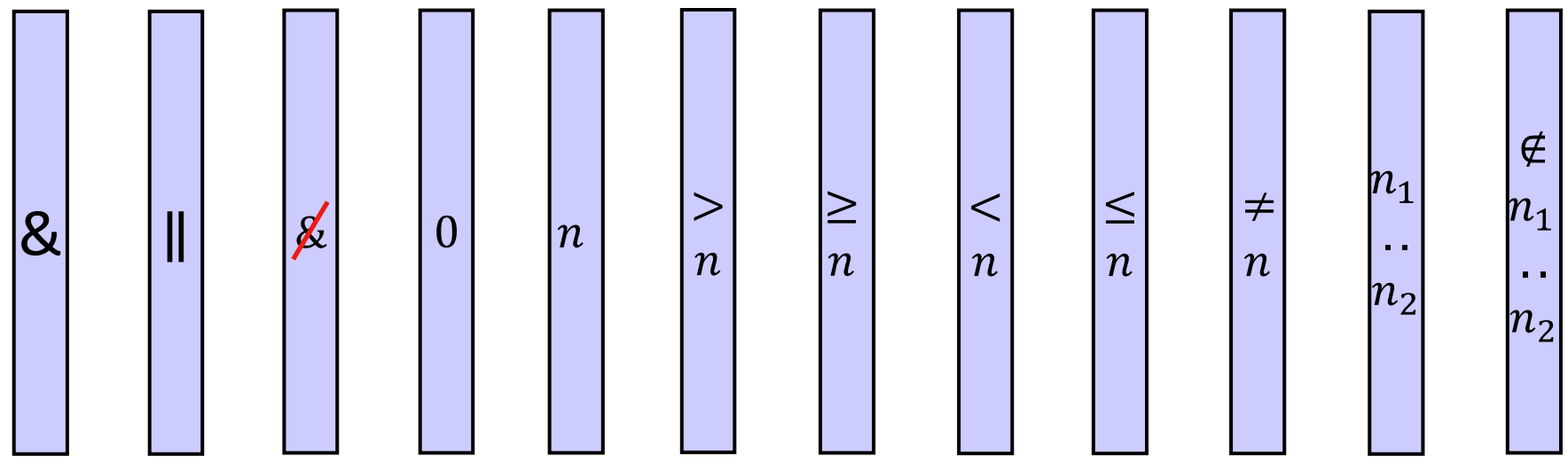

- **All** (denoted '&')

  If $b$ is zero, an assignment matches the pattern if and only if it matches the left component. Otherwise – an assignment matches the pattern if and only if it matches the whole pattern.

- **Some** (denoted '|')

  If $b$ is zero, no assignment matches the pattern. Otherwise – an assignment matches pattern $P$ if and only if it matches pattern $Q$ where

  - *Q's* left component is identical to *P's*
  - $Q$ has $i$ right components identical to *P's*, $1 \le i \le b$, and no other right components
  - The quantifier is replaced with an *All* quantifier

- **Not all (but more than 0)** (denoted by an '&' with stroke)

  If $b$ is zero, no assignment matches the pattern. Otherwise – an assignment $A$ matches pattern $P$ if and only if it matches pattern $Q$ where

  - *Q's* left component is identical to *P's*
  - $Q$ has $i$ right components identical to *P's*, $1 \le i < b$, and no other right components
  - The quantifier is replaced with an *All* quantifier

  and there is no assignment $B$ with a similar left component as *A's* that matches pattern $R$ where

  - *R's* left component and all its right components are identical to *P's*

- The quantifier is replaced with an *All* quantifier

- **None** (denoted ‘0’)

If $b$ is zero, an assignment matches the pattern if and only if it matches the left component. Otherwise – an assignment $A$ matches pattern $P$ if and only if it matches pattern $Q$ where

- *Q’s* left component is identical to *P’s*
- The quantifier and the right components are removed

and there is no assignment $B$ with a similar left component as *A’s* that matches pattern $R$ where

- *R’s* left component is identical to *P’s*
- $R$ has $i$ right components identical to *P’s*, $1 \leq i \leq b$, and no other right components
- The quantifier is replaced with an *All* quantifier

The *None* quantifier may not start a pattern.

- **= n**; $b \geq 1$, $n \in [1, b]$

An assignment $A$ matches pattern $P$ if and only if it matches pattern $Q$ where

- *Q’s* left component is identical to *P’s*
- $Q$ has $n$ right components identical to *P’s*, and no other right components
- The quantifier is replaced with an *All* quantifier

and, if $n \neq b$, there is no assignment $B$ with a similar left component as *A’s* that matches pattern $R$ where

- *R’s* left component is identical to *P’s*
- $R$ has $i$ right components identical to *P’s*, $i > n$, and no other right components
- The quantifier is replaced with an *All* quantifier

- **$>$ n**; $b \geq 2$, $n \in [0, b\text{-}1]$

An assignment $A$ matches pattern $P$ if and only if it matches pattern $Q$ where

- *Q’s* left component is identical to *P’s*
- $Q$ has $i$ right components identical to *P’s*, $n < i \leq b$, and no other right components
- The quantifier is replaced with an *All* quantifier

- **≥ n**; $b \geq 2$, $n \in [1, b]$

An assignment $A$ matches pattern $P$ if and only if it matches pattern $Q$ where

- *Q’s* left component is identical to *P’s*
- $Q$ has $i$ right components identical to *P’s*, $n \leq i \leq b$, and no other right components
- The quantifier is replaced with an *All* quantifier

- **$<$ n (but more than 0)**; $b \geq 2$, $n \in [2, b]$

An assignment $A$ matches pattern $P$ if and only if it matches pattern $Q$ where

- *Q’s* left component is identical to *P’s*
- $Q$ has $i$ right components identical to *P’s*, $1 \leq i < n$, and no other right components

- The quantifier is replaced with an *All* quantifier

and there is no assignment *B* with a similar left component as *A's* that matches pattern *R* where

- *R's* left component is identical to *P's*
- *R* has *i* right components identical to *P's*, $i \geq n$, and no other right components
- The quantifier is replaced with an *All* quantifier

- **≤ n (but more than 0)**; $b \geq 2$, $n \in [1, b]$

  An assignment *A* matches pattern *P* if and only if it matches pattern *Q* where

  - *Q's* left component is identical to *P's*
  - *Q* has *i* right components identical to *P's*, $1 \leq i \leq n$, and no other right components
  - The quantifier is replaced with an *All* quantifier

  and there is no assignment *B* with a similar left component as *A's* that matches pattern *R* where

  - *R's* left component is identical to *P's*
  - *R* has *i* right components identical to *P's*, $i > n$, and no other right components
  - The quantifier is replaced with an *All* quantifier

- **≠ n (but more than 0)**; $b \geq 2$, $n \in [1, b]$

  $\equiv (< n) \lor (> n)$

- **n1..n2**; $b \geq 2$, $n1 \in [1, b]$, $n2 \in [2, b]$, $n1 < n2$

  An assignment *A* matches pattern *P* if and only if it matches pattern *Q* where

  - *Q's* left component is identical to *P's*
  - *Q* has *i* right components identical to *P's*, $n1 \leq i \leq n2$, and no other right components
  - The quantifier is replaced with an *All* quantifier

  and there is no assignment *B* with a similar left component as *A's* that matches pattern *R* where

  - *R's* left component is identical to *P's*
  - *R* has *i* right components identical to *P's*, $i > n2$, and no other right components
  - The quantifier is replaced with an *All* quantifier

- **∉ n1..n2 (but more than 0)**; $b \geq 4$, $n1 \in [2, b-1]$, $n2 \in [3, b]$, $n1 < n2$

  $\equiv (< n1) \lor (> n2)$

The order of the branches does not affect the evaluation result.

**Q8:** *Any person born before 970 and passed away or whose father was born not later than January 1, 950*

The following data semantics is assumed: *null* death date means that the person is still alive.

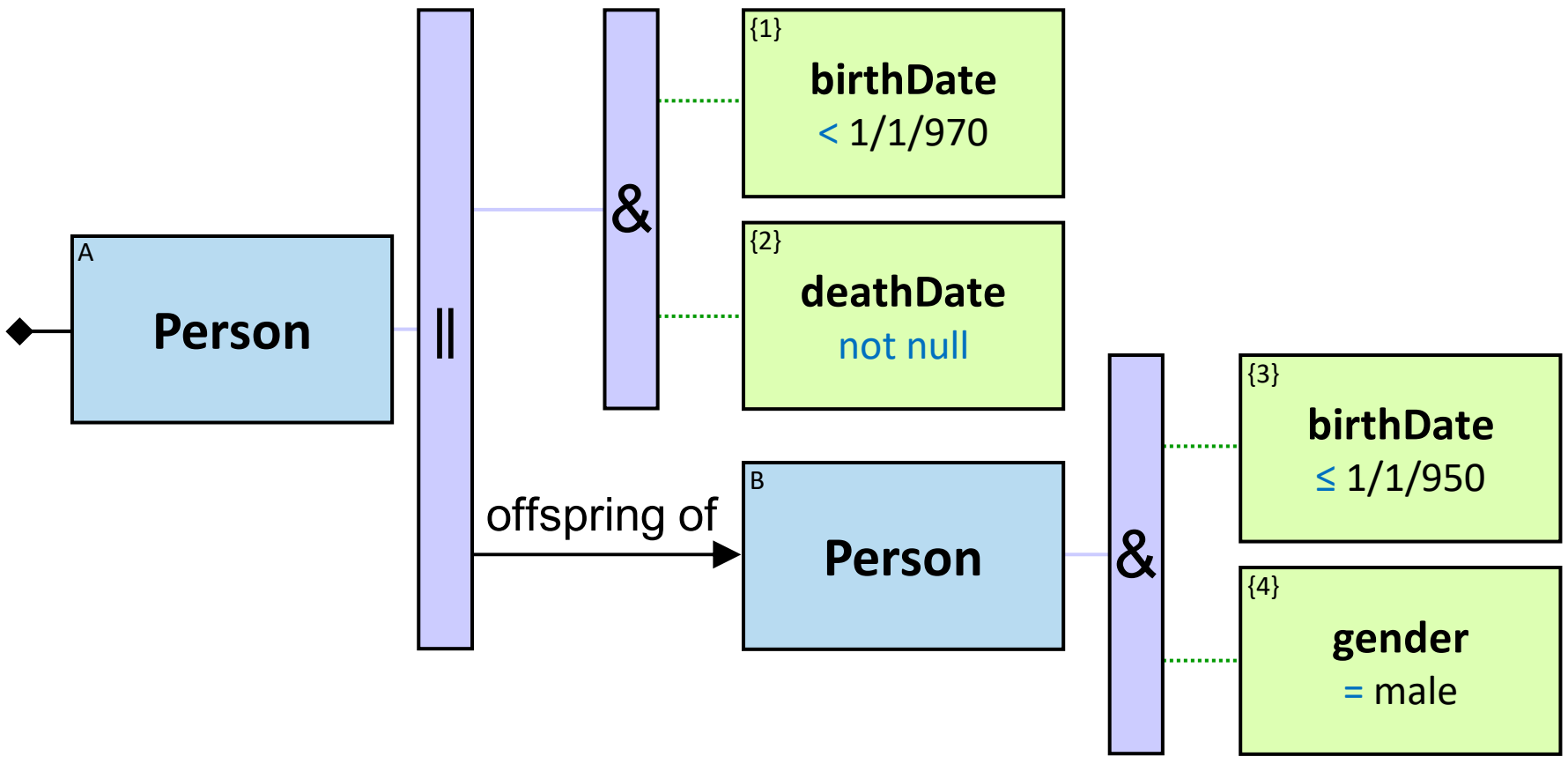

**Q11:** *Any current member of the Masons Guild who on or after January 1, 1011, befriended someone who had left the Saddlers guild or the Blacksmiths guild in June 1010 or later*

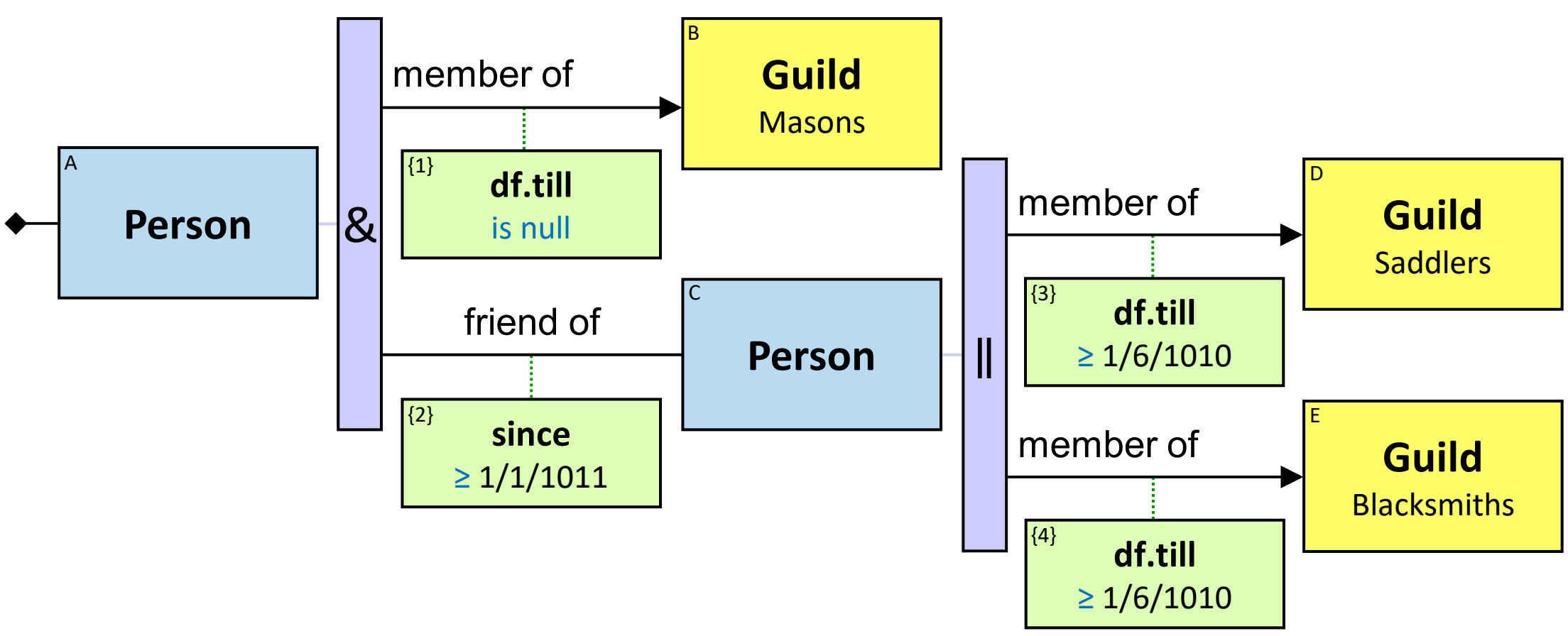

Note that the constraint '*member of df.till is null*' is based on the assumption that a *null* value means that the person is still a member. Alternatively, it could have meant that *df.till* is unknown. It depends on the semantics of the *till* sub-property.

## 9. ENTITY-TAGS

The letter in the top-left corner of each entity rectangle (concrete, typed, and untyped) is called an *entity-tag*.

Entity-tags appear in query results as well: any graph-entity in a query result is annotated with the same tag as the pattern-entity to which it was assigned so that the query poser can understand why any given entity is part of the result. As part of the result, a graph-entity may be annotated with more than one entity-tag, as it may be assigned to several pattern-entities (in the same assignment or in different assignments – when assignments are merged).

Entity-tags may be referenced:

- in *identicality*, *nonidenticality*, and *order constraints*
- in an aggregator *per* clause
- in an A1/M1/M2/M3 "*et × et ×* ..." clause
- in an M1 aggregator "with min/max ..." clause (see Q196)

*Identicality constraints* can be used when the same graph-entity should be assigned to:

- Several typed pattern-entities of the same type
- Several untyped pattern-entities

**Q4:** *Any person A whose dragon was frozen by a dragon owned by (at least) one of A's parents*

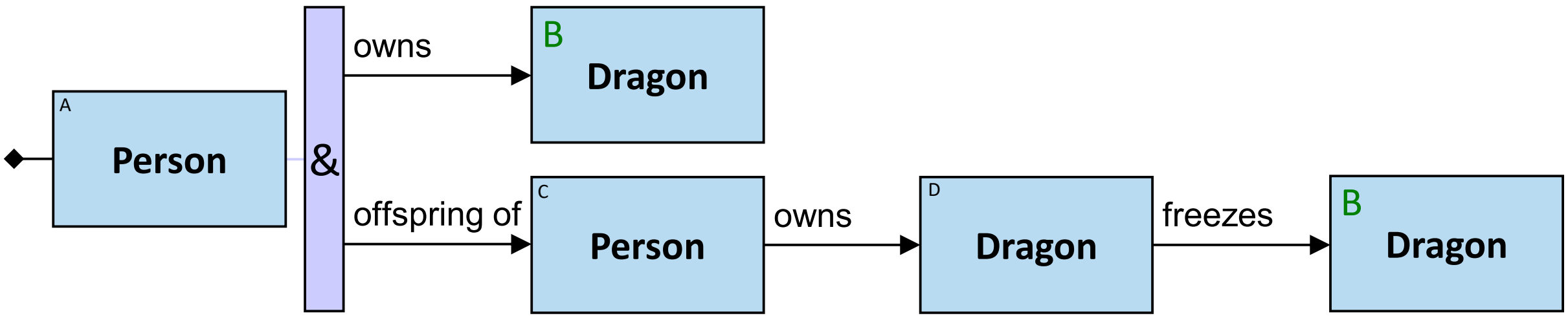

Entity-tag 'B' is used to enforce identical assignment to two *Dragon* pattern-entities.

Entity-tags with identicality constraints are depicted in green.

**Q9:** *Any pair of dragons (A, B) where A froze B in both 980 and 984*

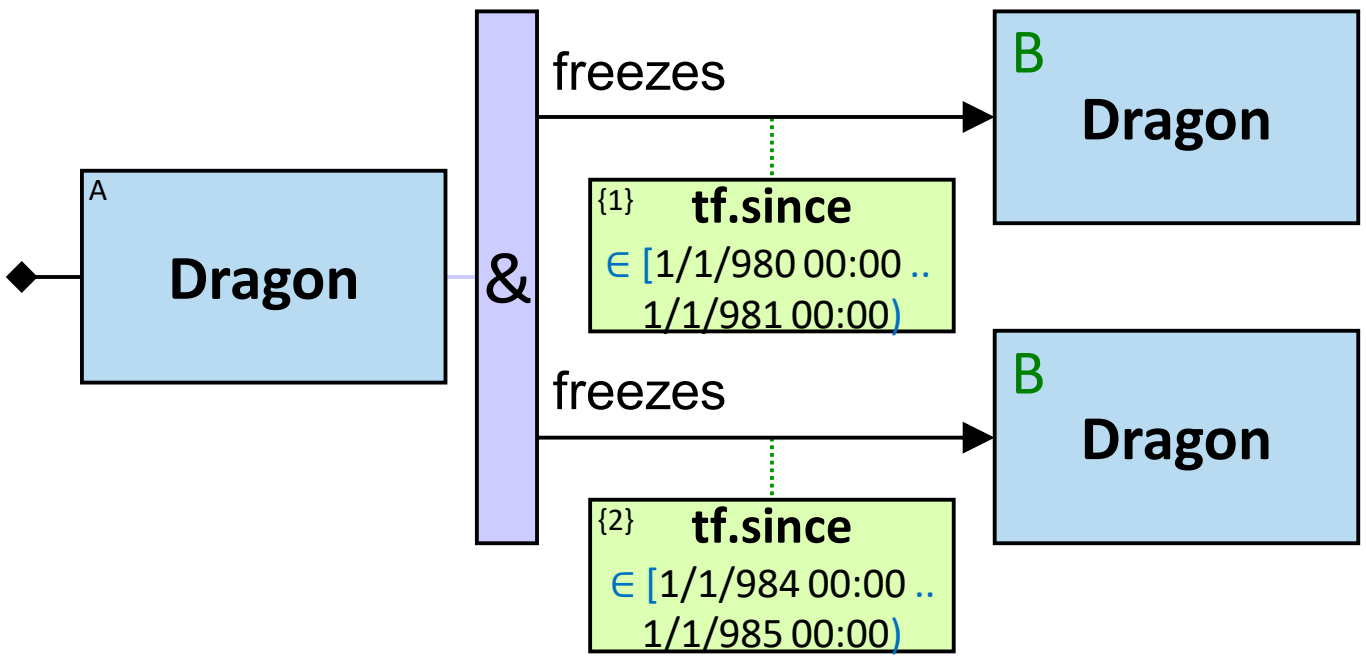

The same visual notation is also used when the same concrete entity appears more than once (see Q25v2, Q26v2)

*Nonidenticality constraints* can be used when different graph-entities should be assigned to typed pattern-entities of the same type or to untyped pattern-entities.

**Q5:** *Any person A whose dragon was frozen by a dragon owned by two of A's parents* (version 1)

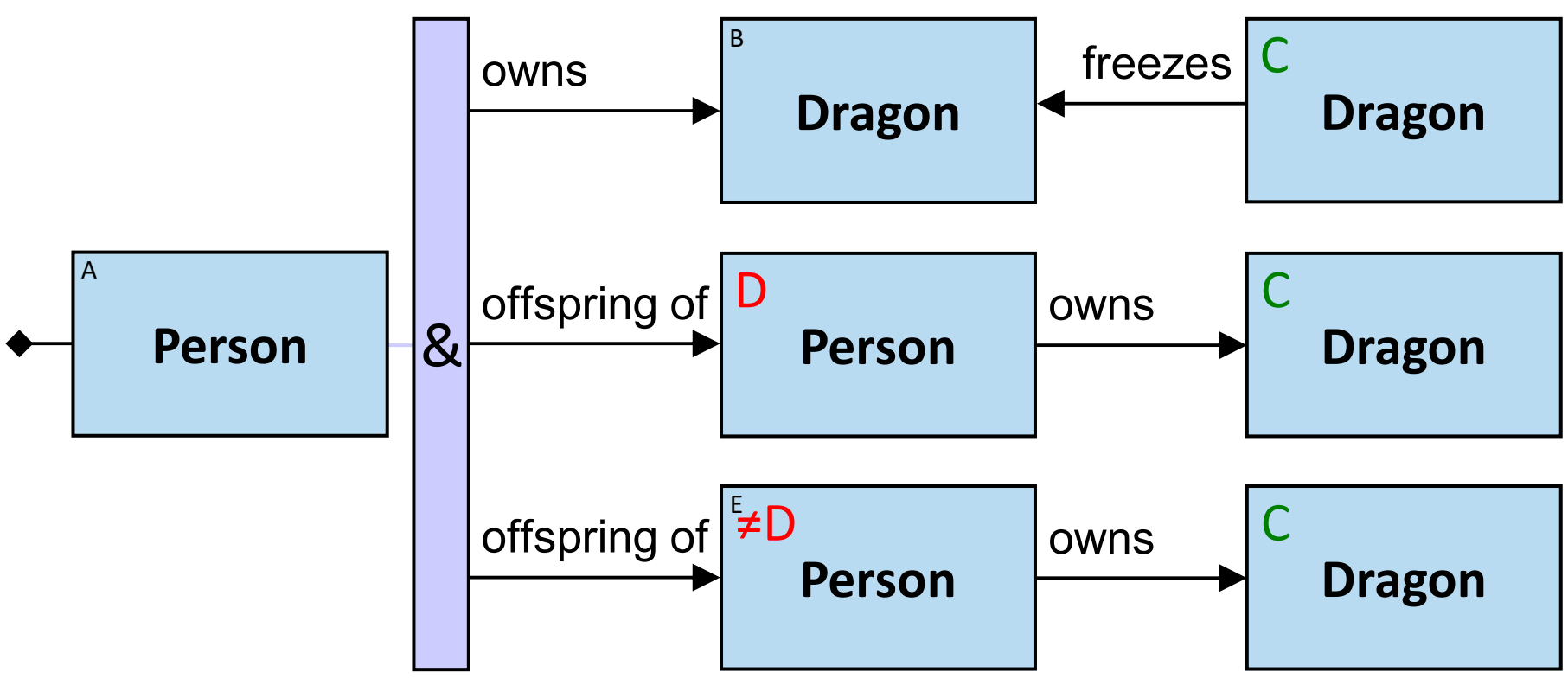

Without the nonidenticality constraint, the same parent could be assigned to both D and E.

Nonidenticality constraints are depicted in red ('≠X'), where X is another entity-tag. Several nonidenticality constraints may be defined for the same pattern-entity, e.g., '≠A,≠C' (see Q57).

**Q6:** *Any person A whose dragon was frozen by two dragons – one owned by one of A's parents, the other owned by another parent (none, one, or both dragons may be owned by both parents)*

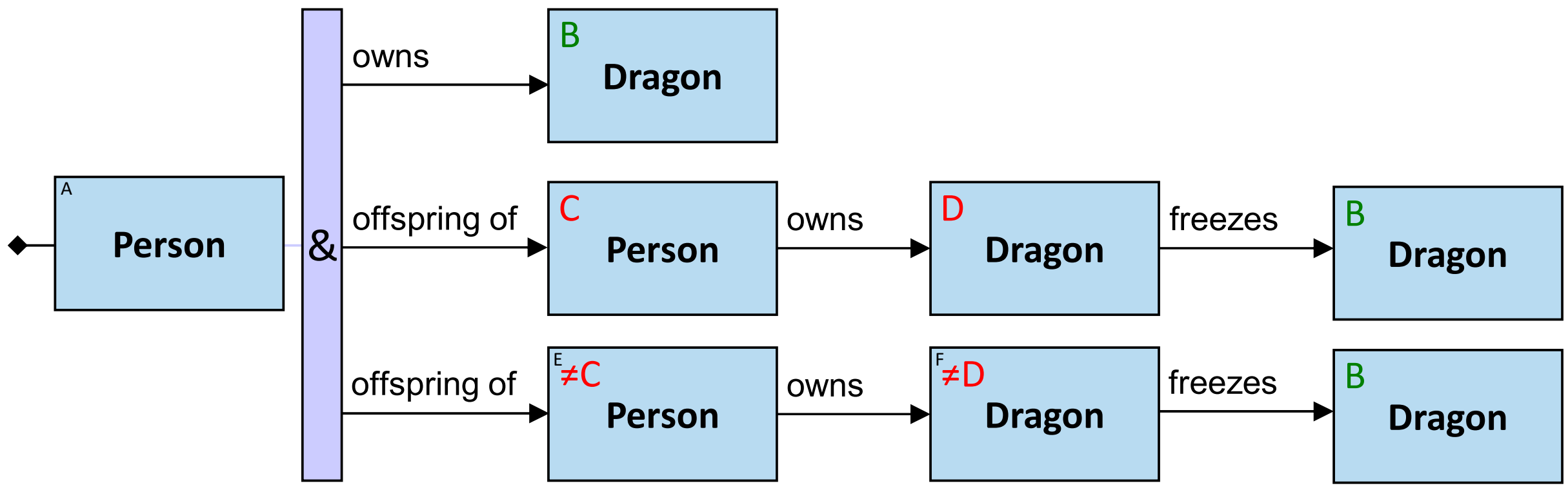

**Q7:** *Any person A whose dragon was either (i) frozen by a dragon owned by two of A's parents or (ii) frozen by two dragons – one owned by one of A's parents and the other owned by A's other parent*

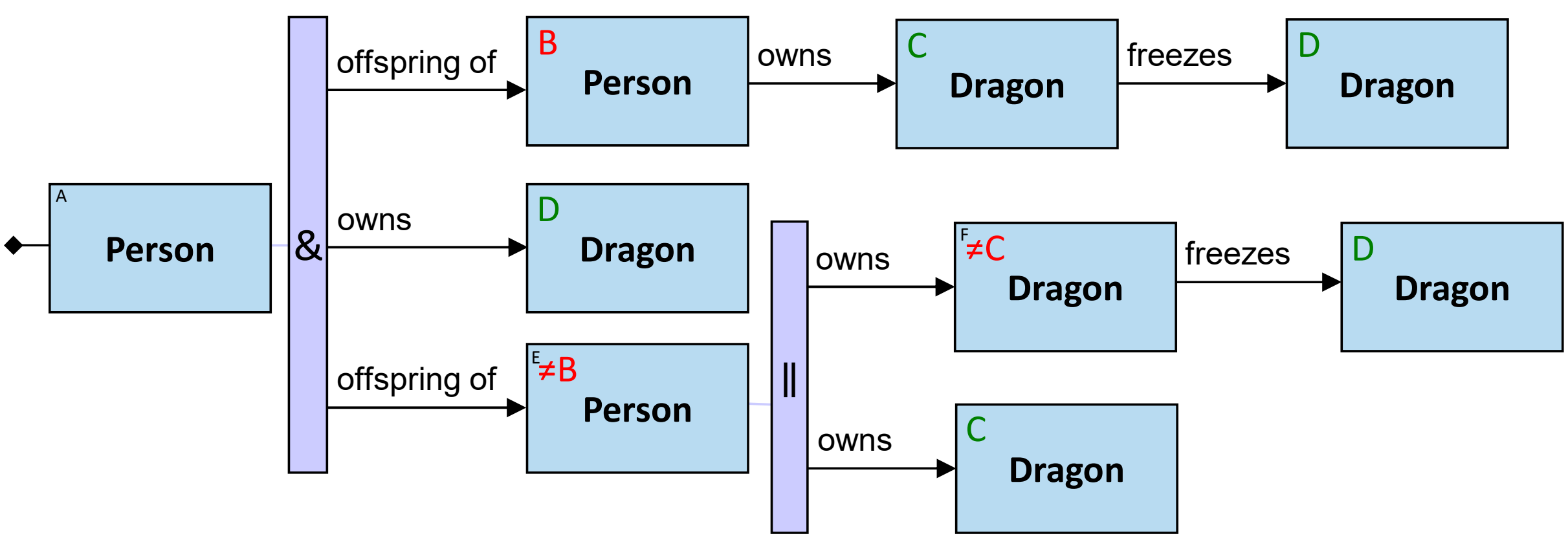

**Q24:** *Any person A having (at least) two parents and owns a dragon that was frozen by a dragon neither of A's parents owns*

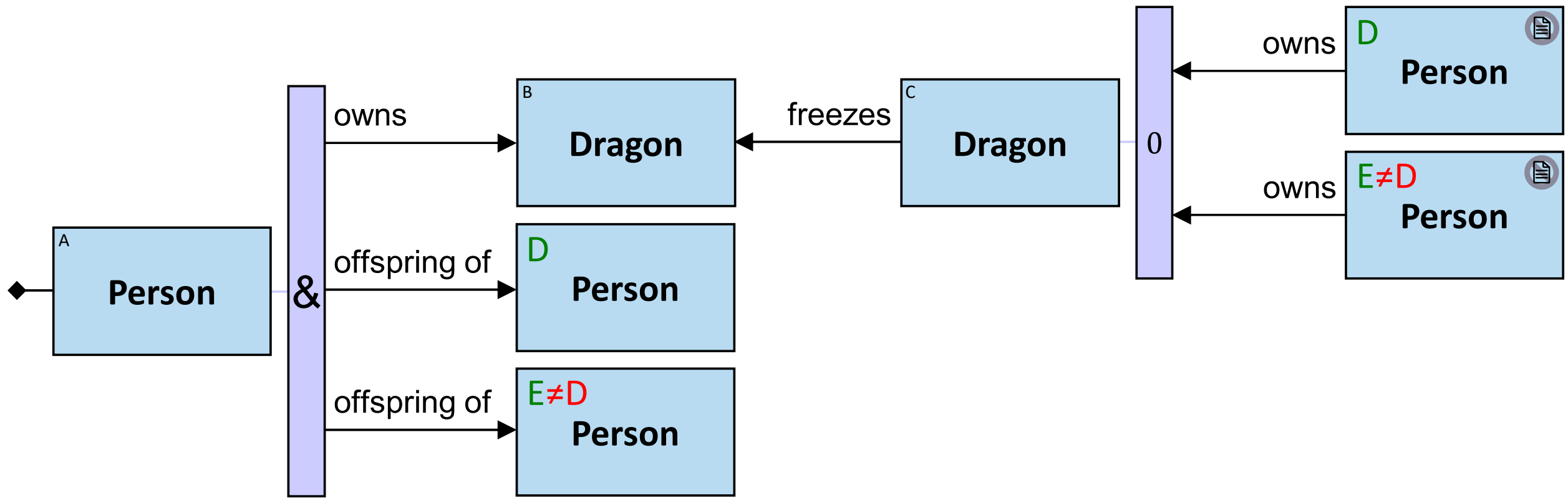

Q24 demonstrates the usage of both identicality constraints and nonidenticality constraints for the same pattern-entity.

Consider Q5v1, Q6, Q7, and Q24. For any given assignment, there is another assignment where the two parents are switched (for example, in Q5v1, the assignments to D and E are switched). Such redundant assignments are usually undesired. Using *order constraints*, we can avoid such redundancies (See Q5v2).

Also, consider the following pattern: *Any three persons A, B, and C, who are pairwise friends.* If persons ($A_1$, $B_1$, $C_1$) compose an assignment, so do ($A_1$, $C_1$, $B_1$), ($B_1$, $A_1$, $C_1$), and all other permutations. Such a factorial increase in the number of assignments is usually undesired. Using *order constraints*, we can express patterns such as *Any three persons A<B<C, who are pairwise friends.*

*Order constraints* can be used when graph-entities should be orderly assigned to either typed pattern-entities of the same type or untyped pattern-entities.

**Q31:** *Any pair of dragons (A, B) where A froze B, A fired at B, B froze A, and B fired at A* (version 1)

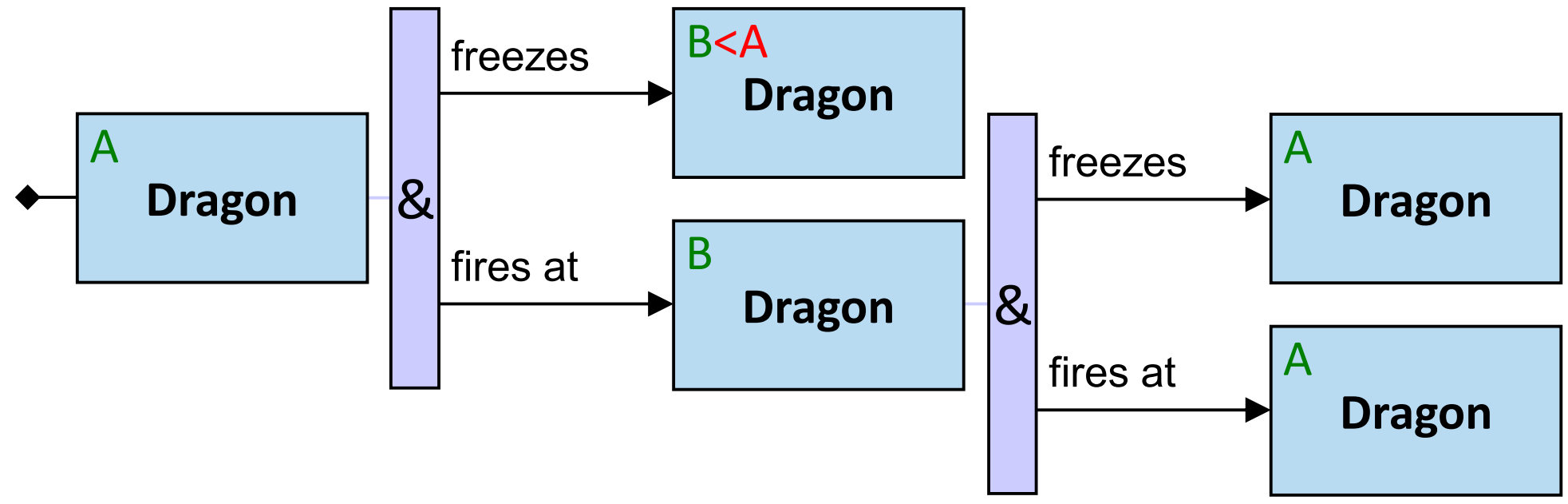

Without the order constraint, any reported pair of dragons would be reported twice: ($D_1$, $D_2$), ($D_2$, $D_1$).

Property graph data models require that each entity and each relationship would be *uniquely identifiable*. To support order constraints, V1 further requires that graph-entities (except *Null* entities) would also be *ordered*.

Order constraints are depicted in red ('<X' or '≤X') where X is another entity-tag. Several nonidenticality or order constraints may be defined for the same pattern-entity, e.g., '<A,<B', '≠A,<C' (see Q83).

See also Q49, Q88, Q115v1, Q272, Q337, Q342, Q347, and Q350.

If, for some assignment, an identicality, nonidenticality, or order constraint is not relevant (e.g., identicality constraint between a pair of entities in two branches of a *Some* quantifier, where the assignment includes only one branch) – the constraint is ignored. Otherwise – the assignment is valid only if the constraint is satisfied.

## 10. NEGATOR

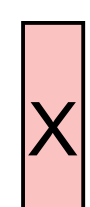

Sometimes we need to express a pattern that contains elements for which there is no assignment. For example, *Any person whose first name is Brandon and who does not own a white horse*. Such patterns are composed of:

- A 'positive' component – *any person whose first name is Brandon*
- A negator – *does not*
- One or more 'negative' component – *own a white horse*

An assignment matches the pattern *only if*:

- It matches a pattern composed only of the positive component: there is an assignment to *a person whose first name is Brandon*.
- The assignment has no superset that matches a pattern composed of both the positive and the negative components: *the person whose first name is Brandon does not own a white horse.*

Such patterns can be composed using *negators*. The pattern starts with the 'positive' component, and negators separate it from the 'negative' components.

A negator is depicted with a **pink 'X' rectangle**. It has one connection on its left side and one connection on its right side. On its left, there is an entity or a quantifier. On its right side – a relationship or a path. A negator may not appear directly left of a relationship or a path with an aggregator (see section 25 – Aggregators).

Query results do not include assignments to entities/relationships/paths right of a negator. Any pattern elements right of a negator is depicted with a gray 'no report' icon on its top-right, indicating that the query result does not include an assignment for it (see section 14 – Latent Pattern-Entities).

**Q12:** *Any person who does not own a horse*

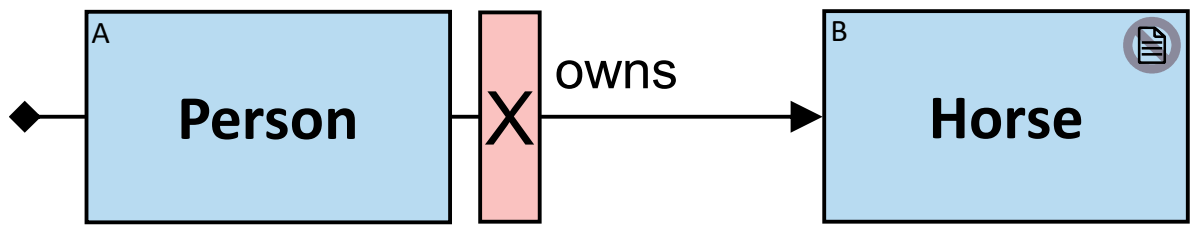

**Q13:** *Any horse not owned by a person*

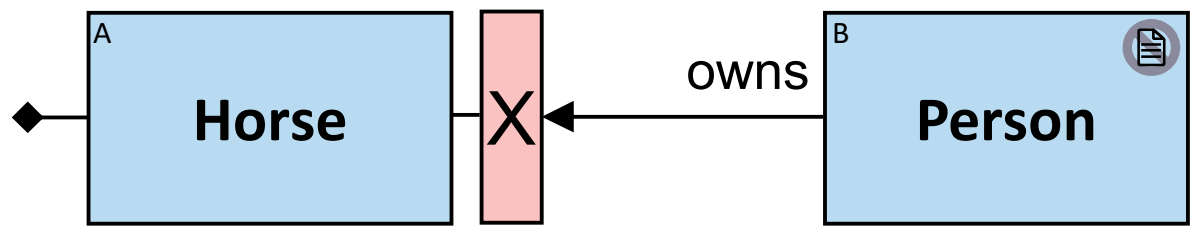

**Q14:** *Sweetfoot – if a person does not own it*

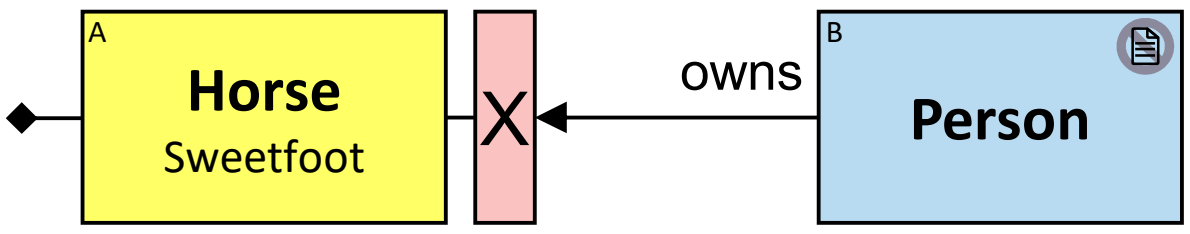

**Q15:** *Brandon Stark – if he does not own a horse*

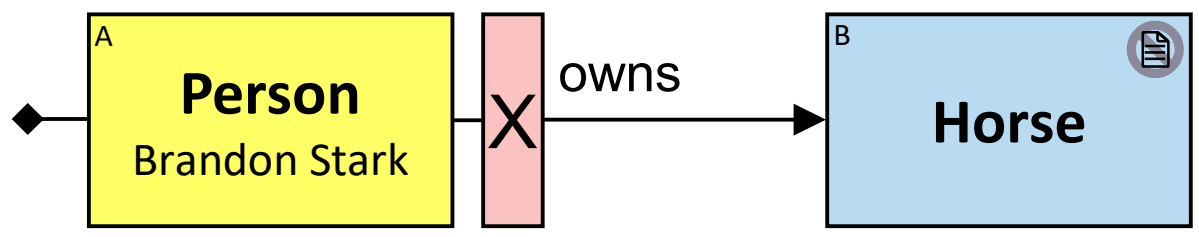

**Q16:** *Any person who does not own Sweetfoot*

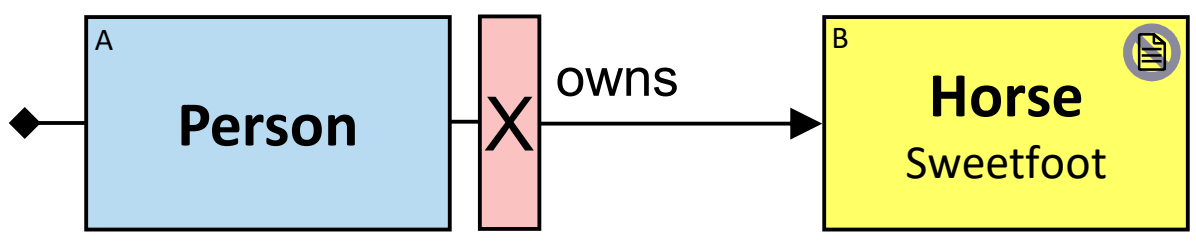

**Q17:** *Any horse not owned by Brandon Stark*

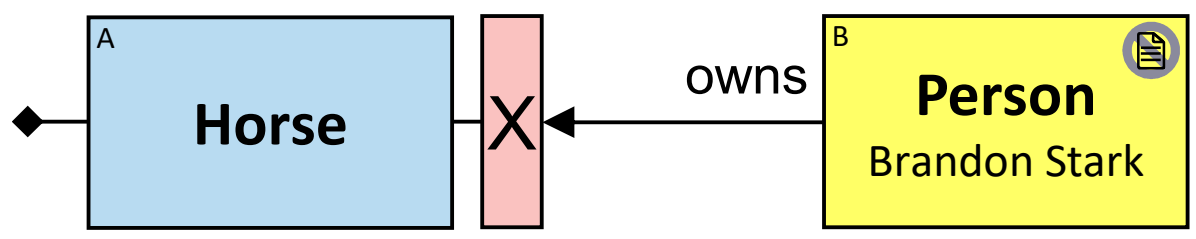

**Q18:** *Brandon Stark – if he does not own Sweetfoot*

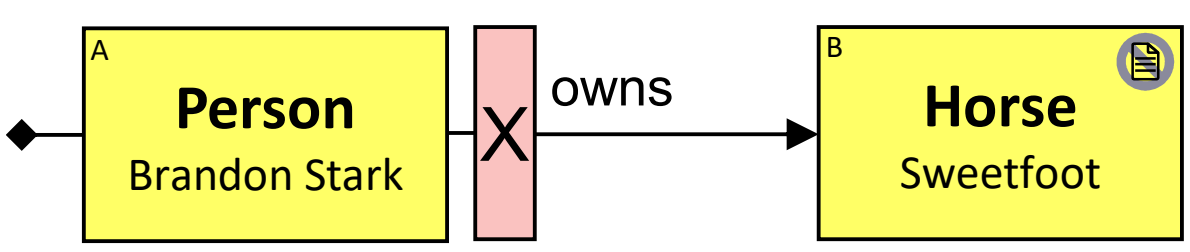

**Q19:** *Sweetfoot – if Brandon Stark does not own it*

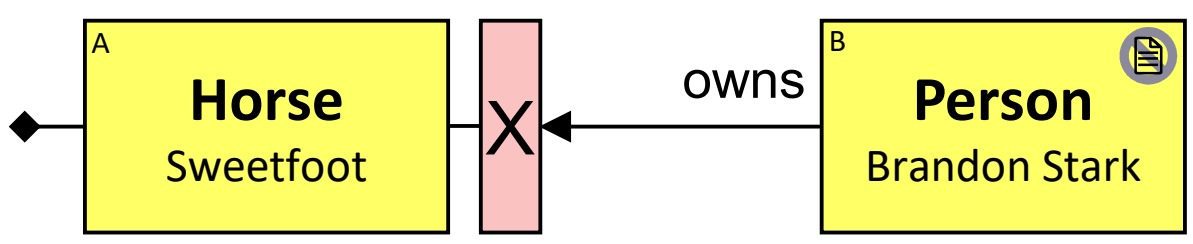

**Q256:** *Any person who does not own a white horse*

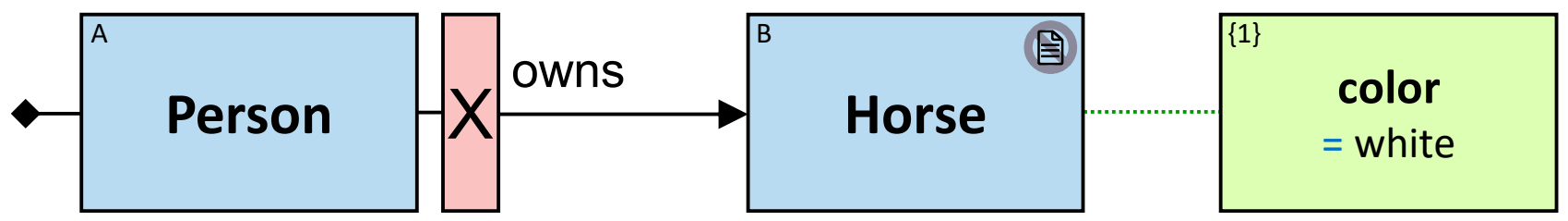

**Q257:** *Any person who did not become a horse owner in or after 1011*

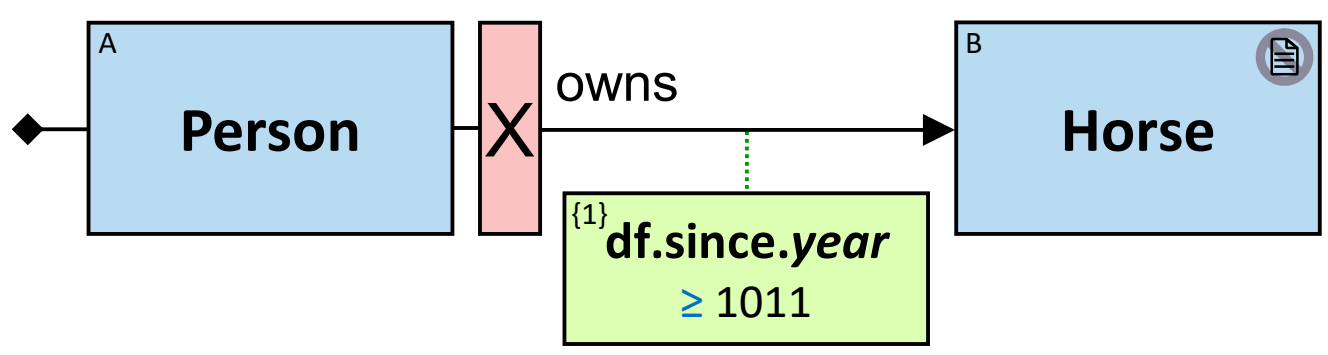

**Q258:** *Any person who did not become a white horse owner in or after 1011*

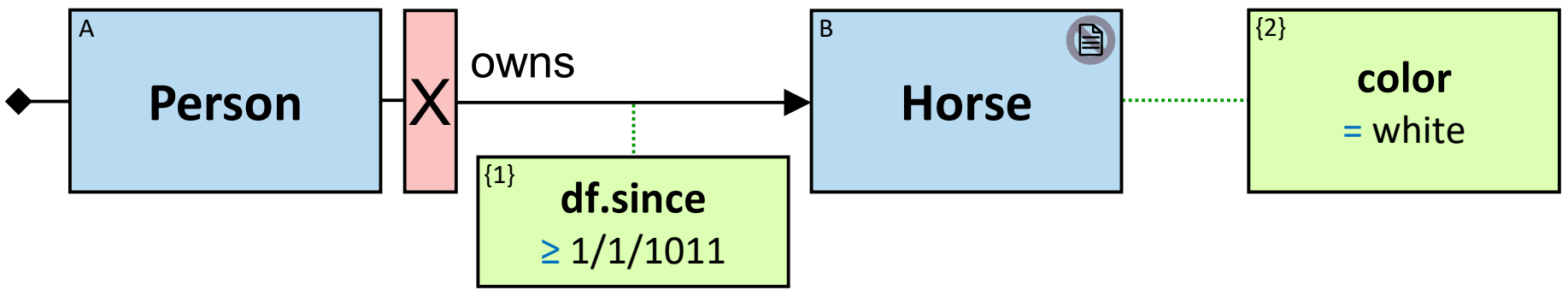

**Q22:** *Any horse not owned by a person who owns a dragon*

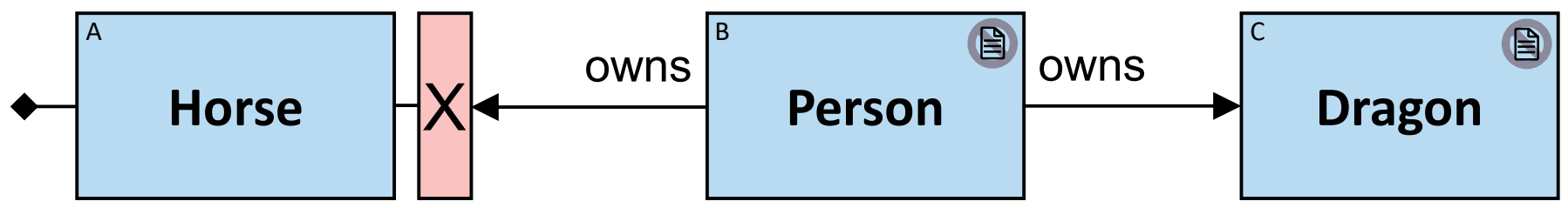

The positive component is *a horse*, while the negative component is *owned by a person who owns a dragon*. The right component is anything that follows the negator – up to the end of the branch.

Valid assignments:

- Any horse that is not owned
- Any horse that none of its owners is a person (e.g., a horse owned by a guild)
- Any horse that each person who owns it – does not own a dragon

**Q333:** *Any person who does not own a dragon that both fired at and froze the same dragon*

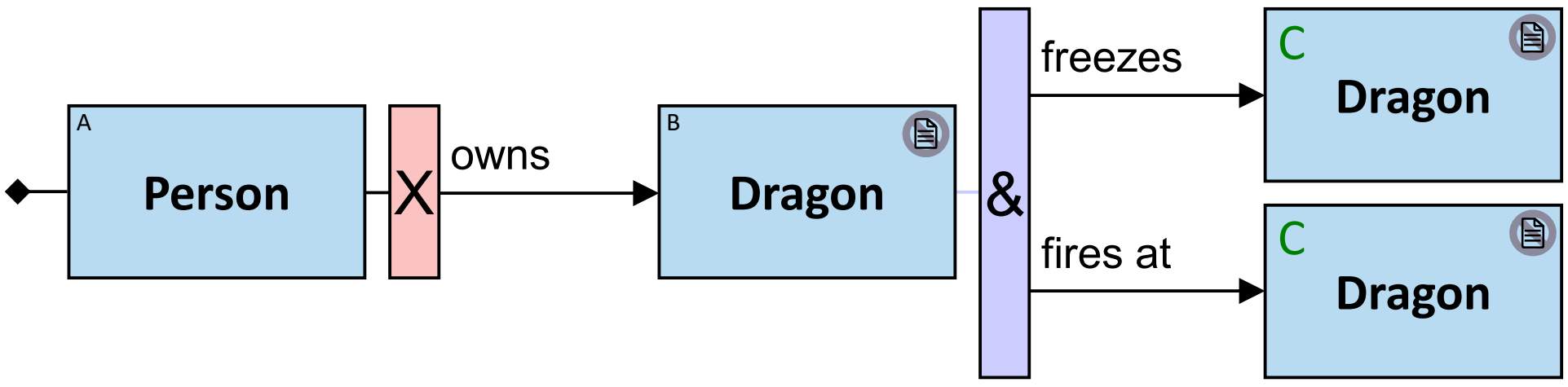

A negator may also appear left of a relationship that directly follows a quantifier's branch:

**Q25:** *Any dragon (C) not fired at by Balerion but fired at by a dragon that Balerion fired at* (two versions)

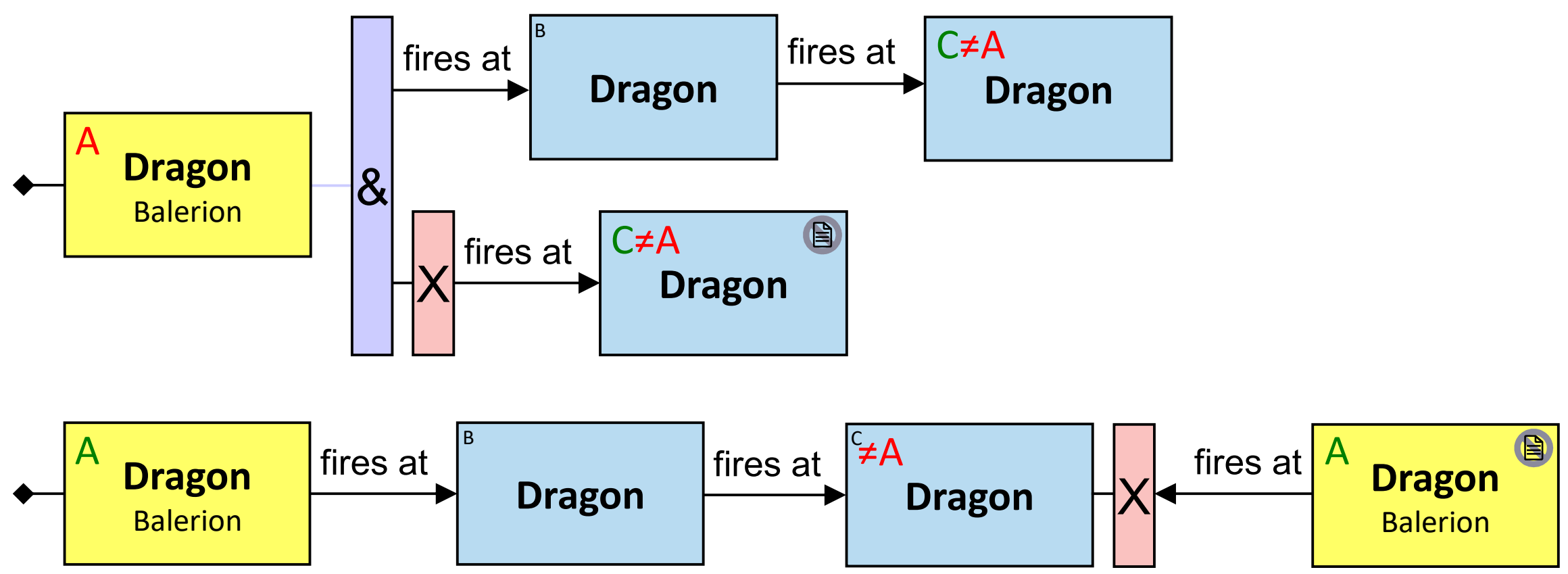

**Q26:** *Any person who is a member of at least one guild that Brandon Stark is a member of, and at least one guild that Brandon Stark is not a member of* (four versions)

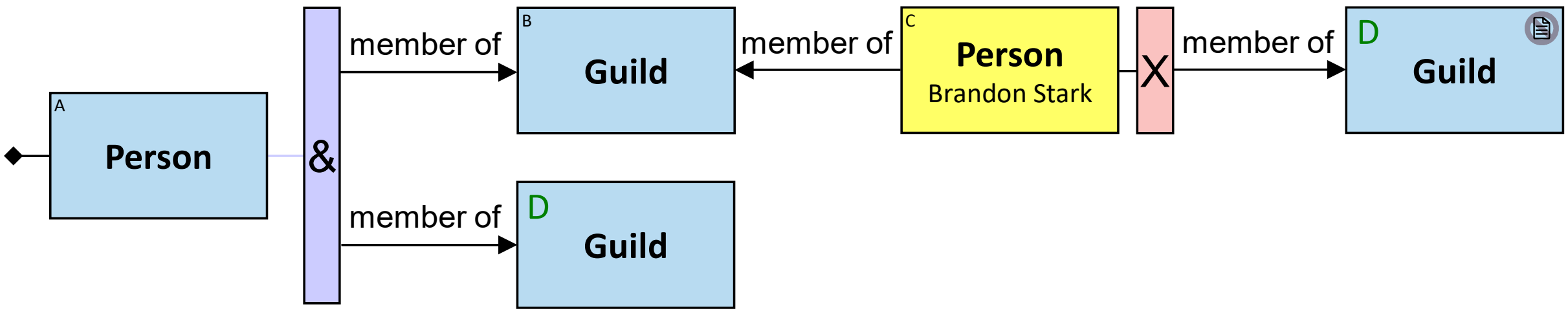

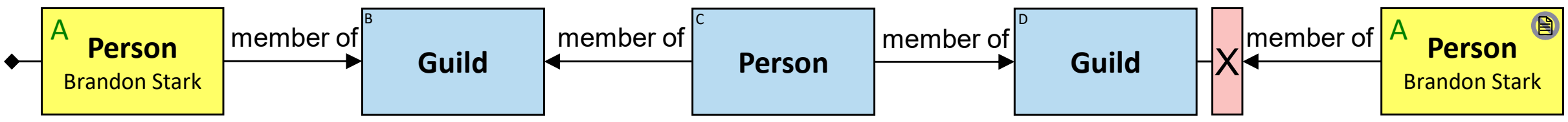

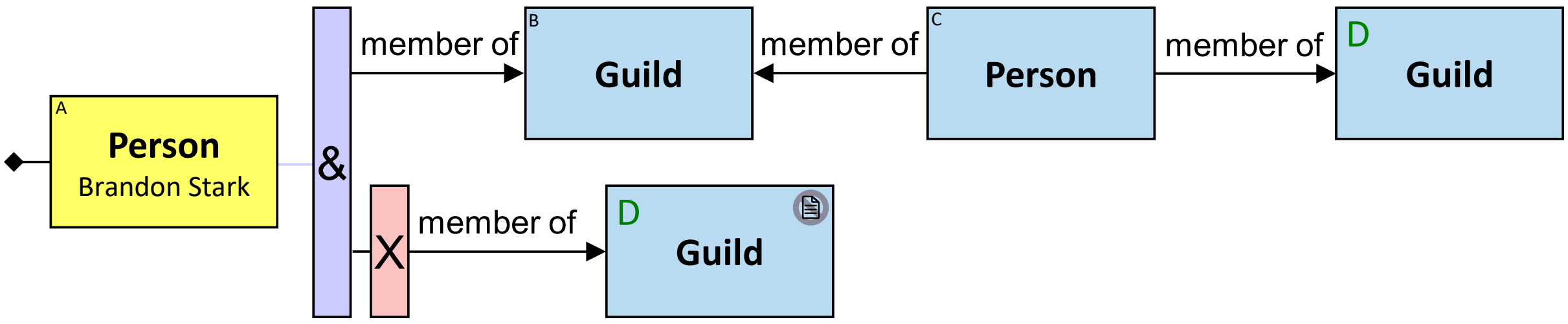

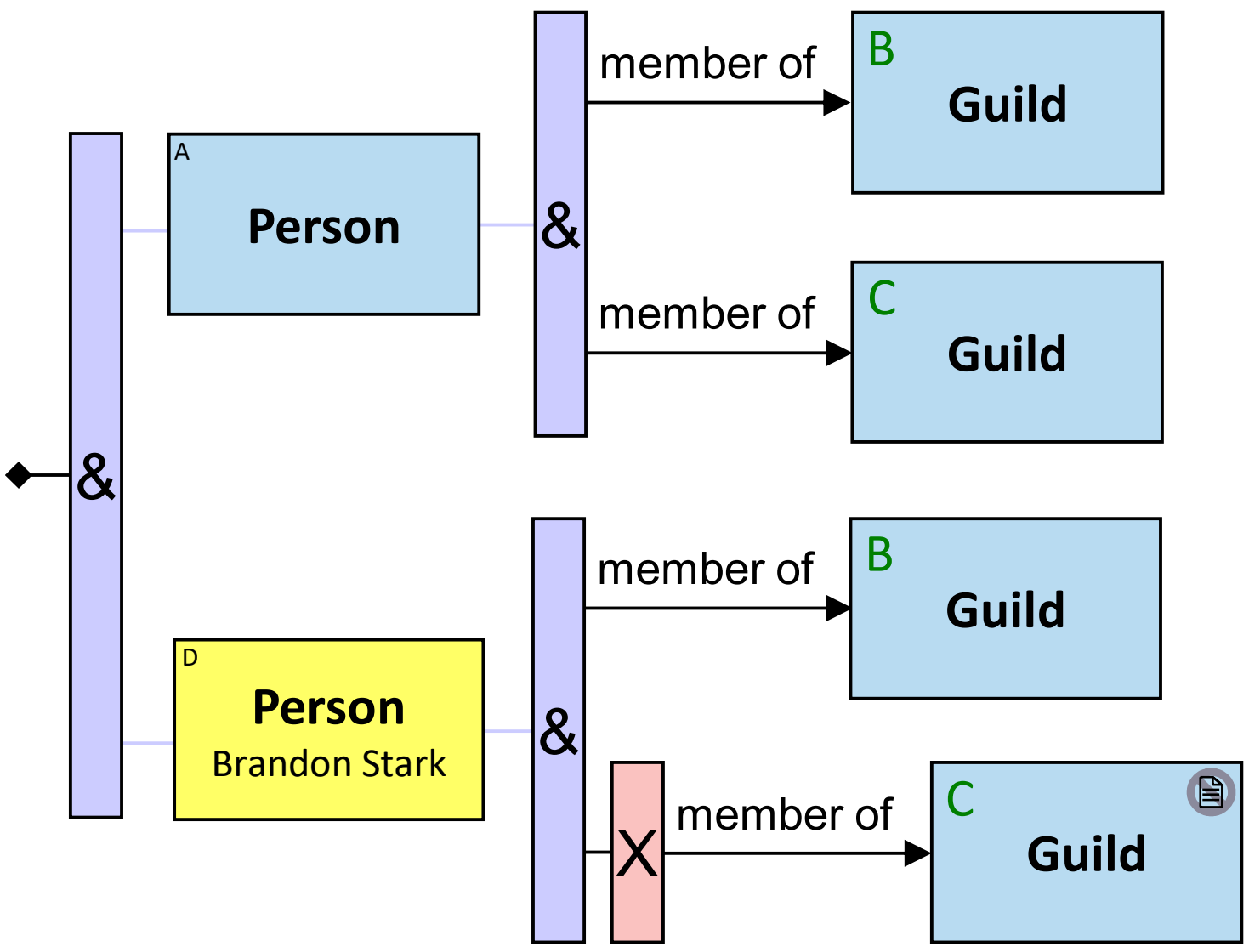

Negators may appear in more than one branch:

**Q20:** *Any horse that is neither owned by Rogar Bolton nor by Robin Arryn* (two versions)

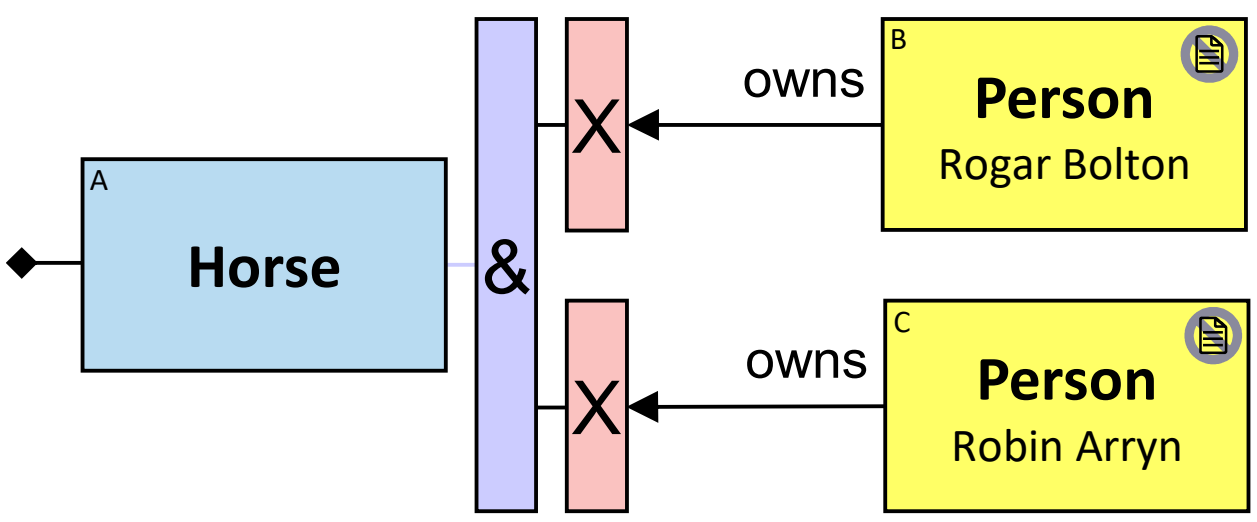

Note that there are two negative components. This pattern can also be represented using the *None* quantifier:

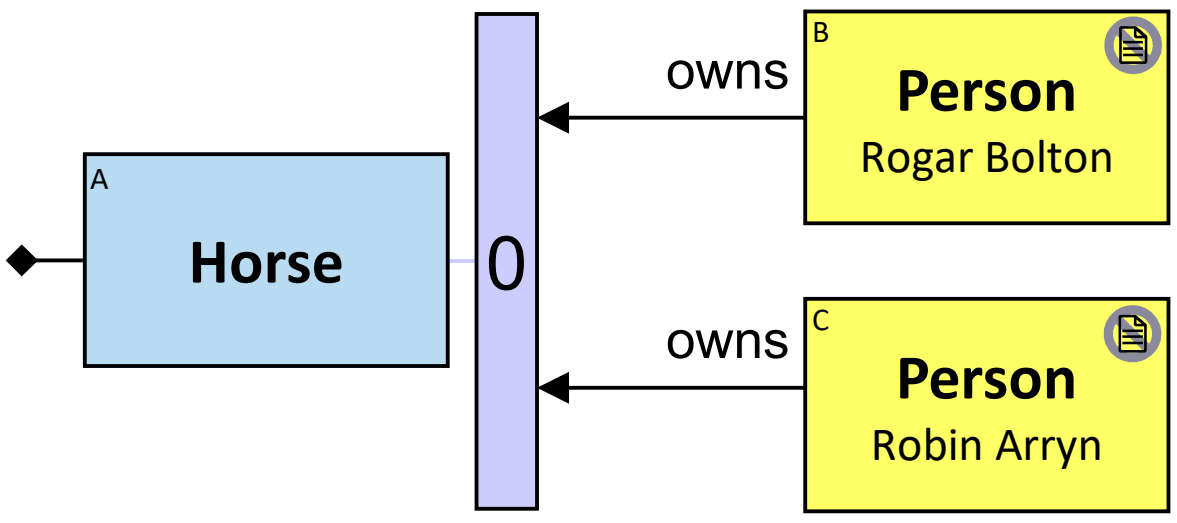

**Q21:** *Any horse not owned by both Rogar Bolton and Robin Arryn*

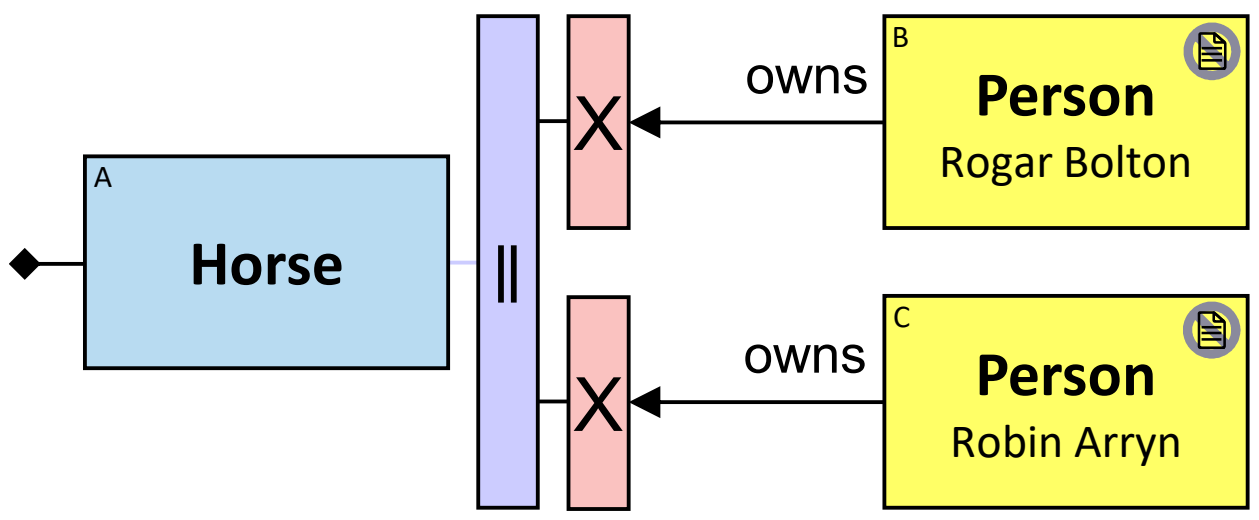

If either Rogar Bolton or Robin Arryn owns the horse – the owner will not be part of the reported assignments.

**Q362:** *Any horse owned either by Rogar Bolton or Robin Arryn but not by both*

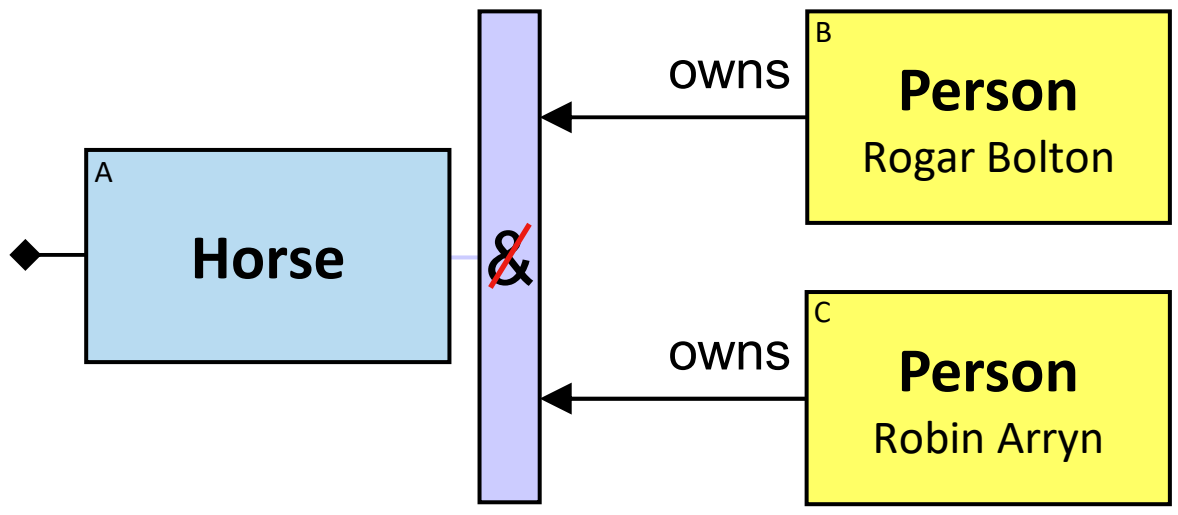

If either Rogar Bolton or Robin Arryn owns the horse – the owner will be part of the reported assignments.

**Q335:** *Any person and that person's parent for whom there is no horse that both are its owners*

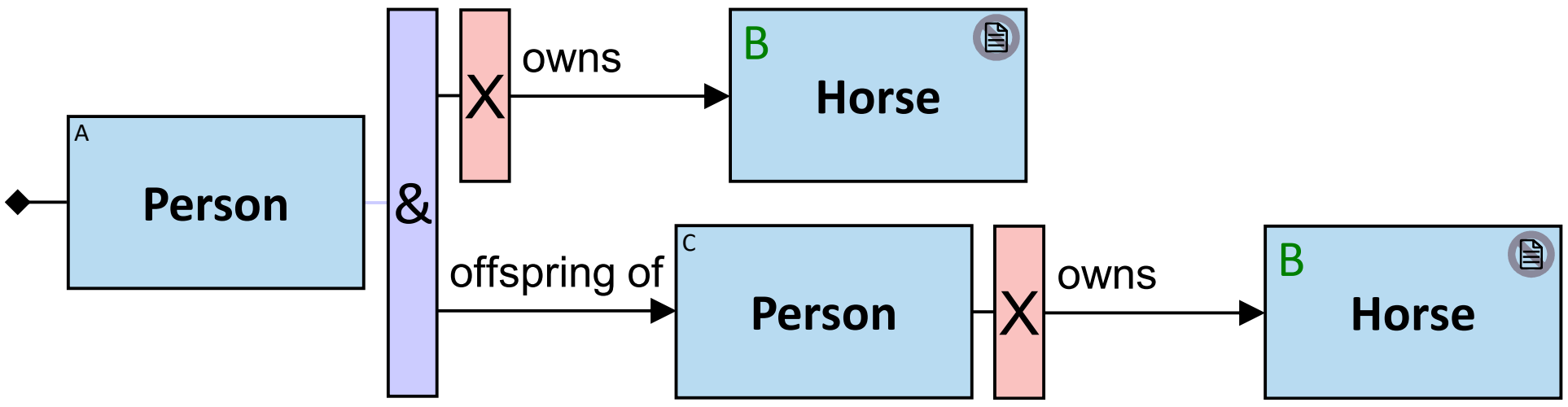

The two negative components refer to the same graph entity.

A sequence of negators:

**Q23:** *Any horse not owned by a person who does not own a dragon*

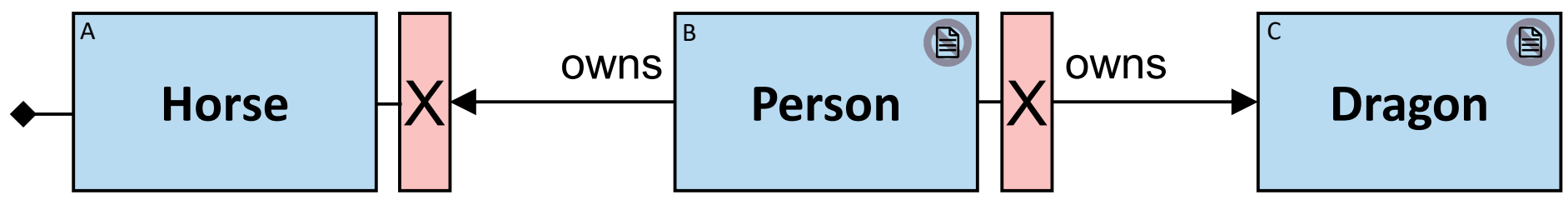

The positive component is *a horse*, while the negative component is *owned by a person who does not own a dragon*. The right component is anything that follows the left negator – up to the end of the branch.

Valid assignments:

- Any horse that is not owned
- Any horse that none of its owners is a person (e.g., a horse owned by a guild)
- Any horse that each person who owns it – also owns a dragon

## 11. RELATIONSHIP/PATH-NEGATOR

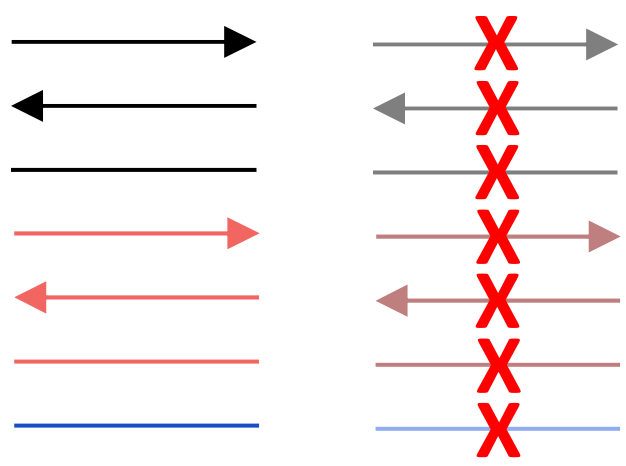

The pattern *a person whose first name is Brandon and who does not own a white horse* has assignments in two cases:

- *There is a person whose first name is Brandon,* and *there are no white horses*
- *There is a person whose first name is Brandon*, *there is at least one white horse*, but *there is no person whose first name is Brandon who owns a white horse*

Sometimes we want an assignment to match the pattern only in the second case – we want only assignments that match the whole pattern except given relationships/paths. Hence, an assignment matches the pattern *only if*:

- It matches a pattern composed of the left component and the right component without the relationship/path:
  - *There is a person whose first name is Brandon*
  - *There is a white horse*
- The assignment has no superset that matches a pattern composed of the left component, the right components, and the relationship/path
  - *There is no person whose first name is Brandon who owns a white horse*

A *relationship/path-negator* is depicted with a **red** '✕' mark at the center of the relationship/path arrow/line. The arrow/line and the associated text are faded.

Components located right of a relationship/path-negator are required to have assignments. These assignments are included in the reported assignments.

**Q357**: *Any dragon for which there is at least one dragon (besides itself) it did not freeze*

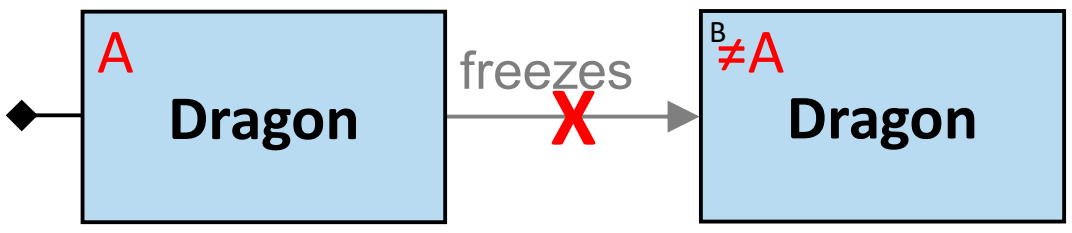

**Q83:** *Any dragon for which there are at least two dragons (besides itself) it did not freeze*

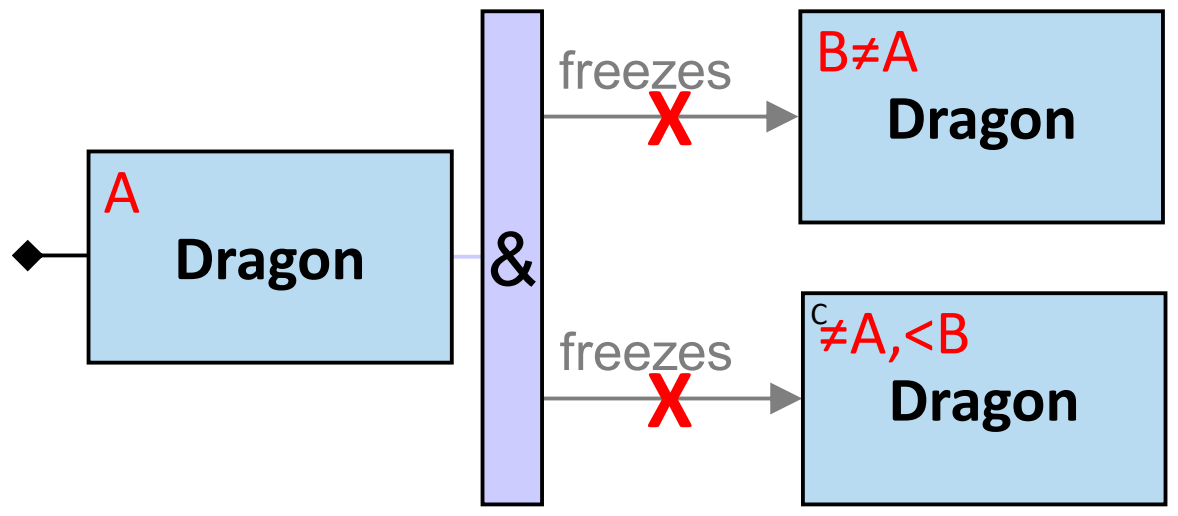

**Q204:** *Any dragon for which there is a dragon not owned by a Sarnorian that did not freeze it*

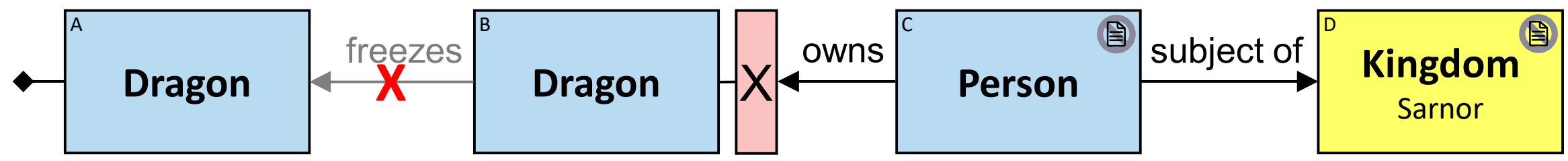

**Q205:** *Any dragon A and Sarnorian subject C for which there is a dragon not owned by C that did not freeze A*

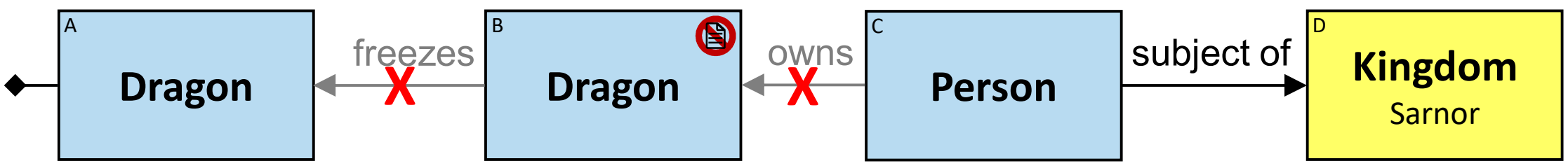

Relationship/path-negators are often used with aggregators (see Q126, Q63-Q66, Q165, Q298, Q345).

A relationship/path with a relationship/path-negator may be wrapped with a negator.

**Q82:** *Any dragon that froze all other dragons.* Alternative phrasing: *Any dragon for which there is no dragon (besides itself) it did not freeze* (version 1)

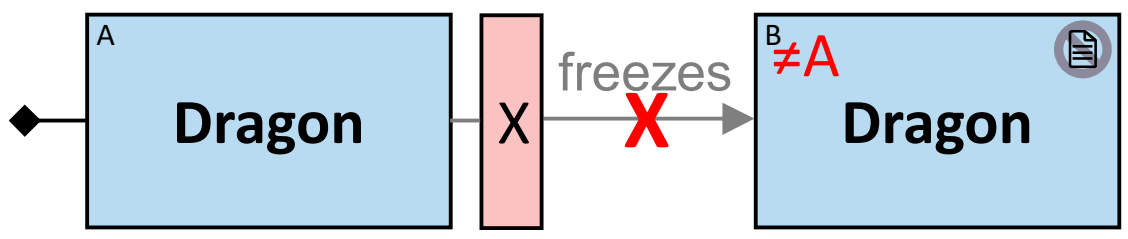

## 12. COMBINER

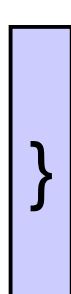

A **vertical purple rectangle with a closing curly brace** represents a *combiner*. A combiner combines two or more consecutive branches of the same quantifier.

On the combiner's left side are relationships and paths, and on its right side – an entity. A sequence of combiners is valid as well. The entity-type on a combiner's right side must match all the relationship-types and paths on its left side.

**Q30:** *Any pair of dragons (A, B) where A both froze B and fired at B* (two versions)

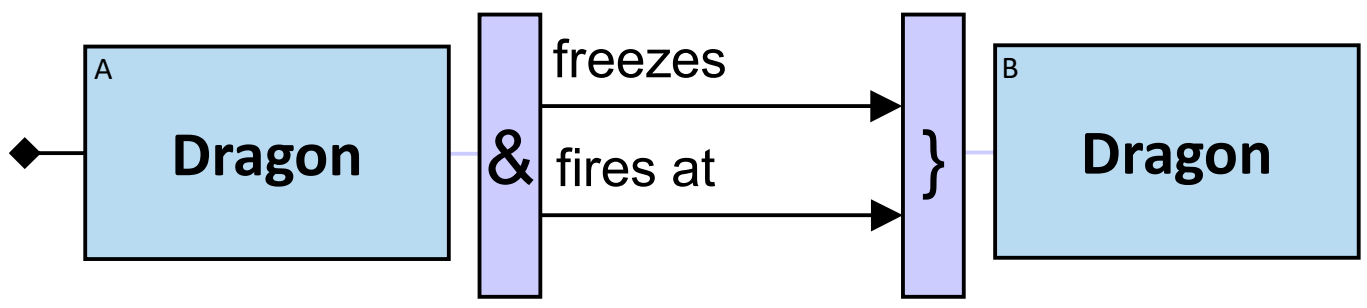

A combiner has the same semantics as the duplication of anything right of it to each branch.

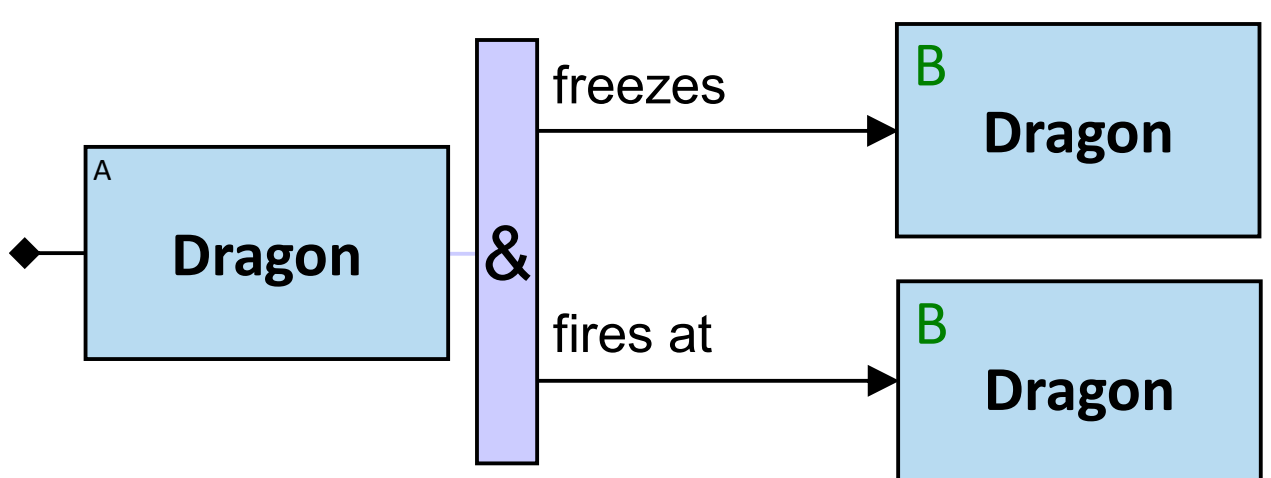

**Q360:** *Any dragon that (froze or fired at) all other dragons.* Alternative phrasing: *Any dragon for which there is no dragon (besides itself) it did not (freeze or fire at)* (version 1)

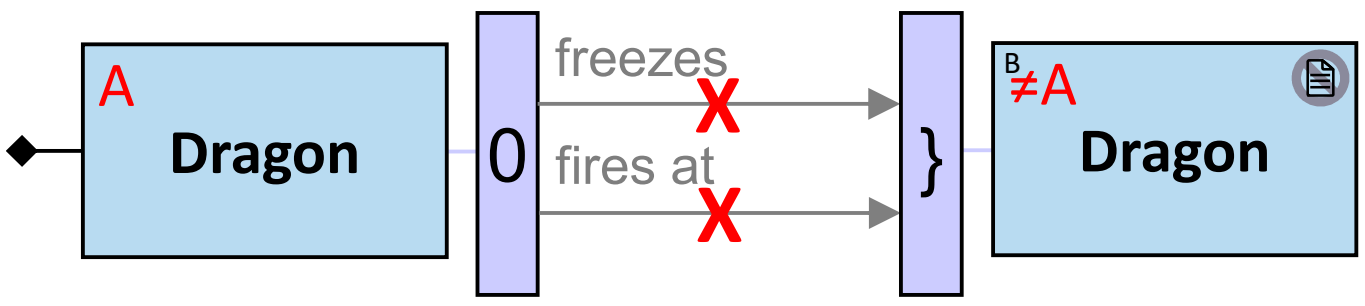

**Q187:** *Any dragon Balerion froze at least once on or after January 1, 1010,* ***and*** *at least once for less than ten minutes*

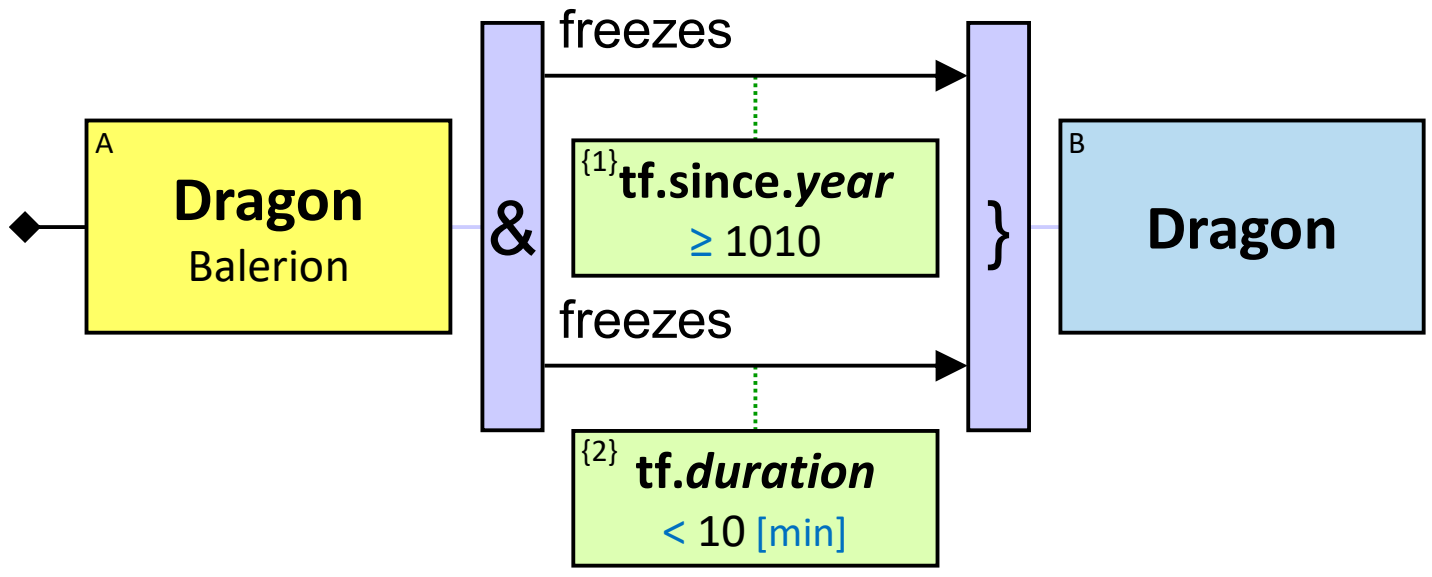

Note that the same graph-relationship may be assigned to both *freezes* pattern-relationships. Compare with Q185.

**Q189:** *Any dragon Balerion froze at least once on or after January 1, 1010,* ***or*** *at least once for less than ten minutes*

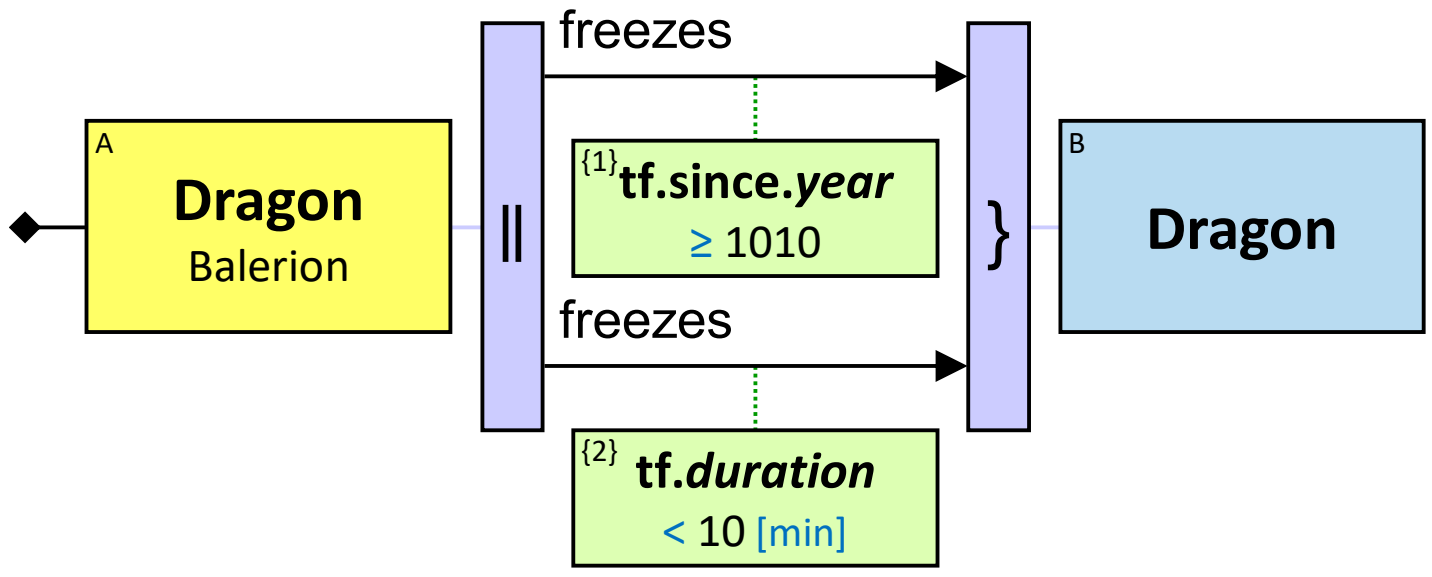

**Q29:** *Any dragon that froze or fired at some dragon* (versions 1-3)

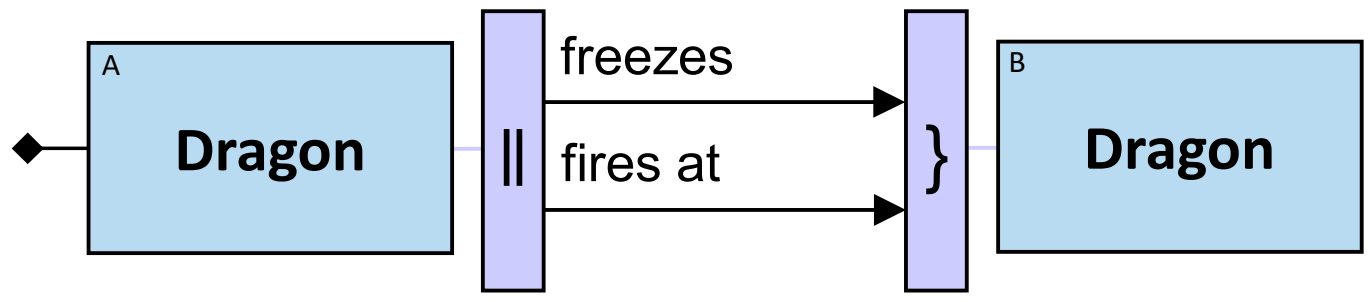

Note that the implied identicality is redundant.

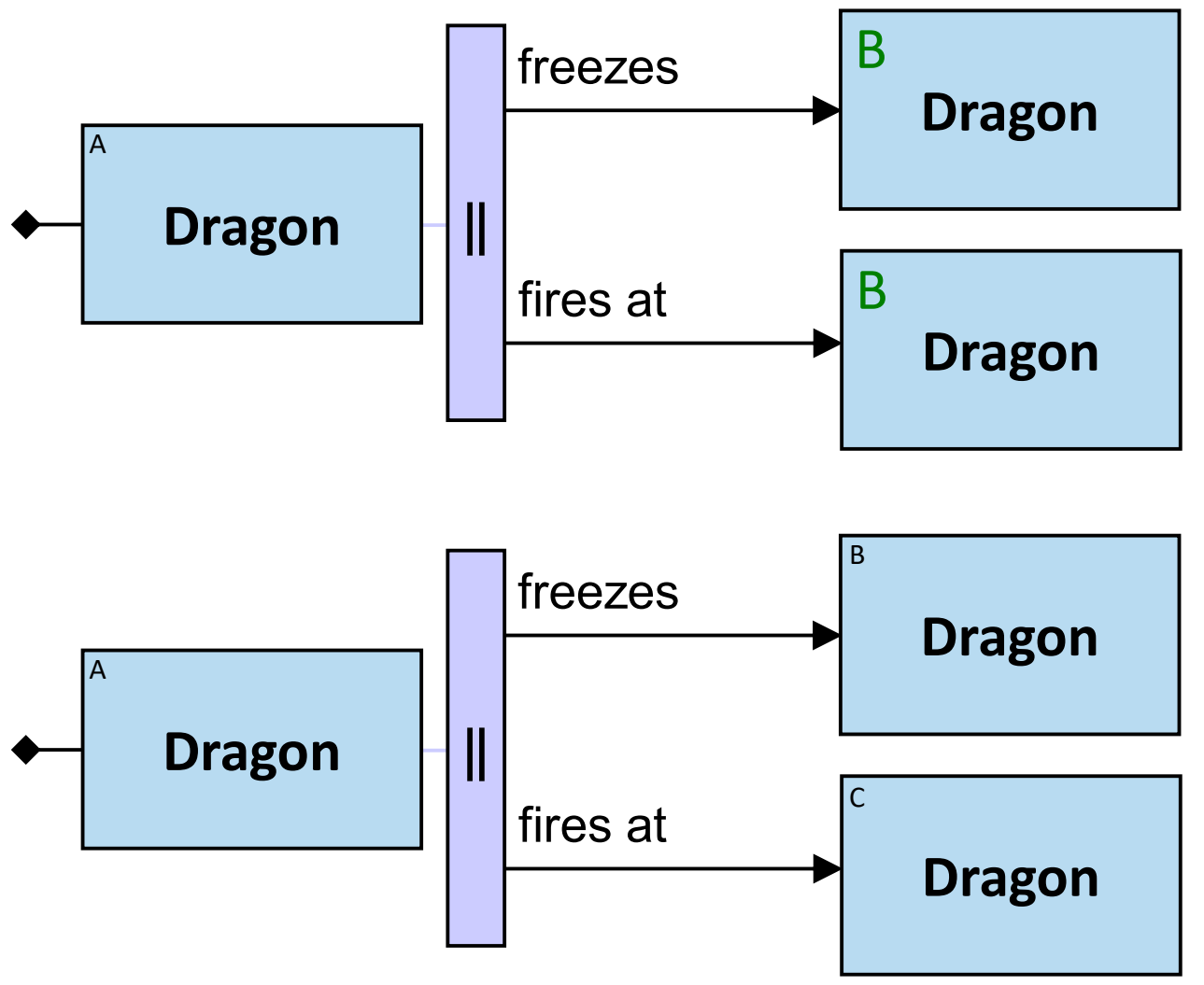

See also note under Q121 – combiner right of an A1 aggregator.

**Q31:** *Any pair of dragons (A, B) where A froze B, A fired at B, B froze A, and B fired at A* (version 2)

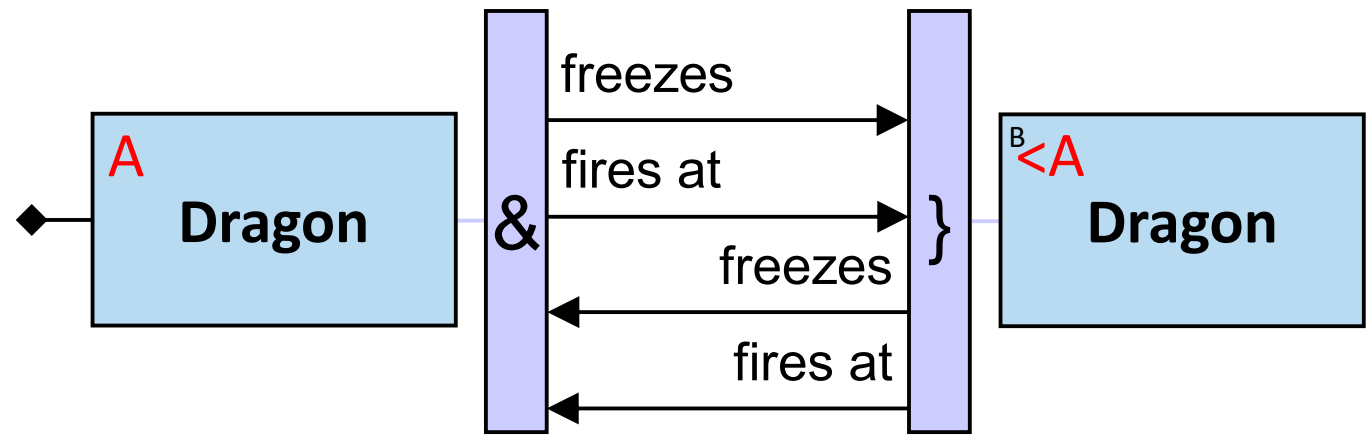

Without the order constraint, any reported pair of dragons would be reported twice: (D1, D2), (D2, D1).

**Q32:** *Any pair of dragons (A, B) where A fired at both B and some dragon that fired at B*

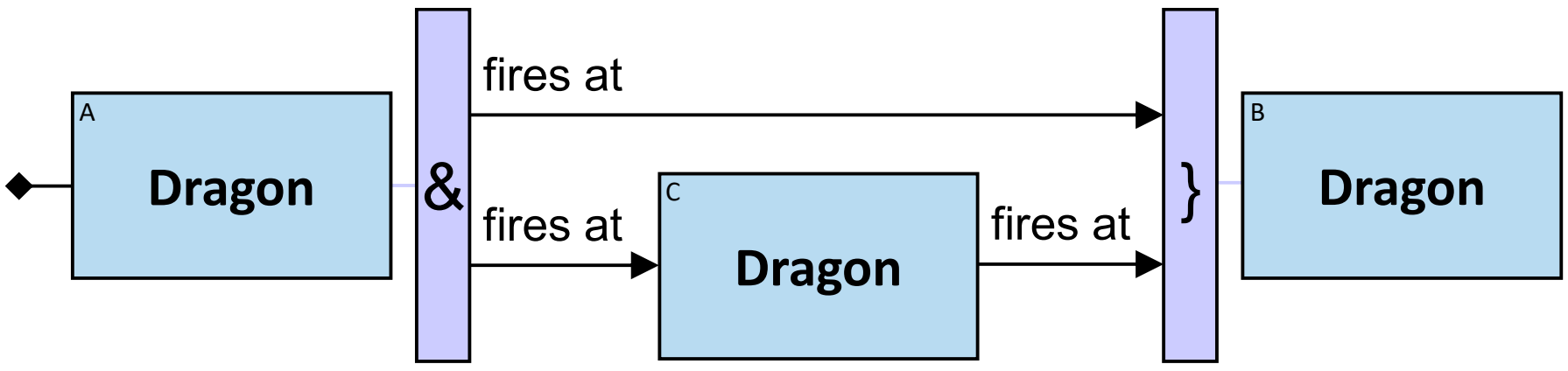

**Q5:** *Any person A whose dragon was frozen by a dragon owned by two of A's parents* (version 2)

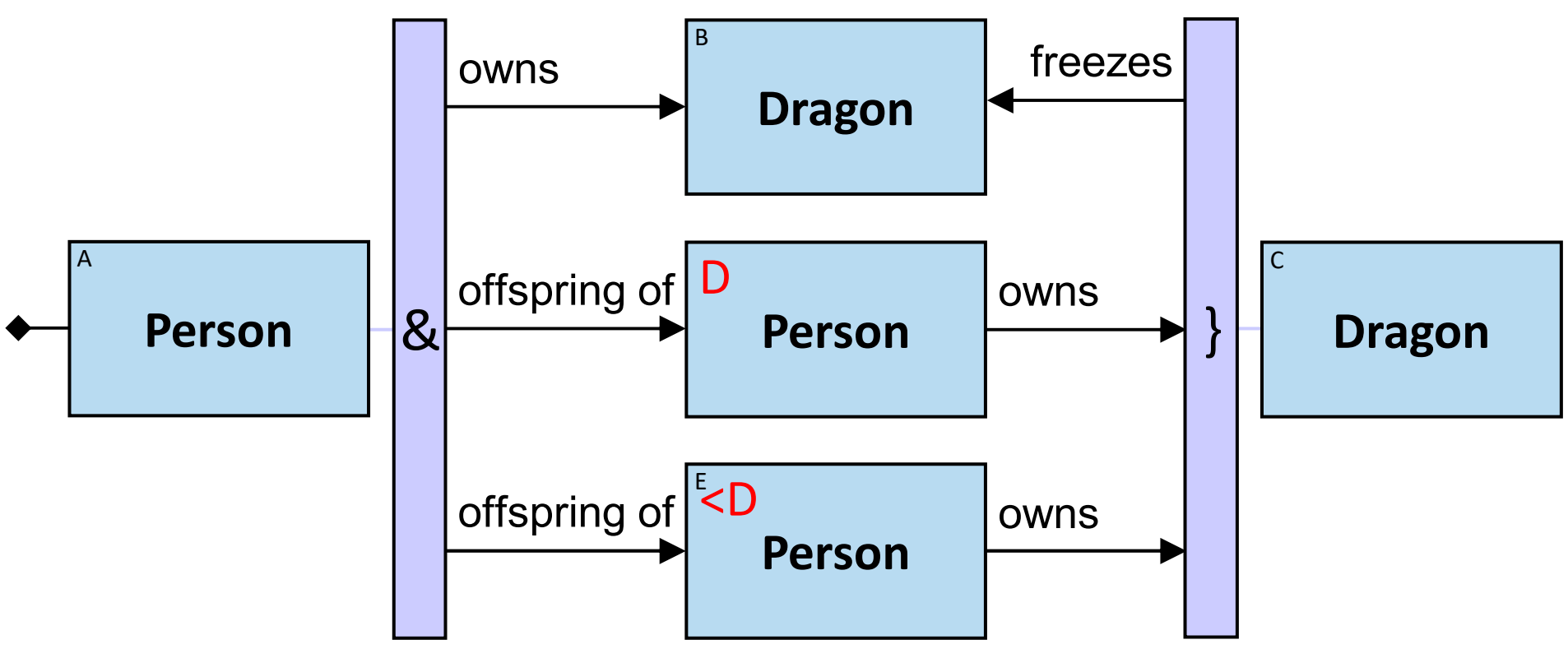

**Q170:** *Any three dragons (A, B, D) where A fired at B, A fired at some dragon that fired at B, B froze D, and B froze some dragon that froze D*

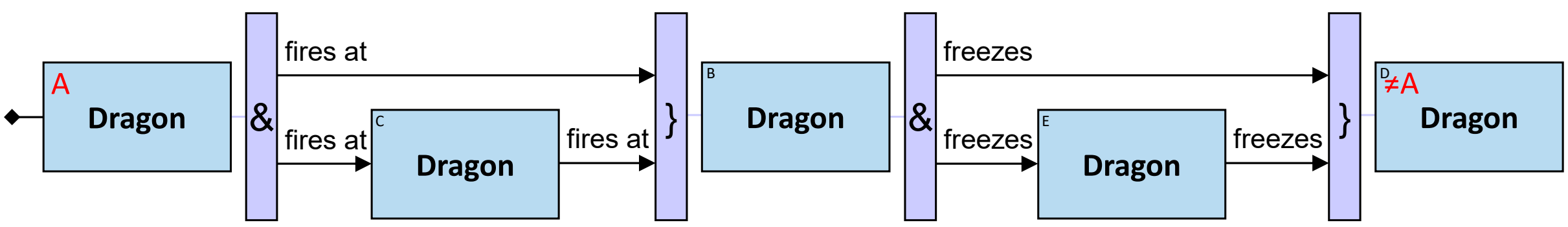

**Q33:** *Any dragon A that froze some dragon B, froze some dragon that froze B, and fired at some dragon*

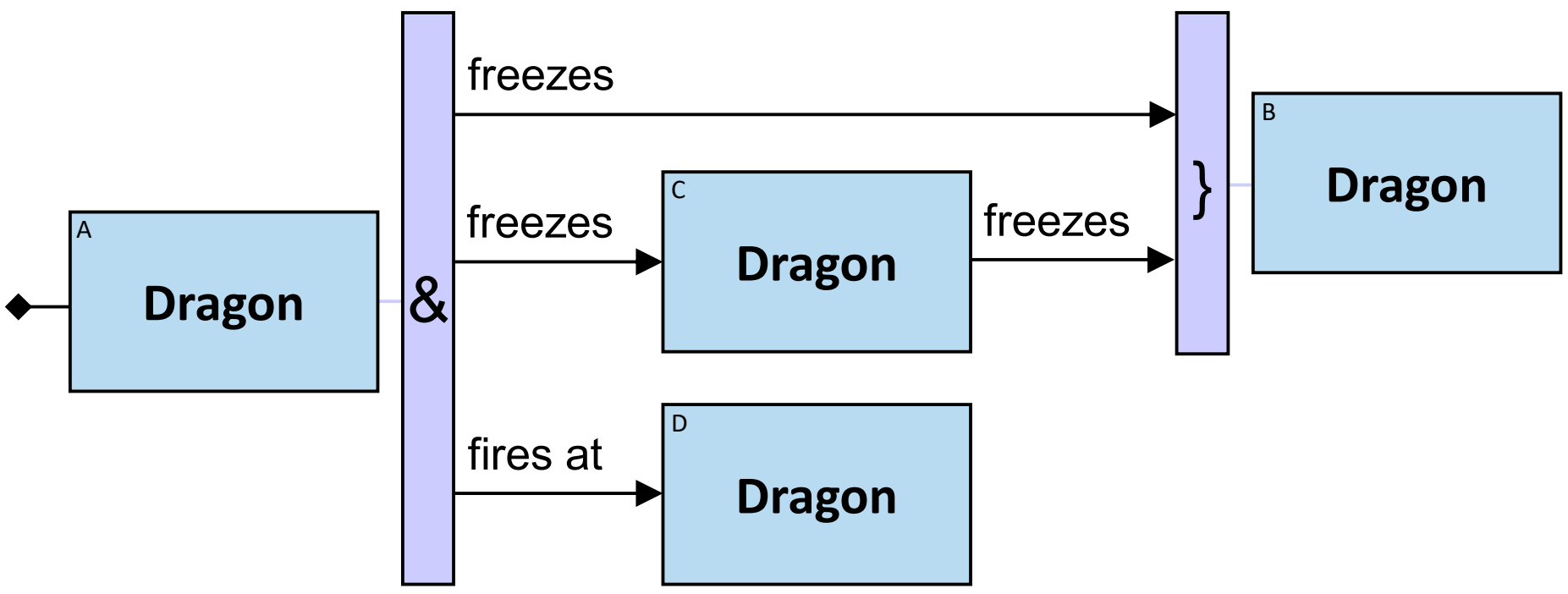

Different branches may be combined with different quantifiers.

**Q34:** *Any dragon A that froze some dragon B, froze some dragon that froze B, fired at some dragon D, and fired at some dragon that fired at D (B and D may be the same dragon or different dragons)*

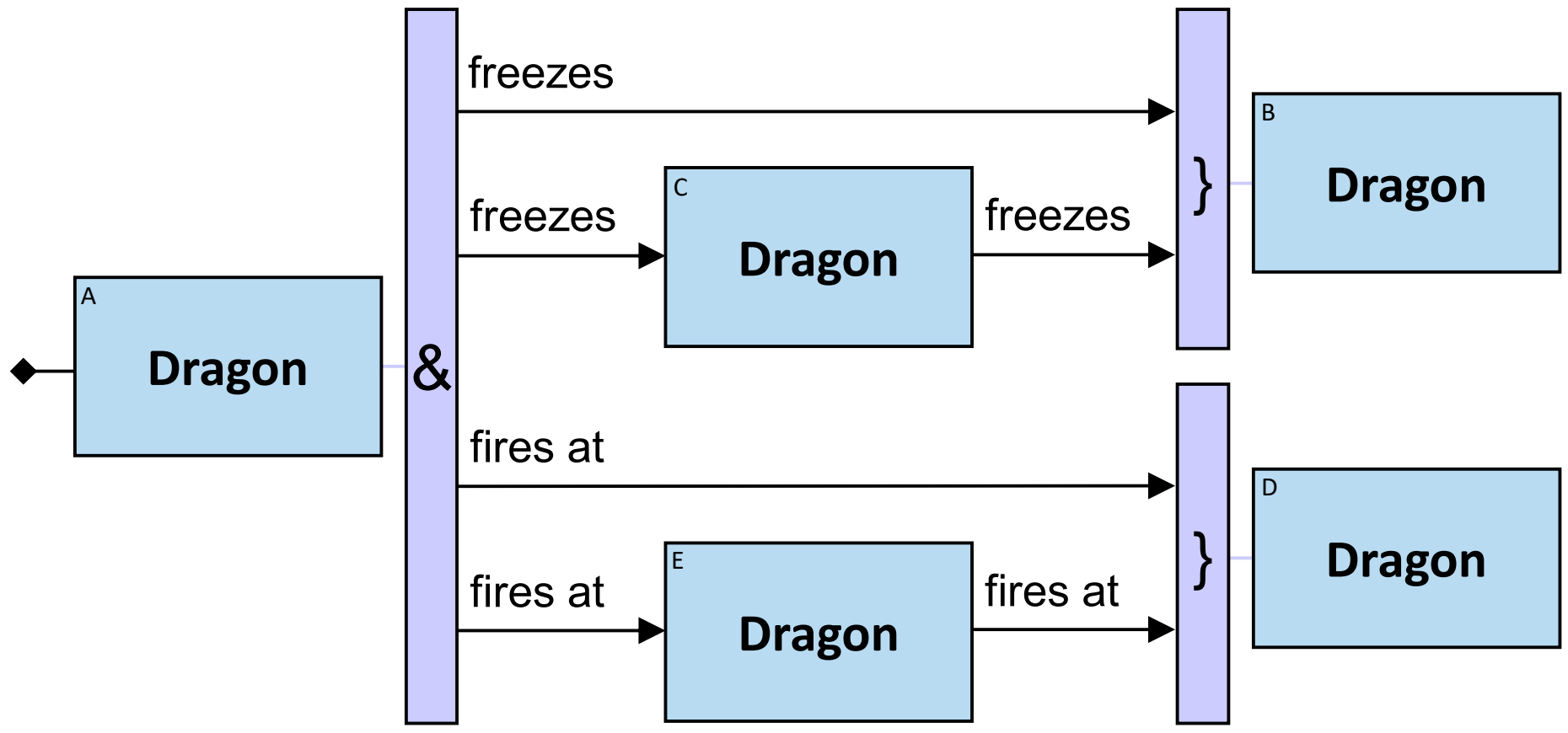

**Q35:** *Any pair of dragons (A, B) where A froze B but did not fire at B*

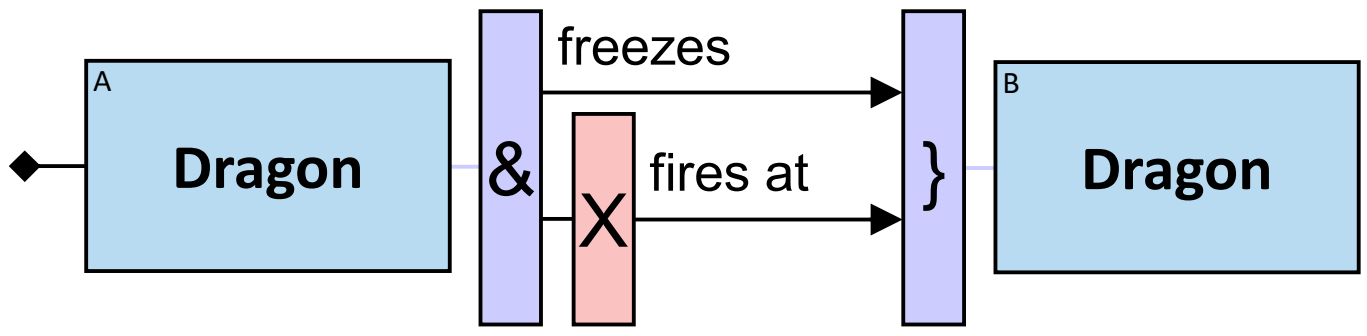

**Q334:** *Any dragon A for which there is no dragon B that A both froze and fired at*

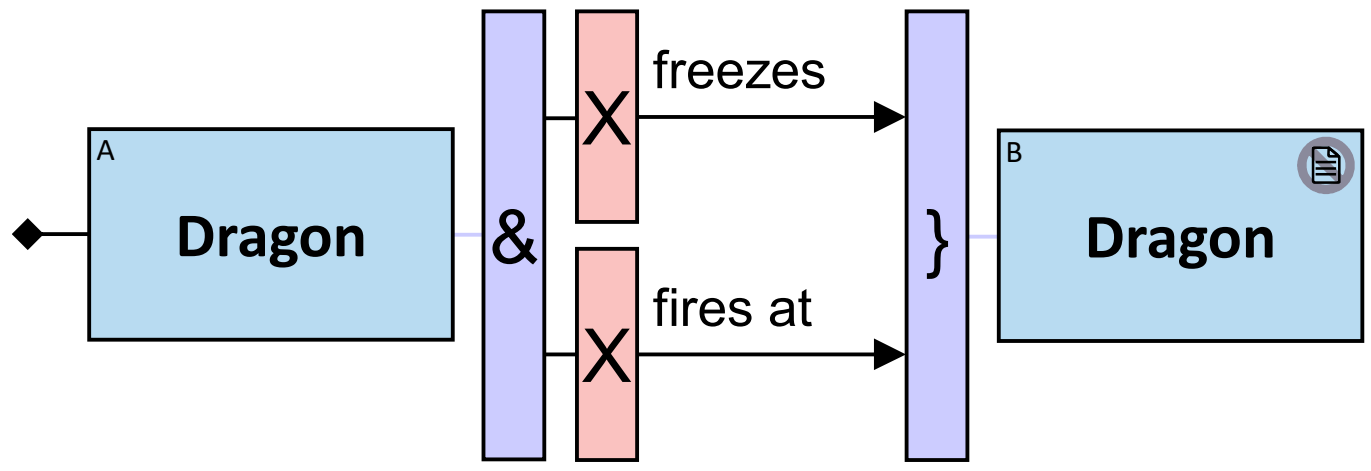

---

## 13. CHAINS, HORIZONTAL QUANTIFIERS, AND HORIZONTAL COMBINER

Relationship's expressions can be *chained*. When chained, each expression with a constraint is a *filtering stage*. An assignment is valid only if all the chained constraints are satisfied, or in other words – if it passed all filtering stages.

**Q188:** *Any dragon Balerion froze at least once – on or after January 1, 1010, for less than ten minutes*

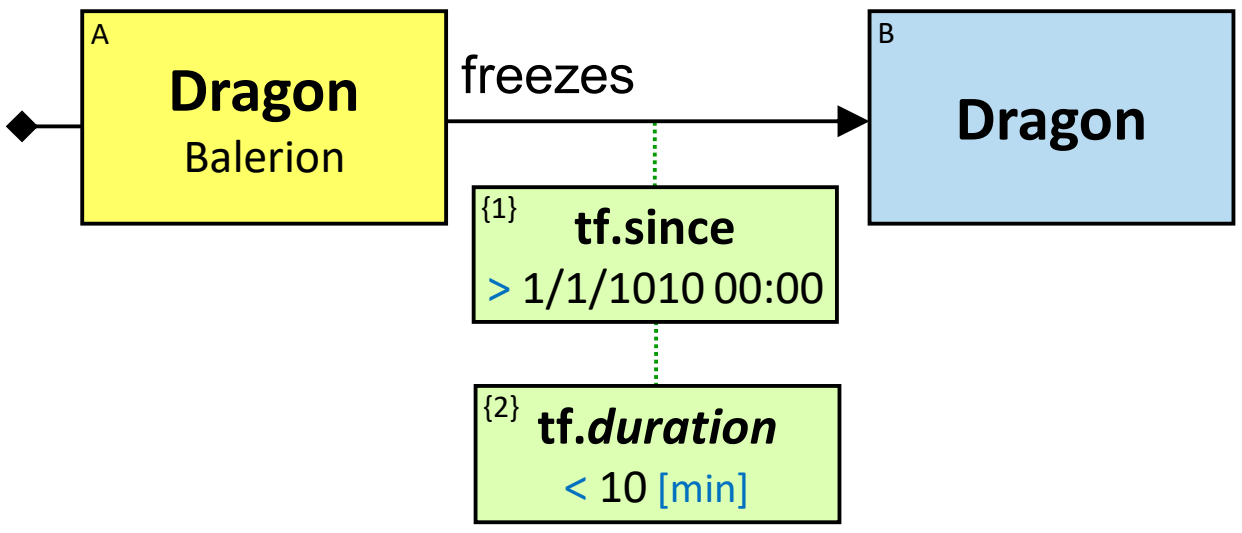

**Q10:** *Any person whose first name is Brandon, who owns some dragon B which froze a dragon C that (i) belongs to an offspring of Rogar Bolton, and (ii) froze a dragon that belongs either to Robin Arryn or to Arrec Durrandon. B froze C at least once in or after 1010 – for longer than 100 seconds*

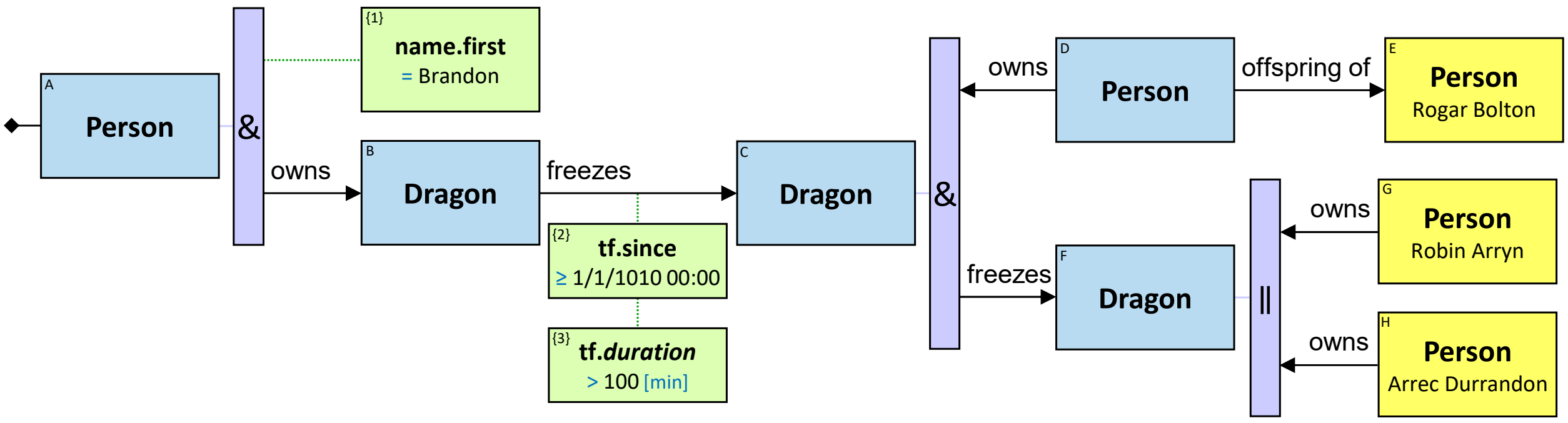

A **horizontal purple rectangle** represents a *horizontal quantifier*. It contains a quantifier type.

Eleven horizontal quantifier types are defined (as for vertical quantifiers, except the *All* quantifier). Their semantics are similar to those of vertical quantifiers. *All* can be implemented by chaining relationship's expressions or by using vertical quantifiers (see Q100).

A horizontal quantifier may appear

- below a relationship
- below a relationship's expression
- below a horizontal quantifier (in a branch)

On its bottom, there zero or more branches. Each branch starts with either

- a relationship's expression, or
- a horizontal quantifier

A horizontal combiner may be used to combine two or more consecutive branches of the same horizontal quantifier. Each branch ends with either

- A relationship's expressions, or
- A horizontal combiner

Below a horizontal combiner

- Another chained stage that starts with
    - a relationship's expression,
    - a horizontal quantifier, or
    - an aggregator

or

- Another horizontal combiner

**Q300:** *Any pair of dragons (A, B) where A froze B, and at least two of the following conditions are satisfied: (i) the freeze duration was longer than ten minutes (ii) the freeze started after January 1, 980 (iii) the freeze ended before February 1, 980*

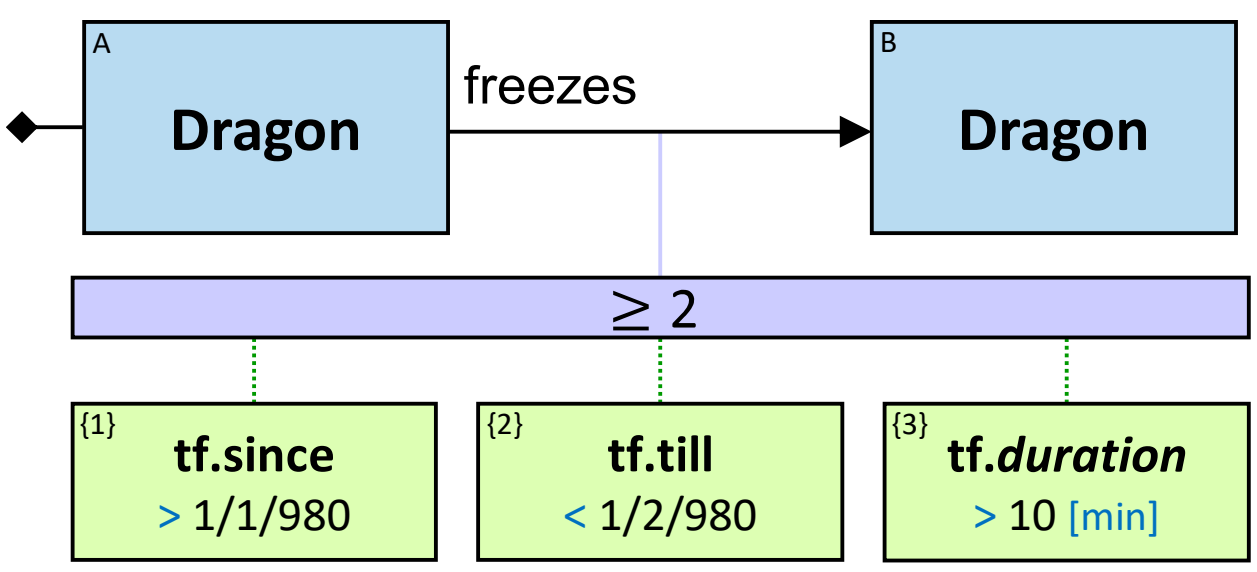

**Q301:** *Any pair of dragons (A, B) where A froze B for more than ten minutes, and either the freeze started after January 1, 980, or the freeze ended before February 1, 980* (two versions)

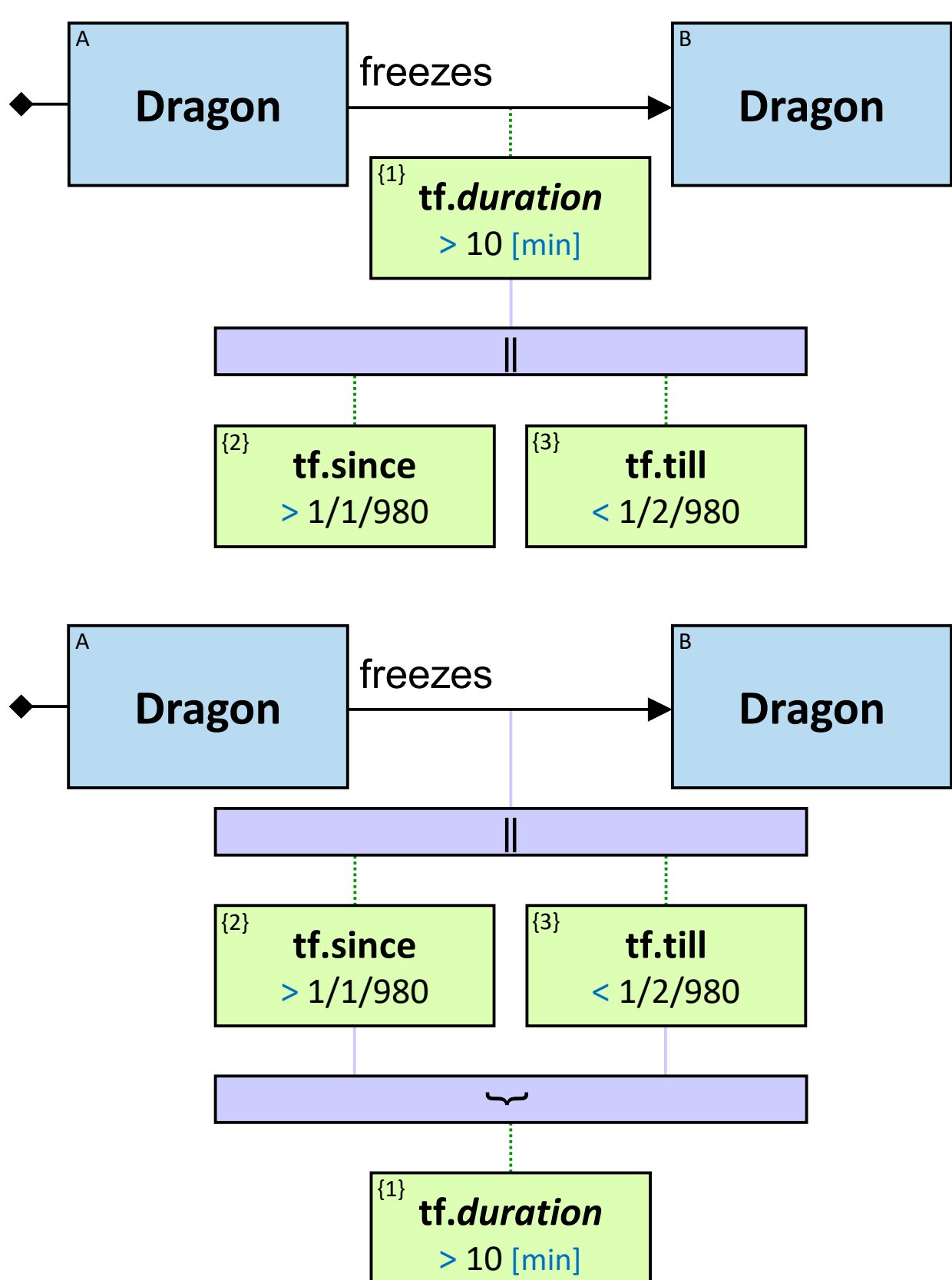

## 14. LATENT PATTERN-ENTITIES

Implicit Latent    Explicit Latent

Pattern entities may be annotated as *implicit latent* or *explicit latent*.

Assignments to *latent pattern-entities* and assignments to pattern-relationships and pattern-paths connected to latent pattern-entities are not included when pattern assignment is reported.

*Implicit latent pattern-entities* are pattern-entities that appear

- right of a negator (see Q12, Q22, Q288)
- right of a *None* quantifier, except branches (or sub-branch of a sequence of quantifiers that follows the *None* quantifier) that starts with an '0' (see Q359)

Such typed and untyped pattern-entities are required to have no assignments; hence, trivially, no assignments would be reported. Such concrete pattern-entities would not be reported as well (see Q20).

Implicit latent pattern-entities are depicted with a **gray 'no report' icon** on their top-right. These icons are automatically added by the interactive pattern builder/visualizer tool and may not be added/removed by the pattern composer.

An entity right of a combiner is depicted as implicit latent if it is implicit latent according to each of the combined branches (see Q334). If it is not implicit latent according to one or more branches – it will not be depicted as implicit latent (see Q35).

*Explicit latent pattern-entities* are non-implicit latent pattern-entities, which the pattern composer marks as latent. Though they may have assignments – those assignments are required not to be reported. Any non-implicit latent pattern-entity may be set as explicit latent.

Explicit latent pattern-entities are depicted with a **red 'no report' icon** on their top-right.

**Q142:** *Any person who owns a white or a black horse and has a parent who owns a dragon. The parent and its dragon are not to be included in the reported assignment*

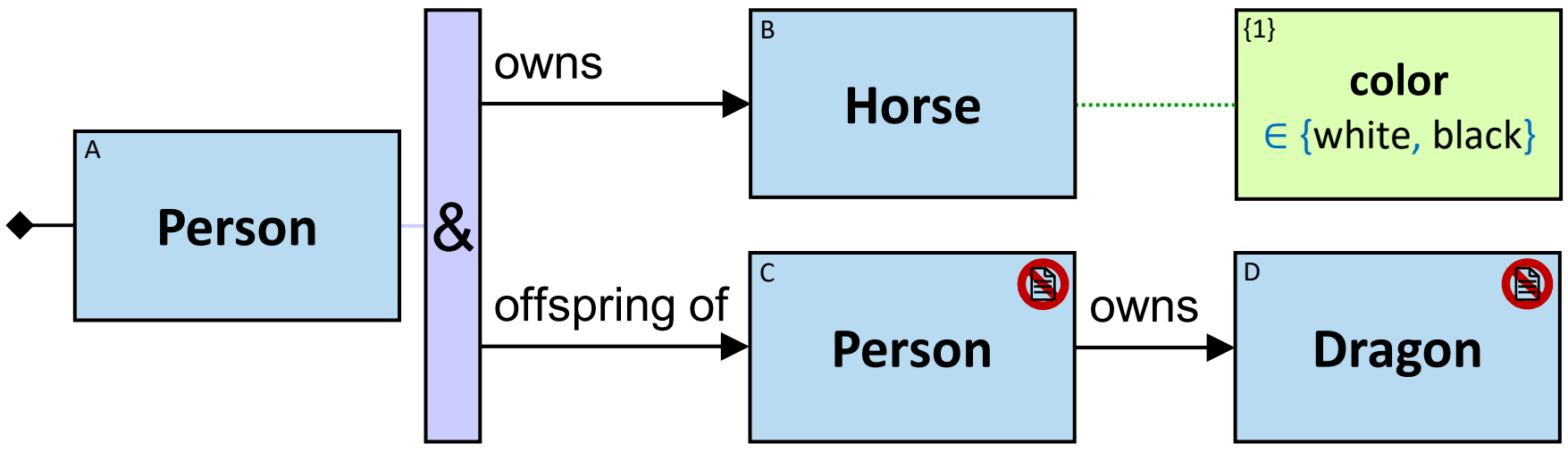

Suppose person $A_1$ has two parents, each with four *owns* relationships to dragons. Without the latent annotations, there would be eight pattern assignments per person and *owns* relationship to a horse. With the latent annotations, there would be only one assignment per person and *owns* relationship to a horse.

**Q143:** *Any person who owns a horse that is neither white nor black and has a parent that does not own a horse. The parent is not to be included in the reported assignment*

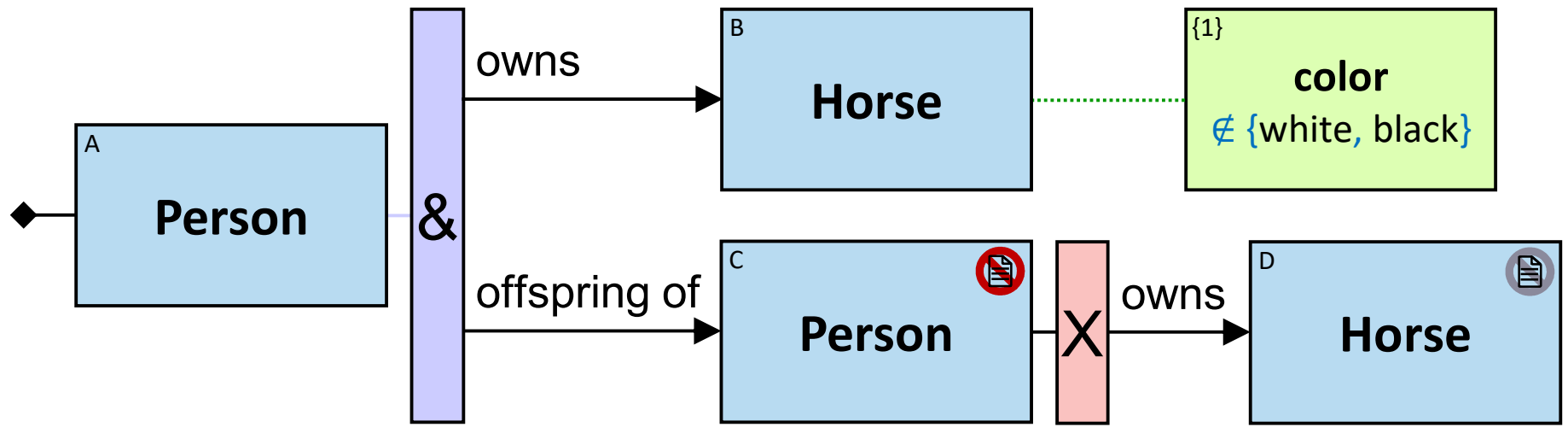

When there are several pattern-entities with the same entity-tag – some may be latent (implicit or explicit) while others are not. See Q337, Q339.

Concrete, typed, and untyped entities may all be latent – implicit or explicit.

At least one pattern-entity should be non-latent.

---

## 15. OPTIONAL COMPONENTS

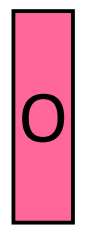

An *optional* is depicted with a **magenta 'O' rectangle**. Anything right of it is optional: if it has a valid assignment – it will be reported (unless it is latent). Otherwise – it will not.

An 'O' may be located directly left of

- a relationship or a path, with or without a relationship/path-negator
- a quantifier, excluding quantifier at the start of the pattern

**Q147:** *Any person A. If A owns both a horse and a dragon – they should be included in the assignment*

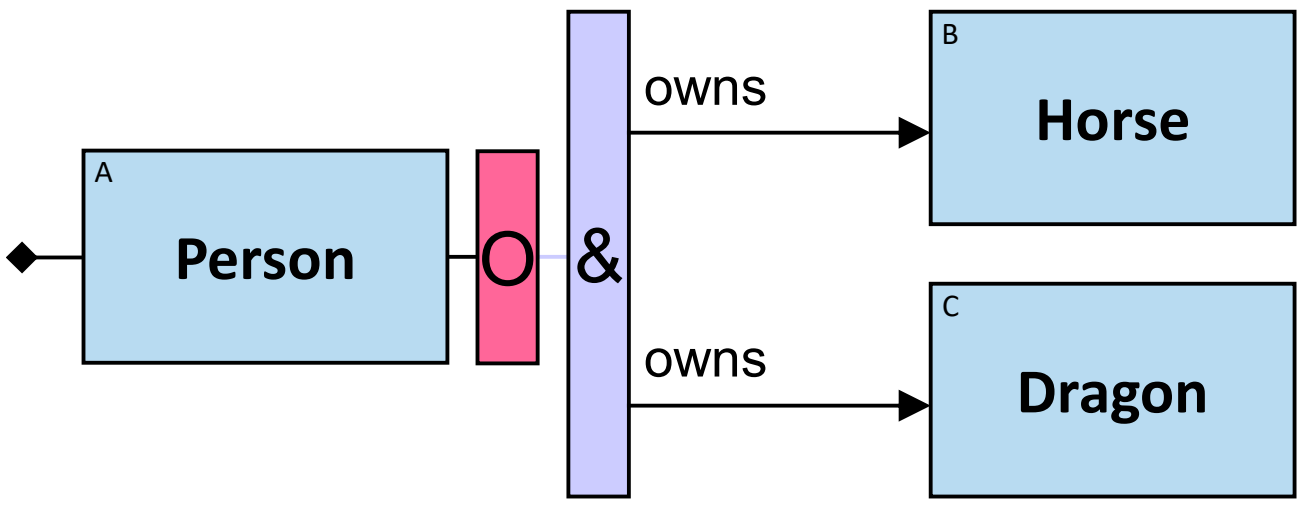

**Q149:** *Any person A. If A owns a horse – it should be included in the assignment. If A owns a dragon – it should be included in the assignment*

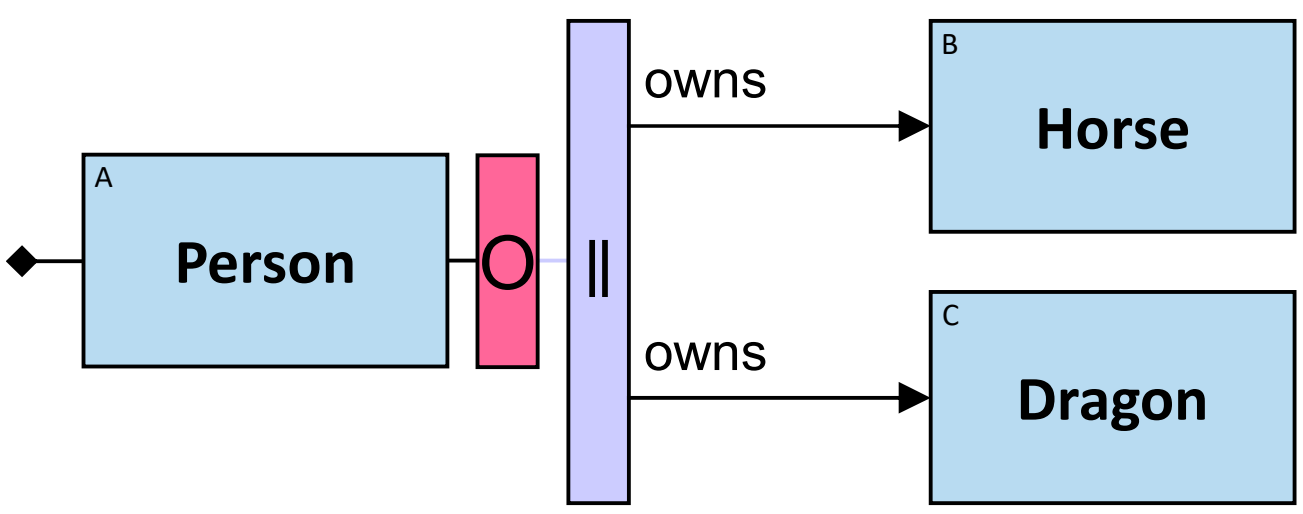

**Q81:** *Balerion. If there is a dragon it did not freeze – it should be included in the assignment*

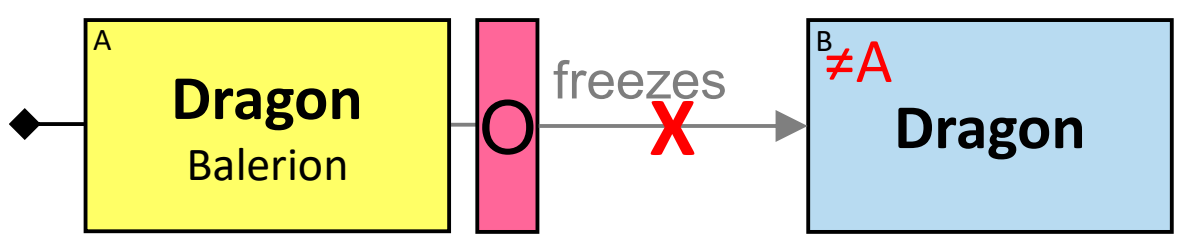

Quantifier's branches that start with an 'O' do not affect the quantifier's evaluation.

**Q144:** *Any person A who owns a white horse. If A has a parent who owns a horse – the parent and its horse should be included in the assignment*

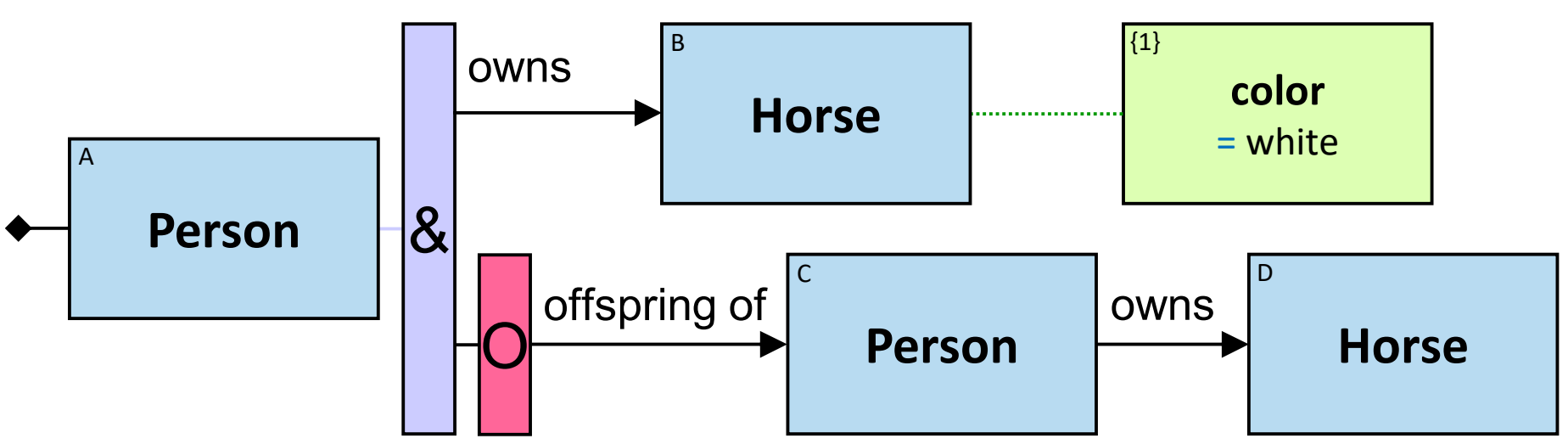

**Q145:** *Any person A who owns a white horse. If A has a parent that does not own a horse – the parent should be included in the assignment*

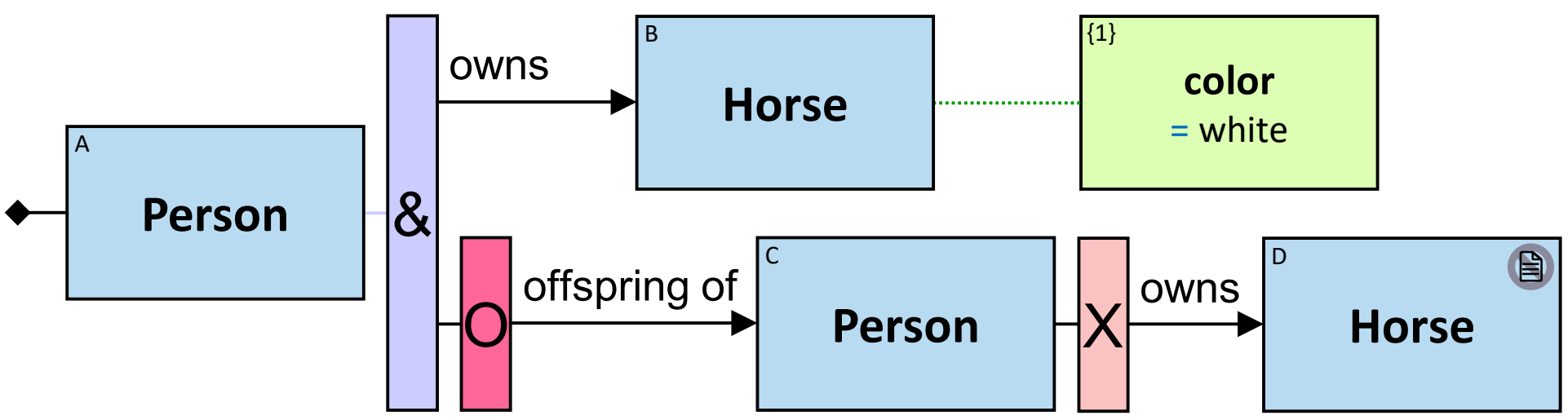

**Q146:** *Any person A who owns a white horse. If A has a parent – the parent should be included in the assignment. If A's parent owns a horse – this horse should be included in the assignment as well*

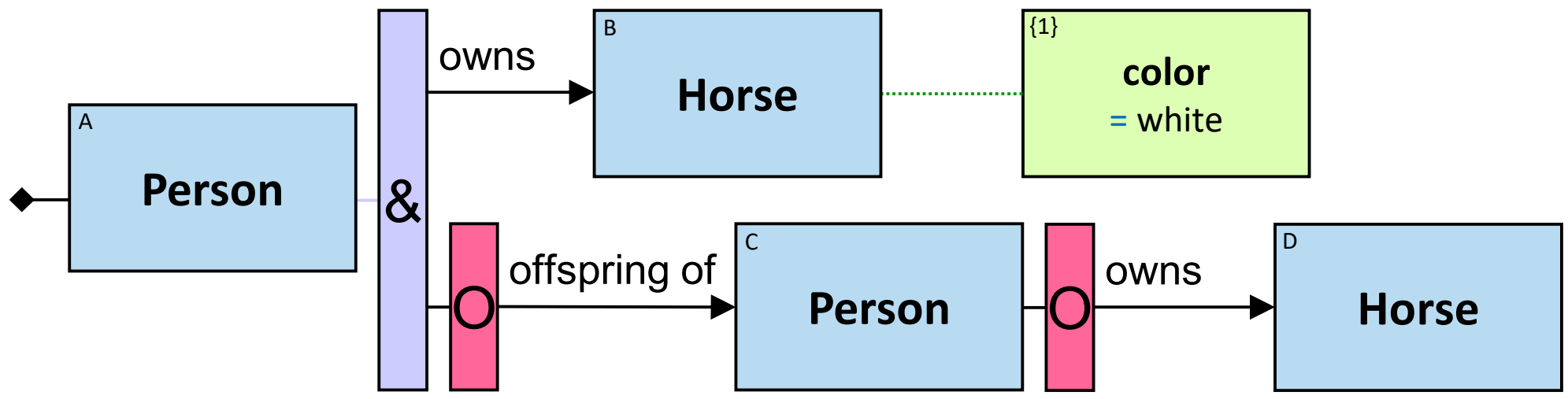

**Q148:** *Any person A who owns both a horse and a dragon. If A has a parent – the parent should be included in the assignment*

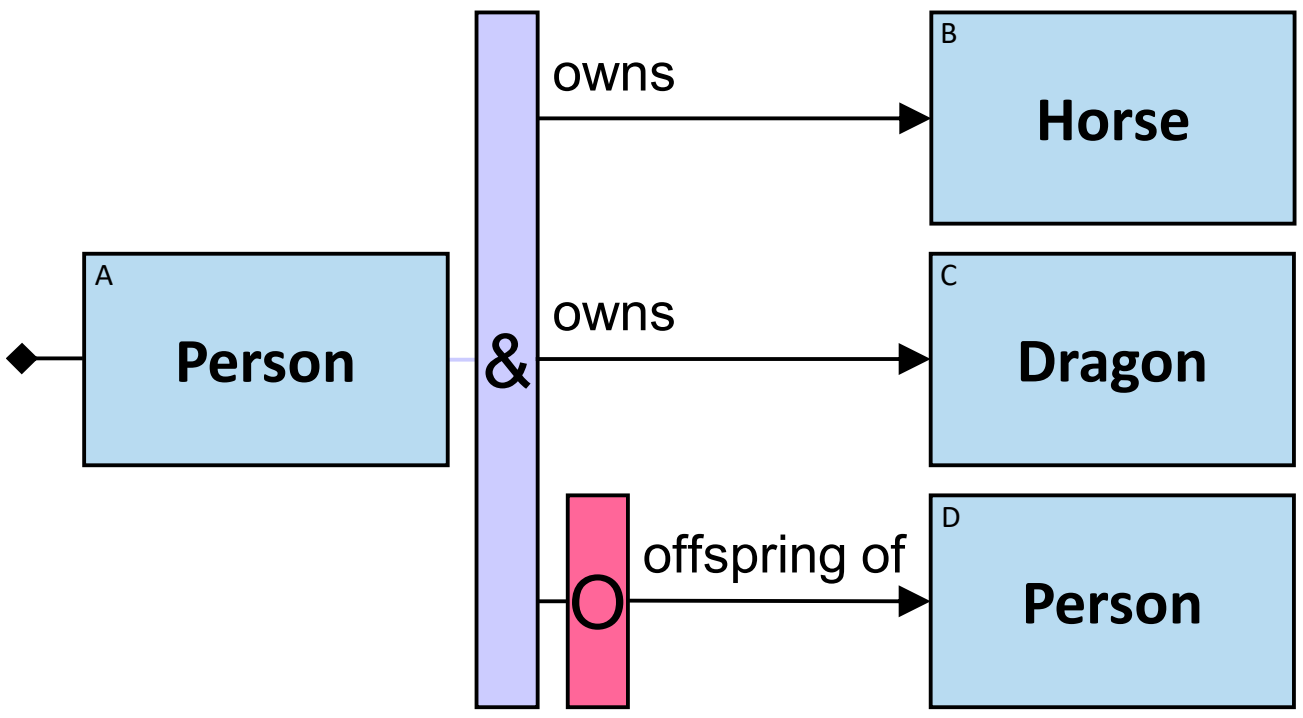

**Q150:** *Any person A who owns a horse or a dragon. If A has a parent – the parent should be included in the assignment*

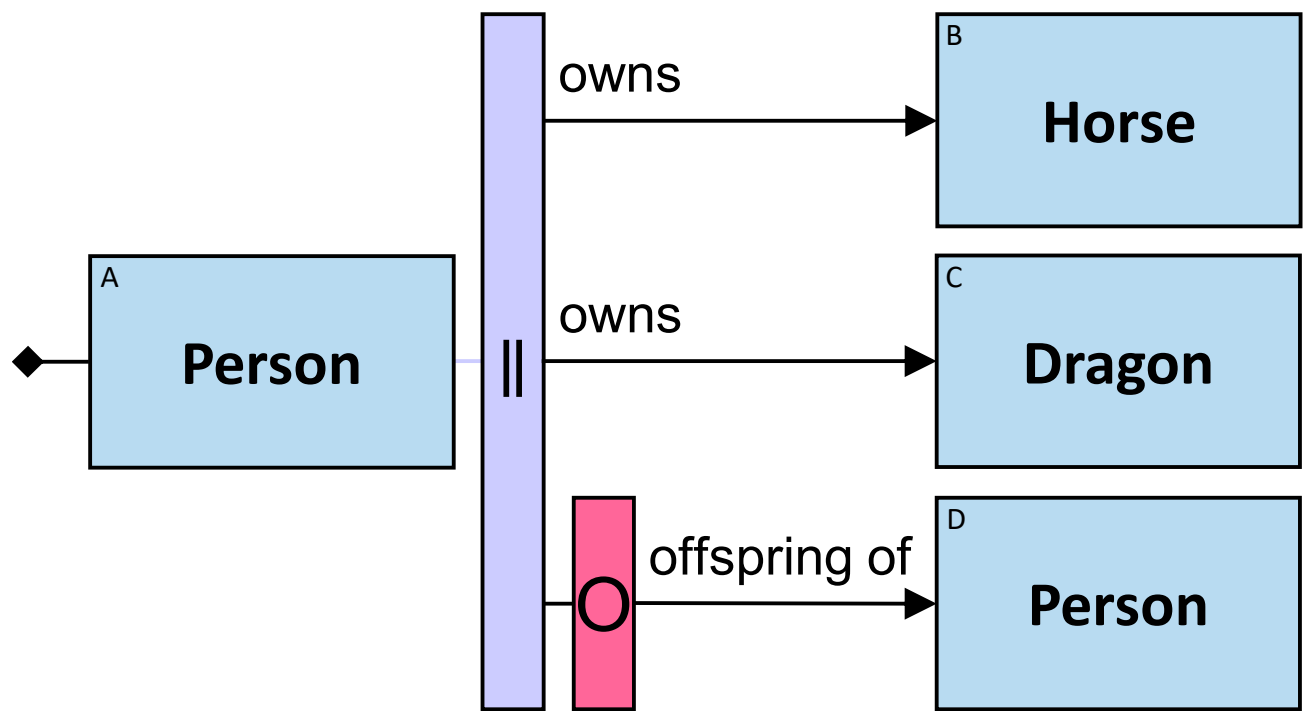

**Q203:** *Any person A. If A owns both a horse and a dragon – they should be included in the assignment. If A owns both a horse and a dragon and also has a parent – the parent should be included in the assignment as well*

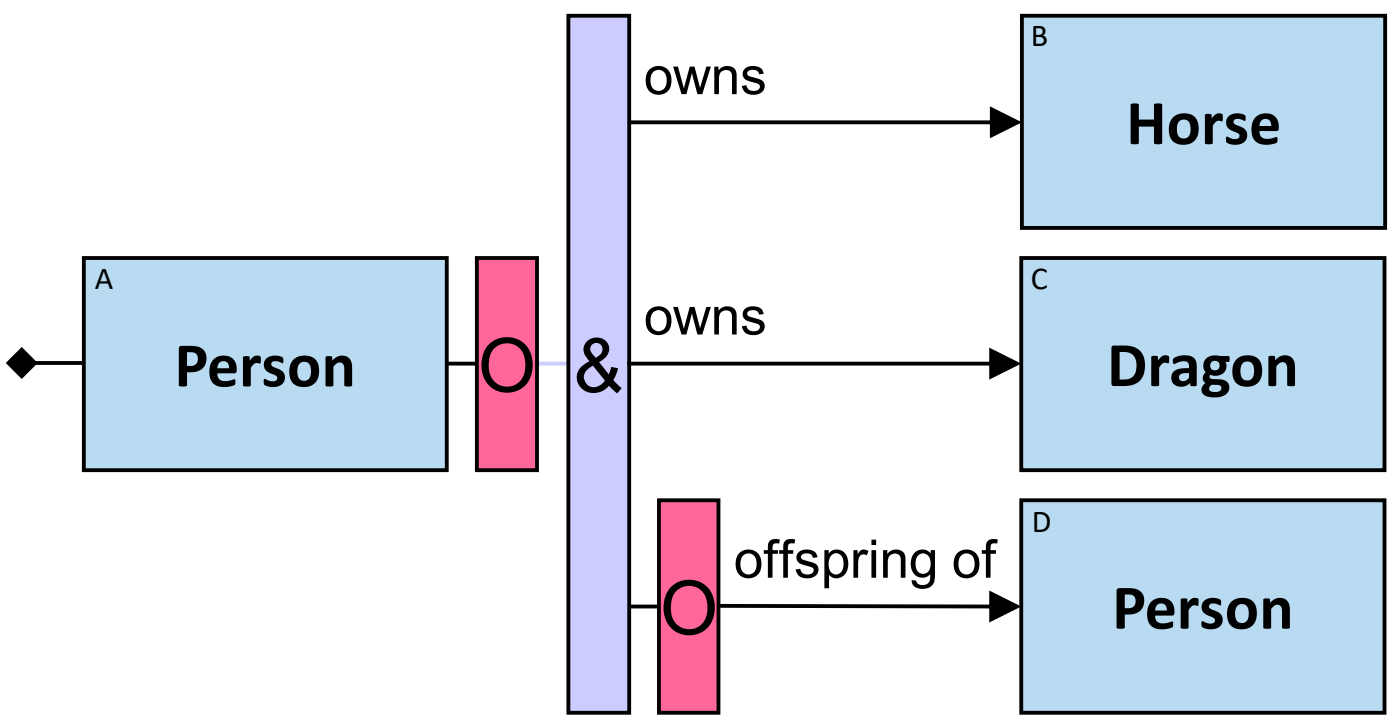

When tags right of an 'O' have no assignments:

- An expression-tag defined right of an 'O' is evaluated to *null* (see Q140, Q141)
- An A1/A2 aggregation-tag defined right of an 'O' is evaluated to zero (see Q125, Q347)
- An A3 aggregation-tag defined right of an 'O' is
  - evaluated to *null* when *aggop* is *min*/*max*/*avg*/*sum* (see Q317, Q318)
  - evaluated to zero when *aggop* is *distinct*
  - evaluated to {} / [] when *aggop* is *set*/*bag*/*union*/*intersection*

- The number of assignments to an entity-tag / Cartesian product of entity-tags defined right of an ‘0’ is zero (see Q295, Q320, Q321)
- The number of assignments to an entity type-tag defined right of an ‘0’ is zero
- The number of assignments to a relationship type-tag defined right of an ‘0’ is zero

***Q359:** Any dragon A where (i) there is no black dragon A froze (ii) there is no white dragon A did not freeze, (iii) there is at least one gold dragon A froze, and (iv) there is at least one silver dragon A did not freeze*

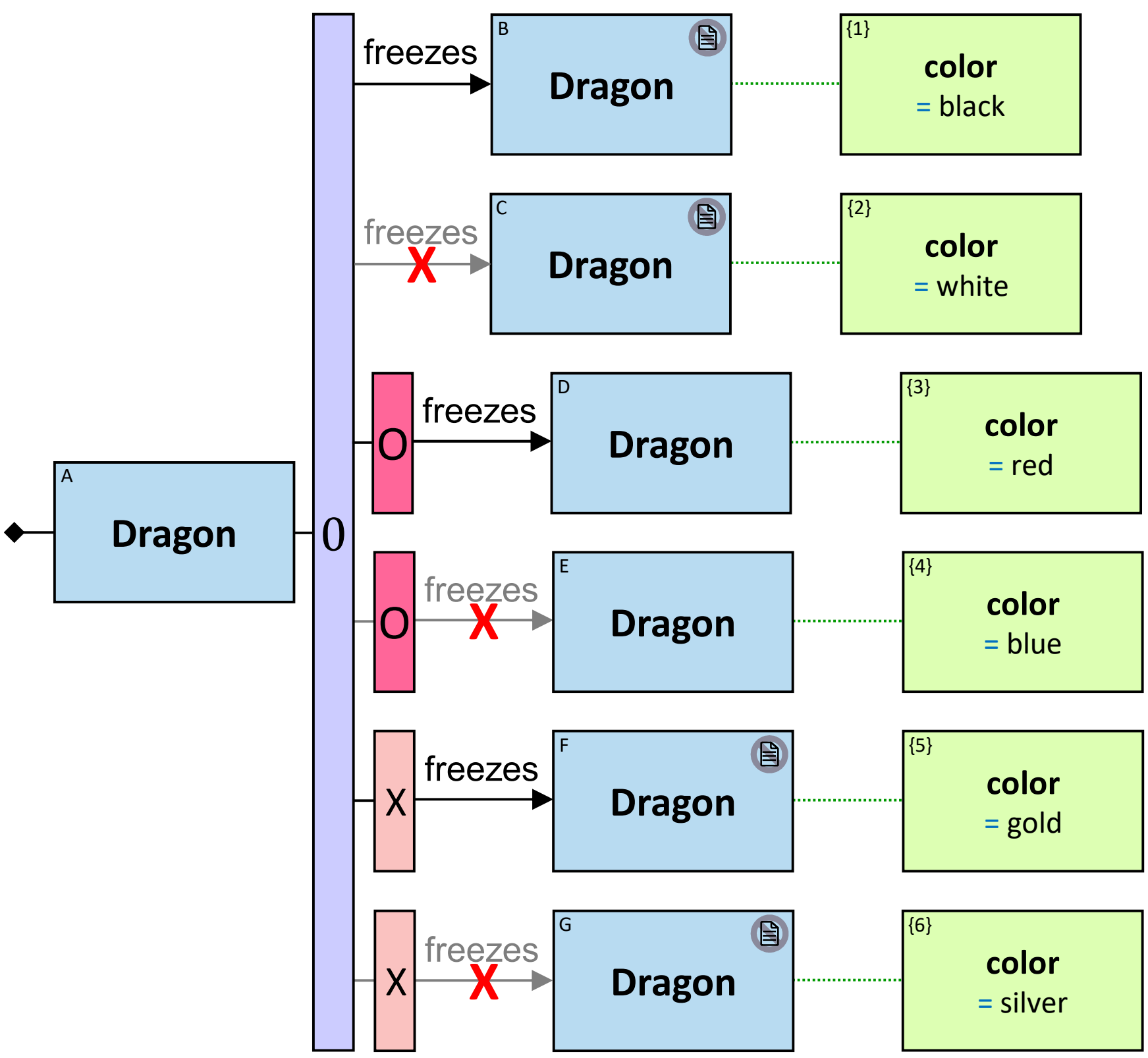

Note that any red dragon that A froze and any blue dragon that A did not freeze is reported. Assignments to optional branches are reported regardless of the quantifier type.

## 16. UNTYPED ENTITIES

A relationship-type may hold between different pairs of entity-types (e.g., *owns*: {(*Person*, *Dragon*), (*Guild*, *Dragon*), (*Null*, *Dragon*)}). *Untyped entities* can be used to express patterns such as *Any dragon and its owners*, where the owner can be either a person, a guild, or *Null*.

A **red rectangle** represents an *untyped entity*. Graph-entities of different types may be assigned to an untyped pattern-entity.

An **empty red rectangle** represents an entity with no explicit type constraints:

**Q36:** *Any person who owns something*

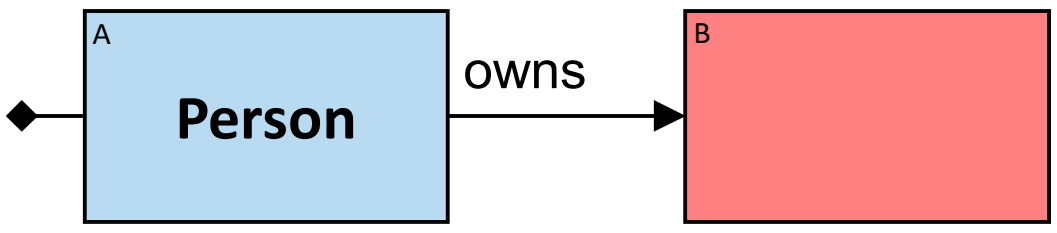

B assignments are *implicitly constrained* to things that a person can own.

*Implicit entity-type constraints* are inferred from the pattern (including from identicality / nonidenticality / order constraints).

**Q49:** *Any three dragons with a cyclic freeze pattern and their owners (if any)*

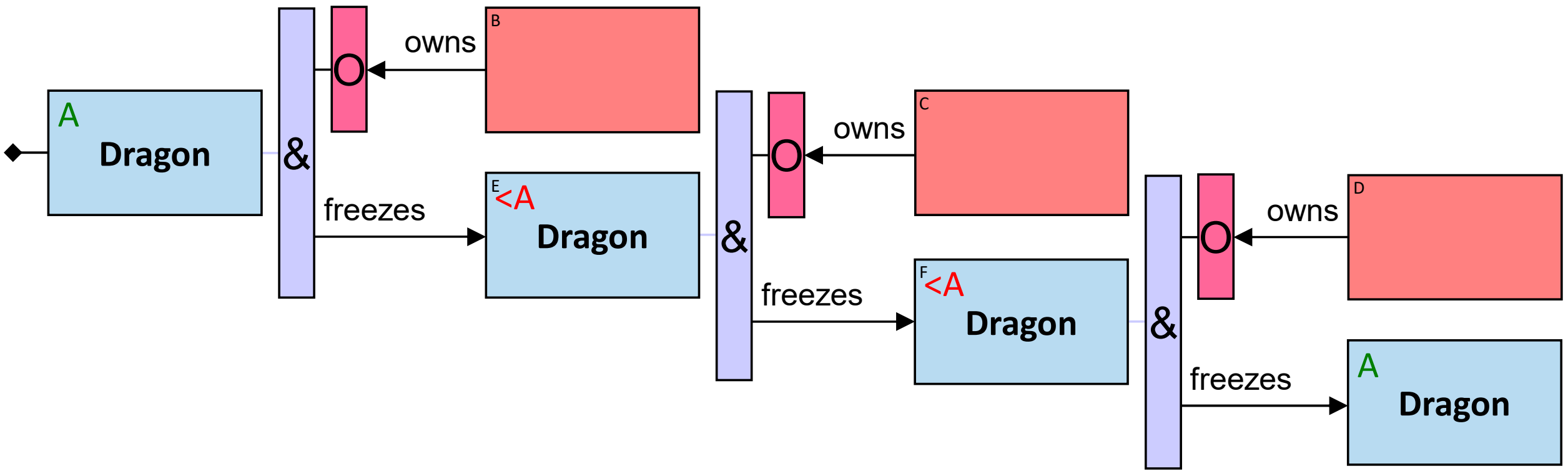

The entity-types of all owners are implicitly constrained to things that can own a dragon.

Entities are implicitly type-constrained also when untyped entities are on either side of a negator or a relationship-negator:

**Q39:** *Any entity of a type that can own a dragon but does not*

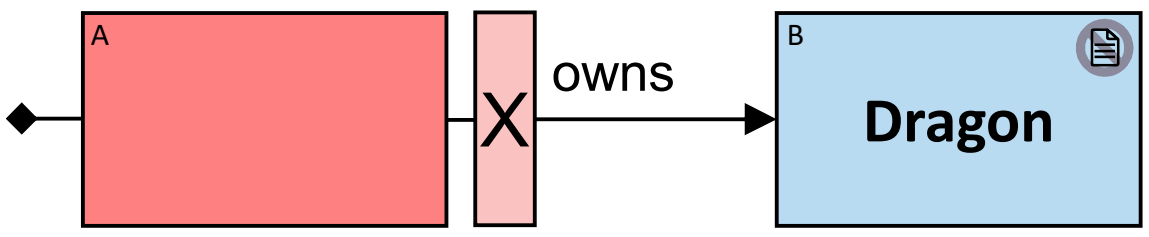

**Q40:** *Any dragon that is not owned*

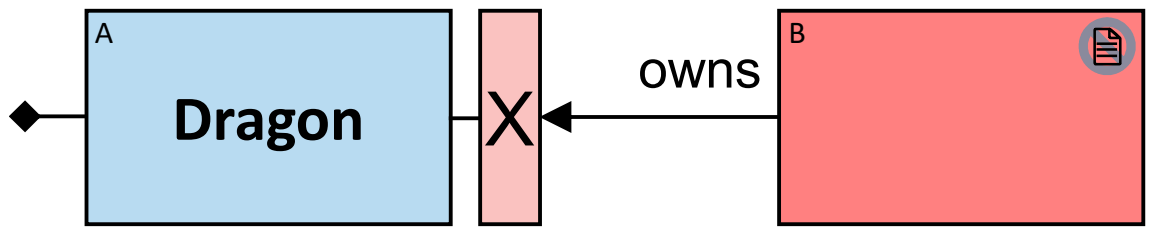

**Q41:** *Any entity of a type that can be owned but is not owned*

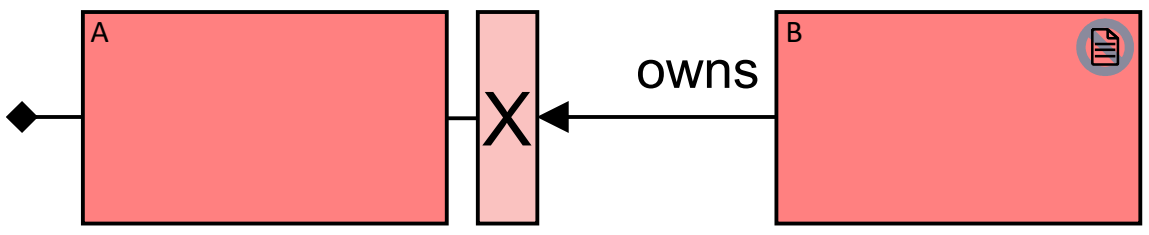

**Q42:** *Any entity of a type that can own something but owns nothing*

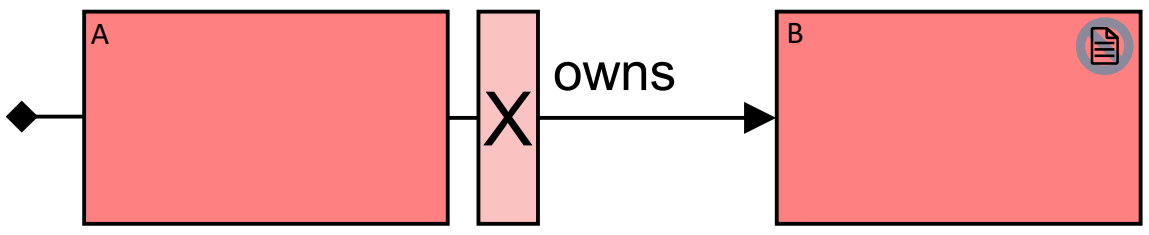

In addition to the implicit type constraints, *explicit type constraints* can be enforced using a type equality constraint ('='), a type inequality constraint ('≠'), a set of allowed types ('∈ {...}'), or a set of disallowed types ('∉ {...}').

**Q43:** *Any dragon that all of its owners (if any) are people*

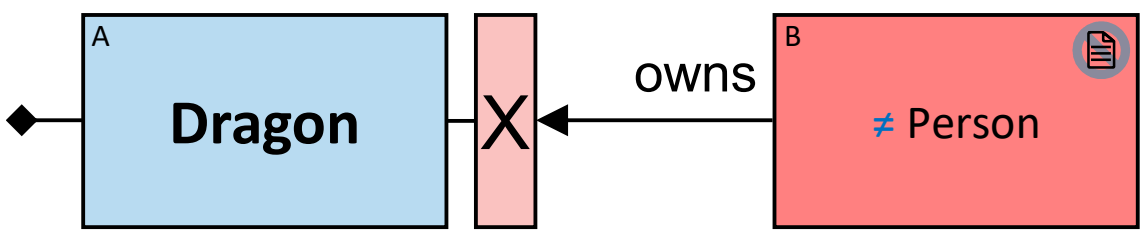

**Q37:** *Any person who owns a horse or a dragon*

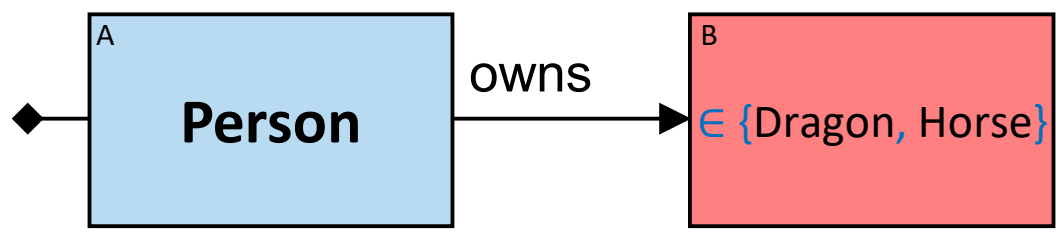

**Q38:** *Any person who owns something which is neither a horse nor a dragon*

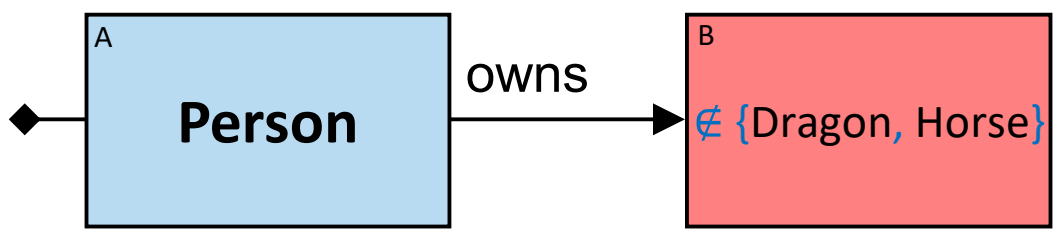

For each untyped entity, implicit and explicit type constraints must not nullify the list of valid entity-types. Explicit type constraints must not contradict implicit type constraints. This can be asserted before executing the query.

Since both *horse* and *dragon* entity-types have a *name* property of the same data type (string) – the following pattern is valid:

**Q291:** *Any person who owns a horse or a dragon whose name starts with an 'M'*

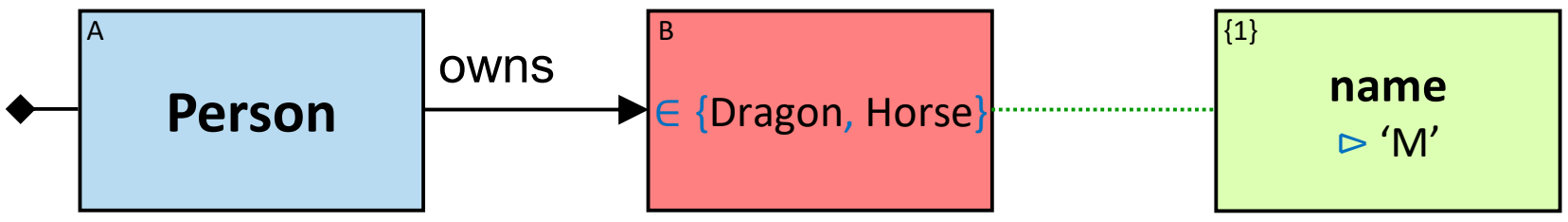

## 17. ENTITY TYPE-TAGS

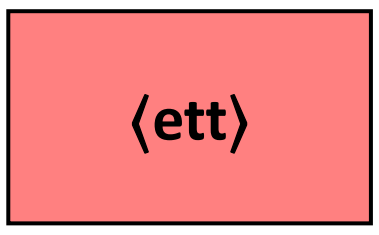

A red rectangle (denoting an untyped entity) may contain an *entity type-tag*, depicted by a **numeric index wrapped in angle brackets**. An entity type-tag represents the entity-type of the entity in a given assignment.

Entity type-tags may be referenced:

- in a type constraint of an untyped entity (see Q50, Q52)
- in an extended aggregator *per* clause (see Q218v2, Q350)
- in an A3 aggregator "aggop ..." clause (see Q167v2)
- in an extended M1/M2/M3 aggregator "[all but] *k* ..." clause (see Q209, Q210)

**Q50:** *Any person who owns (at least) two things of the same type*

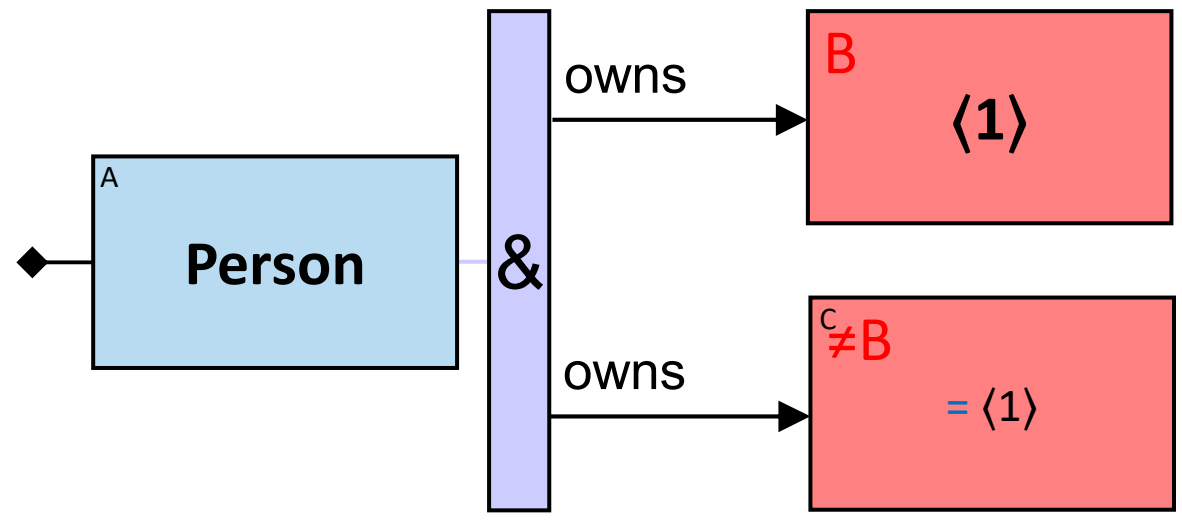

**Q51:** *Any person who owns (at least) two things of different types*

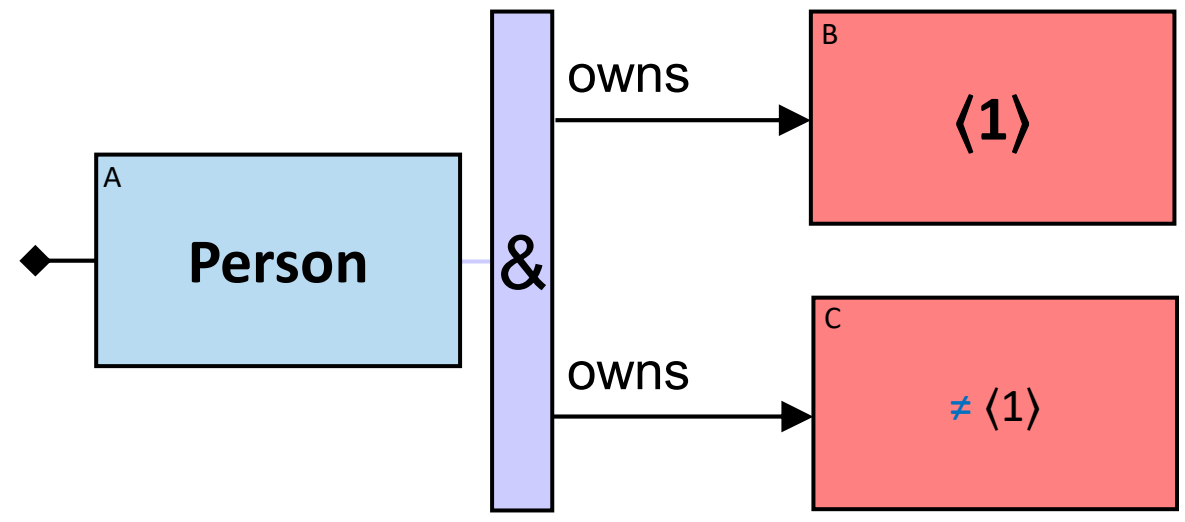

**Q52:** *Any person who owns (at least) two things of different types, neither is a horse*

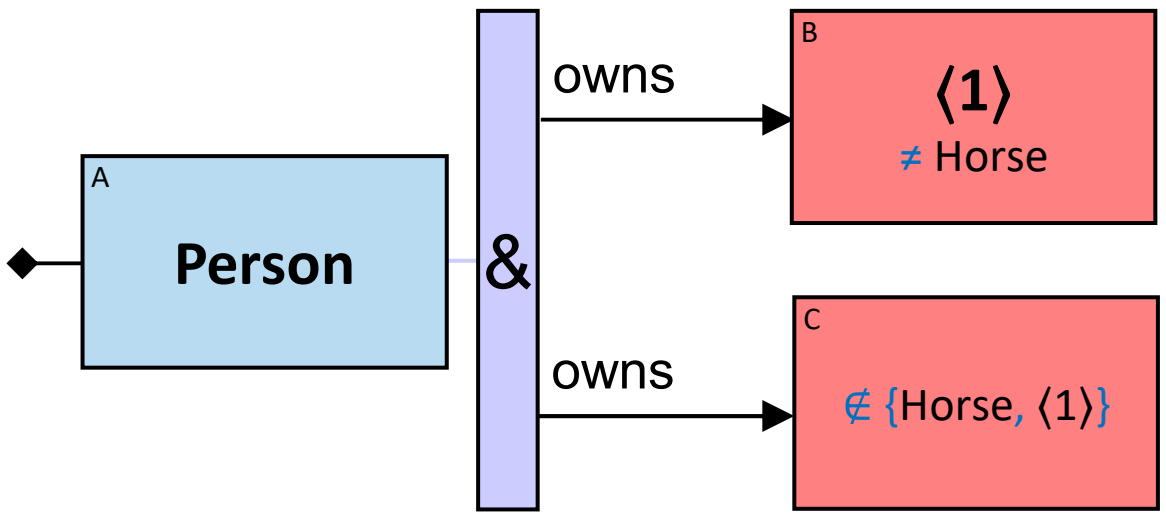

***Q167:*** *Any person whose owned entities are all of the same type* (version 1)

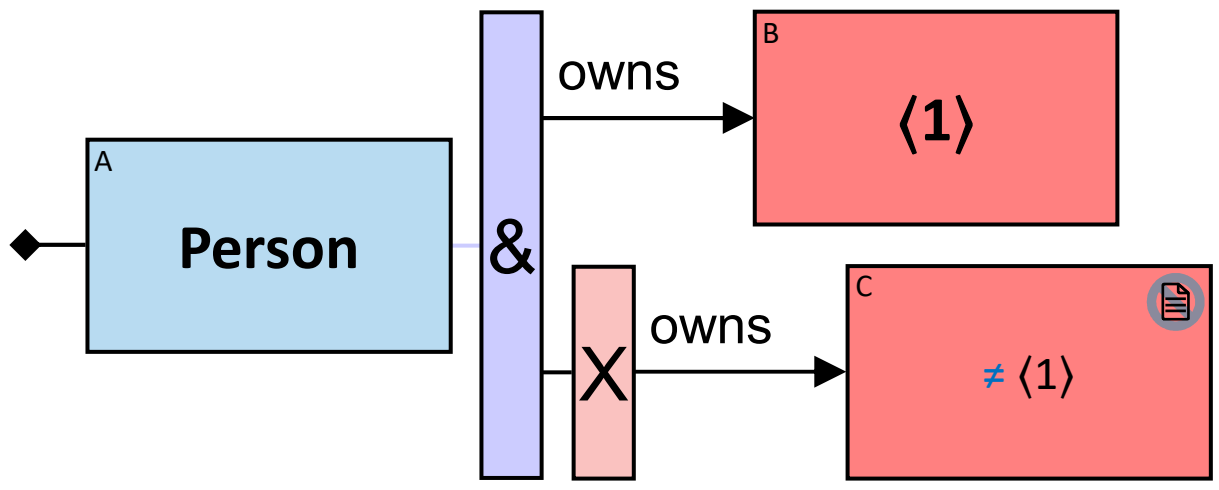

For any quantifier except *all* – an entity type-tag defined in a branch can only be referenced in that branch.

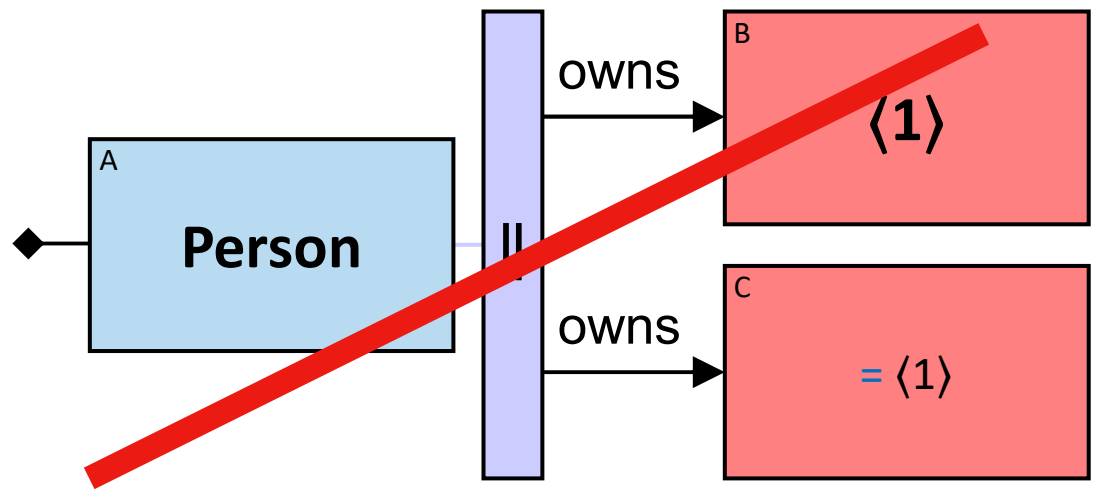

An entity type-tag defined right of a negator can only be referenced right of that negator.

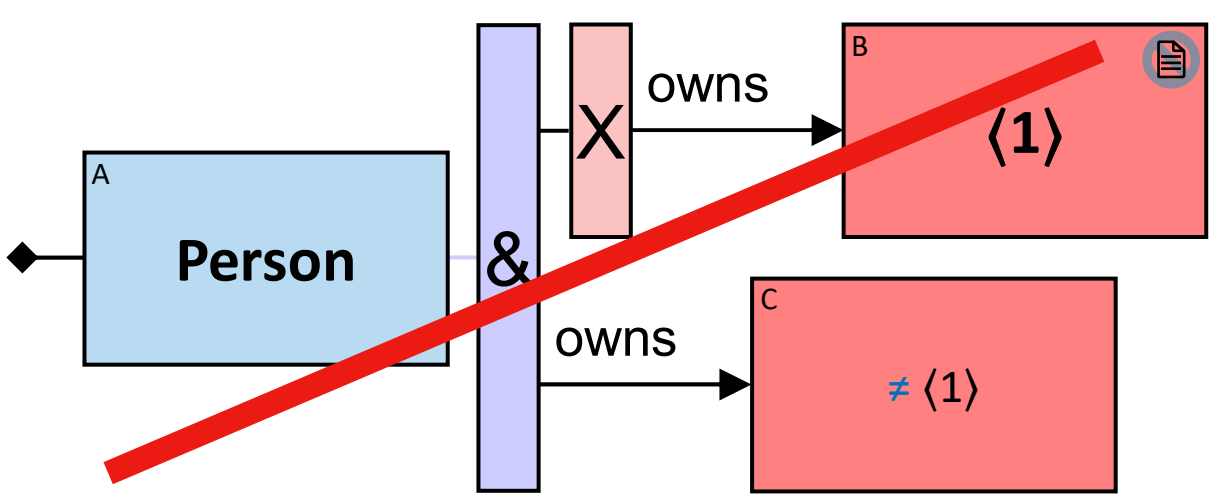

An entity type-tag defined right of an 'O' can only be referenced right of that 'O', except when referenced in a *distinct* ⟨*ett*⟩ A3 aggregator (see Q167v2).

## 18. UNTYPED RELATIONSHIPS

Multiple relationship-types may hold between a pair of entity-types (e.g., *freezes*: {(Dragon, Dragon)}, *fires at*: {(Dragon, Dragon)}. Untyped relationships can be used to express patterns such as *Any pair of dragons with at least one relationship*, where the relationship type can be either *freezes* or *fires at*.

A **red arrow/line** with no type label or with type constraints represents an *untyped relationship*. Graph-relationships of different types may be assigned to an untyped pattern-relationship.

A red arrow/line with no label represents a relationship with no explicit type constraints:

**Q29:** *Any dragon that froze or fired at some dragon* (version 4)

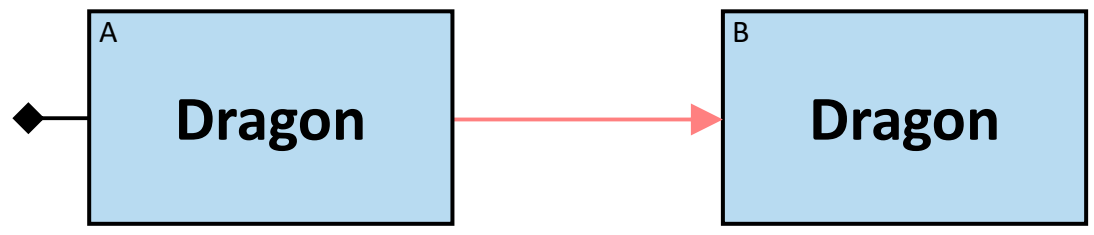

The relationship is *implicitly type-constrained* to directional relationship-types between two dragons.

**Q53:** *Any person with a path of length up to three to Rogar Bolton* (version 1)

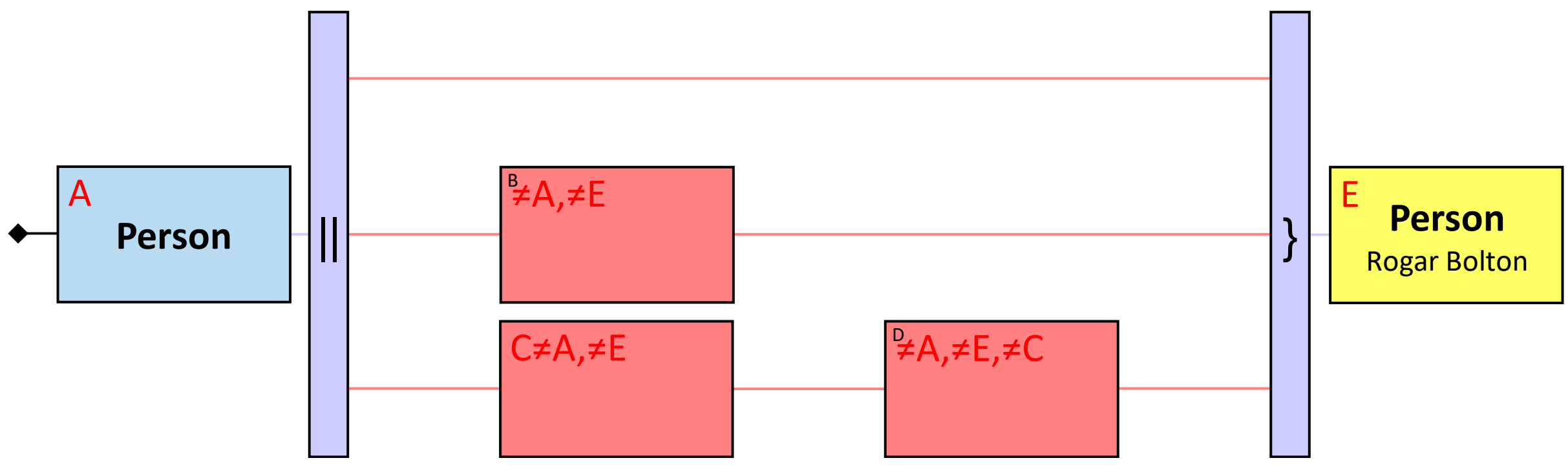

Relationships are implicitly type-constrained also when a negator or a relationship-negator wraps an untyped relationship:

**Q360:** *Any dragon that (froze or fired at) all other dragons.* Alternative phrasing: *Any dragon for which there is no dragon (besides itself) it did not (freeze or fire at)* (version 4)

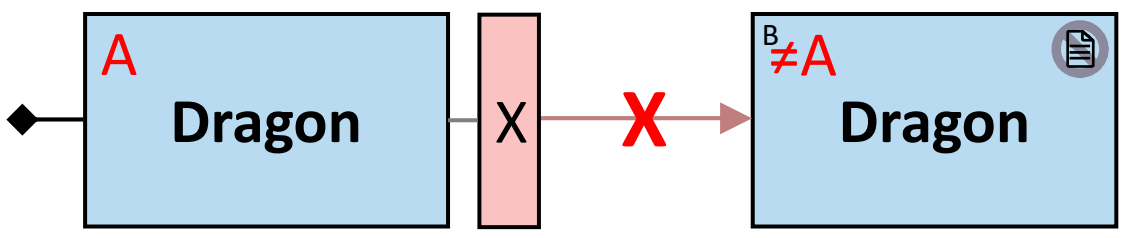

In addition to the implicit relationship-type constraints, *explicit relationship-type constraints* can be enforced using a type equality constraint ('='), a type inequality constraint ('≠'), a set of allowed types ('∈ {...}'), or a set of disallowed types ('∉ {...}').

**Q364:** *Any person and an entity that have a relationship of a type other than 'friend of'*

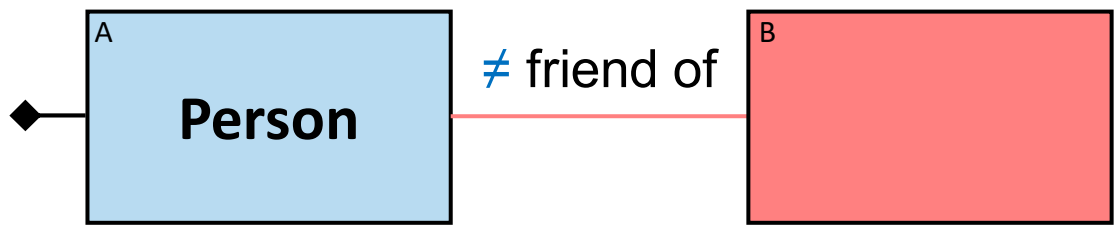

*Implicit relationship-type constraints* are inferred from the pattern (including from relationships directionality and from entities identicality/nonidenticality/order constraints).

For each untyped relationship, implicit and explicit type constraints must not nullify the list of valid relationship-types. Explicit type constraints must not contradict implicit type constraints. This can be asserted before executing the query.

## 19. RELATIONSHIP TYPE-TAGS

⟪1⟫

Instead of a relationship-type, a relationship label may denote a *relationship type-tag*, depicted by a **numeric index wrapped in double angle brackets.** A relationship type-tag represents the relationship-type of the relationship in a given assignment.

Relationship type-tags may be referenced:

- in a type constraint of an untyped relationship (see Q365)
- in an extended aggregator *per* clause (see Q369)
- in an A3 aggregator "aggop …" clause (see Q366v2)
- in an extended M1/M2/M3 aggregator "[all but] *k* …" clause (see Q369)

***Q365:** Any pair of dragons that have relationships of at least two types*

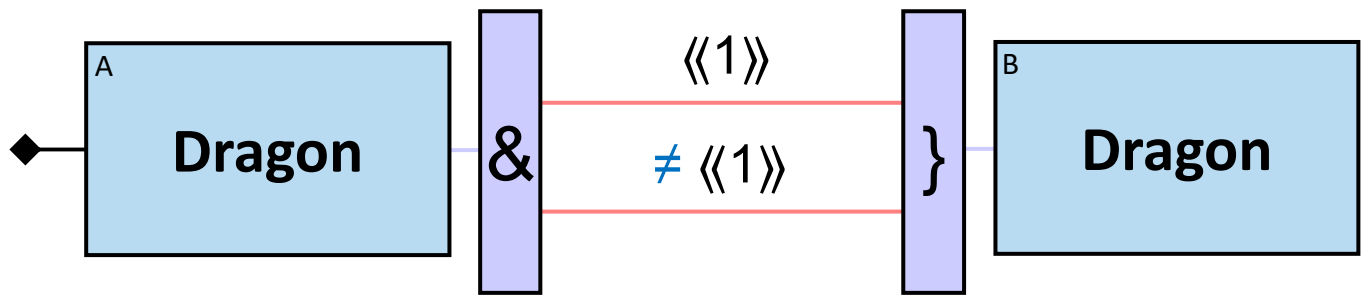

***Q366:** Any pair of dragons that have relationships of at least three types* (version 1)

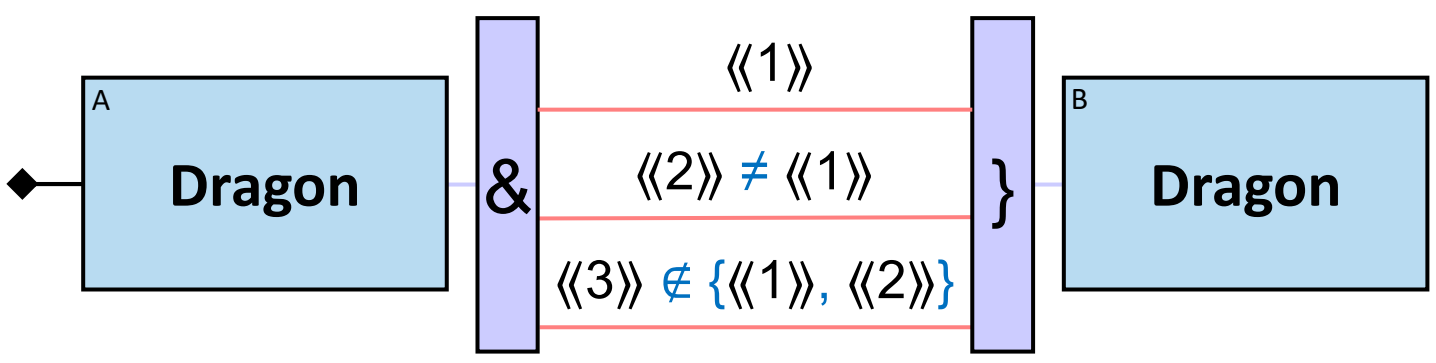

Since there are only two relationship-types between a pair of dragons (*freezes*, *fires at*), this pattern would never have assignments. This can be asserted before executing the query.

***Q367:** Any pair of dragons that have a directional relationship of the same type in both directions*

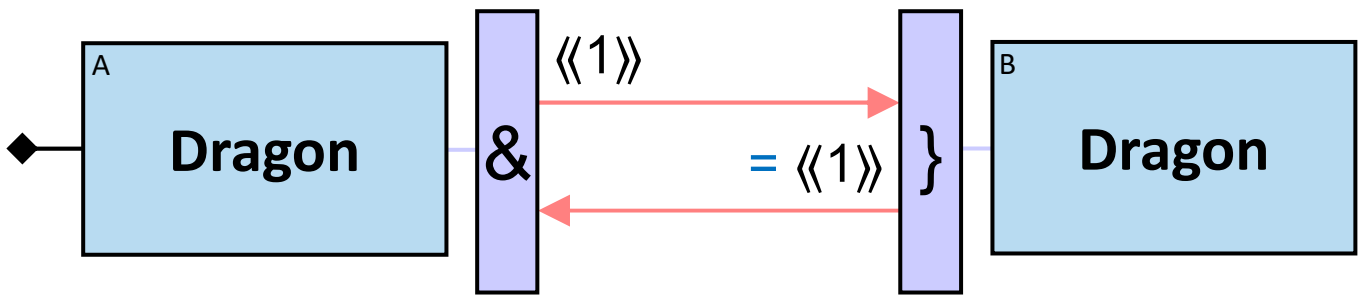

***Q368:** Any pair of dragons that have relationships of at least two types, neither is 'freezes'*

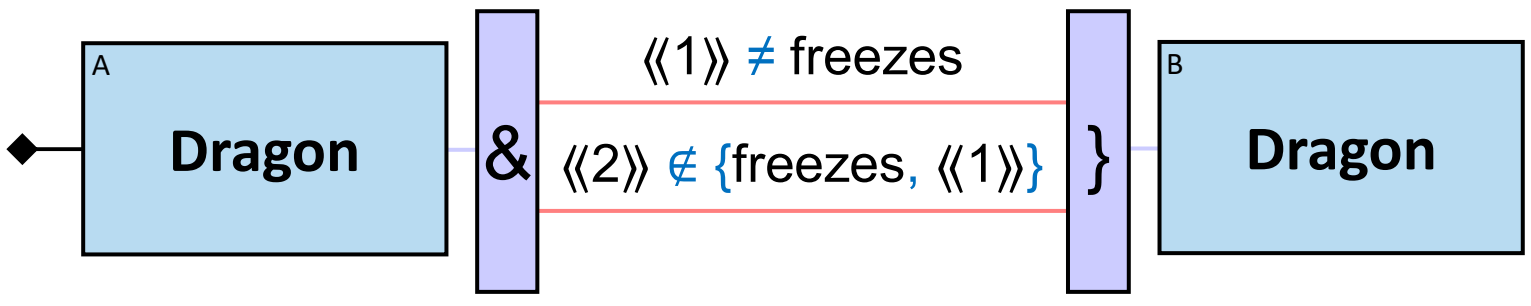

Again, the fact that this pattern would never have assignments can be asserted before executing the query.

For any quantifier except *all* – a relationship type-tag defined in a branch can only be referenced in that branch.

A relationship type-tag defined right of a negator can only be referenced right of that negator.

A relationship type-tag defined right of an 'O' can only be referenced right of that 'O', except when referenced in a *distinct* ⟪*rtt*⟫ A3 aggregator.

---

## 20. NULL ENTITIES

As described above, *graph-entities of type Null* realize action-types and relationships to unknown/unimportant entities.

Since each graph-entity of type *Null* is connected by exactly one relationship, there are additional implicit constraints on each typed entity of type *Null* and on each untyped entity that explicitly allows type *Null*:

- There is no relationship/path on its right (it is a terminal node), or there is no relationship/path on its left (it starts the pattern)
- It is not directly right of a combiner nor directly left of a quantifier
- It is not being aggregated
- Its single connection is either a relationship of a type that supports *Null* entities on this side or a path that may end with such a relationship
- Its entity-tag is not reused (no identicality, nonidenticality, nor order constraints)

As part of a pattern:

- Concrete pattern-entities may not be of type *Null*
- Typed entities may be of type *Null* if there are no implicit type constraints that disallow it
- For untyped entities, type *Null* may be explicitly allowed or disallowed only when there are no implicit type constraints that disallow it

As part of an assignment:

- For an untyped entity – graph entities of type *Null* are valid assignments only if
    - There are no implicit type constraints that disallow it, and
    - There are no explicit type constraints that disallow it, or there are explicit type constraints that allow it

See G1-G20.

---

## 21. PATHS

A *graph-path* is a sequence of graph-entities and graph-relationships that starts with a graph-entity and ends with a graph entity. A *simple graph-path* is a graph-path where all entities are distinct, except possibly the first and last.

Each graph-path has a length. The *path length* is equal to the number of graph-relationships along the path; hence, a relationship is a path of length 1.

A *pattern-path* connects two pattern-entities – like a pattern-relationship. However, while pattern-relationship are assigned with graph-relationships, pattern-paths are assigned with simple graph-paths minus the first and last entities, which are not considered a part of the assignment. Thus, pattern-paths are assigned with at least one graph-relationship.

A **blue line** between two pattern-entities represents a path. Above the line is a *constraint on the path length*. An upper bound on the path length must be defined, hence the constraint must be defined using one of the following constraint types: = *expr*, < *expr*, ≤ *expr*, in *set-expr*, in *bag-expr*, in *list-expr* or in *interval-expr*, where all expressions or interval bounds evaluate to positive integers.

**Q53:** *Any person with a path of length up to three to Rogar Bolton* (version 2)

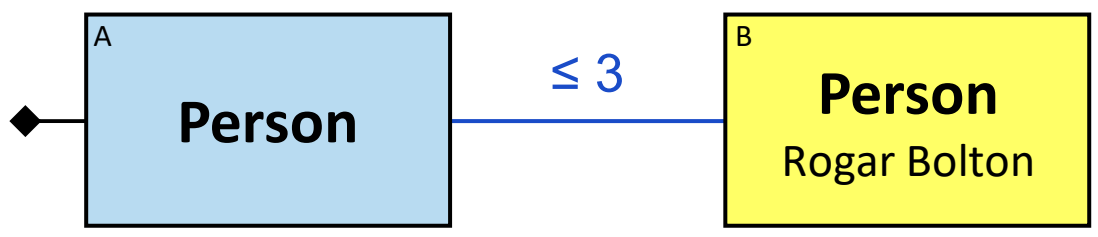

***Q84:*** *Any dragon with no paths of length up to three to other dragons*

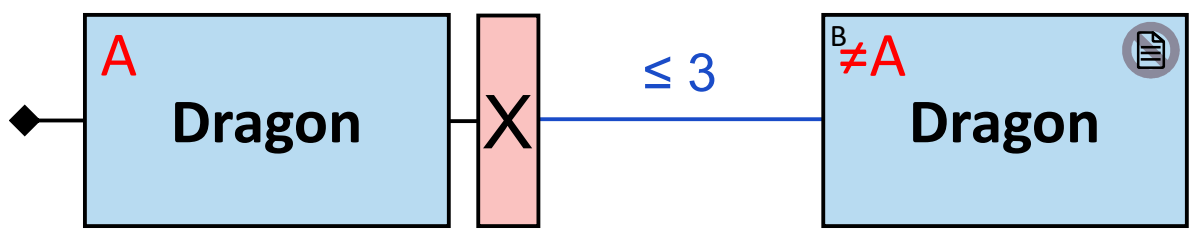

**Q55:** *Any entity with paths of length up to three to Rogar Bolton, Robin Arryn, and Arrec Durrandon*

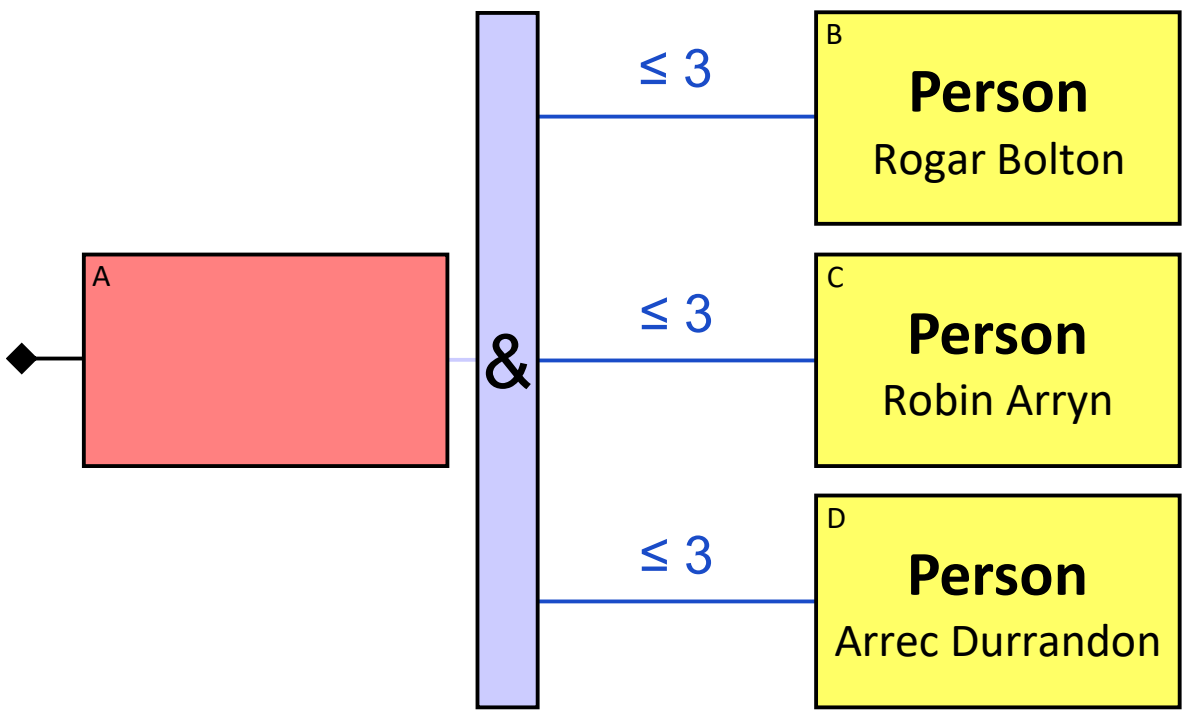

Constraints may be defined for both the entities and the relationships along the path:

*Constraints on relationships along the path* are listed in **blue curly brackets above the path's line**. The brackets may list either:

- Allowed relationship-types – e.g., {fires at, freezes}. Any unlisted relationship-type is disallowed.
- Constraints on the number of relationships of given types, with optionally – a given direction – e.g., {freezes < 2, fires at = 2}, {freezes = 0}, {→ freezes = 2, ← freezes = 1}. Any unlisted relationship-type / direction is allowed.

*Constraints on entities along the path* are listed in **blue curly brackets below the path's line**. The brackets may list either:

- Allowed entity-types – e.g., {*Dragon*}. Any unlisted entity-type is disallowed.
- Constraint on the number of entities of each allowed type – e.g., {Dragon = 0}, {Dragon ≥ 1, Horse ≥ 1}. Any unlisted entity-type is allowed.

A path cannot be composed of *Null* entities since they are terminal nodes. Therefore, the list of allowed entity-types may not contain *Null*. Similarly, the list of disallowed entity-types may not contain *Null* as it is implicitly disallowed.

**Q44:** *Any path of length up to four between Vhagar and Balerion, which is composed only of 'freezes' relationships*

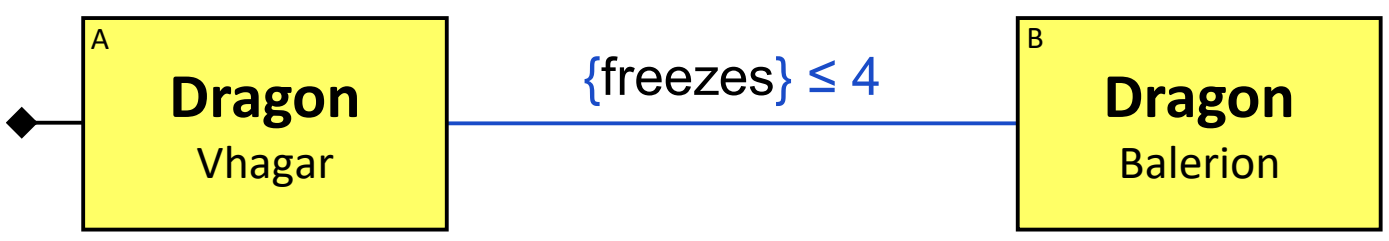

**Q45:** *Any path of length up to four between Vhagar and Balerion composed only of 'freezes' and 'fired at' relationships*

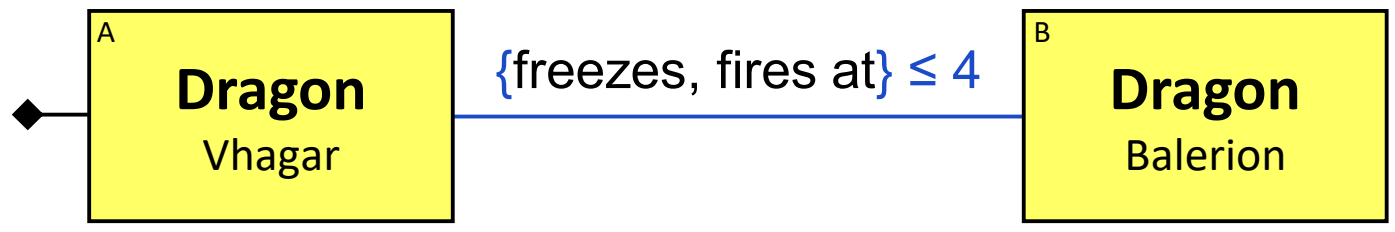

**Q46:** *Any path of length up to four between a dragon owned by Rogar Bolton to a dragon owned by Robin Arryn, which is composed of up to two 'freezes' relationships and only of 'Dragon' entities*

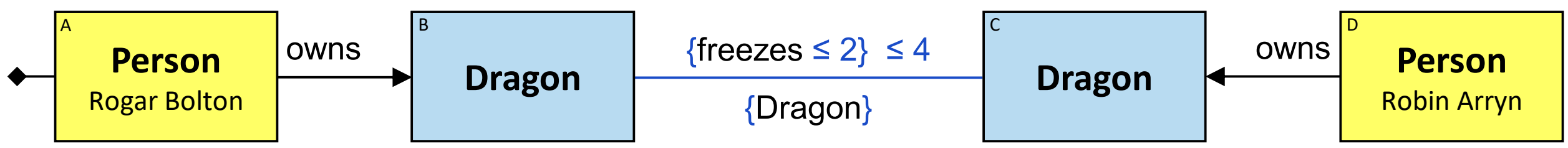

**Q54:** *Any person with paths of length up to three to Rogar Bolton, Robin Arryn, and Arrec Durrandon. The paths composed only of 'friend of' relationships*

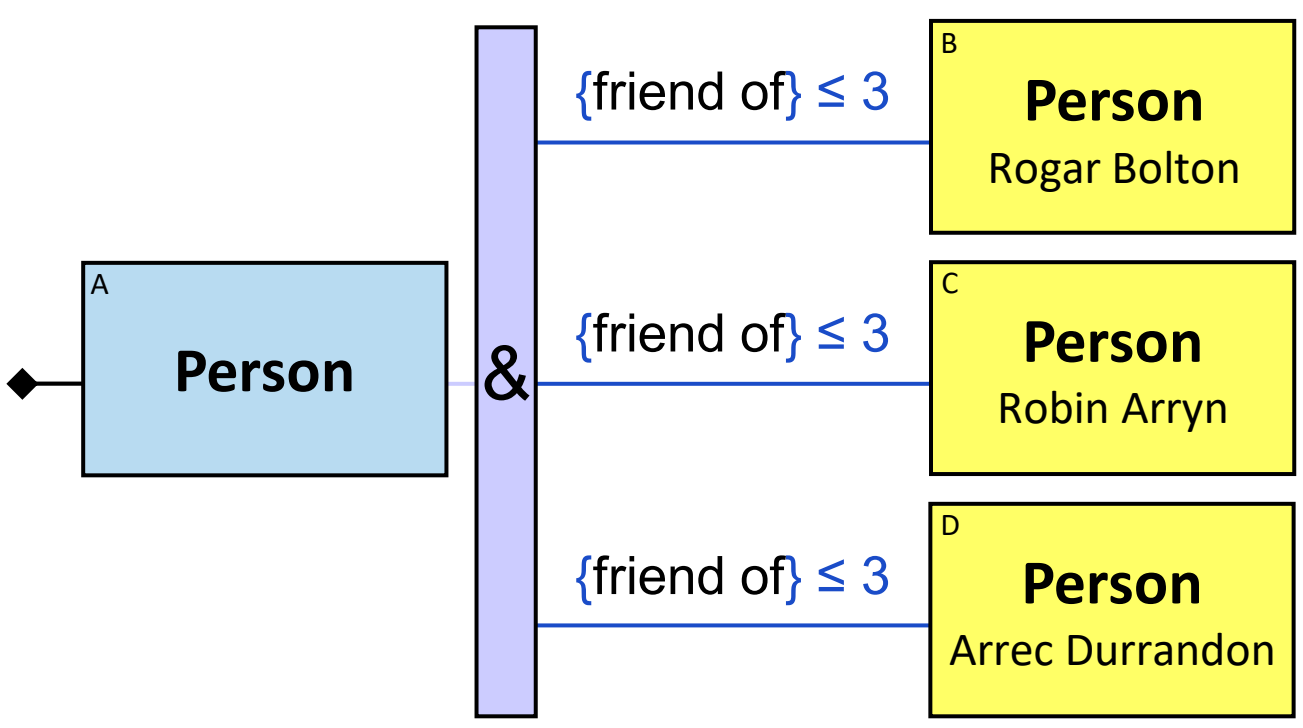

## 22. SHORTEST PATHS

Instead, or in addition to specifying a constraint on the path length – a *shortest path constraint* can limit paths-assignments to the shortest ones that subject to the constraints on the entities/relationships along the path. If, for example, the length of the shortest path that subjects to the constraints is three – only paths of length three are valid assignments.

**Q47:** *All shortest paths between Vhagar and Balerion*

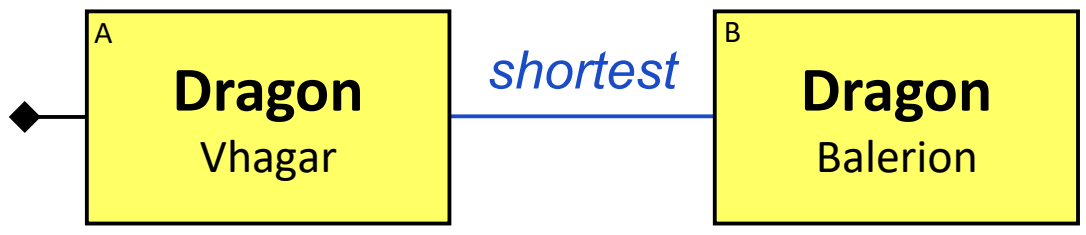

**Q48:** *All shortest paths between Vhagar and Balerion, which are neither composed of 'freezes' relationships nor 'Dragon' entities* (version 1)

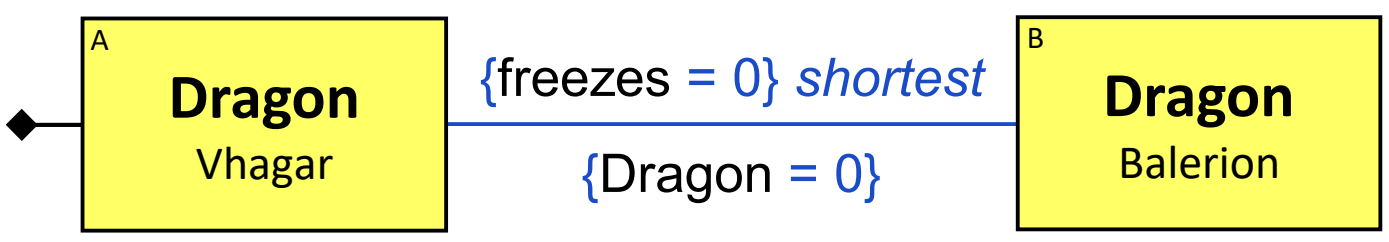

*shortest* may not appear directly right of a negator or a path-negator.

## 23. PATH PATTERNS

An alternative to constraints on the entity-types and the relationship-types along a path are constraints on the patterns which assignments to a path are composed of. A *path pattern* is a pattern that has one entity marked as left-terminal and one entity, possibly the same one, marked as right-terminal.

A *path-assignment* is composed of chained subgraphs; each subgraph is assigned to a path pattern. There is an overlap between assignments to successive path patterns:

- The graph-entity assigned to the left-terminal of the first path pattern of a path is also assigned to the entity preceding the path
- The graph-entity assigned to a right-terminal of a path pattern is also assigned to the left-terminal of the successive path pattern
- The graph-entity assigned to the right-terminal of the last path pattern of a path is also assigned to the entity following the path

A **blue table** below the path defines constraints on the path's path pattern types. The table has two columns:

- A constraint on the number of allowed path pattern instances of this type along the path. The following constraint types can be used: n, < n, ≤ n, > n, ≥ n, $n_1$ .. $n_2$, where the values are positive integer constants.
- A path pattern

The last row defines whether a path-assignment may be composed of other graph-elements ('✔' if yes, '✗' if no).

**Q48:** *All shortest paths between Vhagar and Balerion, which are neither composed of 'freezes' relationships nor 'Dragon' entities* (version 2)

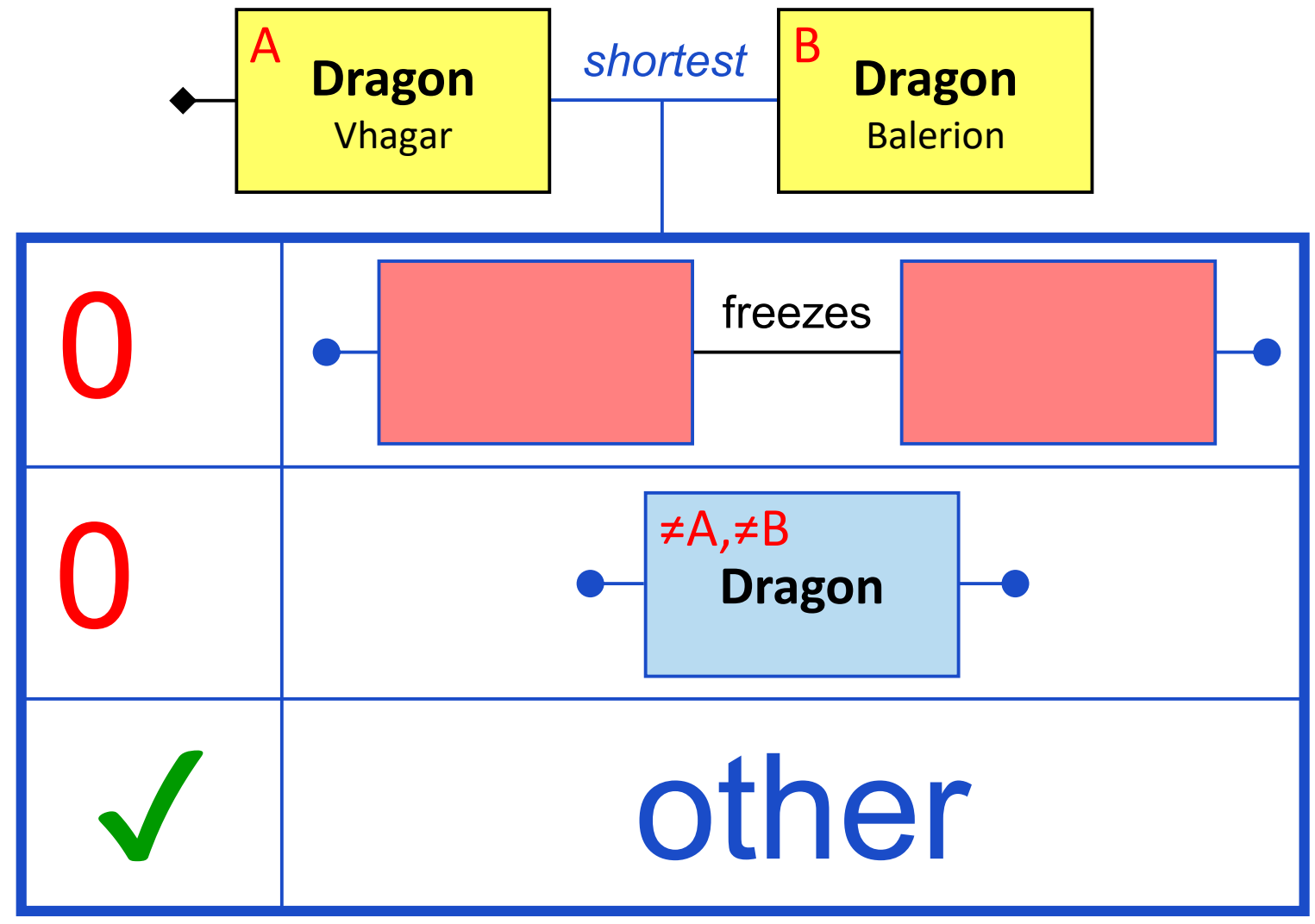

**Q58:** *Any Sarnorian who has a path of length up to six to an Omberian. The path passes through Rogar Bolton*

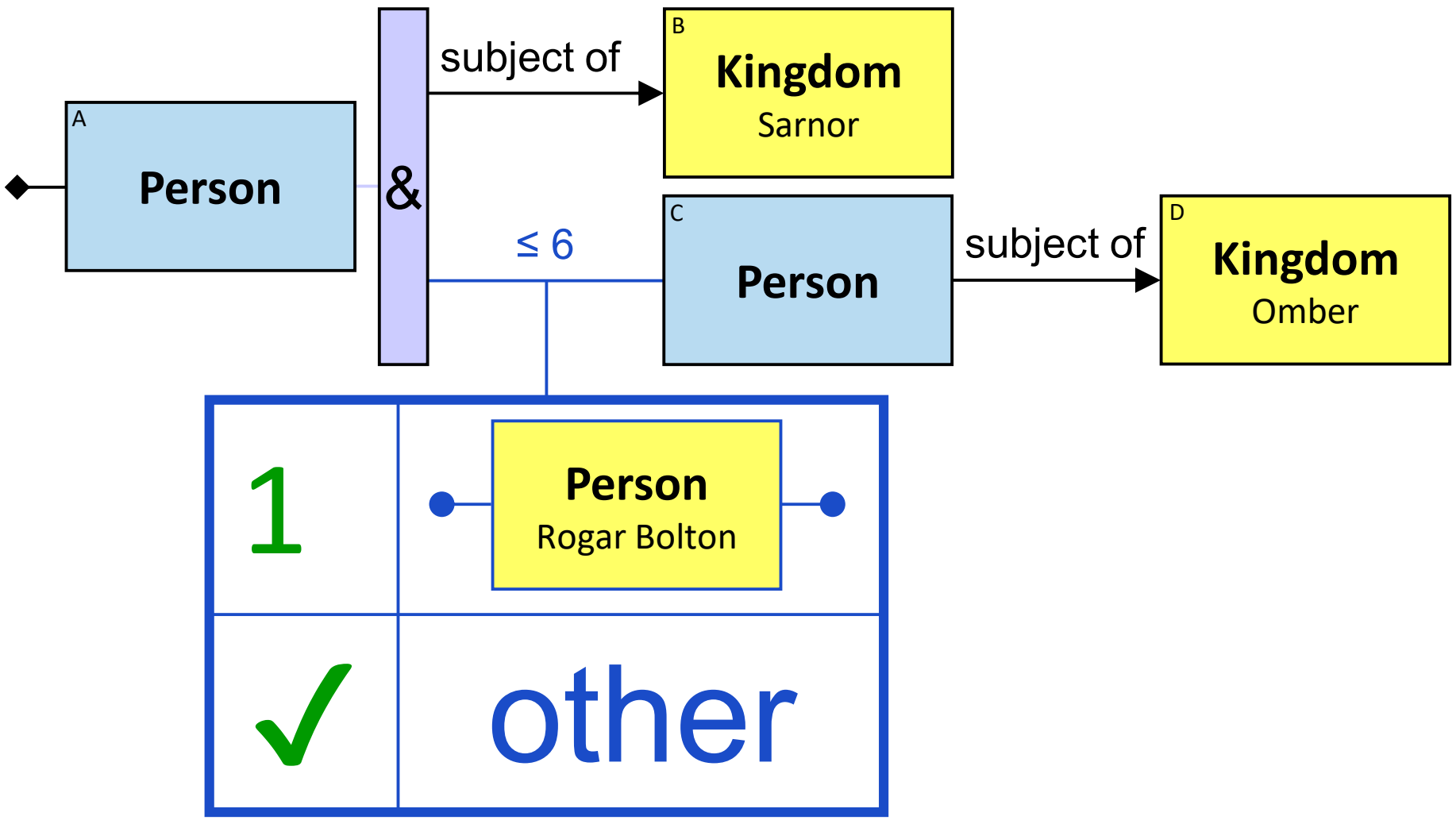

**Q56:** *constraints on path patterns*

In this example, path assignments must be composed of assignments to four path patterns

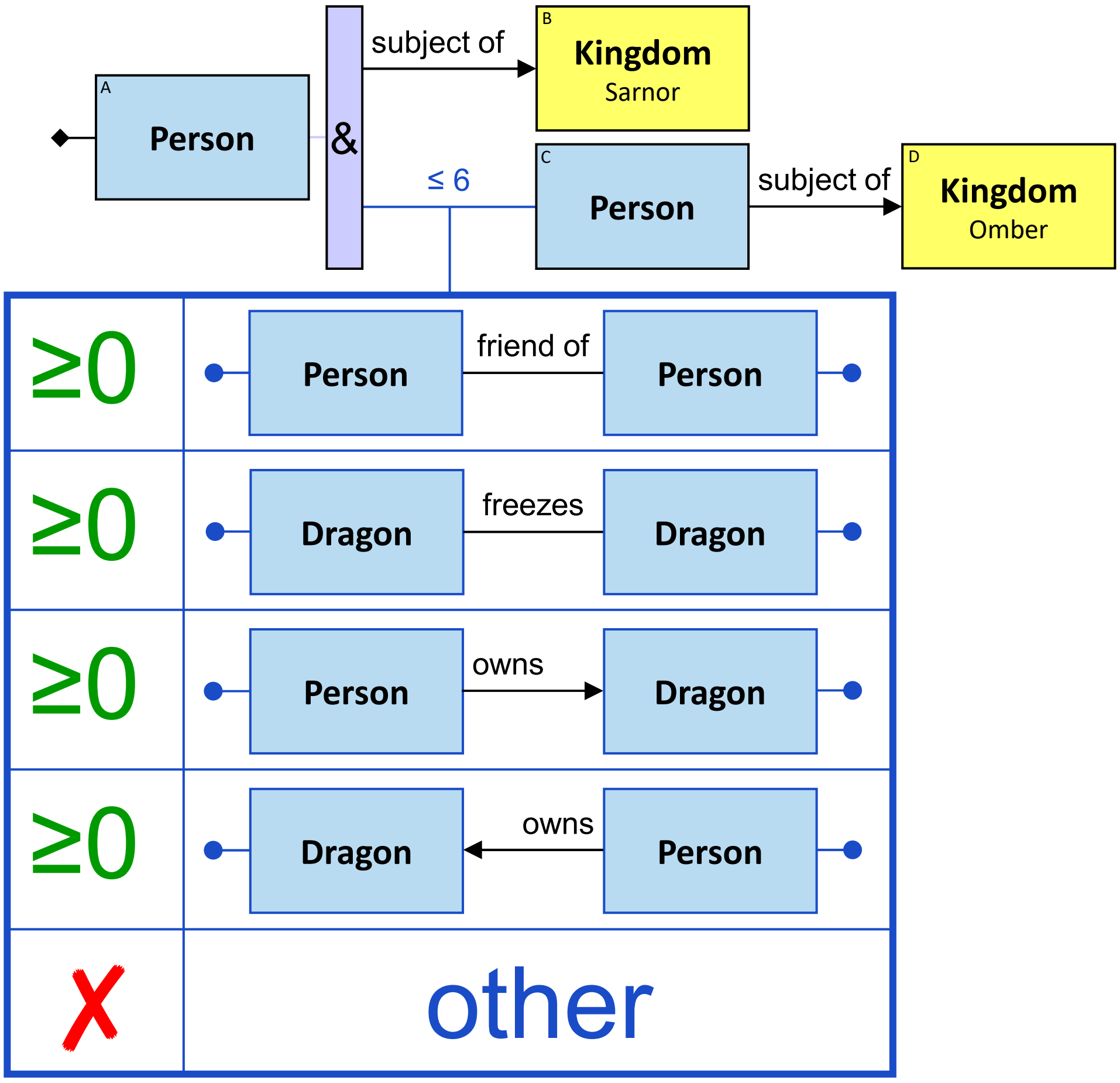

**Q57:** *constraints on path patterns*

- Path assignments must not contain people whose first name starts with 'M' (note that assignments to A and C are not considered part of the path assignment)
- There must be between two and three assignments to any of two path patterns
- The path may be composed of additional graph-elements

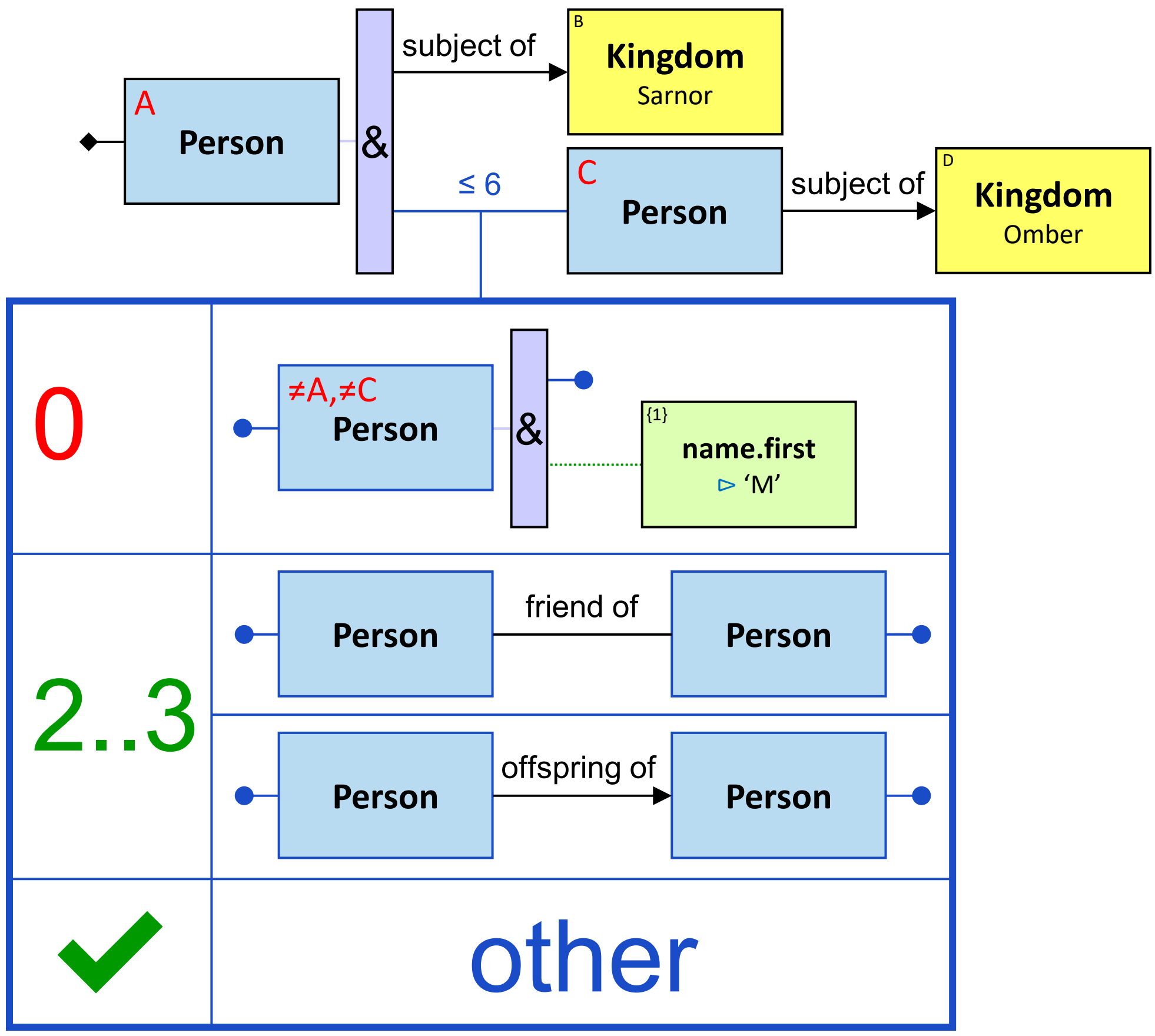

**Q322:** *Any dragon owned by at least five consecutive generations of the same dynasty*

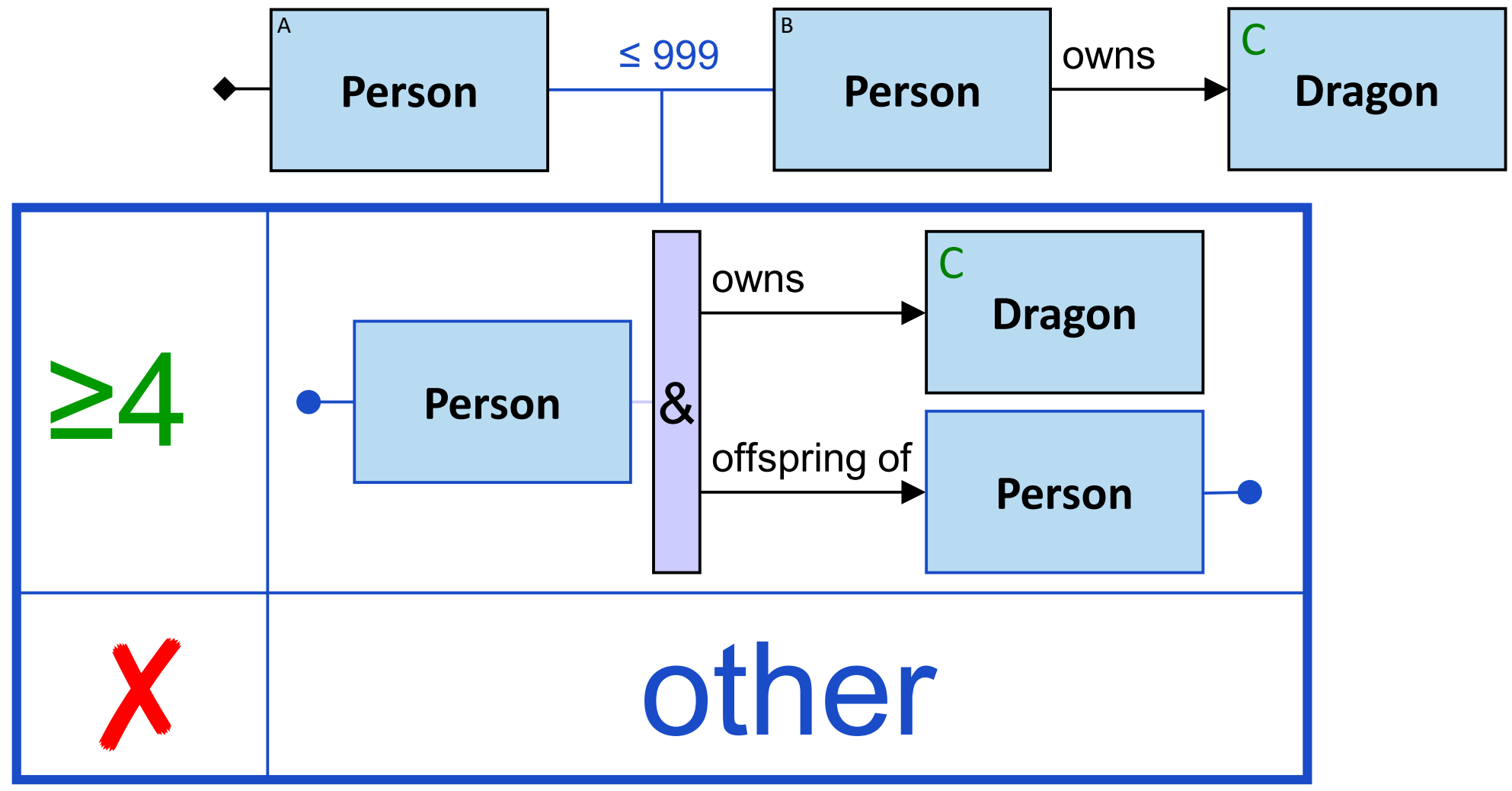

**Q323:** *Any dragon owned by at least five (not necessarily consecutive) generations of the same dynasty*

See also Q290 and Q329.

---

## 24. REFERENCING EXPRESSION-TAGS

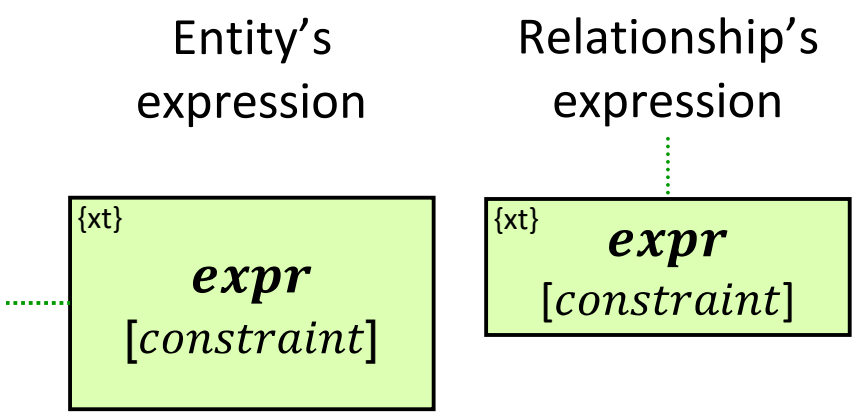

Each green rectangle has an *expression-tag* on its top-left corner, depicted by an **index wrapped in curly brackets** ('{xt}'). An expression-tag represents the result of the evaluation of an entity's expression, relationship's expression, or Cartesian product's expression – for a given assignment.

Expression-tags and aggregation-tags are jointly called *EA-tags*. Each EA-tag has a unique positive integer index. If an EA-tag is referenced at least once – it is depicted in **bold purple**. Otherwise – it is depicted in **black**.

Expression-tags may be referenced:

- in an entity, relationship, or Cartesian product's expression (see Q267v2, Q308, Q349)
- in an entity, relationship, or Cartesian product's expression constraint (see Q108, Q109)
- in a path length constraint

- in an extended aggregator *per* clause (see Q114v2, Q270)
- in an A3 aggregator "aggop ..." clause (see Q116)
- in an extended M1/M2/M3 "all but *k* ..." clause (see Q220, Q262)
- in an M3 aggregator "with min/max ..." clause (see Q130)
- in an A1/A2/A3 aggregator constraint (see Q120, Q255)

**Q108:** *Any person who has the same birth date as Brandon Stark*

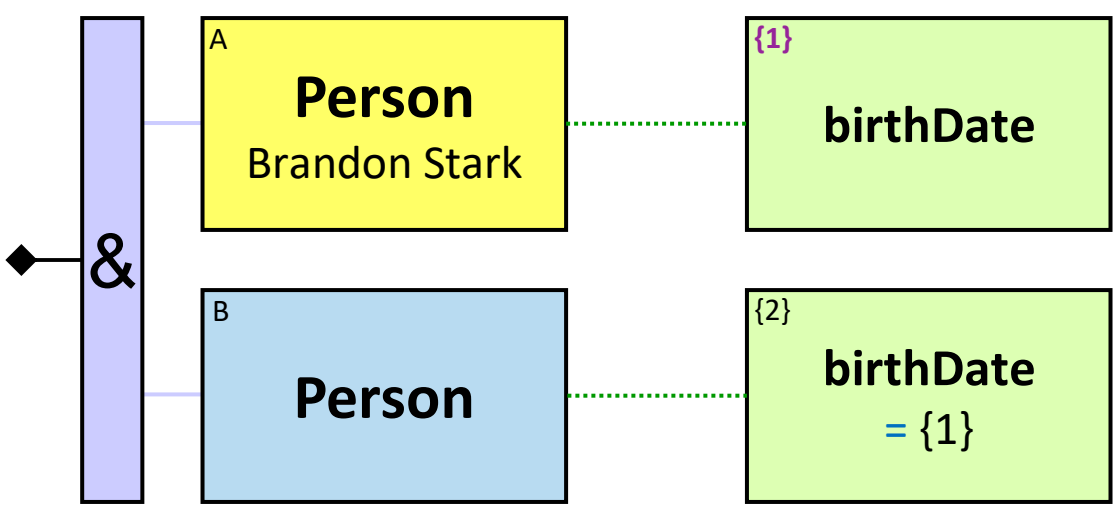

{1} is a property of the only assignment to A. {2} is a property of each unique assignment to B.

For any quantifier except *all* – an EA-tag defined in a branch can only be referenced in that branch.

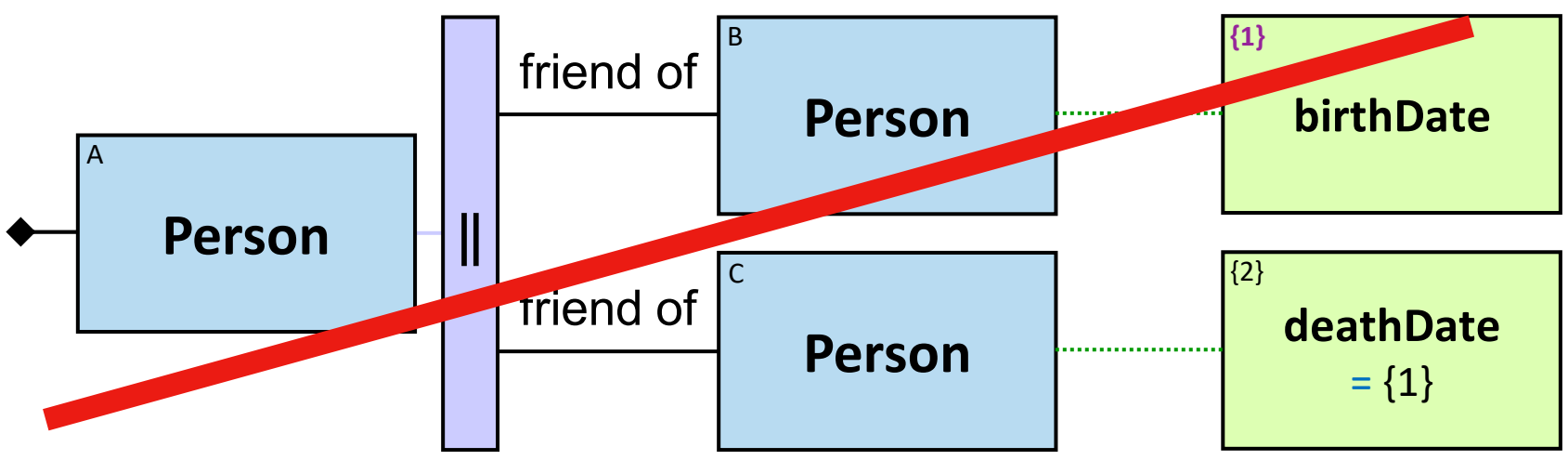

An EA-tag defined in a branch of a horizontal quantifier can only be referenced in that branch.

For each entity/relationship/Cartesian product – if the same expression is used more than once – only one expression-tag will be assigned (see also Q311):

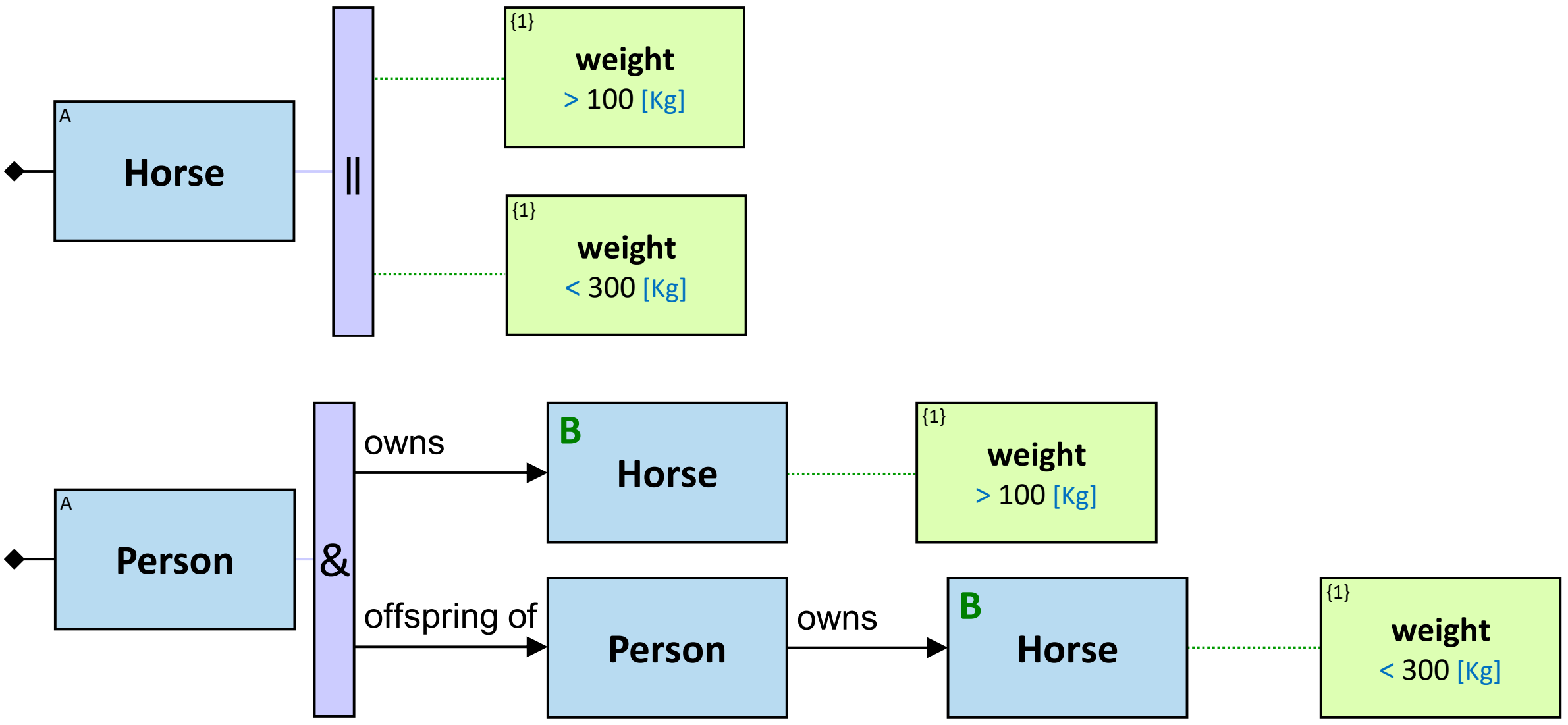

Self or circular references are invalid:

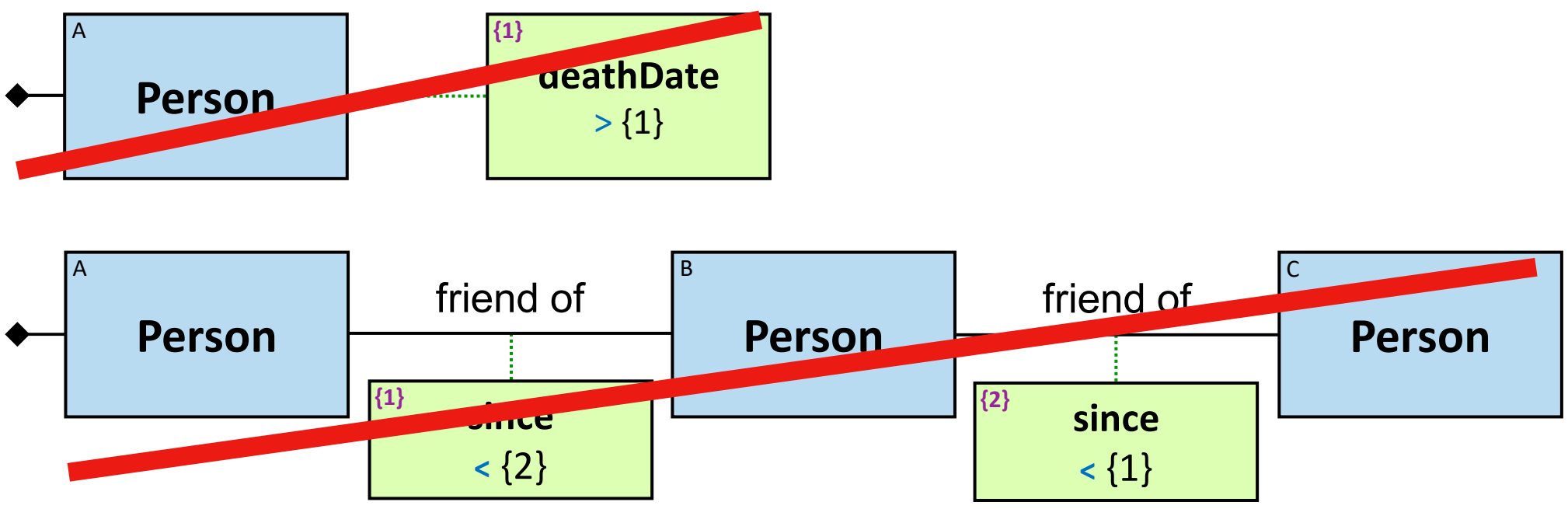

**Q111:** *Any person A who has no friends with whom A shares a birthdate*

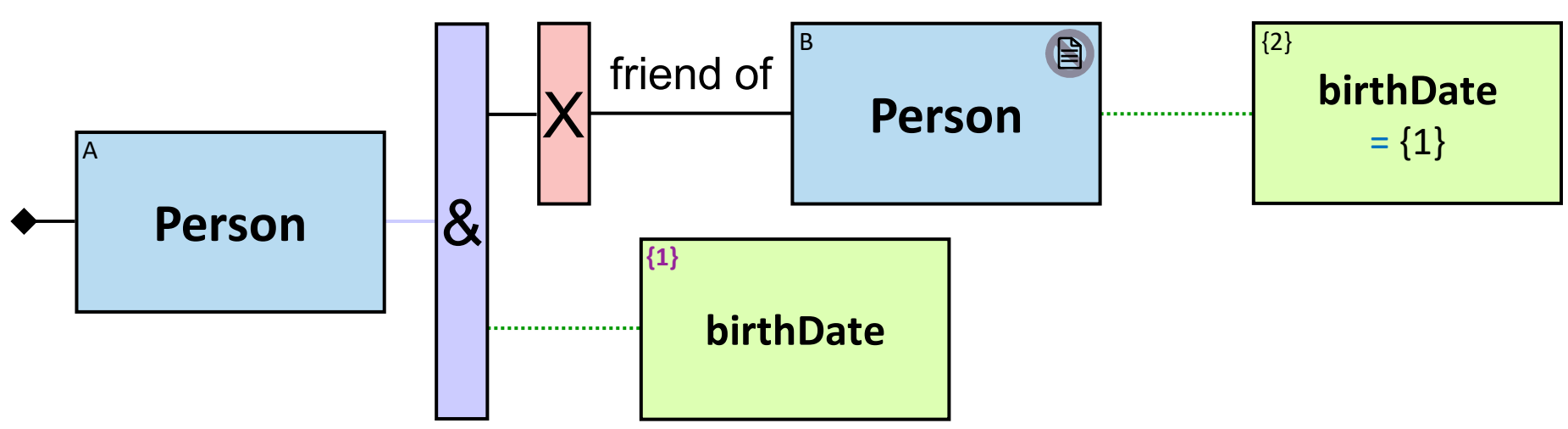

An EA-tag defined right of a negator can only be referenced right of that negator.

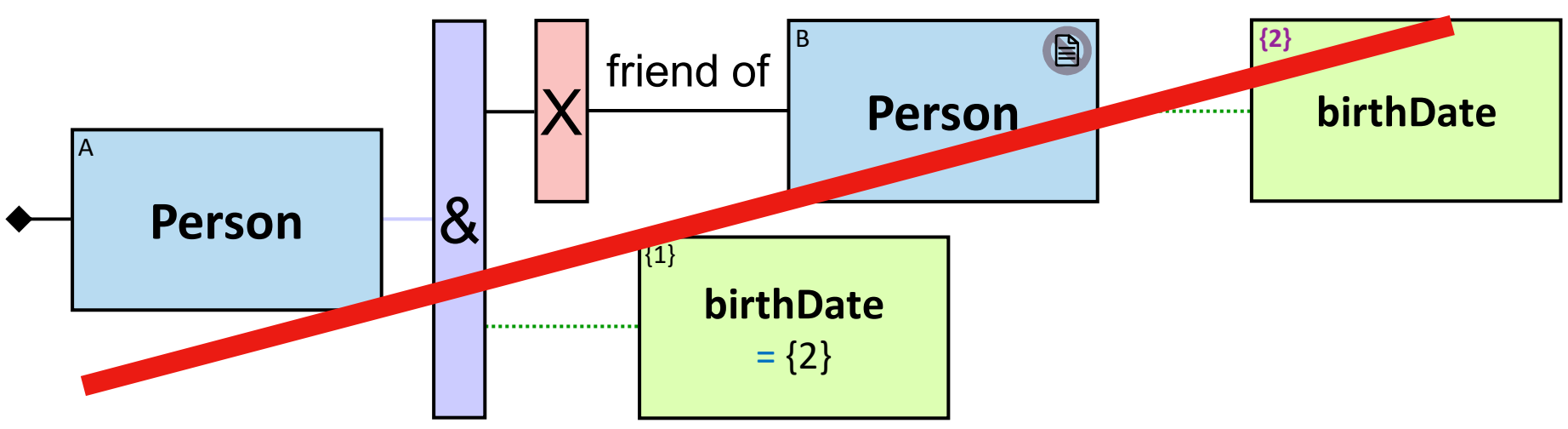

**Q353:** *Any person A that in the day A turned two years old – at least one person was born, but there is no such person A became a friend of since A reached 20*

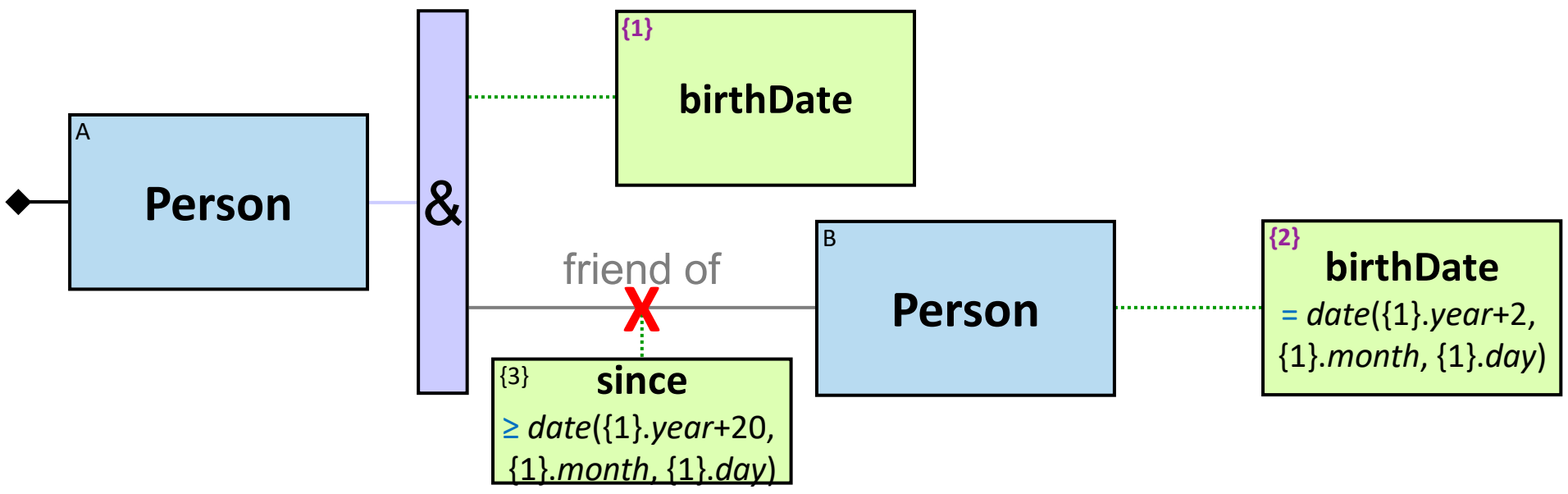

A relationship's expression defined directly right of a relationship-negator can only be referenced in that chain.

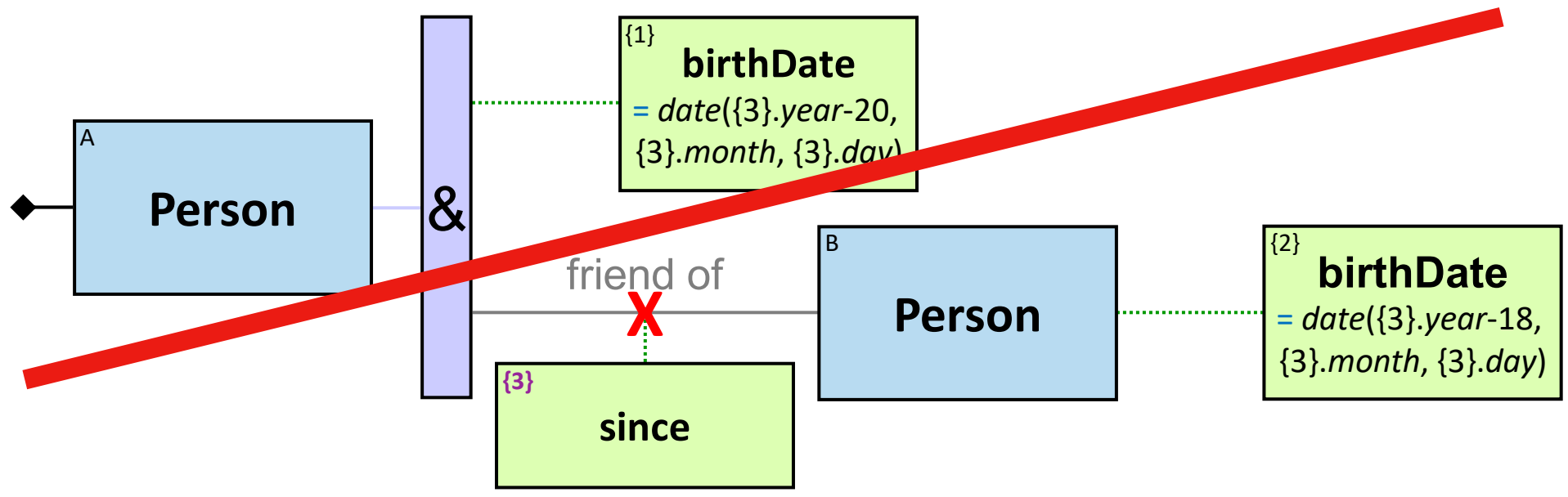

Composite properties and sub-properties are tagged and can be referenced similar to ordinary properties:

**Q109:** *Any person A whose parent owned a horse or a dragon before A's birth* (two versions)

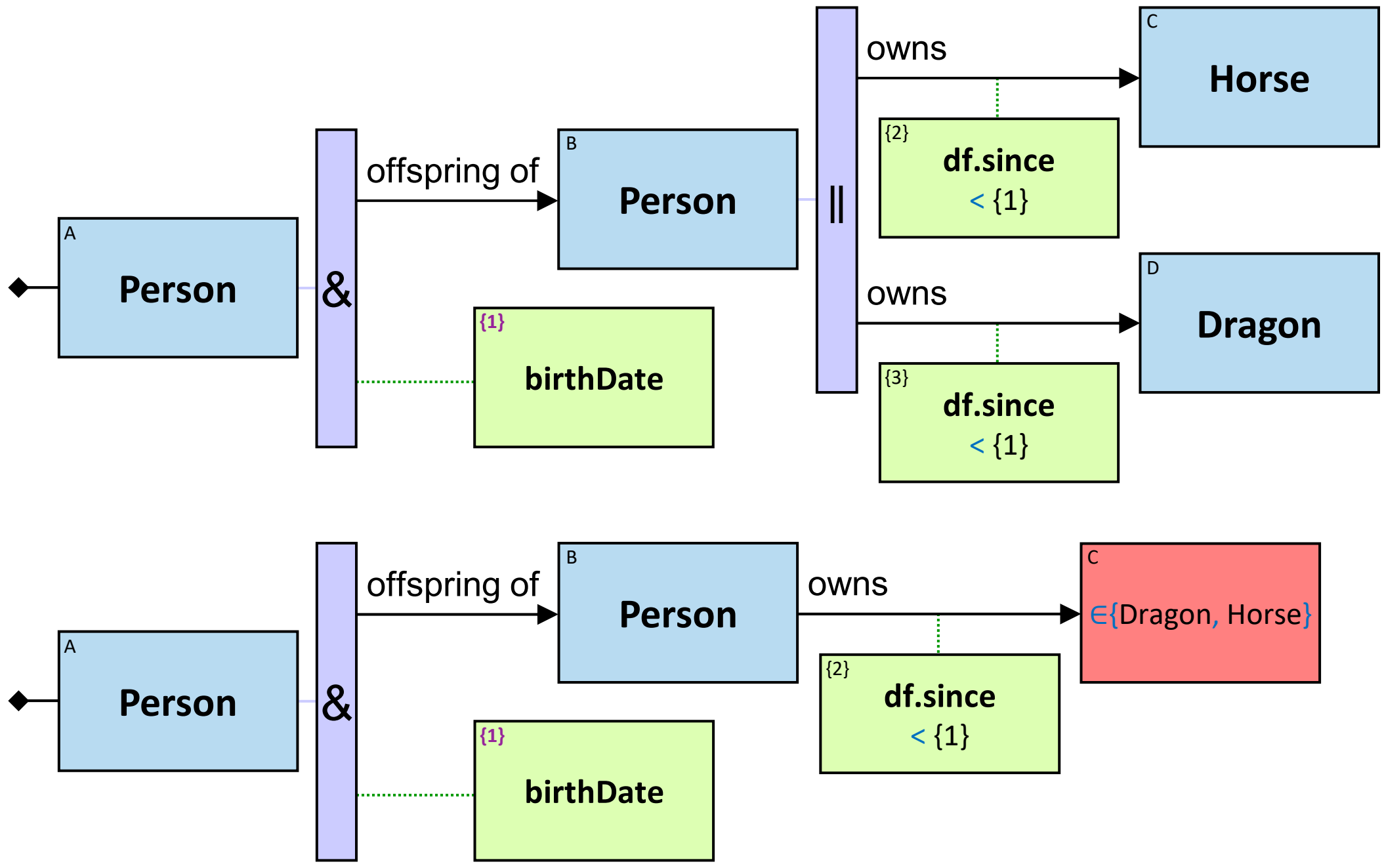

**Q110:** *Any three dragons with cyclic freezes of more than 100 minutes, in chronological order, within six months, and their owners (if any)*

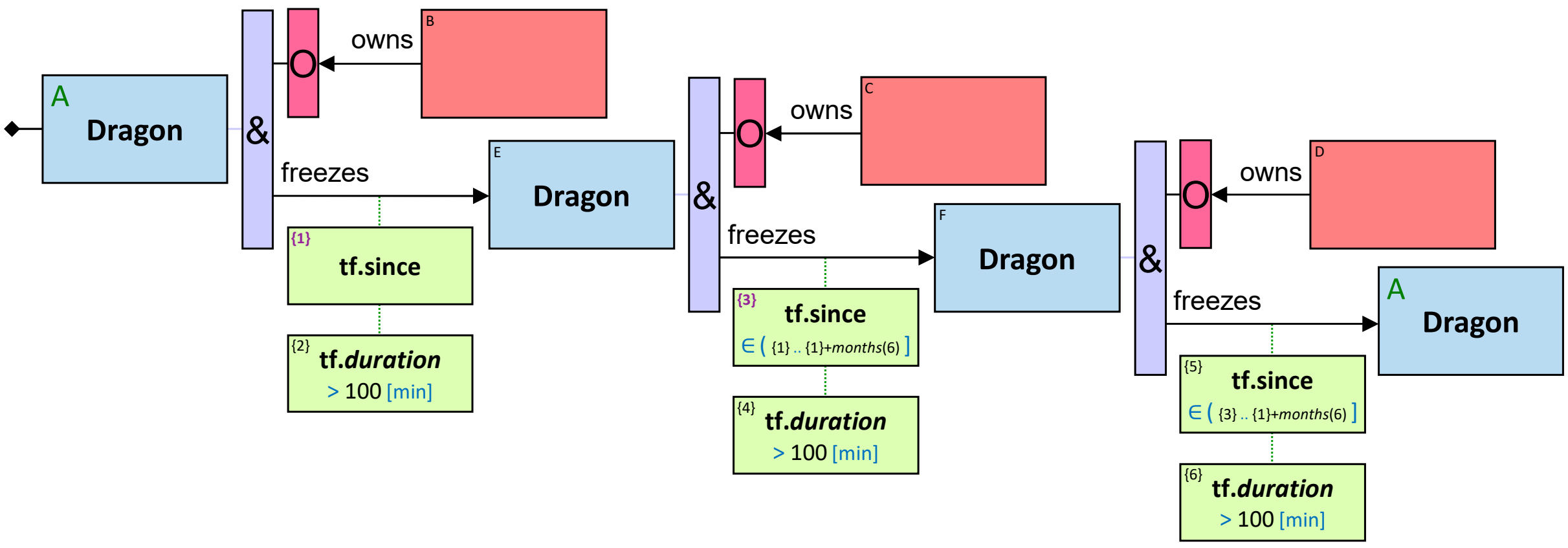

**Q112:** *Any person who owned a horse and a dragon at the same time frames* (three versions)

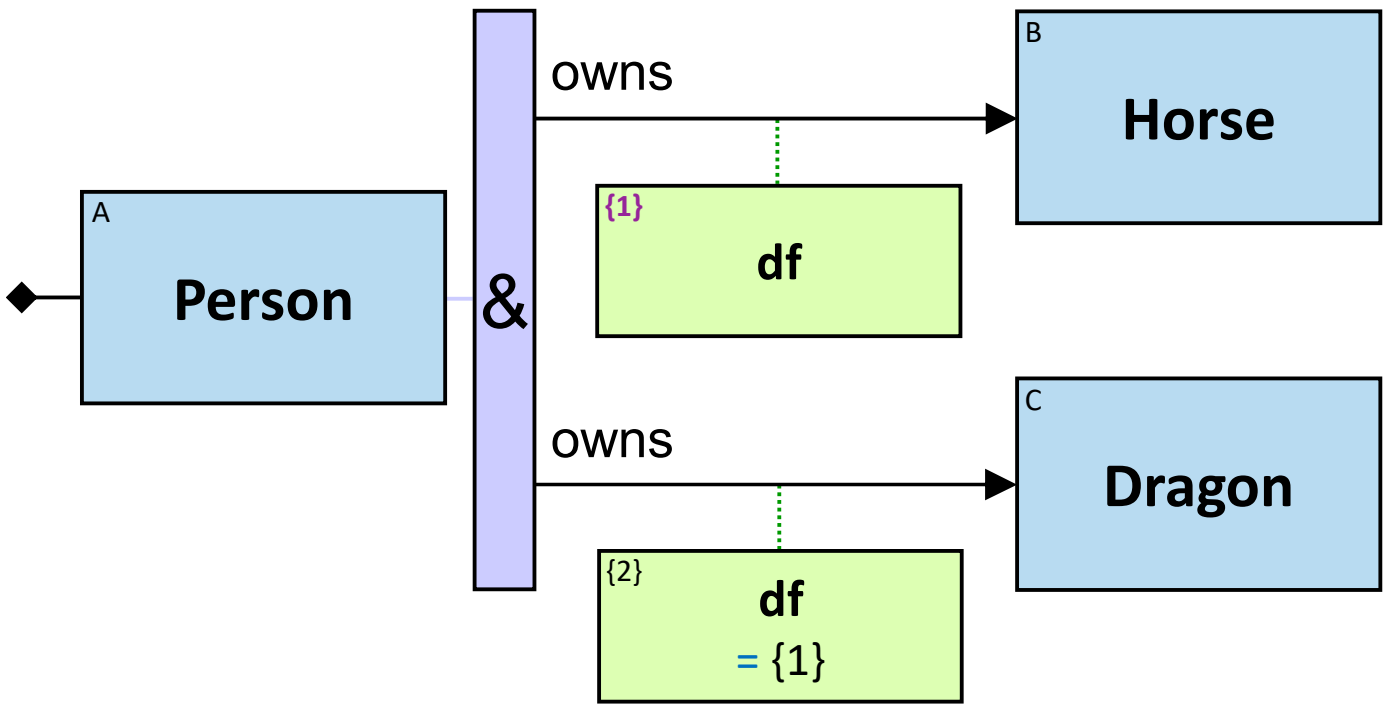

The sub-properties can be compared one-by-one:

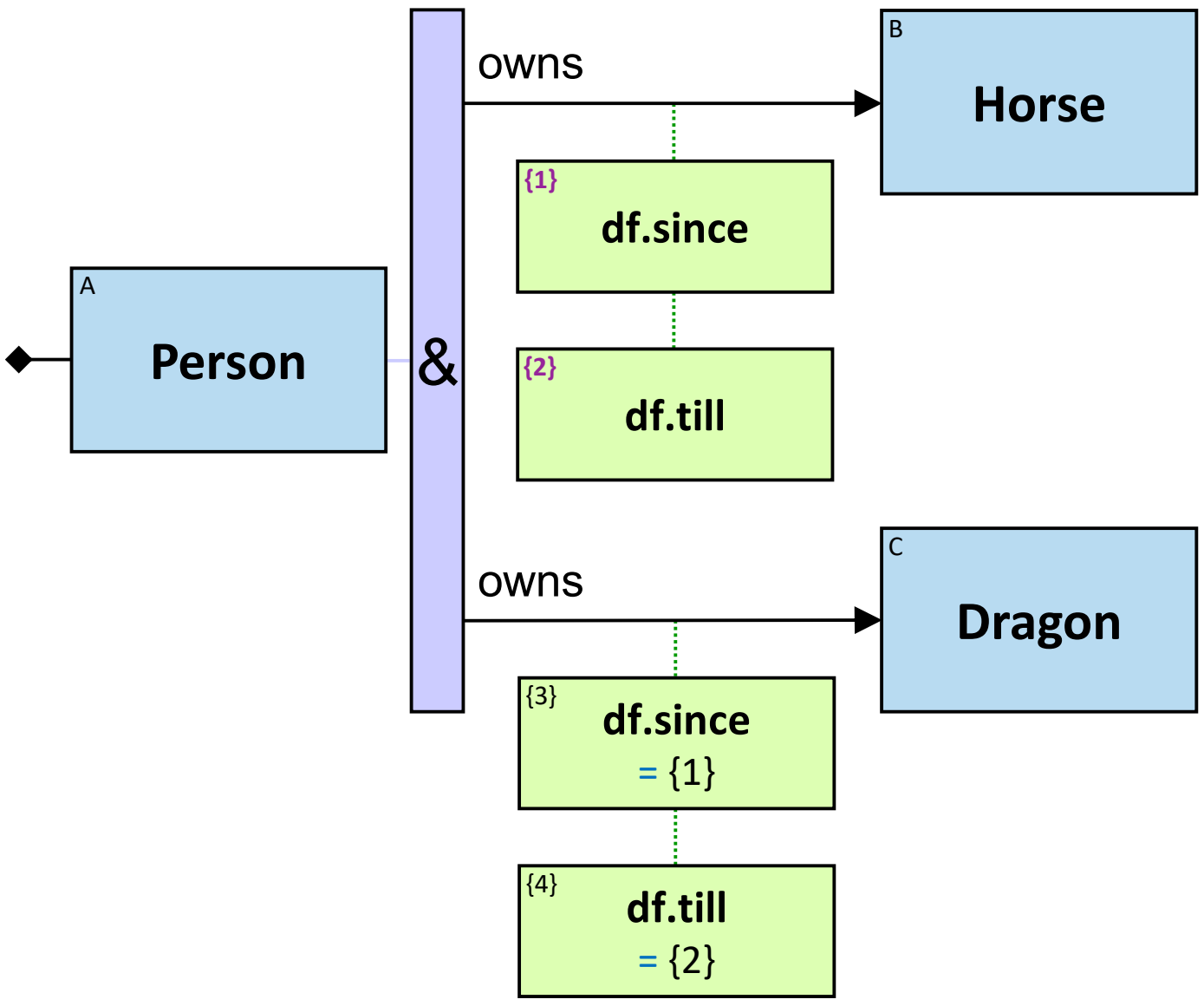

The order of the expressions along a chain does not matter. The following representation is valid, as well:

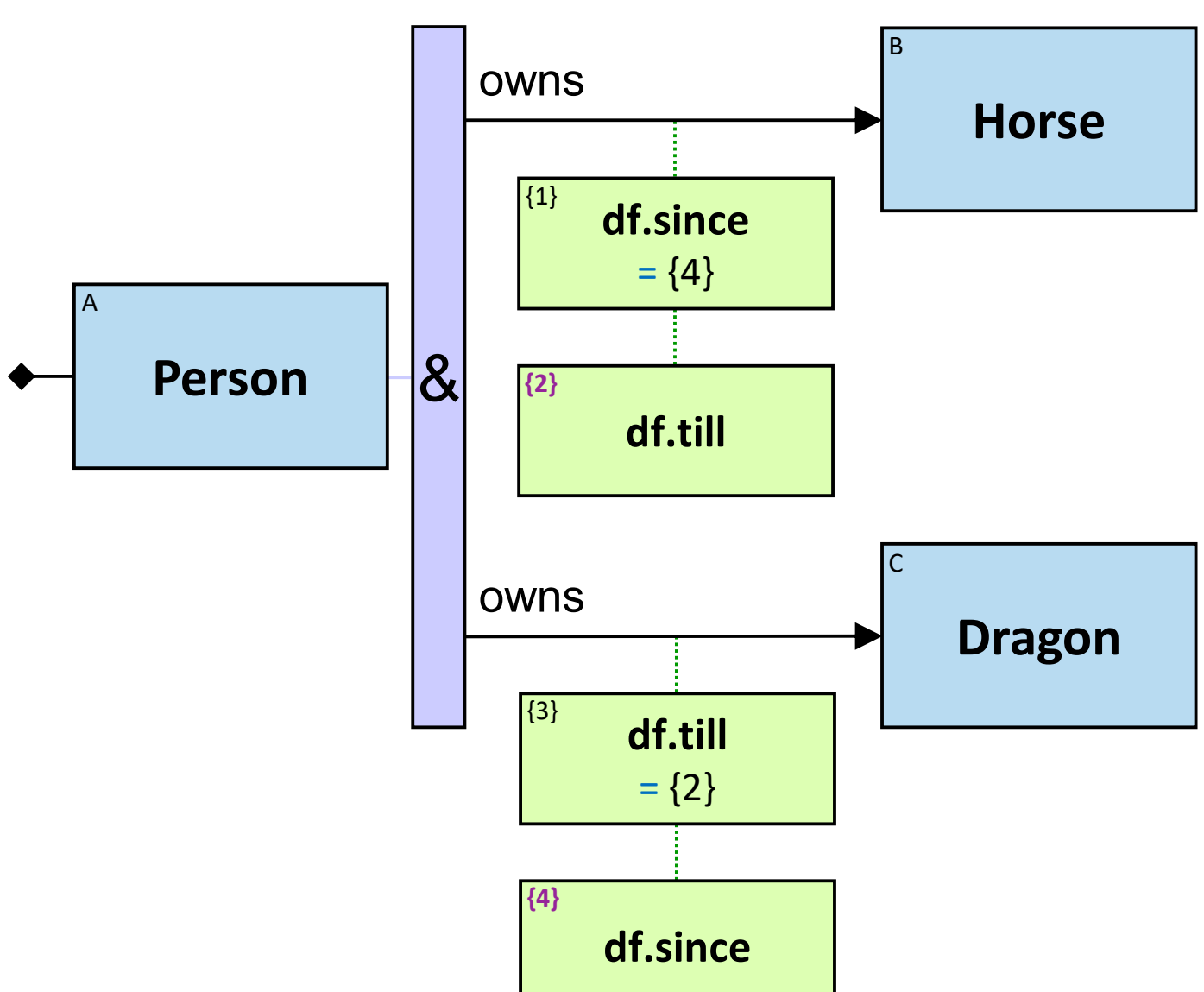

**Q266:** *Any person A who has the same name (first and last) as A's parent* (two versions)

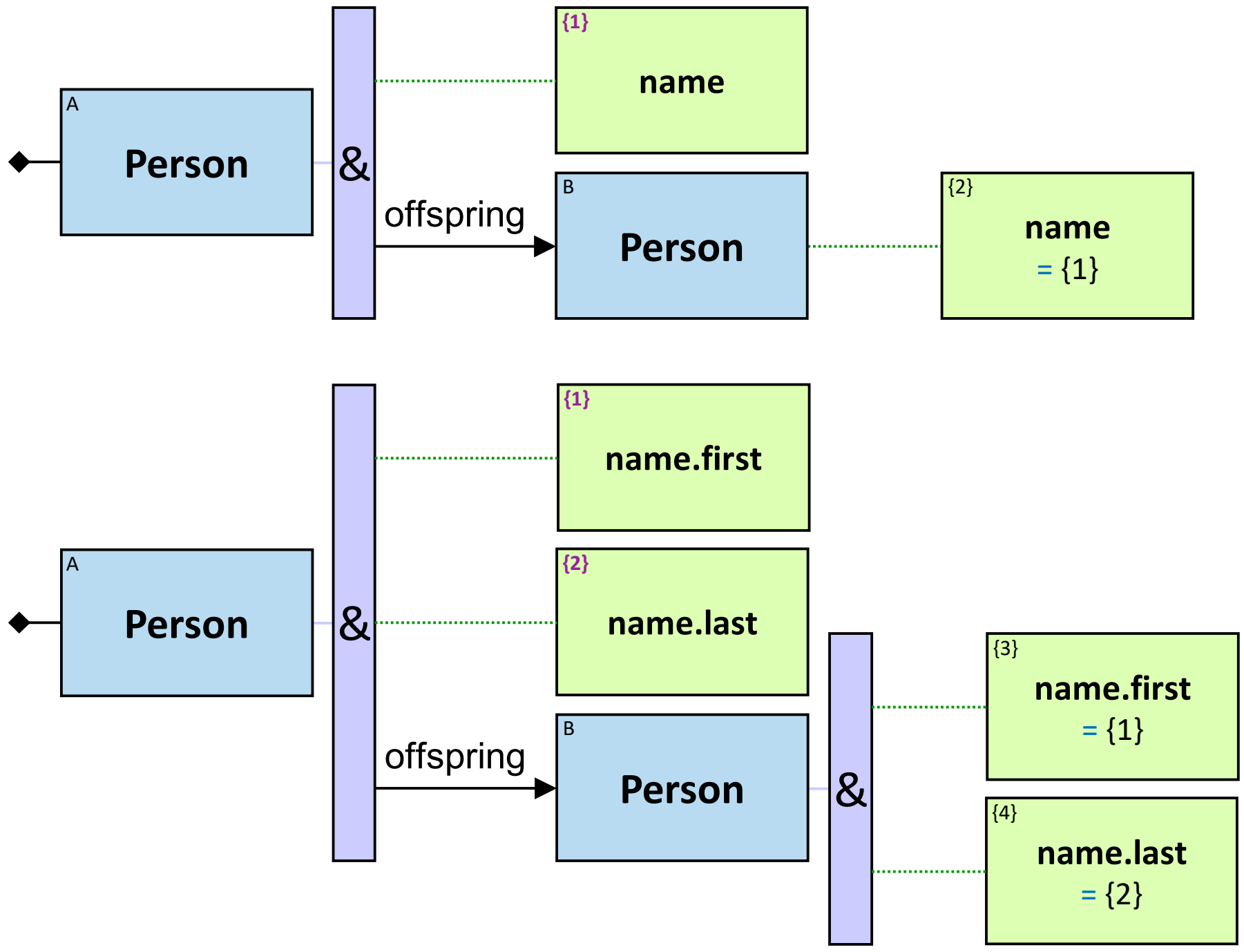

**Q267:** *Any person who was a member of two guilds at overlapping timeframes* (two versions)

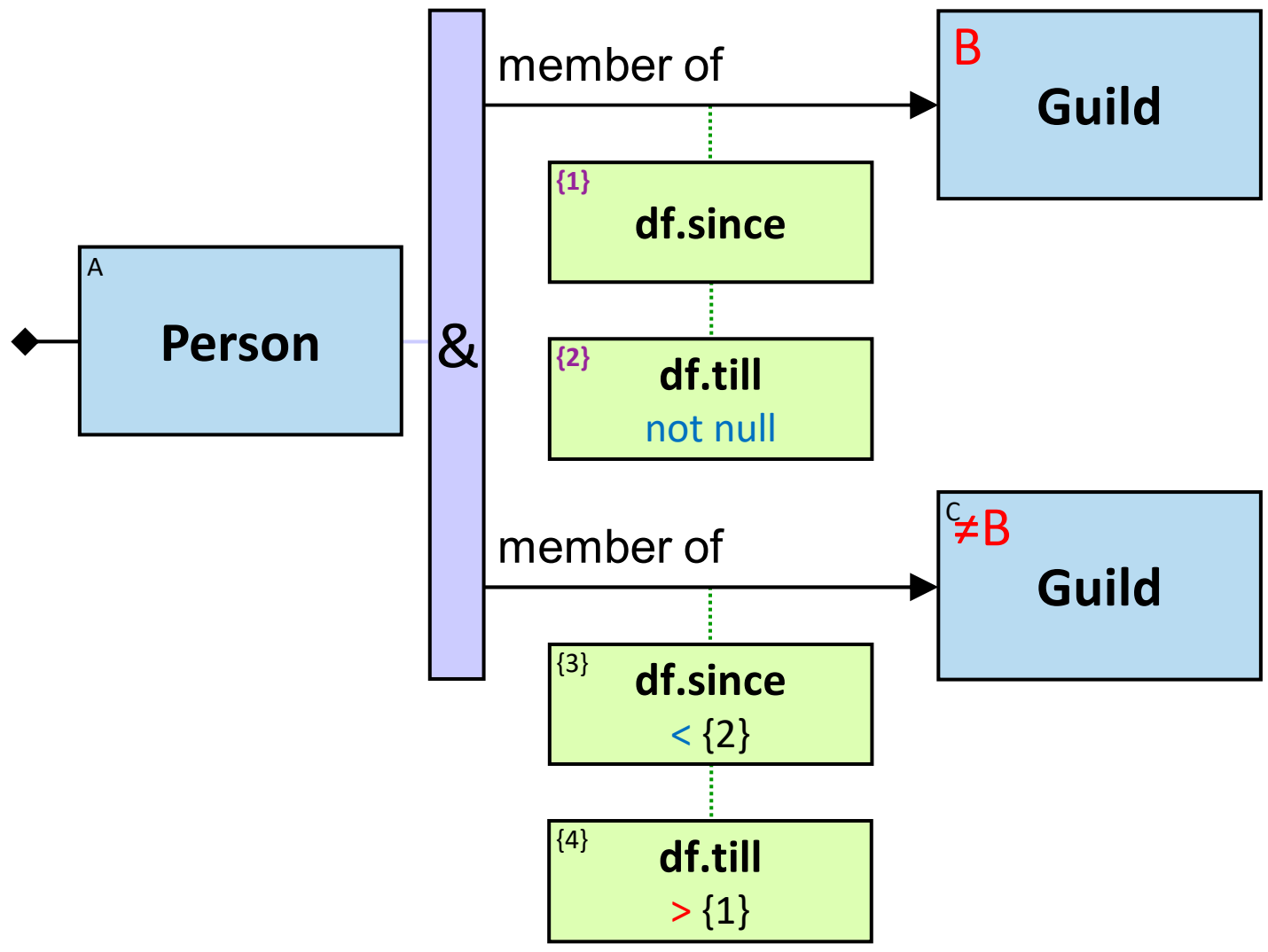

(Assuming that at least one of the *df.till* values is not *null*. See also note under Q11). Note the red constraint operator.

Here is another option, but with a different meaning, as the *overlap* function returns *null* if either *since* or *till* of *df* or of {1} is *null*:

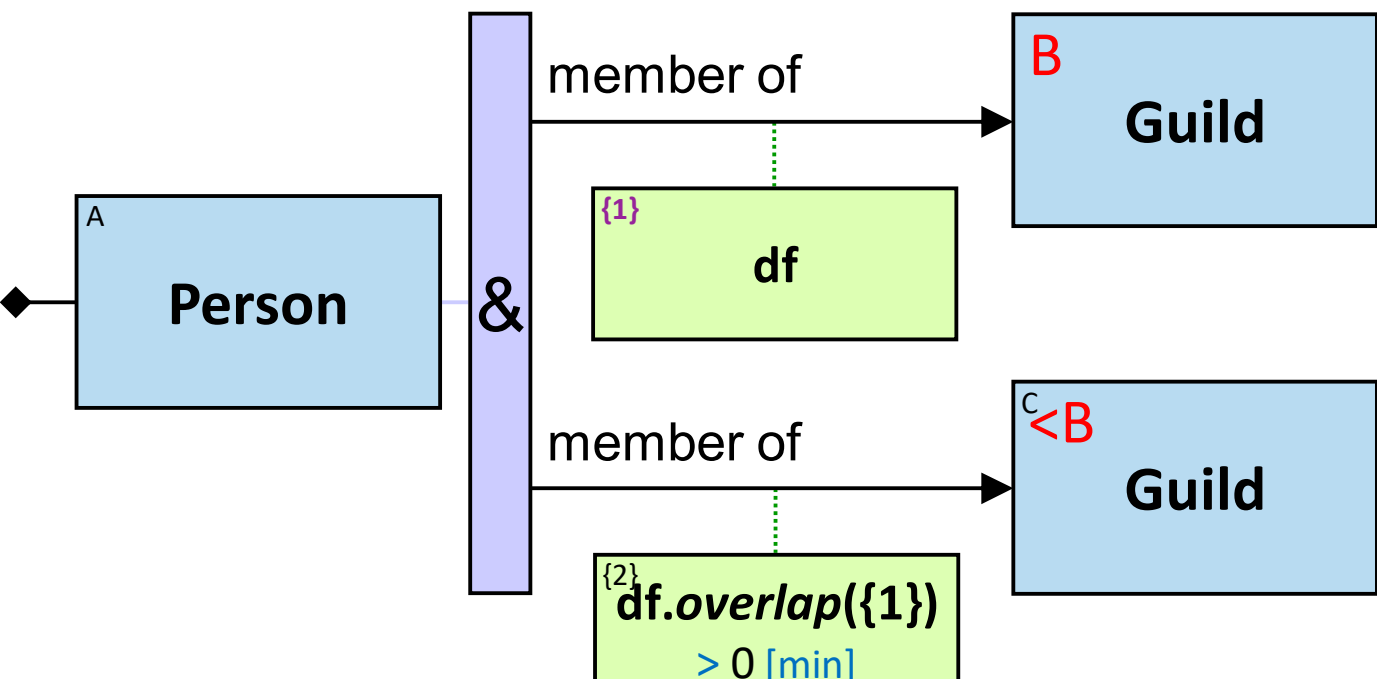

## 25. AGGREGATORS

One can use an order constraint to represent patterns such as *Any person who owns at least two horses*, but it is quite cumbersome for patterns such as *Any person who owns at least 20 horses*. Patterns such as *Any dragon that froze dragons more than ten times* or *Any dragon that froze dragons for a cumulative duration longer than 100 minutes* cannot be expressed without aggregators.

**An orange rectangle** represents an *aggregator*. It is composed of two parts:

- The upper part (dark orange) defines how assignments are split into disjoint sets (*per* clause)
- The lower part (light orange) defines an aggregate expression that is evaluated per each set and may contain a constraint on the result of the evaluation of the aggregate expression

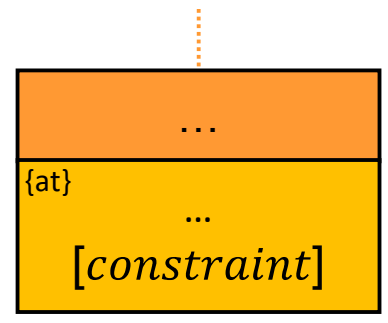

Here are three examples:

First example: *Any person who owns at least 20 horses* (Q59)

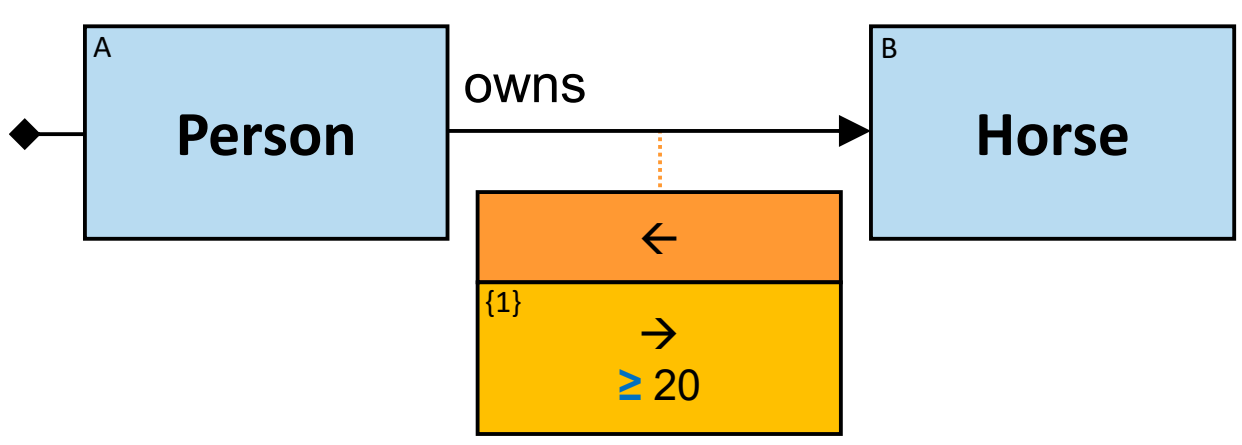

All assignments to the pattern, ignoring the aggregator, are found and split into sets. In each set, all the assignments to A are identical. Only sets with more than two assignments to B are valid pattern assignments.

Second example: *Any dragon that froze dragons more than ten times* (Q71)

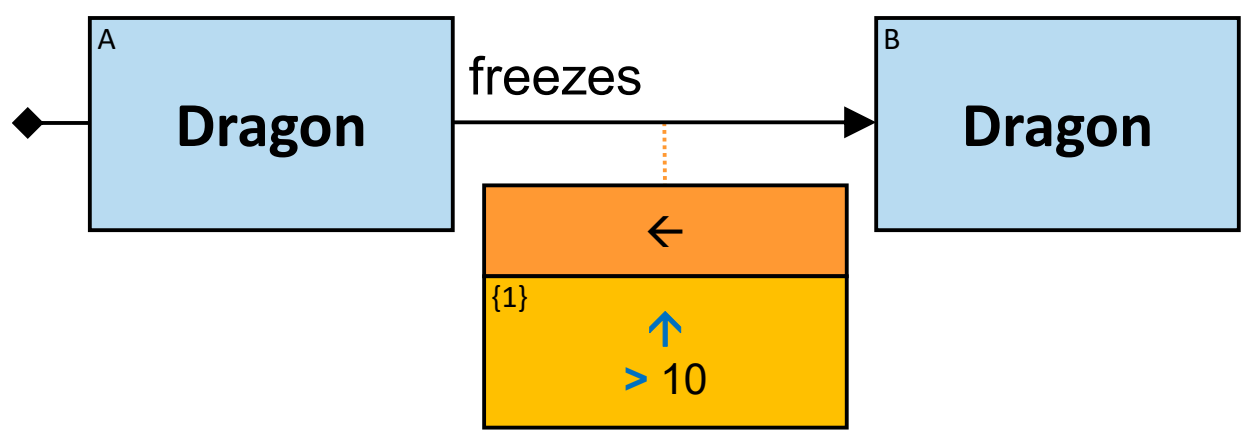

All assignments to the pattern, ignoring the aggregator, are found and split into sets. In each set, all the assignments to A are identical. Only sets with more than ten assignments to the relationship are valid pattern assignments.

Third example: *Any pair of dragons (A, B) where A froze B for a cumulative duration longer than 100 minutes* (Q86)

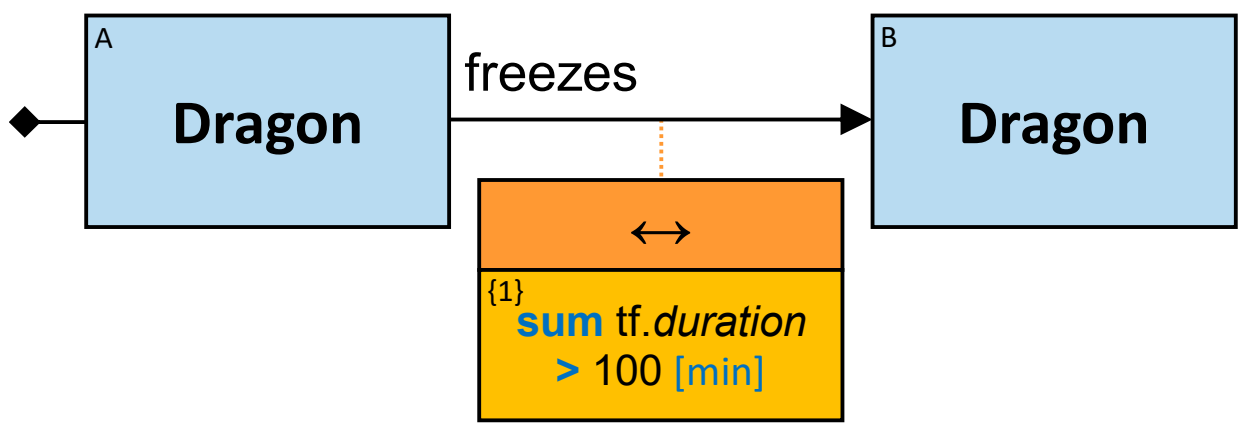

All assignments to the pattern, ignoring the aggregator, are found and split into sets. In each set, all the assignments to both A and B are identical. Only sets where the sum of the duration of all *freezes* relationships is greater than 100 minutes are valid pattern assignments.

**Evaluating an aggregator:**

- Step 1: All assignments to the pattern excluding this aggregator and without constraints based on this aggregation-tag ('at') are found
  - In the first example: all (Person, Horse) pairs (A, B) where A owns B
  - In the second and third examples: all dragons pairs (A, B) where A froze B
- Step 2: The assignments are split into disjoint sets. In each set – the assignment to the entities listed in the *per* clause is identical. Each set is to be aggregated separately.
  - In the first and second examples: In each set – all assignments to A are identical
  - In the third example: In each set – all assignments to both A and B are identical
- Step 3: Per each set of assignments – *at* is evaluated
  - In the first example: per person A: *at* = number of horses A owns
  - In the second example: per dragon A: *at* = cumulative number of its *freezes* relationships to all dragons
  - In the third example: per dragon pair (A, B): *at* = sum of the duration of all *freezes* relationships from A to B
- Step 4: Per each set of assignments: the aggregation constraint (if given) and any other constraint that is based on this aggregation-tag are evaluated
  - In the first example: the aggregation constraint is *at* ≥ 20
  - In the second example: the aggregation constraint is *at* > 10
  - In the third example: the aggregation constraint is *at* > 100 [min]
- Step 5: Each set of assignments for which the constraint is satisfied – is a valid assignment to the pattern

Next, we will define three aggregator types: A1, A2, and A3. Steps 1, 2, 4, and 5 are identical for all types. Step 3 is different, as sets are aggregated in different ways.

Let ***S*** denote the set of all assignments to the pattern excluding this aggregator (subject to aggregators evaluation order) (that is step 1 above).

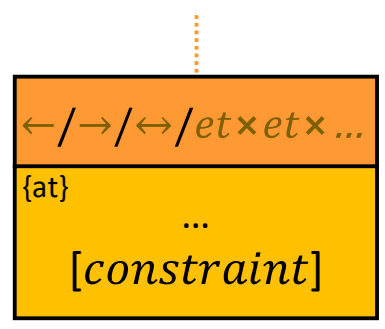

**Upper part**

Let ***T*** denote a Cartesian product of one or more entity-tags.

For all aggregator types (A1, A2, A3, M1, M2, M3, and R1), the *per* clause has one of the following formats:

- '*et1* × *et2* × ...':
  *T* = *et1* × *et2* × ...
- '←':
  *T* = entity-tag directly left of the aggregator
- '→':
  *T* = entity-tag directly right of the aggregator not wrapped with a negator nor preceded by a *None* quantifier (valid only when there is exactly one such entity-tag) (A1: see Q246)
- '↔':
  *T* = *et1* × *et2* where *et1* is directly left of the aggregator, and there is a single entity-tag directly right of the aggregator not wrapped with a negator nor preceded by a *None* quantifier – *et2* (A2: see Q75)

Wherever possible, the notations '←', '→' and '↔' are used instead of entity-tags.

Next, let ***TA*** denote a list of all unique assignments to *T* in *S*. *TA[n]* is the $n^{th}$ unique assignment to *T*.

Let ***S(n)*** denote a subset of *S*: the set of all assignments composed of *TA[n]* (that is step 2 above).

The upper part is optional. When there is no upper part – *S* is not split.

**Lower part**

The lower part contains an aggregation-tag, an aggregate expression, and optionally a constraint on the result of the evaluation of the aggregate expression.

The *aggregation-tag* is on the top-left corner and is depicted by an **index wrapped in curly brackets** ('{at}'). An aggregation-tag represents the result of the evaluation of the aggregate expression for a given set of assignments.

For each entity/relationship/Cartesian product – if the same expression/aggregation is used more than once – only one EA-tag will be assigned (see Q64, Q73v2, Q82v2).

*at* is evaluated per each *TA[n]*. In the first example above, {1} is evaluated per each unique assignment to A in *S*. *at* is a *calculated property* of *TA[n]*. When there is no upper part – *at* is a global property (see {1} in Q82v3, {1} in Q372, {2} in Q356).

Aggregation-tags may be referenced:

- in an entity, relationship, or Cartesian product's expression (see Q315, Q317)

- in an entity, relationship, or Cartesian product's expression constraint (see Q337, Q338v2)
- in a path length constraint
- in an extended aggregator *per* clause (see Q211)
- in an A3 aggregator "[all but] *k* ..." clause (see Q129)
- in an M3 aggregator "with min/max ..." clause (see Q91)
- in an extended M1/M2/M3 aggregator "[all but] *k* ..." clause (see Q212)
- in an A1/A2/A3 aggregator constraint (see Q125, Q127, Q332v2)

---

## 26. A1 AGGREGATOR

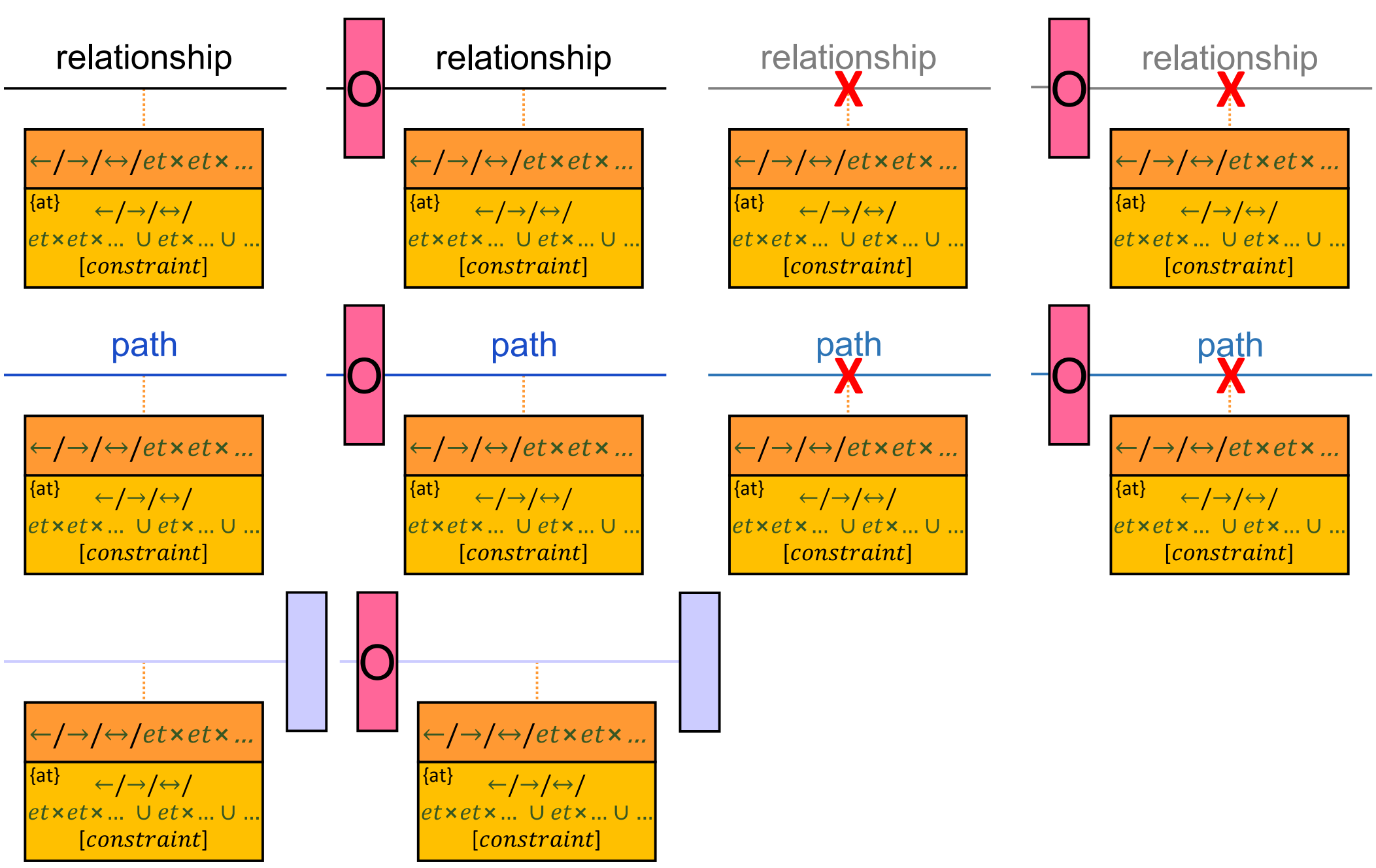

Let ***B*** denote a list of Cartesian products of entity-tags. *B* must be nonempty and must not be composed of entity-tags composing *T*.

The lower part of an A1 aggregator contains the following elements:

- One of the following:
  - '$et11 \times et12 \times \ldots \cup et21 \times et22 \times \ldots \cup \ldots$' (all products have the same arity):
    $B = [et11 \times et12 \times \ldots, et21 \times et22 \times \ldots, \ldots]$ ('∪' – see Q295)
  - '←':
    $\mathrm{card}(B) = 1$; $B[1] = [et]$ where *et* is directly left of the aggregator (see Q249, Q250)
  - '→':
    $B = [et1, et2, \ldots]$ where *et1*, *et2*, ... are directly right of the aggregator and not wrapped with a negator nor preceded by a *None* quantifier (equivalent to '$et1 \cup et2 \cup \ldots$') (see Q294, Q175, Q176 where $\mathrm{card}(B)>1$)
  - '↔':
    $B = [et \times et1, et \times et2, \ldots]$ where *et* is directly left of the aggregator and *et1*, *et2*, ... are directly right of the aggregator and not wrapped with a negator nor preceded by a *None* quantifier (equivalent to '$et \times et1 \cup et \times et2 \cup \ldots$')

  Let ***BA(n, o)*** denote the list of all assignments to *B[o]* in *S(n)*

- **aggregation-tag**: $at(n) = | BA(n, 1) \cup BA(n, 2) \cup \ldots |$

We are using *card(union(all assignments to all elements in B))* instead of *sum(card(assignment to one element in B))* since two elements in *B* may have the same assignment (see Q175), and we are counting *distinct* assignments to all elements in *B* per *n*.

When an optional part has no assignments, A1 aggregation-tags defined right of the 'O' are evaluated to zero (see Q125).

- an optional constraint on *at(n)*

  **For each *n*: *S(n)* is reported only if *at(n)* satisfies the constraint**

  Note that a '*= 0*' constraint cannot be satisfied unless the entities composing *B* are right of an 'O', as such constraint means that there are no assignments to the pattern excluding this aggregator. Similarly:

  - no constraint is equivalent to '*> 0*' constraint
  - '*≠ expr*', '*< expr*', and '*≤ expr*' constraints are not satisfied when *at(n)* is zero
  - '*= expr*', '*≥ expr*', and '*∈ [expr1, expr2]*' constraints are not satisfied when *at(n)* is zero even if *expr* is evaluated to zero

A1 location:

- Below a relationship/path

  A relationship/path with an A1 below it may have a relationship/path-negator (see Q63-Q66) and may be wrapped with an 'O' (see Q82v2, Q126).

- Below a quantifier-input (excluding quantifier at the start of the pattern)

  See Q305v1, Q121, Q122. A quantifier-input with an A1 below it may be wrapped with an 'O'.

- Below the pattern's start (even when followed by a quantifier)

  See Q27, Q261, Q332.

- If all entities in $T \cup B$ are in a sequence: if all the entities in *T* appear right of all the entities in *B*: left of the rightmost entity in *T* (see Q247). Otherwise: right of the leftmost entity in *T* (see Q243)
- If entities in $T \cup B$ are in different branches of a quantifier: below the quantifier-input (see Q27)

***Q59:** Any person who owns at least 20 horses*

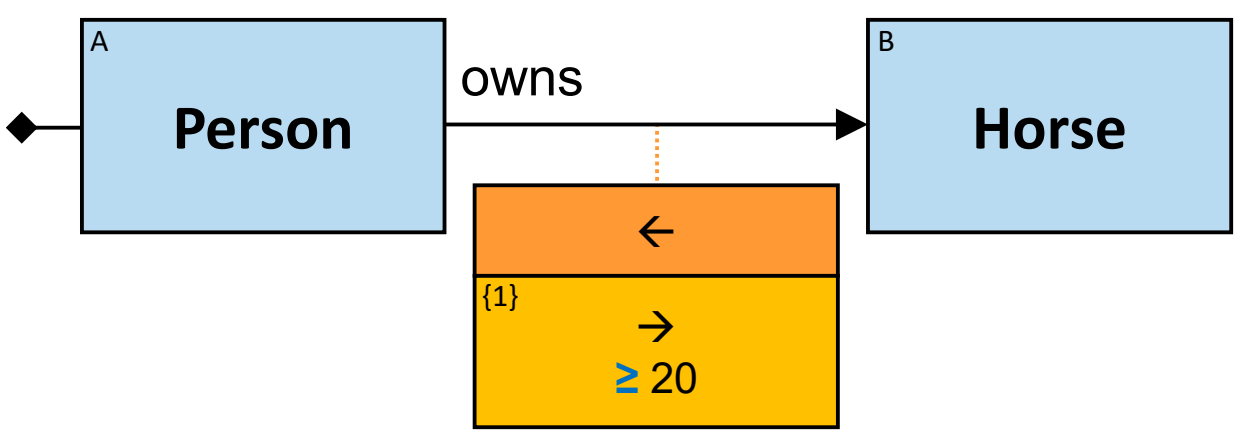

{1} is a property of each unique assignment to A in *S*.

***Q60:*** *Any dragon that was frozen by exactly five dragons*

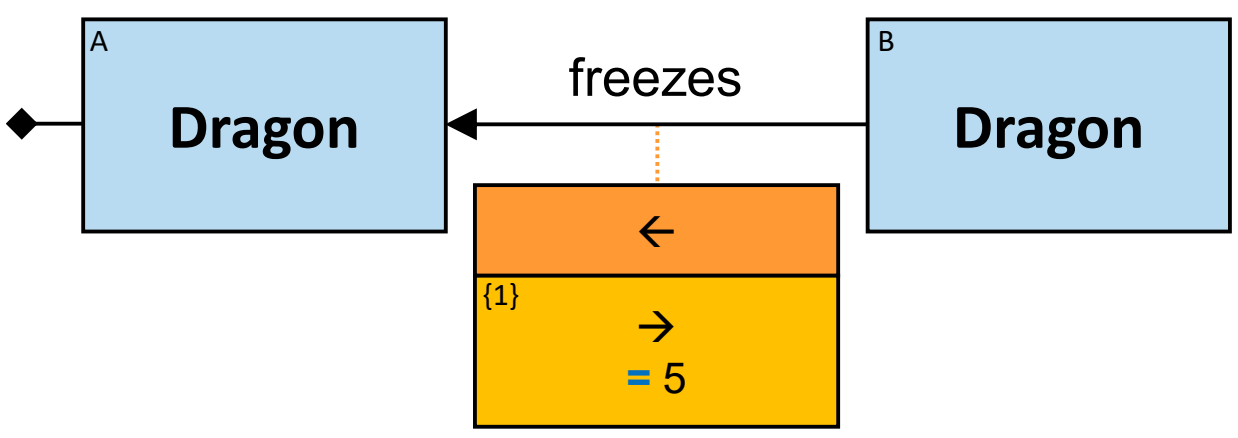

If some dragon B froze some dragon A more than once – B would still be counted only once per A. A1 counts *distinct* entity assignments.

***Q61:*** *Any entity that owns more than two entities*

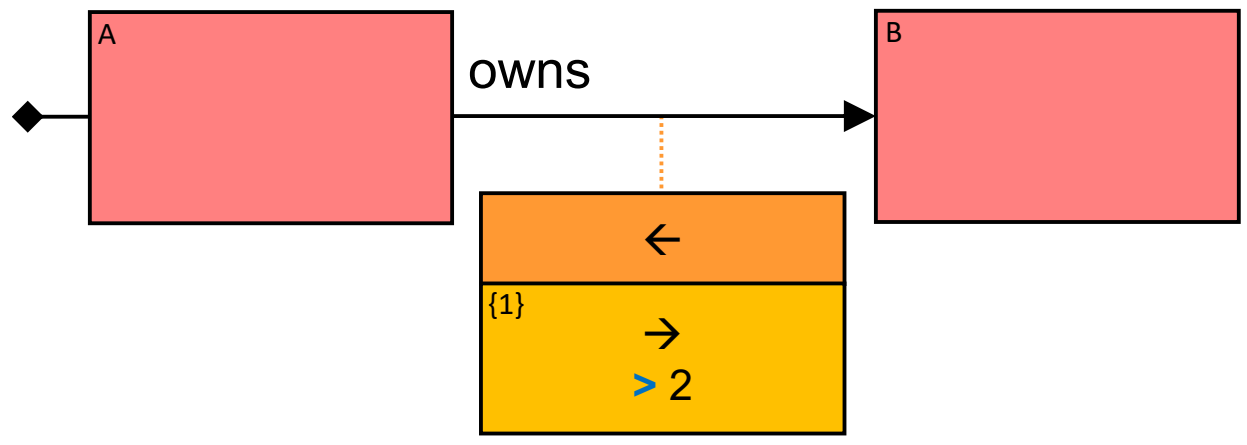

***Q62:*** *Any person who has paths of length up to four to more than five people*

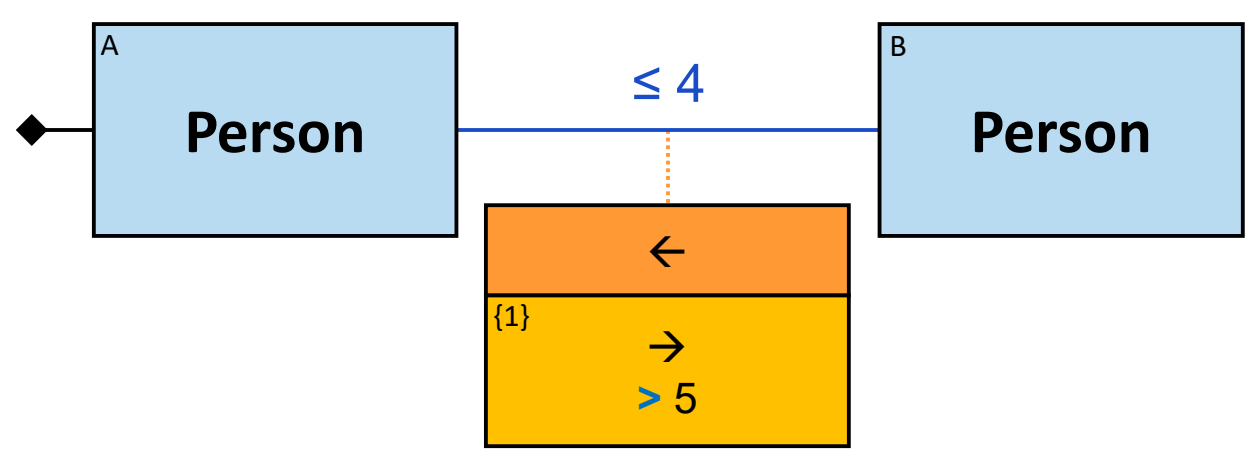

***Q136:*** *Any dragon A that froze (dragons that froze dragons B). The cumulative number of distinct Bs (per A) is greater than 100* (two versions)

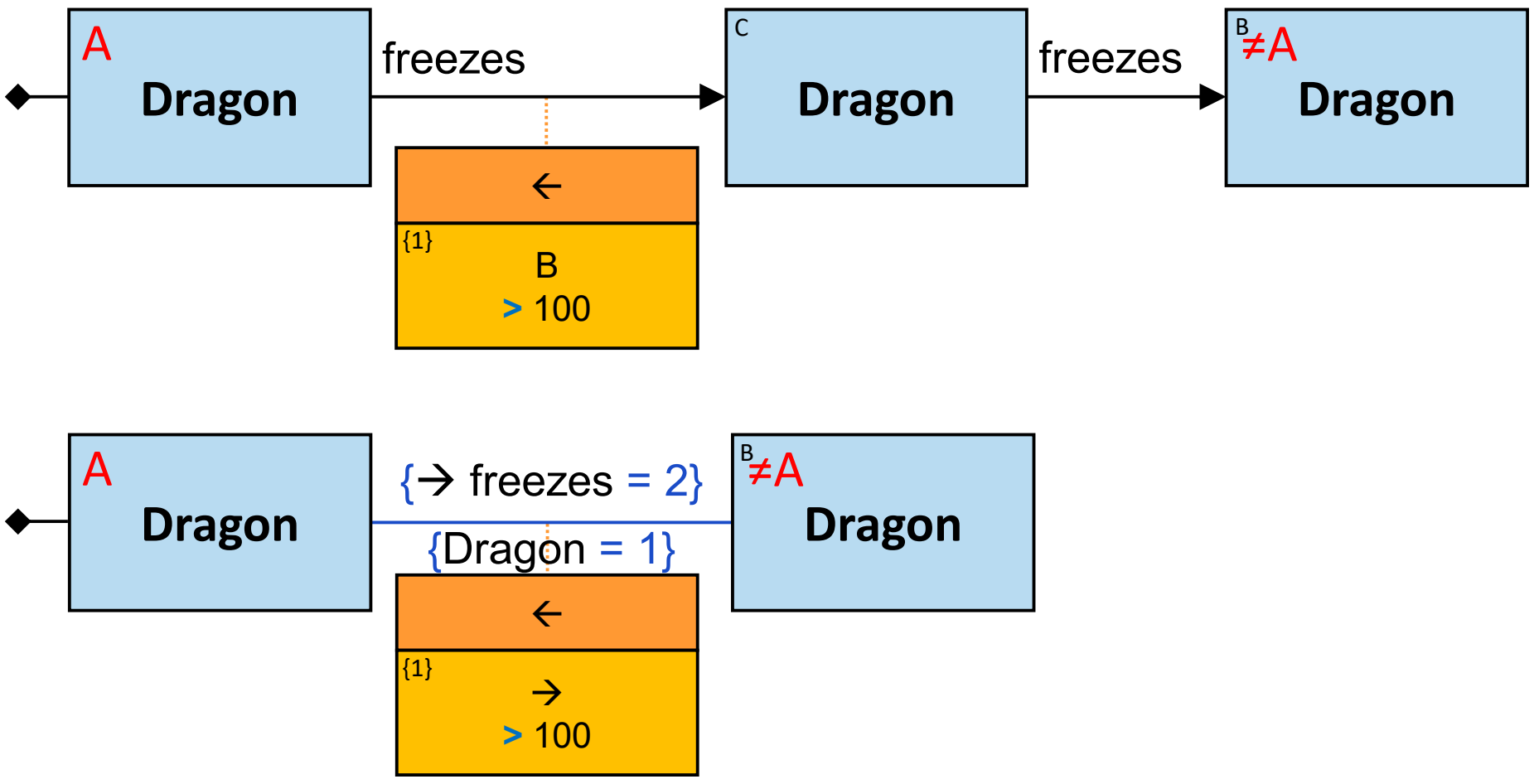

***Q177:*** *Any pair of dragons (A, B) were A was frozen by at least ten B's and froze each one of those* (two versions)

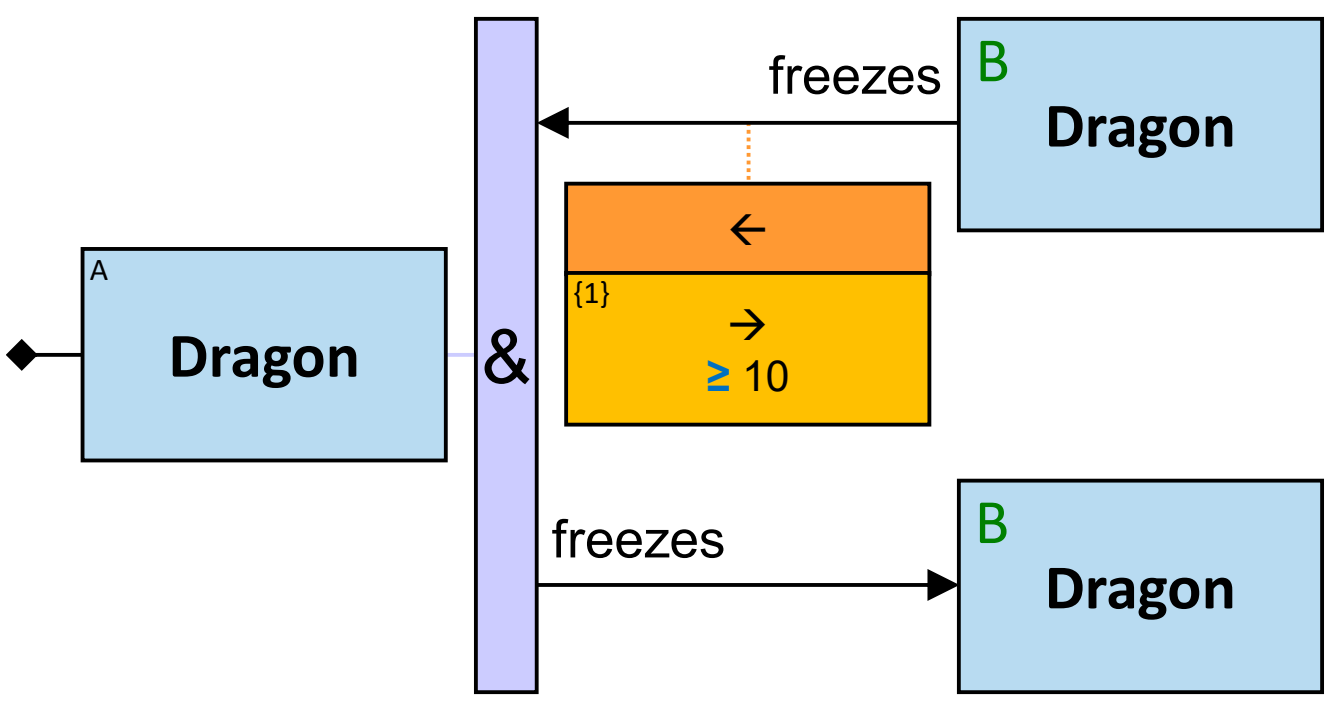

First, any pair (A, B) matching the pattern excluding the aggregator is found. Then, the aggregation constraint is checked:

For each assignment to A, there are at least ten assignments to B such that (B froze A and A froze B)

This second version is for illustrative purposes only:

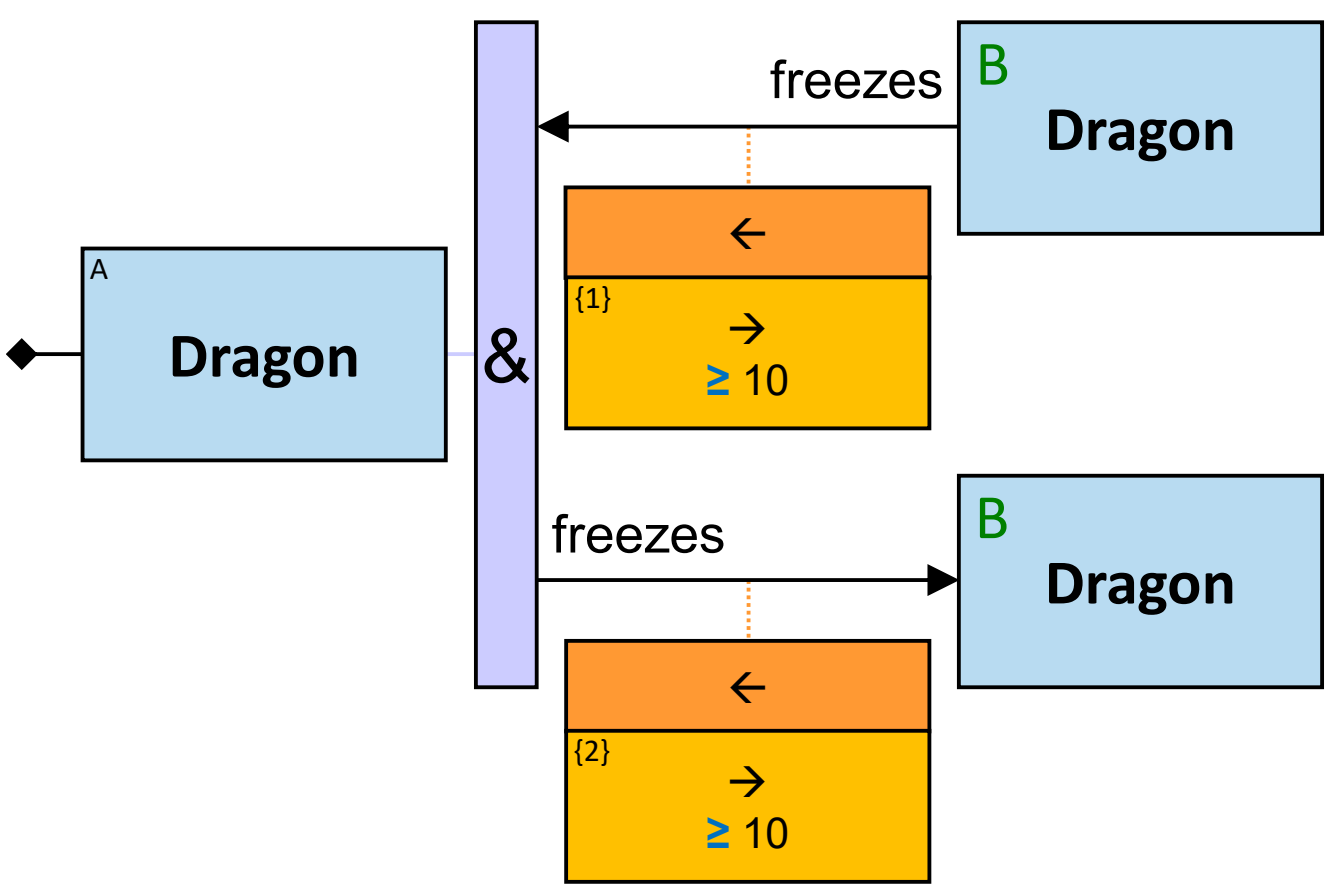

First, any pair matching the pattern excluding the aggregators is found. Then, the aggregator constraints are checked one by one:

For each assignment to A:

- There are at least ten assignments to B such that (B froze A, and A froze B)
- There are at least ten assignments to B such that (A froze B, and B froze A)

***Q178:** Any dragon A frozen by at least ten dragons and either (i) A froze only one dragon – which is not one of those, or (ii) A froze at least two dragons*

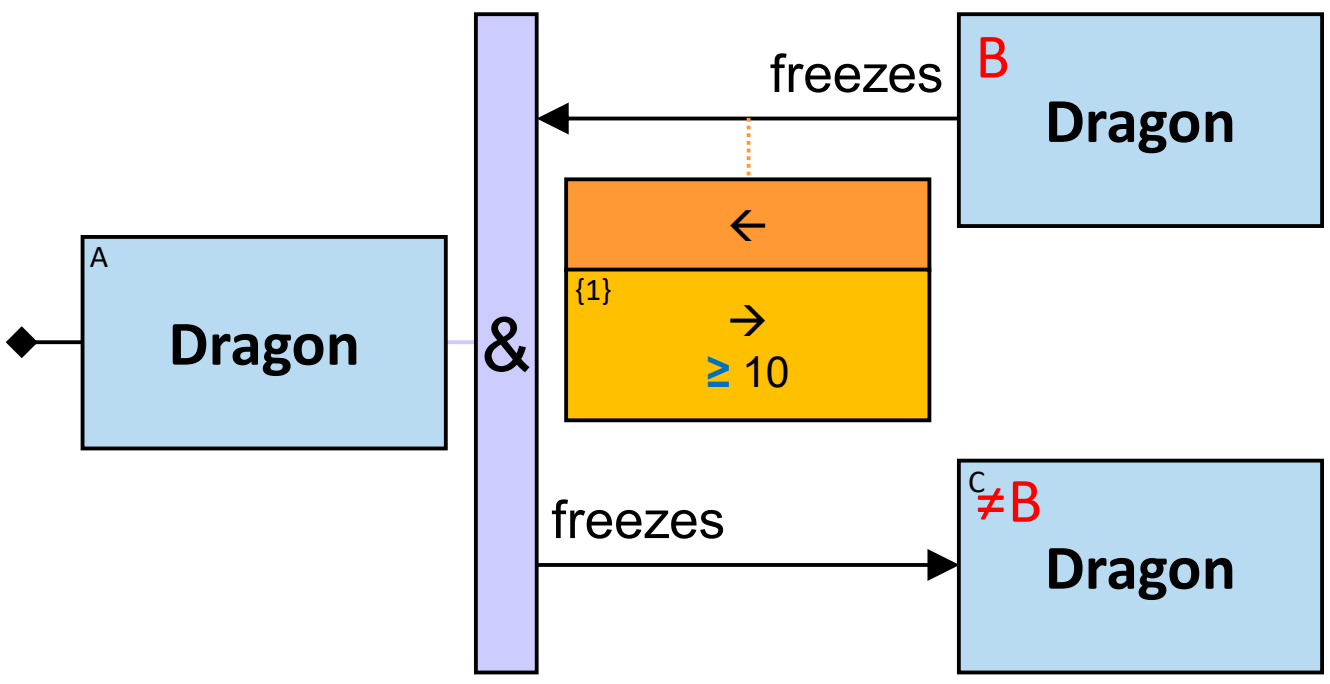

First, any dragon triplet (A, B, C) matching the pattern excluding the aggregator is found. Then, the aggregation constraint is checked:

For each assignment to A:

- There are at least ten assignments to B such that (B froze A, and A froze a dragon that is not B)

Hence, for each assignment to A:

- At least ten dragons froze A, and either (i) A froze only one dragon – which is not one of those, or (ii) A froze at least two dragons

***Q85:** Any dragon that froze at least ten dragons and was frozen by at least ten dragons* (two versions)

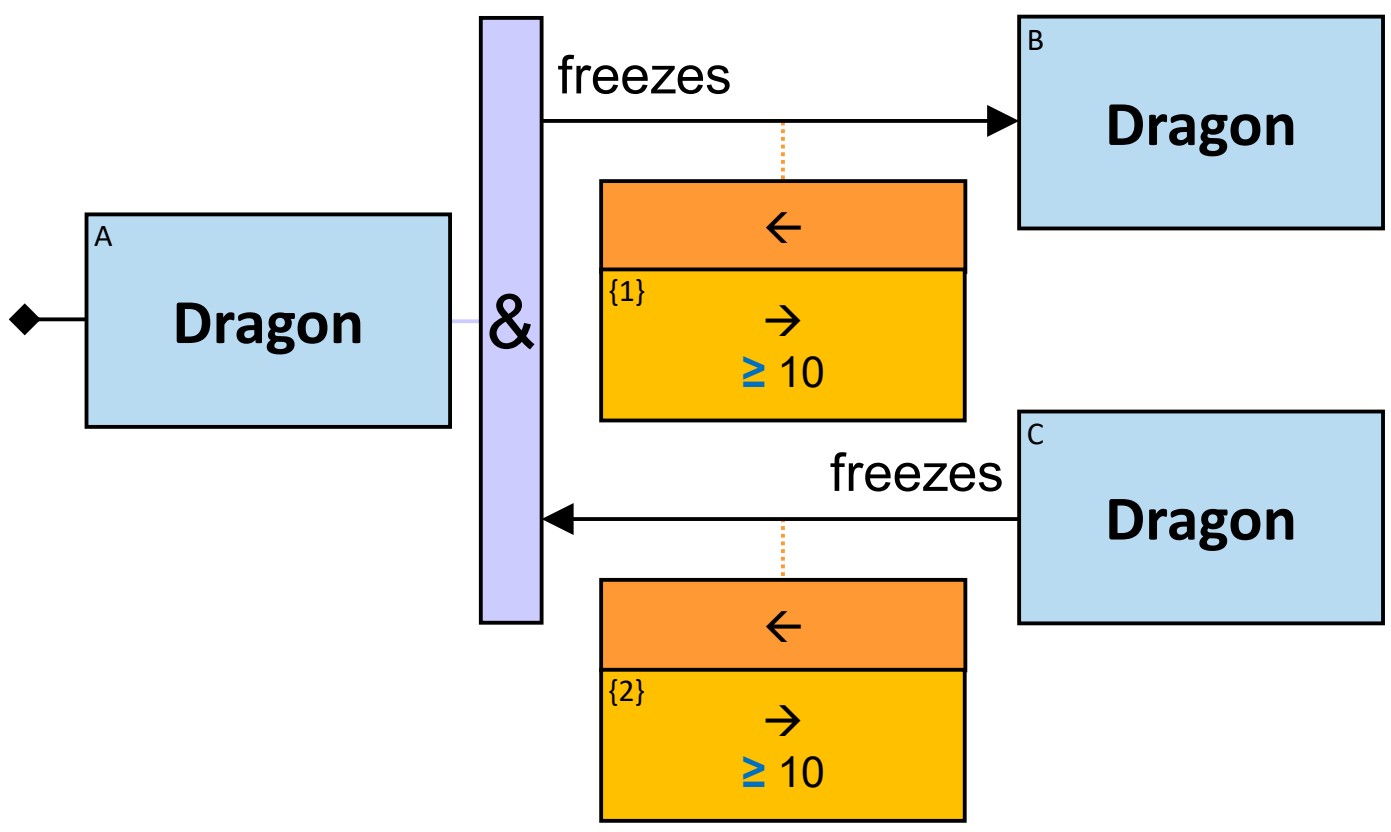

This second version is for illustrative purposes only:

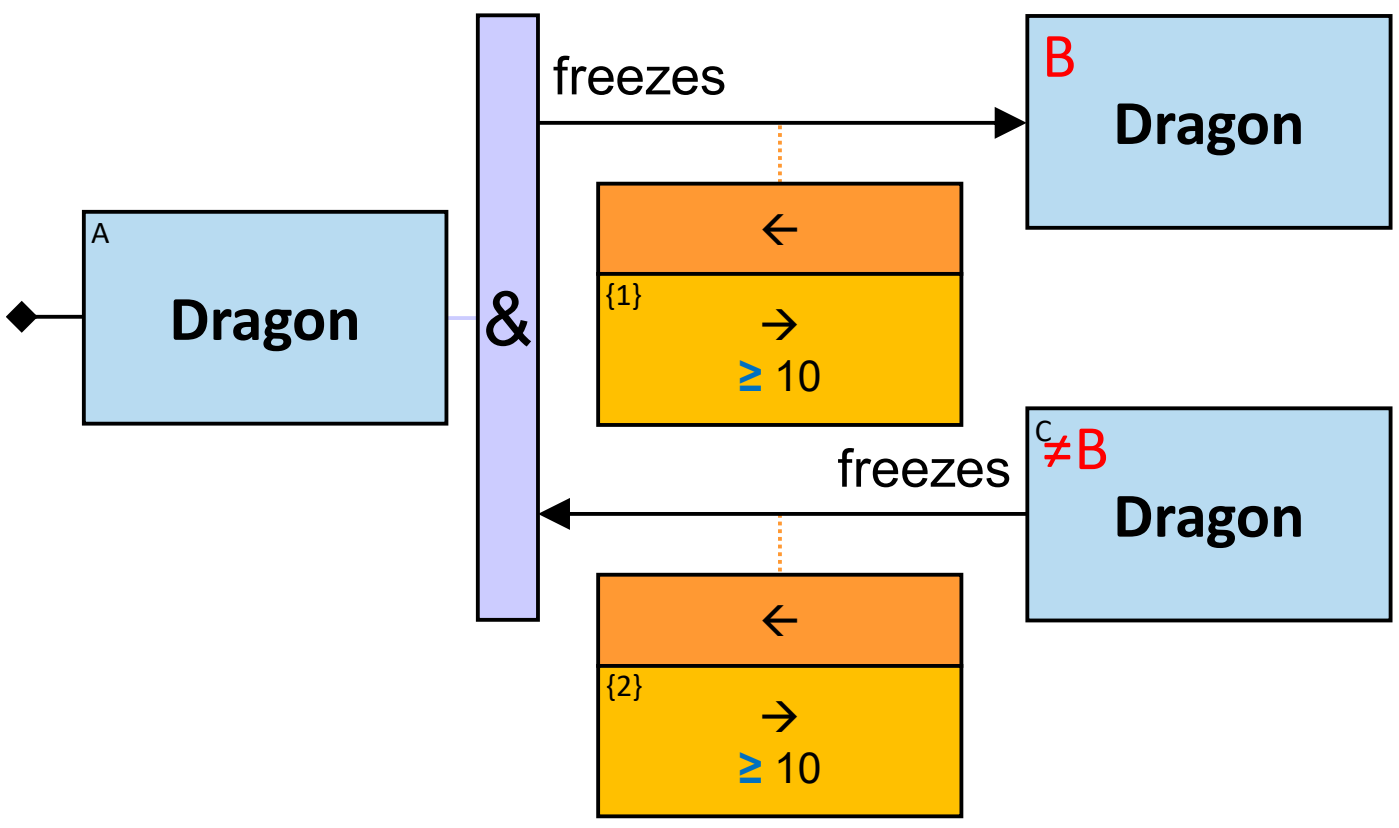

First, any dragon triplet (A, B, C) matching the pattern excluding the aggregators is found. Then, the constraints are checked one by one:

For each assignment to A:

- There are at least ten assignments to B such that A froze B, and a dragon other than B froze A
- There are at least ten assignments to C such that C froze A, and A froze a dragon other than C

Hence, for each assignment to A:

- At least ten dragons froze A, and either (i) A froze only one dragon – which is not one of those, or (ii) A froze at least two dragons

and also

- A froze at least ten dragons, and either (i) only one dragon froze A – which is not one of those, or (ii) at least two dragons froze A

Hence:

- A froze at least ten dragons, and at least ten dragons froze A

***Q101:*** *Any person who owns at least five white horses*

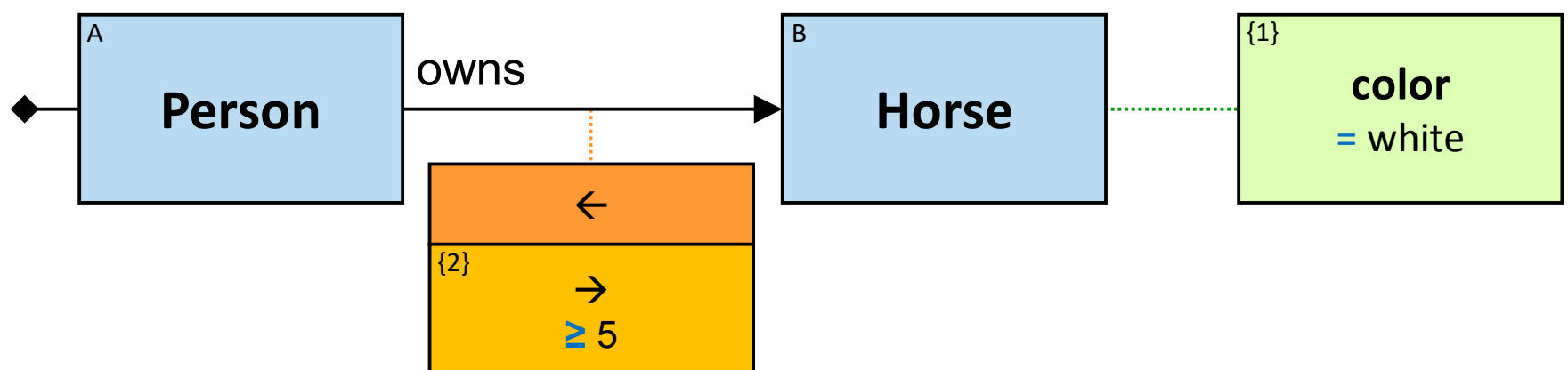

***Q102:*** *Any dragon A that was frozen by exactly two dragons, each of which was frozen at least once. A might have been frozen by additional dragons that were never frozen*

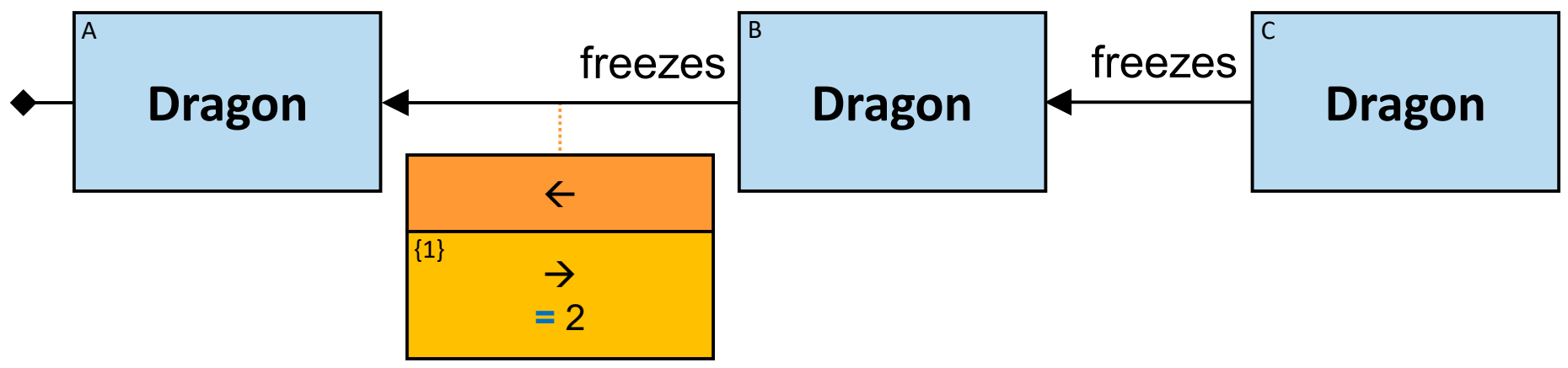

***Q166:*** *Any **dragon** that more than five Sarnorians own a dragon that froze **it***

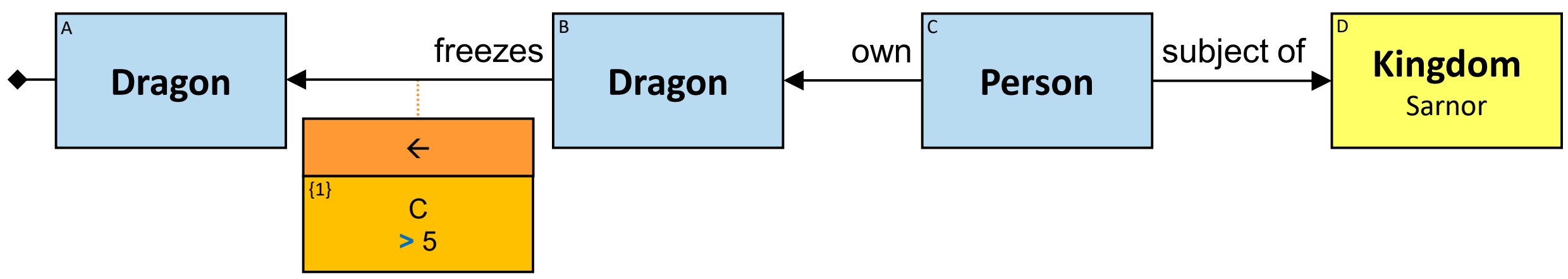

***Q246:*** *Any dragon B that froze at least one dragon and was frozen by more than ten dragons*

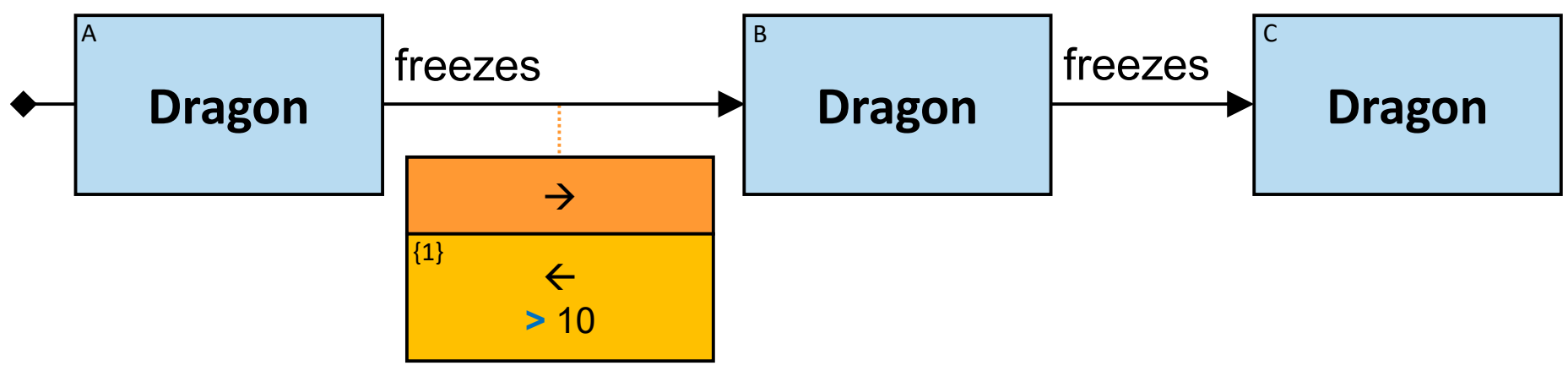

{1} is a property of each unique assignment to B in *S*.

***Q247:*** *Any dragon C that more than ten dragons froze dragons that froze it*

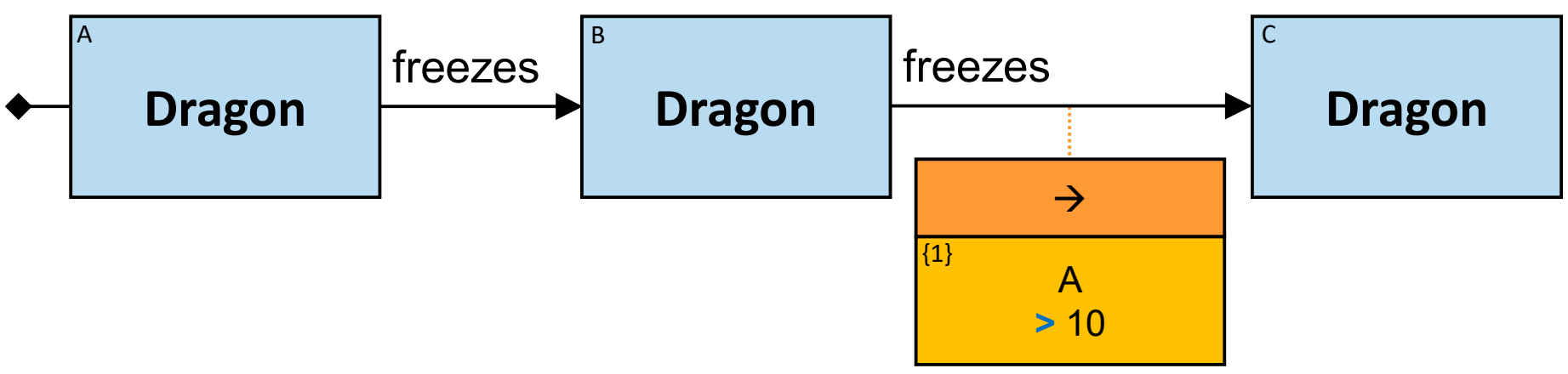

***Q113:*** *Any person A who has at least five friends with whom A shares a birthdate*

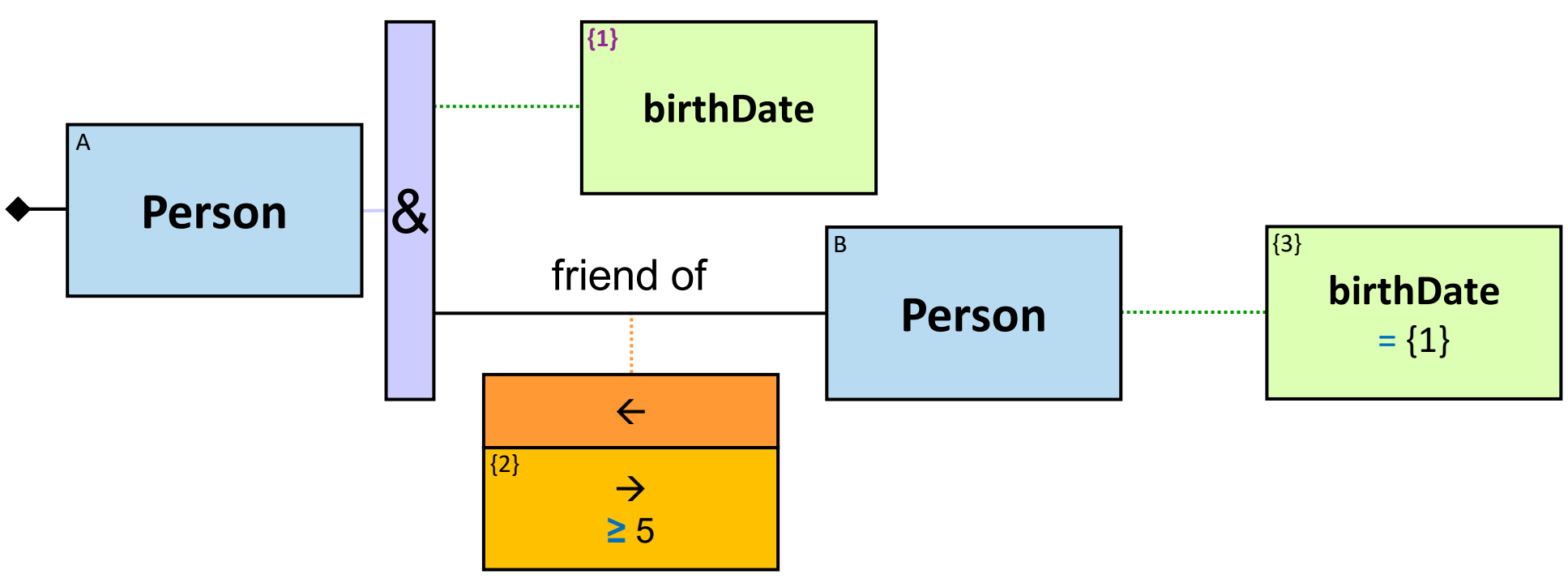

***Q114:** Any person who owns at least five horses of the same color* (version 1)

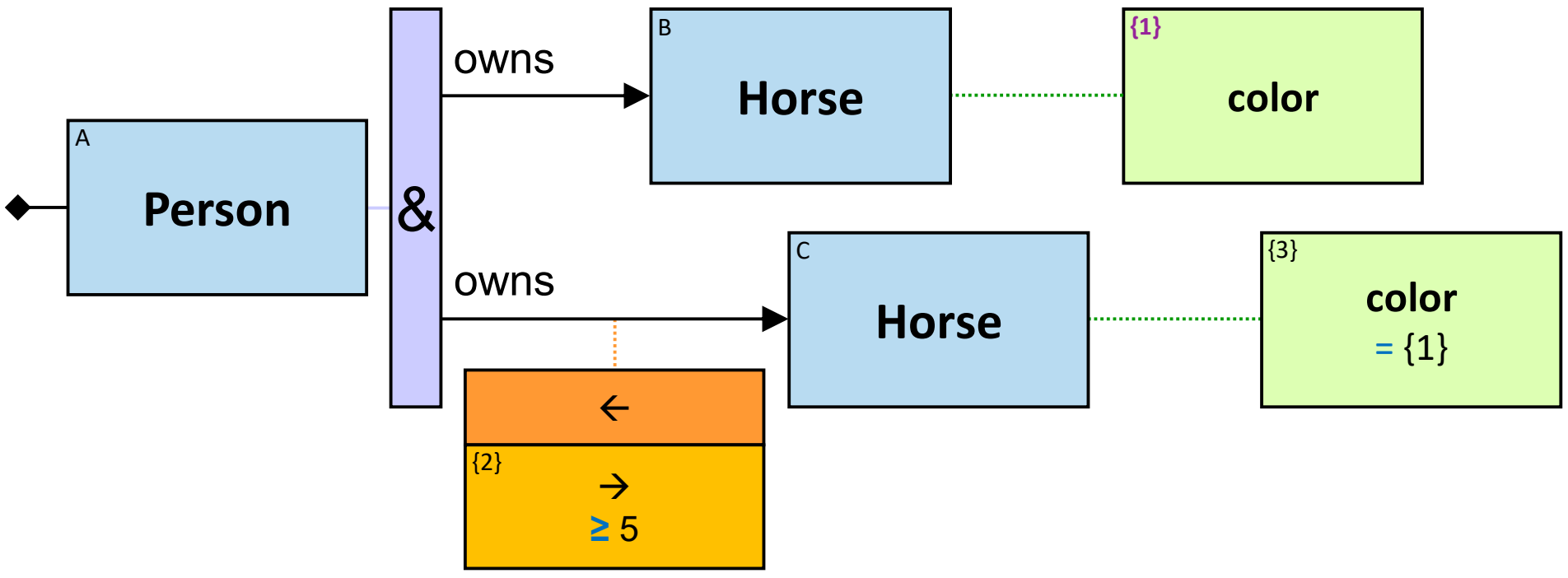

***Q218:** Any person who owns at least five entities of the same type* (version 1)

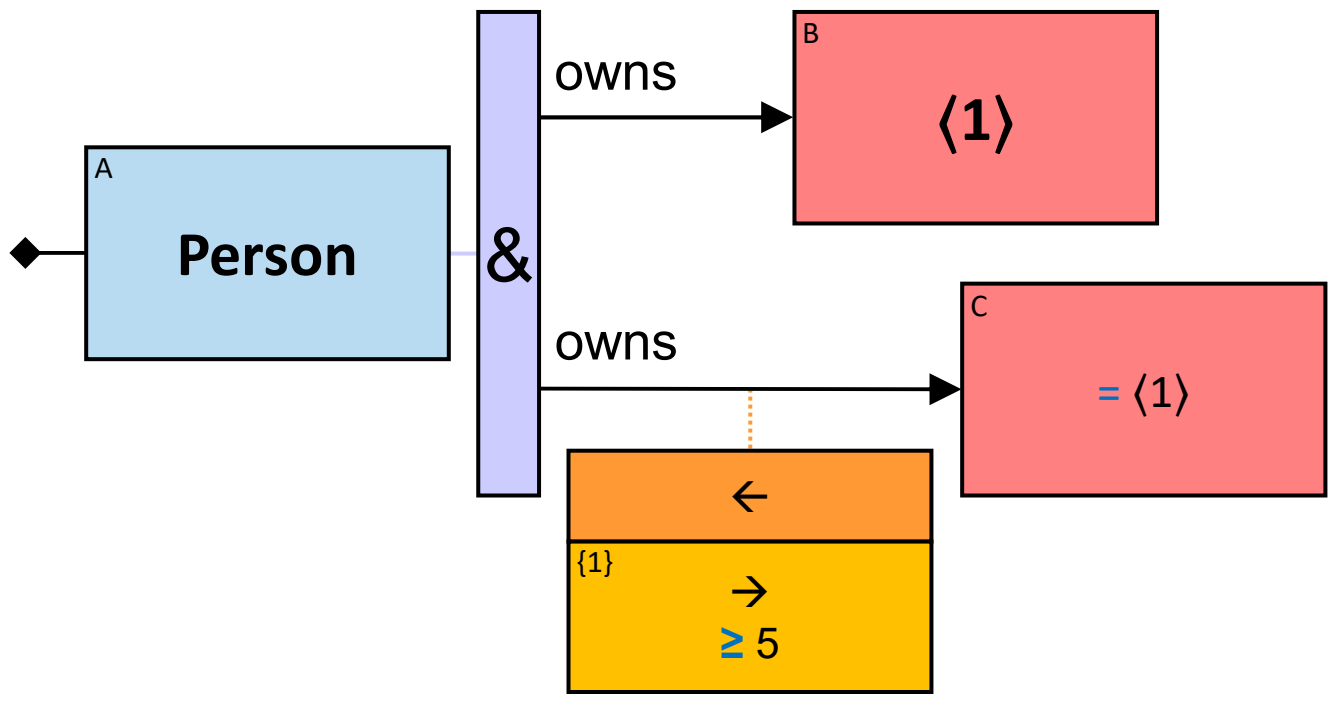

***Q292:** Any person A that at least 80% of A's horses are black*

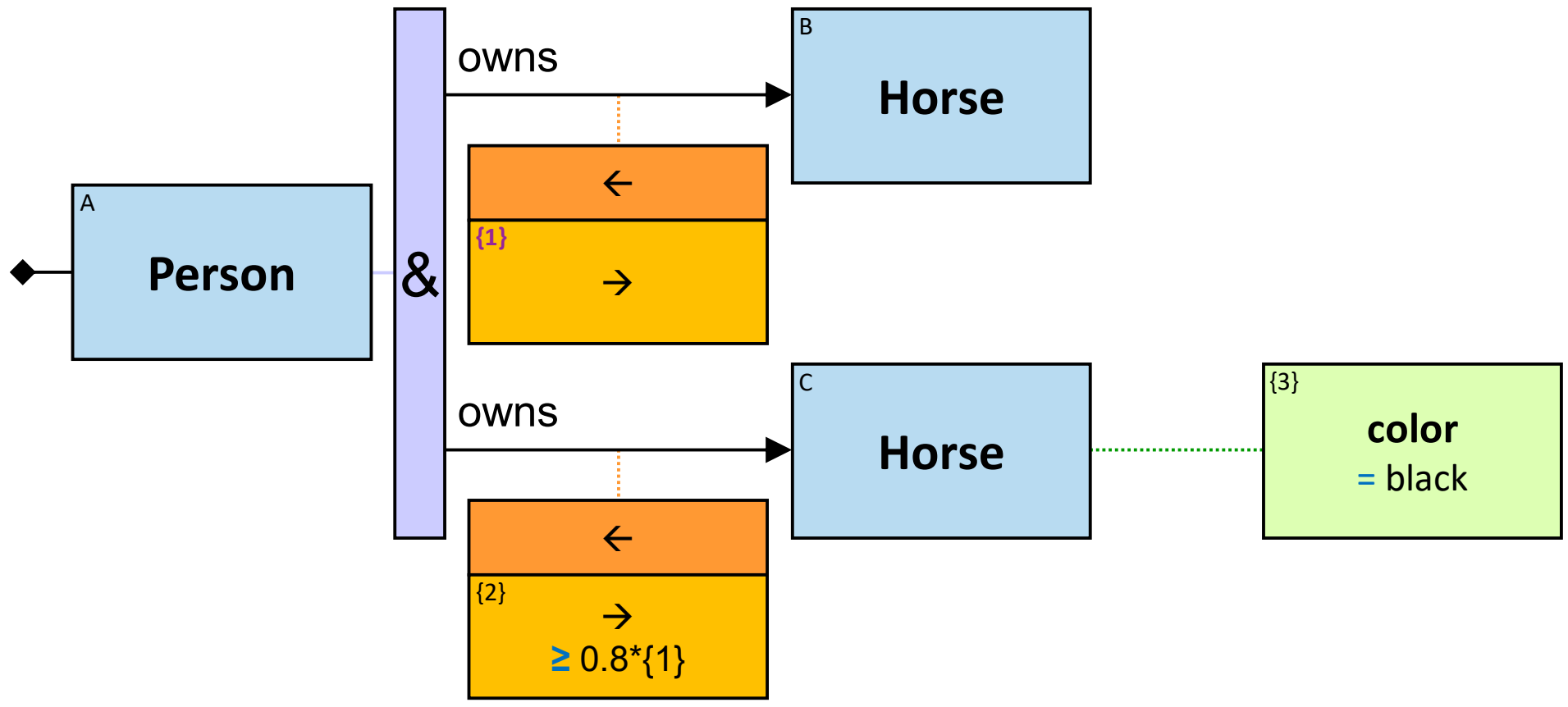

***Q63:*** *Any Masons Guild member who more than five Masons Guild members are not friends of*

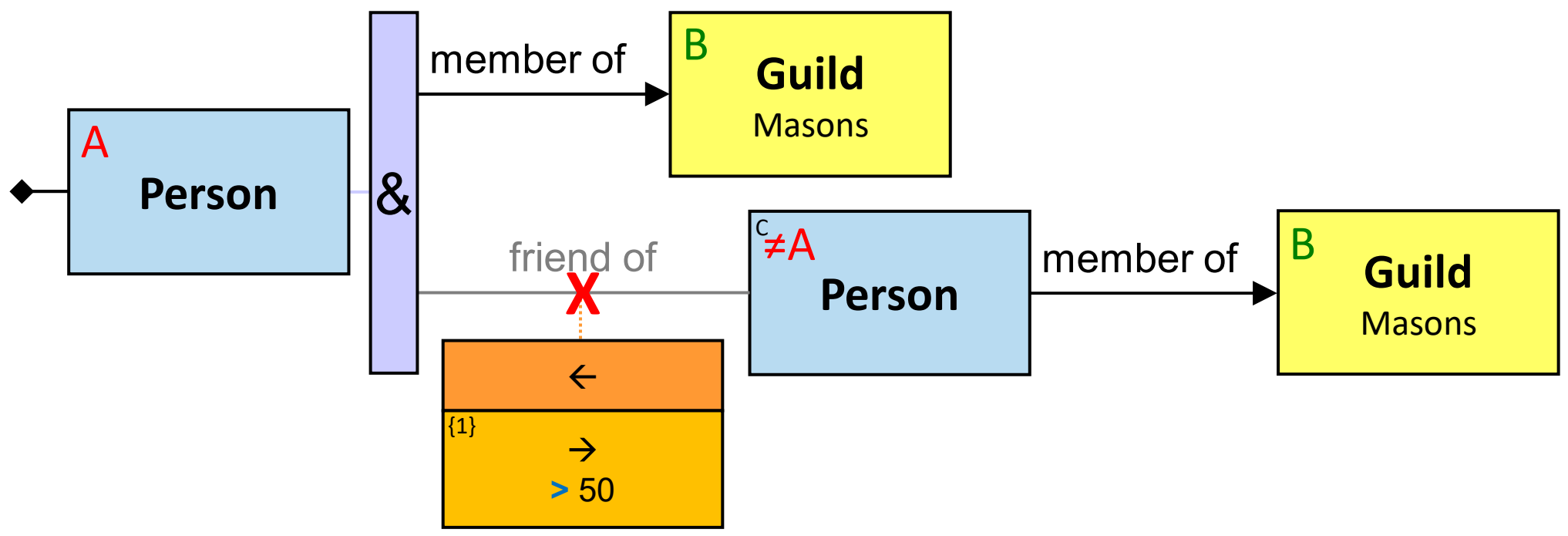

***Q65:*** *Any person A who does not own more than two things that (at least) one of A's parents owns/owned*

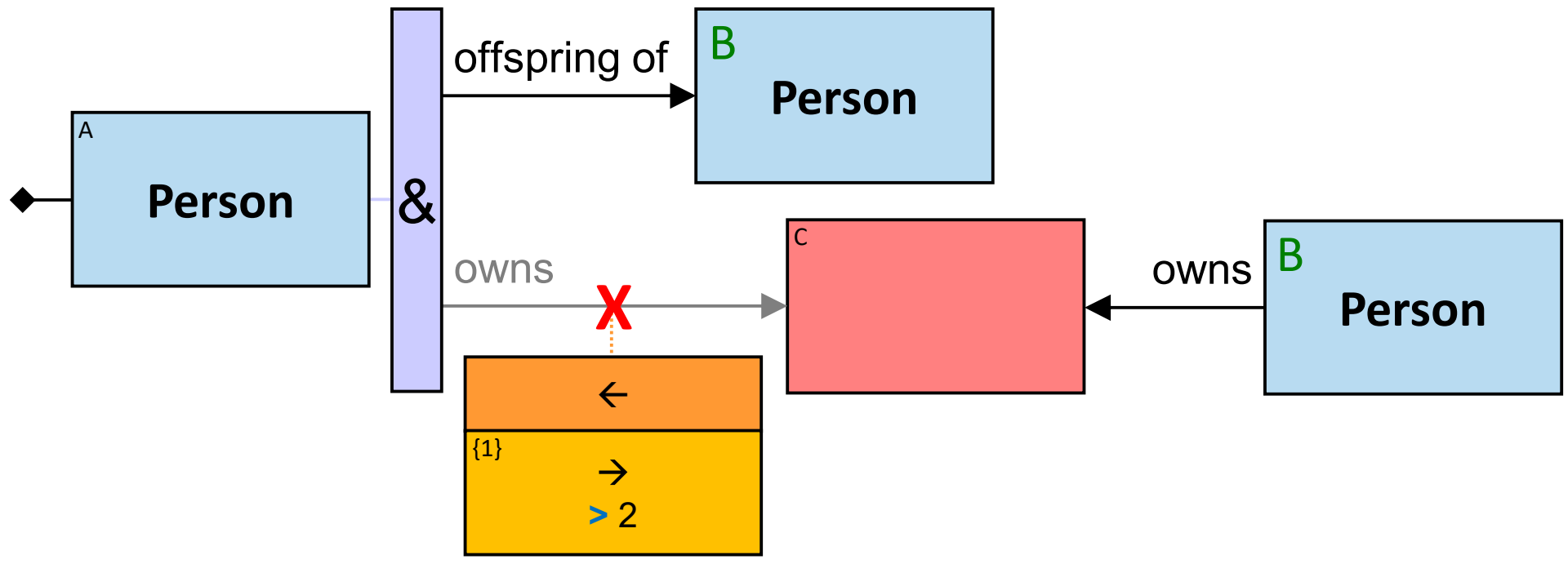

***Q66:*** *Any person from whom more than five people do not have a path of length up to six*

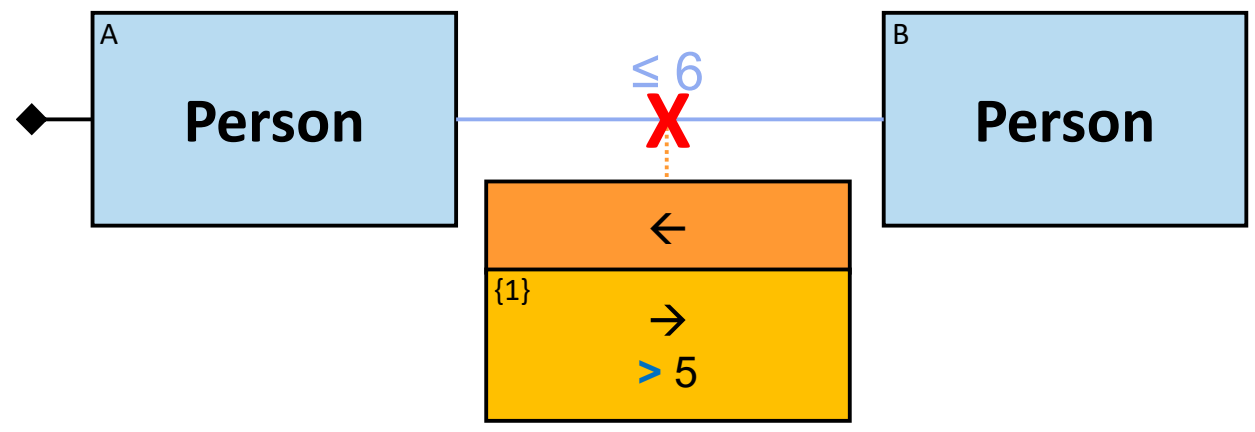

***Q151:*** *Any horse owned by more than ten people, at least one is Sarnorian. Only the Sarnorian owners should be reported*

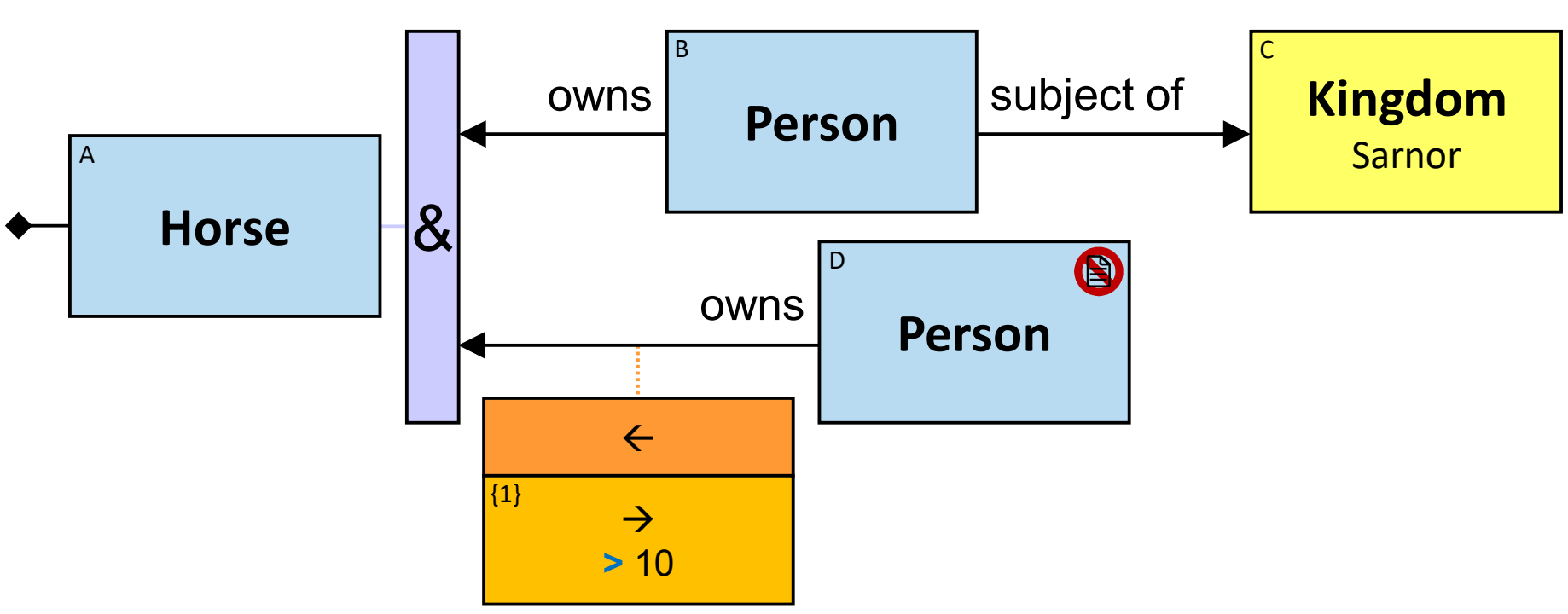

***Q152:** Any horse owned by more than ten people. Only the Sarnorian owners should be reported*

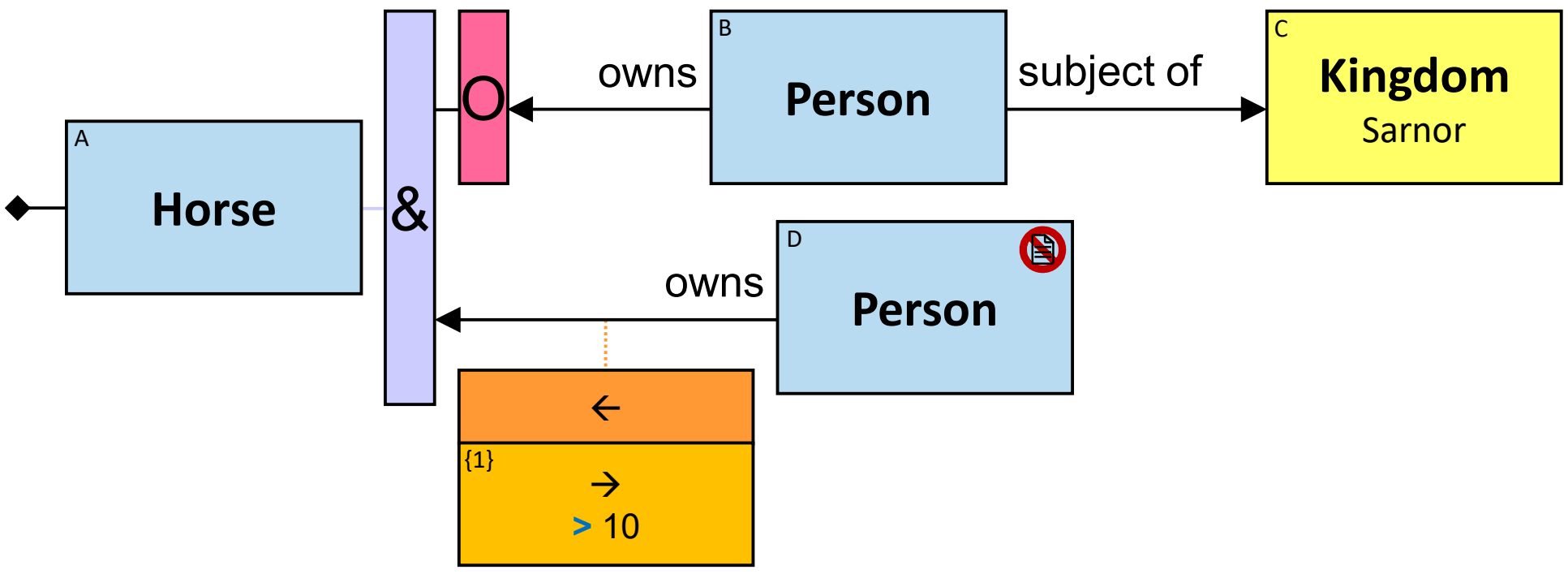

***Q125:** Any dragon that the number of dragons it froze is greater than the number of dragons that froze it*

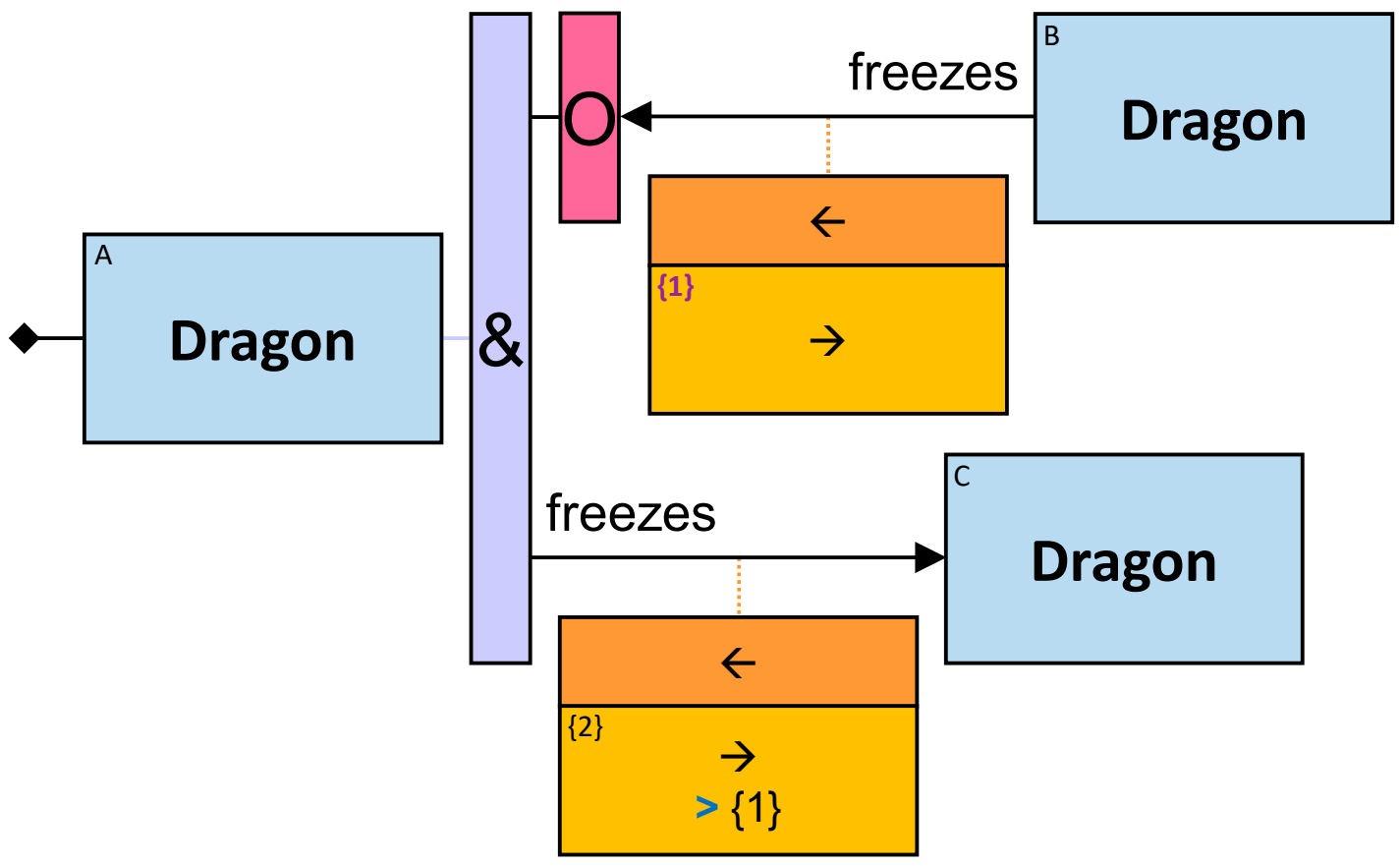

{1} is evaluated to zero when the optional part has no valid assignments.

***Q126:** Any dragon that the number of dragons it froze is greater than the number of dragons it did not freeze*

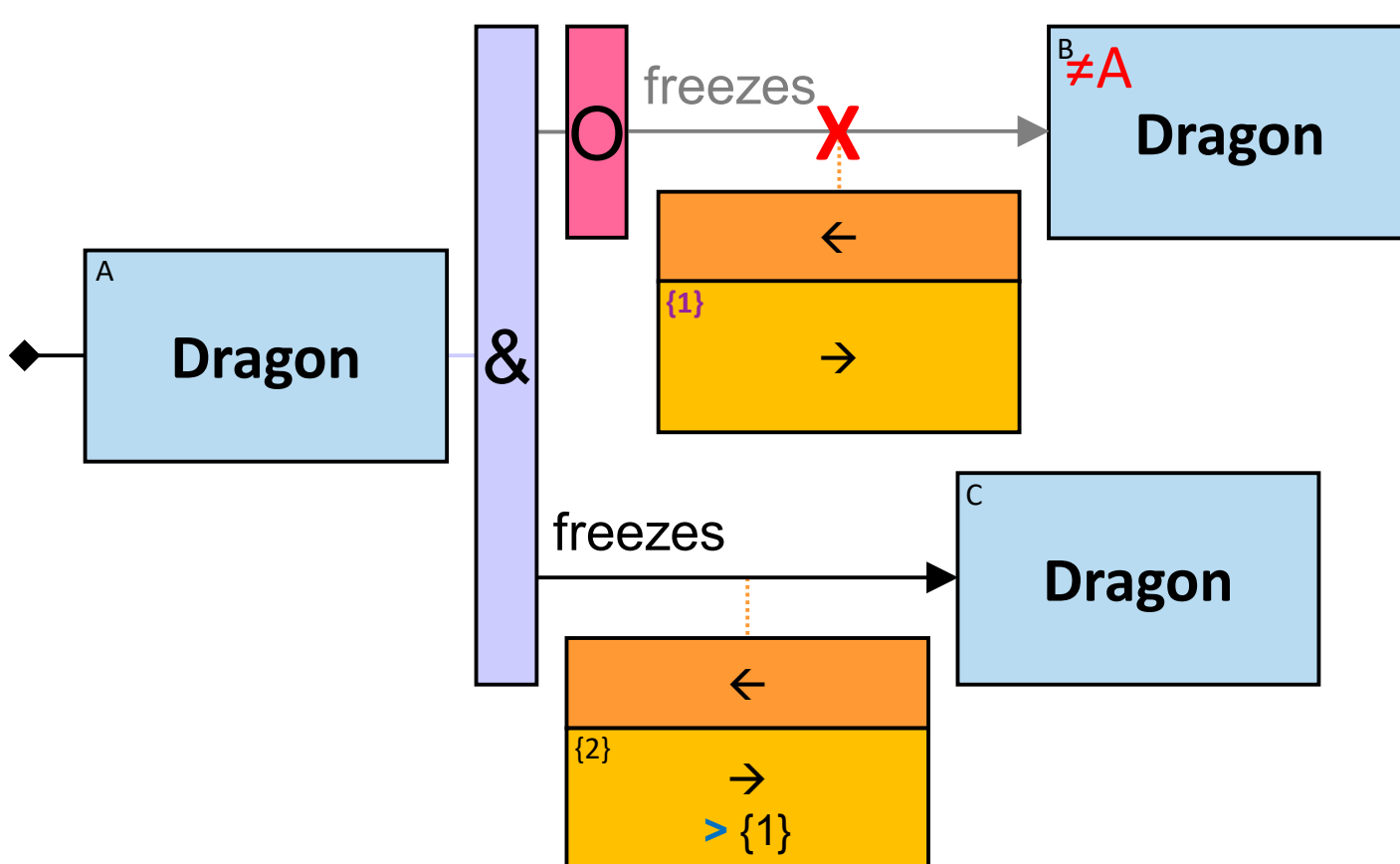

***Q249:*** *Any triplet of sets (people A, dragons B, horses C) where (i) any person in A owns at least one dragon in B and at least one horse in C, (ii) any dragon in B is owned by at least ten people in A, and (iii) any horse in C is owned by at least ten people in A*

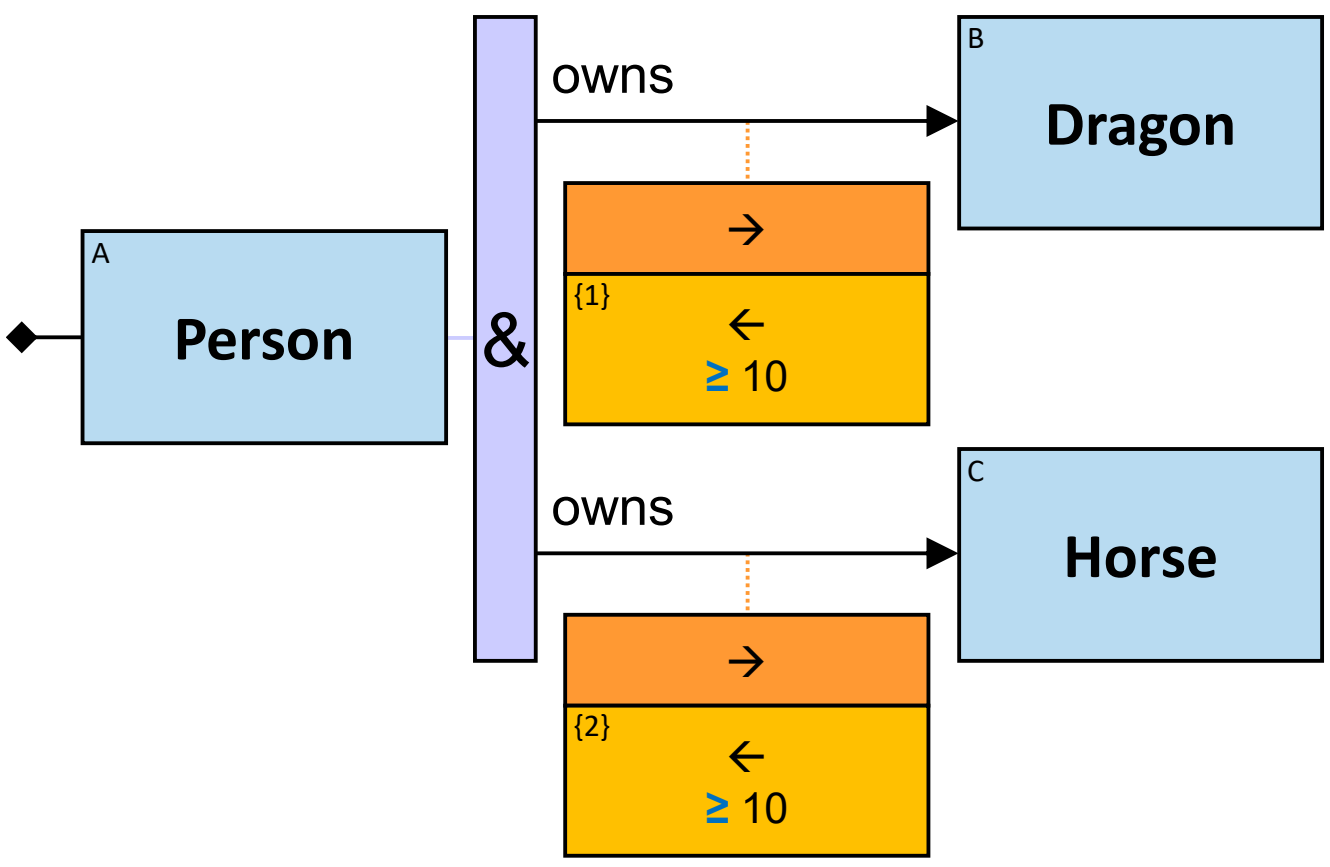

***Q250:*** *Any triplet of sets (people A, dragons B, horses C) where (i) any person in A owns at least one dragon in B or at least one horse in C, (ii) any dragon in B is owned by at least ten people in A, and (iii) any horse in C is owned by at least ten people in A*

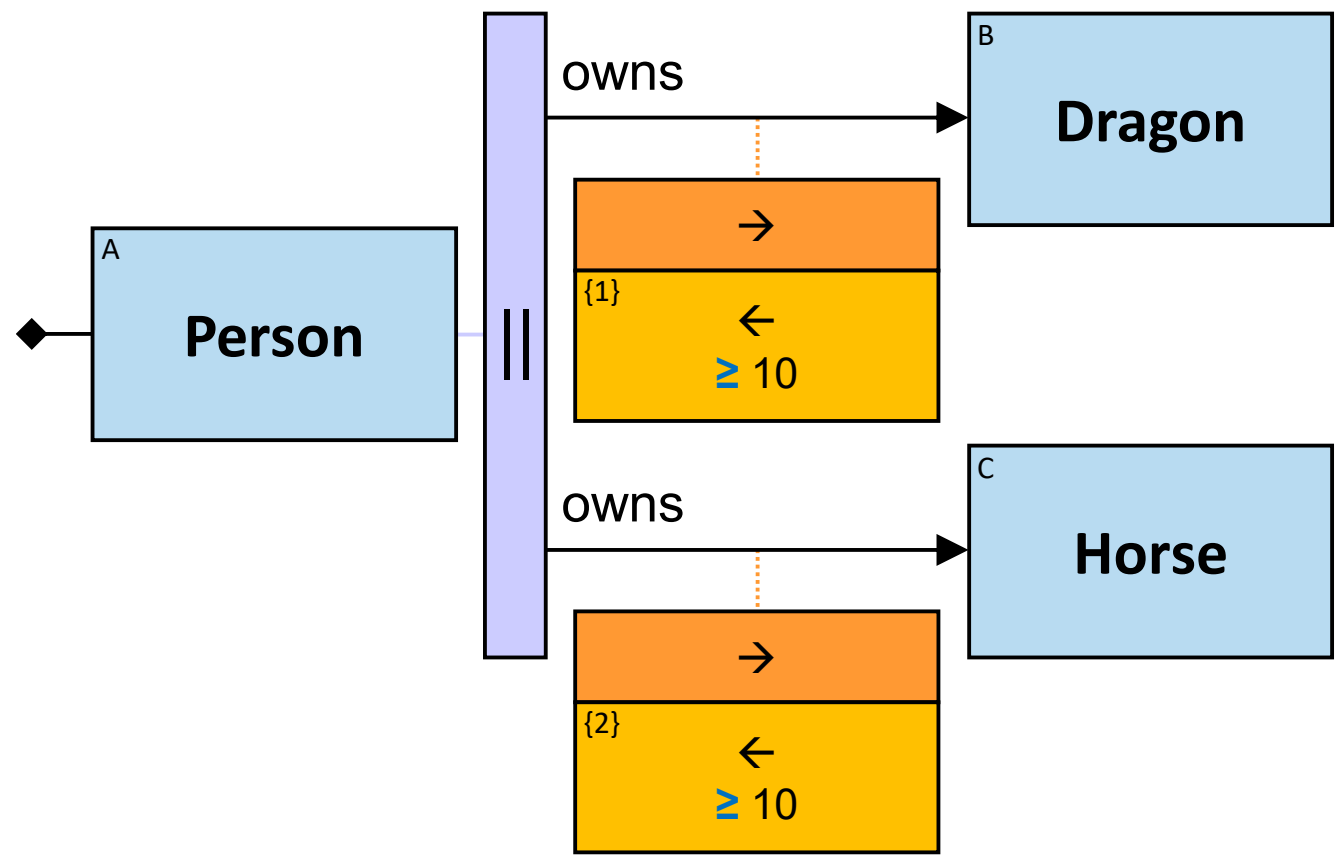

***Q344:*** *Any triplet of sets (people A, dragons B, horses C) where (i) any person in A owns at least one dragon in B and at least one horse in C, (ii) any dragon in B is owned by at least ten people in A, and (iii) any horse in C is owned by less than five people in A*

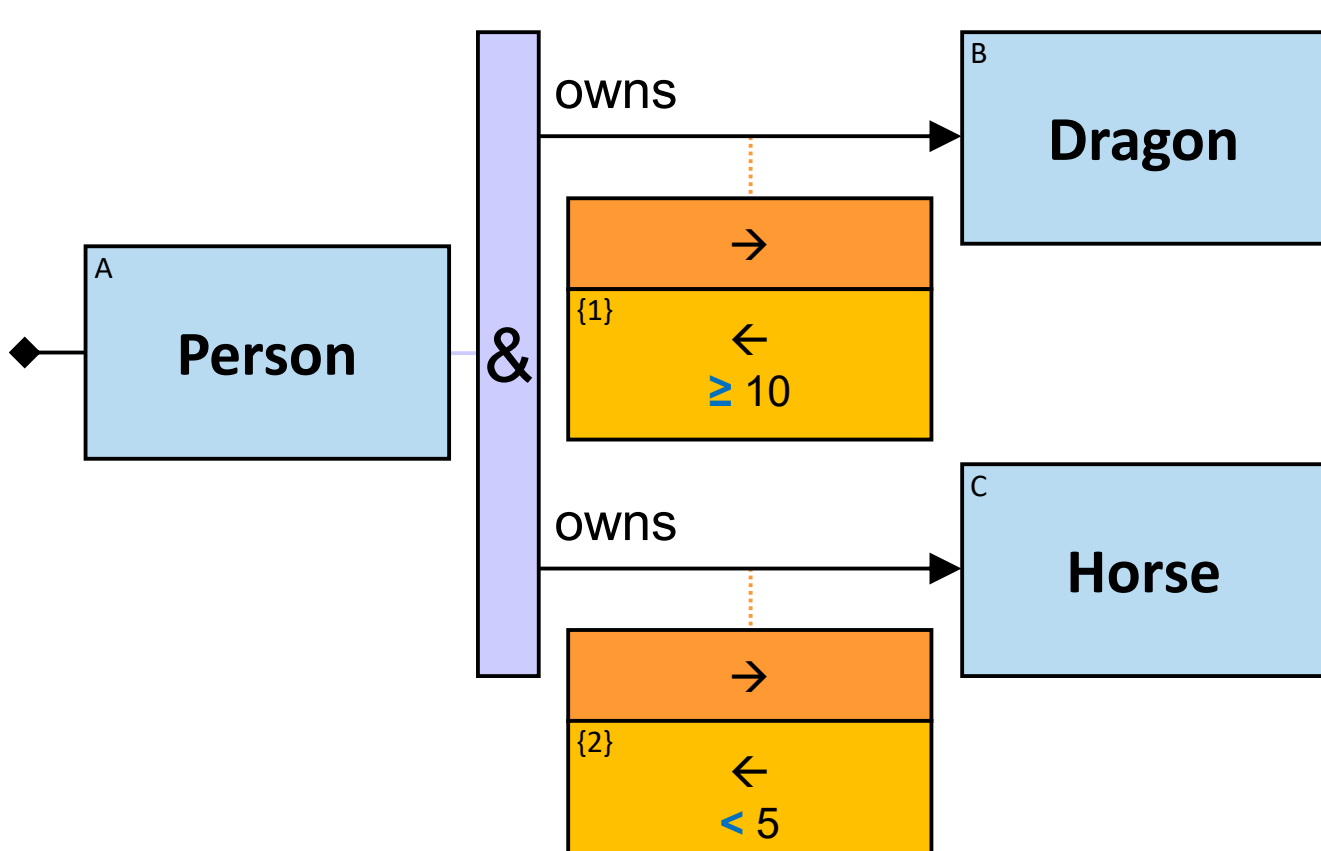

***Q345:*** *Any triplet of sets (people A, dragons B, horses C) where (i) any person in A owns at least one dragon in B and does not own at least one horse in C (ii) any dragon in B is owned by at least ten people in A, and (iii) any horse in C is not owned by at least ten people in A*

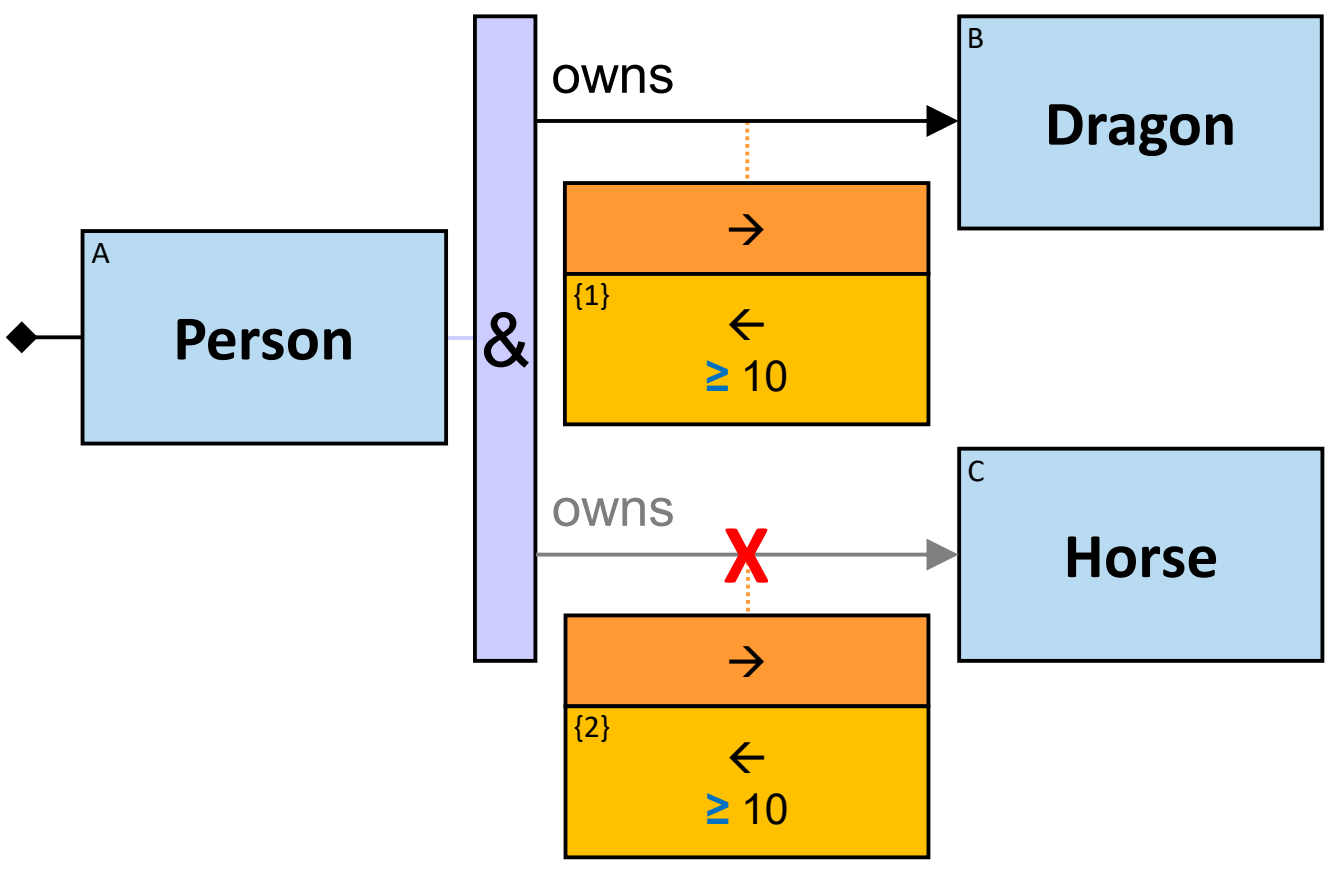

**Q82:** *Any dragon that froze all other dragons.* Alternative phrasing: *Any dragon for which there is no dragon (besides itself) it did not freeze* (version 2)

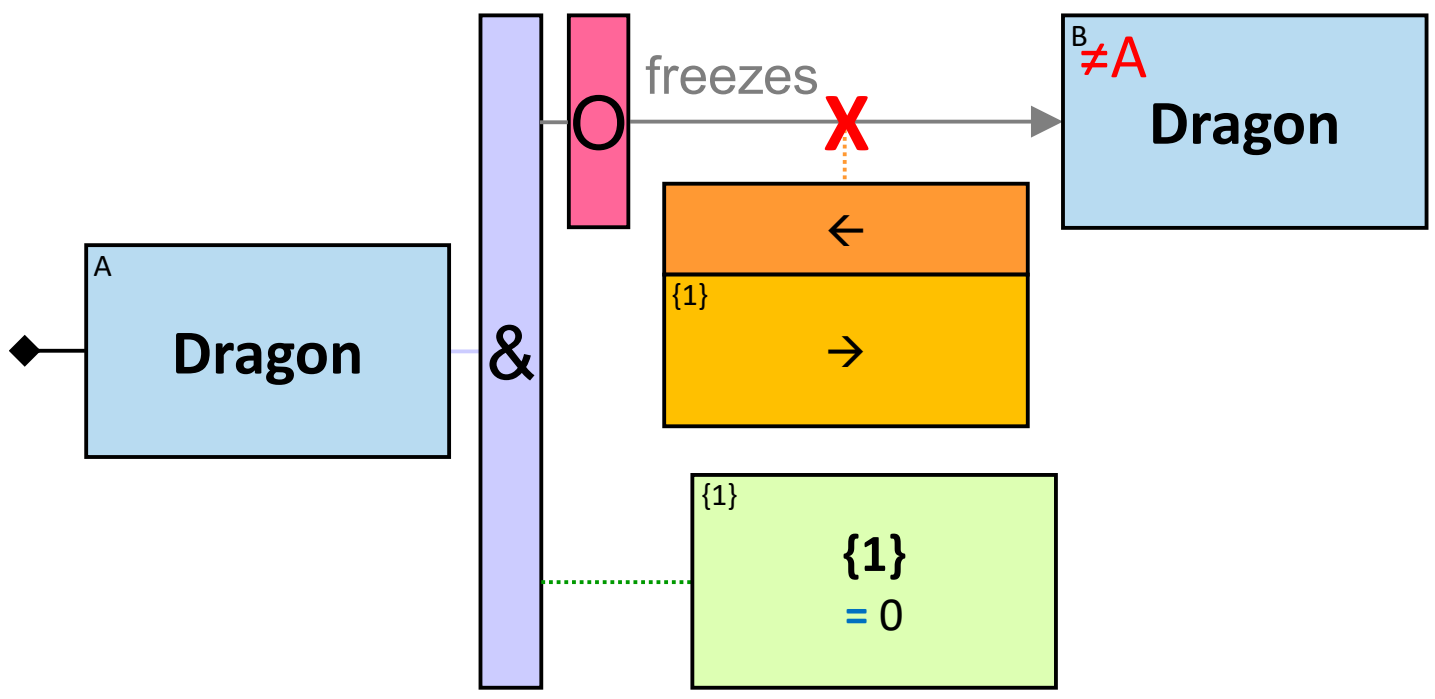

The dragons that were frozen will not be a part of the assignment.

**Q360:** *Any dragon that (froze or fired at) all other dragons.* Alternative phrasing: *Any dragon for which there is no dragon (besides itself) it did not (freeze or fire at)* (version 2)

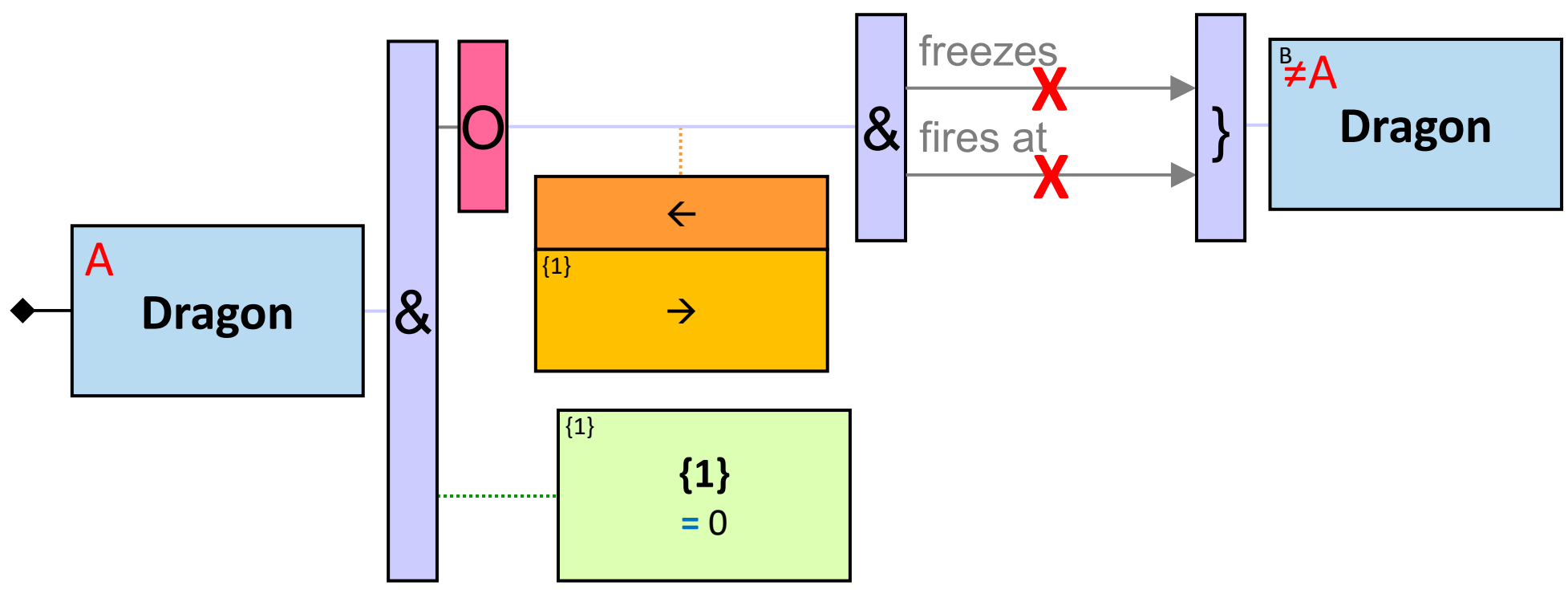

*Q64: Any dragon that no more than four dragons owned by Sarnorians did not freeze it*

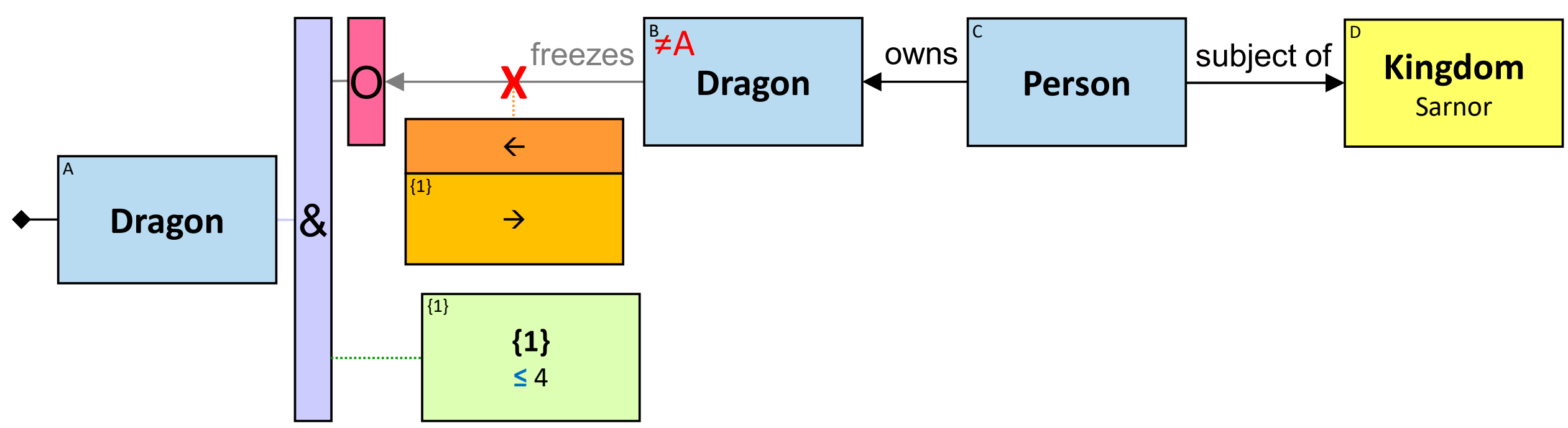

*Q165: Any dragon that no more than four Sarnorians own a dragon that did not freeze it*

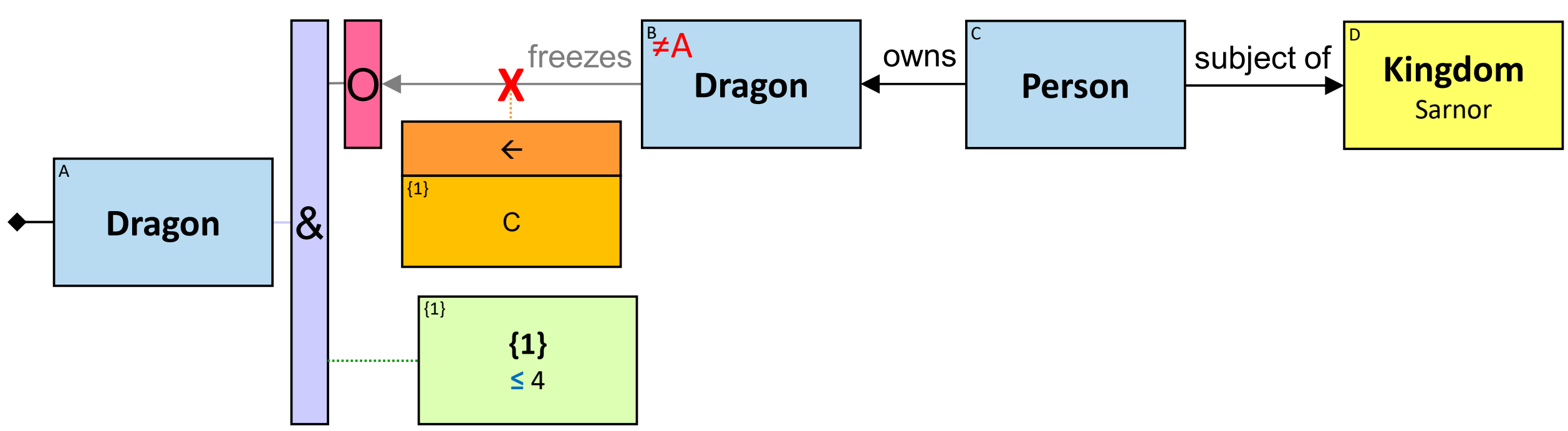

*Q206: Any dragon A that for each of no more than four Sarnorians C there is a dragon C does not own that did not freeze A*

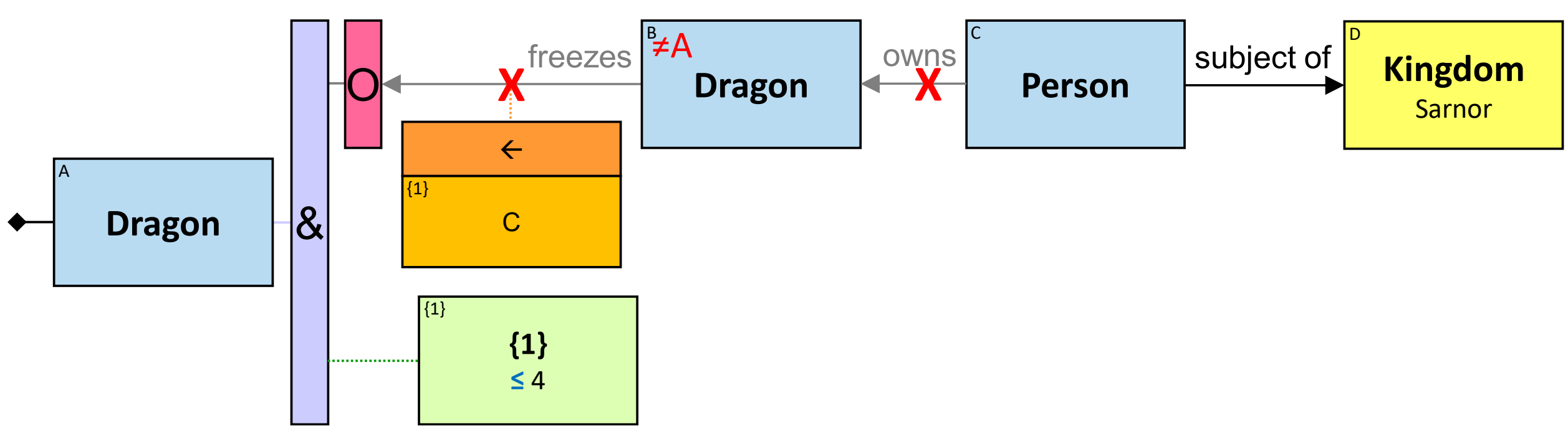

*Q305: Any person A that the number of horses A owns plus the number of dragons A owns – is at least ten* (two versions)

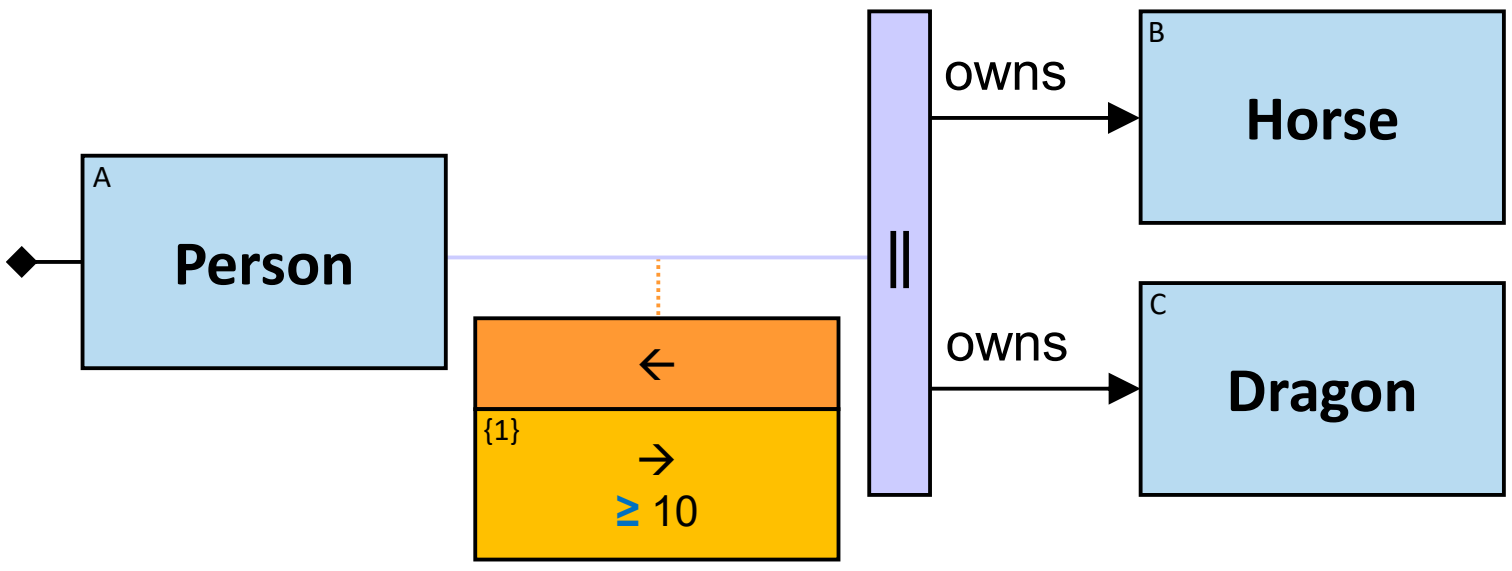

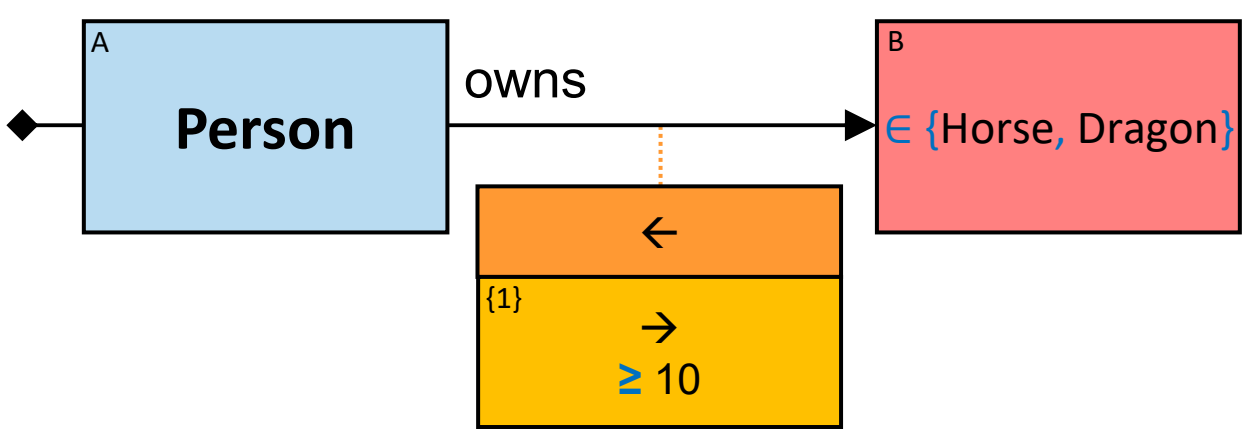

*Q121: Any dragon that froze or fired at at least ten dragons*

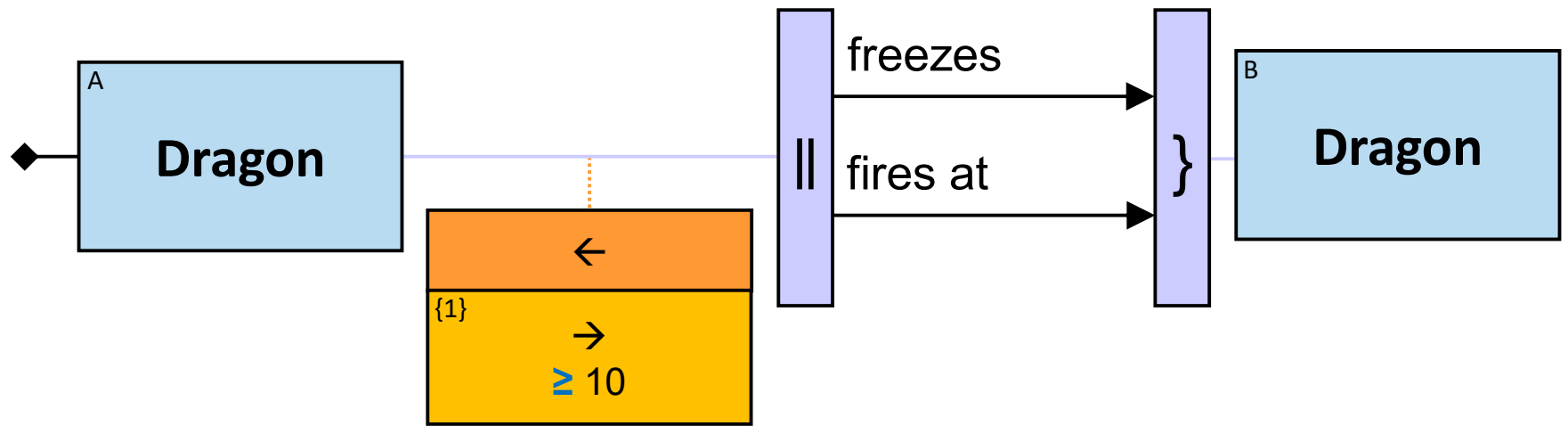

('→' is the entity directly right of the combiner (in this example – B))

Note that any dragon that was both frozen and fired at – would be counted only once.

*Q122: Any dragon that fired at dragon B and fired at a dragon that fired at B – for at least ten different B's*

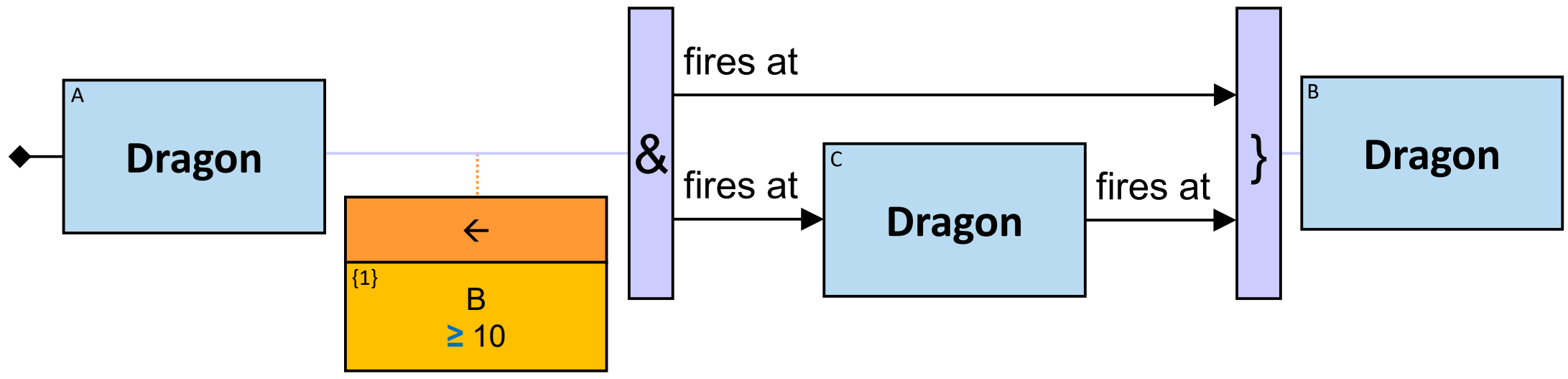

*Q294: Any dragon that fired at Balerion and at least nine other dragons*

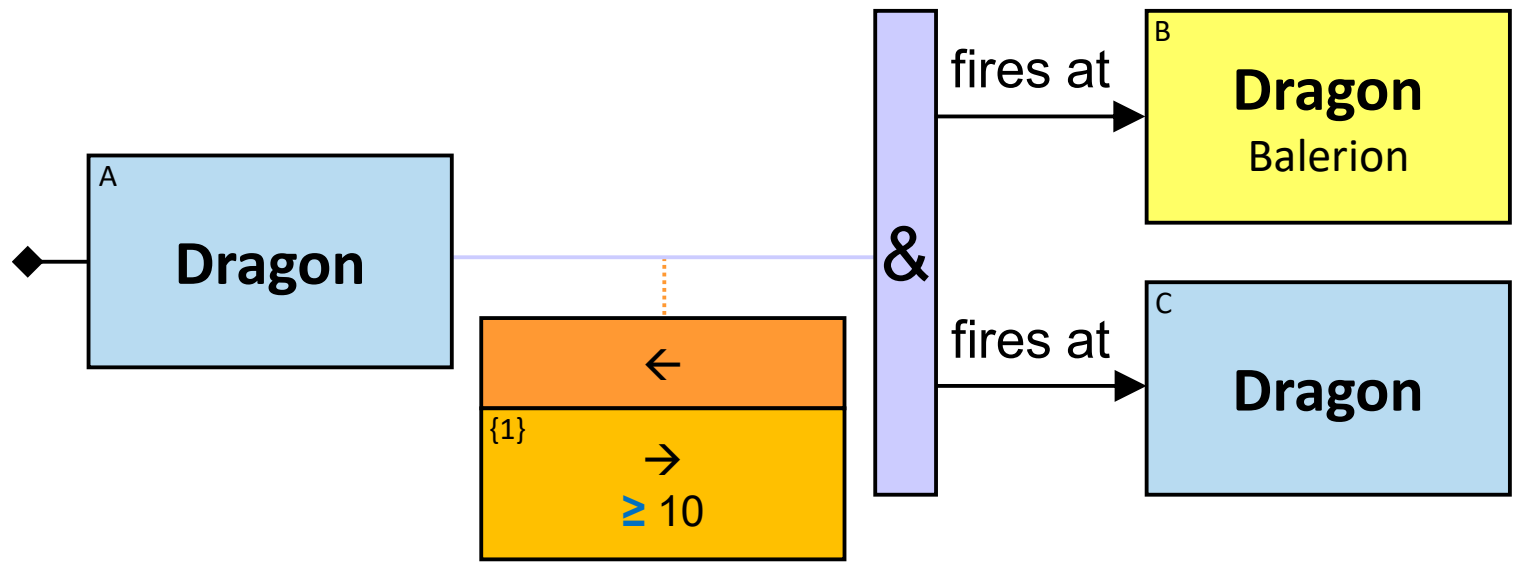

***Q175:** Any dragon that froze some dragon at least once and fired at some dragon at least once. The number of dragons it froze/fired at – is at least ten*

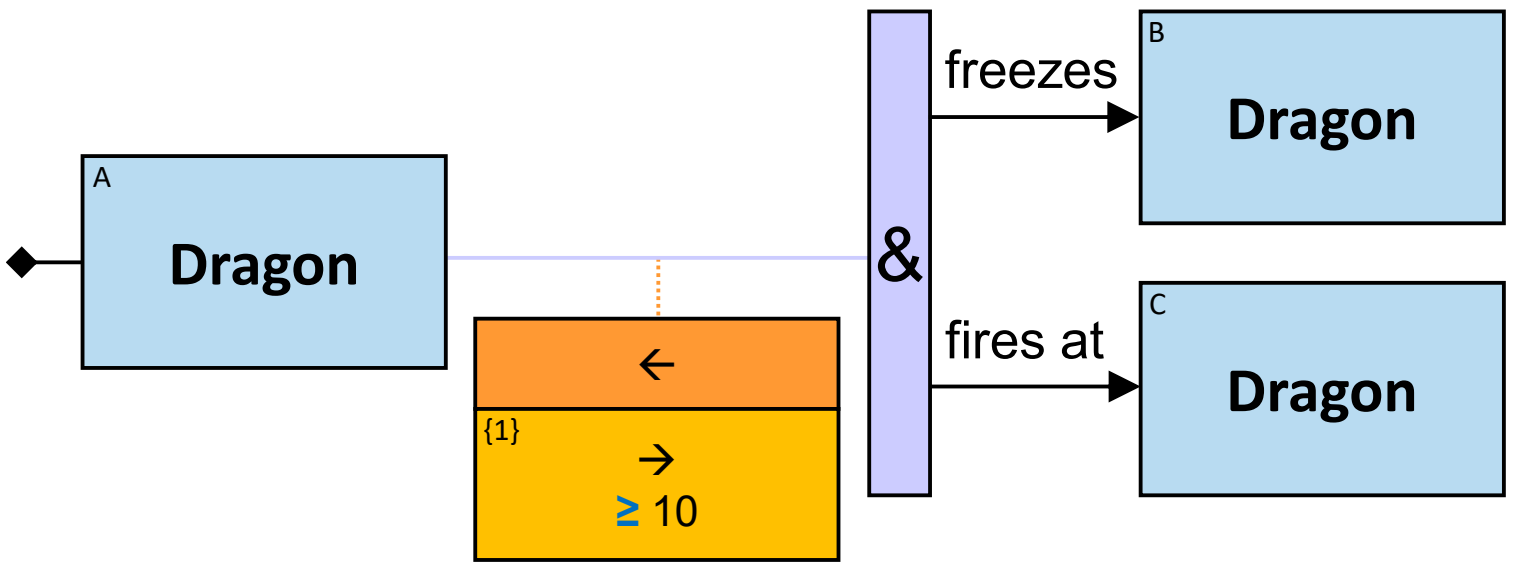

Note that if a dragon were both froze and fired at – it would be counted only once. A1 counts *distinct* entity assignments.

***Q298:** Any dragon A where (i) there is at least one black dragon A froze (ii) there is at least one white dragon A did not freeze, (iii) there is no gold dragon A froze, (iv) there is no silver dragon A did not freeze, and (v) the number of black or red dragons A froze plus the number of white and blue dragons A did not freeze – is at least ten. Report all black and red dragons that A froze and all white and blue dragons that A did not freeze*

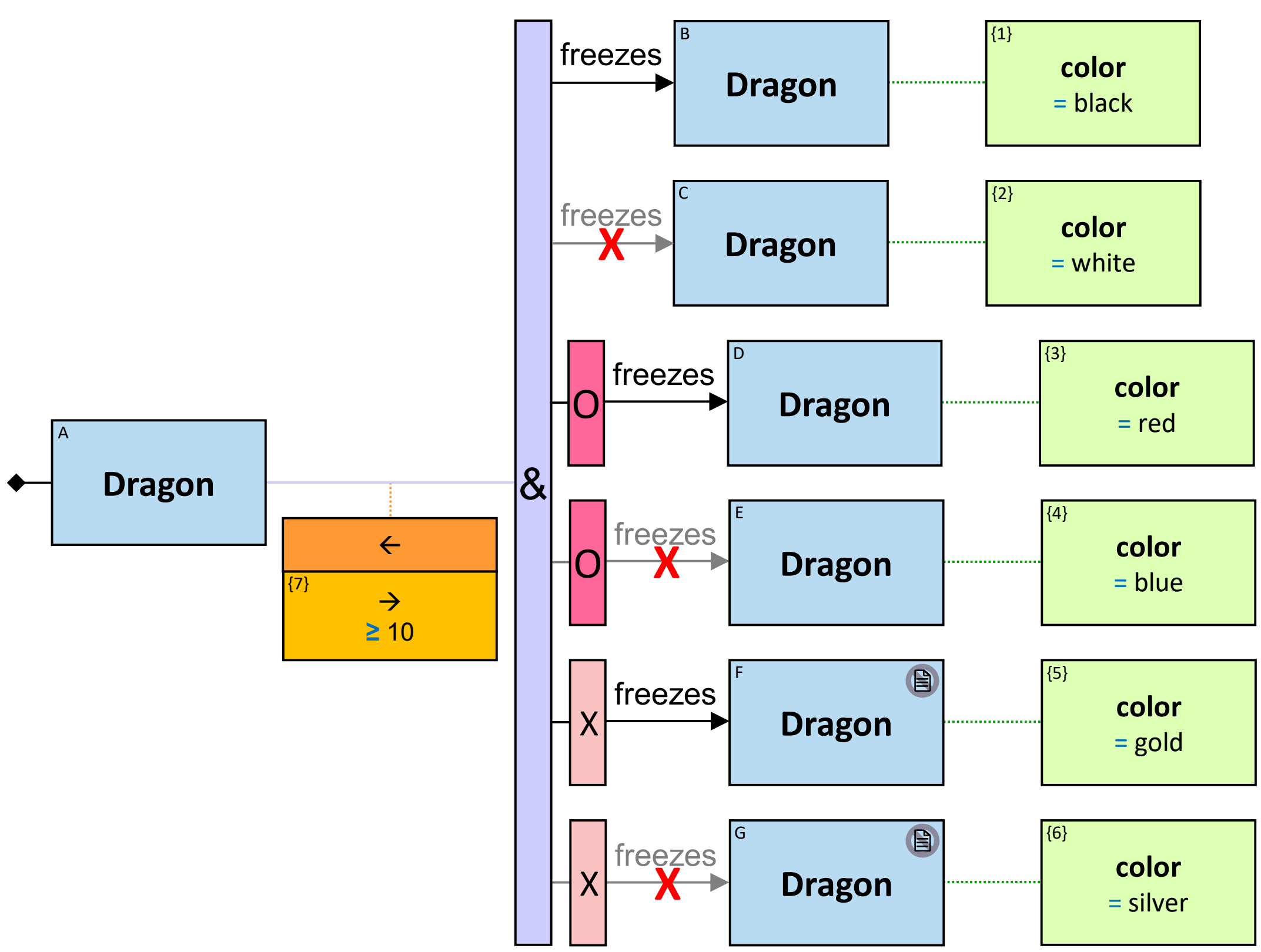

Note that the entities that are right of a negator are not aggregated, but the entities that a right of a relationship/path-negator and no negator are aggregated. Compare with Q358.

***Q176:*** *Any dragon that either (i) froze at least one dragon and fired at at least one dragon it did not freeze. The number of dragons it froze/fired at is at least ten, or (ii) froze at least ten dragons*

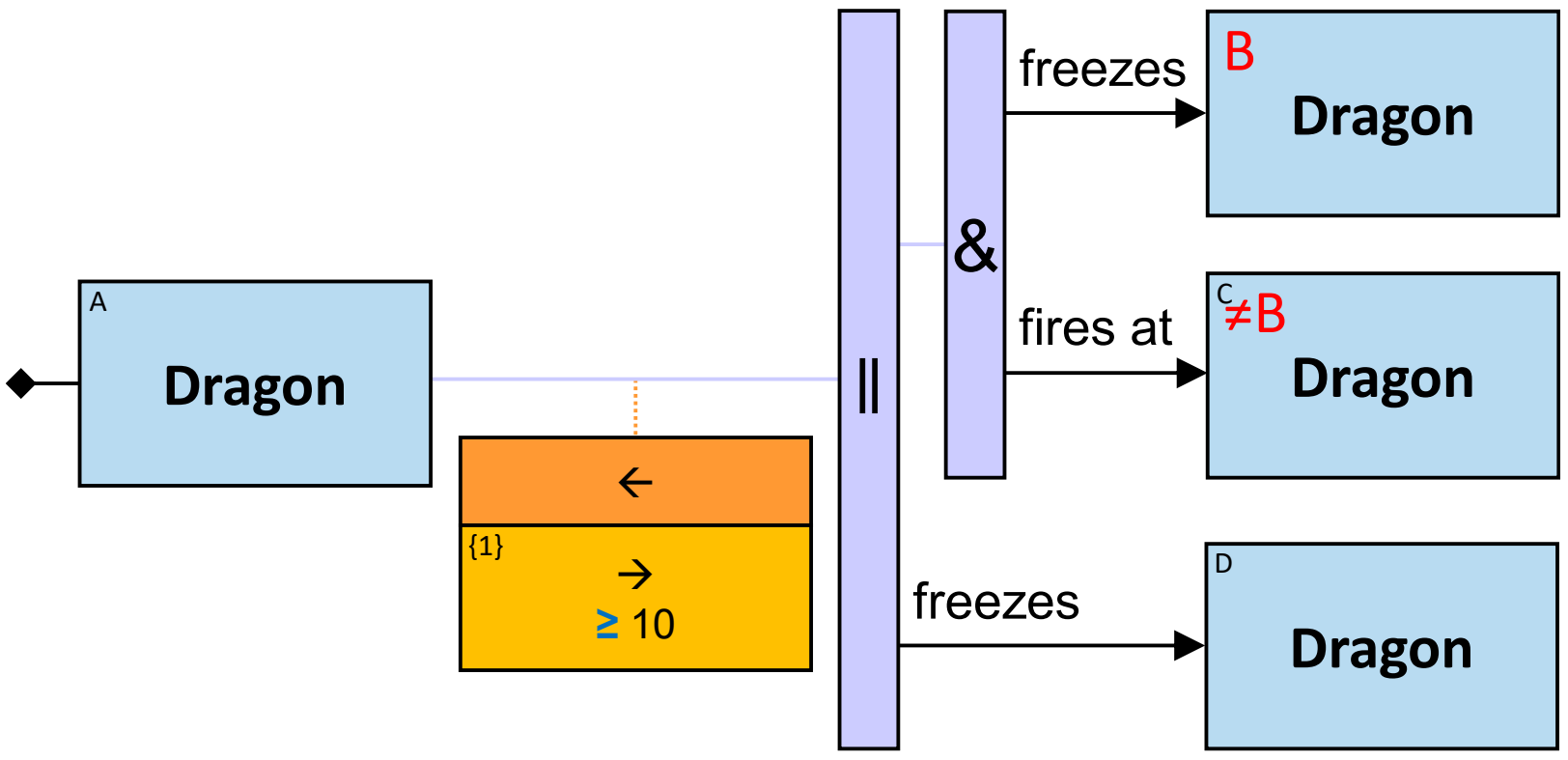

***Q293:*** *Any person A that at least 80% of the horses owned by A and/or by (at least) one of A's parents – are jointly owned by A and by (at least) one of A's parents*

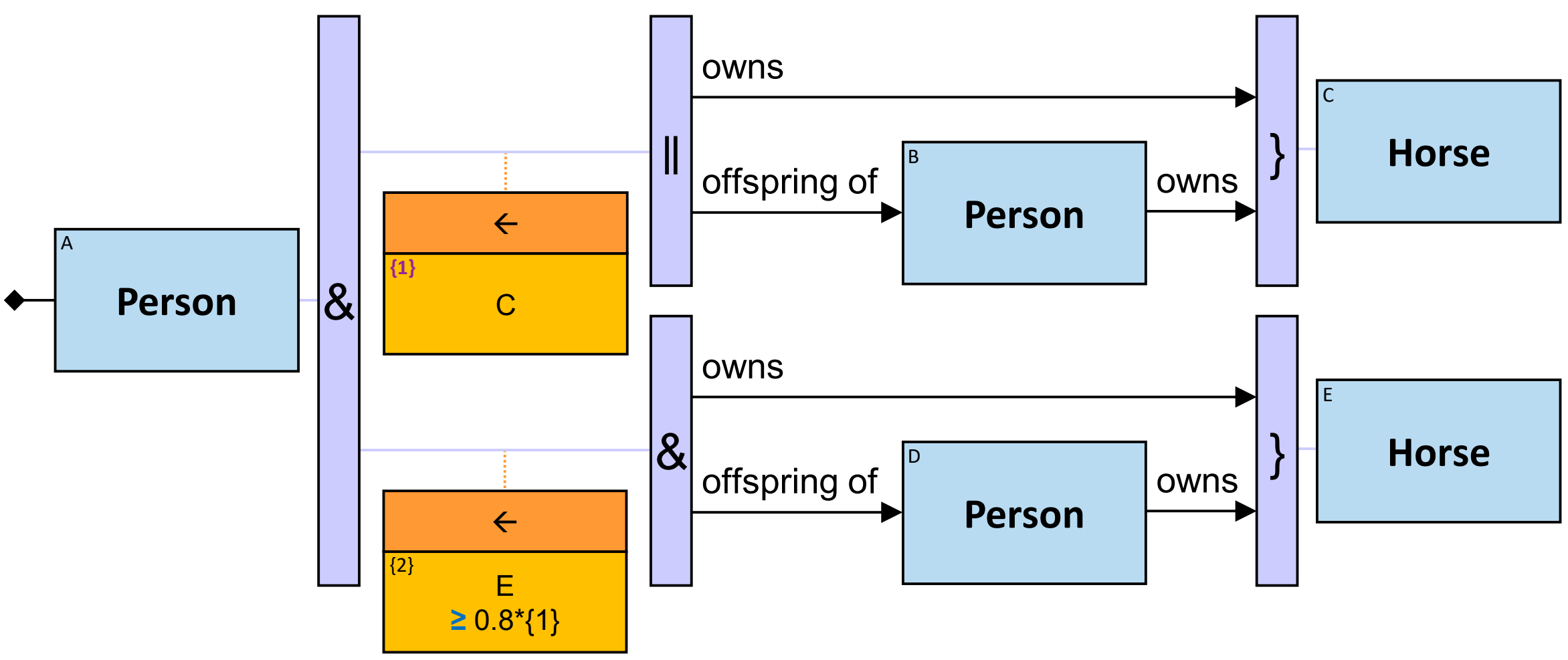

***Q288:*** *Any person A who owns a dragon that froze more dragons than any dragon owned by any of A's ancestors*

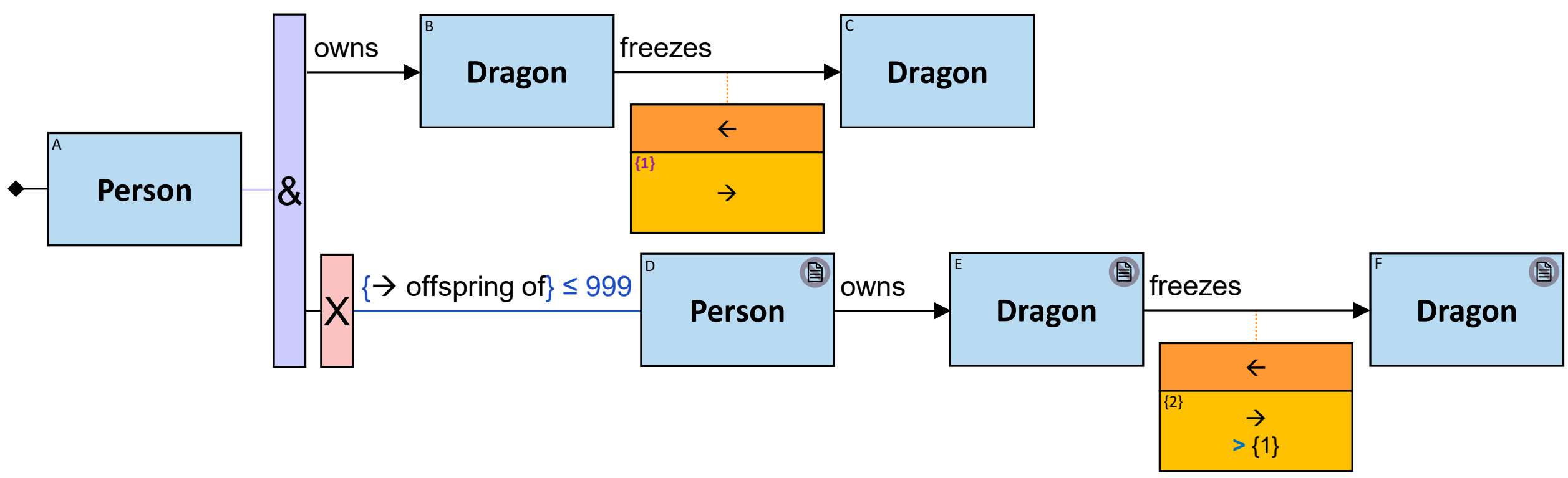

*Q290: Any path of length up to ten between Rogar Bolton and Robin Arryn that does not contain hubs (in this pattern – hubs are entities with degree ≥ 1000)*

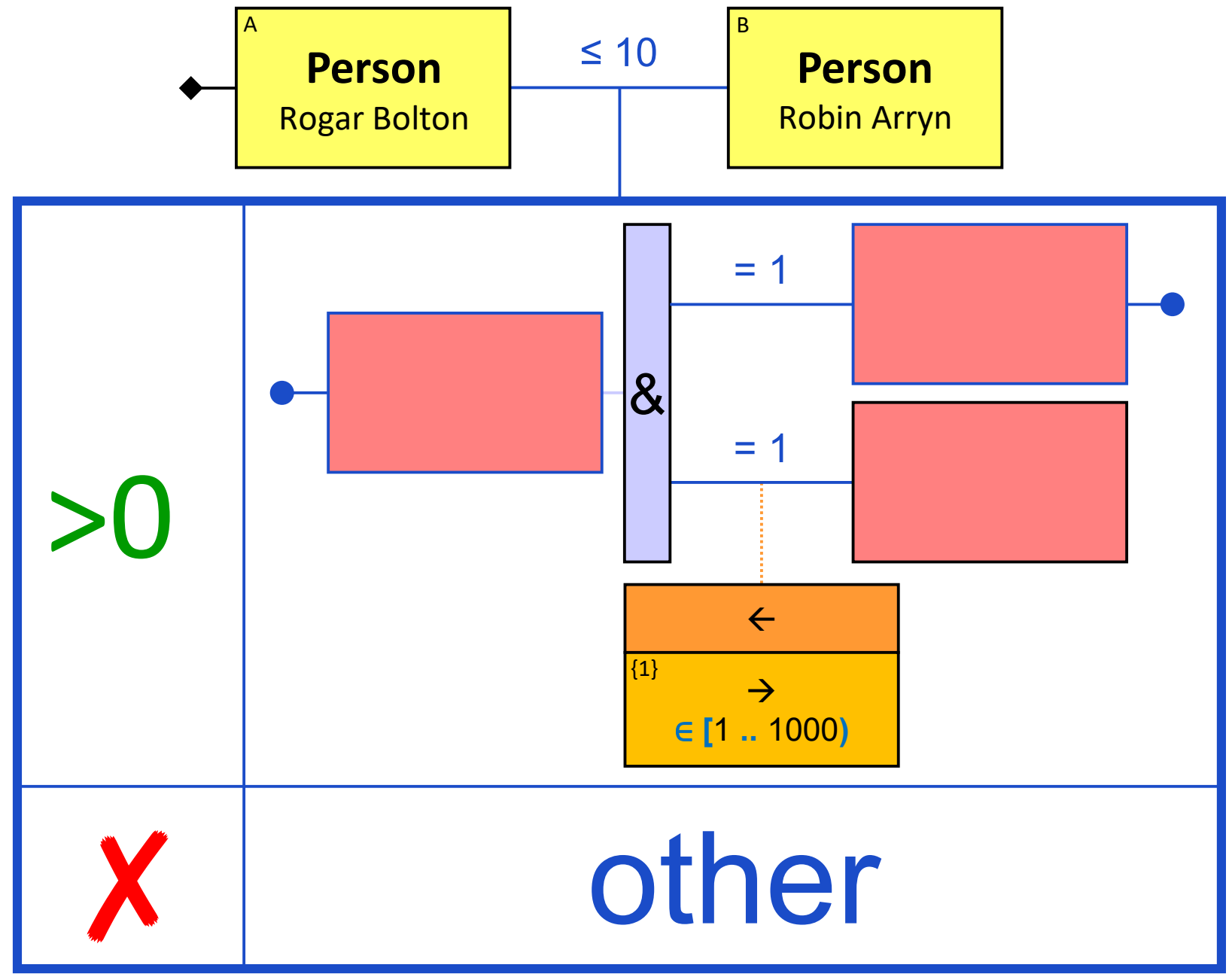

*Q329: Any Sarnorian who has paths of length up to six to more than five Omberians, each of these paths passes through Rogar Bolton*

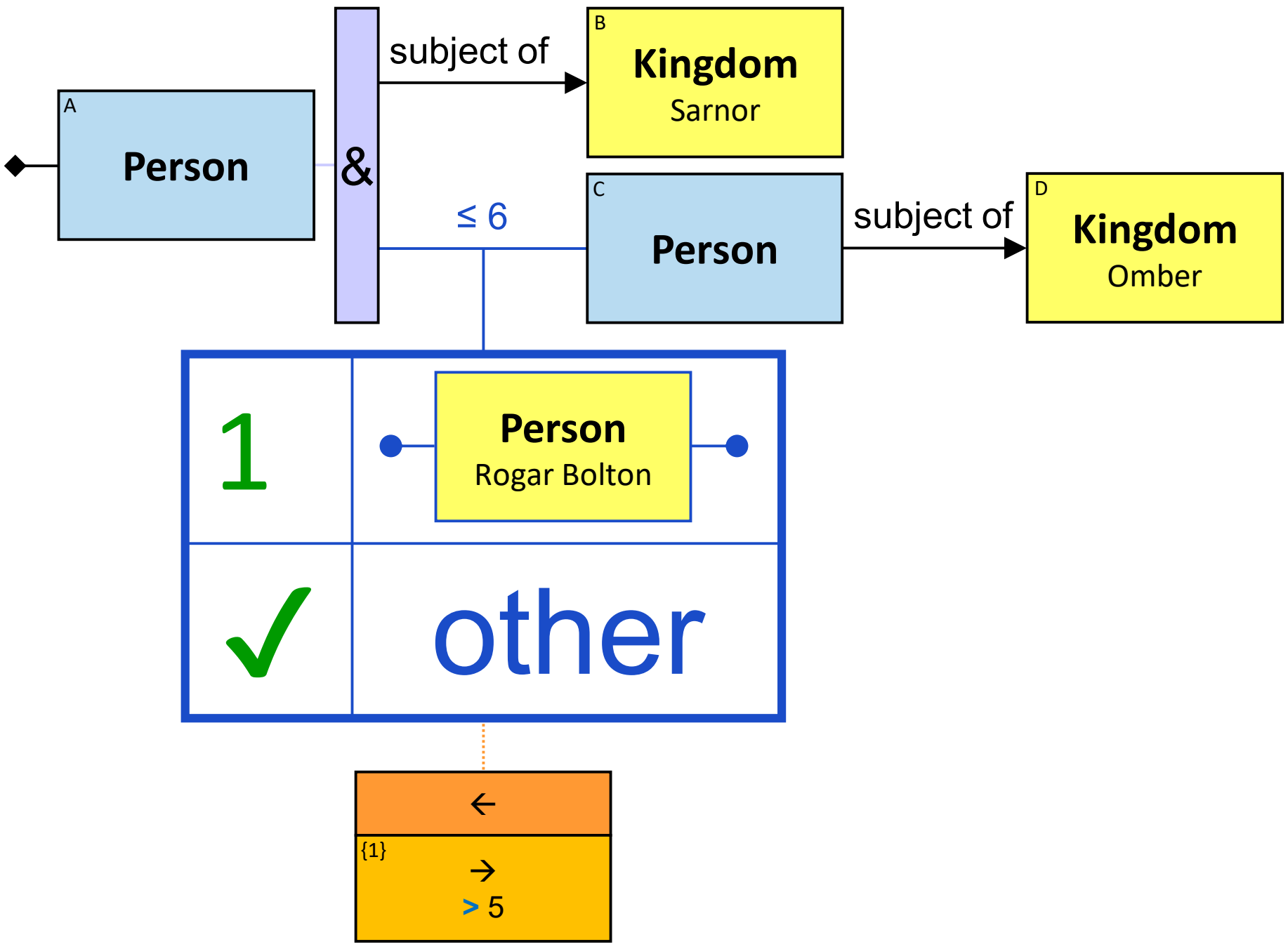

*Q295: Any person A that the number of dragons A owns plus the number of dragons A does not own that A's dragons fired at – is at least ten*

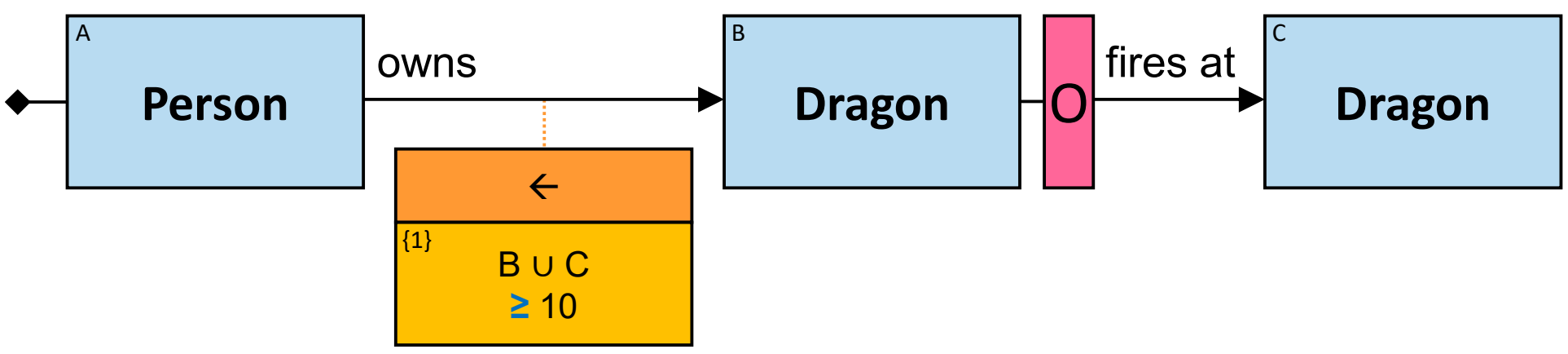

Note that if any of A's dragons were fired at one of A's other dragons - it would not be counted twice.

*Q248: Any pair of dragons (A, C) where A froze more than ten dragons that froze C*

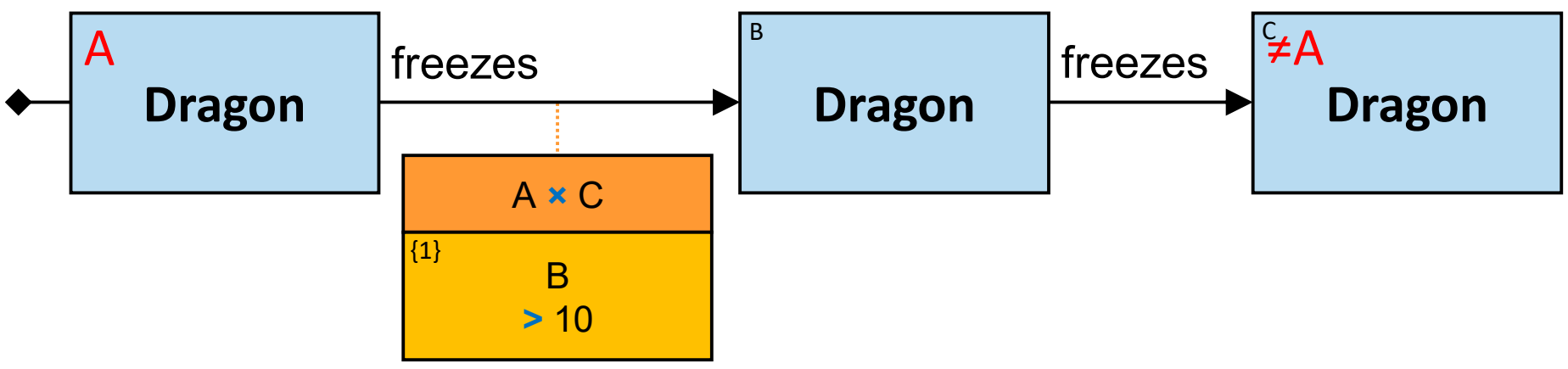

{1} is a property of each unique assignment to A × C in *S*.

*Q243: Any pair of people (A, D) where at least five of A's dragons froze one or more D's dragons*

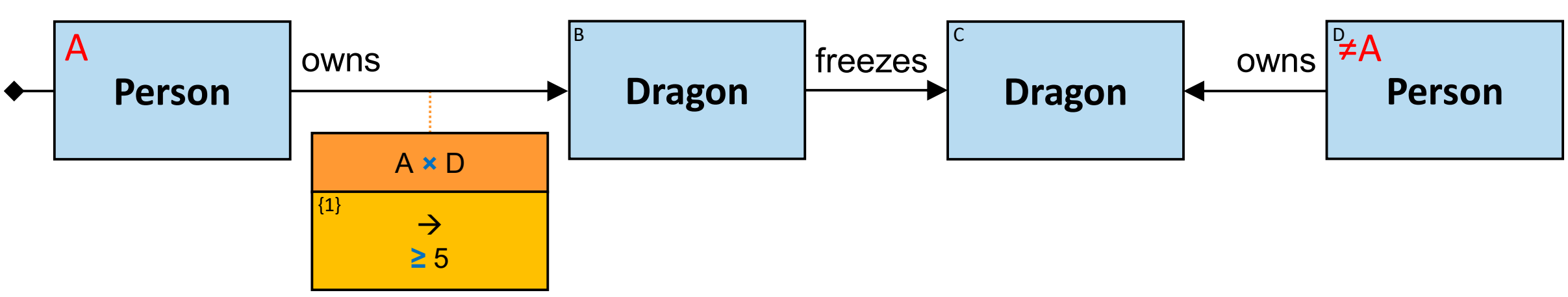

*Q244: Any pair of people (A, D) where at least five of A's dragons froze D's dragons, and at least five D's dragons were frozen by one or more A's dragons*

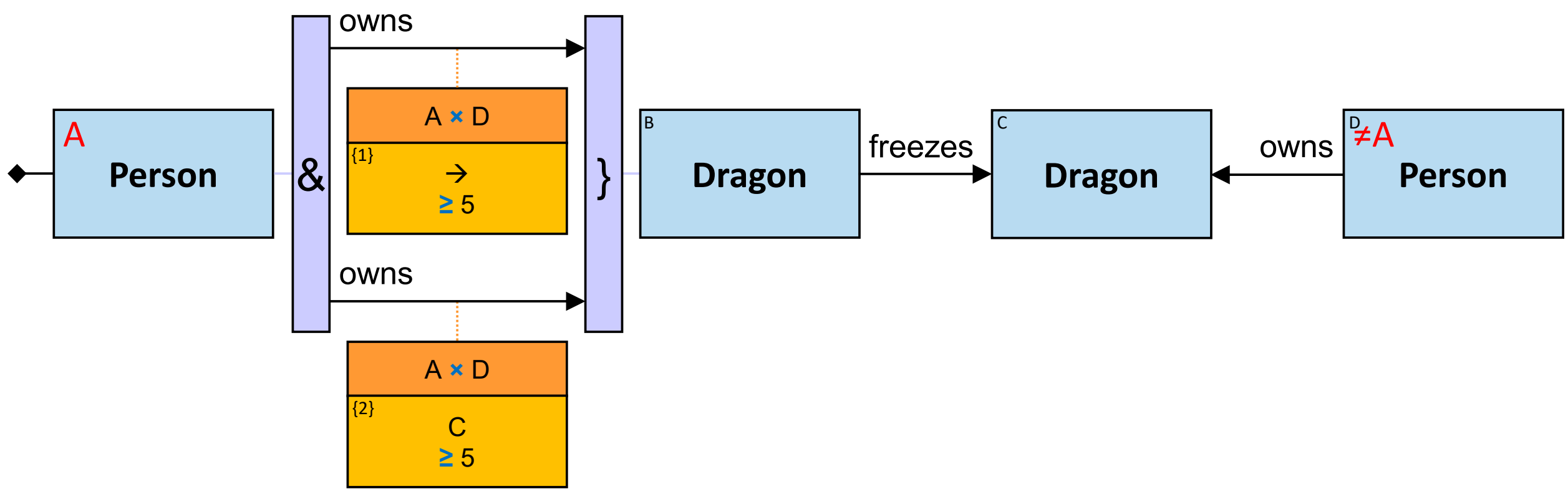

*Q27: Any person A where there are less than 500 horses of the same colors as A's owned horses*

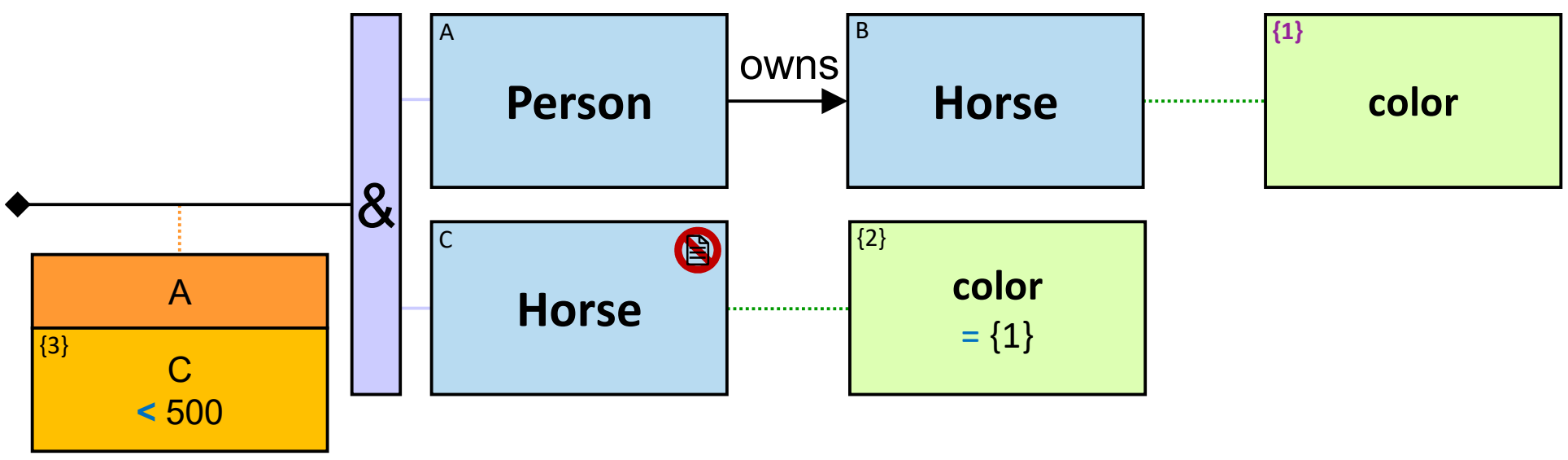

*Q28: Any person who owns a horse of a rare color (there are less than 500 horses of that color)*

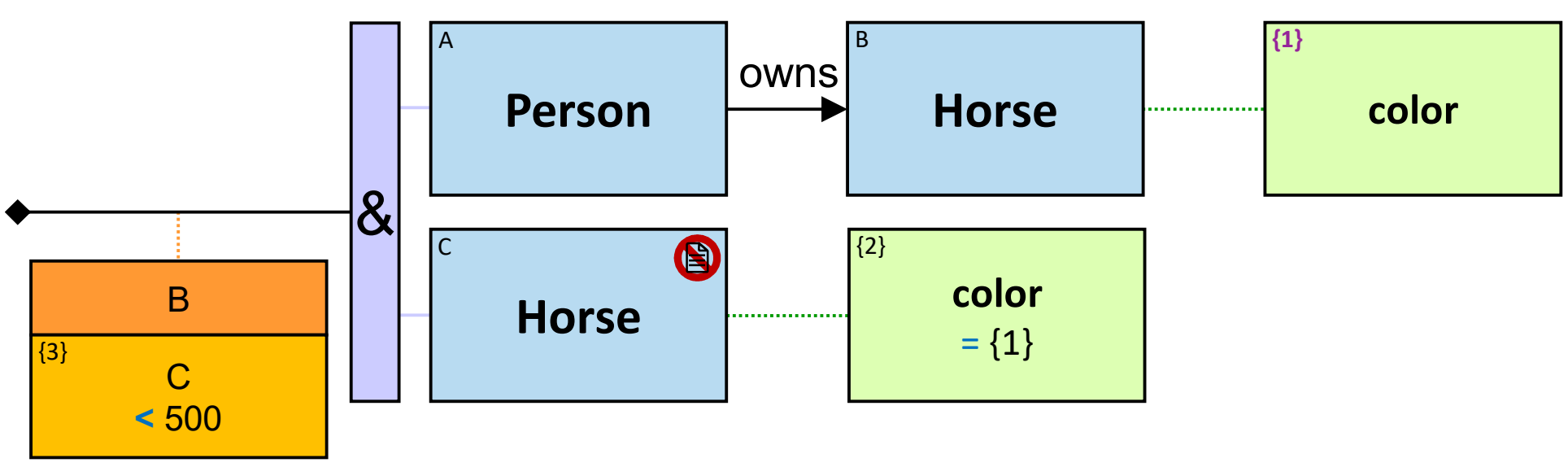

Patterns Q346, Q352, and Q355, are similar in structure, but the aggregation is different:

*Q346: Any dragon that froze at least ten dragons of the colors of horses of its owners*

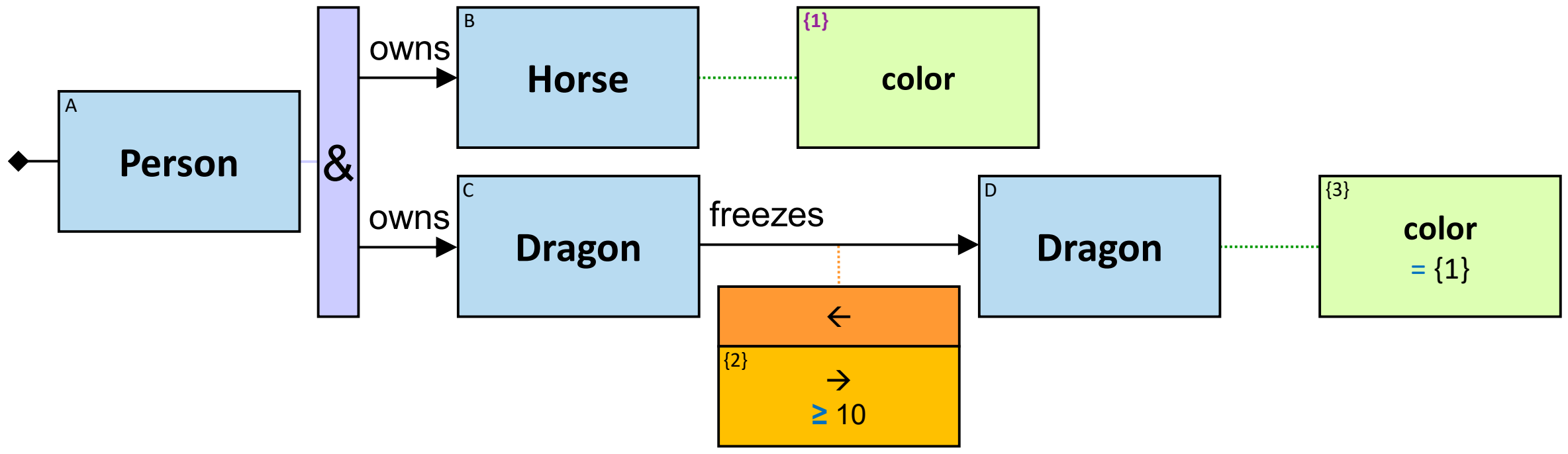

*Q352: Any dragon that froze at least ten dragons of the colors of horses of one of its owners*

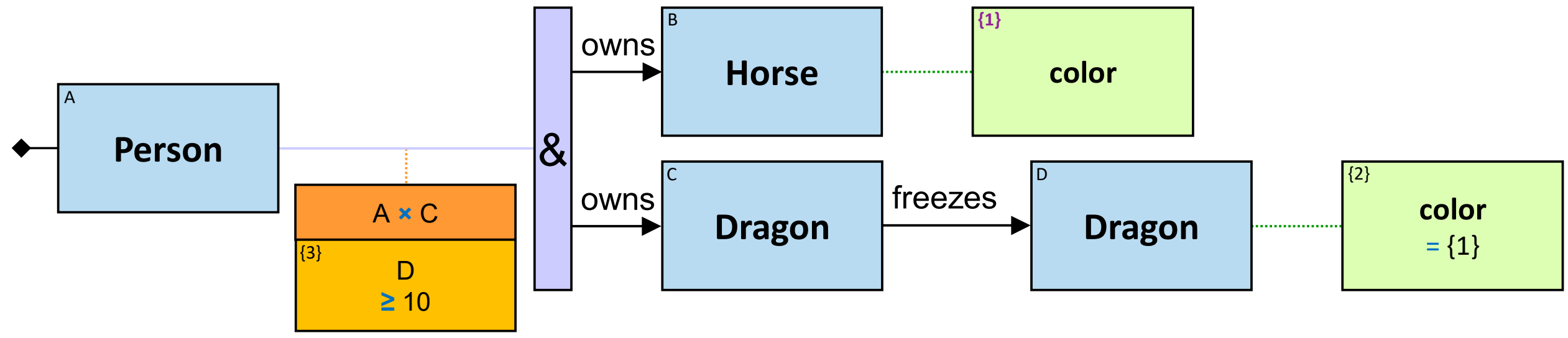

*<b>Q355:</b> Any dragon that froze at least ten dragons of the color of one horse of one of its owners*

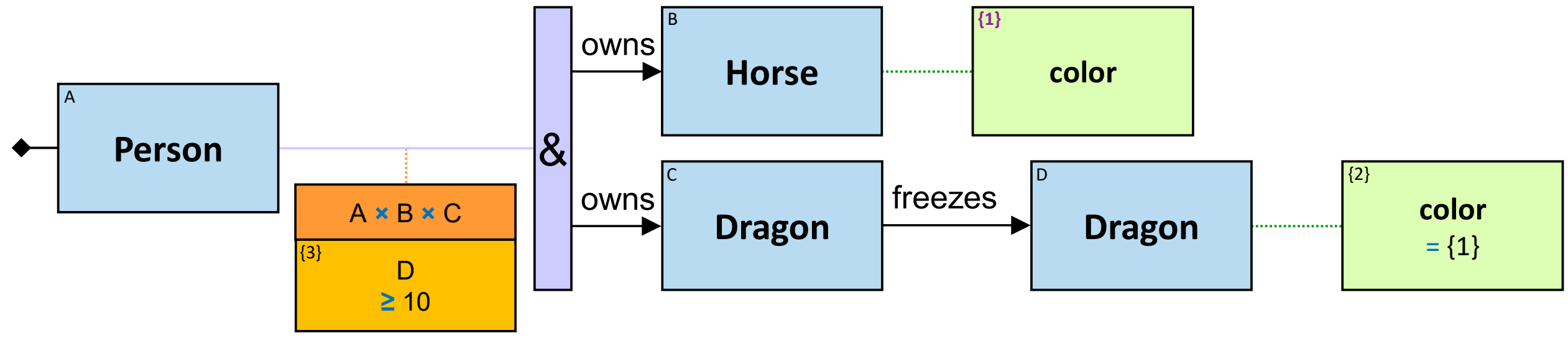

*<b>Q207:</b> Any pair of dragons where the difference between the number of dragons that one of them froze and the number of dragons that the other froze is less than 10*

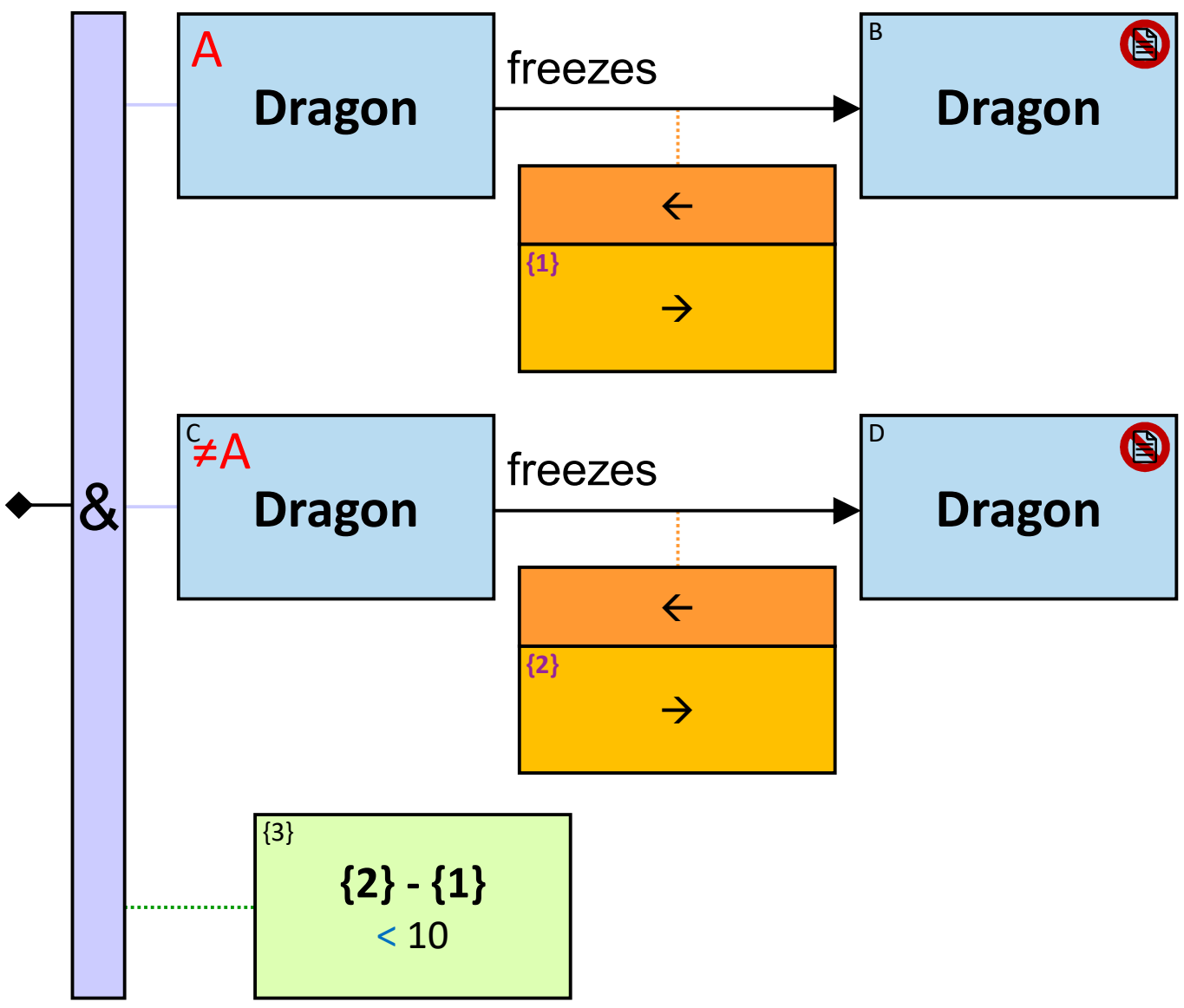

{3} is a property of each unique assignment to A × C in *S*.

*<b>Q279:</b> Any pair of people (A, E), each own dragons, where no A's dragon froze any of E's dragons*

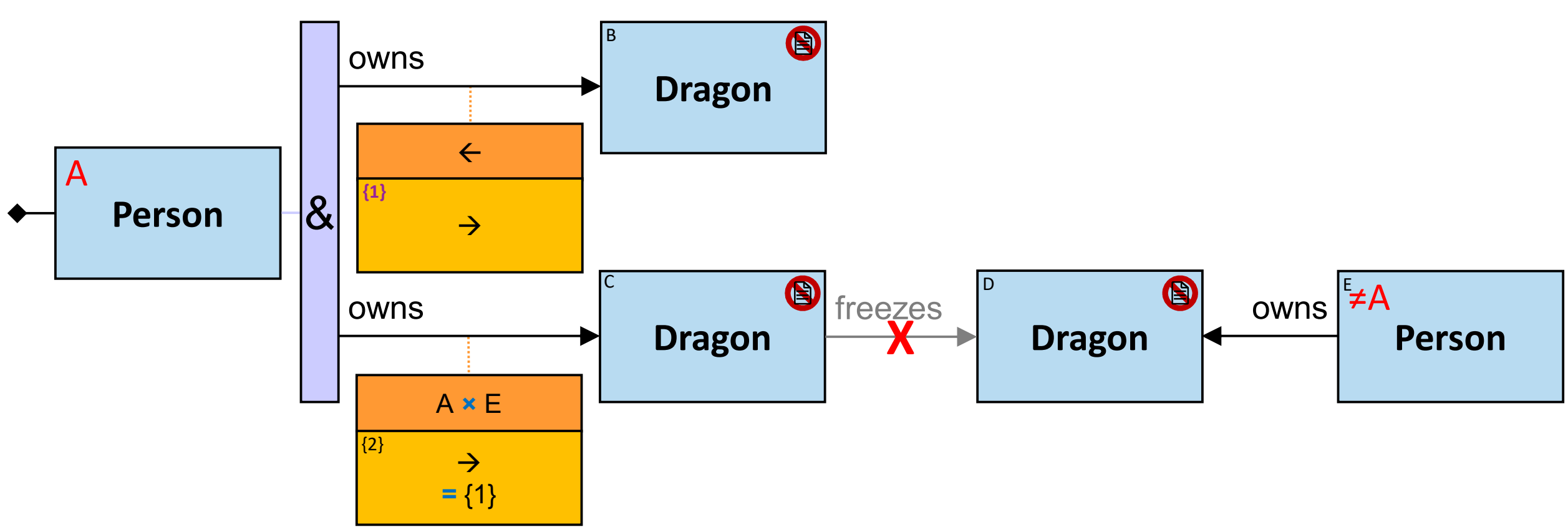

**Q82:** *Any dragon that froze all other dragons.* Alternative phrasing: *Any dragon for which there is no dragon (besides itself) it did not freeze* (version 3)

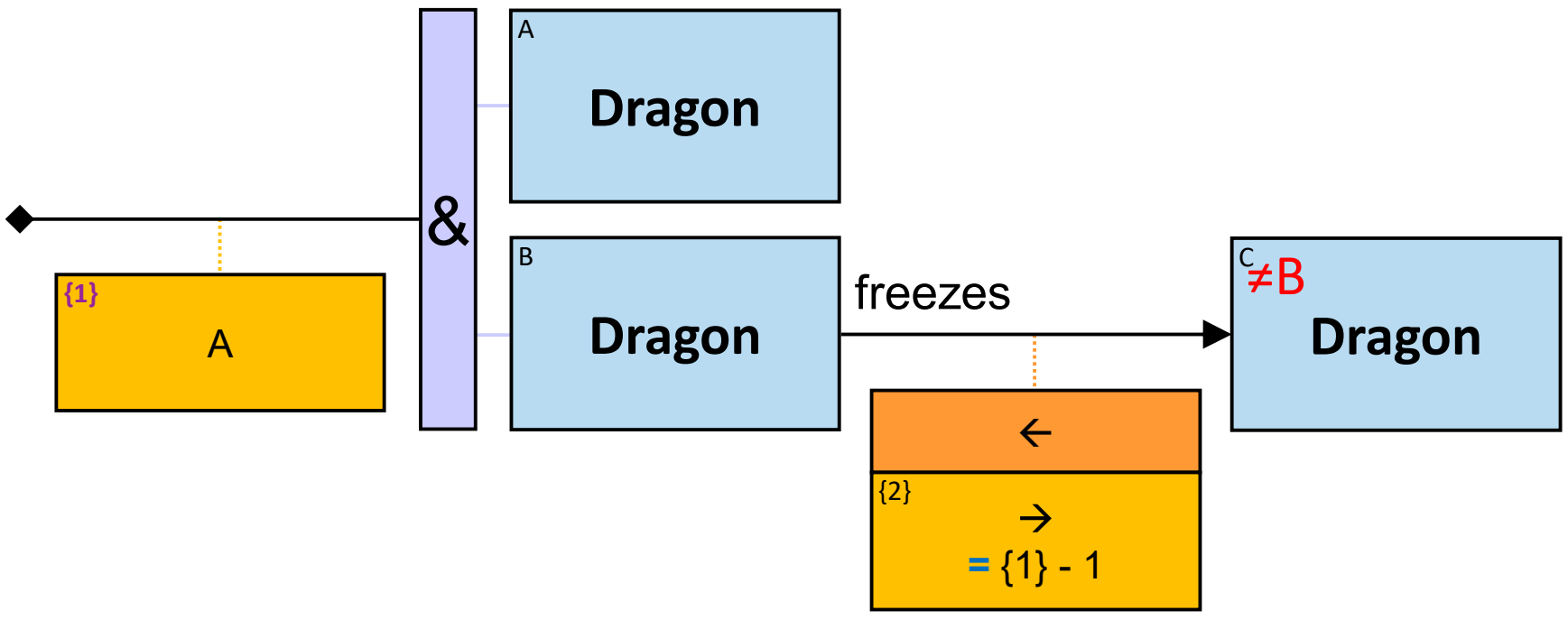

{1} is a global property.

In contrast to Q82v1 and Q82v2, the frozen dragons will be a part of the assignment.

**Q360:** *Any dragon that (froze or fired at) all other dragons.* Alternative phrasing: *Any dragon for which there is no dragon (besides itself) it did not (freeze or fire at)* (version 3)

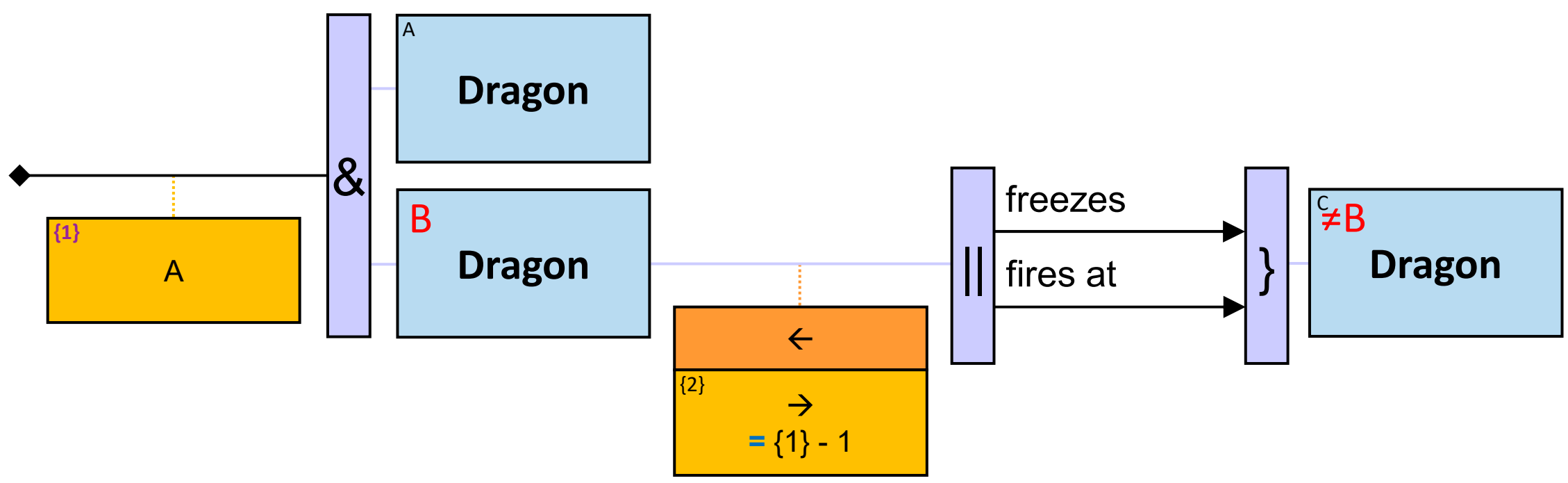

***Q375:** The difference between the number of white dragons and the number of black dragons*

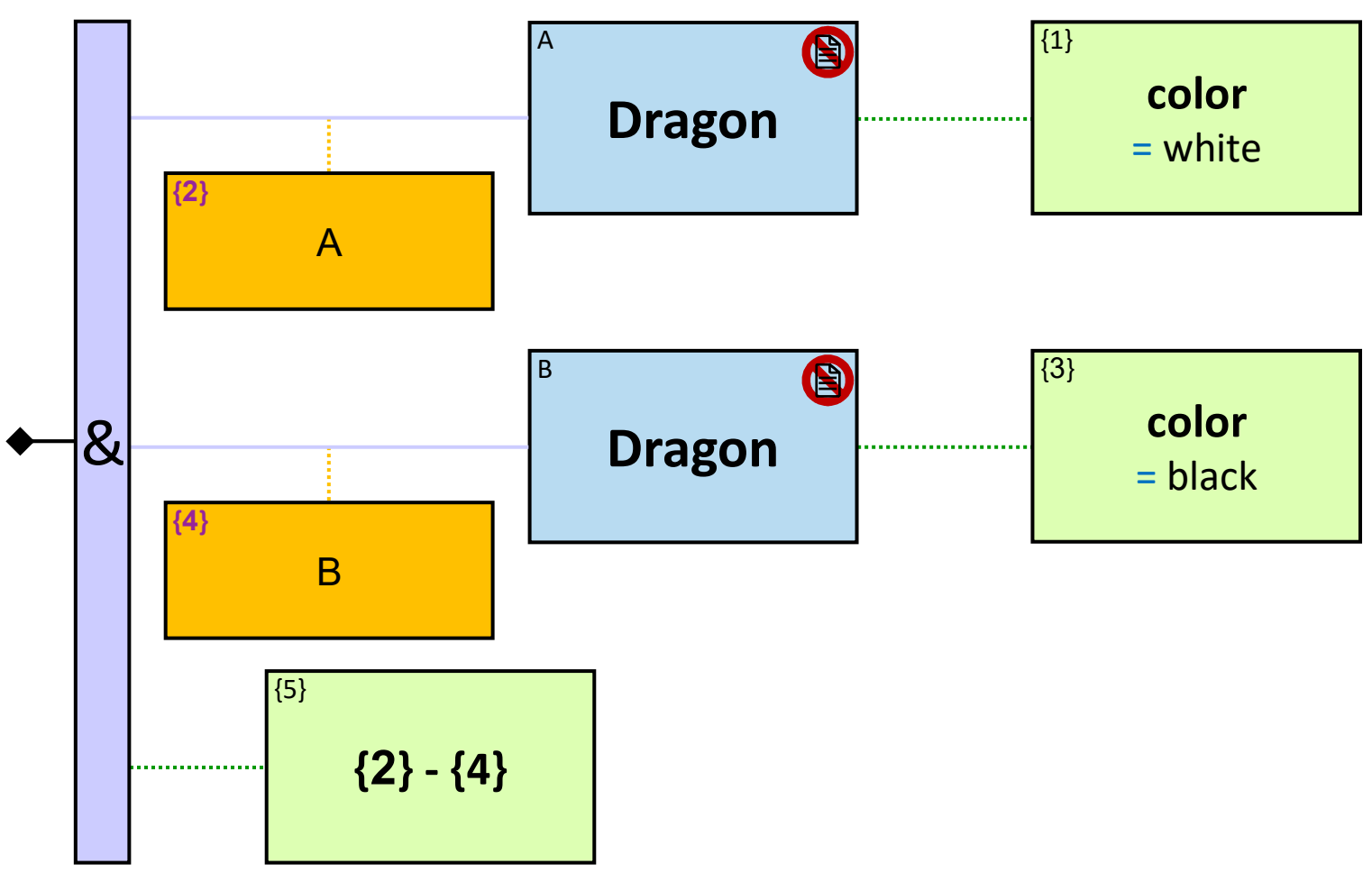

{2}, {4} and {5} are global properties. {5} is a global expression.

## 27. A2 AGGREGATOR

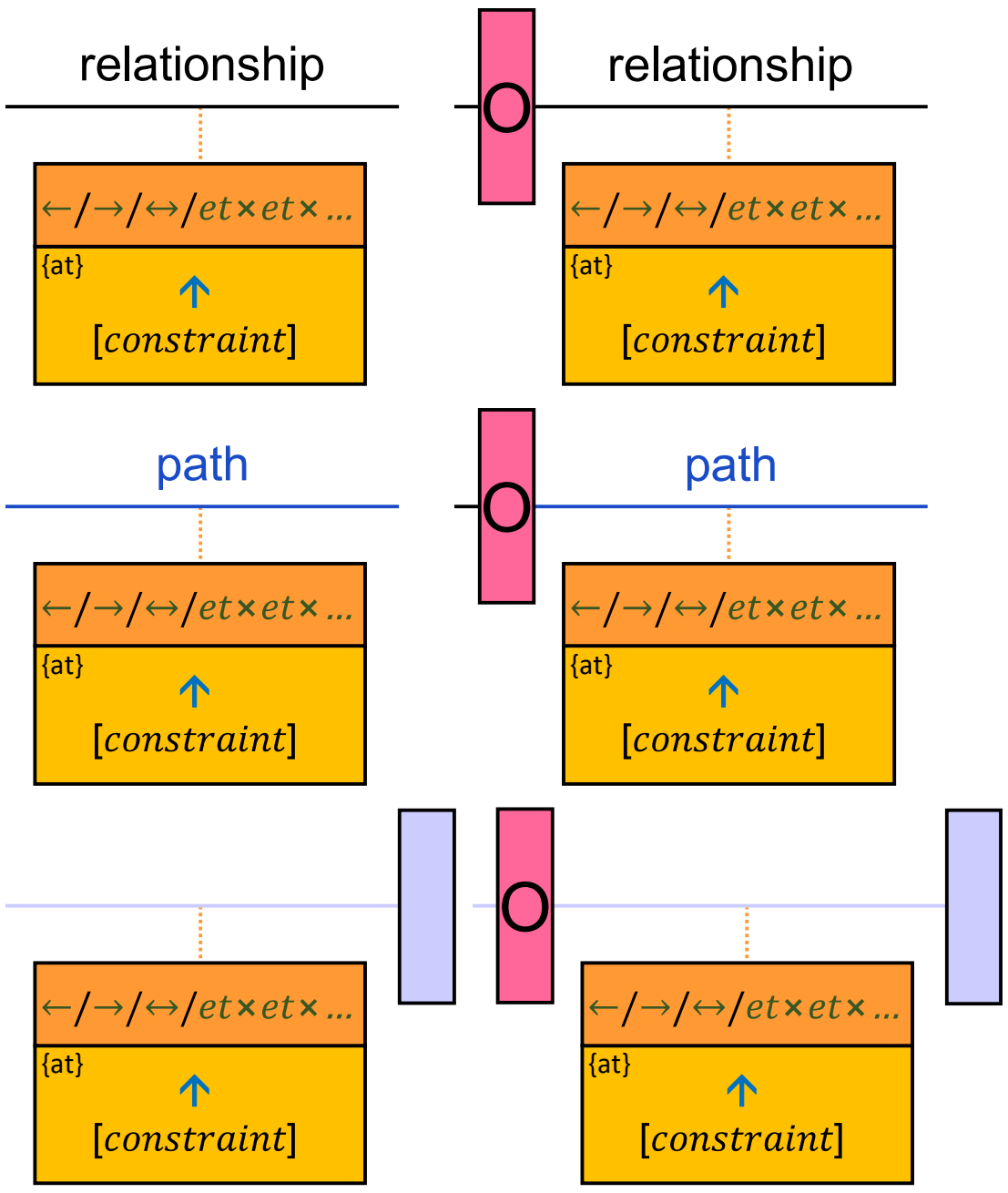

Let ***R*** denote a list where each element is a pattern-relationship/pattern-path. $R$ must be nonempty.

The lower part of an A2 aggregator contains the following elements:

- '↑' (an arrow pointing to the relationship/path on top of the aggregator):

  When A2 appears below a relationship/path – card($R$) = 1; $R[1]$ = the relationship/path

  When A2 appears below a quantifier-input – each element in $R$ is the relationship/path that follows one branch of the quantifier, excluding relationships/paths wrapped with a negator or a relationship/path-negator.

  Let ***RA(n, o)*** denote the list of all assignments to $R[o]$ in $S(n)$.

- **aggregation-tag**: $at(n) = |RA(n, 1) \cup RA(n, 2) \cup \dots |$

  We are using *card(union(all assignments to all elements in R))* instead of *sum(card(assignment to one element in R))* since two elements in $R$ may have the same assignment (see Q100), and we are counting *distinct* assignments to all elements in $R$ per $n$.

  When an optional part has no assignments, A2 aggregation-tags defined right of the 'O' are evaluated to zero (see Q347).

- an optional **constraint** on $at(n)$

  **For each *n*: *S(n)* is reported only if *at(n)* satisfies the constraint**

  Note that a '$= 0$' constraint cannot be satisfied, as such constraint means that there are no assignments to the pattern excluding this aggregator. Similarly:

  - no constraint is equivalent to '$> 0$' constraint

- ‘≠ *expr*’, ‘< *expr*’, and ‘≤ *expr*’ constraints are not satisfied when *at(n)* is zero
- ‘= *expr*’, ‘≥ *expr*’, and ‘∈ *[expr1, expr2]*’ constraints are not satisfied when *at(n)* is zero even if *expr* is evaluated to zero

A2 location:

- Below the relationship/path whose assignments are counted.

  A relationship/path with an A2 below it may be wrapped with an ‘O’.

- Below the quantifier-input (except a *None* quantifier) that assignments to relationships/paths on its right are counted (see Q297, Q174).

  A quantifier-input with an A2 below it:

  - may be wrapped with an ‘O’ (except at the pattern’s start)
  - at least one branch of the quantifier (or sub-branch of a sequence of quantifiers) must start with a relationship/path with no relationship/path-negator that is not wrapped with a negator nor preceded by a *None* quantifier.

***Q72:*** *Any dragon that was frozen exactly ten times* (two versions)

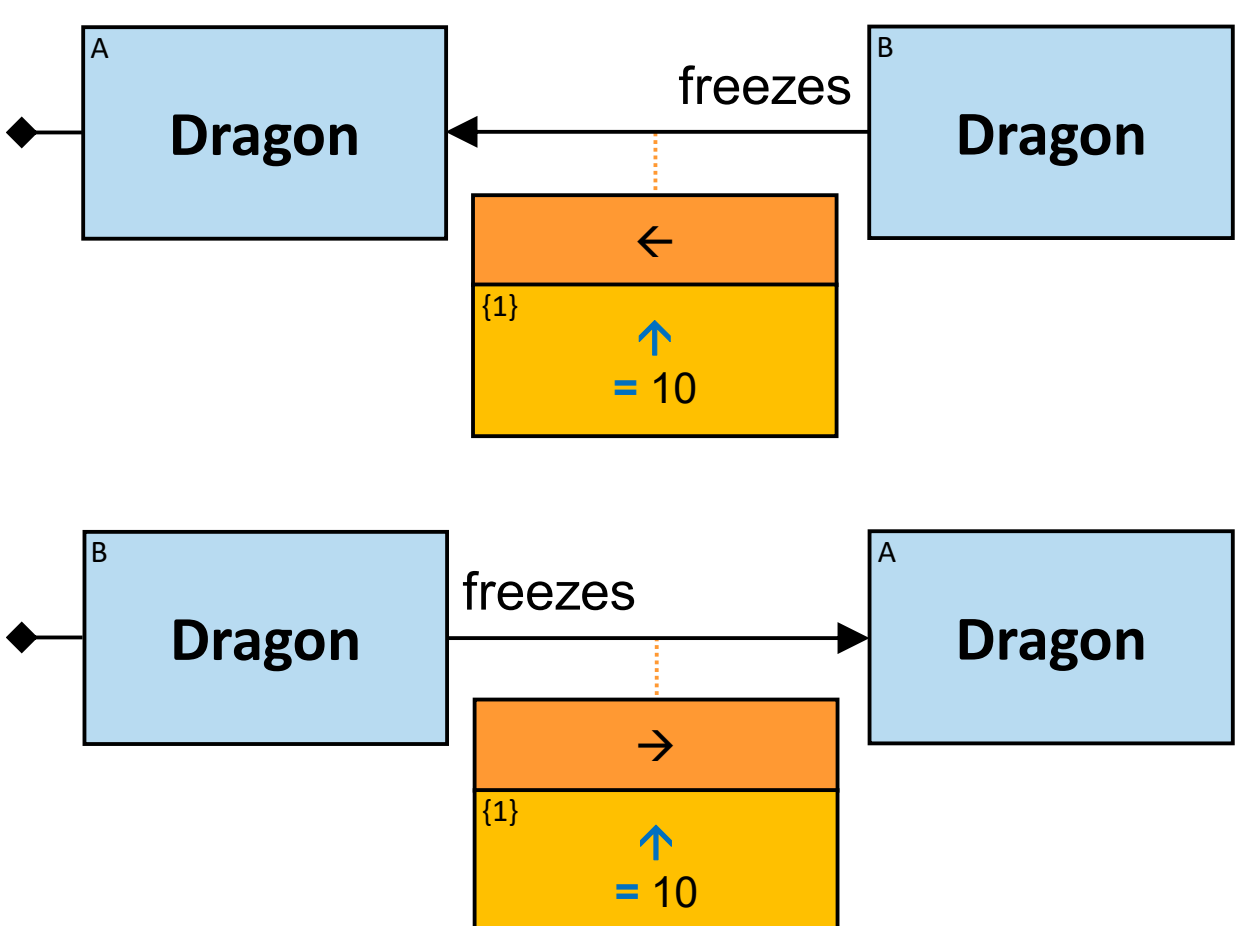

***Q71:*** *Any dragon that froze dragons more than ten times*

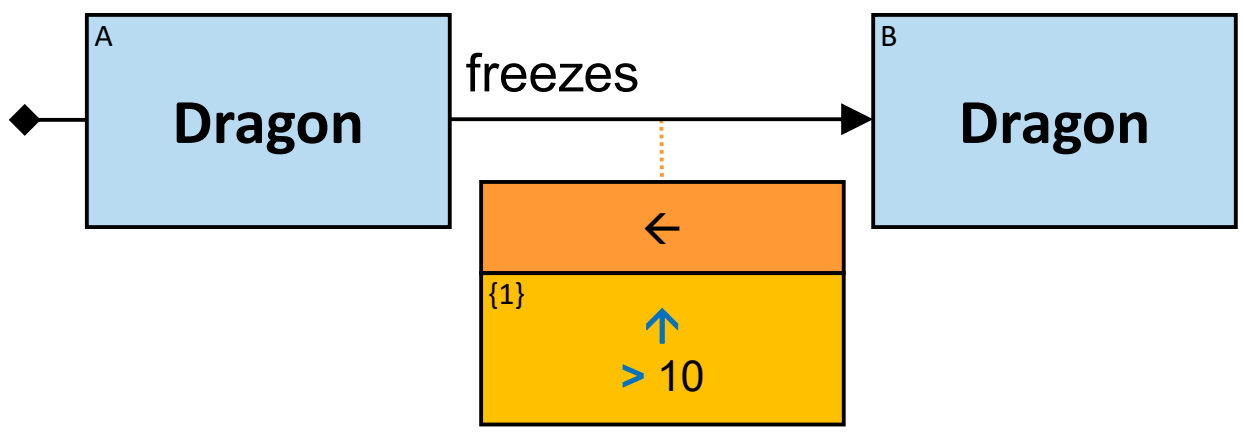

***Q73:*** *Any dragon that did not freeze dragons or froze dragons no more than ten times* (two versions)

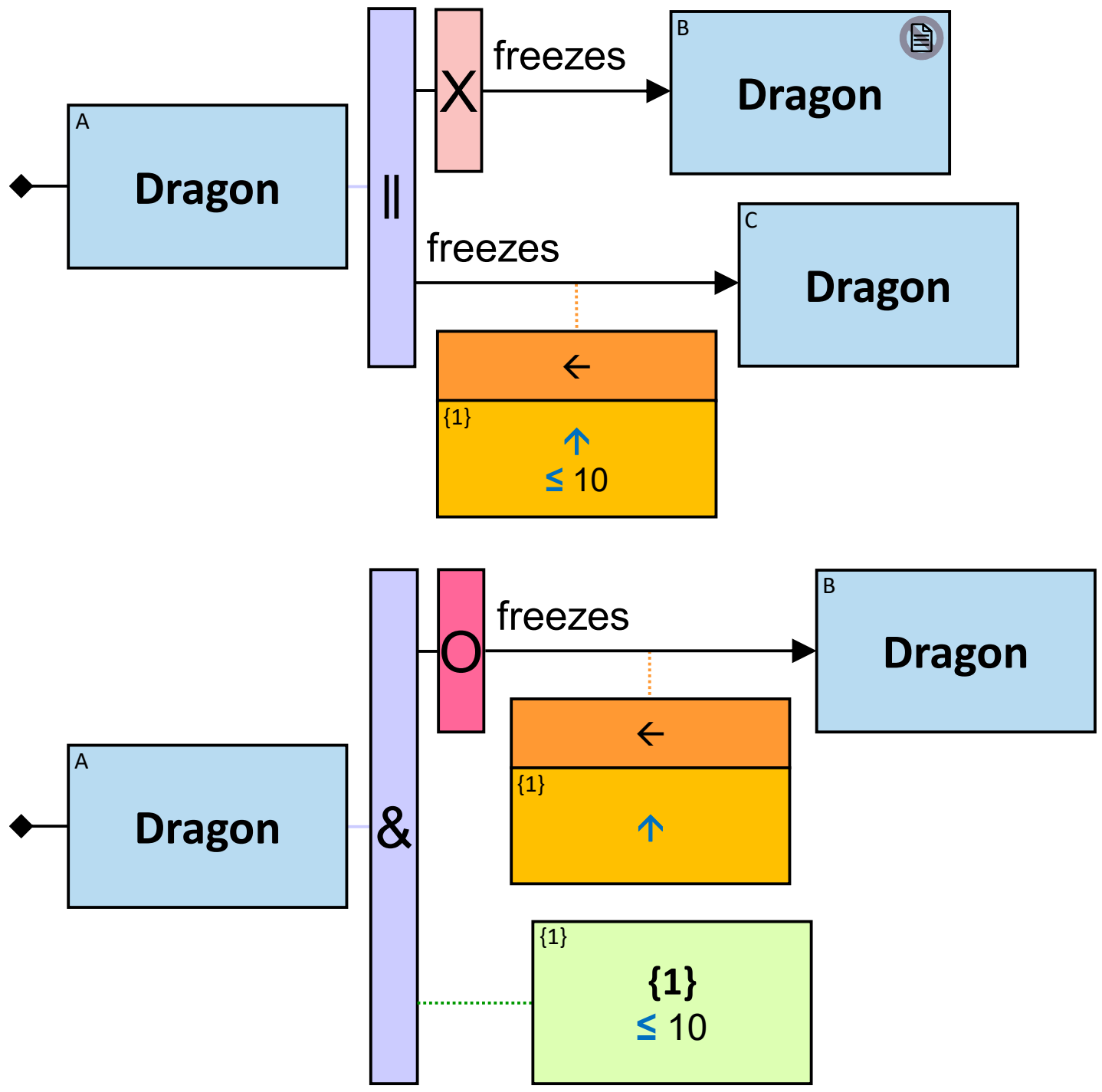

***Q123:*** *Any dragon that either froze or fired at dragons – at least ten times*

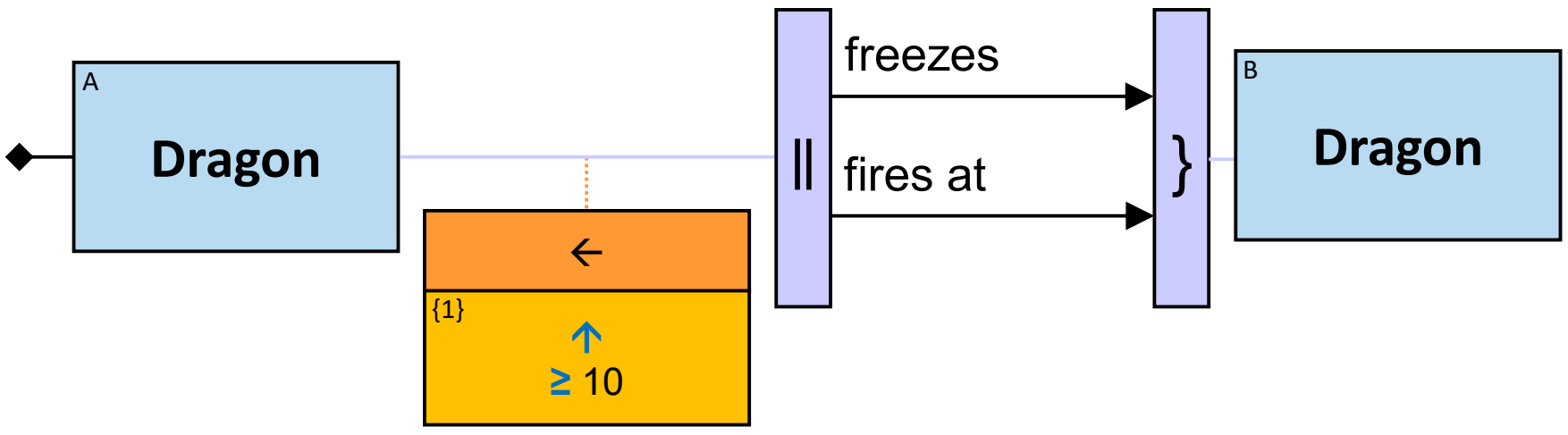

(Per assignment to A – counting the number of assignments to the relationship directly right of the quantifier)

***Q104:*** *Any person who owned white horses at least ten times (same or different horses)*

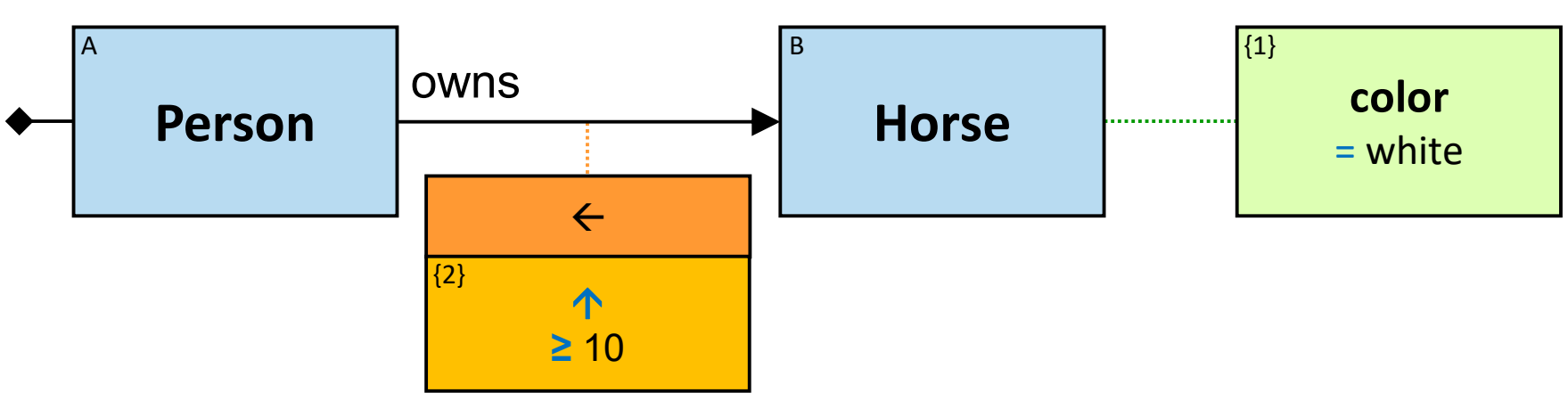

***Q296:*** *Any person who owned horses at least ten times – each horse is either white or weighs more than 100 Kg* (two versions)

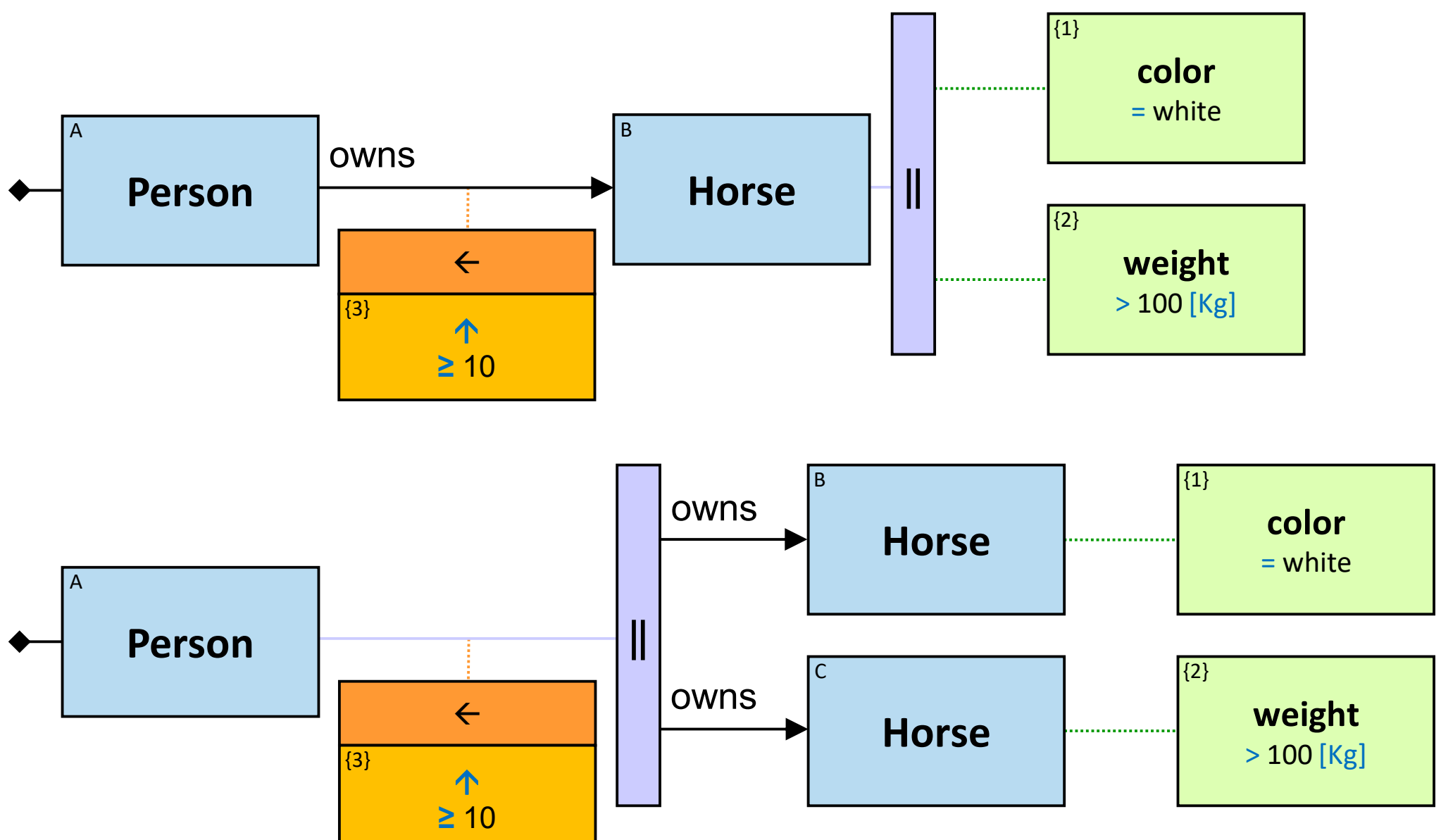

If B and C are assigned with the same graph-entity (a white horse that weighs more than 100 Kg), identical assignments to the two *owns* relationships would be counted only once. A2 counts *distinct* relationship/path assignments.

***Q79:*** *Any person with more than five paths of length up to four to other people*

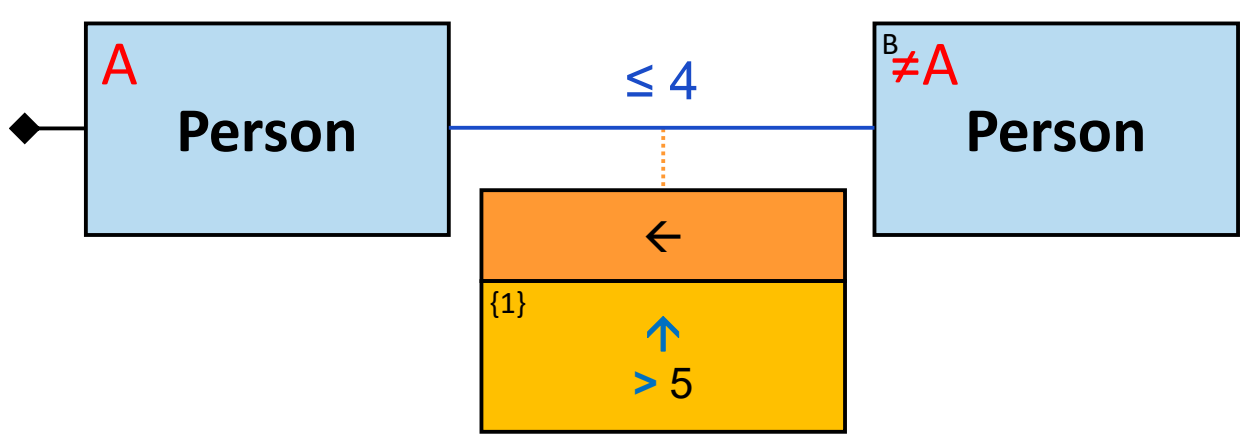

***Q105:*** *Any dragon A that was frozen exactly two times by dragons, each of* which was frozen at least once. *A might have been frozen by additional dragons that were never frozen*

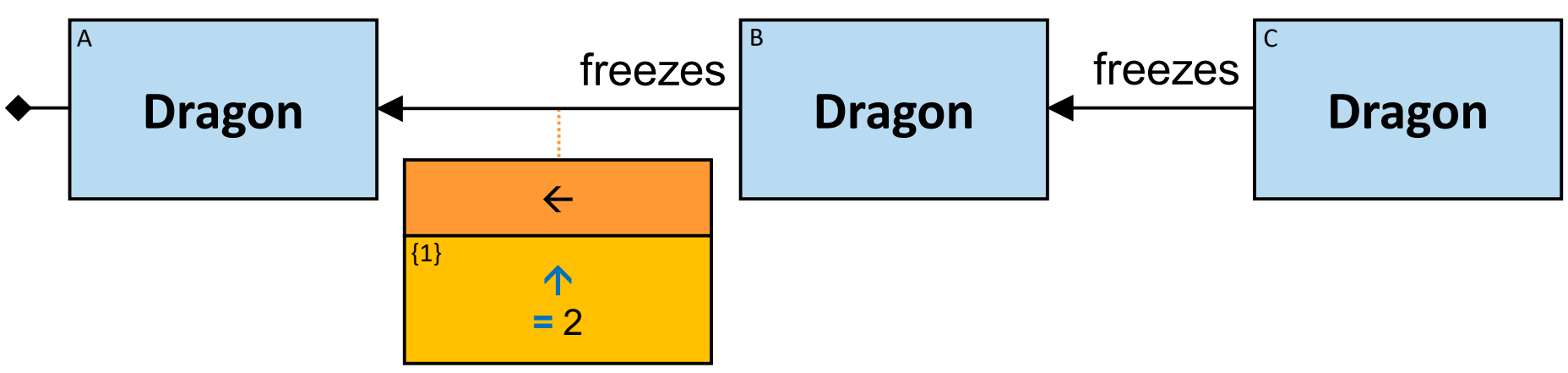

***Q245:*** *Any dragon B that was frozen at least once and froze dragons exactly twice*

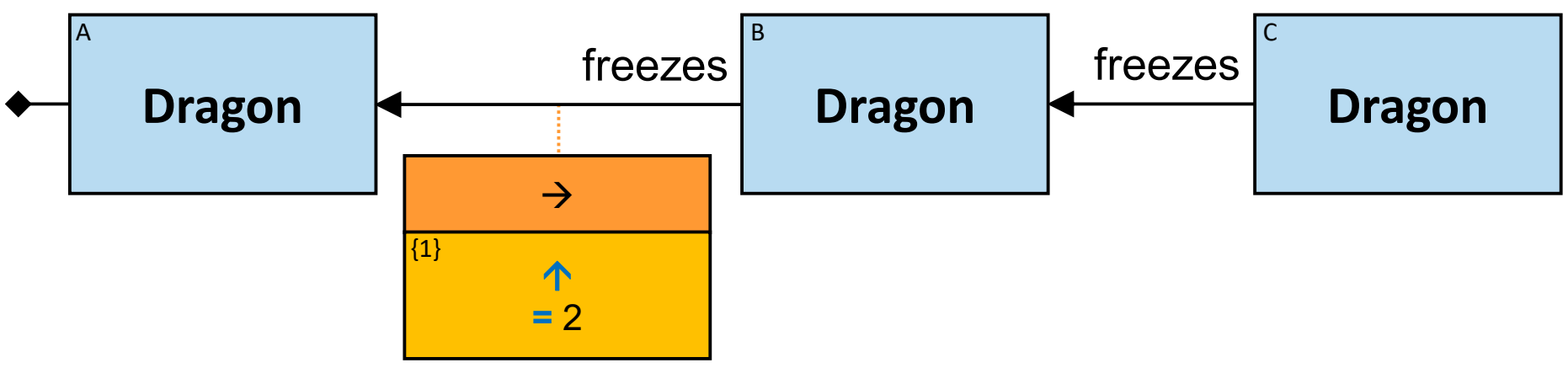

***Q185:*** *Any dragon Balerion froze at least once on or after January 1, 1010, at least once for less than ten minutes, and at least twice (on or after January 1, 1010, or for less than ten minutes)*

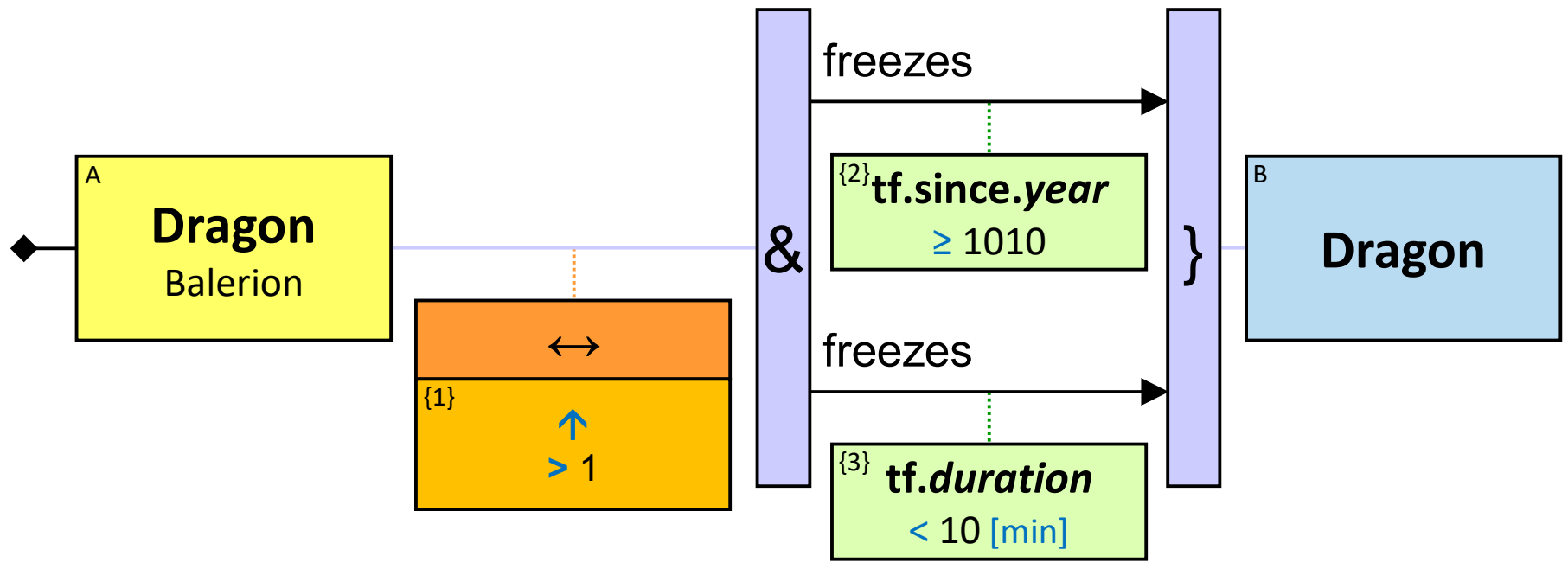

(counting the number of *distinct freezes* relationships)

***Q127:*** *Any dragon that froze dragons owned by Sarnorians more times than dragons owned by Omberians*

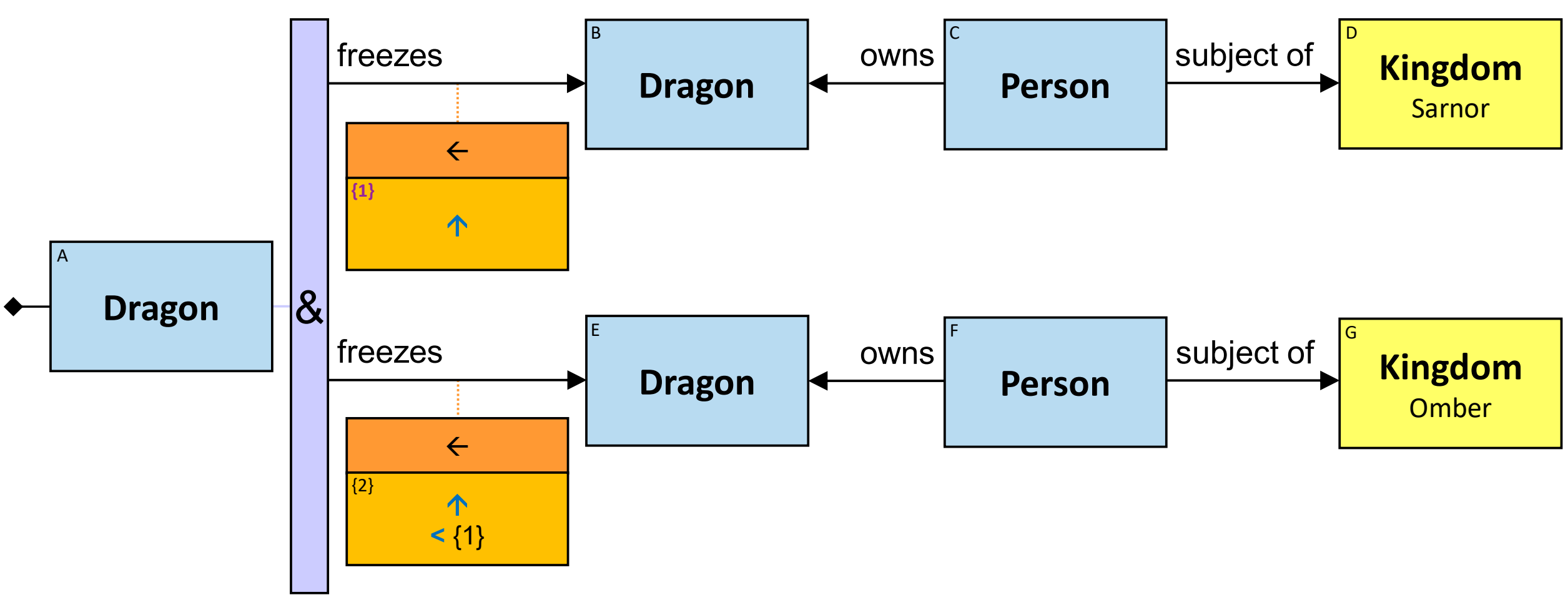

***Q339:*** *Any dragon Balerion froze more than ten times; report only those freezes that are in or after 1010*

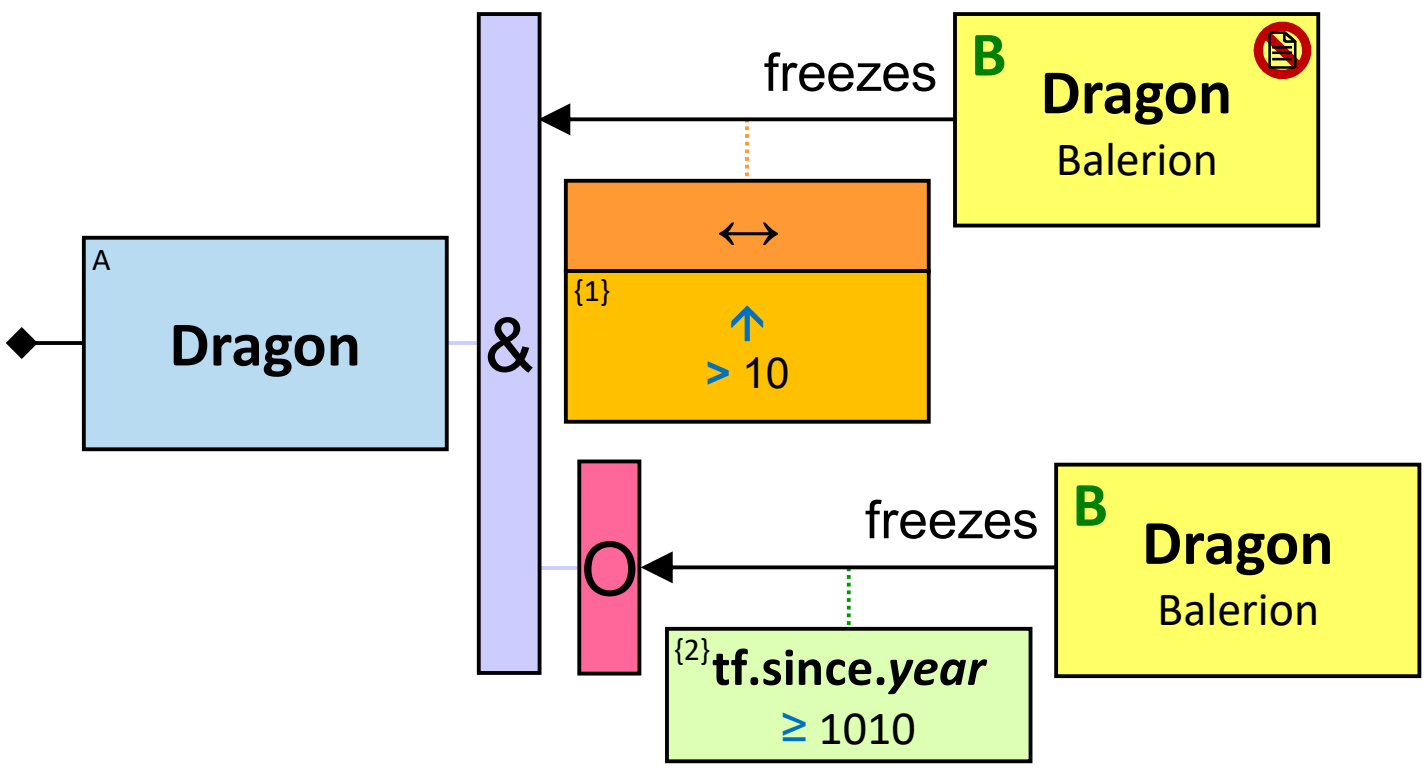

***Q297:*** *Any dragon that the number of times it fired at dragons plus the number of paths of length up to three between it and some horse – is at least ten*

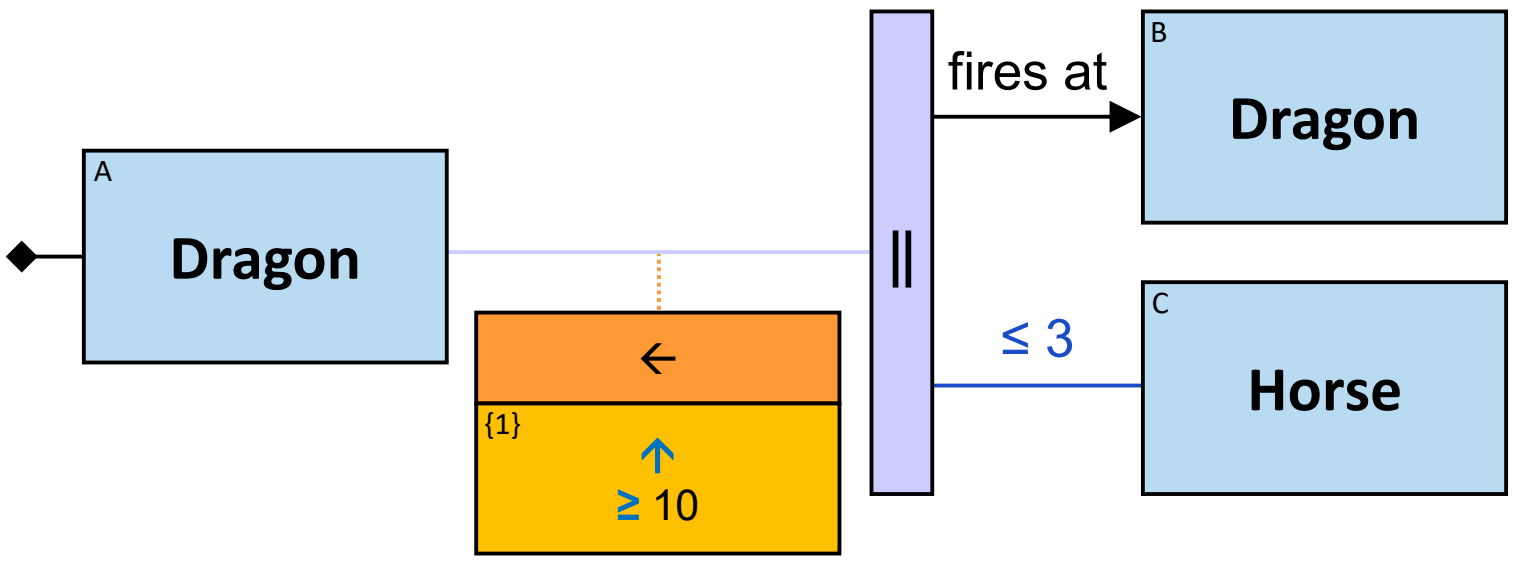

***Q124:*** *Any dragon that either (froze a dragon) or (fired at a dragon that fired at a dragon) – at least ten times*

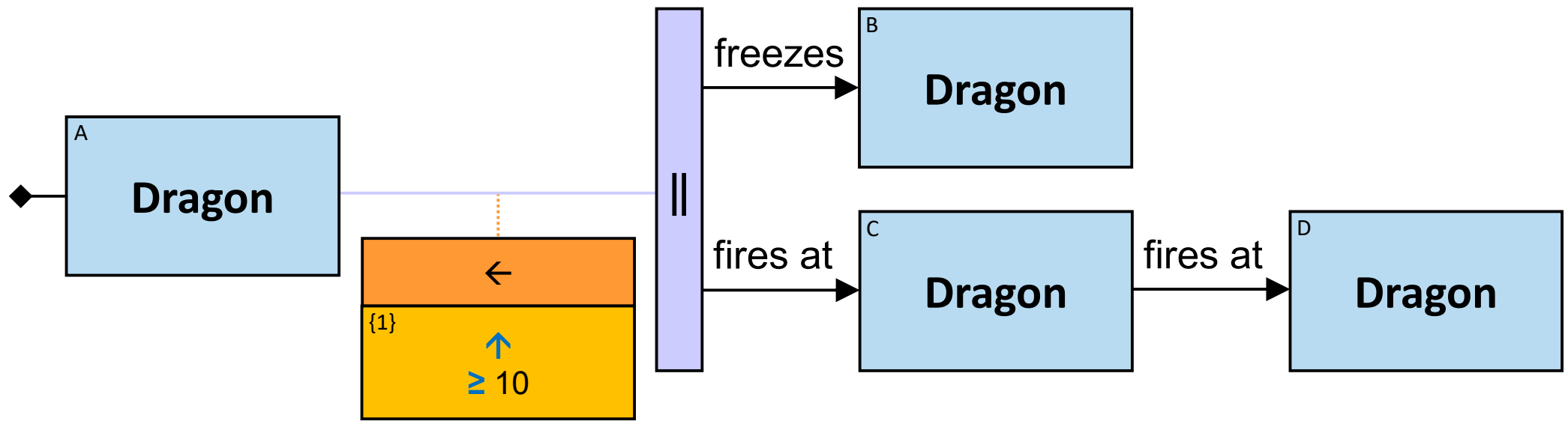

***Q173:*** *Any dragon that fired at at least two dragons and fired at least ten times*

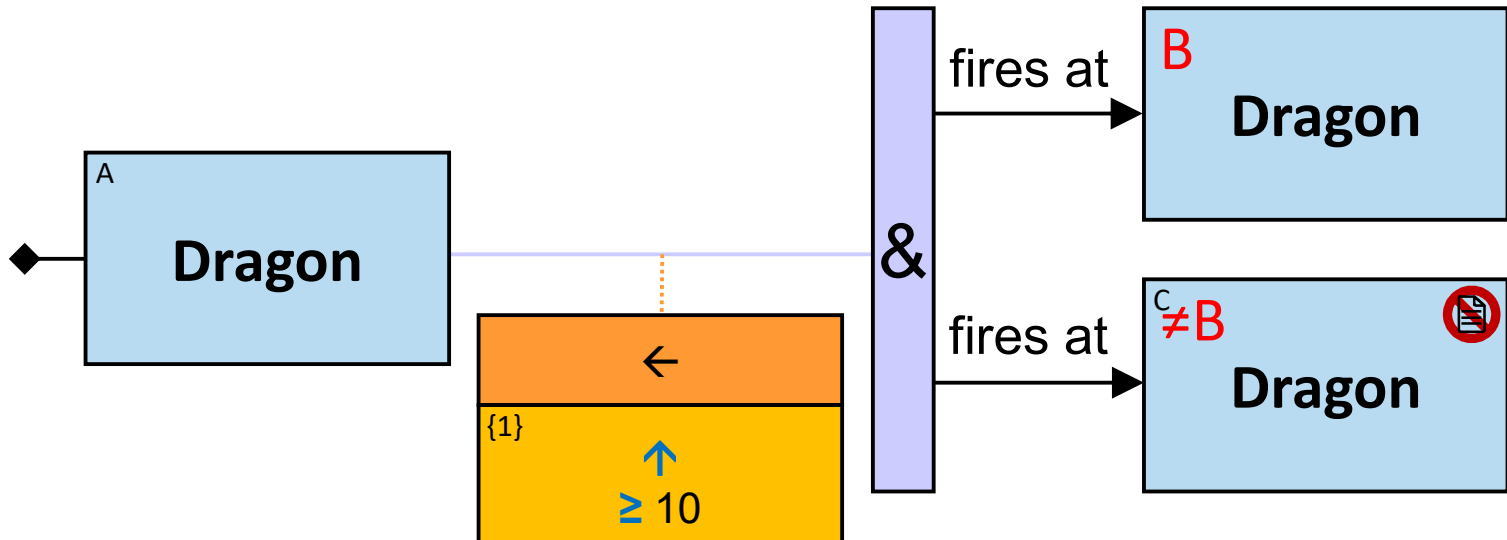

(counting the number of *distinct fired at* relationships). C is explicit latent to avoid redundant results (any dragon fired at would be assigned at least once to B and at least once to C).

***Q174:** Any dragon that either (i) froze at least one dragon and fired at at least one dragon it did not freeze, or (ii) froze at least two dragons. If (i): the number of times it froze/fired at dragons is at least ten; otherwise: the number of times it froze dragons is at least ten*

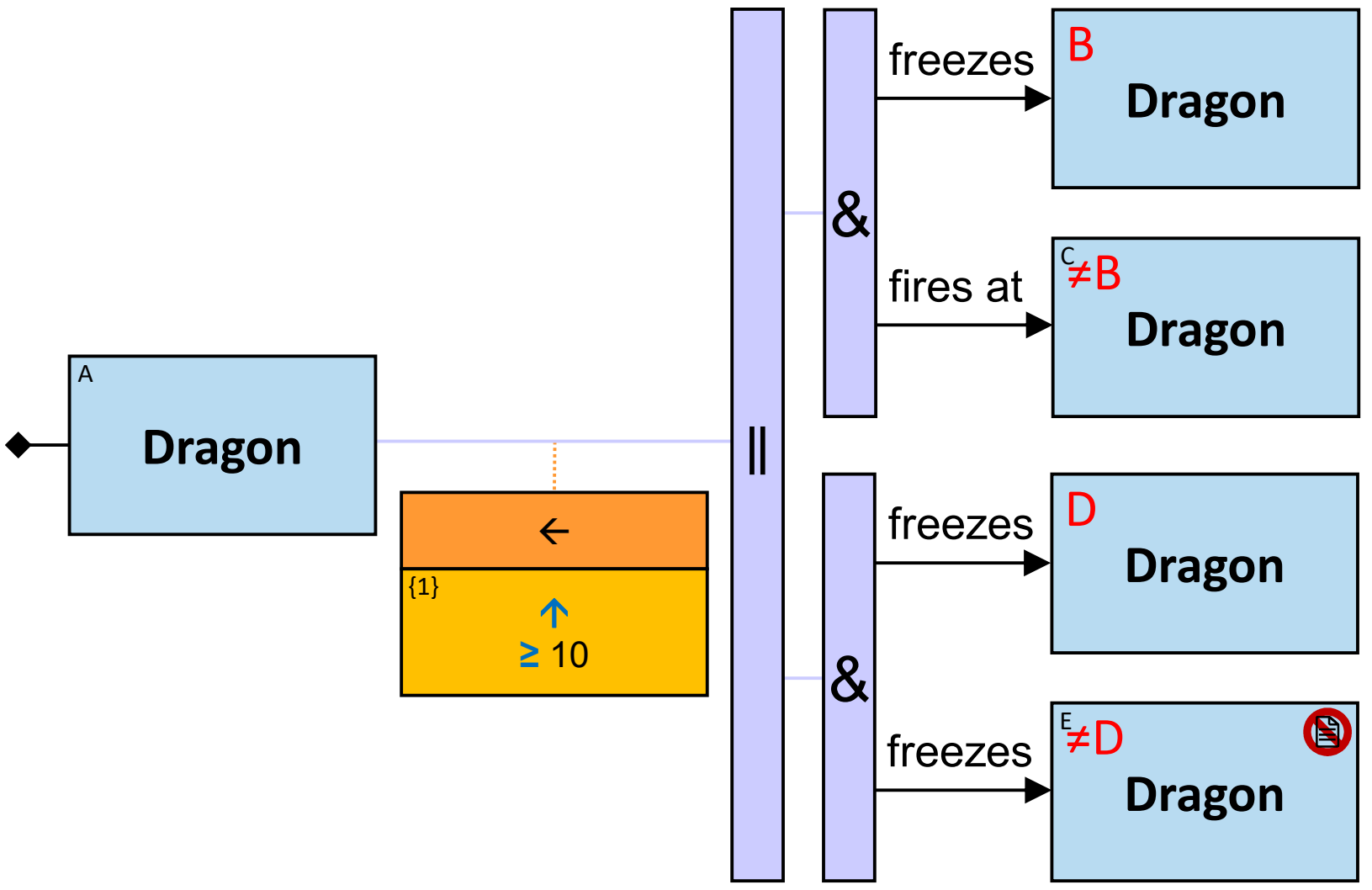

***Q75:** Any pair of dragons (A, B) where B froze A between eight and ten times*

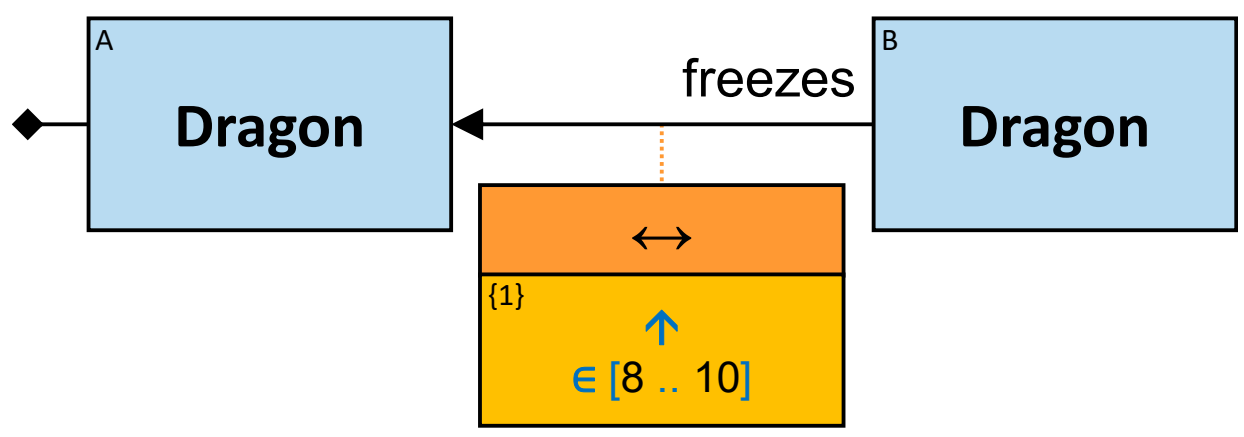

***Q76:** Any dragon that froze Balerion between eight and ten times*

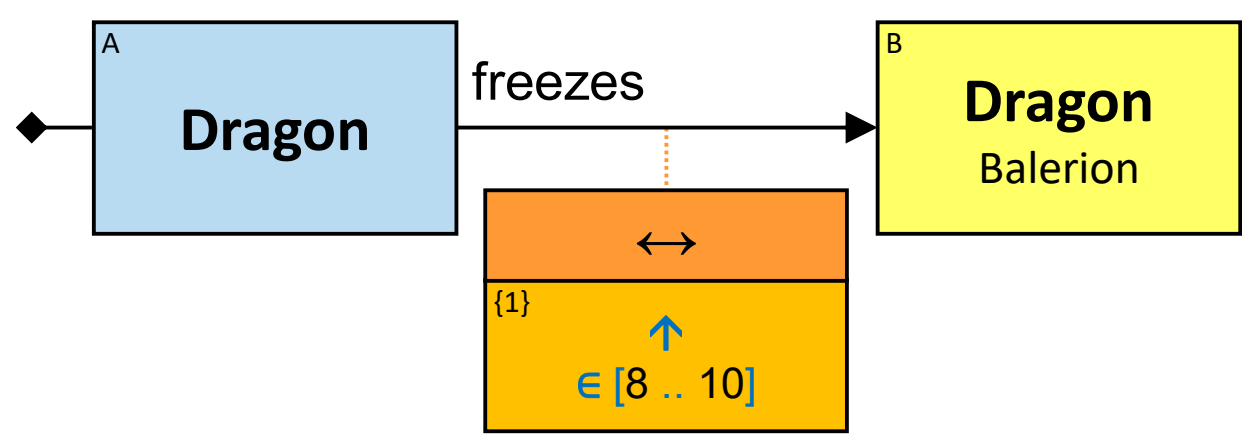

***Q242:** Any pair of people (A, D) where at least five times one of A's dragons froze one of D's dragons*

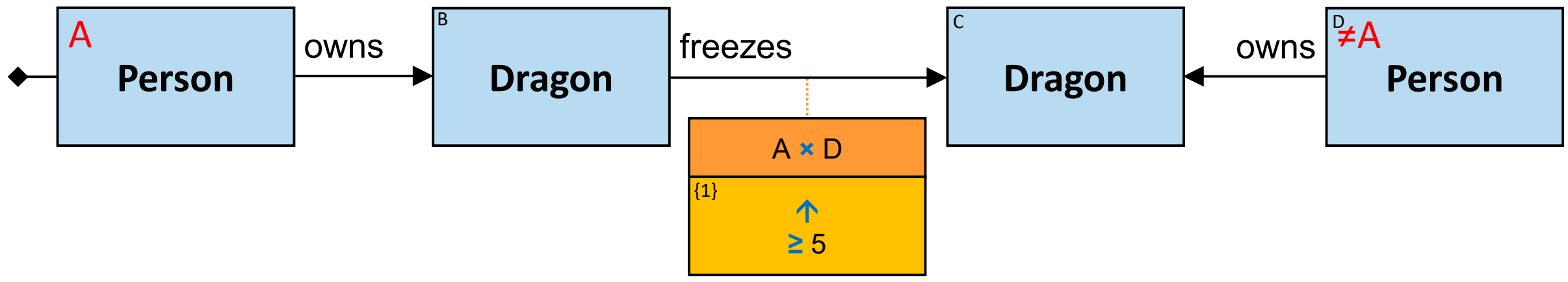

*Q358: Any dragon A where (i) there is at least one black dragon A froze (ii) there is at least one white dragon A did not freeze, (iii) there is no gold dragon A froze, (iv) there is no silver dragon A did not freeze, and (v) the number of times A froze black and red dragons is at least ten. Report all black and red dragons that A froze*

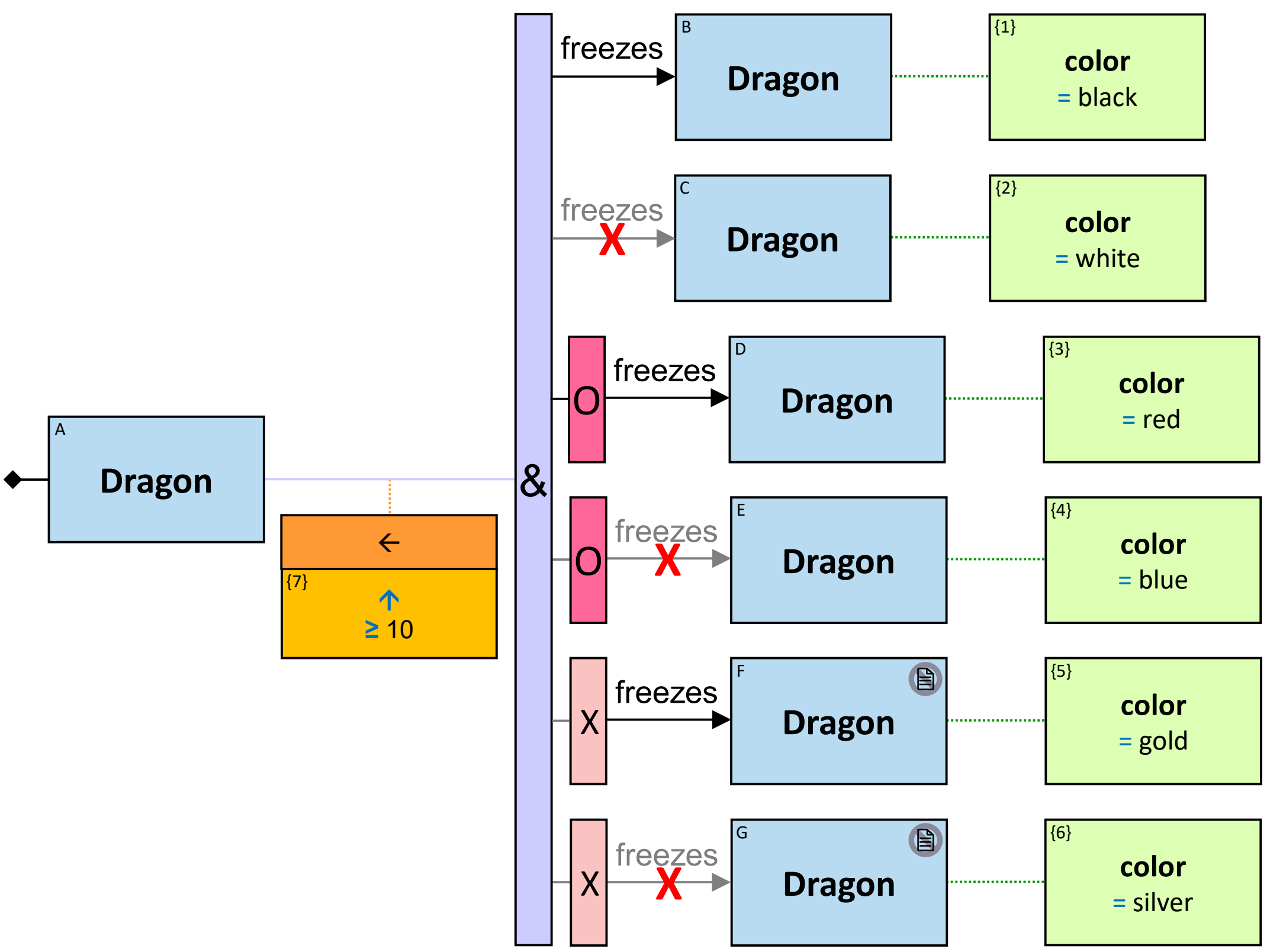

Pattern-relationships that are wrapped with a negator and pattern relationships with a relationship/path-negator are not aggregated. Compare with Q298.

*Q372: The number of times dragons were frozen*

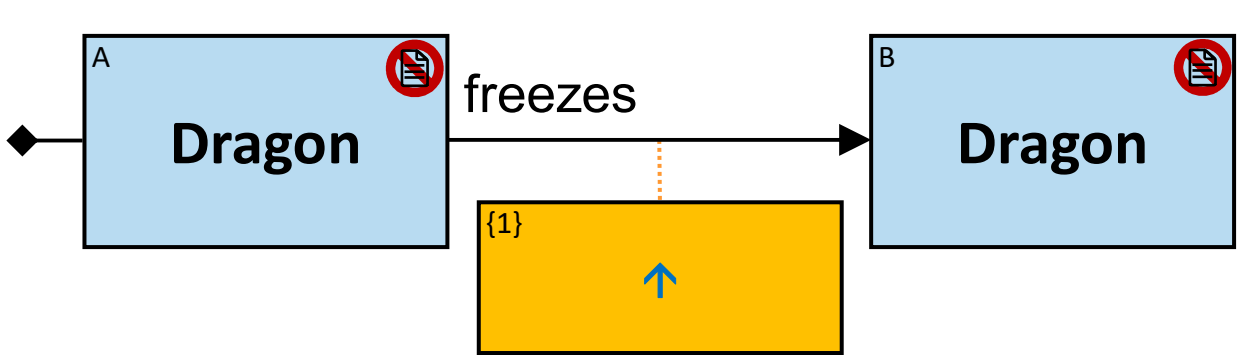

{1} is a global property.

## 28. A3 AGGREGATOR

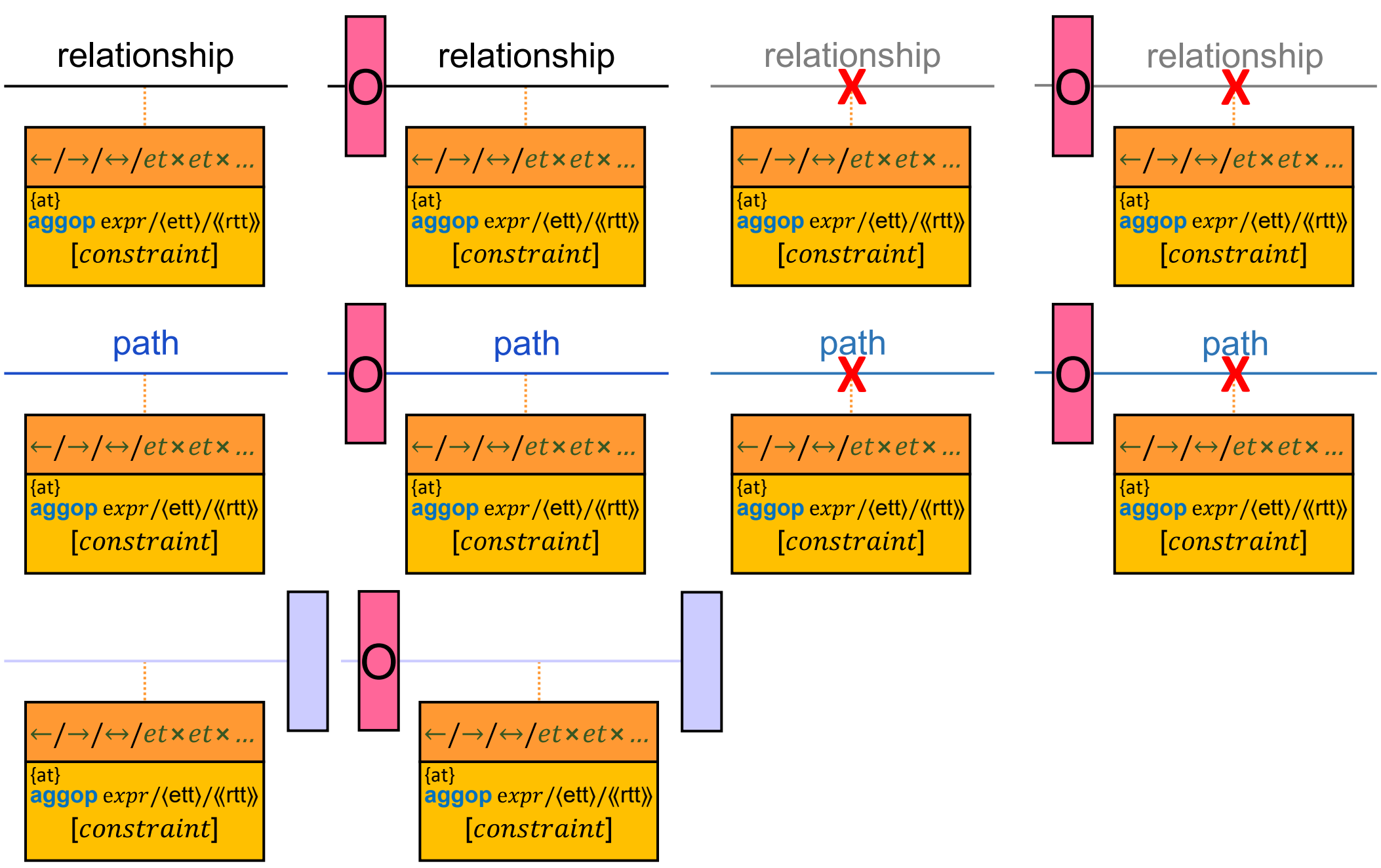

The lower part of an A3 aggregator contains the following elements:

- One of the following:
  - *aggop expr*

    *aggop* is *min/max/avg/sum* – for aggregating values of a supported type, *union/intersection* – for aggregating sets/bags of any type, or *distinct/set/bag* – for aggregating values of any type. *distinct* returns the number of distinct *non-null* evaluation results; *set* returns a set of all *non-null* evaluation results (see Q332v2); *bag* returns a bag of all *non-null* evaluation results (see Q315).

    *expr* is an expression composed of constants, properties, and sub-properties of *R* (when A3 appears below a relationship *R* with no relationship-negator) (see Q86, Q354), EA-tags defined on top of the aggregator (see Q277), and EA-tags defined right of the aggregator (see Q116, Q137, Q169)

  - *distinct ⟨ett⟩*

    *⟨ett⟩* is an entity type-tag defined right of the aggregator (see Q167v2)

  - *distinct ⟪rtt⟫*

    *⟪rtt⟫* is a relationship type-tag defined on top of the aggregator or right of the aggregator (see Q366v2)

  Let ***BA(n)*** denote the list of the evaluation results of *expr/⟨ett⟩/⟪rtt⟫* for all assignments in *S(n)*.

- **aggregation-tag**:
  - if *aggop* is *min/max/avg/sum/distinct/union/intersection*:
    *at(n) = aggop(BA(n)[1], BA(n)[2], ...)*
  - *if aggop* is *set*: *at(n) = {BA(n)[1], BA(n)[2], ...}* (with no duplicate values and no *null* values)
  - *if aggop* is *bag*: *at(n) = [BA(n)[1], BA(n)[2], ...]* (with duplicate values and no *null* values)

When an optional part has no assignments, A3 aggregation-tags defined right of the 'O' are evaluated to *null* when *aggop* is *min/max/avg/sum*, evaluated to zero when *aggop* is *distinct*, and evaluated to {} / [] when *aggop* is set/bag/union/intersection.

- an optional **constraint** on *at(n)*

  **For each *n*: *S(n)* is reported only if *at(n)* satisfies the constraint**

  When *aggop* is *distinct*: a '*= 0*' constraint cannot be satisfied unless *ett* / *rtt* is right of an 'O', as such constraint means that there are no assignments to the pattern excluding this aggregator. Similarly:

  - no constraint is equivalent to '*> 0*' constraint
  - '*≠ expr*', '*< expr*', and '*≤ expr*' constraints are not satisfied when *at(n)* is zero
  - '*= expr*', '*≥ expr*', and '*∈ [expr1, expr2]*' constraints are not satisfied when *at(n)* is zero even if *expr* is evaluated to zero

A3 location:

- Below a relationship/path

  When A3 appears below a relationship *R* and *expr* is composed of at least one property or sub-property of *R* – *R* may be wrapped with an 'O'. Otherwise – a relationship/path with an A3 below it may have a relationship/path-negator and may be wrapped with an 'O'.

- Below a quantifier-input (excluding quantifier at the start of the pattern)

  A quantifier-input with an A3 below it may be wrapped with an 'O'.

- Below the pattern's start (even when followed by a quantifier)

  See Q263.

- If all entities in *T* are in a sequence: A3 appears directly right of the leftmost member of *T*
- If entities in *T* are in different branches of a quantifier: A3 appears directly left of the quantifier

When units of measure are defined for the expression (based on the aggregation operator, on the units of measures of the properties, and on the operators that compose the expression) – they are depicted as well (see Q86, Q117).

***Q86:*** *Any pair of dragons (A, B) where A froze B for a cumulative duration longer than 100 minutes*

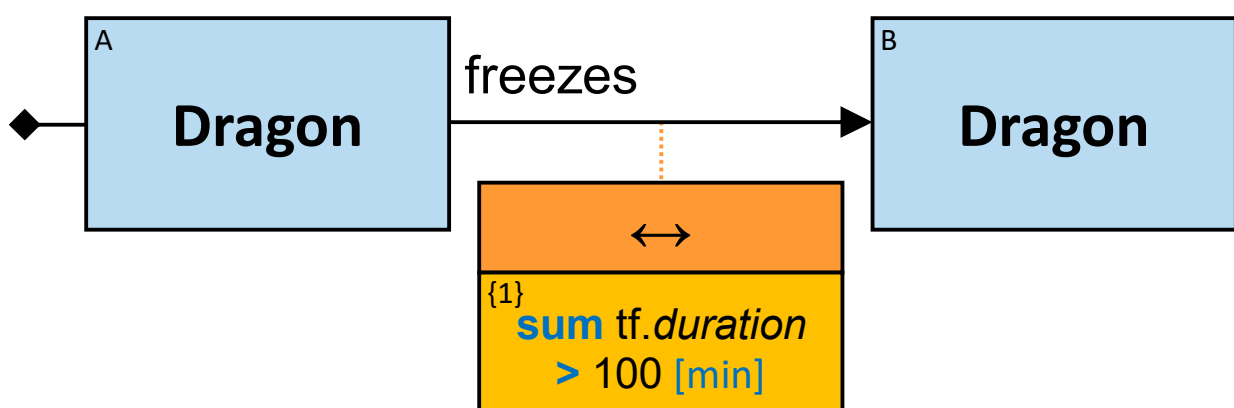

***Q87:*** *Any dragon that was frozen at least once, and the cumulative duration it was frozen for is less than 100 minutes*

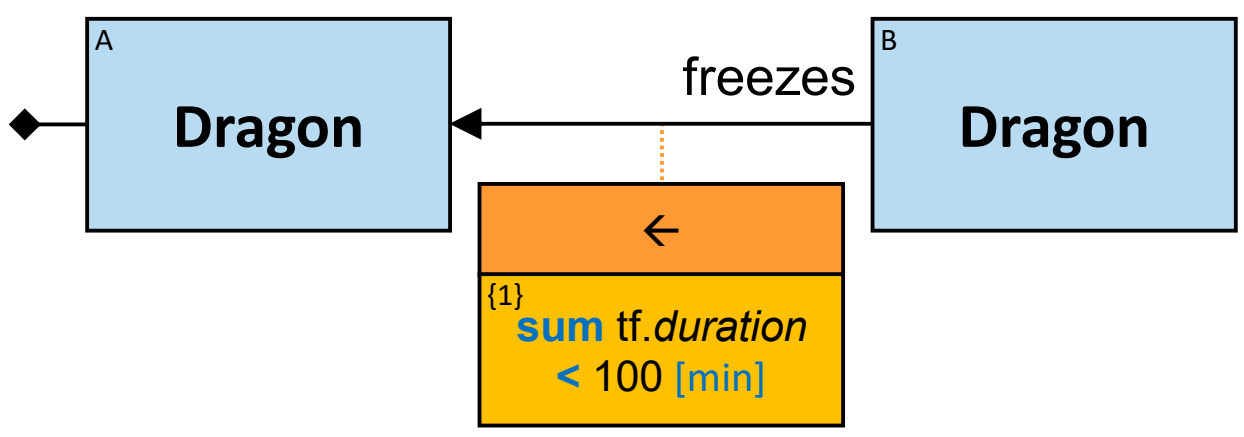

***Q89:*** *Any dragon that froze dragons for more than three different durations*

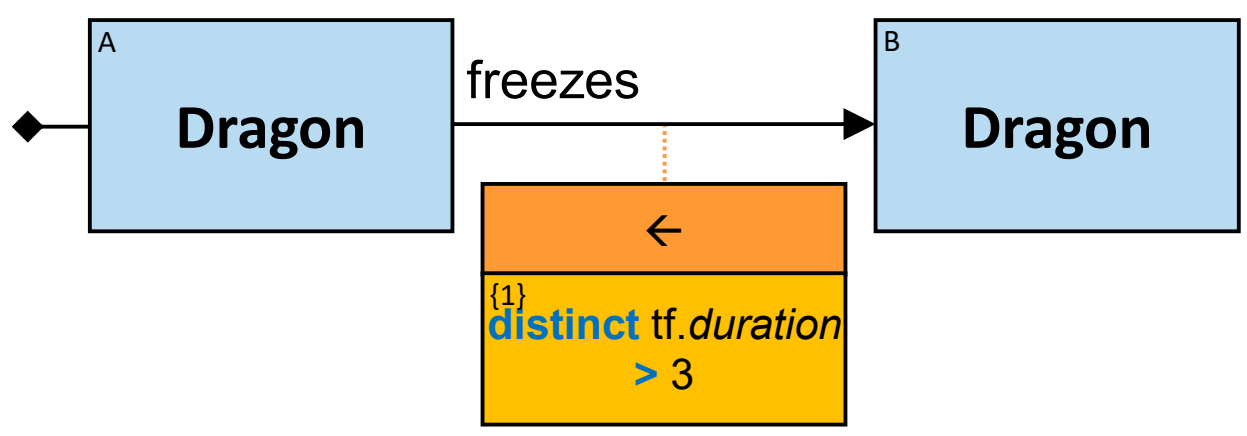

***Q167:*** *Any person whose owned entities are all of the same type* (version 2)

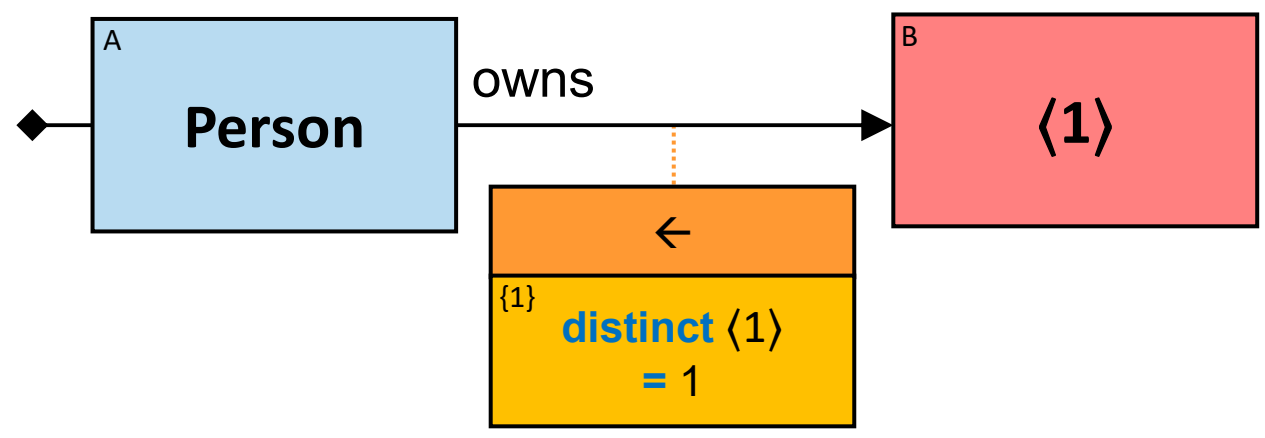

***Q366:*** *Any pair of dragons that have relationships of at least three types* (version 2)

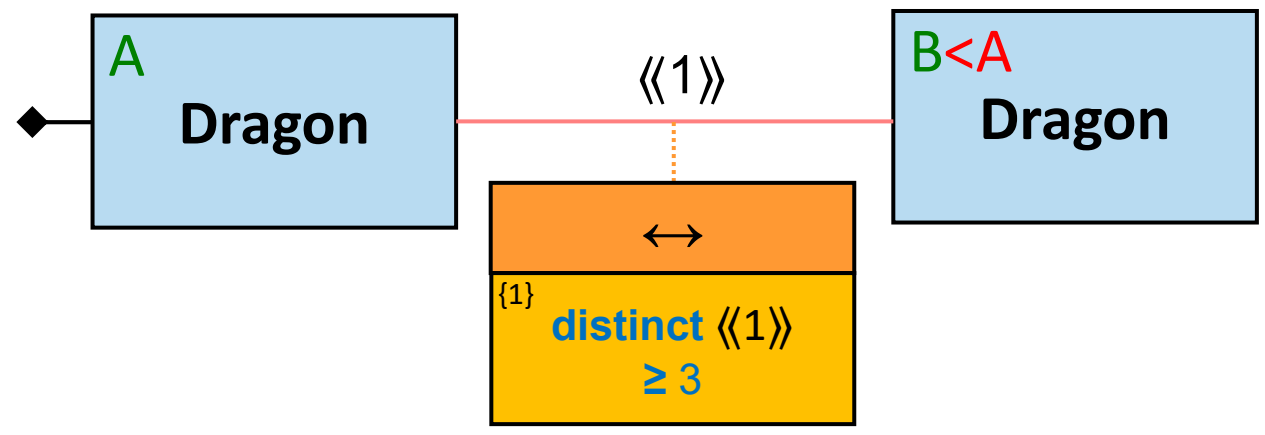

***Q117:*** *Any person whose owned horses' average weight is greater than 450 Kg*

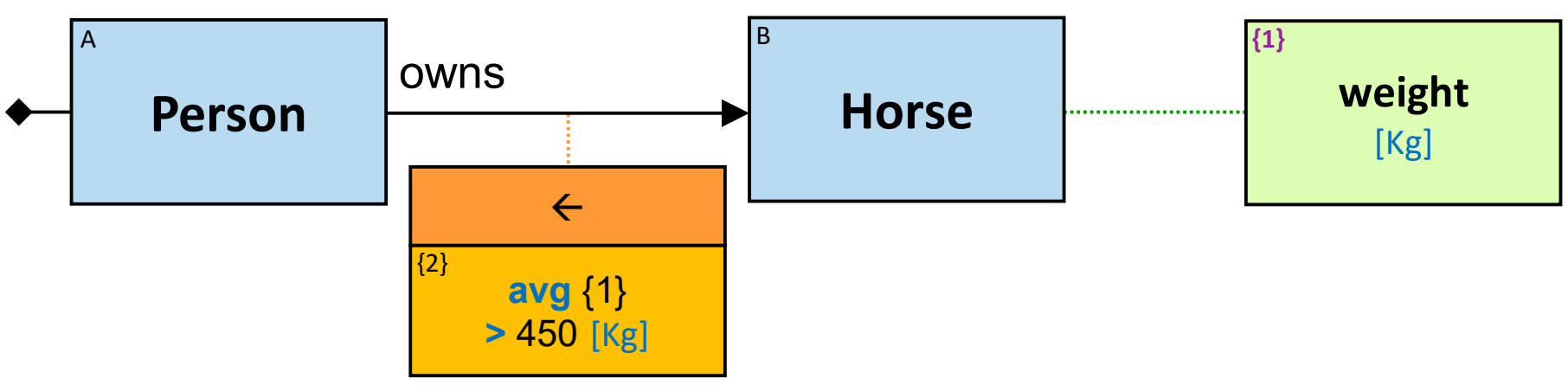

If a horse were owned twice – it would be counted only once.

*Q116: Any person A that the number of distinct colors of A's owned horses – is between one and three*

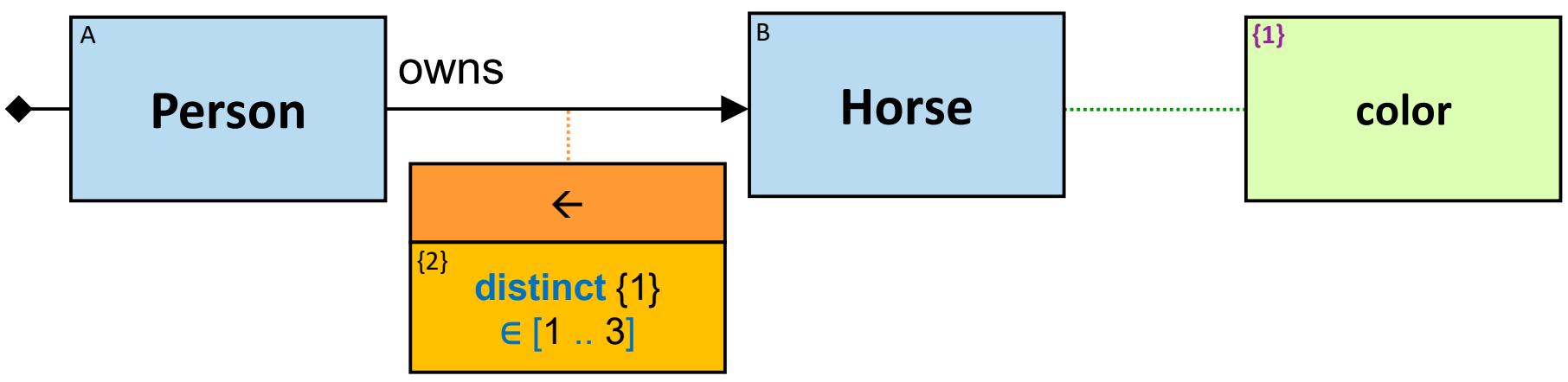

*distinct* does not aggregate *null* evaluation results. If all dragons have a *null* color, {2} would be equal to zero.

*Q134: Any person A that the number of distinct colors of all horses owned by A's friends – is between one and three*

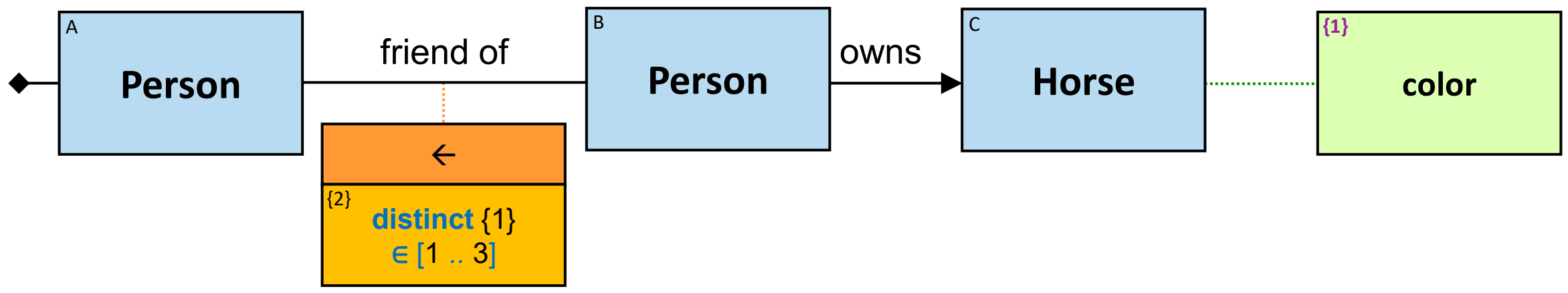

*Q229: Any person A that the number of distinct colors of all horses owned by A's friends and parents – is between one and three*

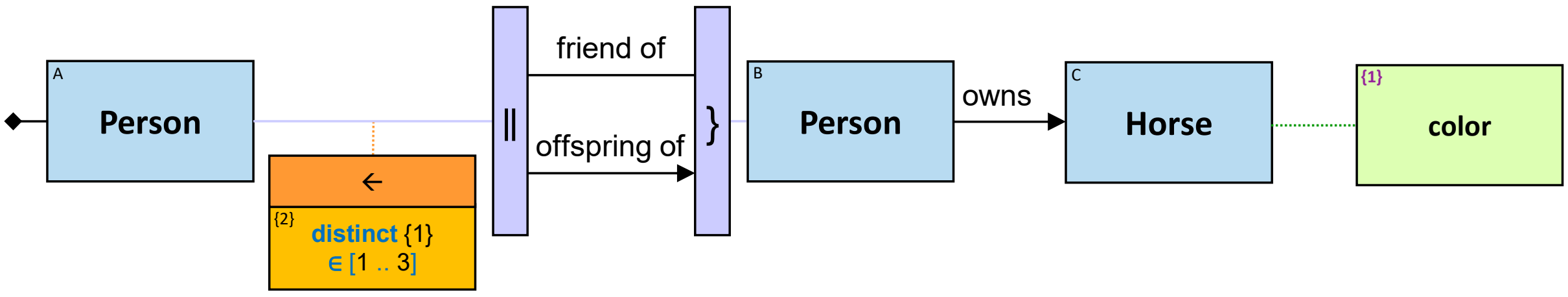

*Q135: Any person A that the average weight of all horses owned by A's friends – is greater than 450 Kg*

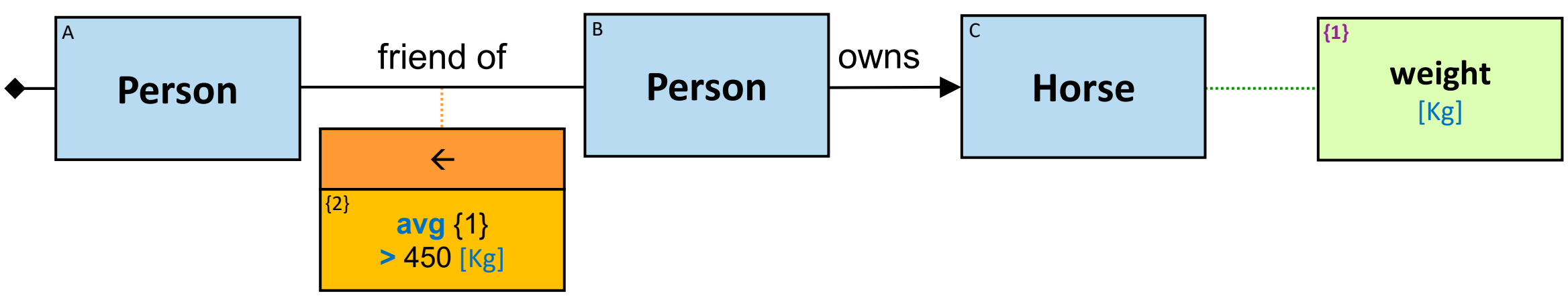

Note that if some person $A_1$ is a friend of two people who jointly own a horse, this horse's weight would be counted once.

***Q137:*** *Any dragon A that froze dragons B. Each B froze at least one dragon, which is not A. All these B's together froze dragons for more than 100 minutes cumulatively*

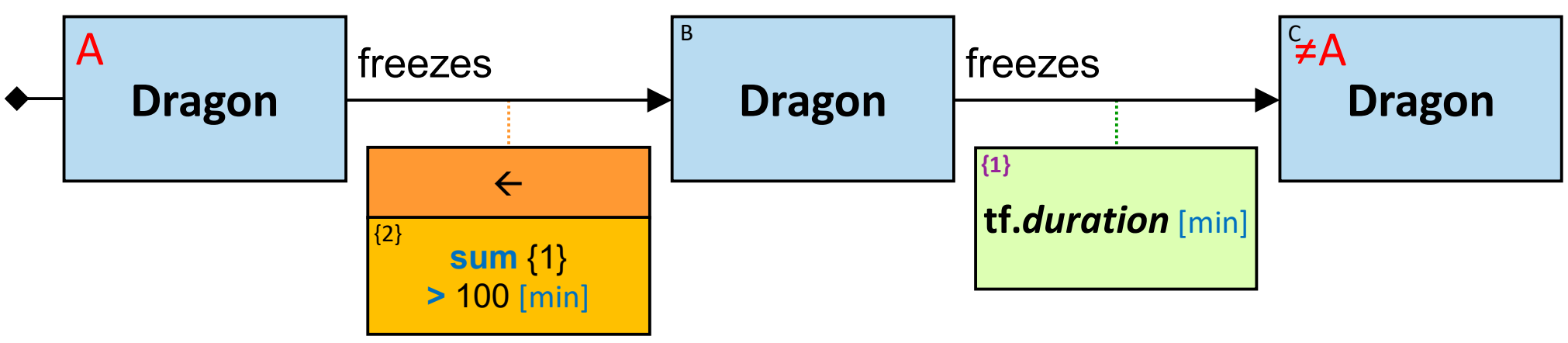

***Q336:*** *For each dragon: each time it froze some dragon for a duration longer than the average duration of all its freezes*

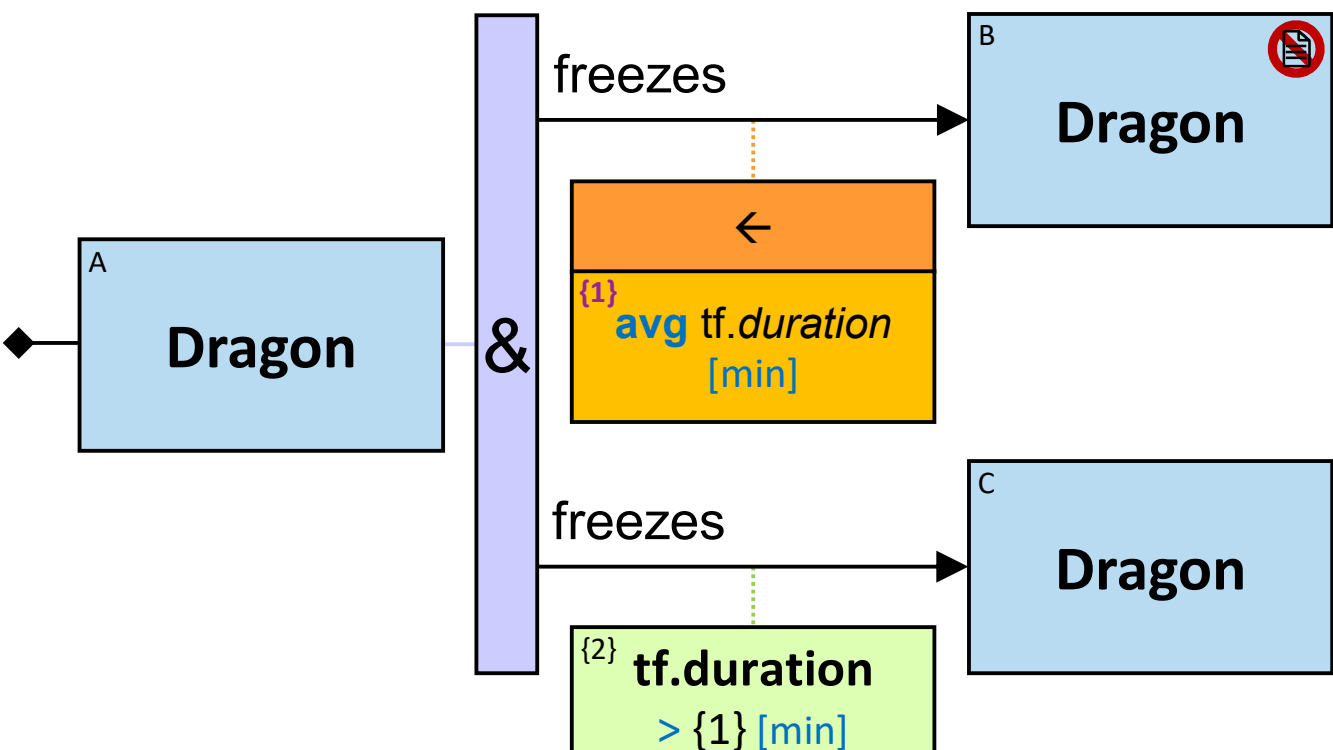

First, any triplet (A, B, C) matching the pattern excluding the aggregator and the constraint in {2} is found. Then, {1} is evaluated per assignment to A, and then the constraint in {2} is evaluated per relationship assignment. See also Q337.

Note that a dragon that always froze dragons for the same duration is not a valid assignment to A.

***Q354:*** *Any dragon that froze one dragon and fired at another dragon, or froze a dragon and fired at more than one dragon, or fired at a dragon and froze more than one dragon, and there is some dragon that the earliest time it fired at it is earlier than the earliest time if froze any other dragon*

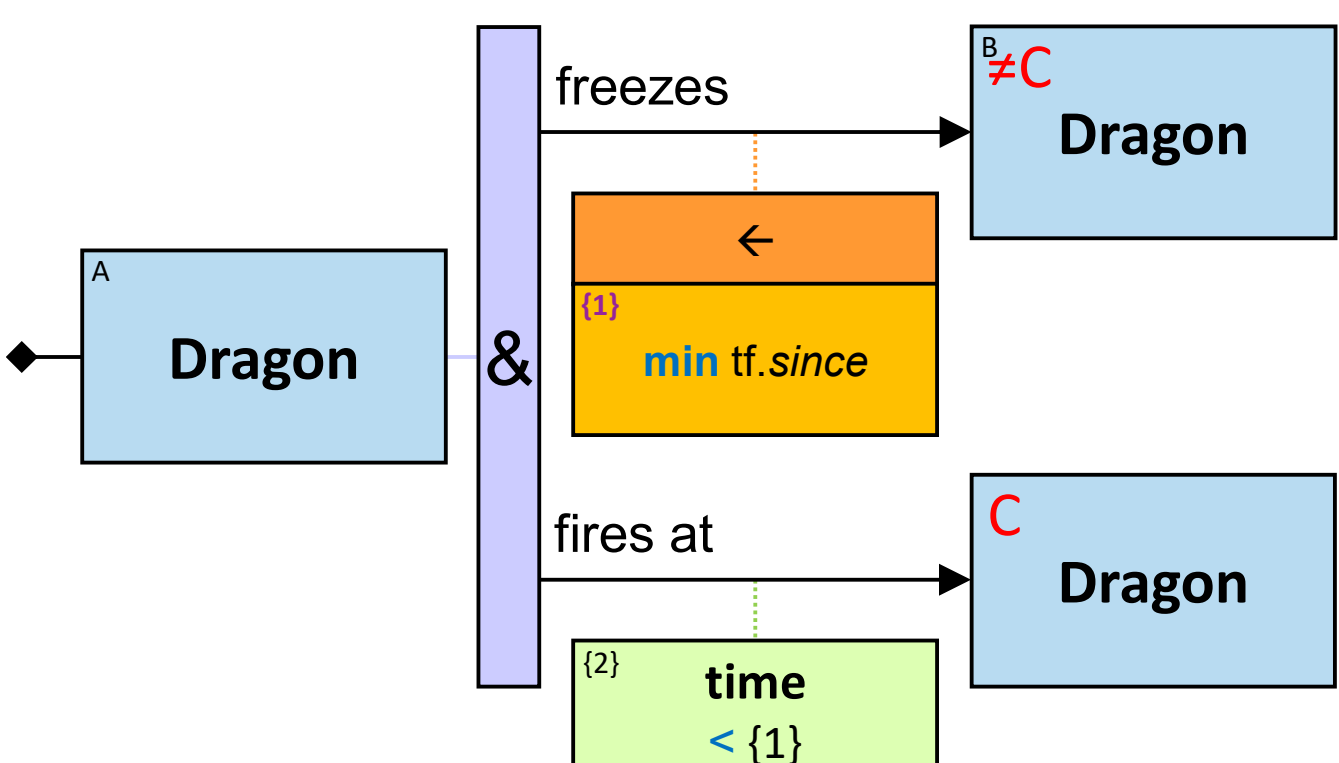

*Q139: Any person A who owns horses of the same number of colors as the number of colors of the horses owned by A's parents cumulatively*

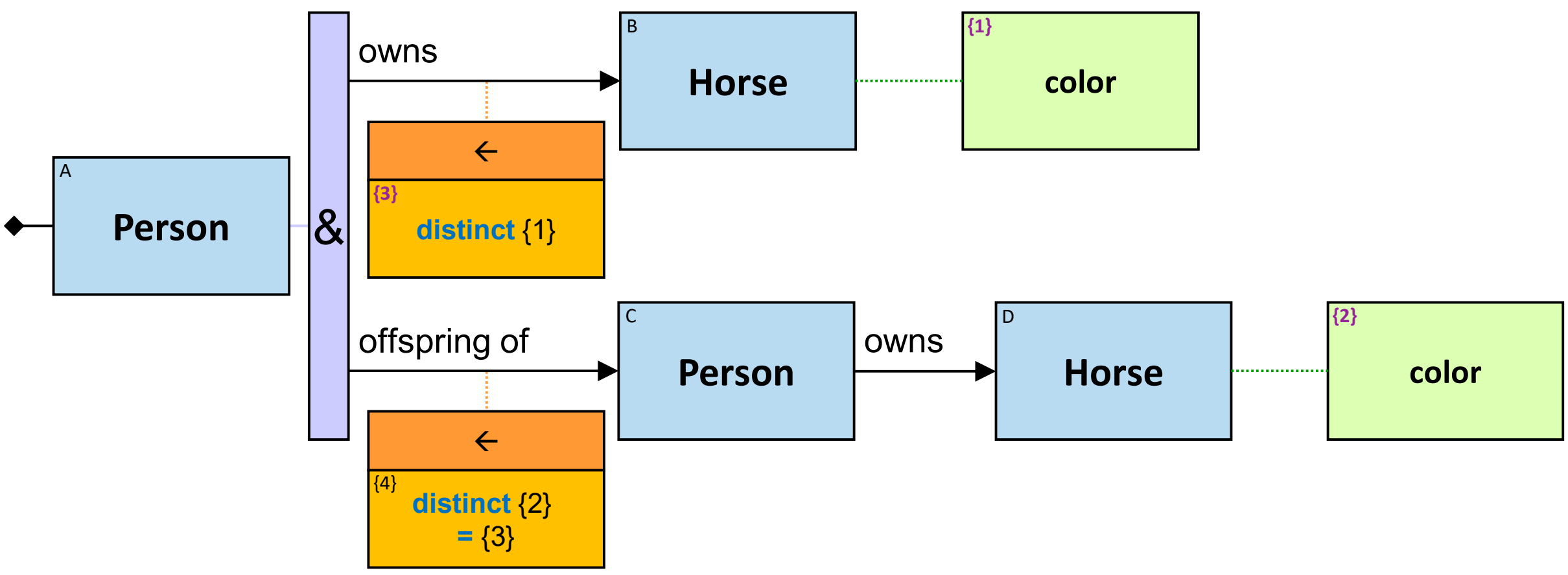

*Q98: Any pair of dragons (A, X) where A froze more than three dragons and (A froze X more than ten times or for a cumulative duration of more than 100 minutes)*

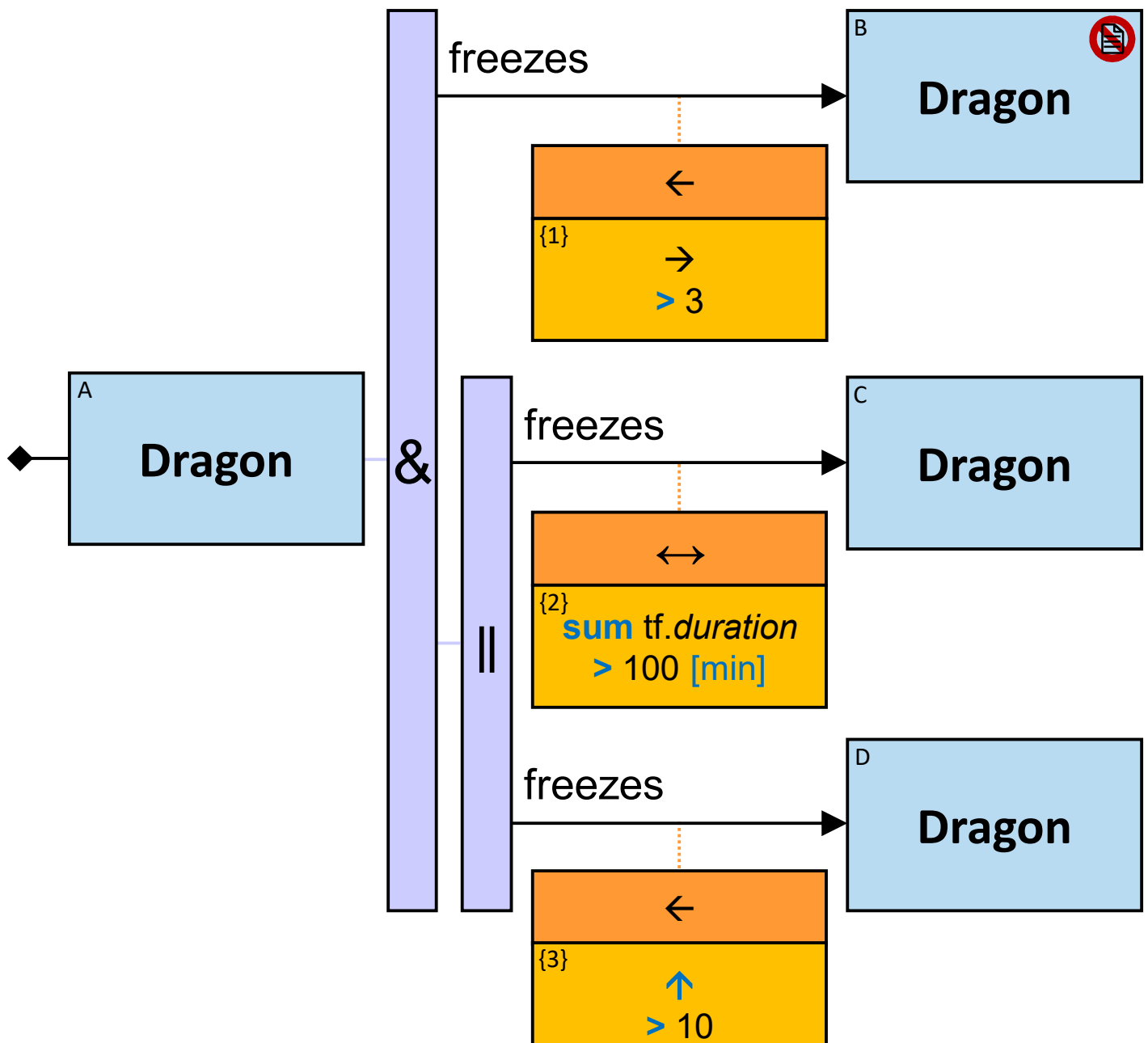

***Q88:*** *Any pair of dragons (A, B) where the cumulative duration A froze B is equal to the cumulative duration B froze A*

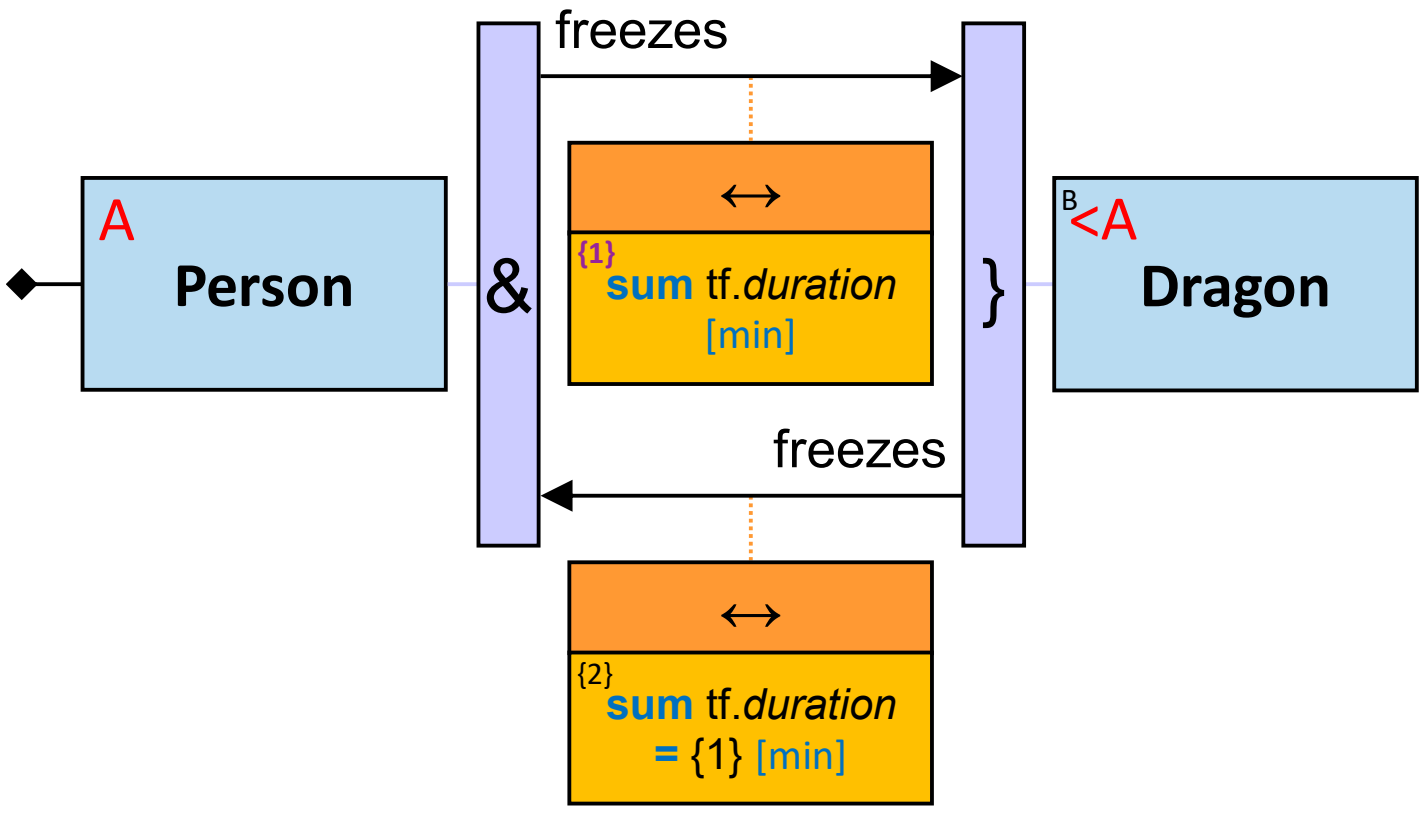

A property of entity *et* cannot be aggregated per *et,* nor per a superset of *et*.

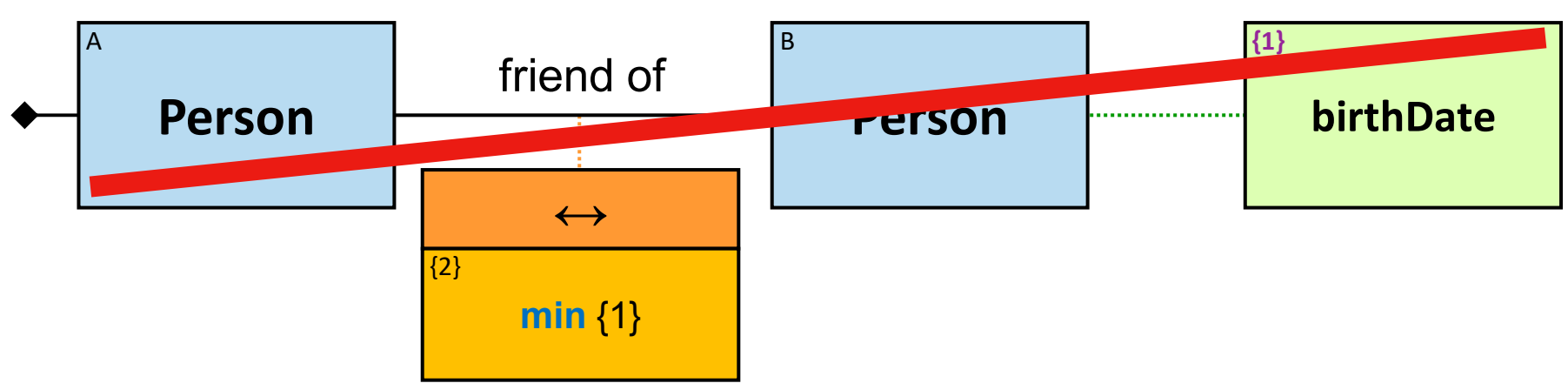

{1} is a property of B and hence cannot be aggregated per A × B.

If entity *et* has a property that references

- an aggregation of a property of a Cartesian product composed of *et1, or*
- an aggregation of a property of a relationship between *et* and *et1*

then

- a property of *et* cannot be constrained based on a property of *et1*
- a property of *et1* cannot be constrained based on a property of *et*

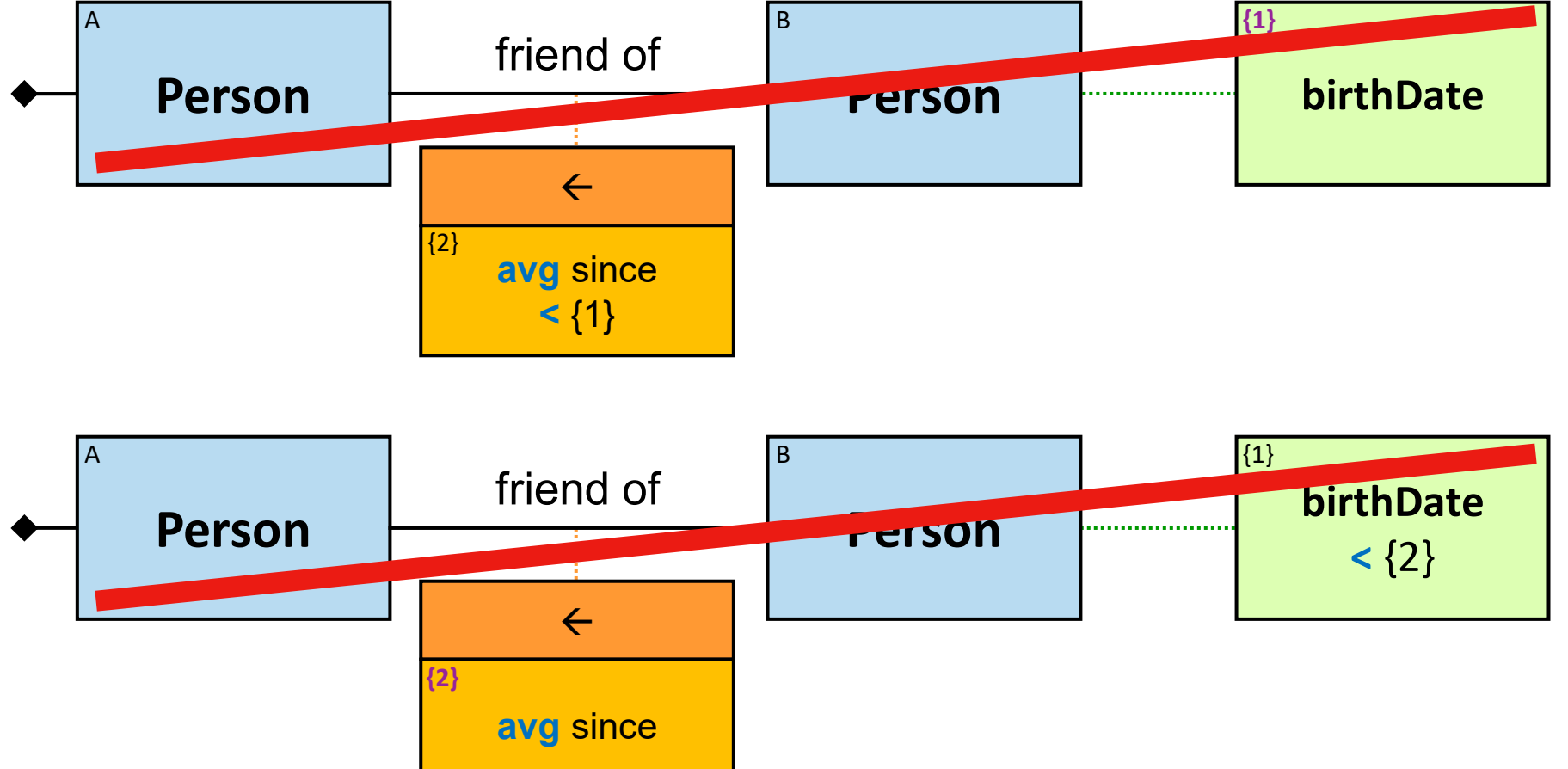

In the two patterns above, {2} – a property of A, is referencing an aggregation of a property of a relationship between A and B. Therefore, a property of A cannot be constrained by a property of B. Similarly, a property of B cannot be constrained by a property of A.

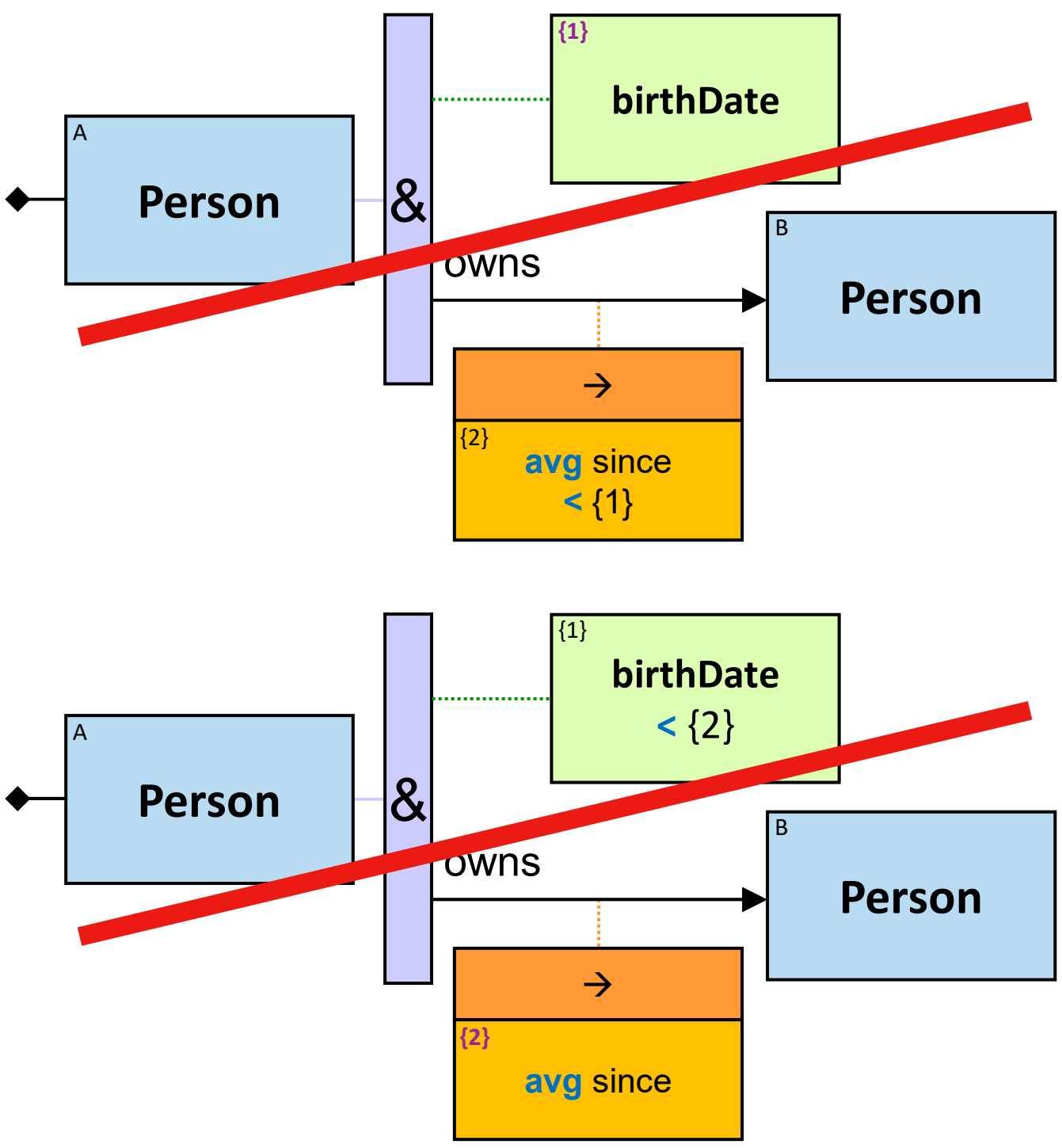

In the two patterns above, {2} – a property of B, is referencing an aggregation of a property of a relationship between A and B. Therefore, a property of B cannot be constrained by a property of A. Similarly, a property of A cannot be constrained by a property of B.

## 29. MIN/MAX AGGREGATORS

Sometimes we need to limit assignments to a Cartesian product of entity-tags, based on some value, to [all but] the *k* assignments for which the value is the lowest/highest. Here are some examples:

- *Any person A and A's five oldest offspring*
- *Any dragon and the three dragons it froze the largest number of times*
- *Any dragon and the four dragons it froze for the maximal cumulative duration*

Three types are M aggregators are presented:

- M1: the value is the number of assignments to a union of Cartesian products of entity-tags
- M2: the value is the number of assignments to relationships/paths
- M3: the value is the evaluation result of an expression

M1 and M2 are syntactic sugar, which, alternatively, can be implemented by chaining A1 or A2 to an M3.

## 30. M1 AGGREGATOR

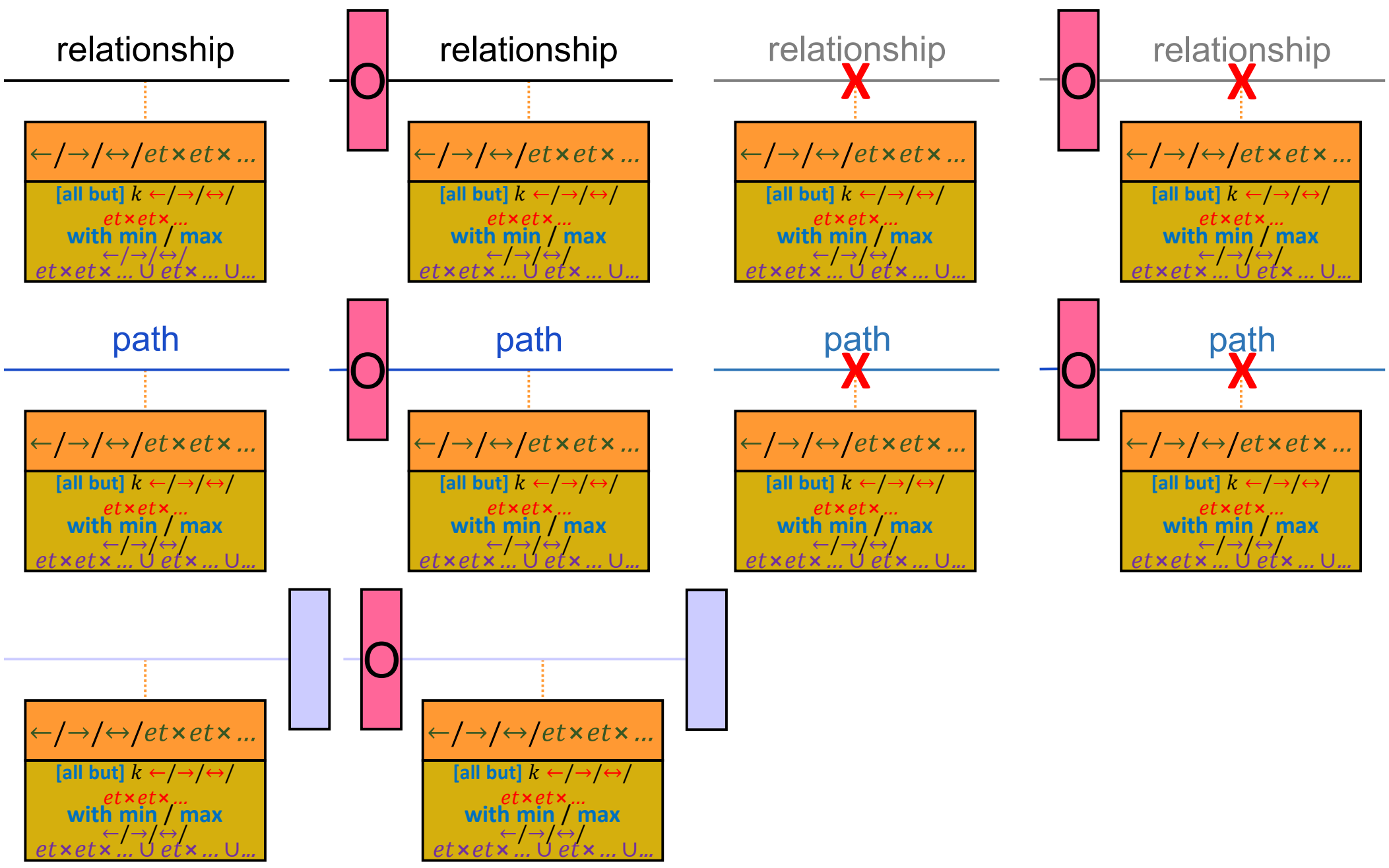

Let ***B*** denote a Cartesian product of one or more entity-tags. $B$ must be nonempty and must not be composed of entity-tags composing $T$.

Let ***M*** denote a list of Cartesian products of one or more entity-tags. $M$ must be nonempty and must not be composed of entity-tags composing $B$ or $T$.

The lower part of an M1 aggregator contains the following elements:

- optional: "all but"
- $k$: positive integer
- One of the following:
  - '$et1 \times et2 \times \ldots$': $B = et1 \times et2 \times \ldots$
  - '←': $B$ = entity-tag directly left of the aggregator
  - '→': $B$ = entity-tag directly right of the aggregator not wrapped with a negator nor preceded by a *None* quantifier (valid only when there is exactly one such entity-tag)
  - '↔': $B = et1 \times et2$, where $et1$ is the entity-tag directly left of the aggregator, and there is a single entity-tag directly right of the aggregator not wrapped with a negator nor preceded by a *None* quantifier – $et2$

  Let ***BA(n)*** denote the list of all unique assignments to $B$ in $S(n)$. $BA(n)[o]$ is the $o^{th}$ assignment.

- One of the following:
  - with min
  - with max

- One of the following (*with min/max …*):
  - '$et11 \times et12 \times \ldots \cup et21 \times et22 \times \ldots \cup \ldots$' (all products have the same arity): $M = [et11 \times et12 \times \ldots, et21 \times et22 \times \ldots, \ldots]$
  - '←': card($M$) = 1; $M[1] = [et]$ where $et$ is directly left of the aggregator

- ‘→’:
  *M* = [*et1*, *et2*, …] where *et1*, *et2*, … are directly right of the aggregator and not wrapped with a negator nor preceded by a *None* quantifier (equivalent to ‘*et1* ∪ *et2* ∪ …’)
- ‘↔’:
  *M* = [*et* × *et1*, *et* × *et2*, …] where *et* is directly left of the aggregator and *et1*, *et2*, … are directly right of the aggregator and not wrapped with a negator nor preceded by a *None* quantifier (equivalent to ‘*et* × *et1* ∪ *et* × *et2* ∪ …’)

Let ***MA(n, o, p)*** denote the list of all assignments to *M[p]* in the subset of *S(n)* that contains *BA(n)[o]*.

***MC(n, o)*** = *MA(n, o, 1)* ∪ *MA(n, o, 2)* ∪ … – the set of unique assignments to elements in *M* in the subset of *S(n)* that contains *BA(n)[o]*.

**For each *n*: from the set of assignments in *S(n)* – [all but] the *k* assignments *BA(n)[k]* with the minimal/maximal positive card(*MC(n, k)*) are reported.**

If there are only *j*<*k* assignments – all those *j* are valid assignments. If there are *j*>*k* assignments with equal extremum – all those *j* are valid assignments.

M1 location:

- Below a relationship/path

  A relationship/path with an M1 below may have a relationship/path-negator and may be wrapped with an ‘O’.

- Before a quantifier-input (excluding quantifier at the start of the pattern)

  A quantifier-input with an M1 below may be wrapped with an ‘O’ (except at the pattern’s start).

- Below the pattern’s start (even when followed by a quantifier)

  See Q299v1, Q299v2, Q262.

- If *T* is empty – M1 appears directly right of the leftmost entity in *B*
- If *T* is not empty and all the entities in *T* appear right of all the entities in *B*: left of the rightmost entity in *T*. Otherwise: right of the leftmost entity in *T*
- If entities in *T*∪*B* are in different branches of a quantifier: directly left of the quantifier

***Q67:** The three people with the largest number of parents*

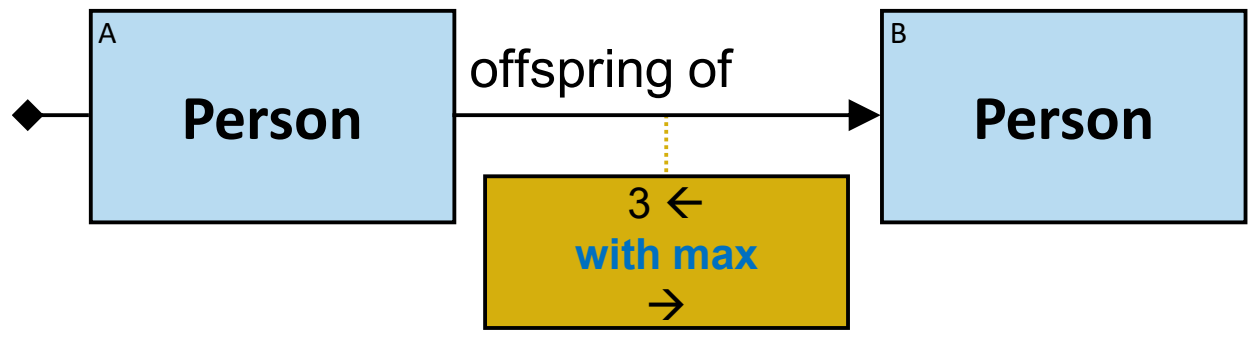

***Q68:*** *The two dragons that were frozen by the largest number of dragons*

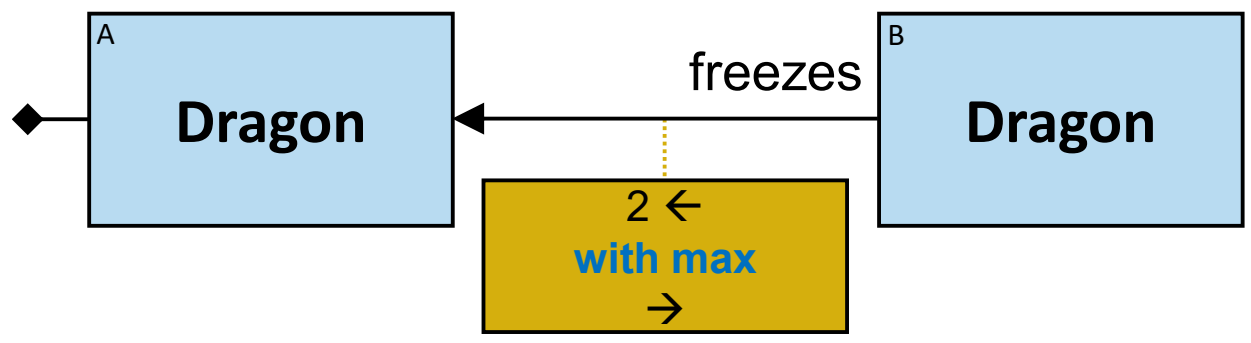

***Q69:*** *The two entities that own the largest number of entities*

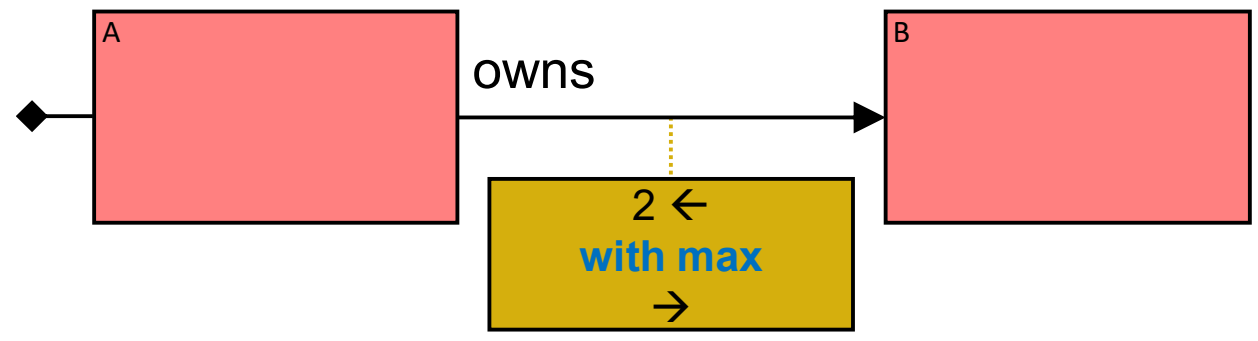

***Q70:*** *The five people who the number of people with a path of length up to four from each of them – is the smallest*

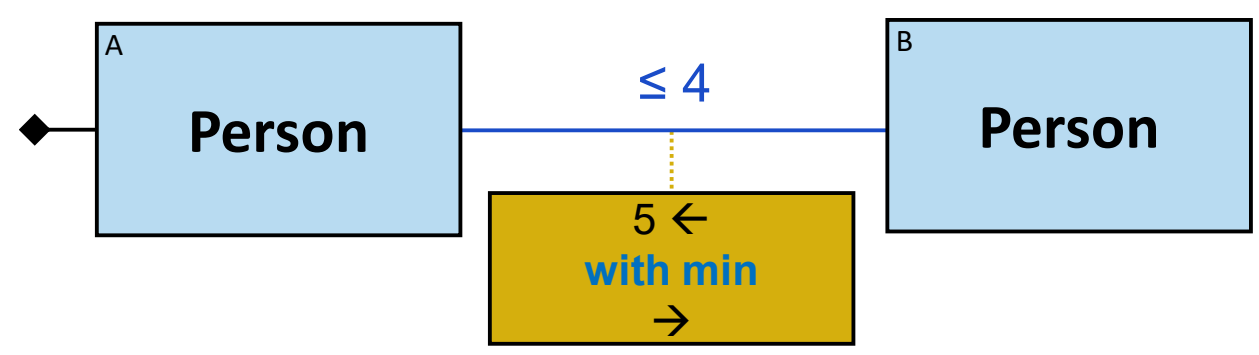

***Q196:*** *Any* ***dragon*** *owned by Brandon Stark and the three dragons* ***it*** *froze that froze the largest number of dragons*

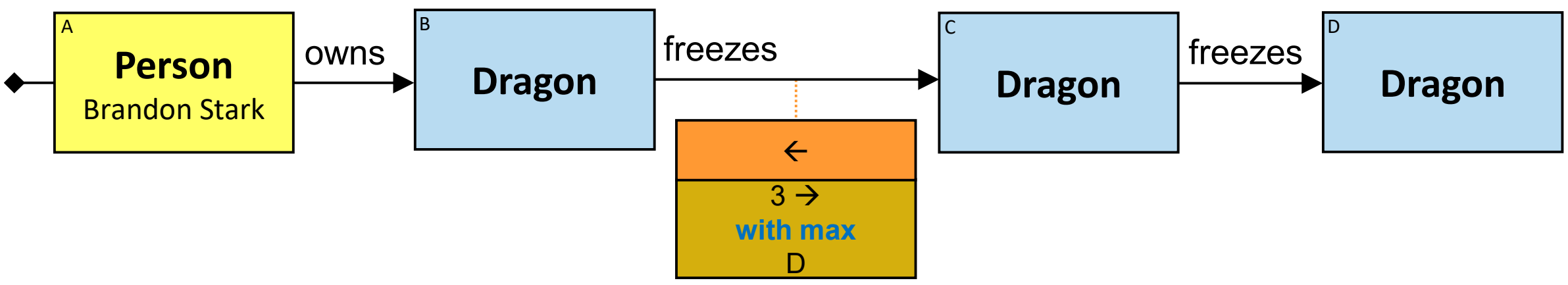

***Q197:*** *Any person A and A's three dragons that the dragons they froze – froze the largest number of distinct dragons cumulatively*

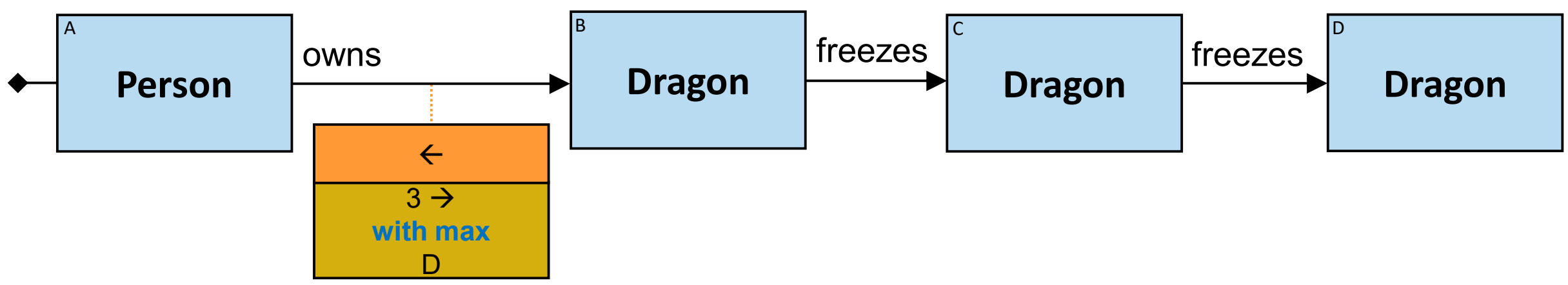

*Q234: Any person and the three dragons whose dragons froze – that froze the largest number of dragons*

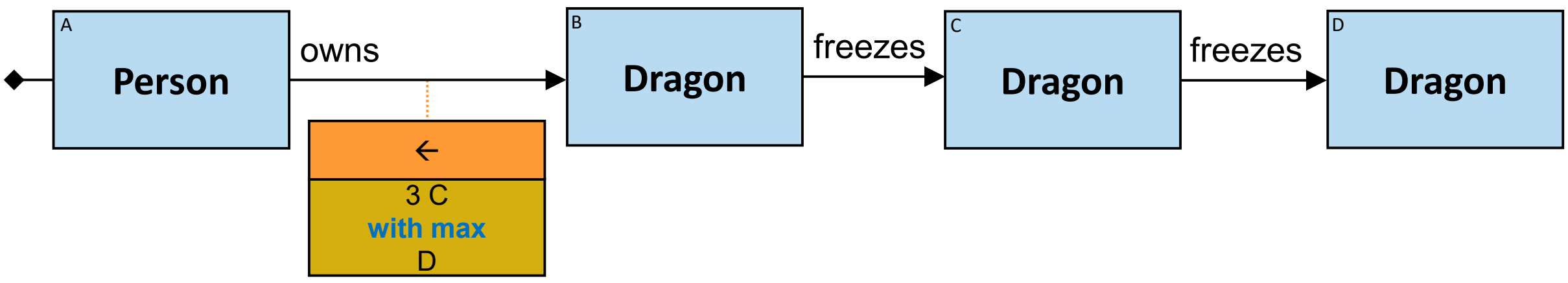

*Q236: Any person A and the three dragons whose dragons froze – that were frozen by the largest number of A's dragons*

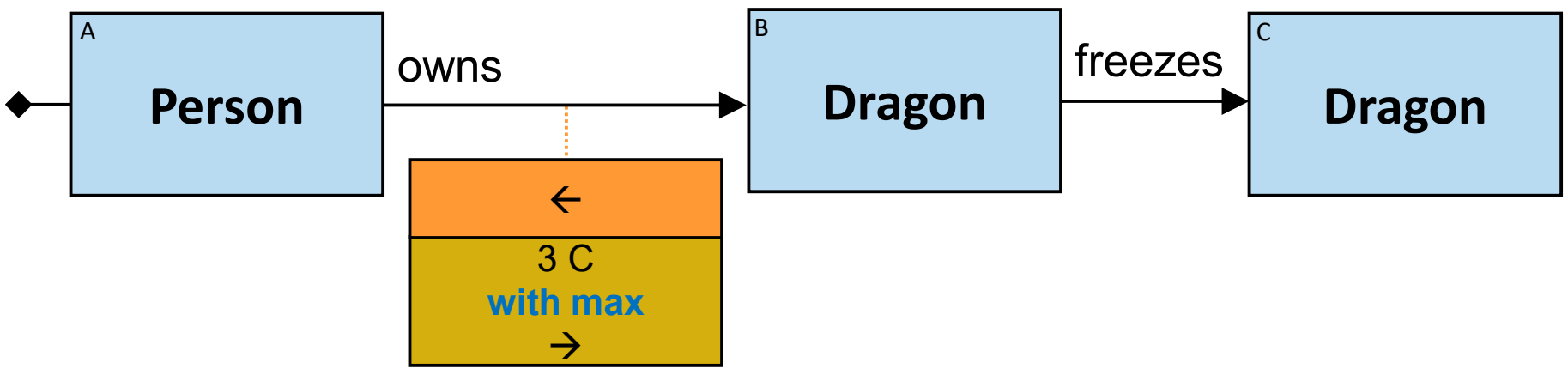

*Q227: Any **dragon** owned by Brandon Stark, and the three dragons **it** froze or fired at – that froze the largest number of dragons*

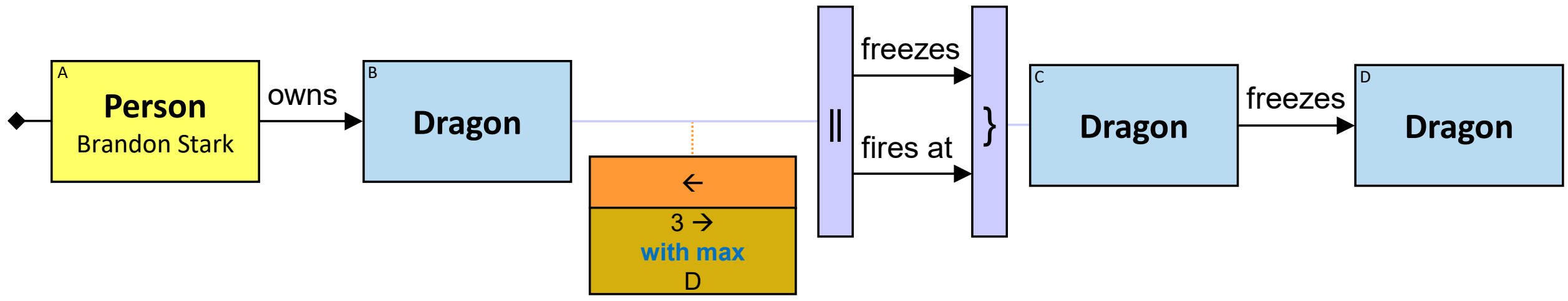

*Q238: For any pair of people (A, D) where A's dragons froze D's dragons – A's three dragons that froze the largest number of D's dragons*

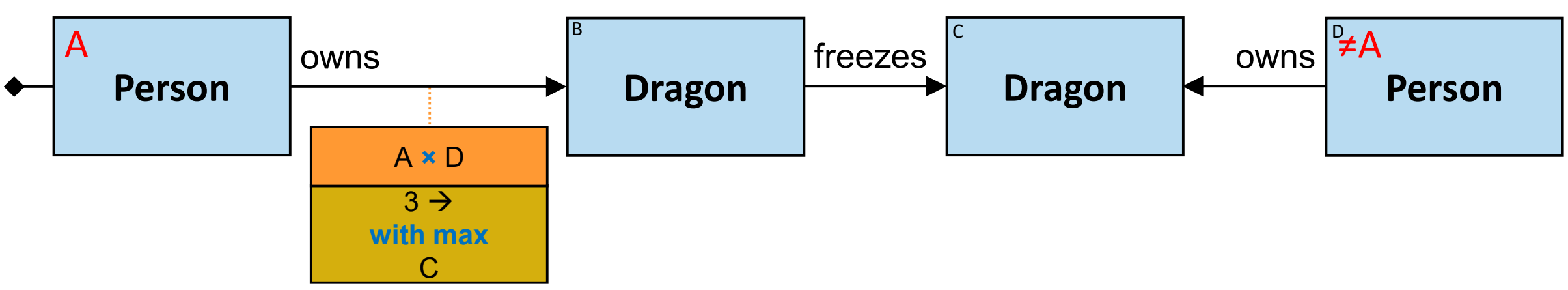

***Q299:*** *For any pair of people (A, E) where at least one of A's dragons froze at least one of E's dragons – A's (up to) three dragons that cumulatively froze the largest number of E's dragons* (two versions)

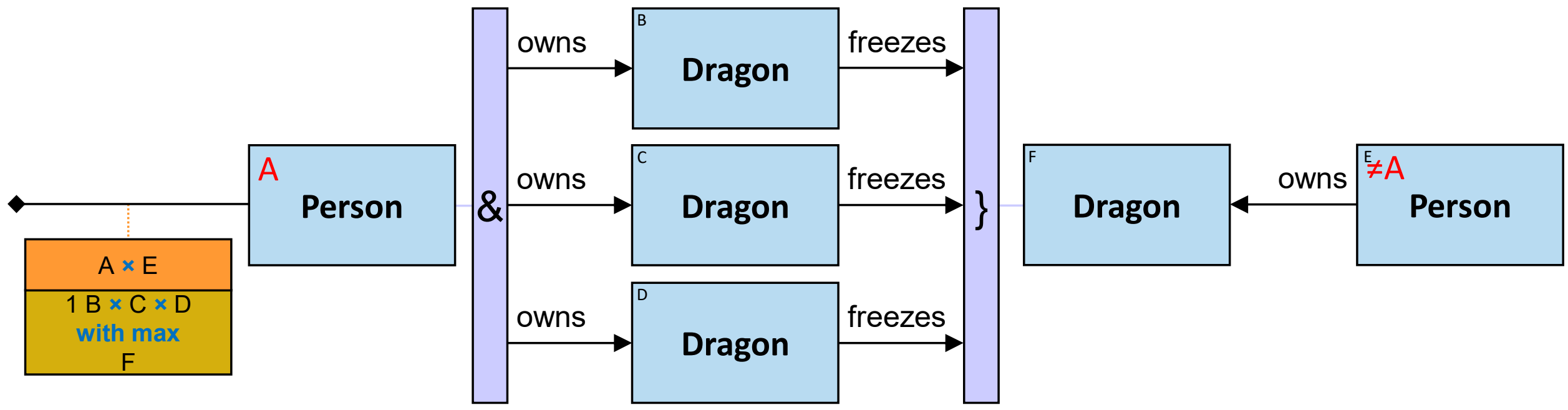

Note that the same graph-entity may be assigned to B, C, and D.

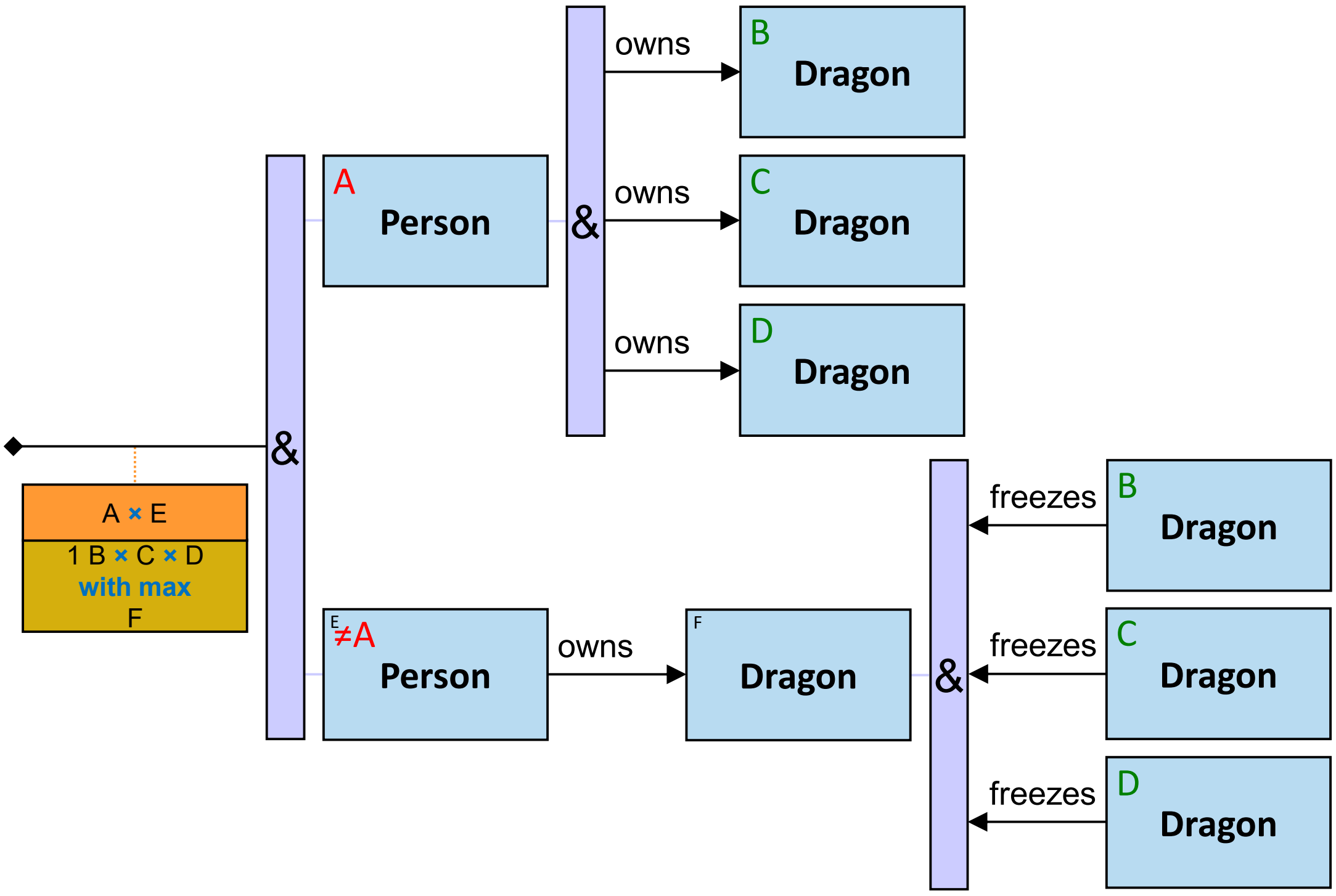

## 31. M2 AGGREGATOR

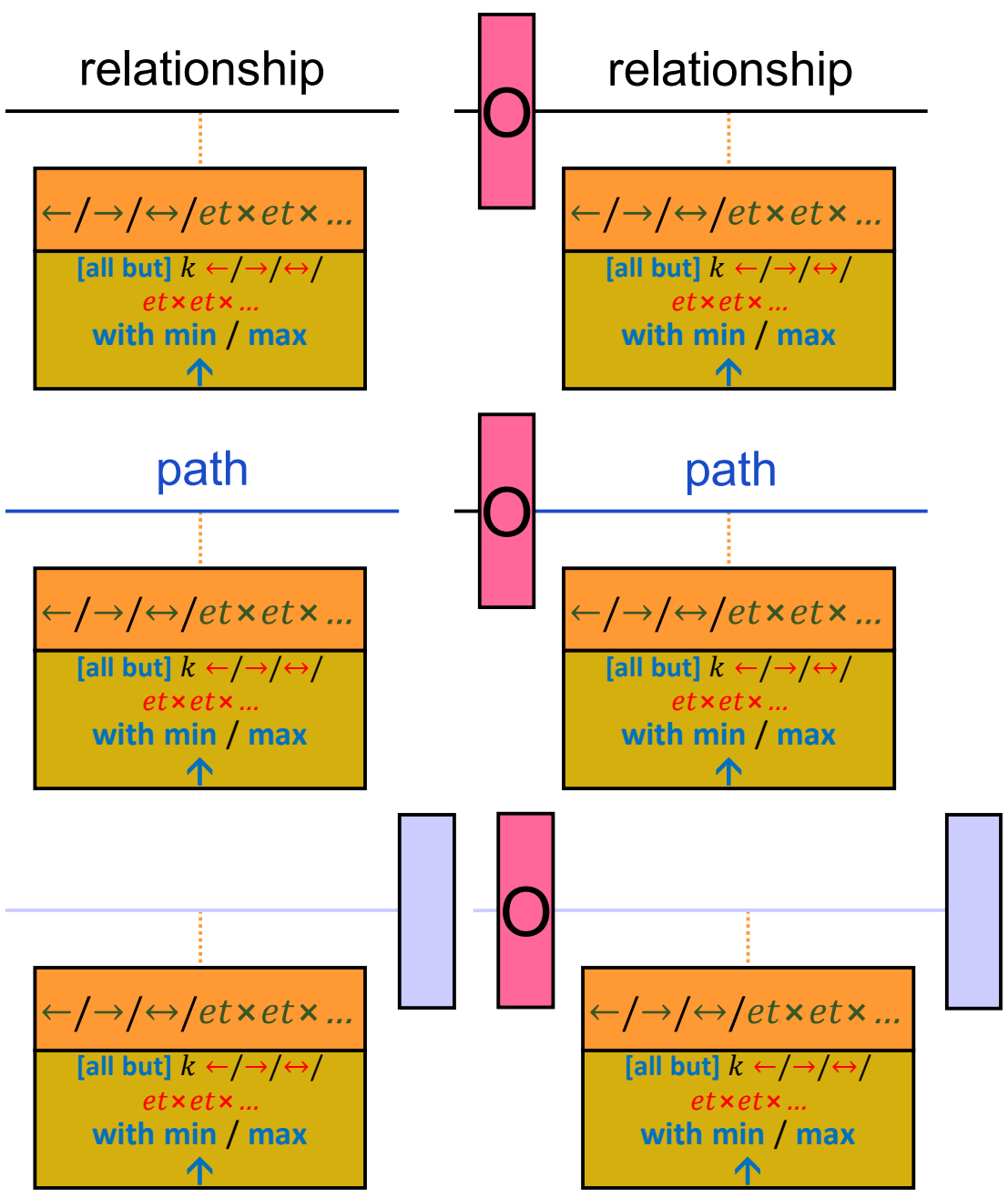

Let ***B*** denote a Cartesian product of one or more entity-tags. *B* must be nonempty and must not be composed of entity-tags composing *T*.

The lower part of an M2 aggregator contains the following elements:

- optional: "all but"
- *k*: positive integer
- One of the following:
  - '*et1* × *et2* × ...': *B* = *et1* × *et2* × ...
  - '←': *B* = entity-tag directly left of the aggregator (see Q78, Q171, Q172)
  - '→': *B* = entity-tag directly right of the aggregator not wrapped with a negator nor preceded by a *None* quantifier (valid only when there is exactly one such entity-tag) (see Q195, Q228)
  - '↔': *B* = *et1* × *et2*, where *et1* is the entity-tag directly left of the aggregator, and there is a single entity-tag directly right of the aggregator not wrapped with a negator nor preceded by a *None* quantifier – *et2* (see Q77, Q80)

  Let ***BA(n)*** denote the list of all unique assignments to *B* in *S(n)*. *BA(n)[o]* is the *o*$^{th}$ assignment.

- One of the following:
  - with min ↑
  - with max ↑

  Let ***R*** denote a list where each element is a pattern-relationship/pattern-path. *R* must be nonempty.

  When M2 appears below a relationship/path – card(R) = 1*; R[1]* = the relationship/path.

  When M2 appears below a quantifier-input – each element in *R* is the relationship/path that follows one branch of the quantifier, excluding relationships/paths wrapped with a negator or a relationship/path-negator.

Let ***RA(n, o, p)*** denote the list of all assignments to *R[p]* in the subset of *S(n)* that contains *BA(n)[o]*.

***RC(n, o)*** = *RA(n, o, 1)* ∪ *RA(n, o, 2)* ∪ … – the set of unique assignments to elements in *R* in the subset of *S(n)* that contains *BA(n)[o]*.

**For each *n*: from the set of assignments in *S(n)* – [all but] the *k* assignments *BA(n)[k]* with the minimal/maximal positive card(*RC(n, k)*) are reported.**

If there are only *j*<*k* assignments – all those *j* are valid assignments. If there are *j*>*k* assignments with equal extremum – all those *j* are valid assignments.

M2 location:

- Below the relationship/path whose assignments are counted

  A relationship/path with an M2 below it may be wrapped with an 'O'.

- Below the quantifier-input (except a *None* quantifier) that assignments to relationships/paths on its right are counted

  A quantifier-input with an M2 below it:

  - may be wrapped with an 'O'
  - at least one branch of the quantifier (or sub-branch of a sequence of quantifiers) must start with a relationship/path with no relationship/path-negator that is not wrapped with a negator nor preceded by a *None* quantifier

***Q78:*** *The four dragons that froze Balerion the largest number of times*

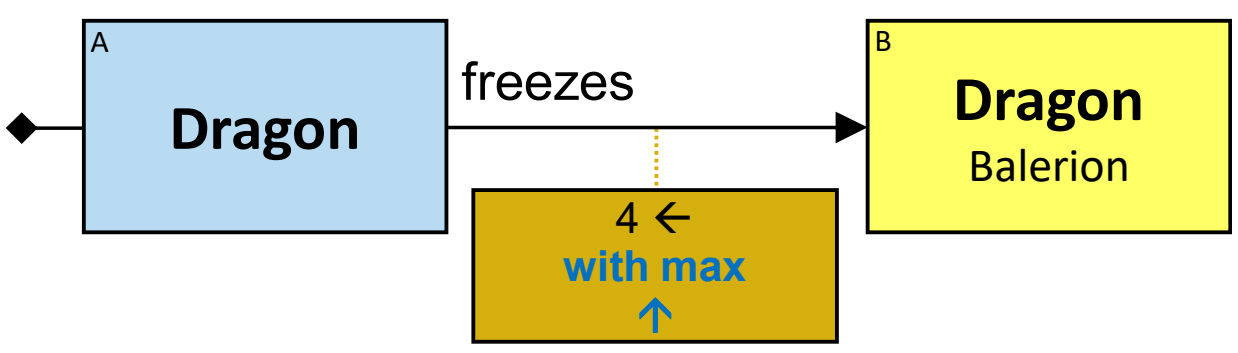

***Q171:*** *The two dragons that were frozen the largest number of times*

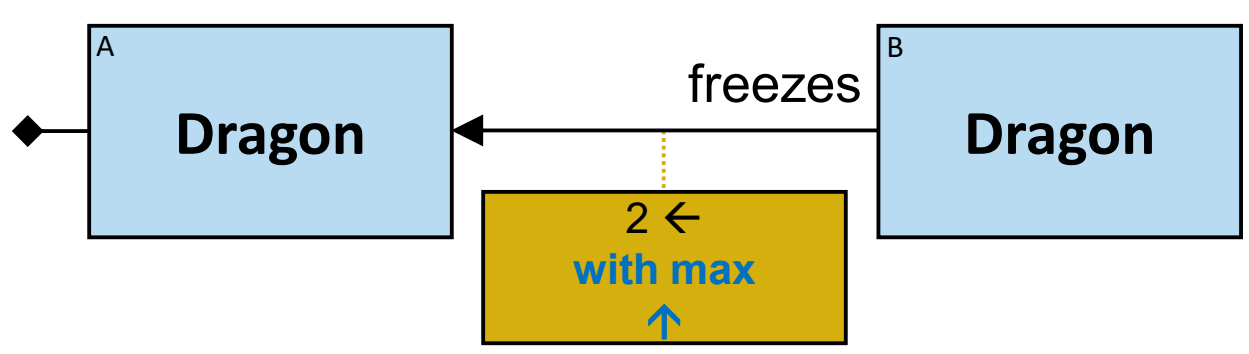

***Q172:*** *The five people with the smallest positive number of paths of length up to four to some person*

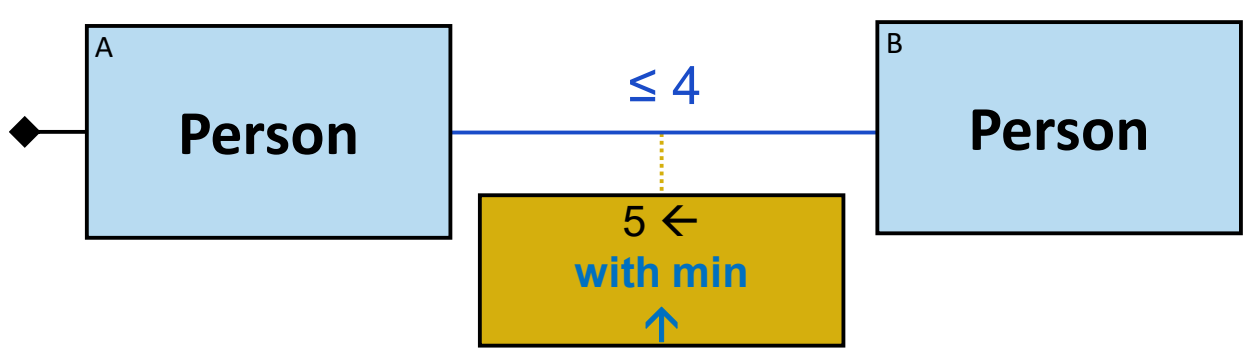

(Compare with Q324)

***Q77:*** *The five pairs of dragons (A, B) with the largest number of times B froze A*

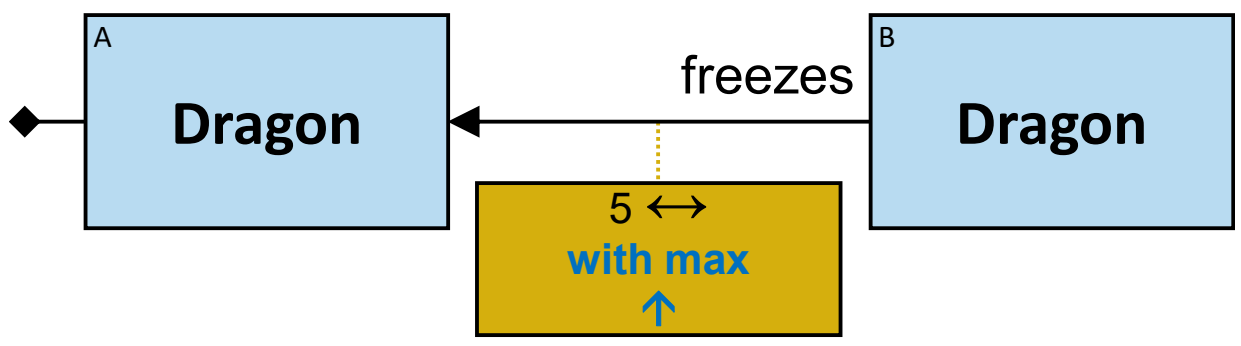

***Q80:*** *The three pairs of people with the largest number of paths of length up to four between them*

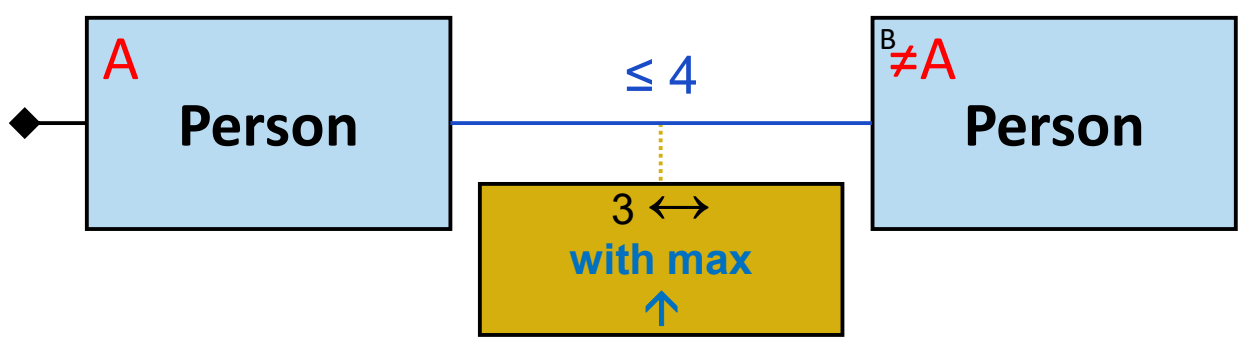

***Q195:*** *Any dragon owned by Brandon Stark and the three dragons it froze the largest number of times*

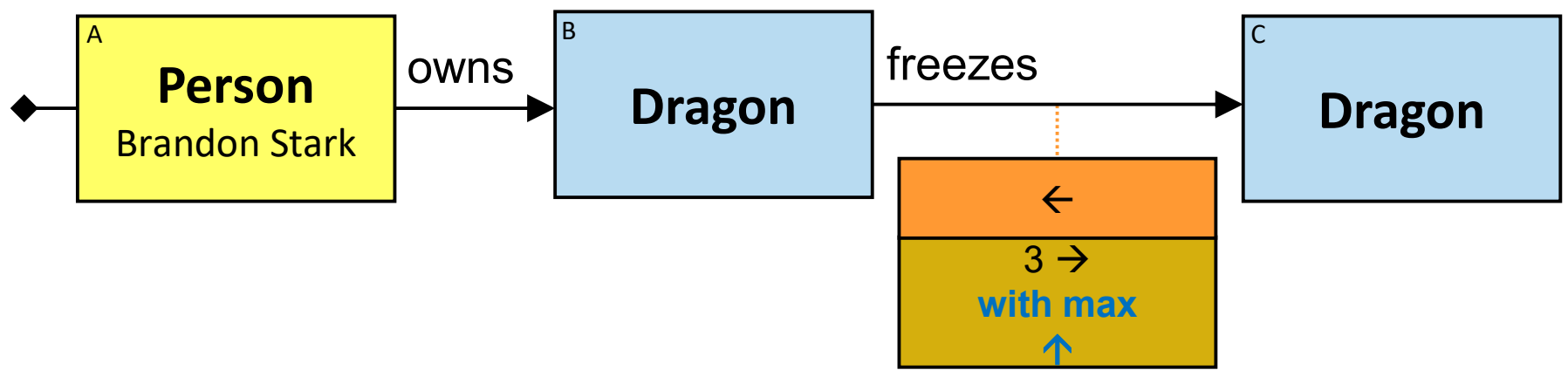

***Q231:*** *Any person A and the three dragons that were frozen by dragons that were frozen the largest number of times by A's dragons*

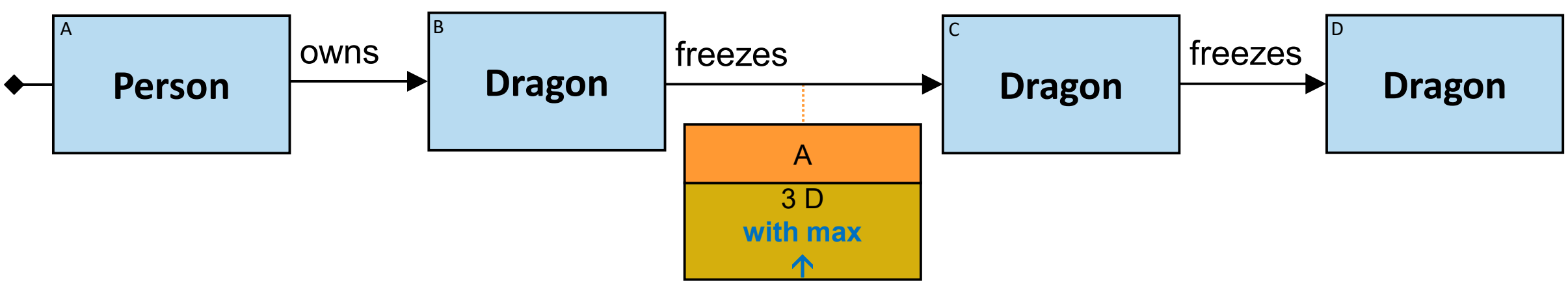

***Q228:*** *Any dragon owned by Brandon Stark that fired at at least two dragons, and the three dragons it fired the largest number of times*

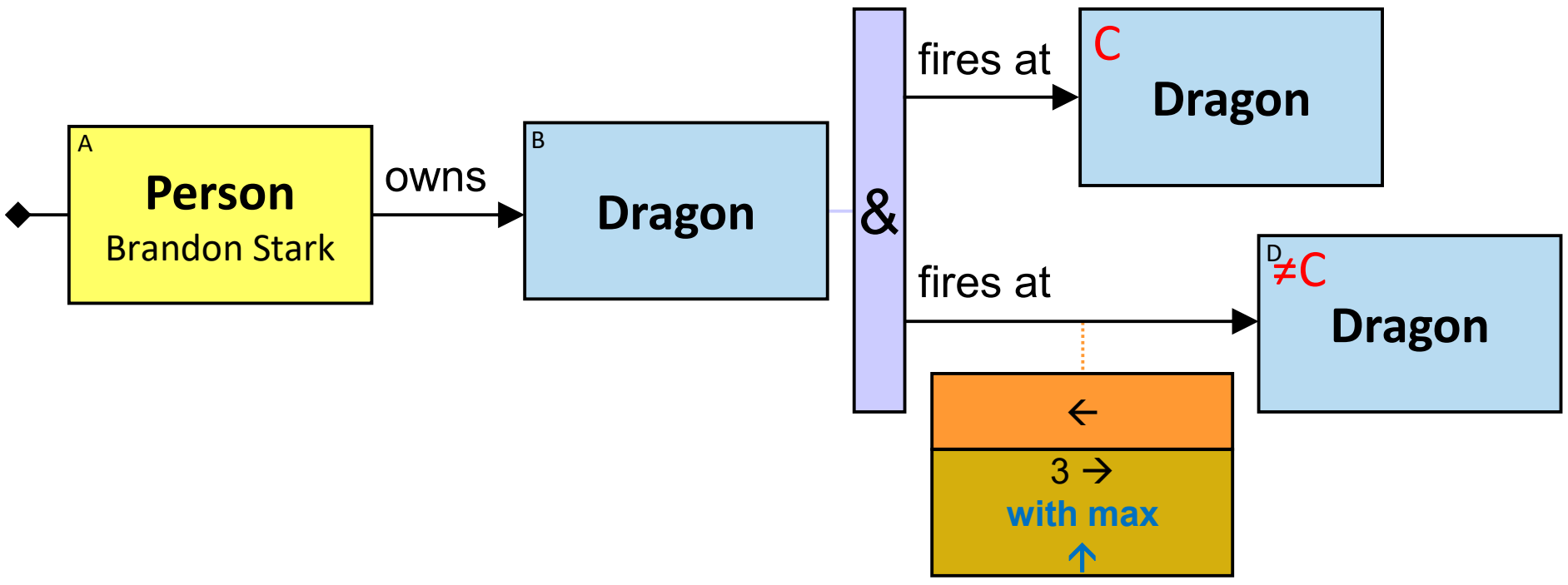

(counting the number of relationships assignments directly right of the quantifier)

***Q237:*** *For any pair of (A – a dragons owner, and C – a dragon frozen by A's dragons) – the three dragons owned by A that froze C the largest number of times*

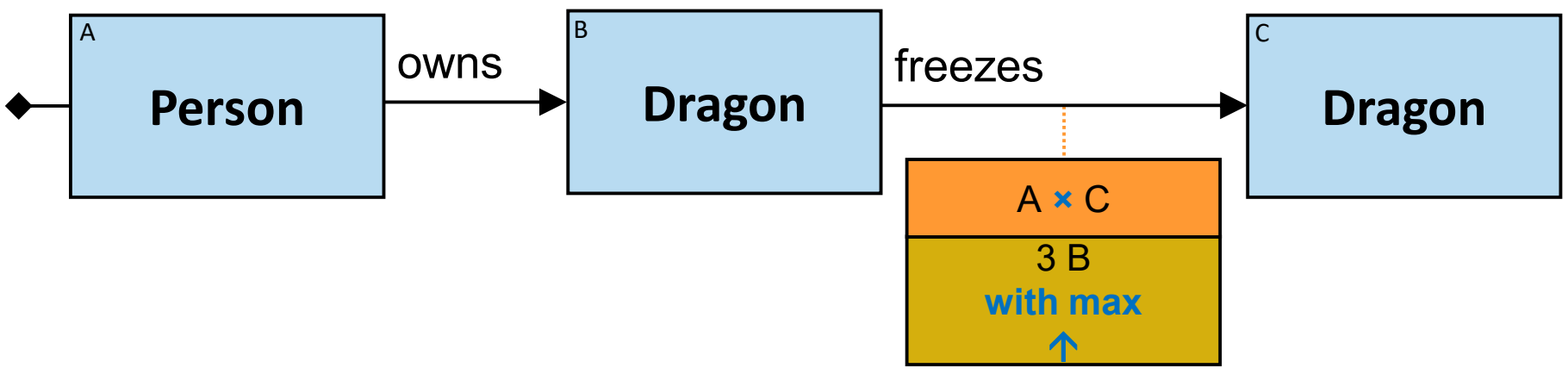

***Q239:*** *For any pair of people (A, D) where A's dragons froze D's dragons – the three pairs of (A's dragon B, D's dragon C) where B froze C the largest number of times*

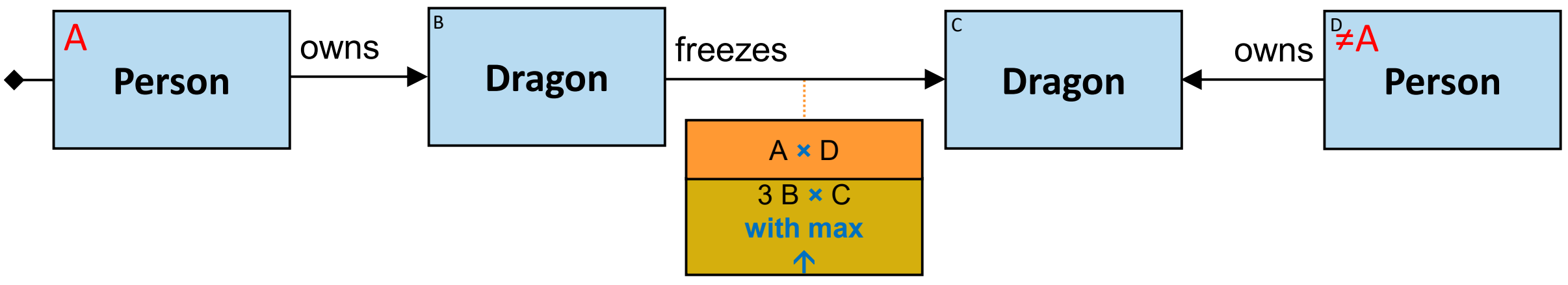

## 32. M3 AGGREGATOR

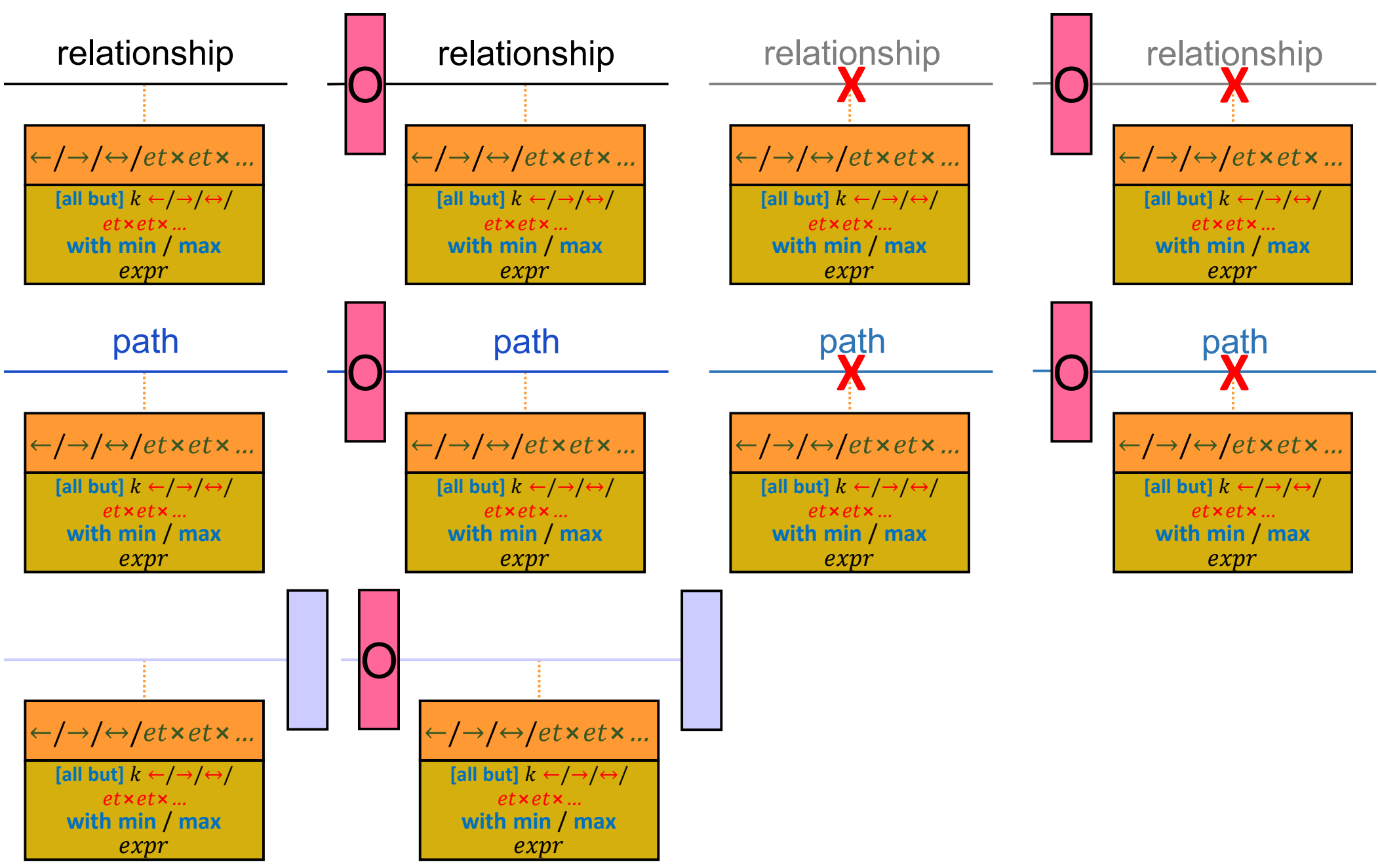

Let ***B*** denote a Cartesian product of one or more entity-tags. *B* must be nonempty and must not be composed of entity-tags composing *T*.

The lower part of an M3 aggregator contains the following elements:

- optional: “all but”
- *k*: positive integer
- One of the following:
    - ‘*et1* × *et2* × ...’: *B* = *et1* × *et2* × ...
    - ‘←’: *B* = entity-tag directly left of the aggregator
    - ‘→’: *B* = entity-tag directly right of the aggregator not wrapped with a negator nor preceded by a *None* quantifier (valid only when there is exactly one such entity-tag)
    - ‘↔’: *B* = *et1* × *et2*, where *et1* is the entity-tag directly left of the aggregator, and there is a single entity-tag directly right of the aggregator not wrapped with a negator nor preceded by a *None* quantifier – *et2*

  Let ***BA(n)*** denote the list of all unique assignments to *B* in *S(n). BA(n)[o]* is the $o^{th}$ assignment.

- One of the following:
    - with min *expr*
    - with max *expr*

  *expr* is an expression composed of constants, properties, and sub-properties of *R* (when M3 appears below a relationship *R* with no relationship-negator) (see Q74), EA-tags defined on top of the aggregator (see Q91, Q274, Q306), and EA-tags defined right of the aggregator (see Q130, Q128)

  Let ***RA(n, o)*** denote the minimal/maximal assignment to *expr* in the subset of *S(n)* that contains *BA(n)[o]*.

**For each *n*: from the set of assignments in *S(n)* – [all but] the *k* assignments *BA(n)[k]* with the minimal/maximal *RA(n, k)*.**

If there are only *j*<*k* assignments – all those *j* are valid assignments. If there are *j*>*k* assignments with equal extremum – all those *j* are valid assignments.

M3 location:

- Below a relationship/path

  When M3 appears below a relationship *R,* and *expr* is composed of at least one property or sub-property of *R*, *R* may be wrapped with an ‘O’. Otherwise – a relationship or path with an M3 below it may have a relationship/path-negator and may be wrapped with an ‘O’.

- Before a quantifier-input (excluding quantifier at the start of the pattern)

  A quantifier-input with an M3 below it may be wrapped with an ‘O’.

- Below the pattern’s start (even when followed by a quantifier)

  See Q130, Q216, Q264.

- If *T* is empty – M3 appears directly right of the leftmost entity in *B*.
- If *T* is not empty and all the entities in *T* appear right of all the entities in *B*: left of the rightmost entity in *T*. Otherwise: right of the leftmost entity in *T*
- If entities in *T*∪*B* are in different branches of a quantifier: directly left of the quantifier

*Q130: The four oldest people*

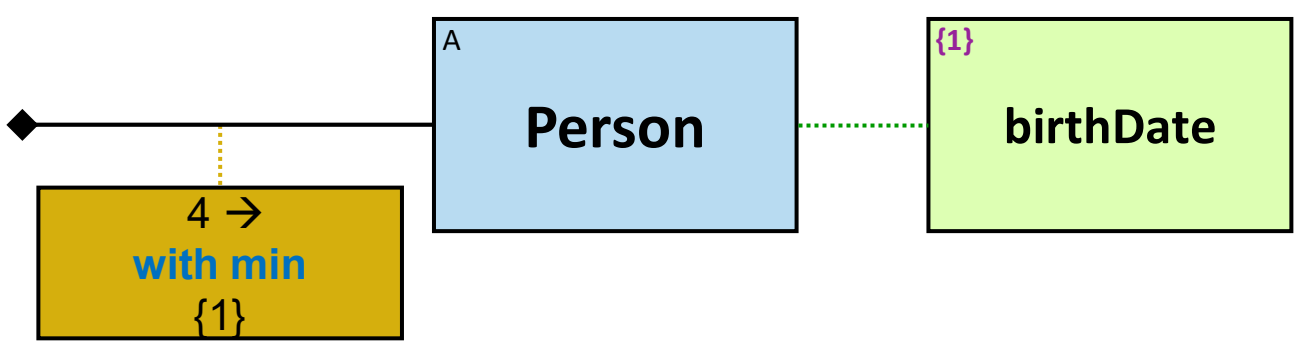

*Q131: The four oldest males*

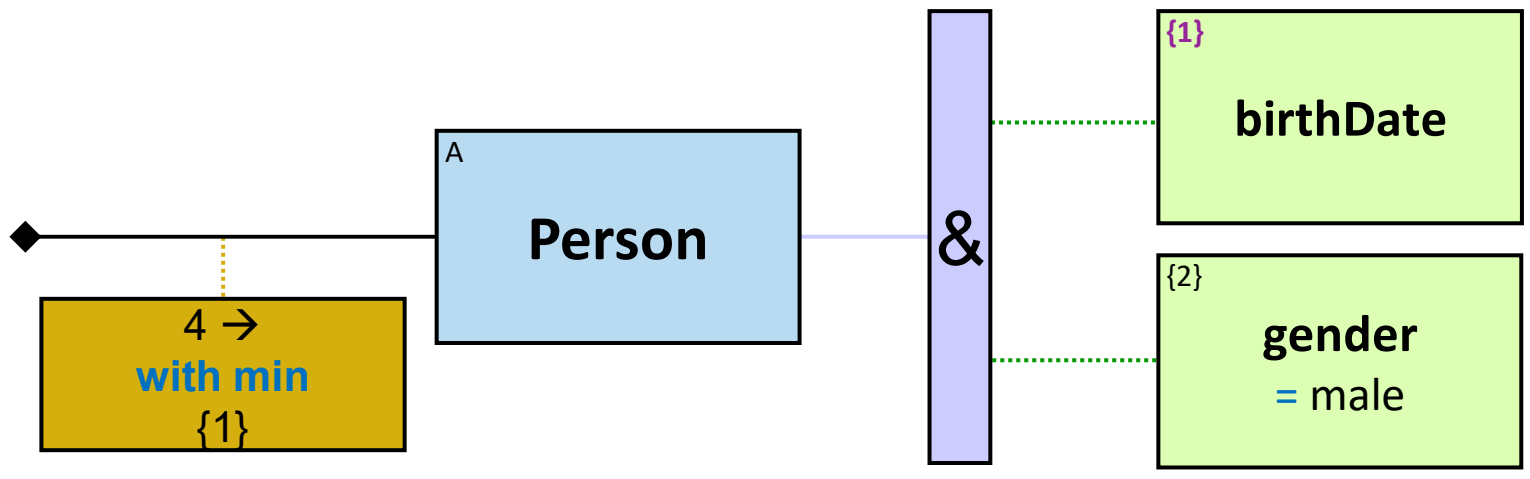

*Q118: Any person A and A's three oldest offspring*

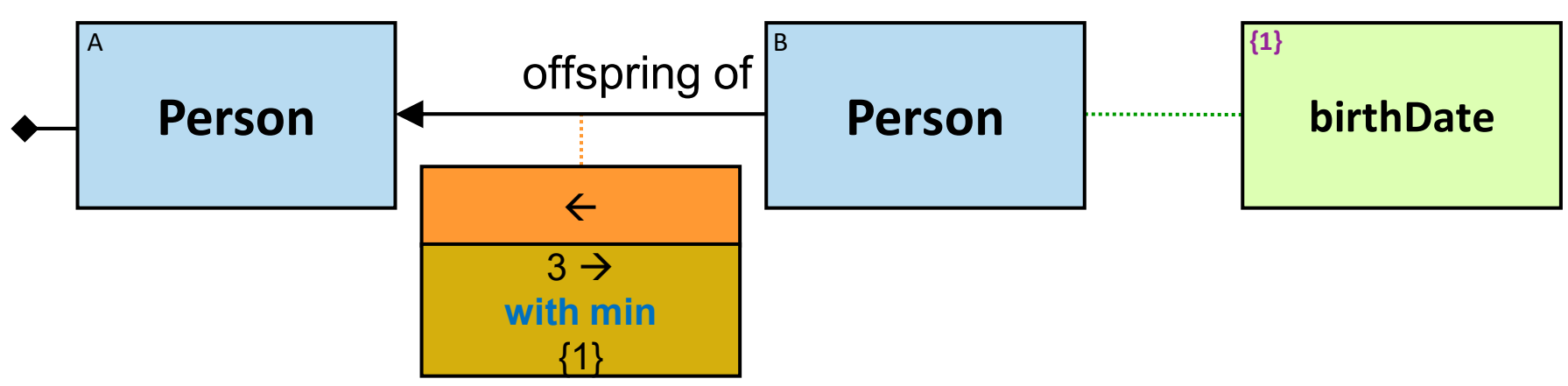

*Q119: Any person A and A's three youngest sons*

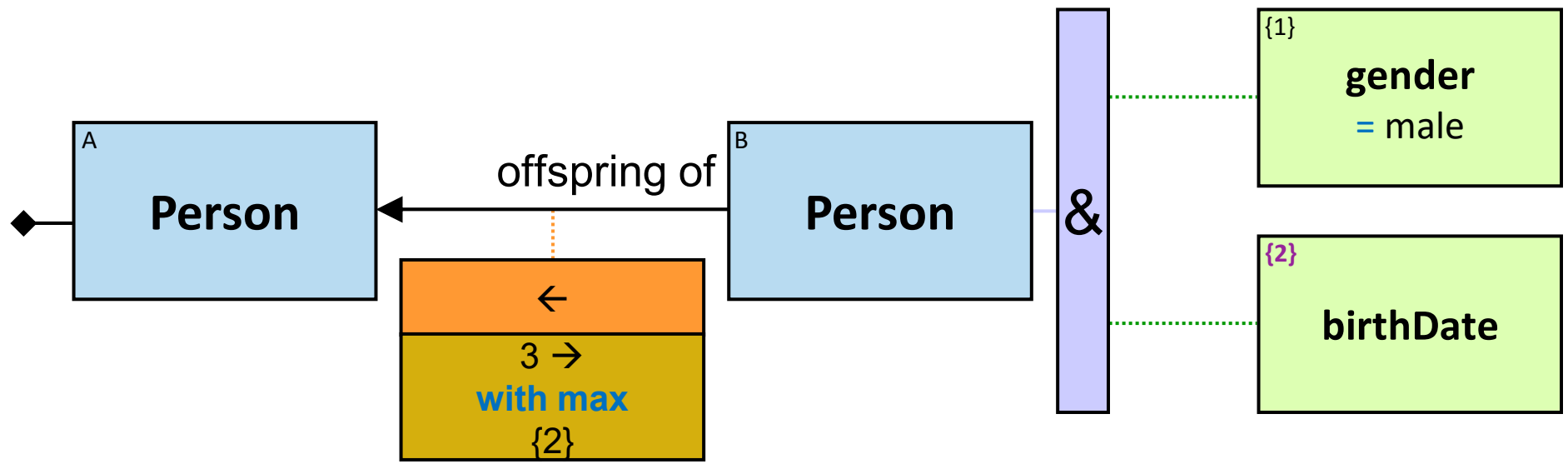

*Q74: Any person A and A's three horses with the minimal first ownership start date*

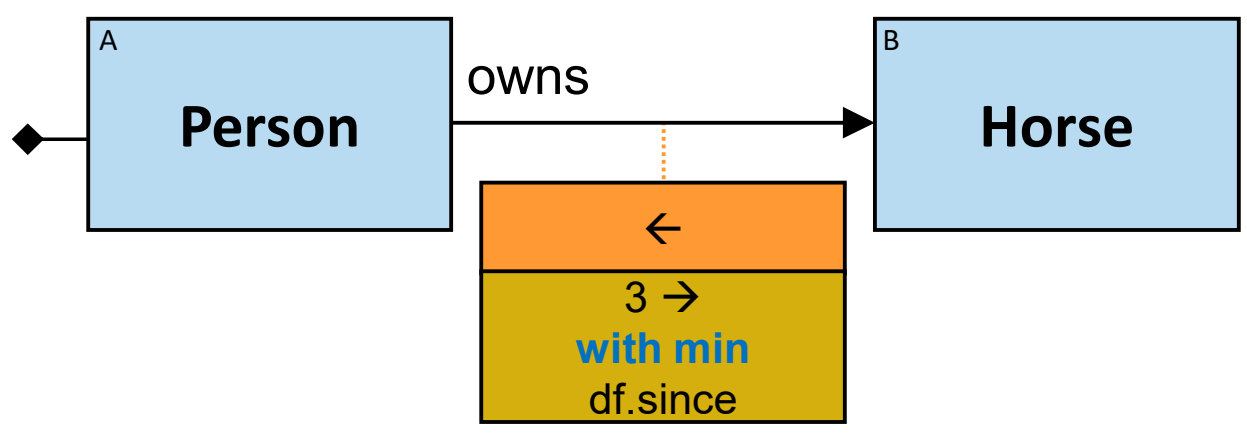

***Q230:*** *Any person A and the three people A is a friend of or friend of an offspring of that own the heaviest horses*

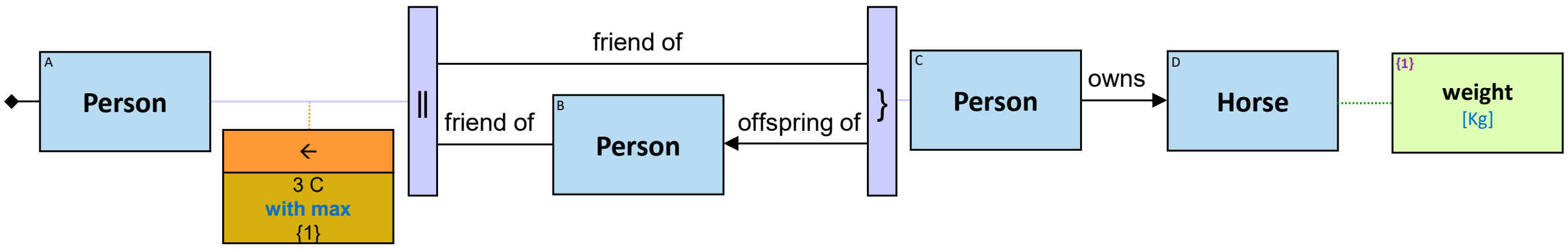

***Q232:*** *Any person A and the three heaviest horses owned by people A is a friend of or a friend of an offspring of*

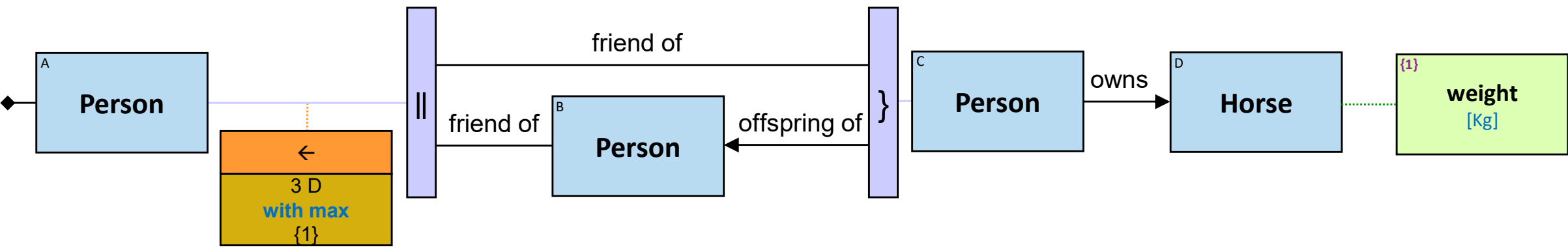

---

## 33. R1 AGGREGATOR

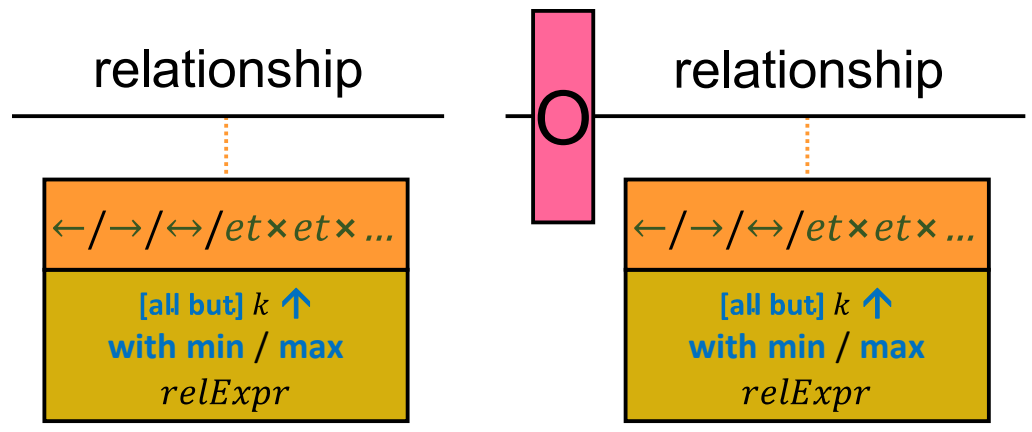

Let ***R*** denote the relationship R1 appears below it.

The lower part of an R1 aggregator contains the following elements:

- optional: "all but"
- *k*: positive integer
- '↑' (an arrow pointing to the relationship on top of the aggregator)
- One of the following:
  - with min *relExpr*
  - with max *relExpr*

  *relExpr* is an expression containing at least one property or sub-property of *R*

  Let ***RA(n)*** denote the list of all unique assignments to *R* in *S(n)*. *RA(n)[o]* is the o[th] assignment.

**For each *n*: from the set of assignments in *S(n)* – [all but] the *k* assignments *RA(n)[k]* with the minimal/maximal *relExpr* are reported.**

If there are only *j*<*k* assignments – all those *j* are valid assignments. If there are *j*>*k* assignments with equal extremum – all those *j* are valid assignments.

R1 location:

- Below the relationship whose property is referenced

  The relationship may be wrapped with an 'O'.

*Q241: The four longest freezes*

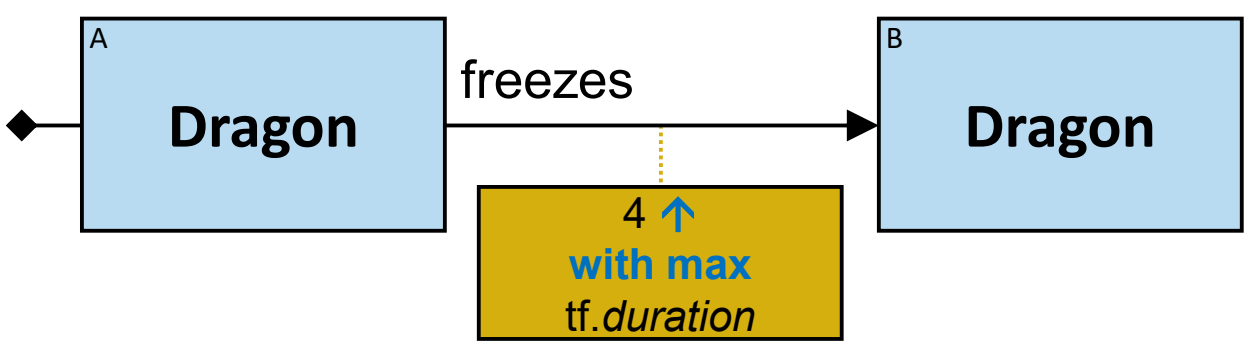

*Q161: For each dragon that froze at least one dragon at least once: the four longest freezes*

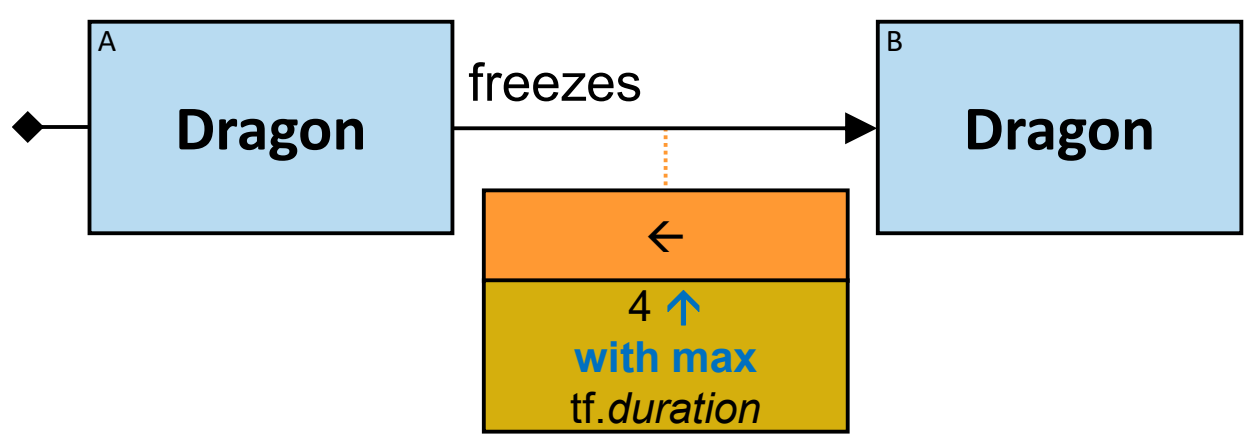

*Q160: For each pair of dragons (A, B) where A froze B at least once: The four longest freezes*

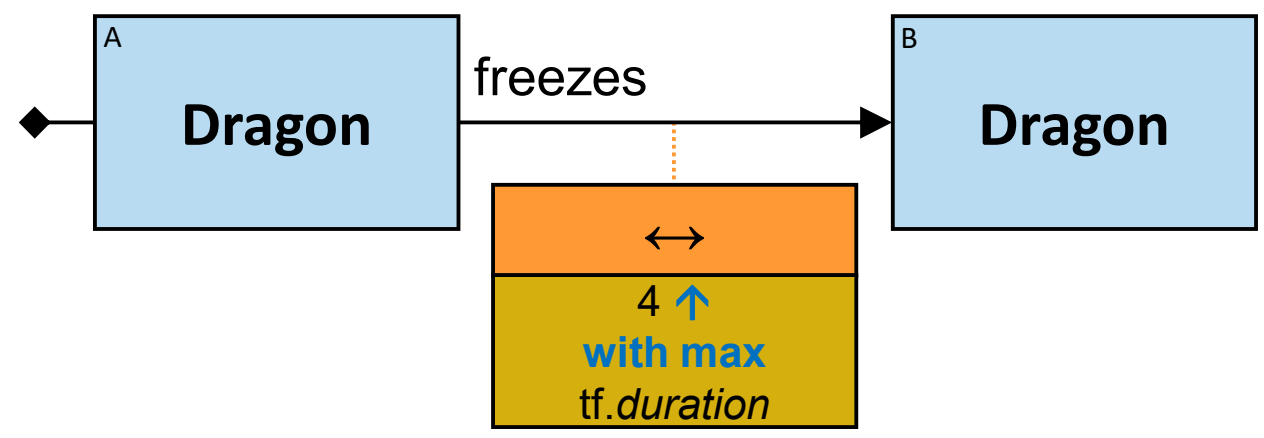

*Q240: For any pair of people (A, D): The four longest freezes where any of A's dragons froze any of D's dragons*

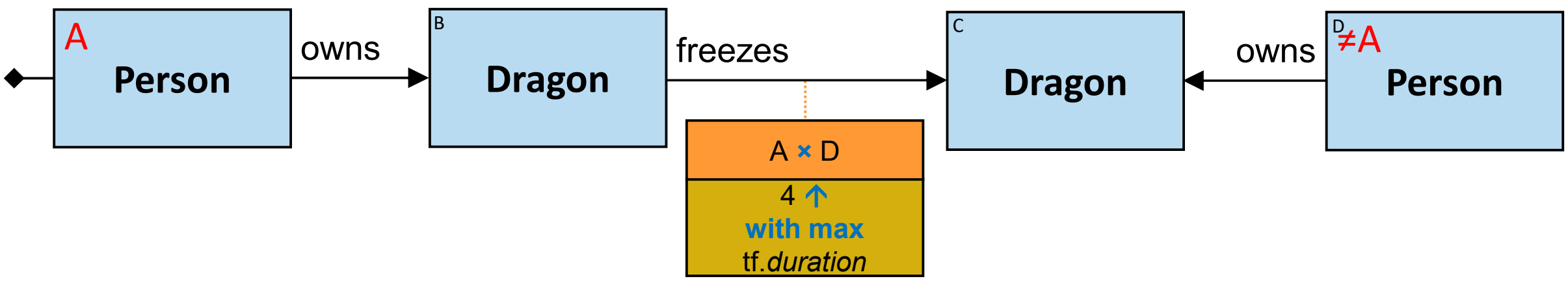

## 34. AGGREGATOR CHAINS

Relationships' expressions and aggregators can be *chained*. When chained, each relationship's expression and each aggregator with a constraint is a filtering stage. An assignment is valid only if all the chained constraints are satisfied, or in other words – if it passed all filtering stages.

Within a chain, relationships' expressions may not appear below aggregators. Also, an aggregator may not appear in a horizontal quantifier's branch. It may appear below a horizontal combiner that combines all the branches (see Q302, Q303).

***Q96:*** *Any dragon Balerion froze more than ten times – each on or after January 1, 1010, and for a duration shorter than ten minutes*

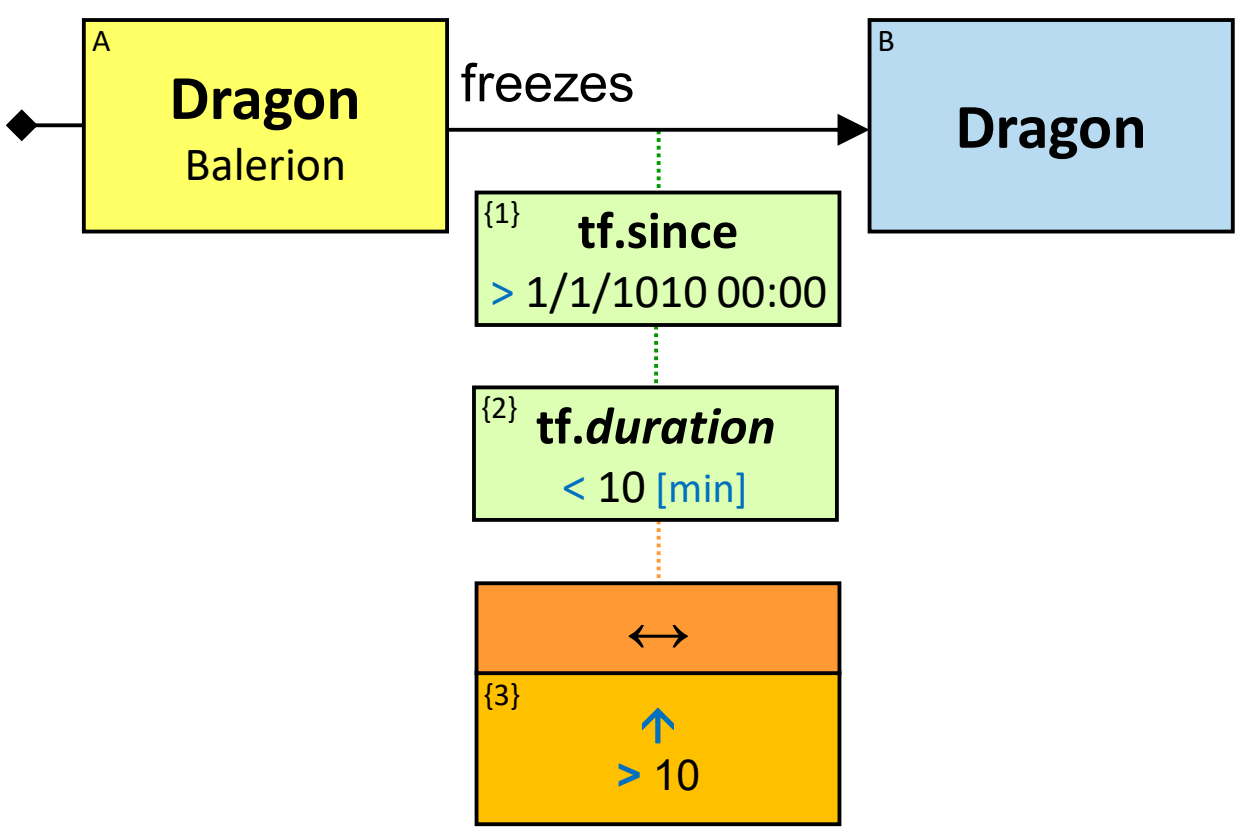

Note that the order of the constraints along the chain matters: the top two constraints filter relationships based on their properties' value. The third constraint is on the number of relationships that passed these filters.

***Q302:*** *Any dragon A that froze at least three dragons – each at least one time where at least two of the following conditions are satisfied: (i) the freeze duration was longer than ten minutes (ii) the freeze started after January 1, 980 (iii) the freeze ended before February 1, 980*

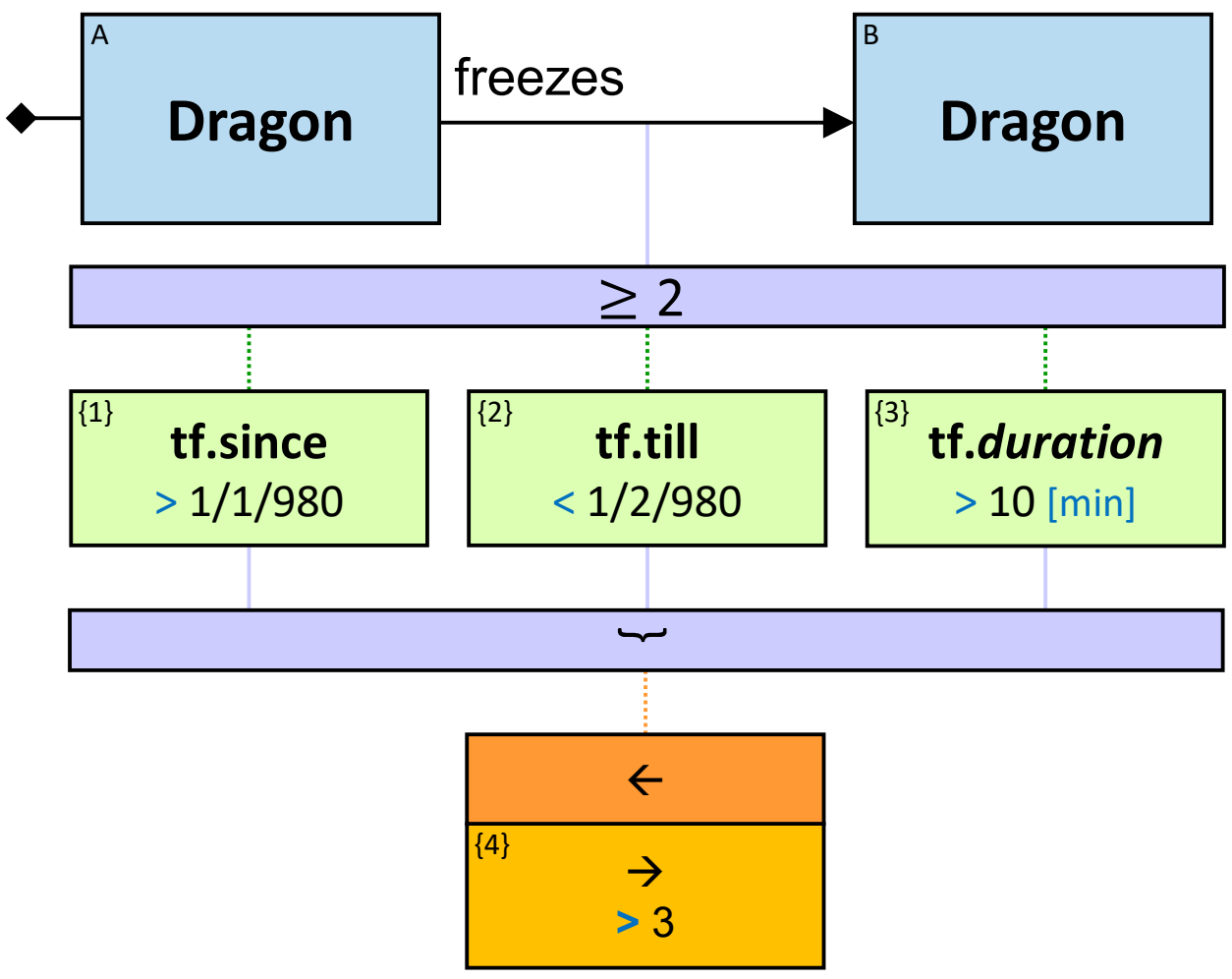

See Q185 for a different pattern structure that could be used.

***Q303:*** *Any pair of dragons (A, B) where A froze B at least three times, each for more than ten minutes, and either the freeze started after January 1, 980 or ended before February 1, 980*

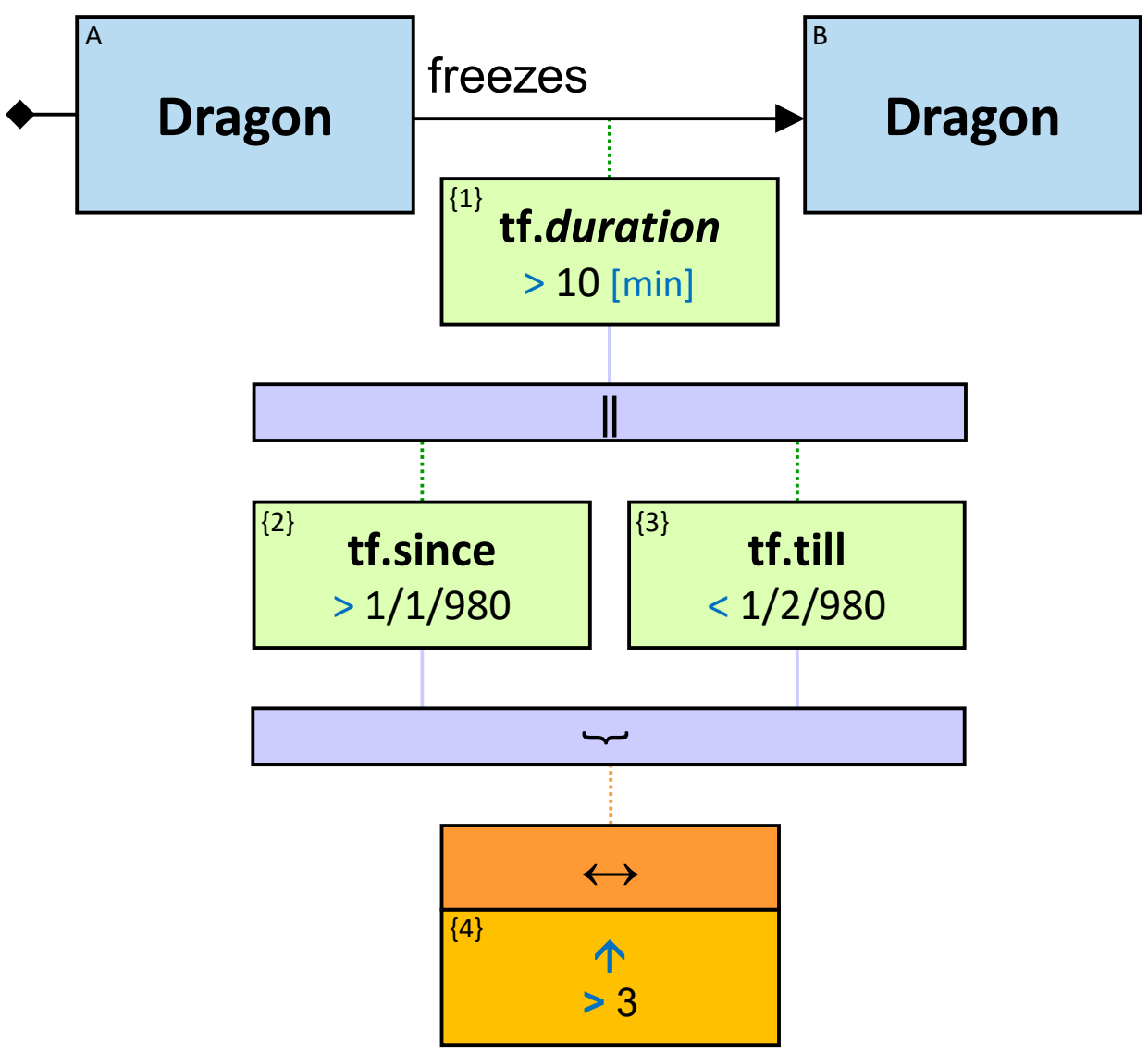

***Q186:*** *Any dragon that Balerion froze more than ten times for less than ten minutes, and at least once for ten minutes or more*

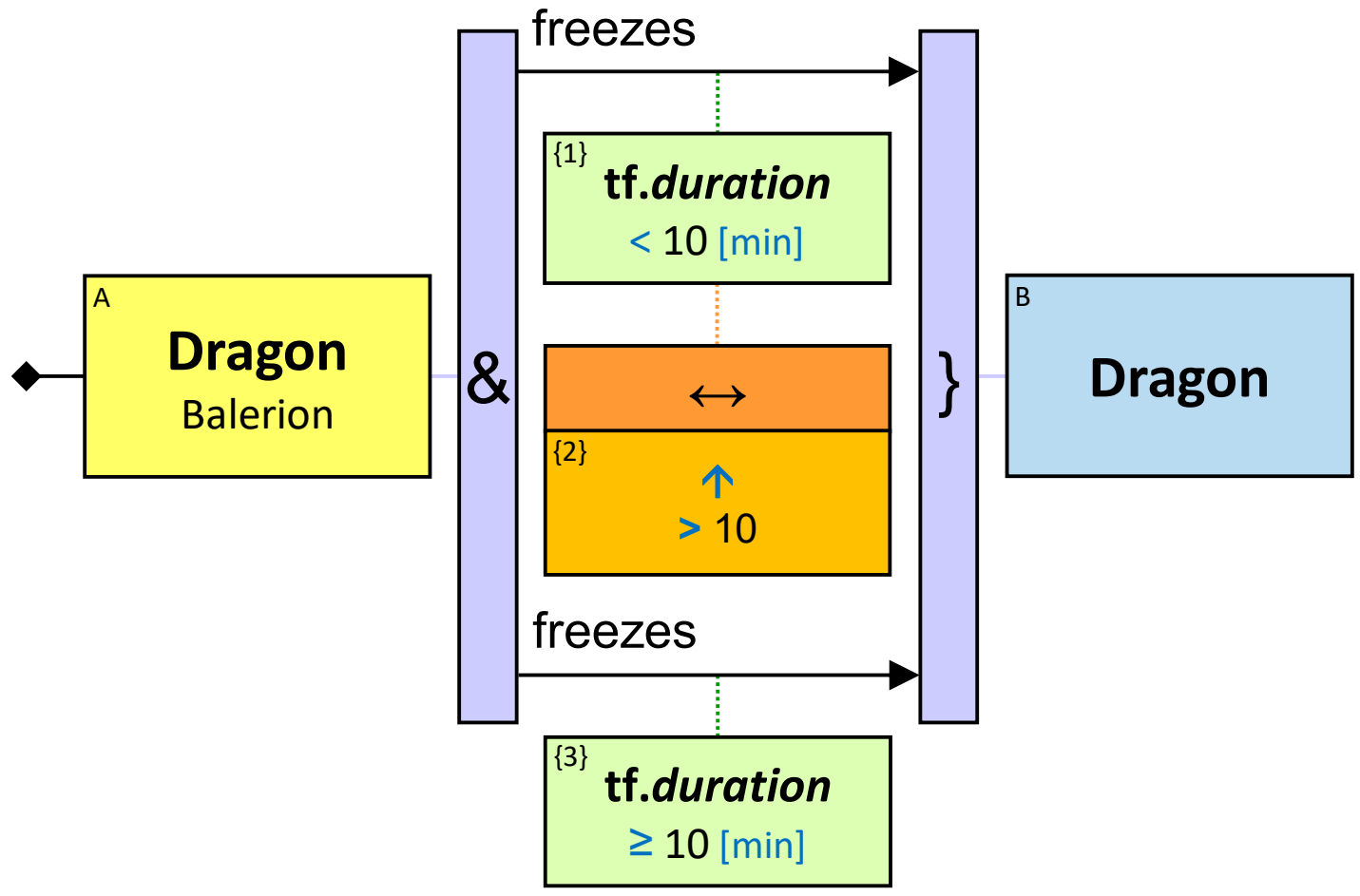

***Q99:** Any dragon that Balerion froze more than ten times for less than ten minutes, and not once for ten minutes or more* (two versions)

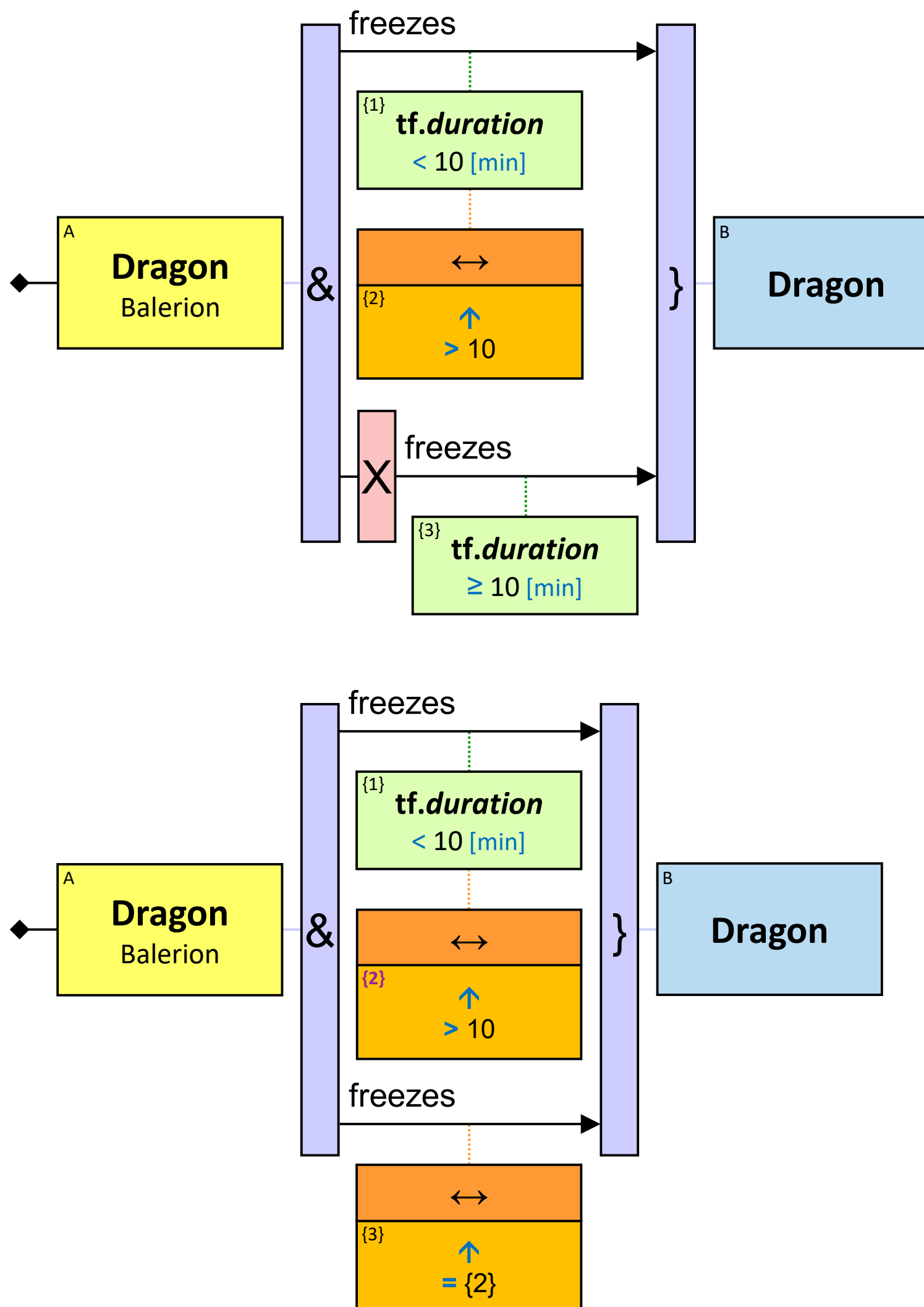

***Q94:** Any dragon that froze more than three dragons: each more than five times for more than ten minutes*

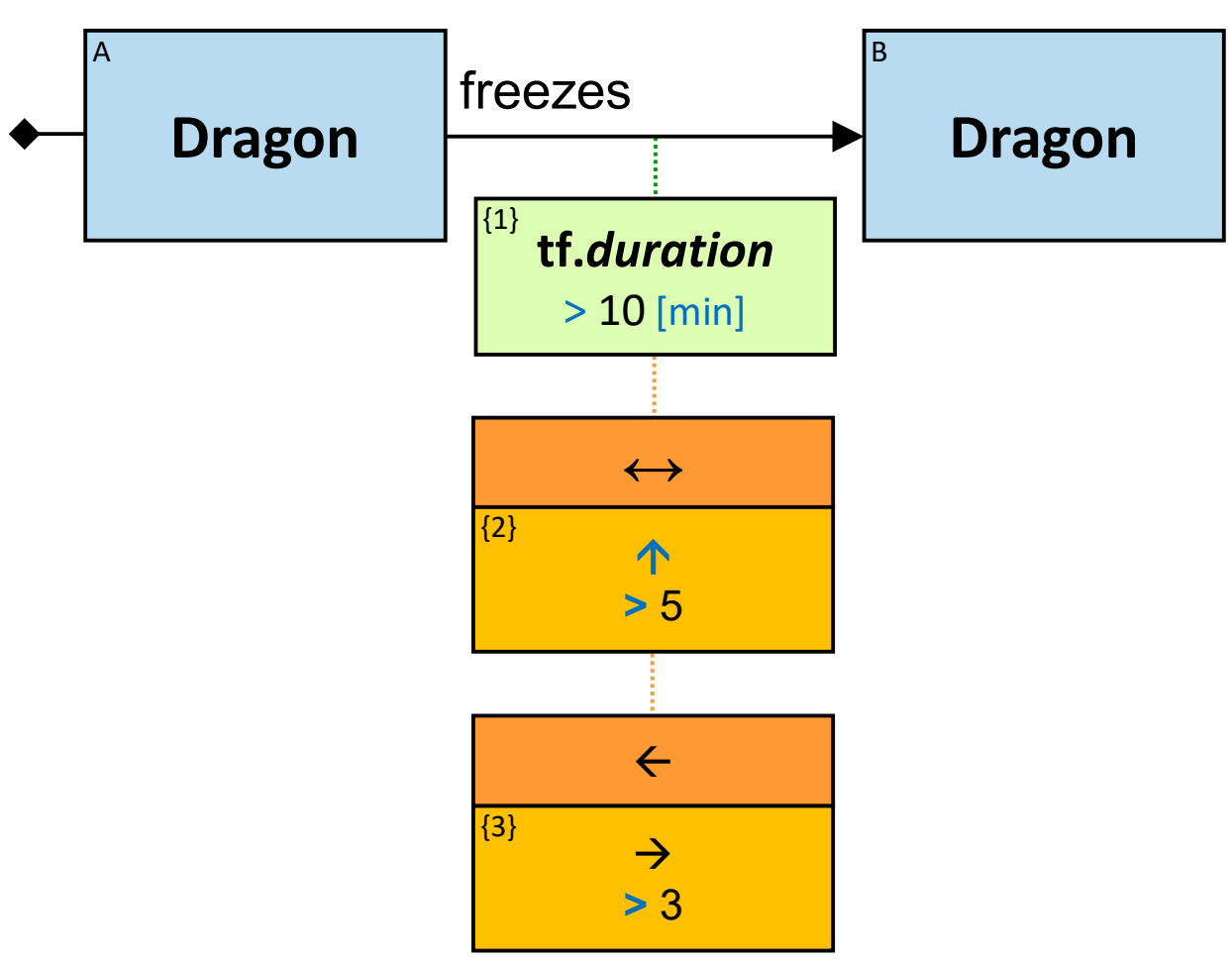

Filtering stages:

- Pass freezes longer than ten minutes
- Pass pairs of dragons (A, B) where A froze B more than five times for more than ten minutes
- Pass dragons A that froze more than three B's each more than five times for more than ten minutes

***Q93:*** *Any dragon A that froze more than three dragons – each at least five times for more than ten minutes*

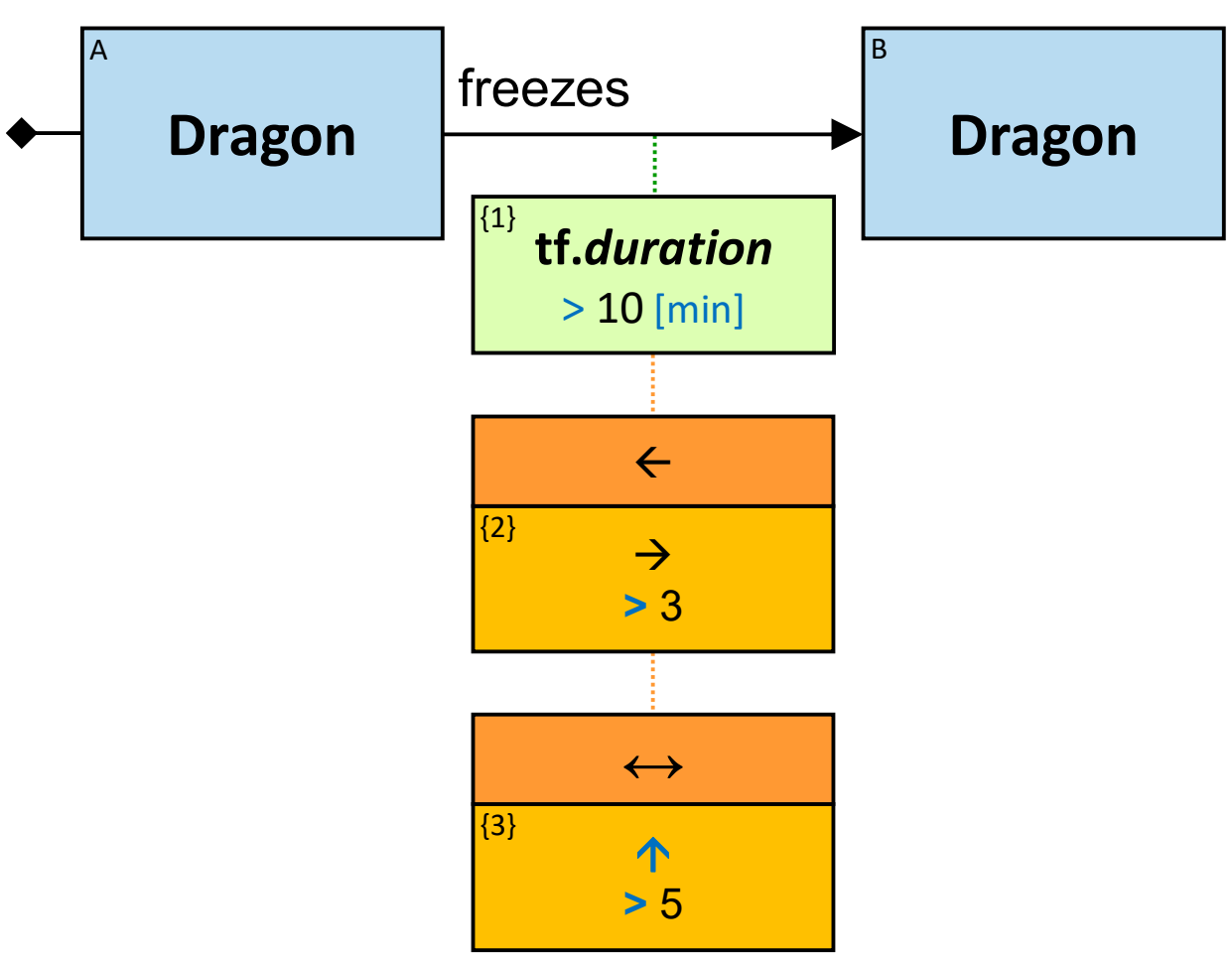

(Similar to Q94, but the order of the aggregators is switched.) Filtering stages:

- Pass freezes longer than ten minutes
- Pass dragons A that froze more than three B's – each at least once for more than ten minutes
- Pass pairs of dragons (A, B) where A froze B more than five times for more than ten minutes

A dragon A that froze more than three dragons, each at least once for more than ten minutes, but froze only a single dragon more than five times for more than ten minutes – is a valid assignment in Q93 but not in Q94.

***Q272:*** *Any pair of dragons (A, B) where the longest freeze duration is at least ten times longer than the shortest freeze duration*

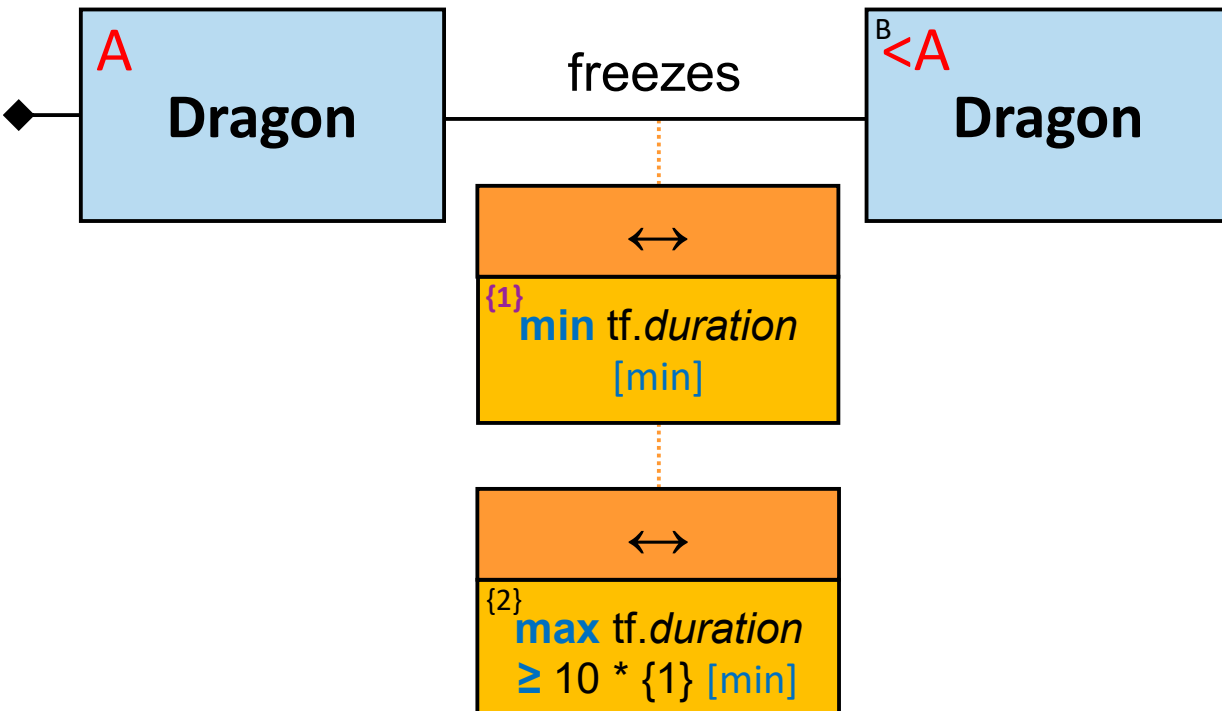

*Q363: Any dragon that the last time it froze some dragon for the first time was in or after 1010*

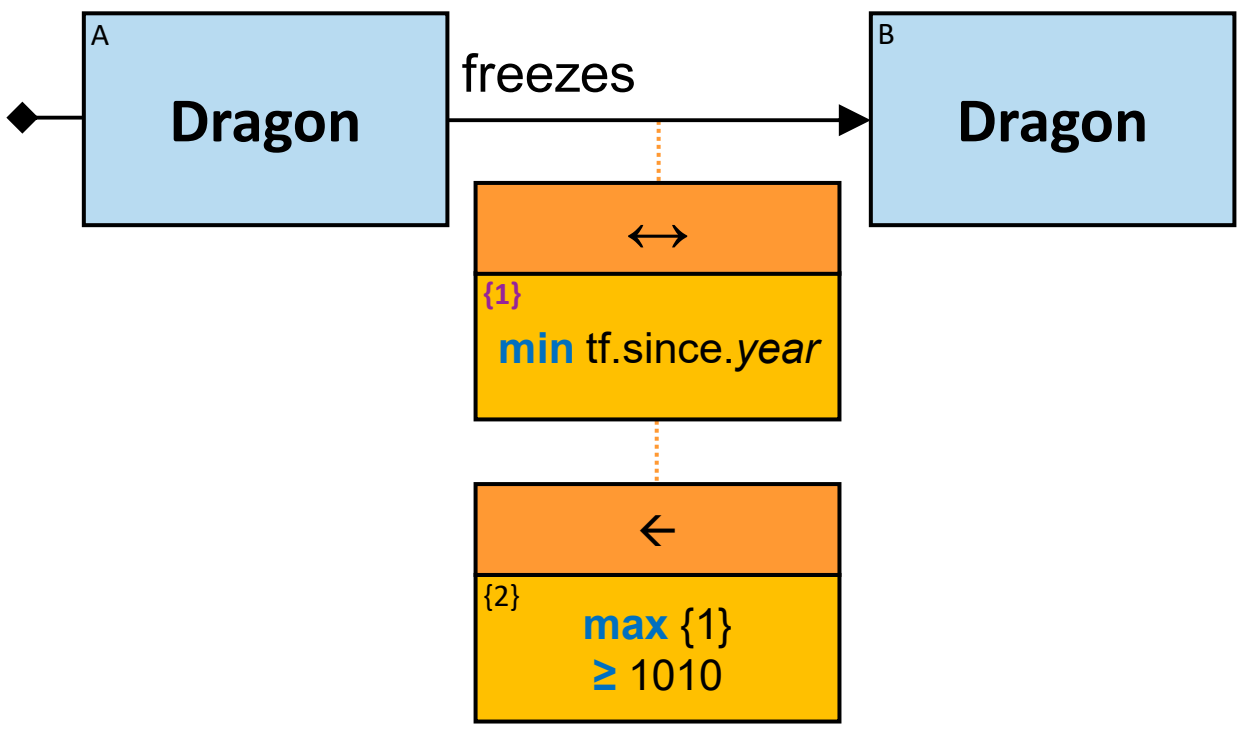

*Q91: The four dragons with the maximal (shortest duration they were frozen for)*

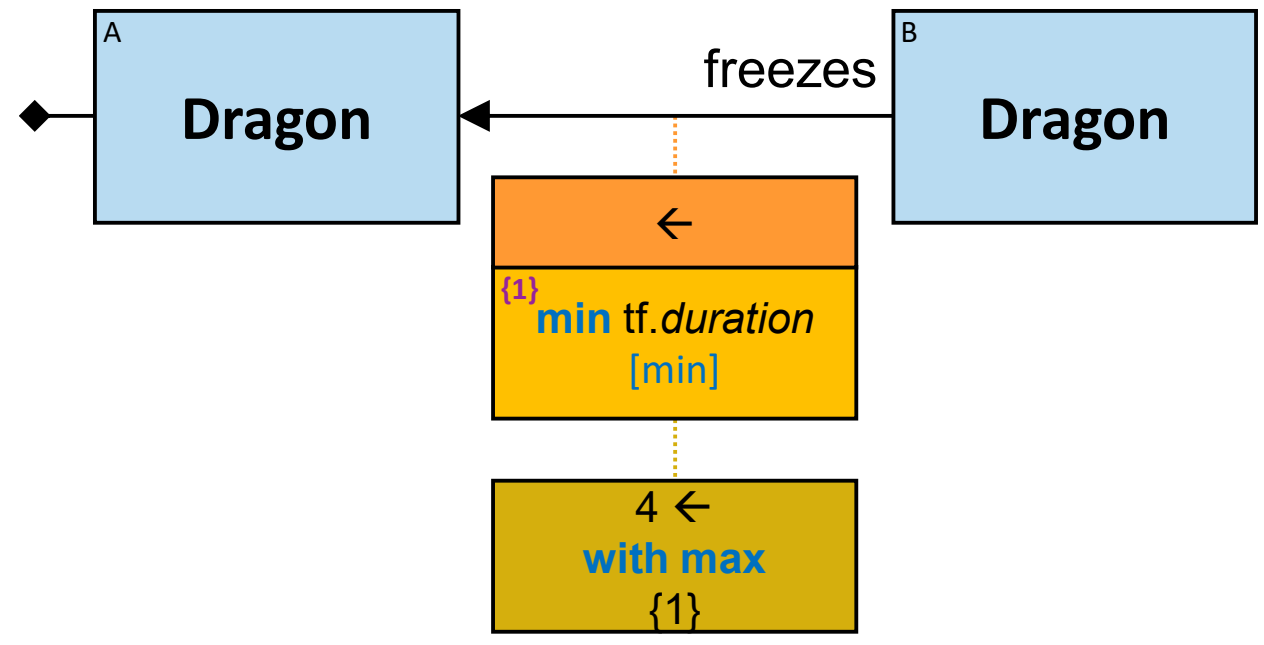

*Q90: The four pairs of dragons (A, B) where A froze B for the maximal cumulative duration*

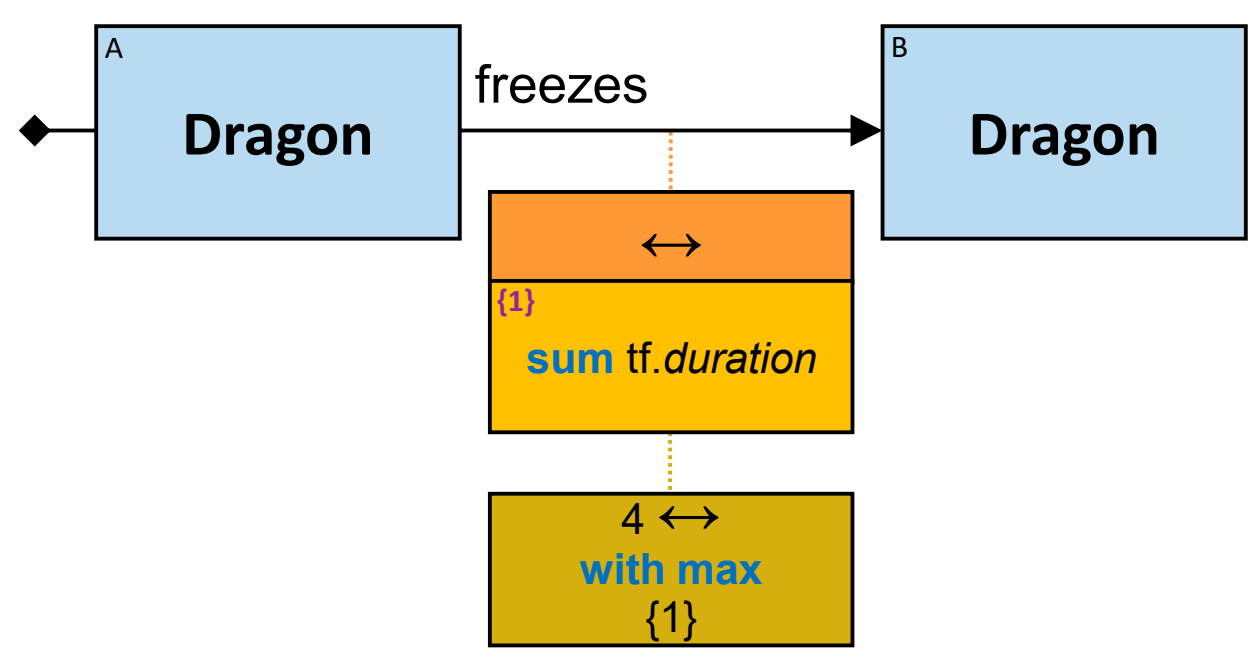

*Q92: The four dragons with the maximal (average duration they froze dragons for)*

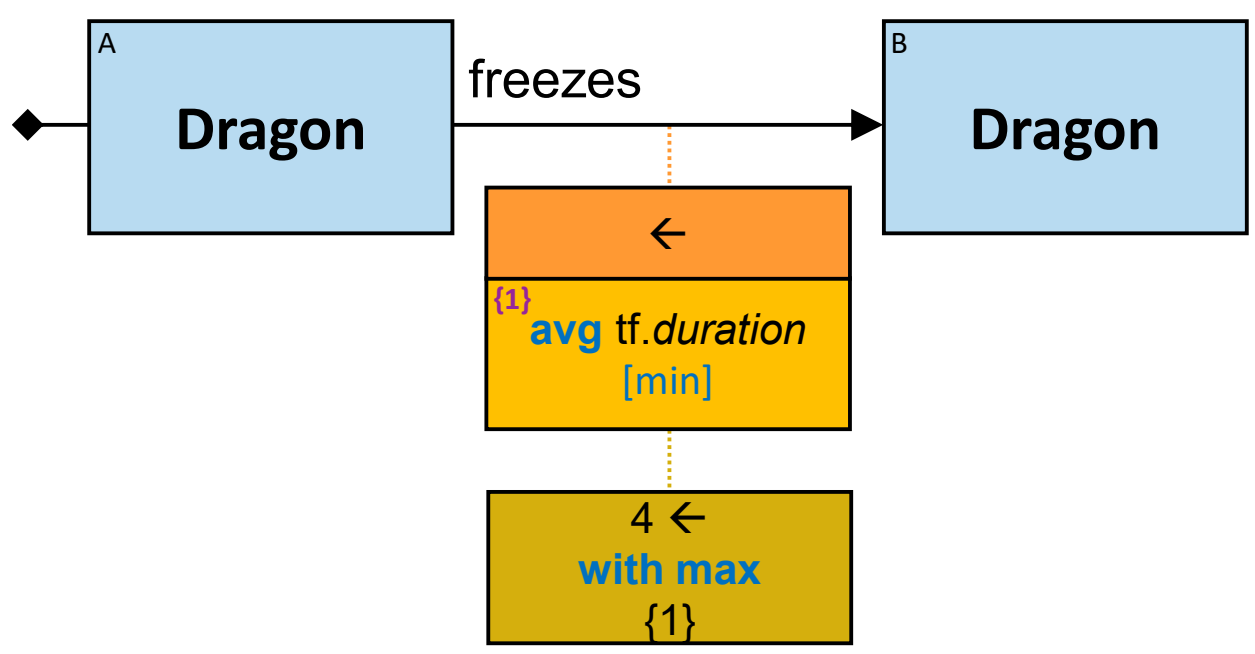

***Q328:*** *The three people with the maximal cumulative horse ownership days*

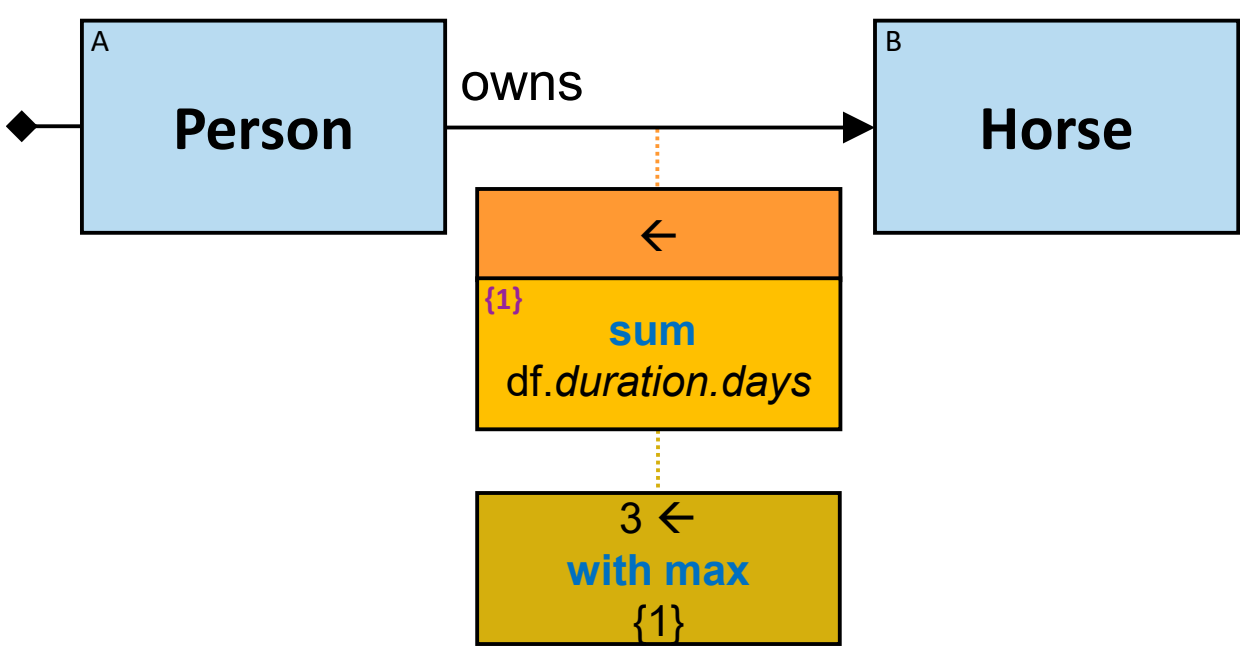

***Q201:*** *For each dragon that froze at least ten dragons: the three dragons it froze for the maximal cumulative duration*

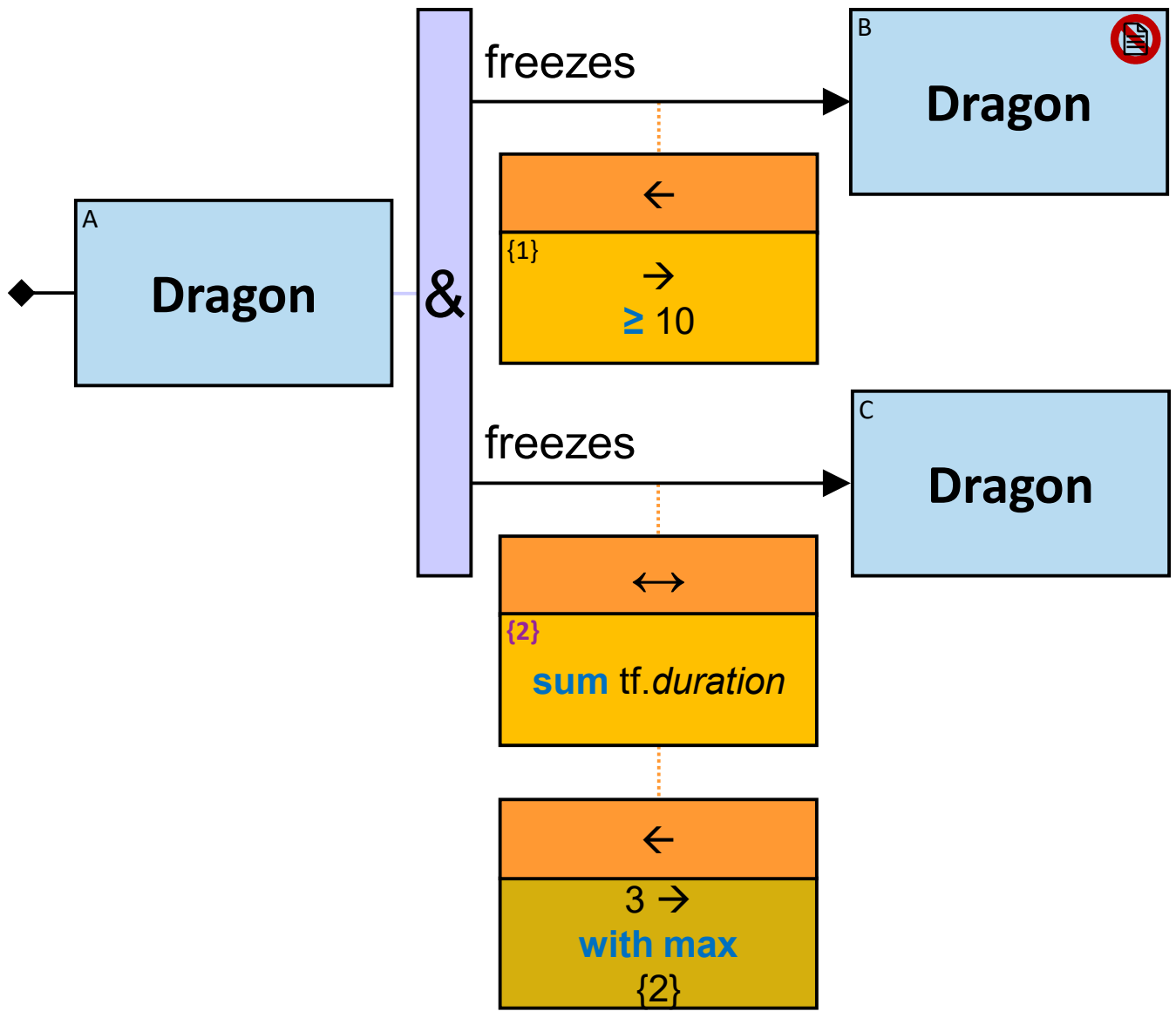

***Q274:*** *The four dragons with the largest difference between (the longest duration they froze dragon for) and (the shortest duration they froze dragon for)*

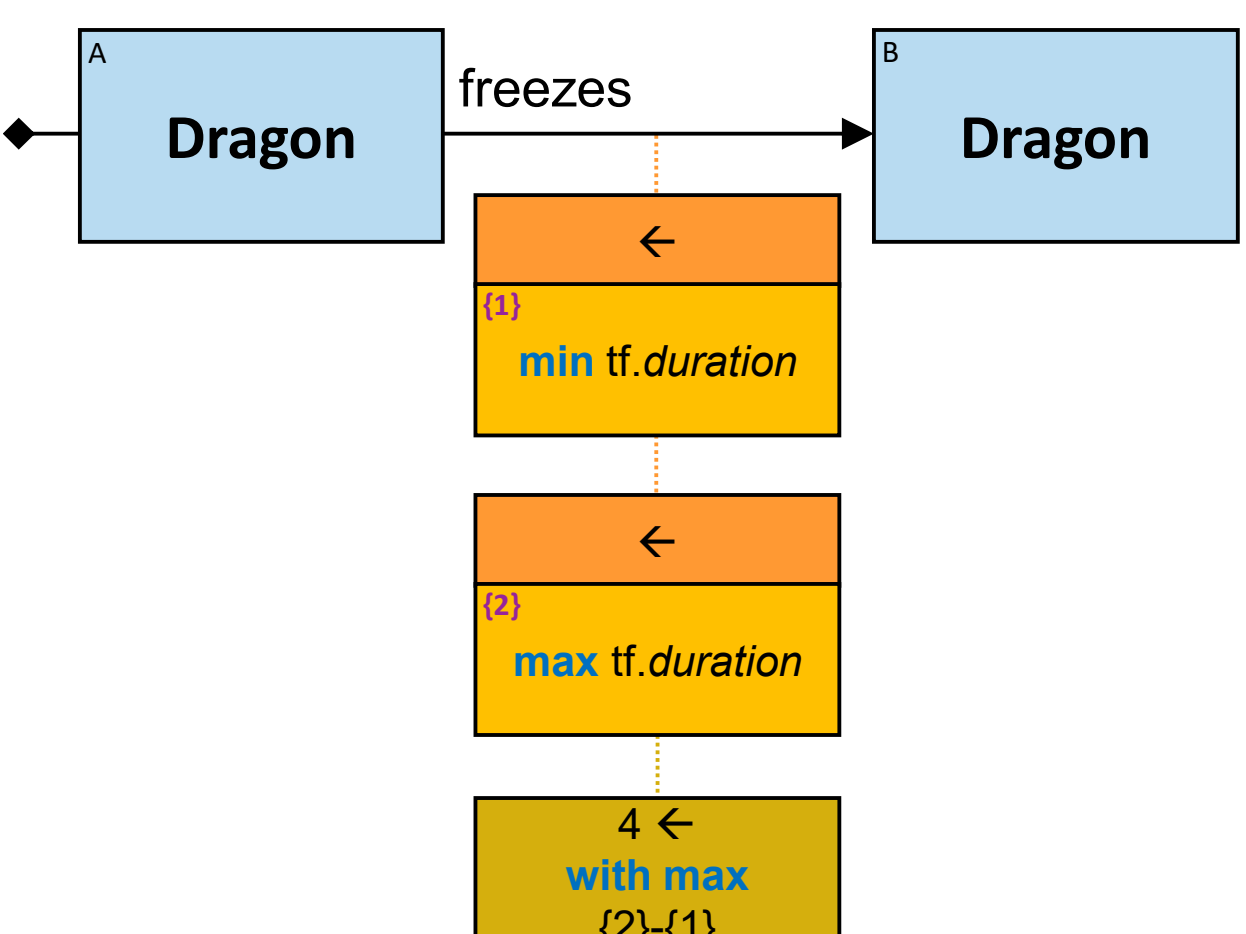

*Q182: Any dragon owned by Brandon Stark and the three dragons it froze for the maximal duration*

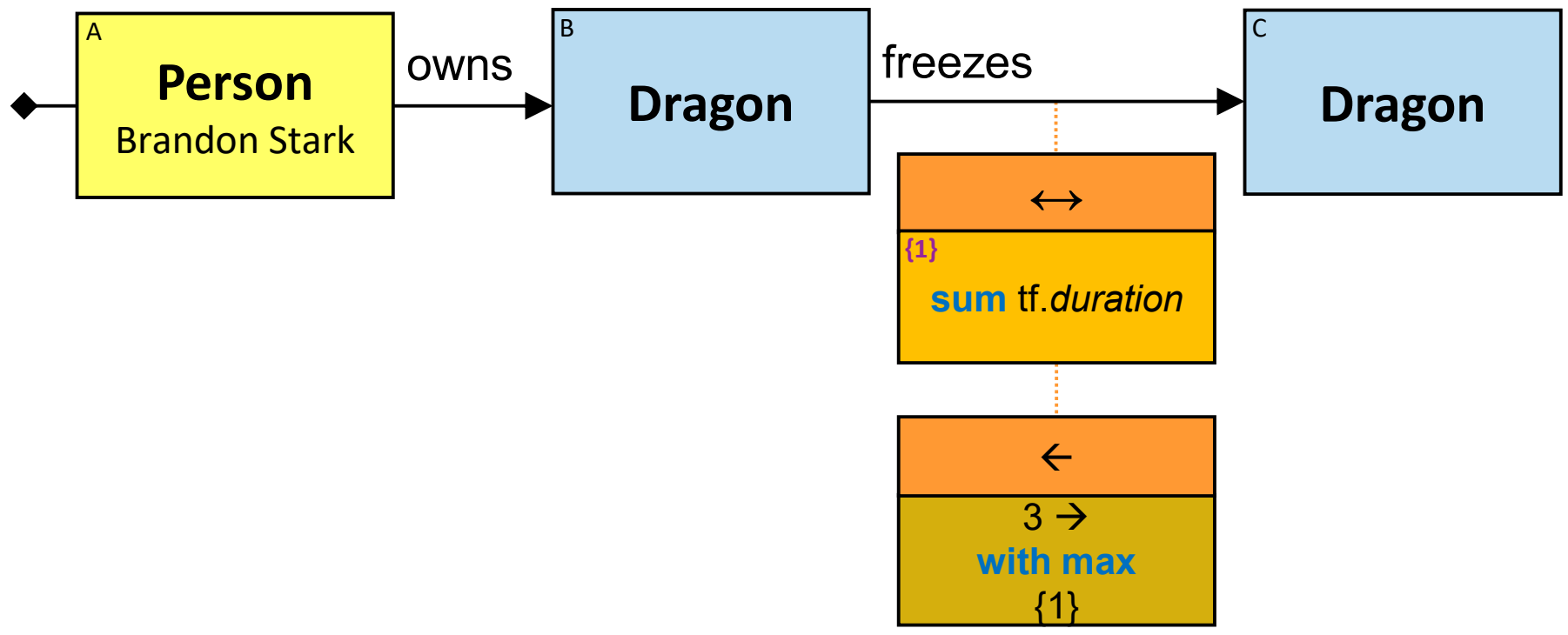

*Q133: The four people that the average weight of their horses is maximal*

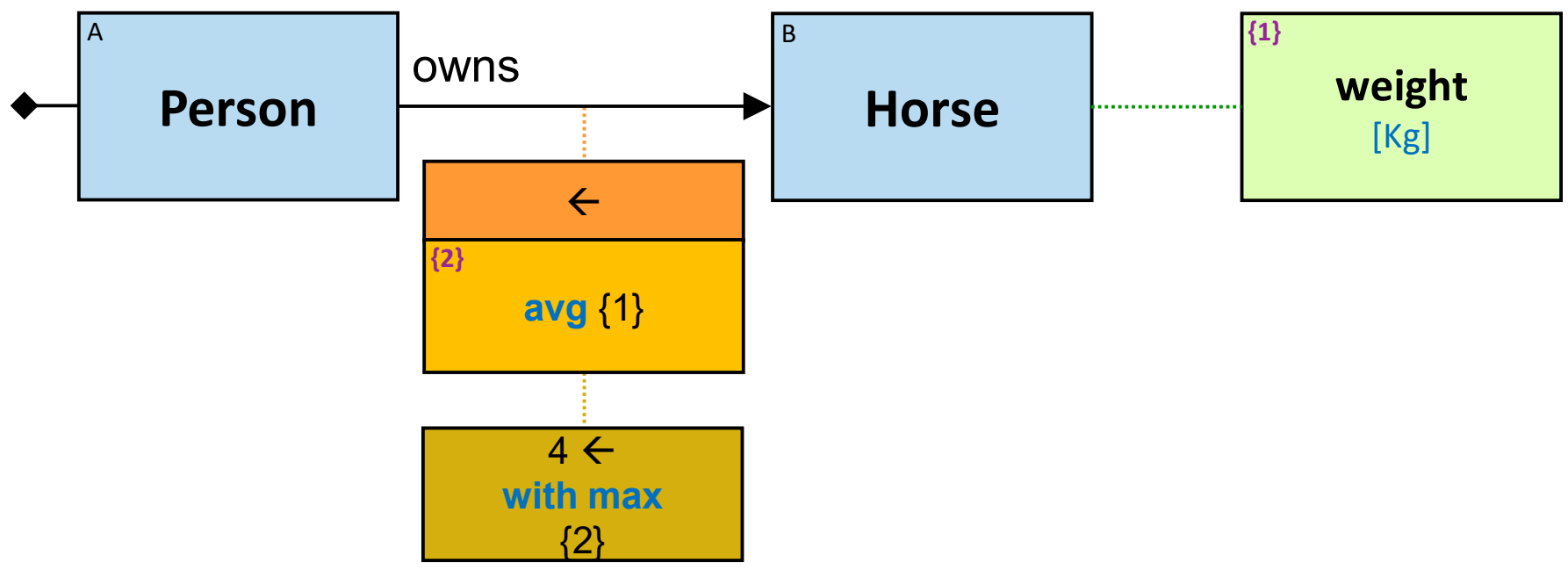

*Q132: The four people who own horses of the largest number of colors*

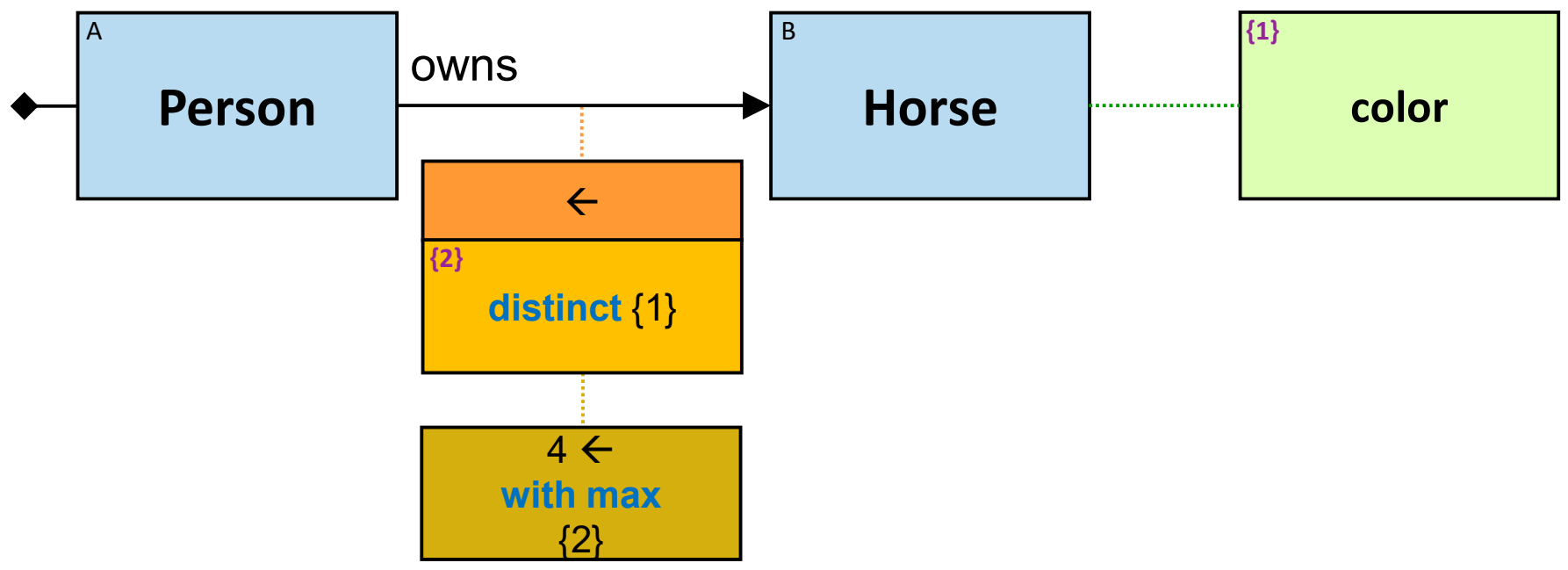

*Q233: Any dragon A than froze dragons Bs that froze dragons Cs, and the three Cs for which A froze Bs for the maximal cumulative duration*

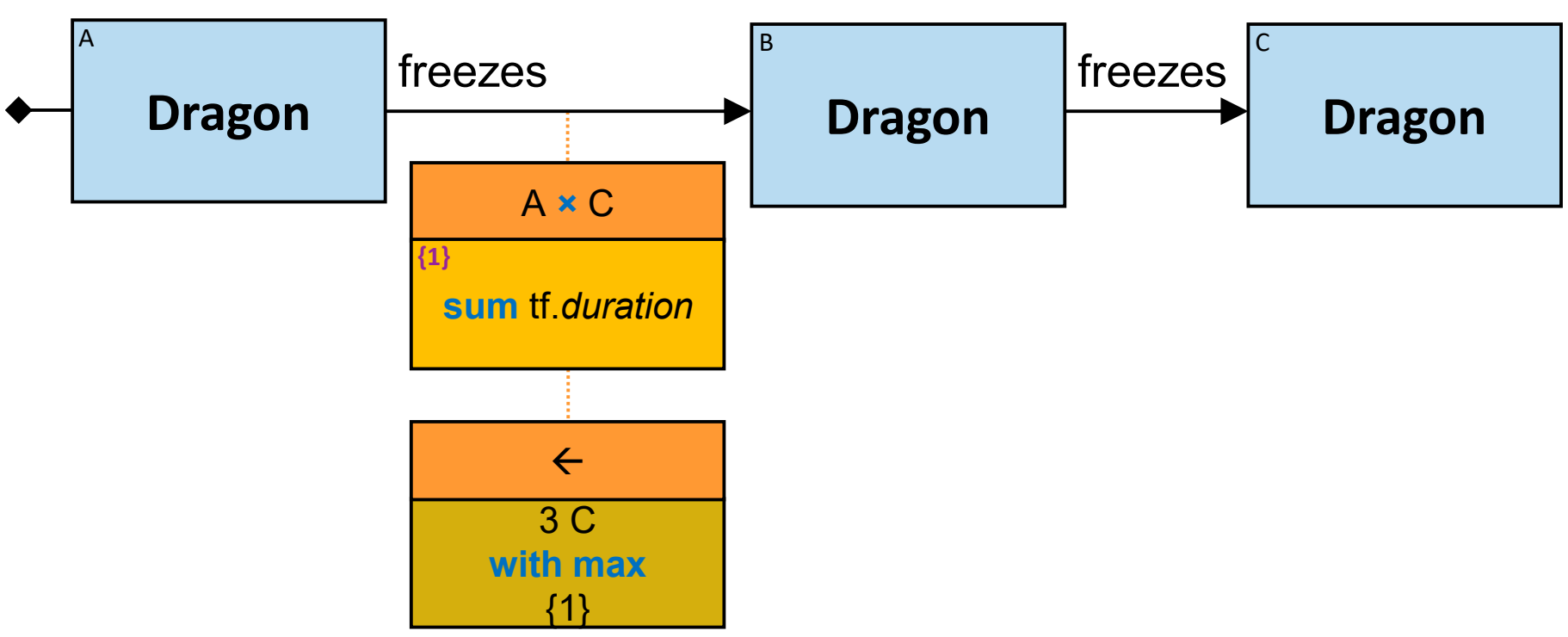

*Q138: The four people who the people each of them is a friend of – cumulatively own horses of the largest number of colors*

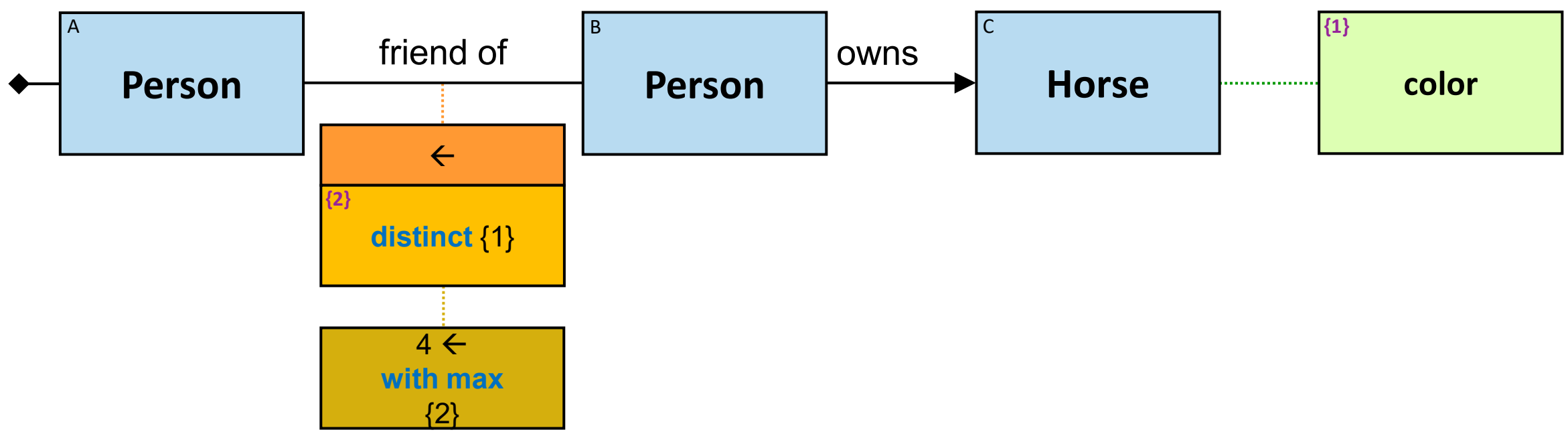

*Q183: The three dragons that dragons owned by Brandon Stark froze for the maximal cumulative duration*

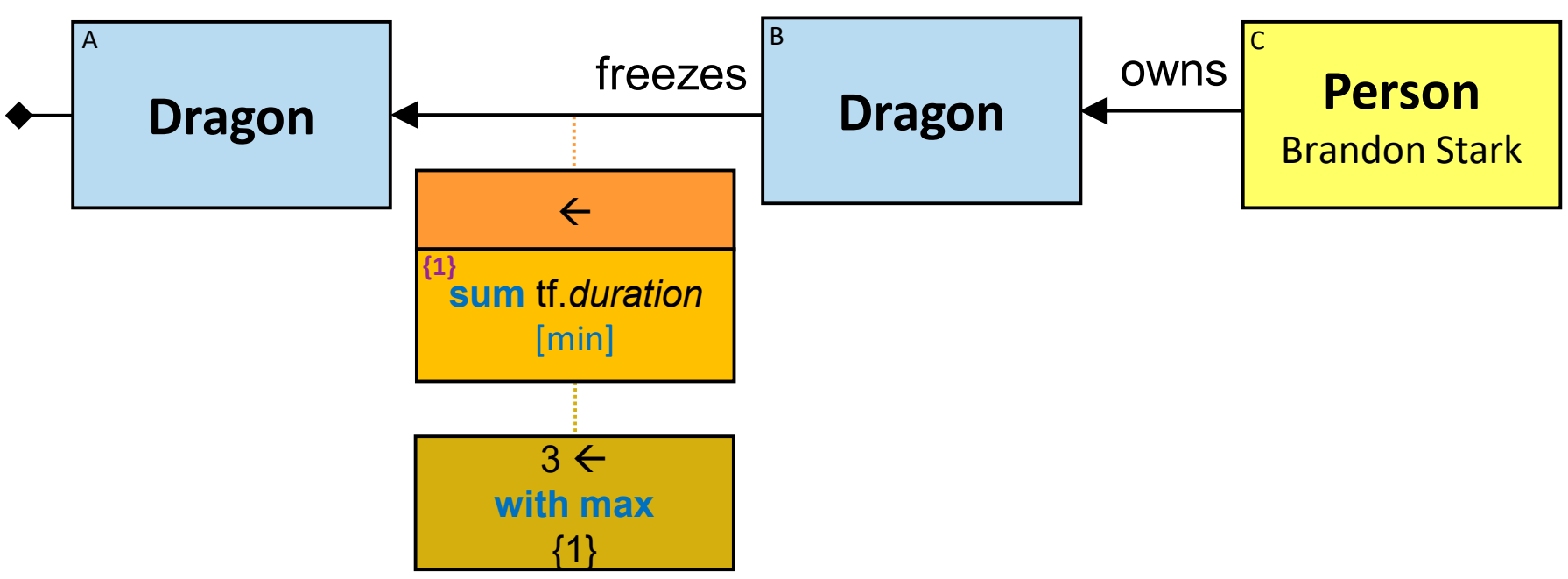

*Q168: The three people who the number of types of entities each of them owns is the largest*

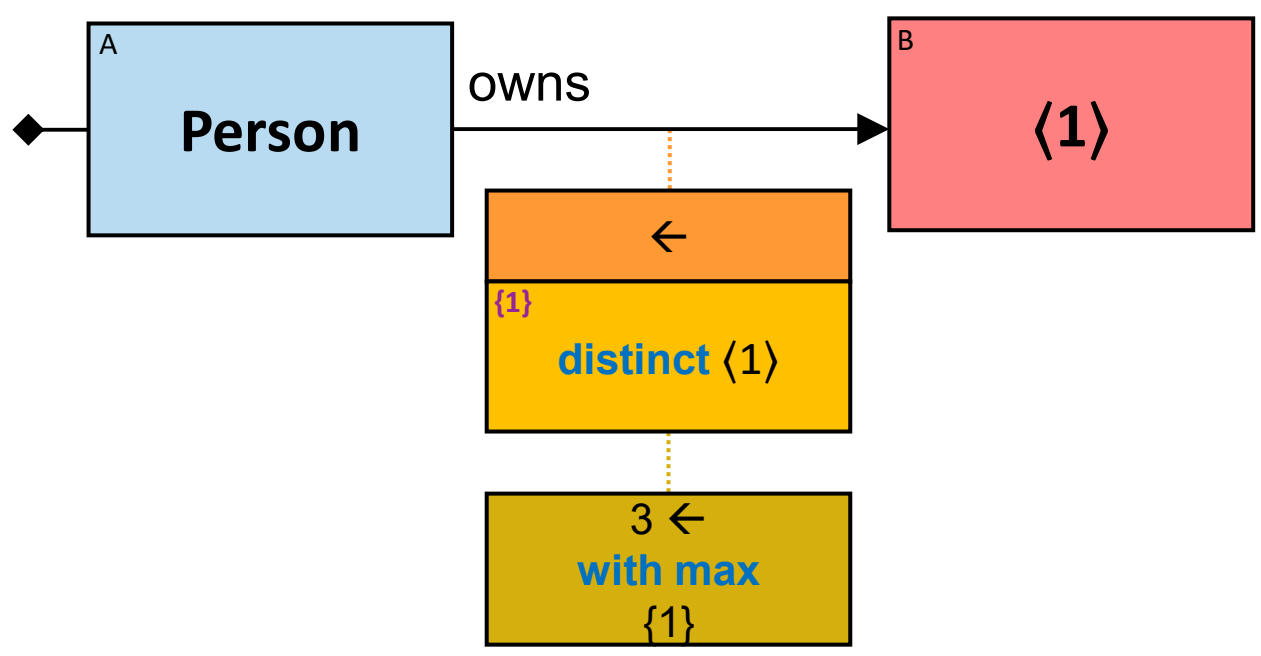

*Q163: Any dragon that the average duration of the ten shortest times it froze dragons is longer than 60 minutes*

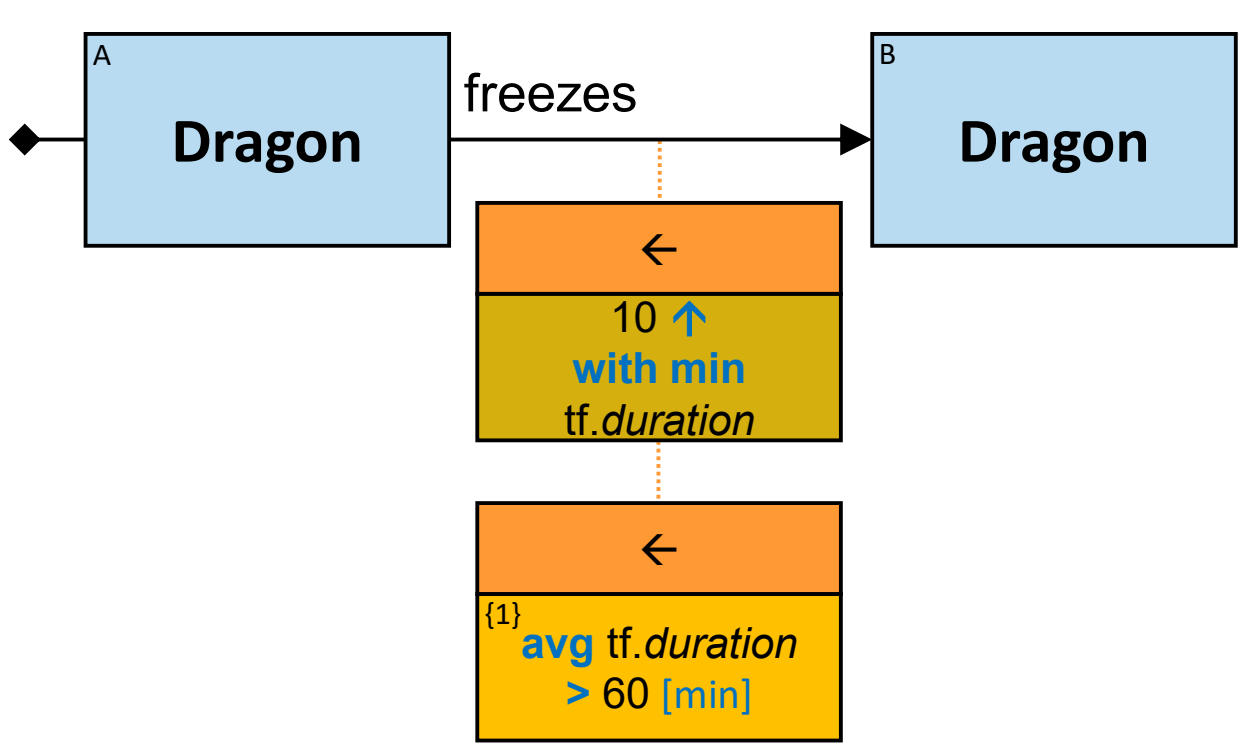

***Q162:*** *Any pair of dragons (A, B) where the second shortest duration A froze B is longer than 60 minutes*

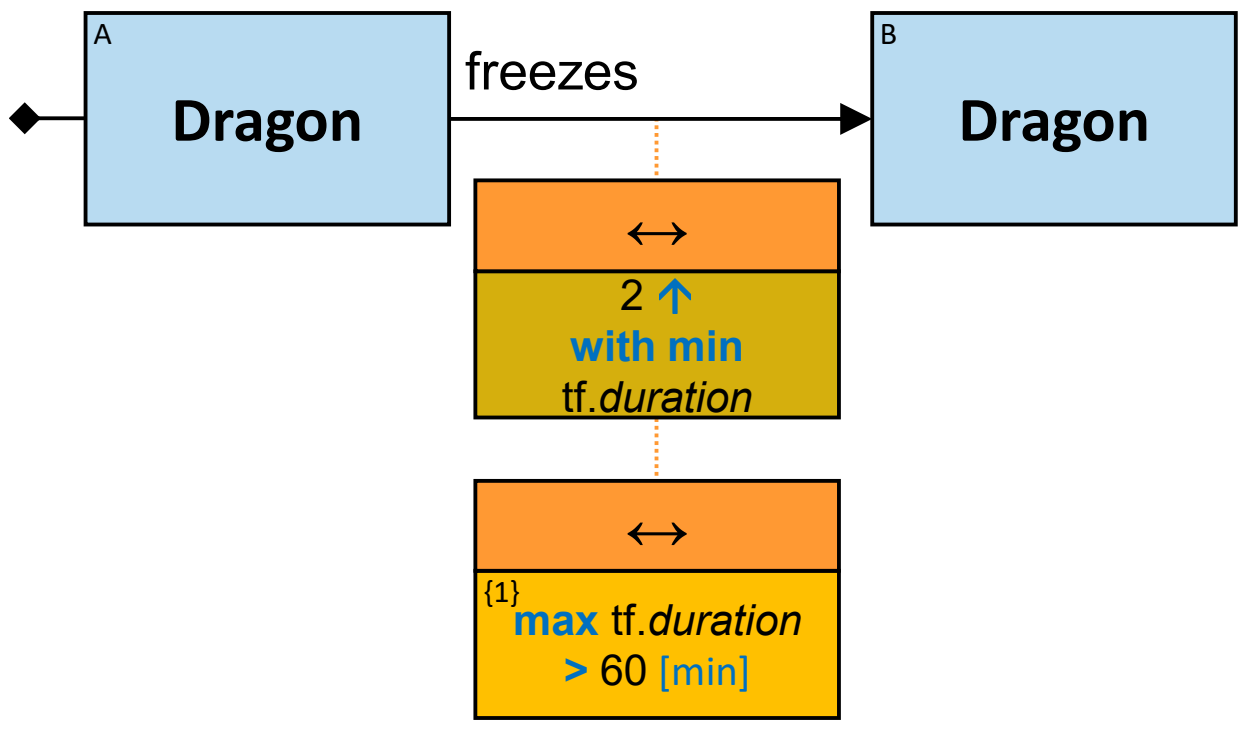

***Q341:*** *For each pair of dragons (A, B): the two shortest freezes A froze B. If there are more than two freezes with equal minimal duration – the two earliest amongst them*

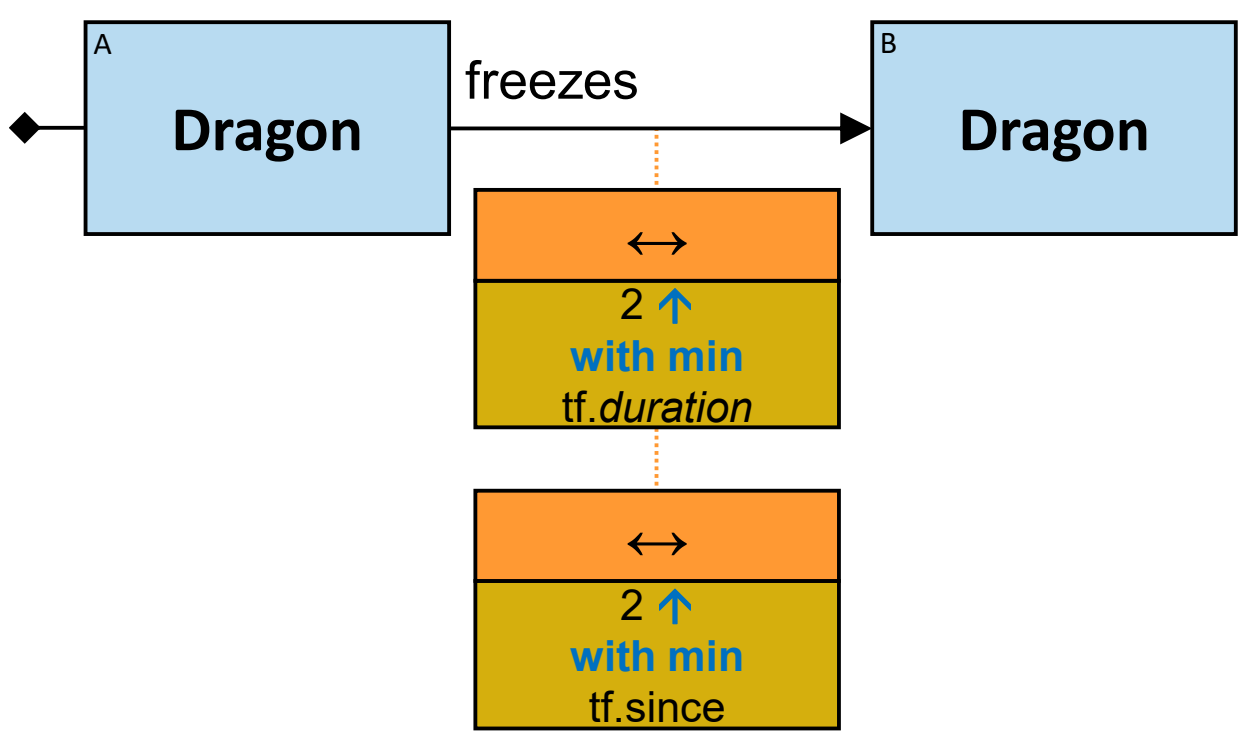

***Q268:*** *Any pair of dragons (A, B) where the average duration of the 11th-20th most prolonged freezes A froze B is longer than 60 minutes* (two versions)

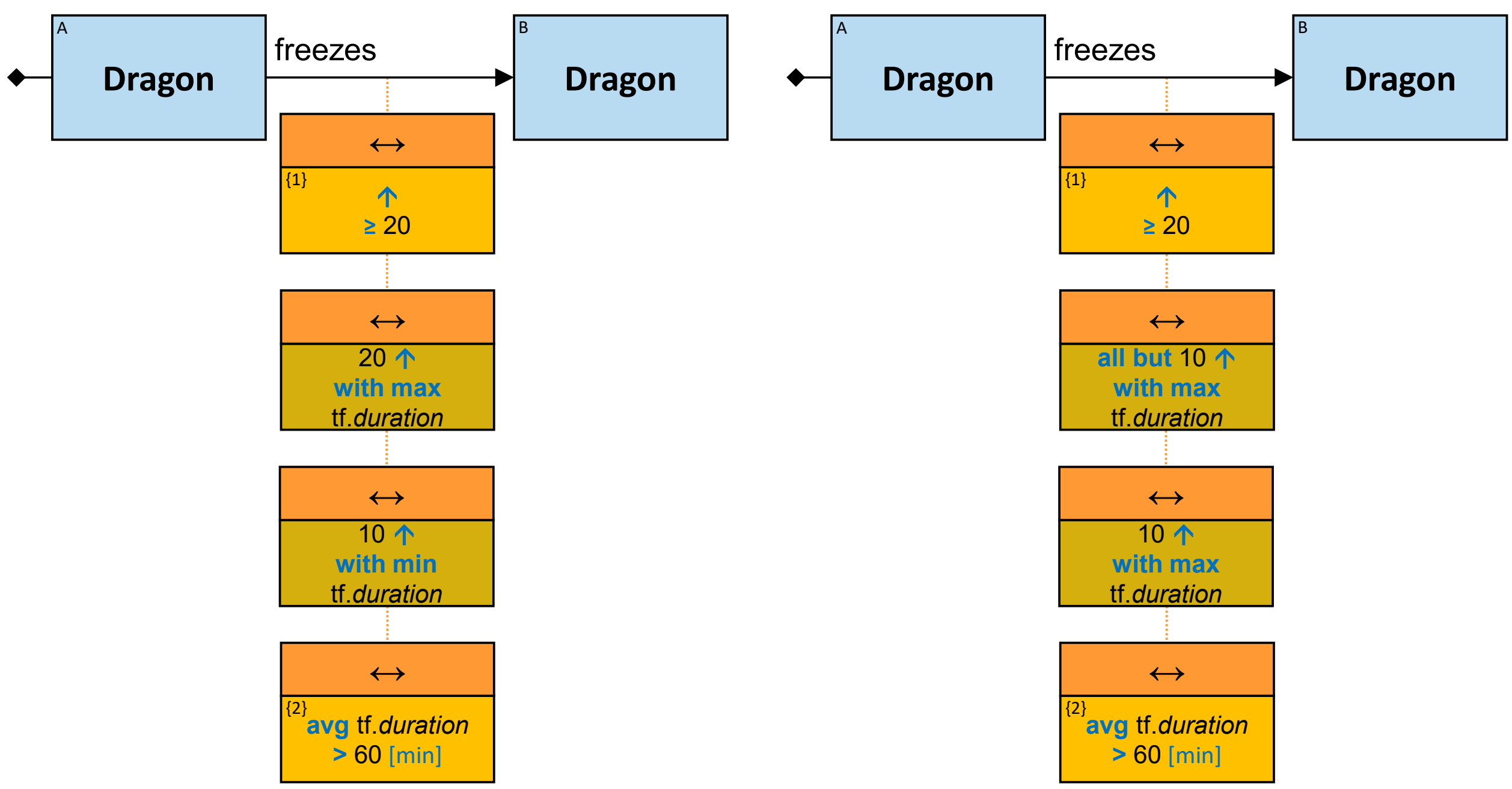

***Q120:** Any person A whose three oldest offspring's cumulative height is lower than A's height*

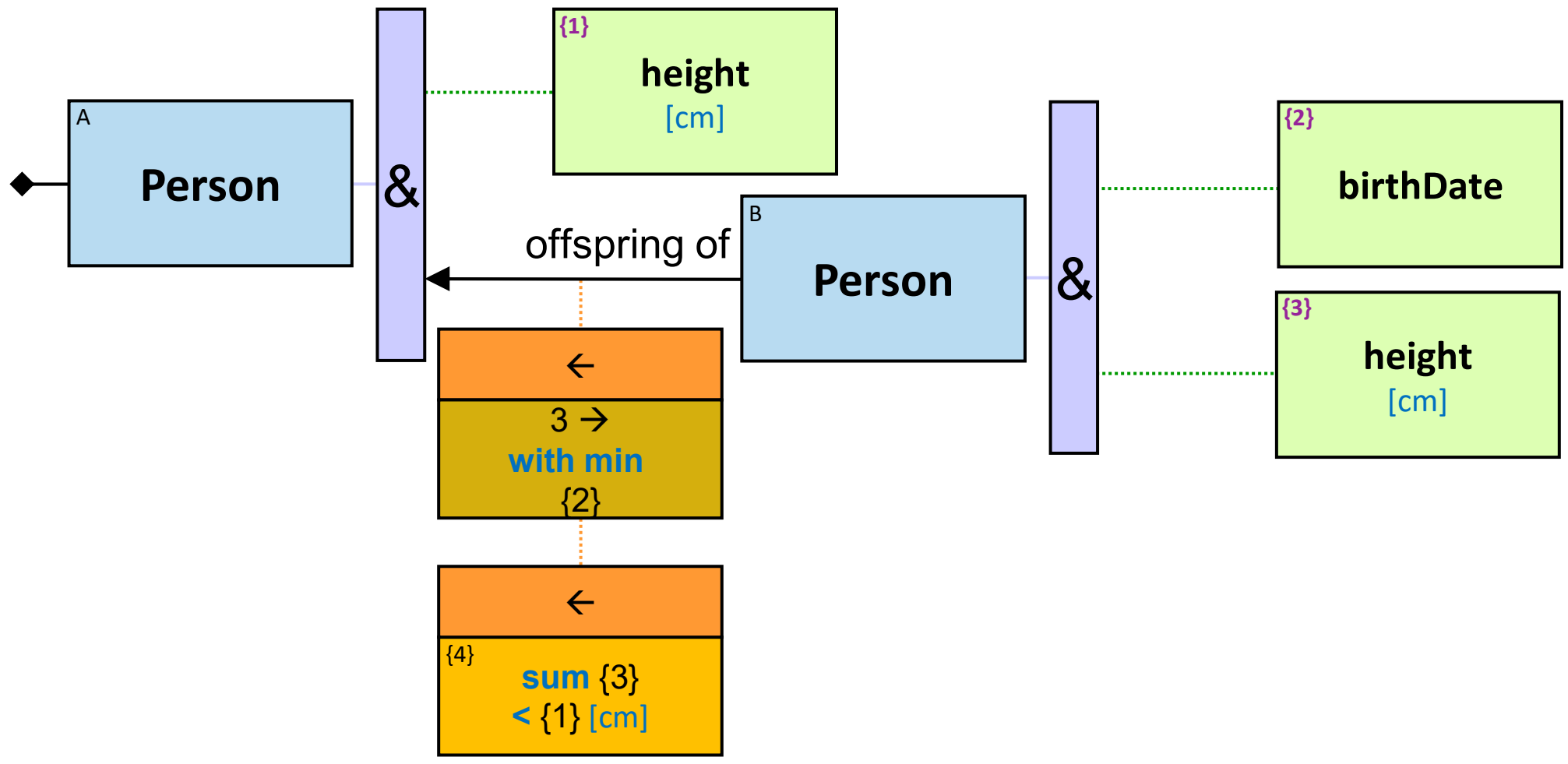

***Q202:** Any person A whose three oldest offspring's average height is lower than the average height of all A's offspring*

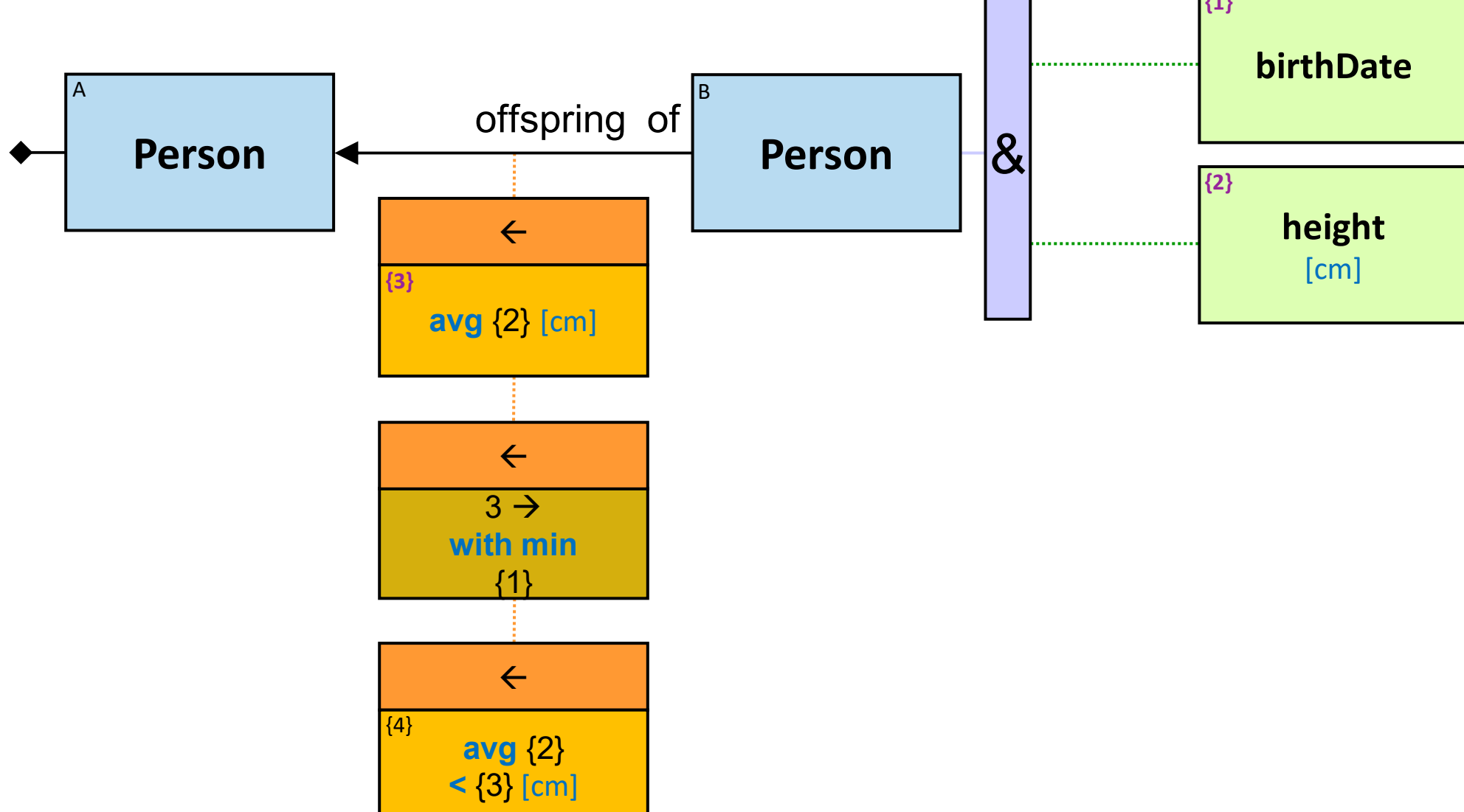

***Q208:*** *Any person whose three oldest horse-owning sons cumulatively own horses of three colors*

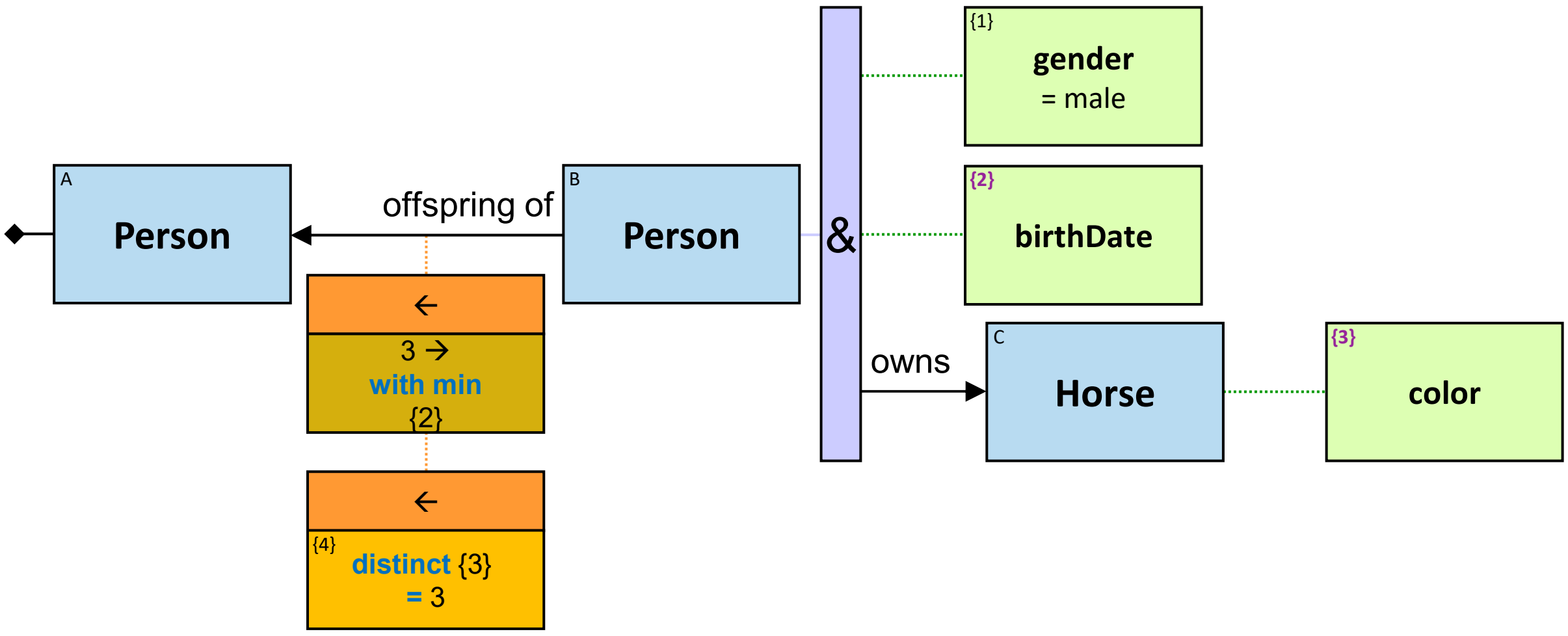

***Q140:*** *Any person whose three oldest sons cumulatively own horses of three colors*

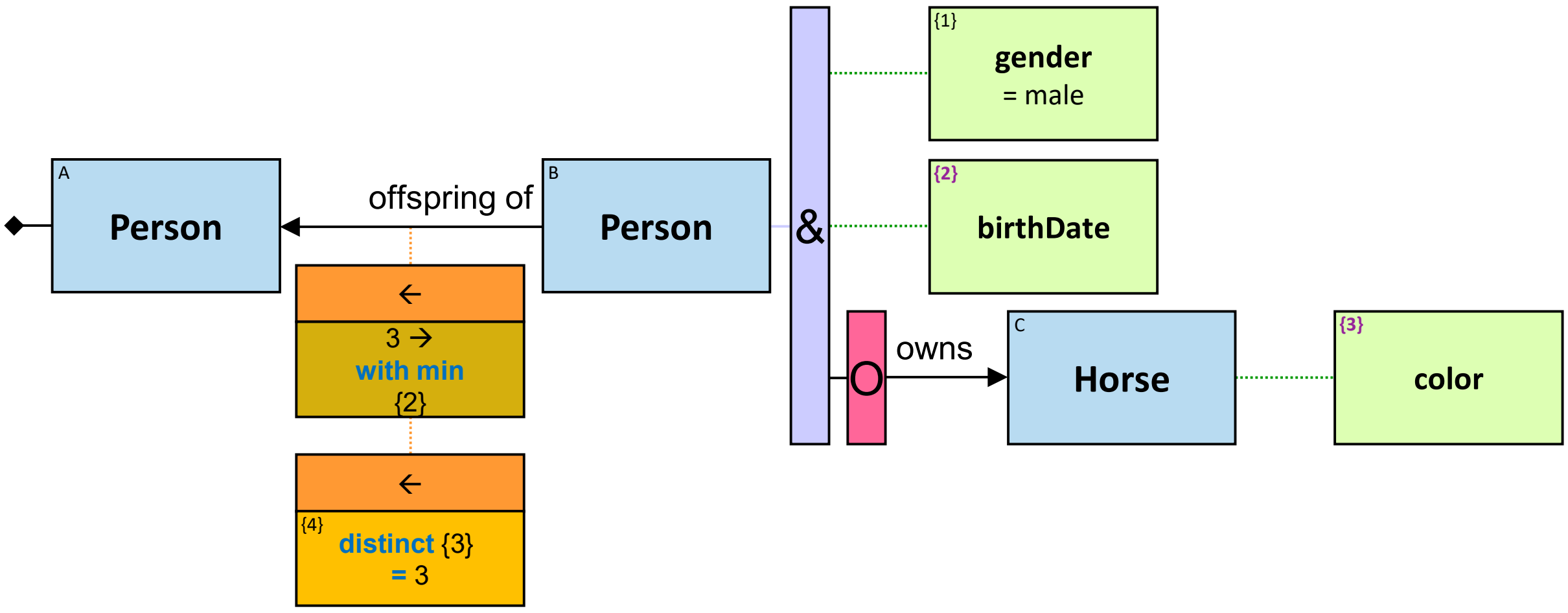

Note that {3} is defined right of an 'O' and is evaluated to *null* when the optional part has no valid assignment. *distinct* does not aggregate *null* evaluation results.

***Q141:** Any person A whose three oldest sons cumulatively own horses of the same number of colors as of those cumulatively owned by A's three youngest daughters*

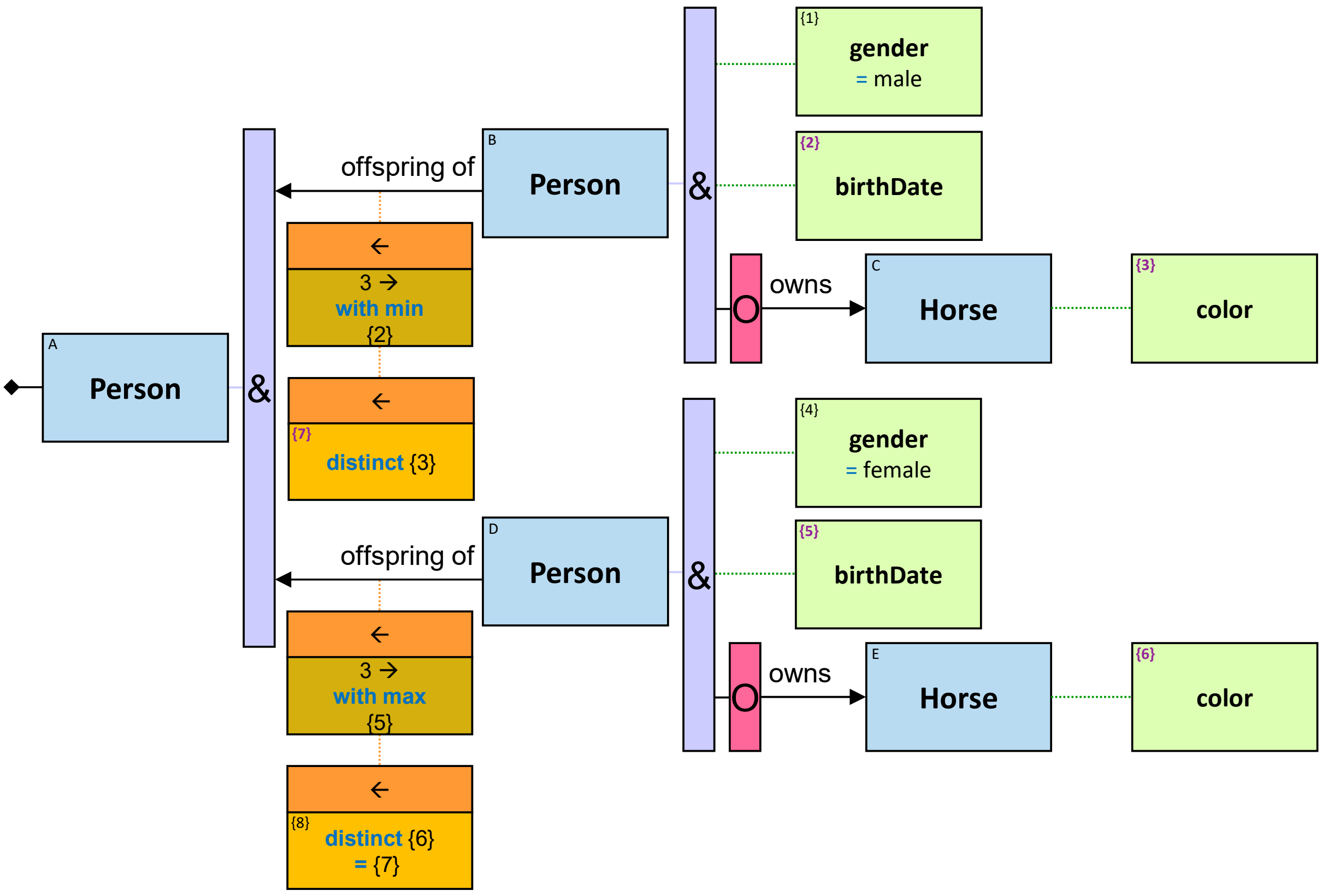

***Q95:** Any dragon Balerion froze more than five times, each for more than ten minutes, and the total duration of those freezes was longer than 100 minutes* (three versions)

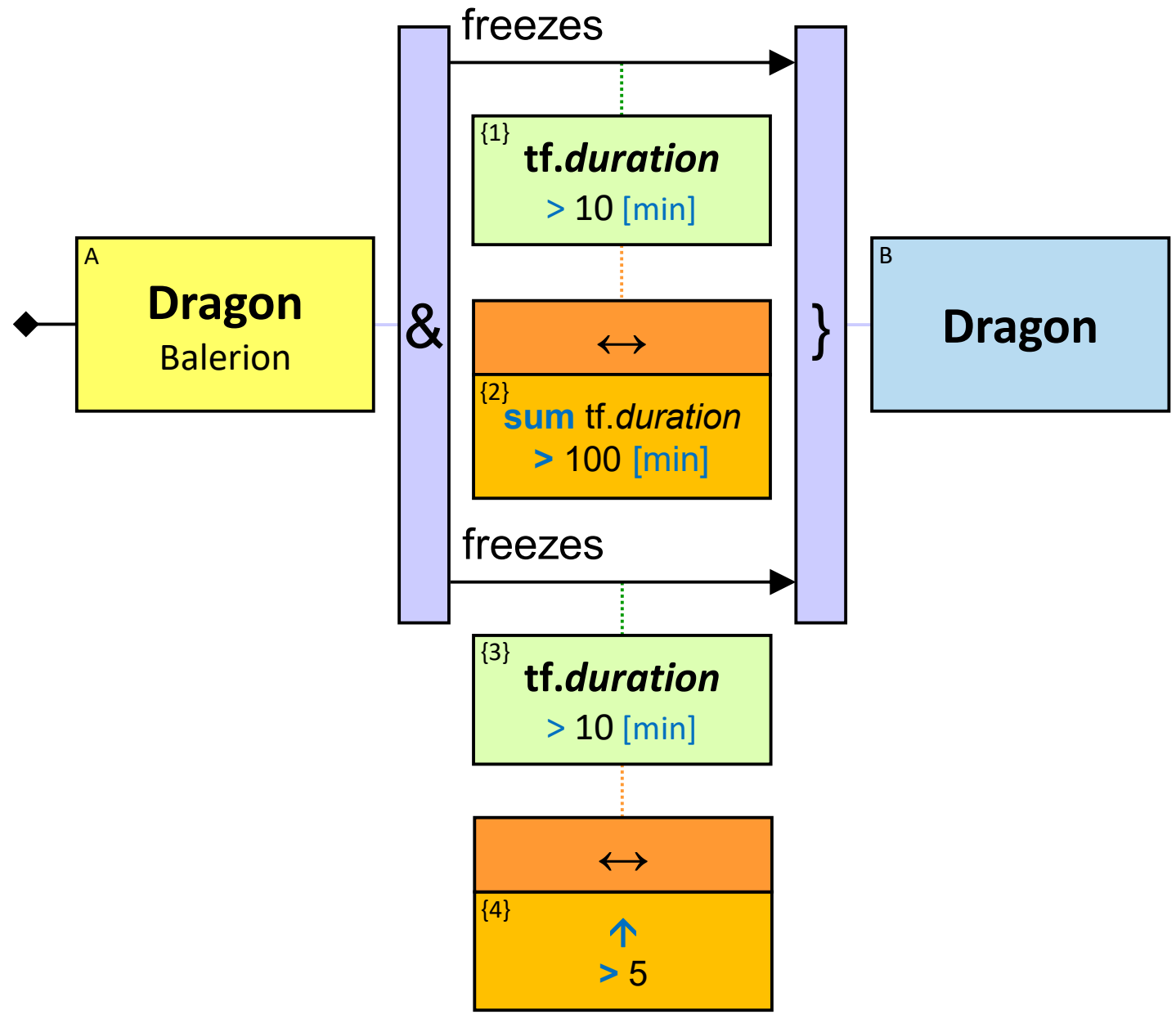

The two *per pair* constraints could be chained instead. The meaning would be similar:

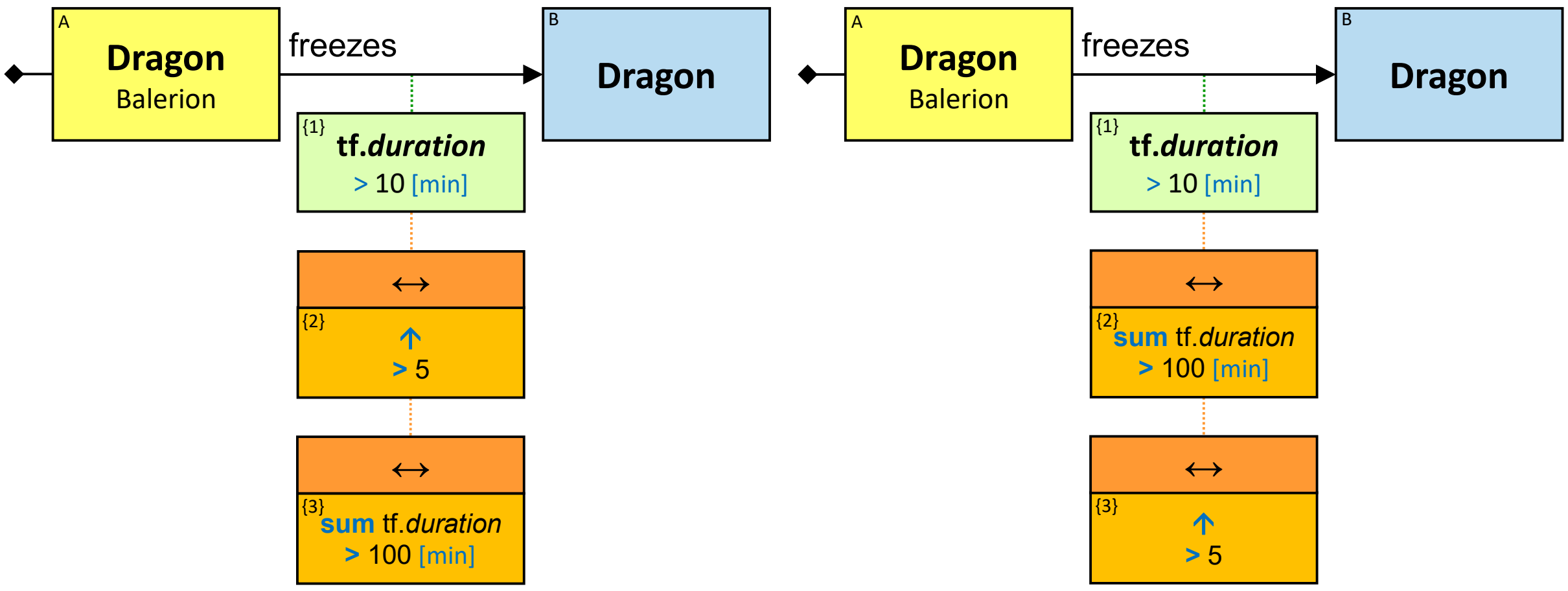

***Q199:** Any dragon that (froze more than ten times each of more than ten dragons) and (froze more than 20 times each of less than ten dragons)*

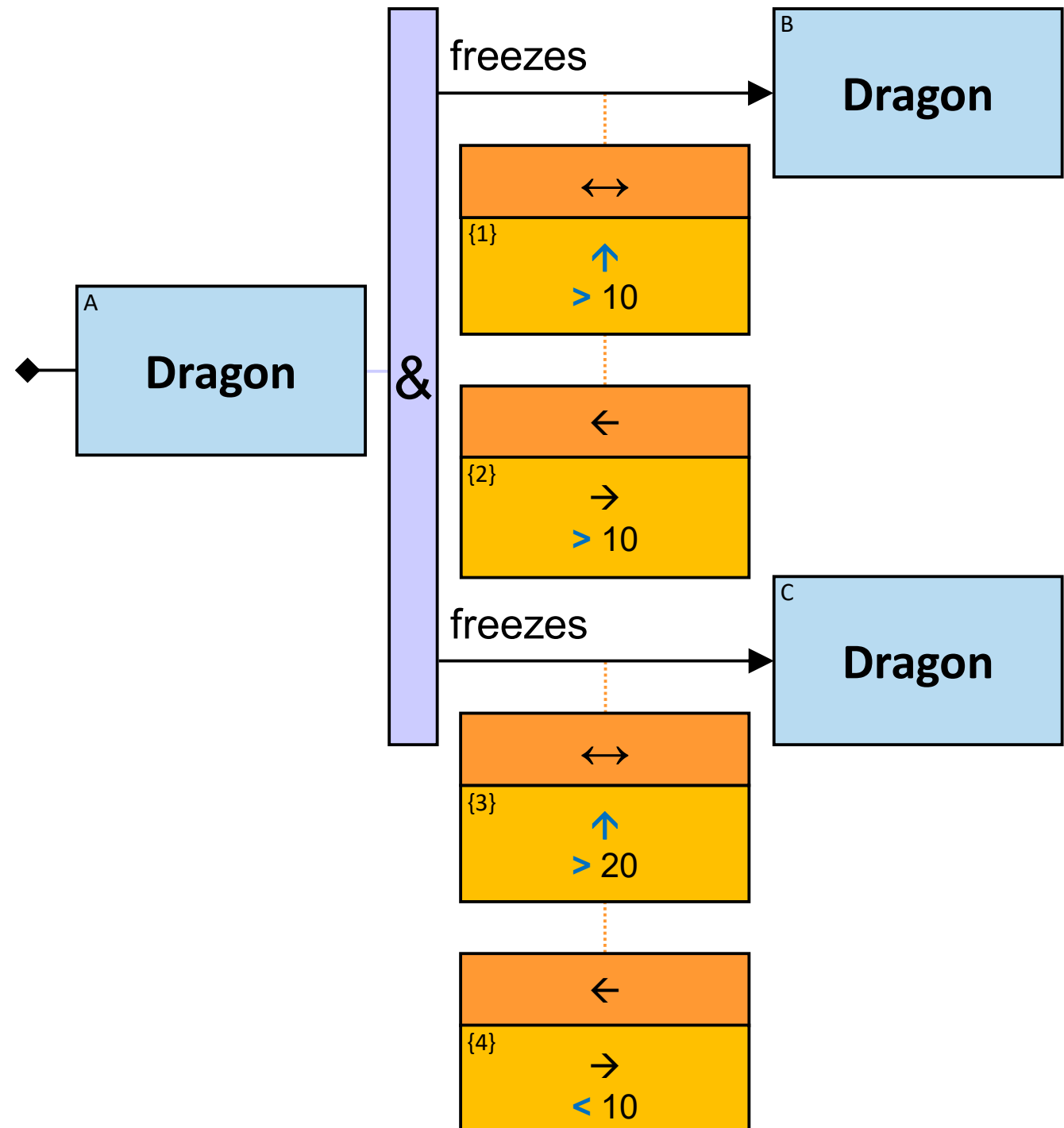

***Q200:*** *Any dragon that (froze more than ten times each of more than ten dragons. For each of these ten dragons – the freezes had exactly two distinct durations) and (froze more than 20 times each of less than ten dragons. The average freeze duration of all these dragons – is longer than three minutes)*

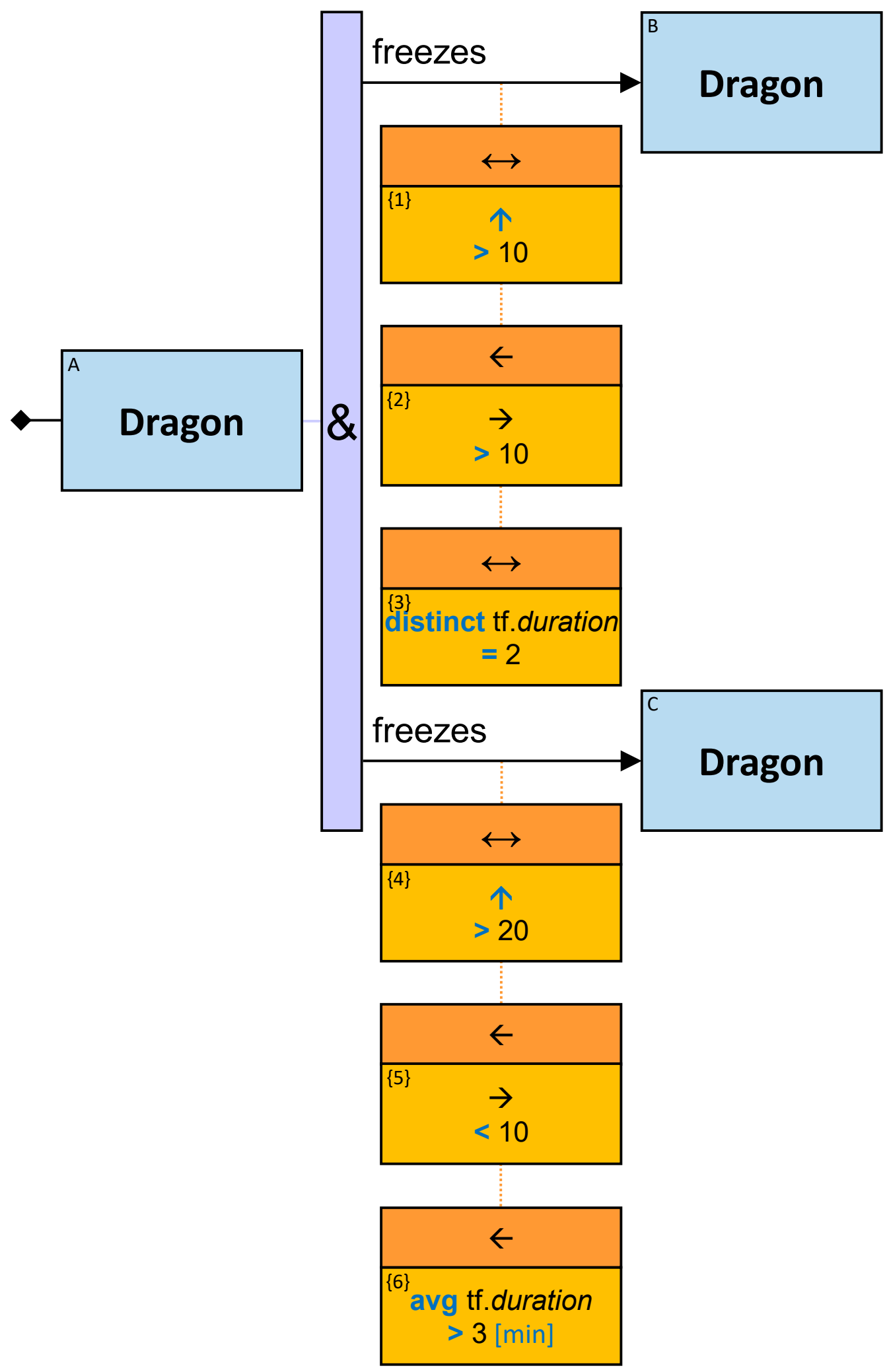

***Q277:*** *Any dragon that the longest (cumulative duration it froze some dragon) is more than ten times longer than the shortest (cumulative duration it froze some dragon)*

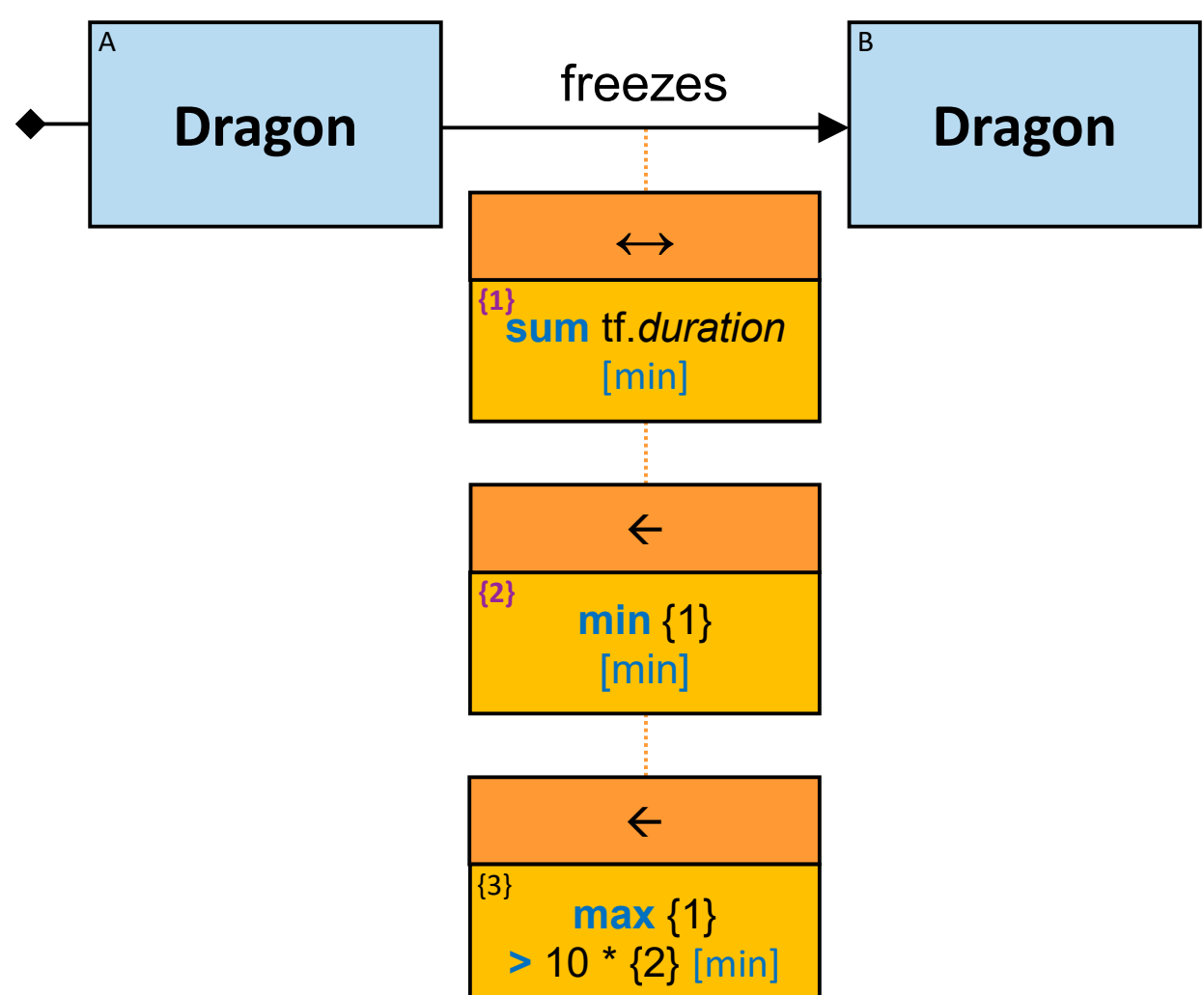

***Q351:** Any person who owns two horses of different colors, a dragon that froze at least ten dragons of the same color as the first horse, and a dragon that froze at least ten dragons of the same color as the second horse*

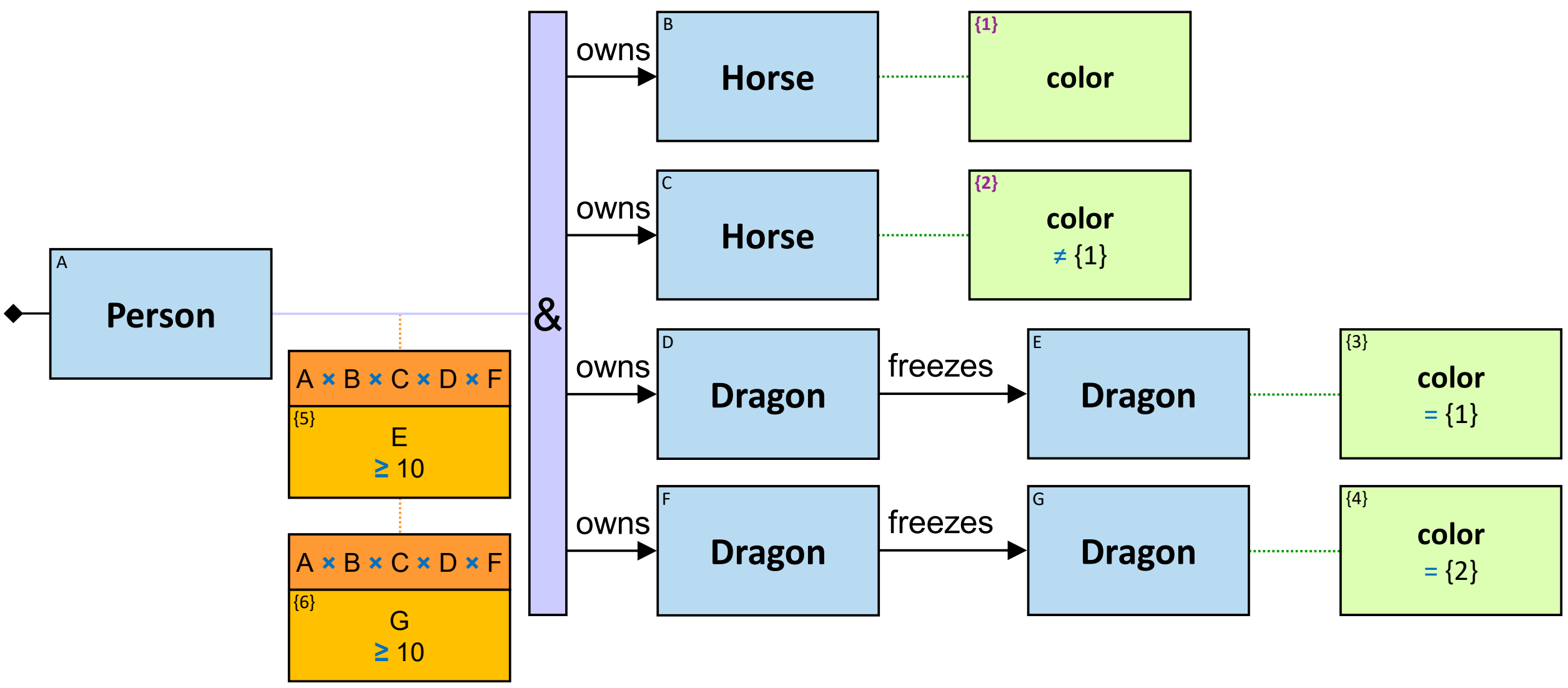

***Q320:** Any person A where there are more horses of the same colors as A's owned horses than dragons of the same colors as A's owned dragons* (version 1)

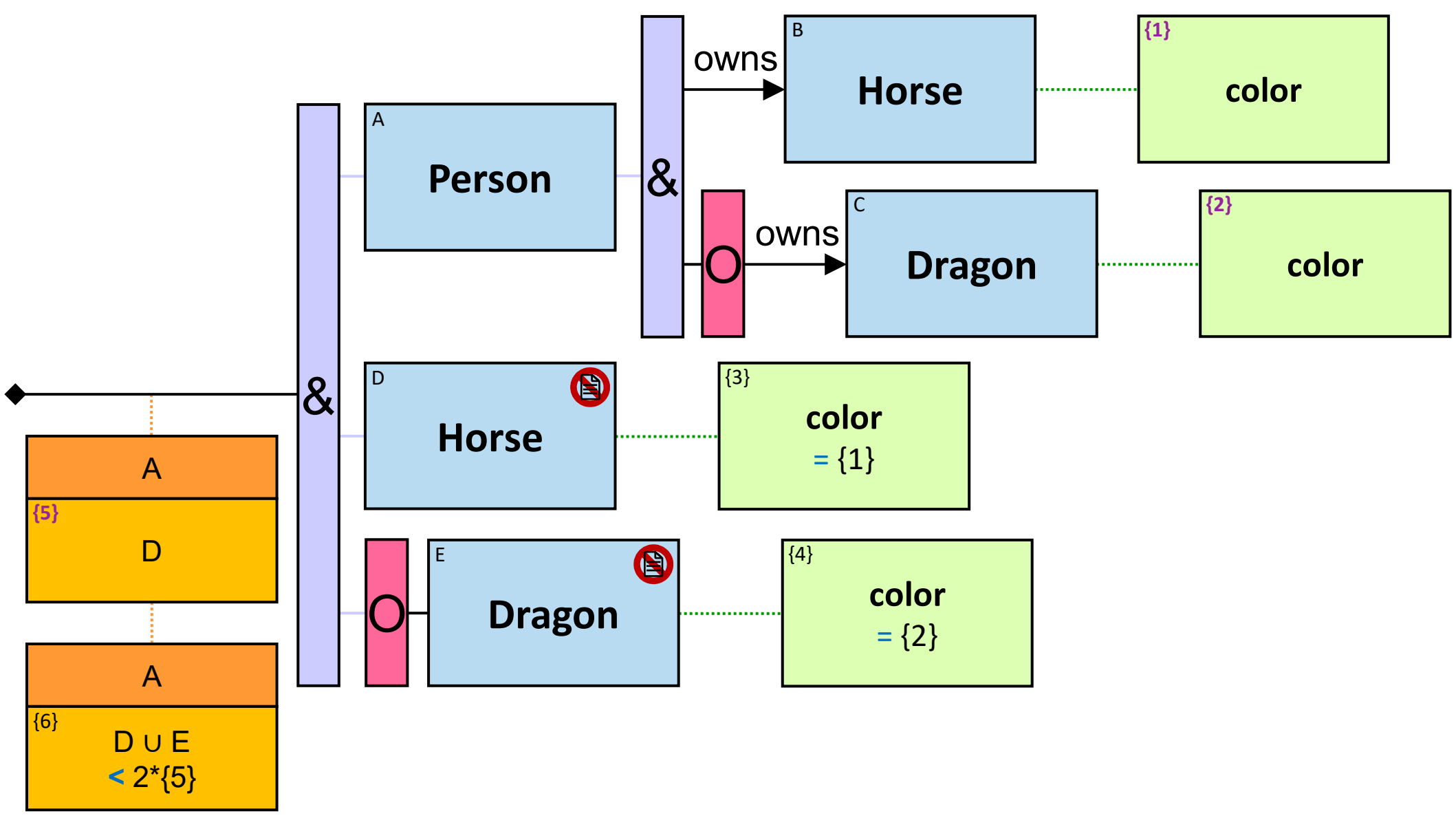

The number of assignments to E is zero when the optional part has no assignment. We cannot simply test if E < {5} since if there are zero assignments to E per A – the constraint is not satisfied.

***Q321:*** *Any person A where more horses than dragons have the same name-length as a horse/dragon A owns* (version 1)

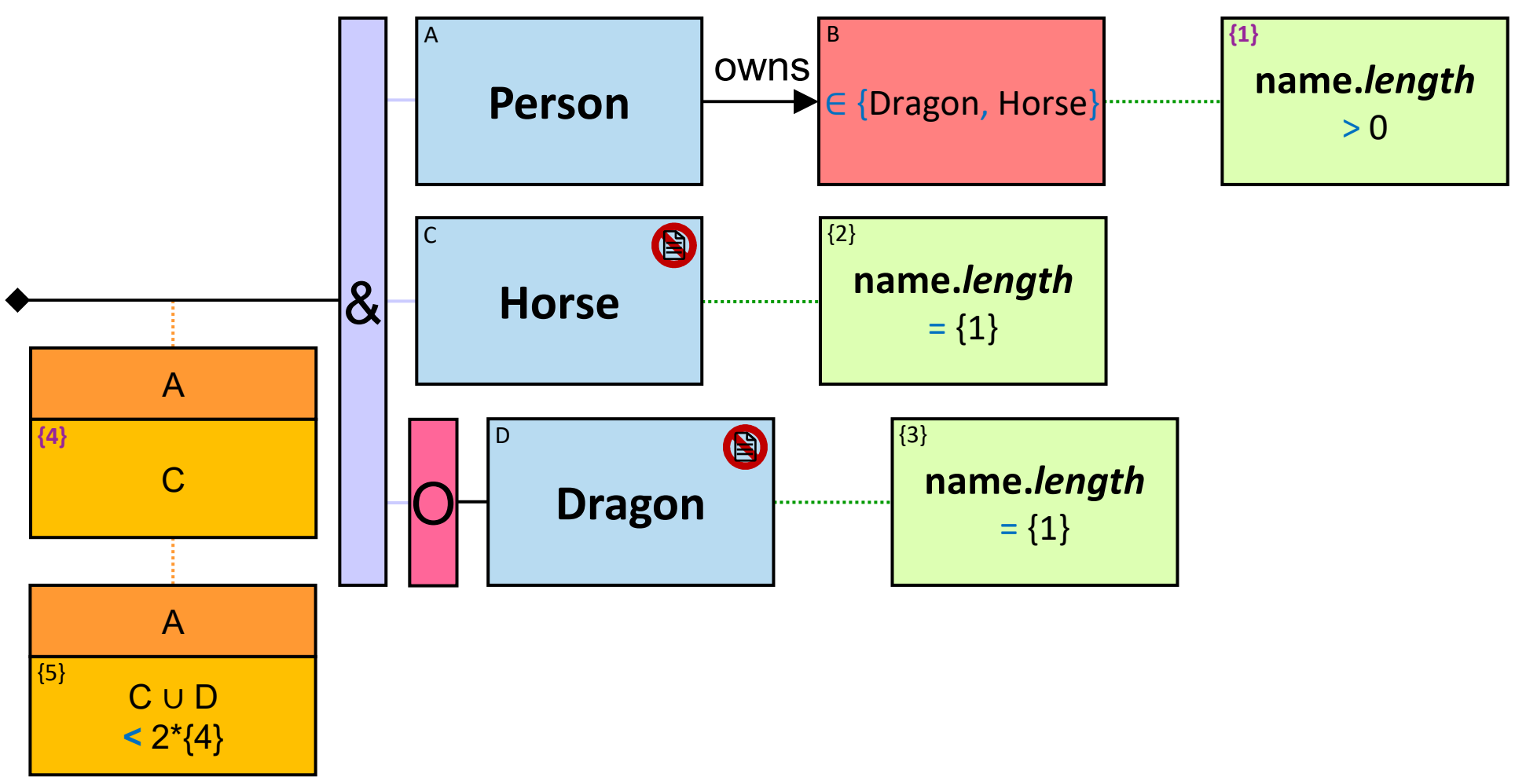

The number of assignments to D is zero when the optional part has no assignment. We cannot simply test if D < {4} since if there are zero assignments to D per A – the constraint is not satisfied.

***Q259:*** *Any person who since 1011 become an owner of no horses or up to four horses* (two versions)

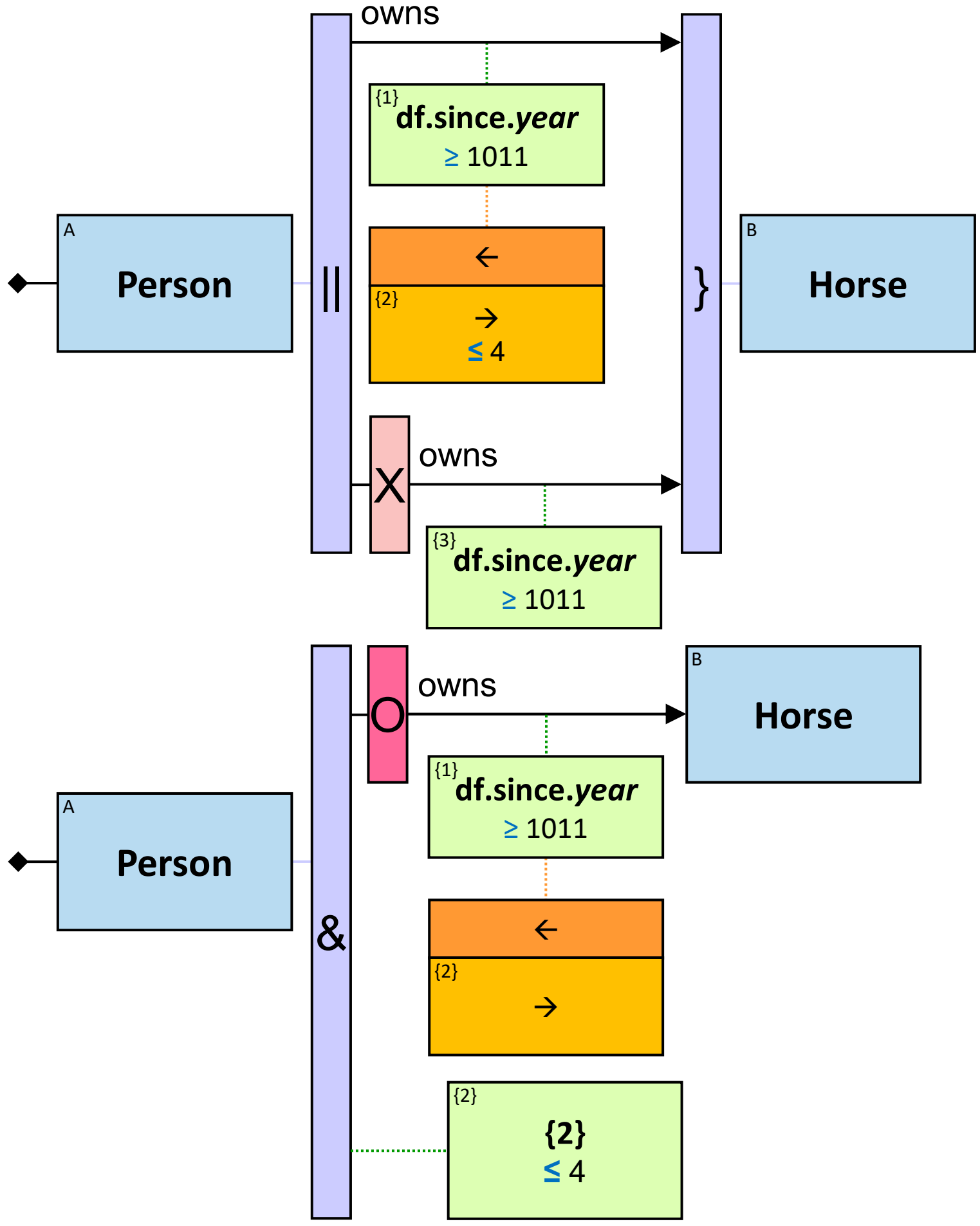

***Q337:*** For each pair of dragons: each time one of them froze the other for a duration longer than the average duration of all the freezes that are longer than the average duration one of them froze the other

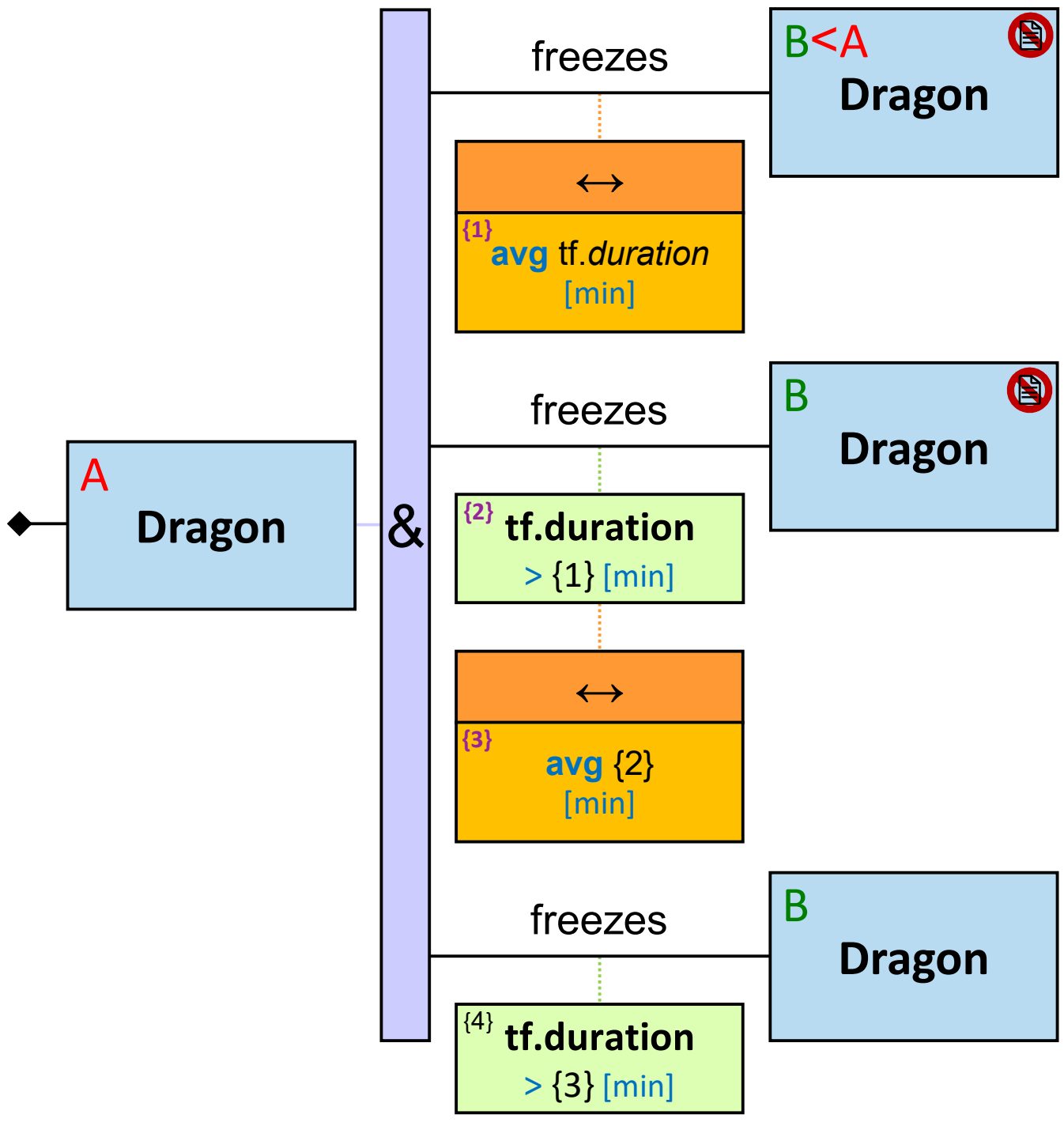

***Q317:*** *Any dragon that the time difference between the earliest time it froze/fired at some dragon and the latest time it froze/fired at some dragon – is at least one year*

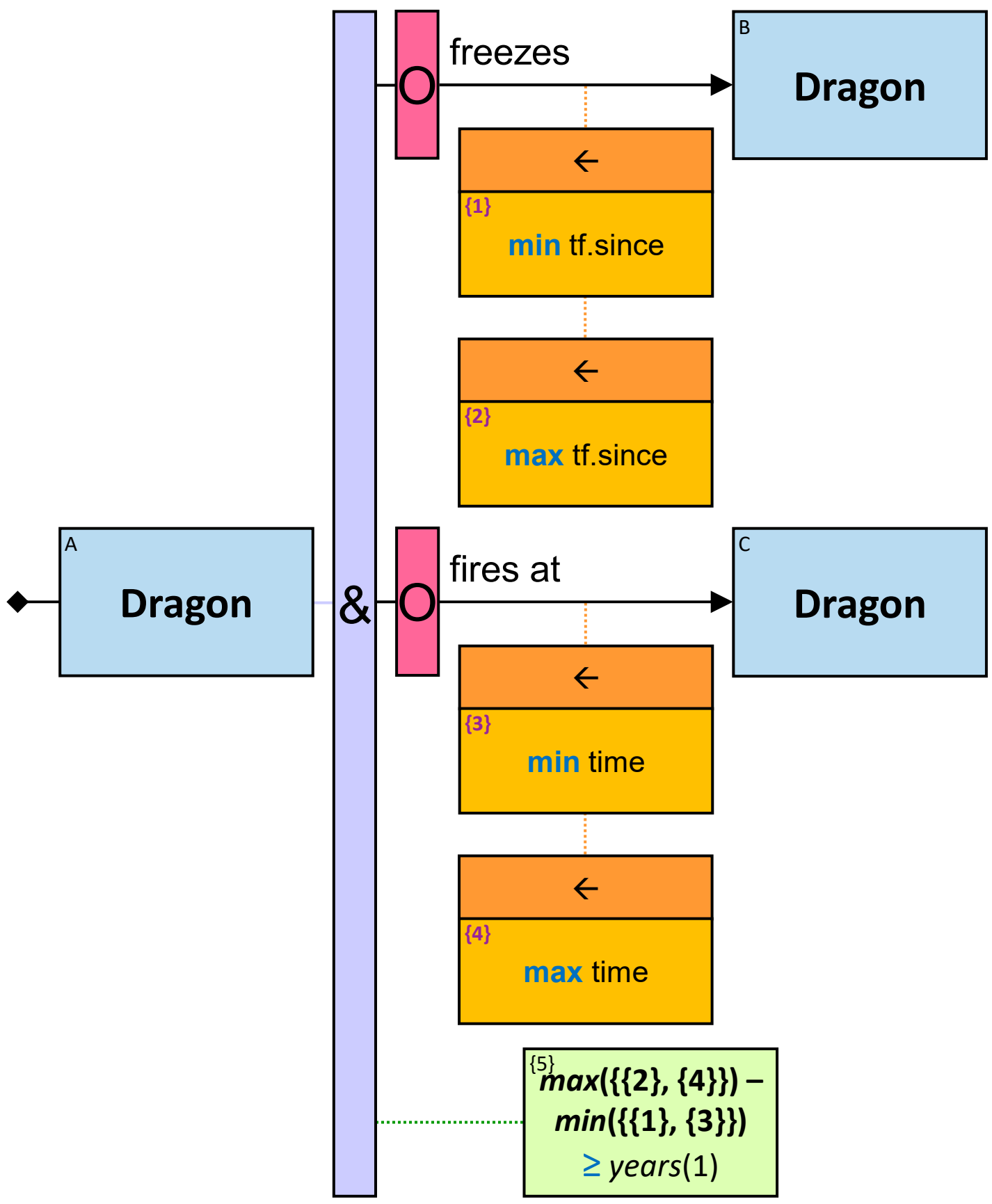

Note that {1}, {2}, {3} and {4} are defined right of an 'O', hence evaluated to *null* when the optional part has no assignment. Since *min*({...}) and *max*({...}) ignore *nulls*, the pattern would yield the expected results even when dragon A froze no dragon or when dragon A fired at no dragon.

## 35. AGGREGATOR SEQUENCES

Multiple aggregators may appear along a *sequence*. Along a sequence, aggregator's constraints are evaluated from right to left, except where a constraint depends on the evaluation result of a constraint on its left (see Q376).

***Q128:*** *Any person A and A's three offspring who own horses of the largest number of colors*

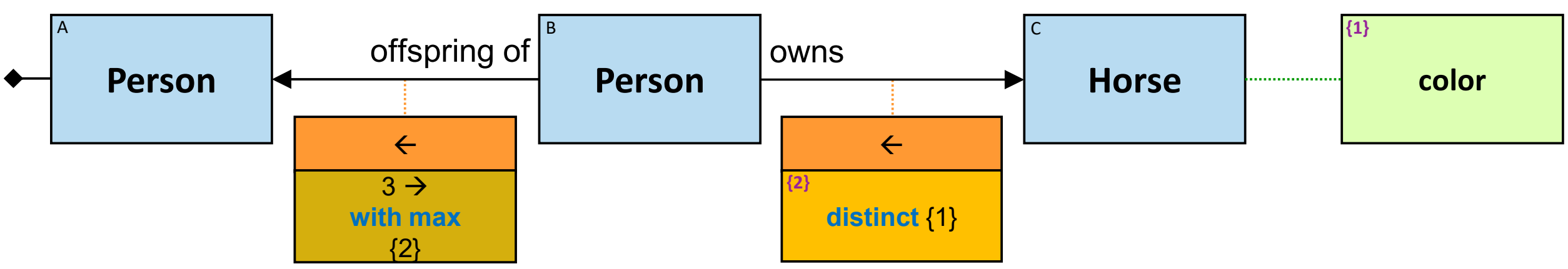

***Q198:*** *Any person A and A's three dragons that (for each of them: the four dragons it froze that froze the largest number of dragons) – froze the largest number of distinct dragons cumulatively*

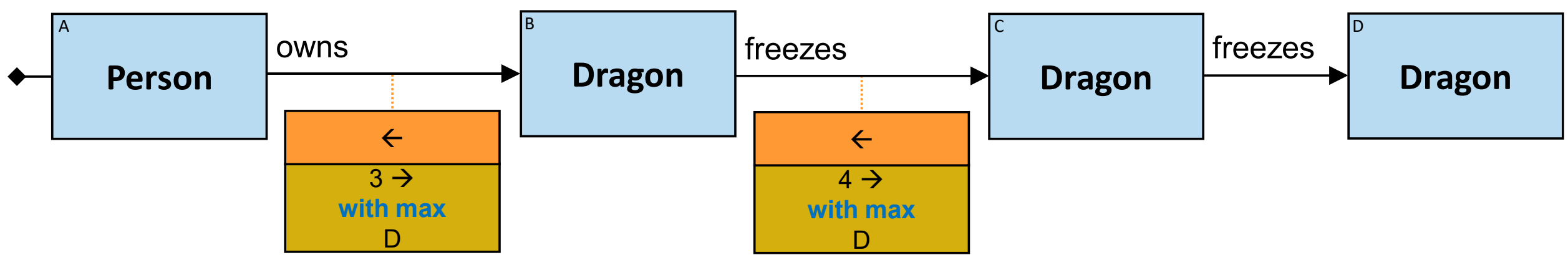

***Q179:*** *Any pair of dragons (A, B) where A froze B for a cumulative duration longer than the cumulative duration B froze dragons* (two versions)

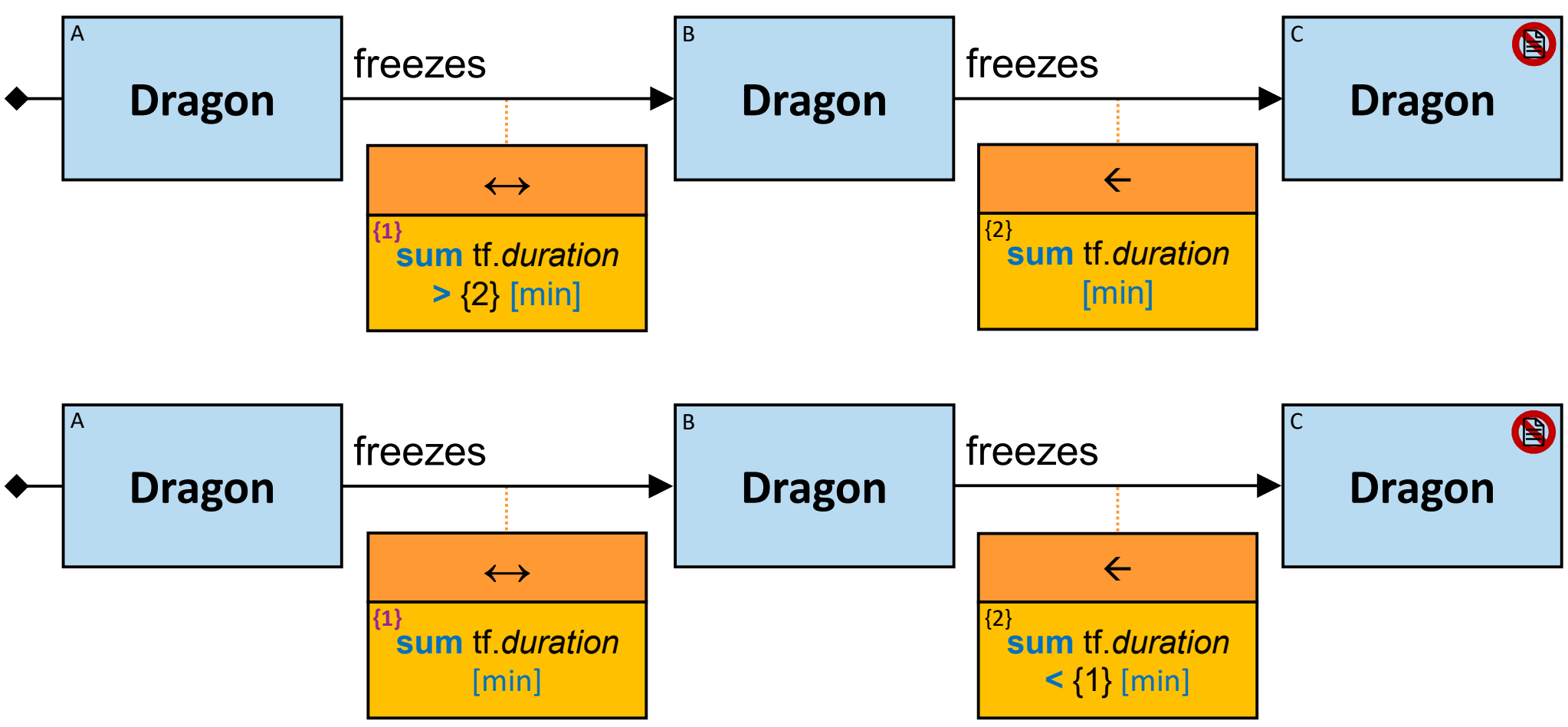

***Q103:*** *Any dragon A that froze at least three dragons, each of which was frozen by at least four dragons other than A*

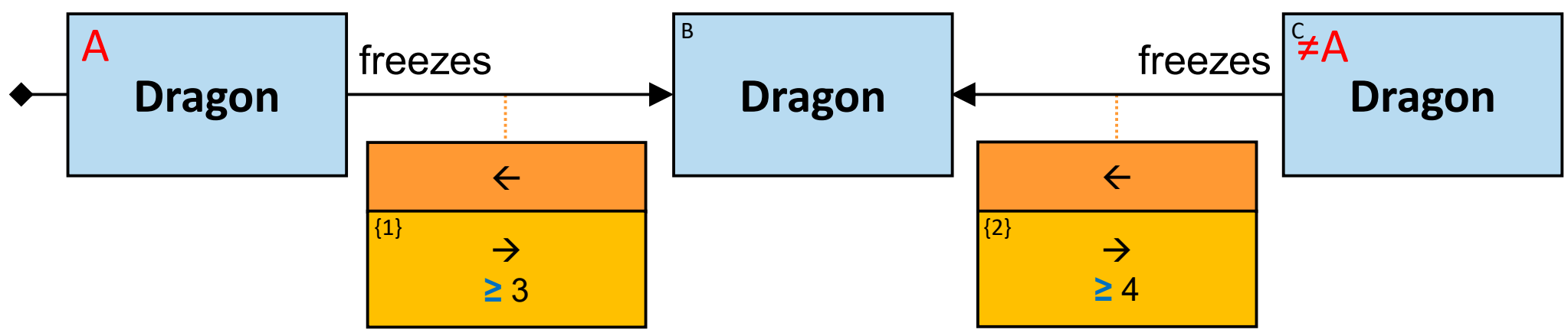

***Q106:*** *Any dragon A that froze dragons at least three times – each was frozen at least four times by dragons other than A*

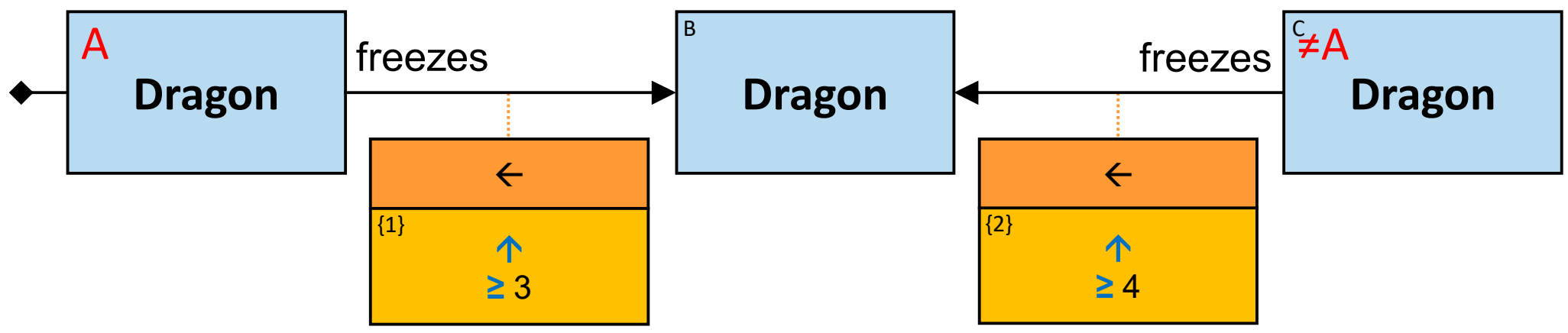

***Q164:*** *Any dragon that the number of times dragons it froze have frozen dragons (cumulatively) – is equal to the number of times dragons it fired at have fired at dragons (cumulatively)*

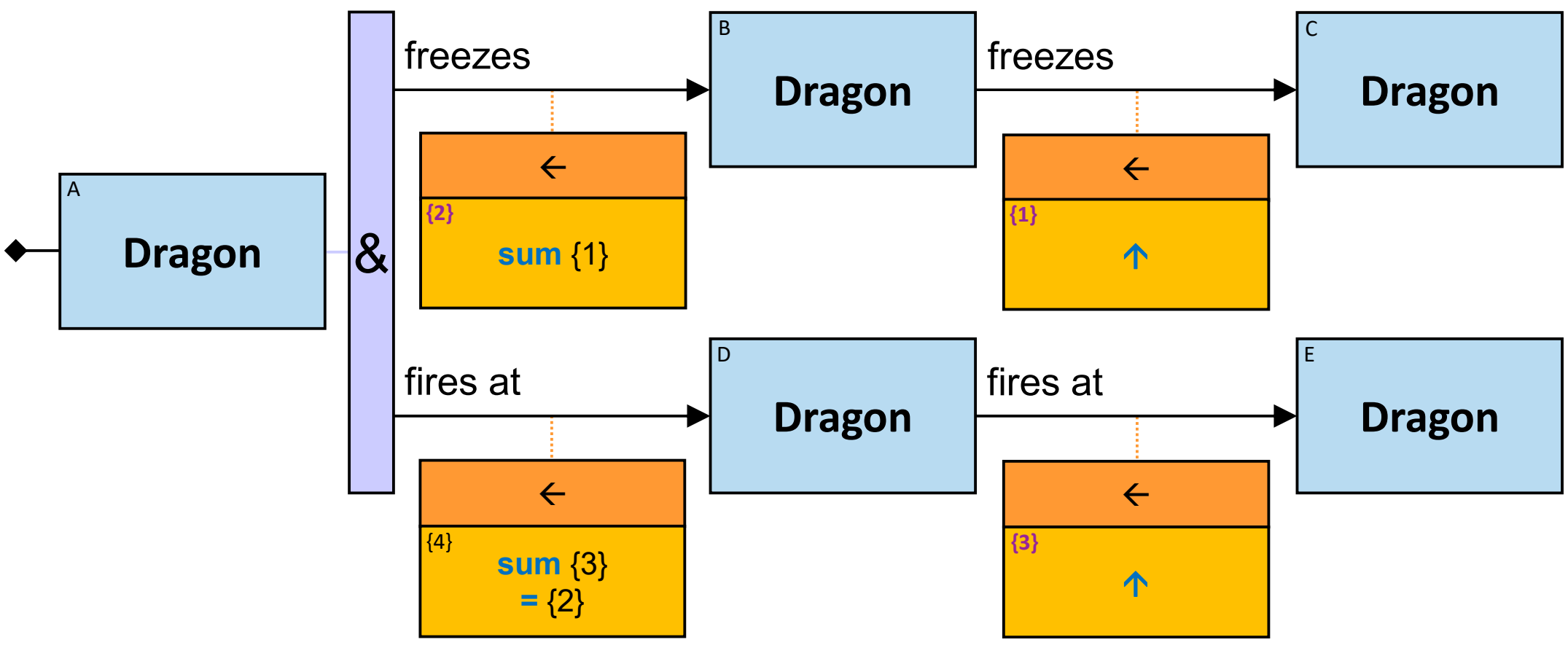

***Q356:*** *The average number of horse ownerships per person*

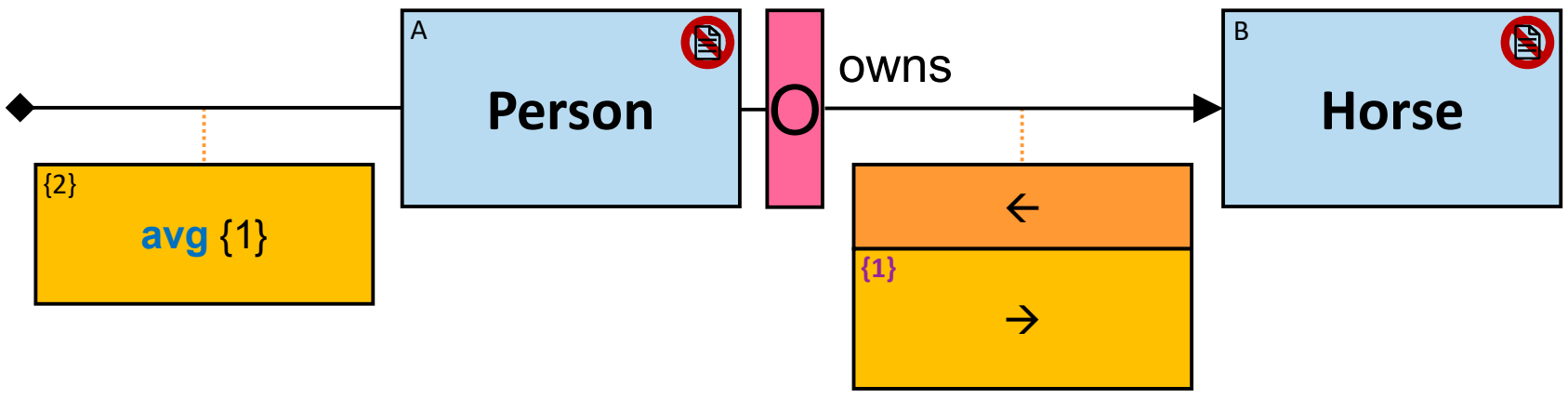

***Q324:*** *The five people with the smallest number (including zero) of paths of length up to four to some person*

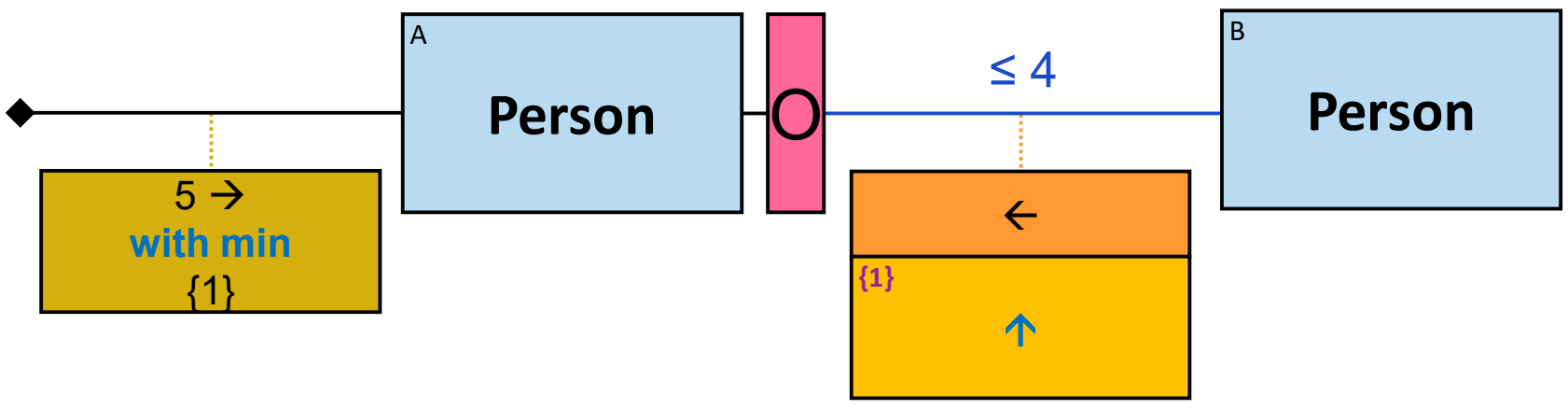

(Compare with Q172)

***Q107:*** *Any* ***dragon*** *that (the number of dragons, each owned by five people, that froze* ***it****) is five, and that the number of times* ***it*** *was frozen by those dragons (cumulatively) is not five*

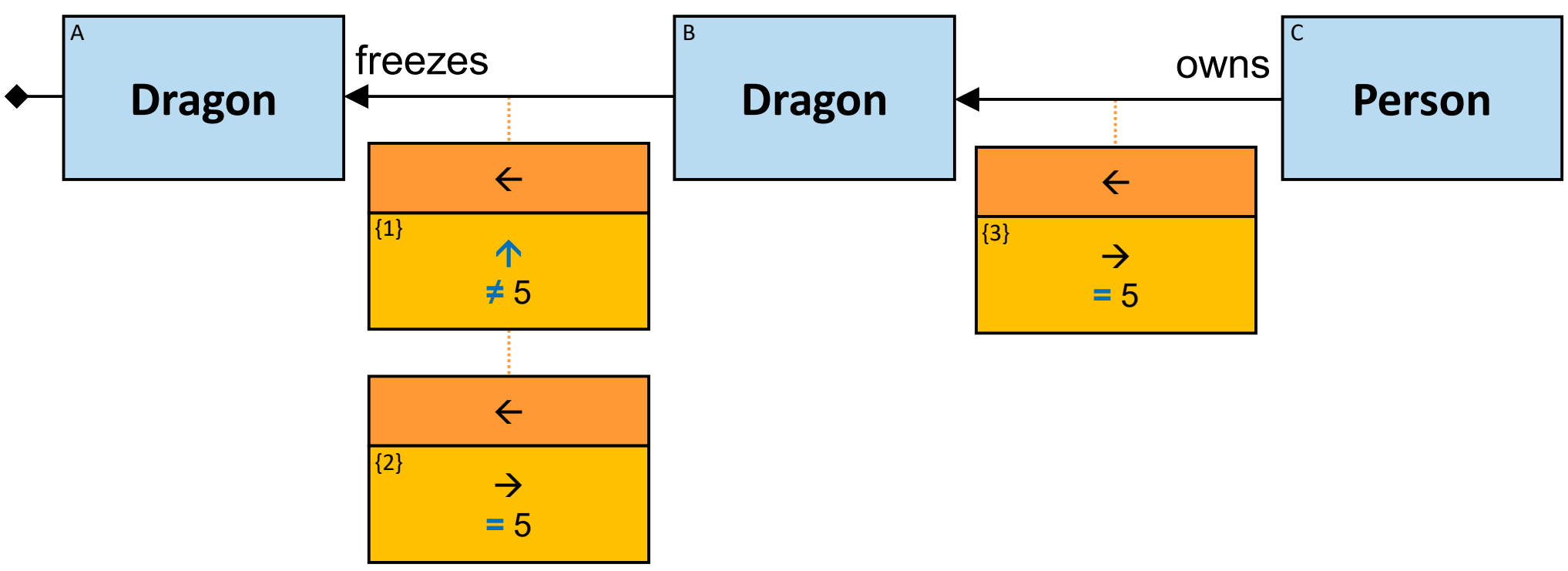

***Q169:*** *Any person A who (has at least one offspring who owns at least three horses) and (each of A's offspring who owns at least three horses – owns horses of at least three colors)*

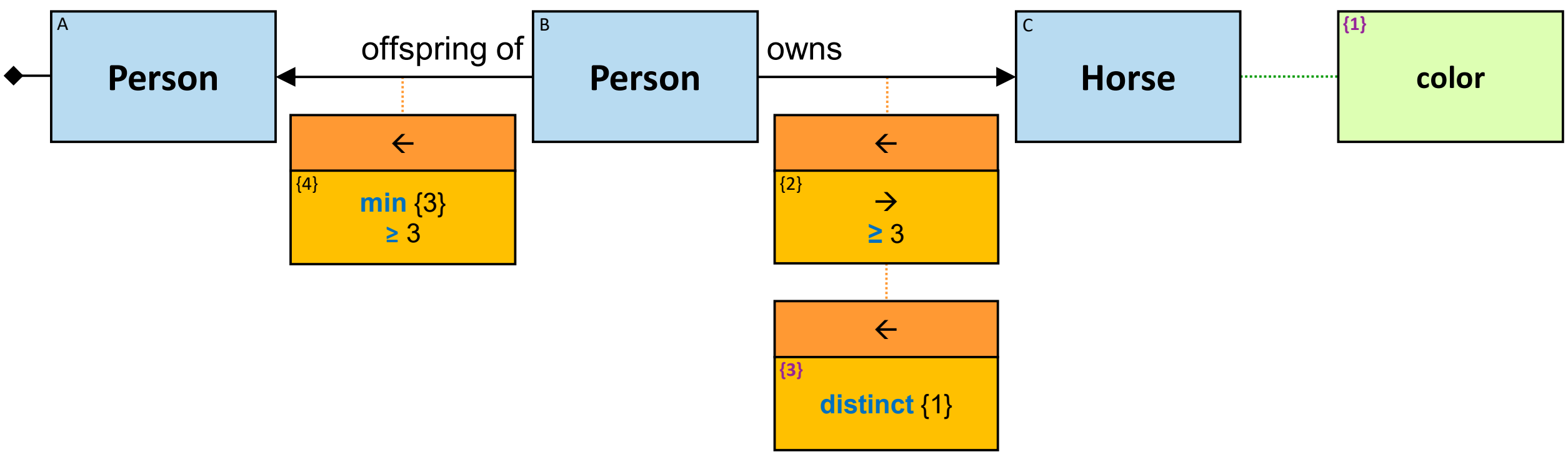

*Q129: Any person A that (each of A's offspring who owns at least one horse – owns a different number of horses)*

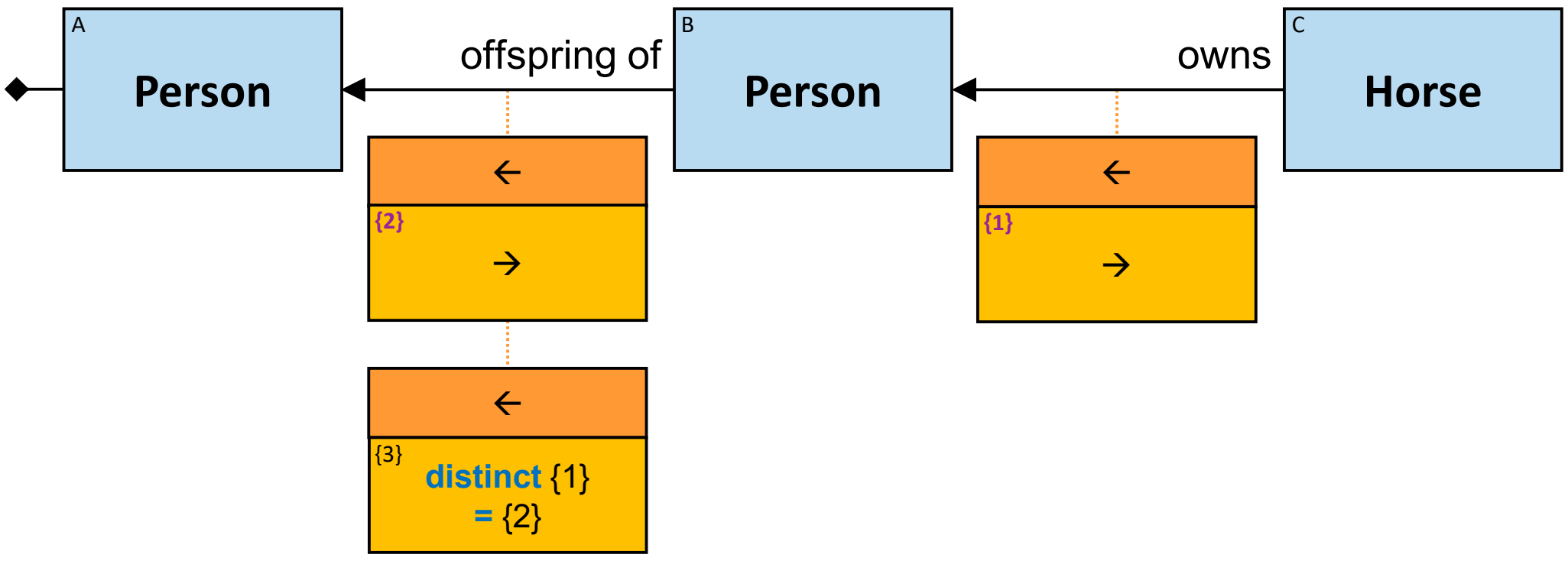

*Q181: Any dragon with no intersection between the groups of dragons frozen by any two dragons it froze*

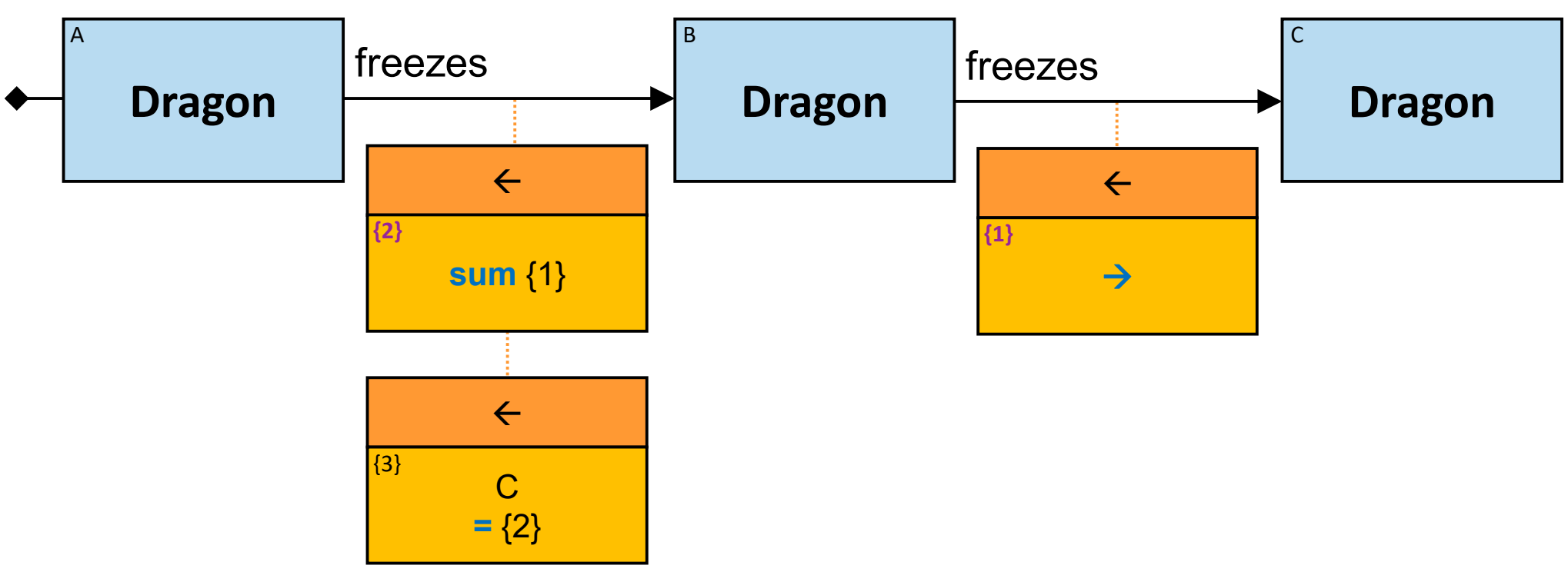

*Q213: Out of the pairs of dragons (A, B) where A froze B for a cumulative duration longer than the cumulative duration B froze dragons – the five pairs with the largest number of times A froze B*

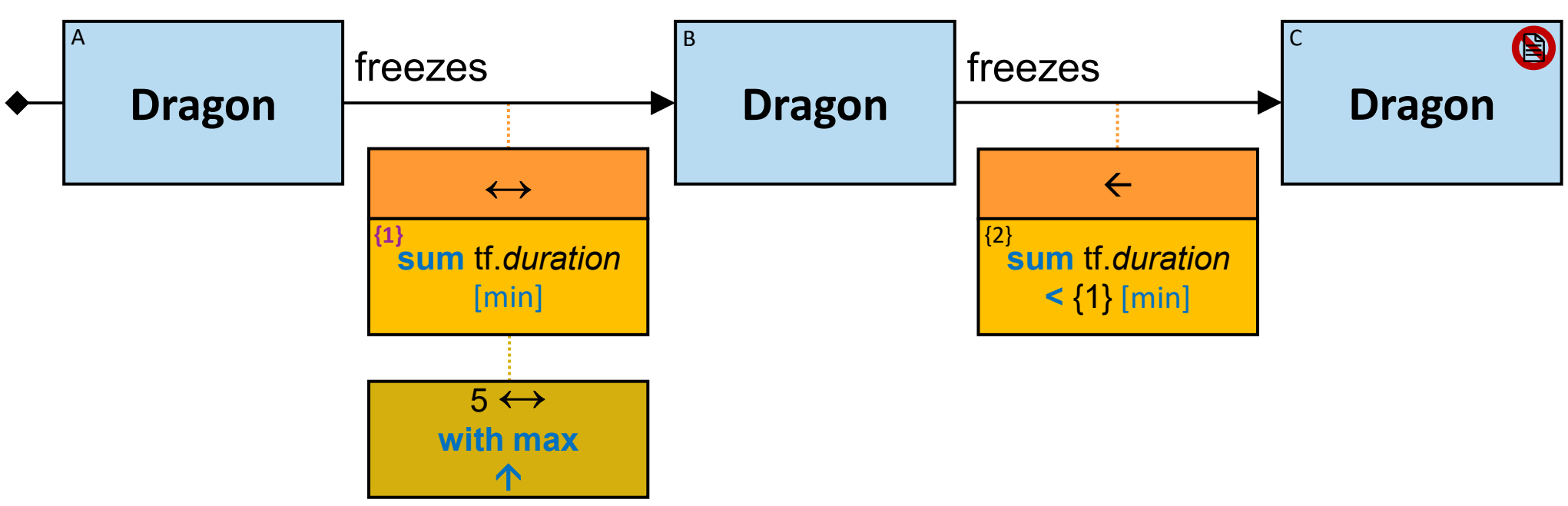

When the order of the aggregators along the chain is switched, the meaning changes:

***Q376:*** *Out of the five pairs of dragons (A, B) with the largest number of times A froze B – the pairs where A froze B for a cumulative duration longer than the cumulative duration B froze dragons*

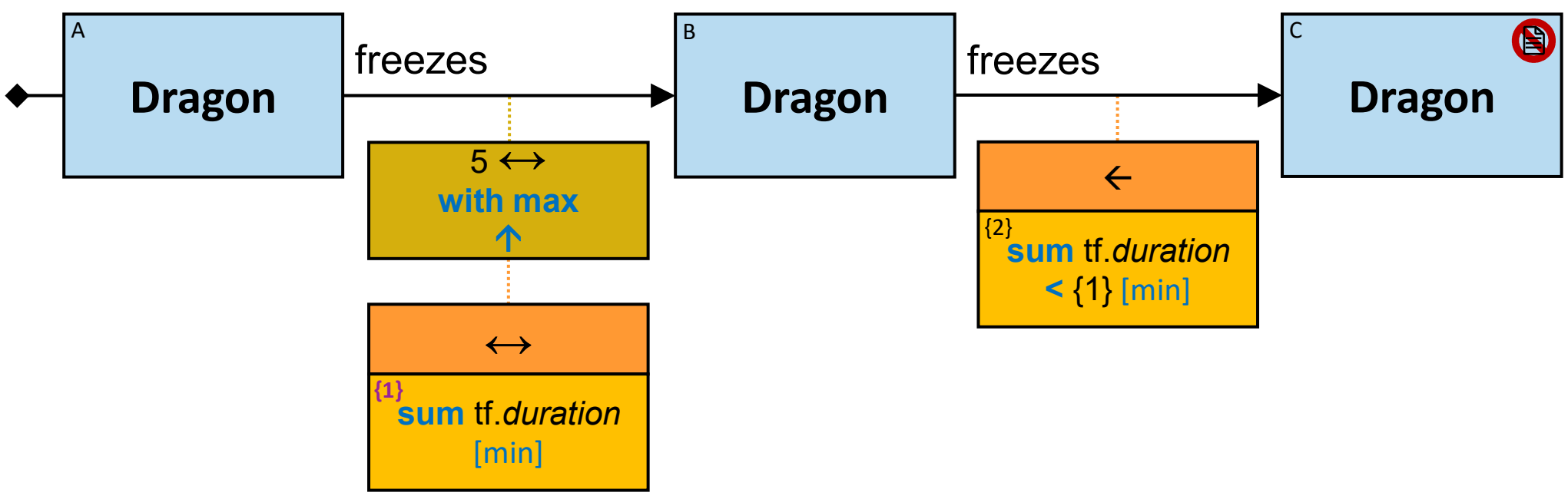

Since {2} has a constraint that depends on the evaluation of {1}, {1} must be evaluated first. However, the M2 aggregator filters (A, B) pairs before {1} is evaluated.

***Q191:*** *Any dragon that froze dragons S and fired at dragons T.* $|S|\geq 3$, $|T|\geq 3$ *and* $|S\cup T|\geq 10$

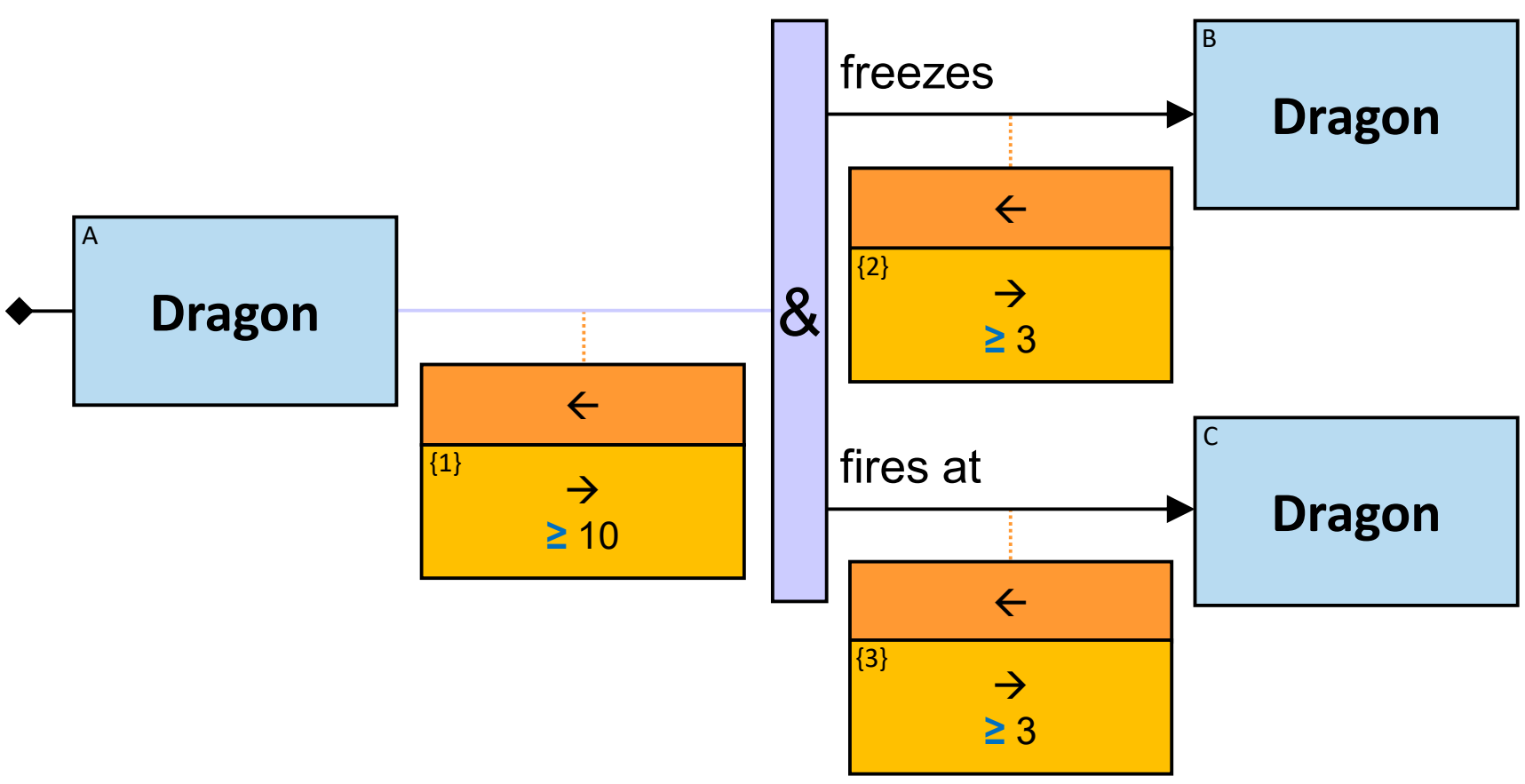

***Q192:*** *Any dragon that froze dragon* $m\geq 3$ *times and fired at dragons* $n\geq 3$ *times.* $m+n\geq 10$

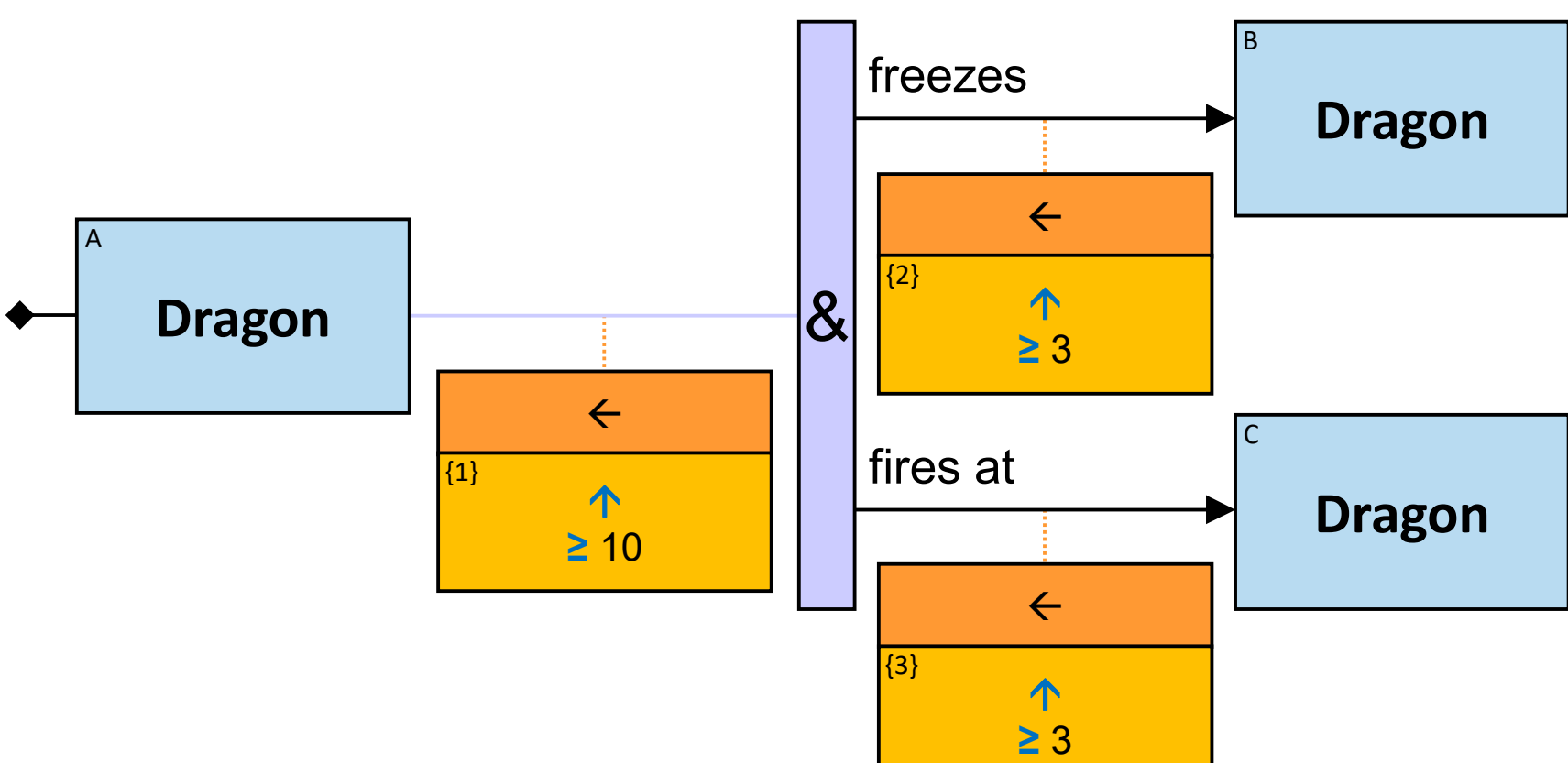

***Q193:*** *Any dragon that froze dragons S, fired at dragons T, and fired ≥3 times.* $|S|\geq 3$ *and* $|S\cup T|\geq 10$

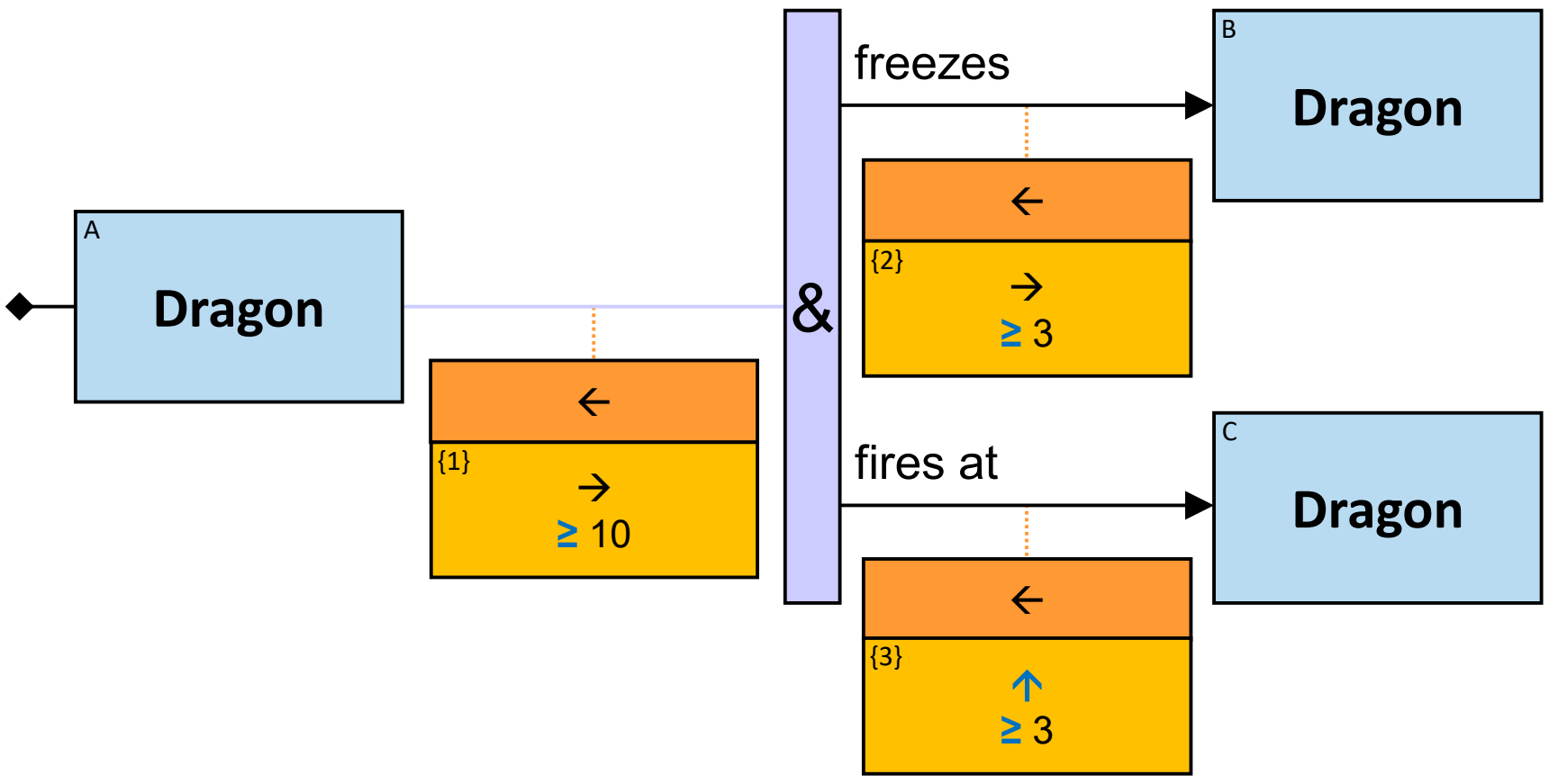

***Q194:*** *Any dragon that froze dragons m times, fired at dragons n≥3 times, and froze ≥3 dragons.* $m+n\geq 10$

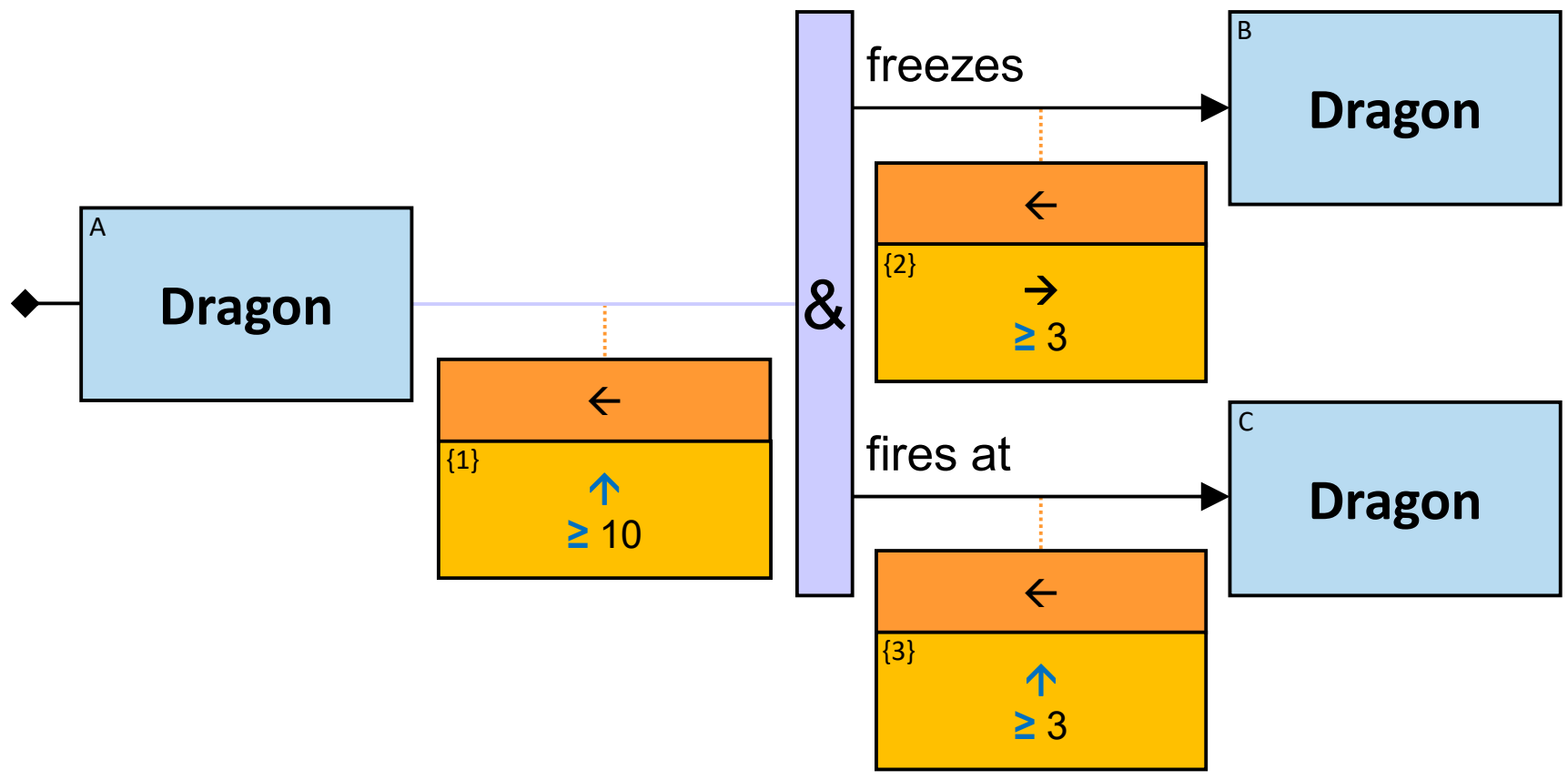

***Q97:*** *Any dragon that froze more than three dragons – each more than ten times or for a cumulative duration of more than 100 minutes*

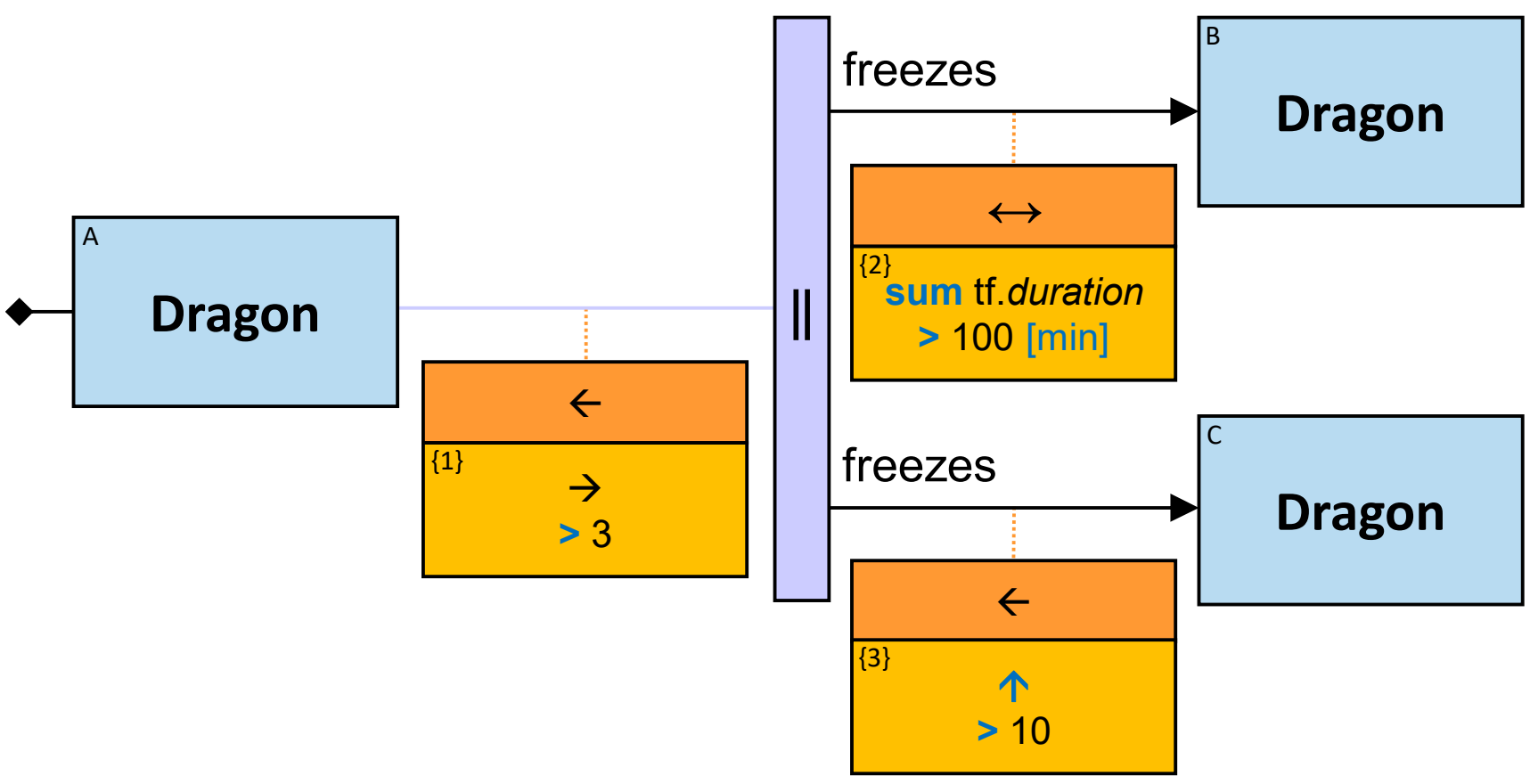

***Q180:*** *Any pair of dragons (A, B) where the cumulative duration A and B froze each other – is longer than both the cumulative duration A froze other dragons and the cumulative duration B froze other dragons*

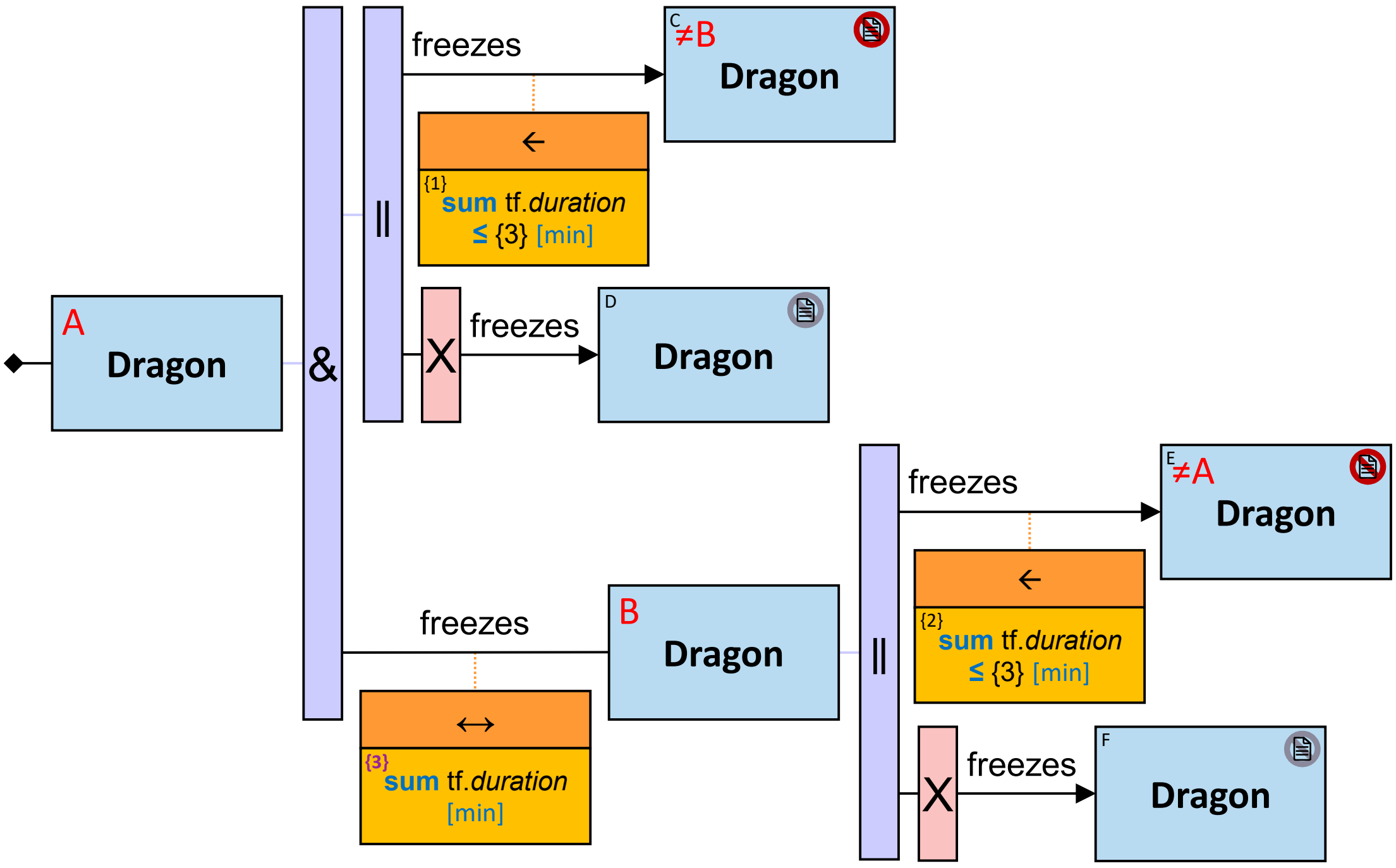

***Q347:*** *The five dragon triplets with the largest number of times triplet-members froze one another*

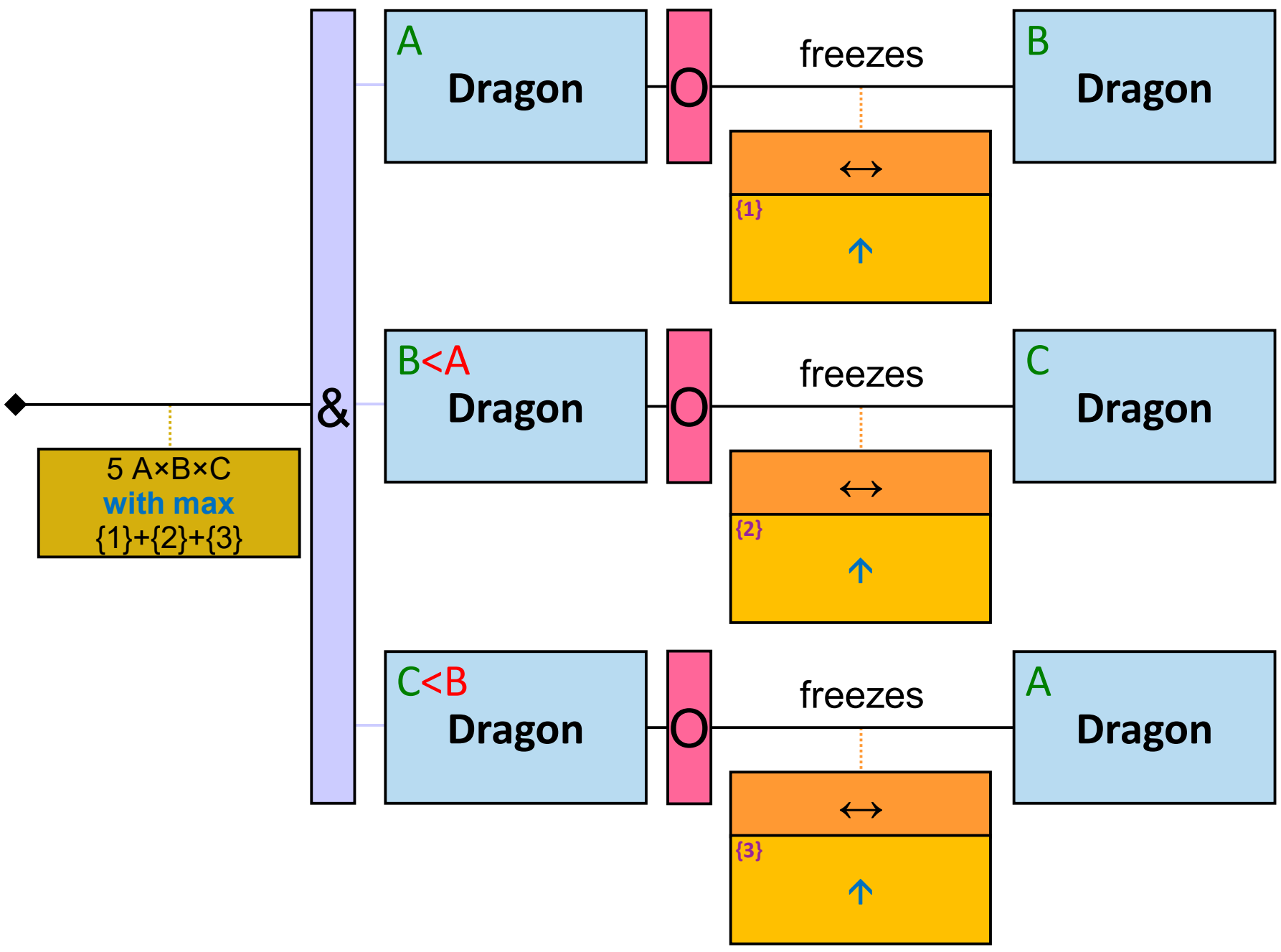

Note that we need to count each triplet only once (ignoring permutations).

***Q100:*** *Any dragon that Balerion froze more than ten times for less than ten minutes, more than ten times on or after January 1, 1010, more than 15 times for less than ten minutes or on or after January 1, 1010, and more than 100 times altogether*

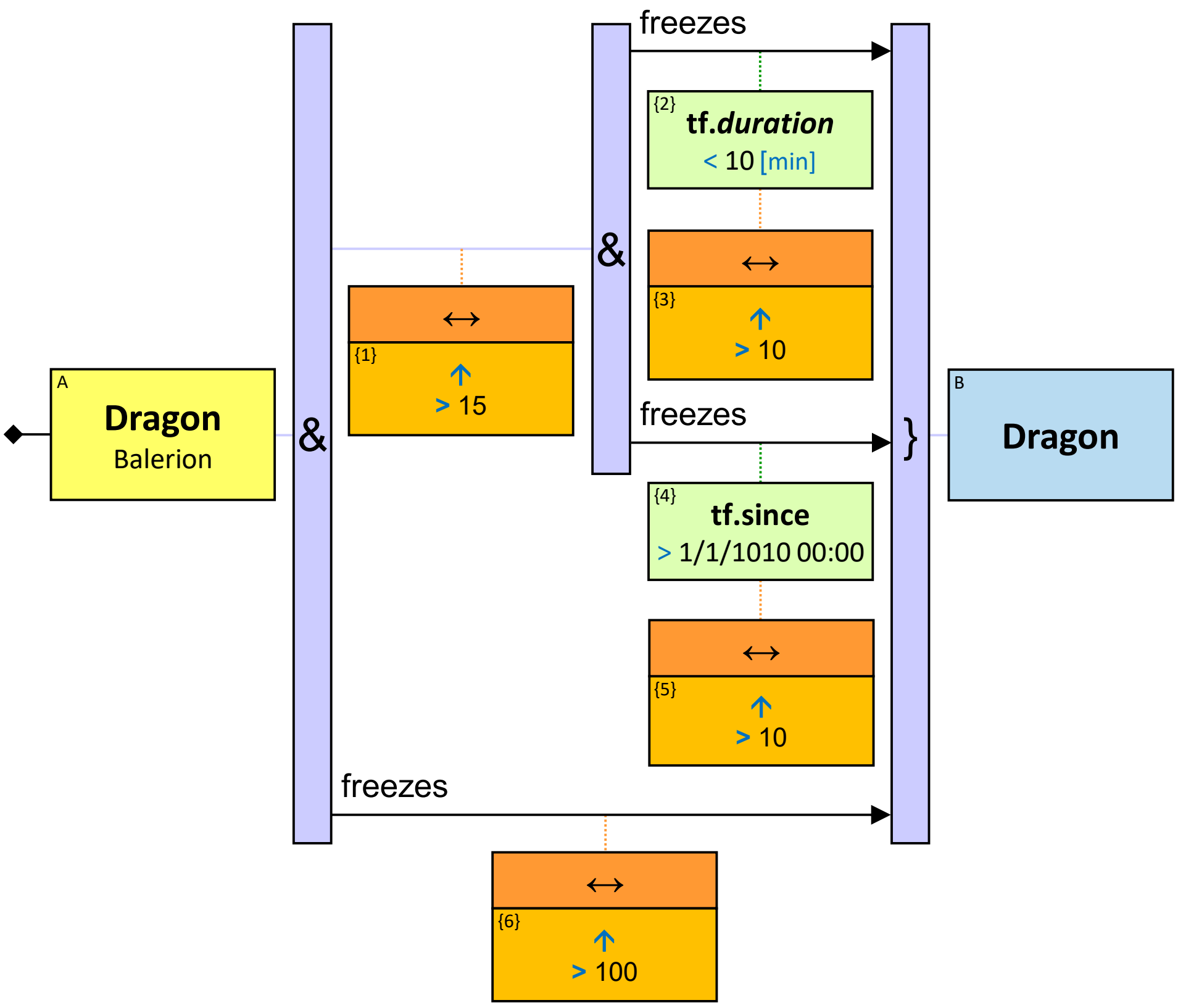

---

## 36. EXTENDED AGGREGATORS

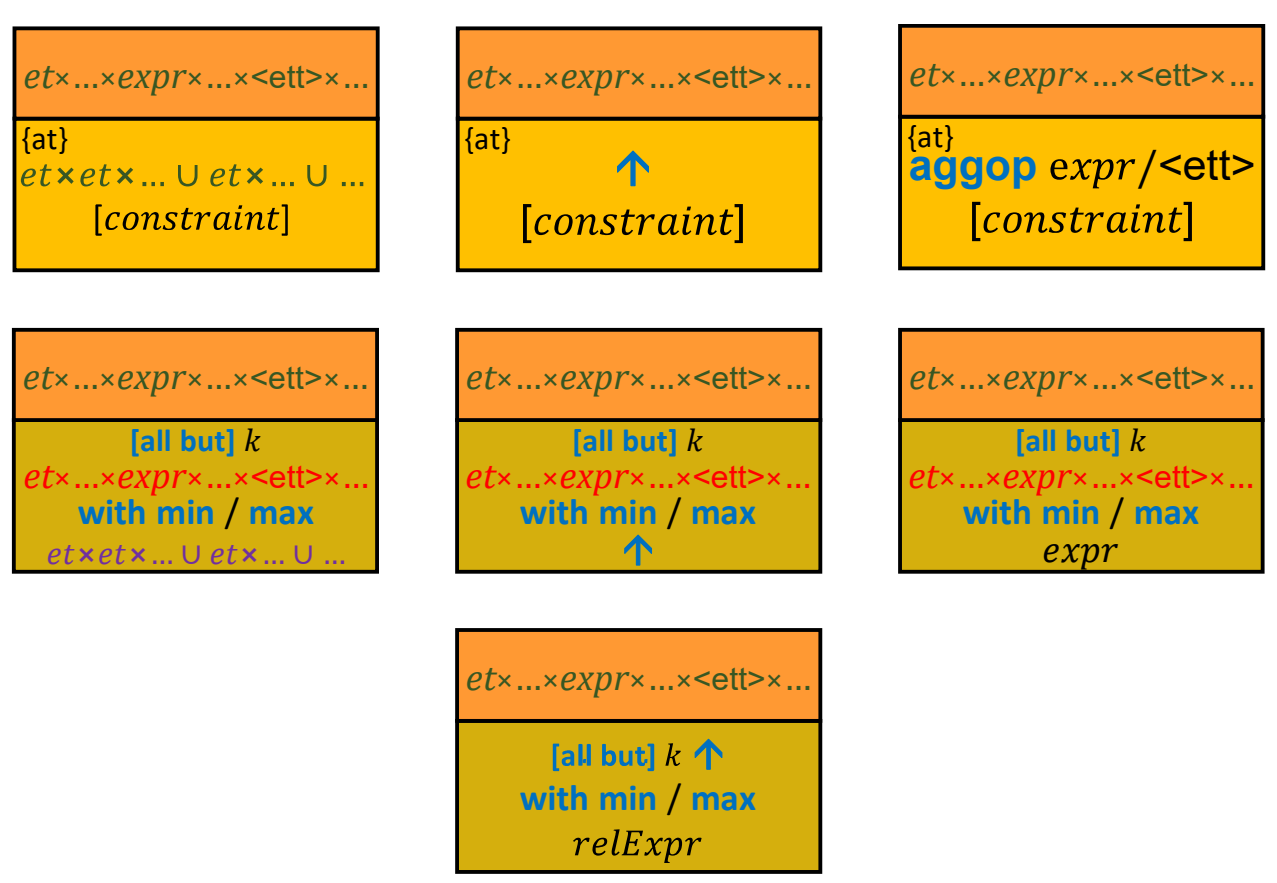

For extended aggregators, $T$ denotes an *extended Cartesian product* – a Cartesian product not just of entity-tags but of

- Zero or more entity-tags
- Zero or more expressions, each can be either
    - an inherent property of the attached relationship

- an EA-tag
- Zero or more entity type-tags
- Zero or more relationship type-tags

Hence, for all extended aggregator types (A1, A2, A3, M1, M2, M3, and R1), the *per* clause has the following format:

- *'et1 × et2 × … × expr1 × expr2 × … × ⟨ett1⟩× ⟨ett2⟩× … × ⟪rtt1⟫× ⟪rtt2⟫× …'*:
  *T = et1 × et2 × … × expr1 × expr2 × … × ⟨ett1⟩× ⟨ett2⟩× …× ⟪rtt1⟫× ⟪rtt2⟫× …*

Wherever possible, the notations '←', '→' and '↔' are used instead of entity-tags.

For all extended M aggregators (M1, M2, and M3), the same is true also for *B.*

Here are two examples:

First example: *Any person who at a certain date became an owner of more than five horses* (Q115v2)

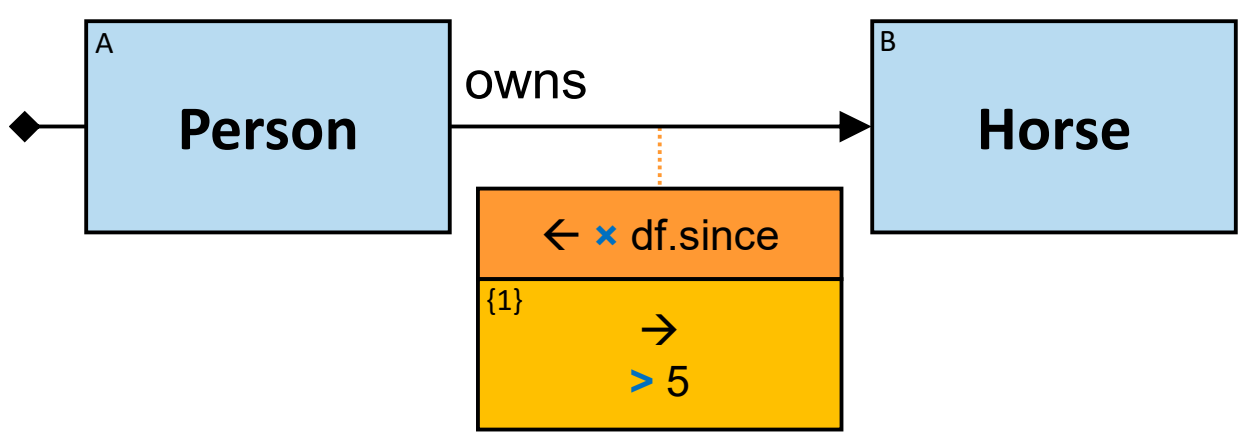

{1} is evaluated per each unique assignment to A × *df.since* in *S*. In other words, {1} is a property of each unique assignment to A × *df.since* in *S*.

Second example: *For each color of dragons that Balerion froze – the three dragons it froze the largest number of times* (Q215)

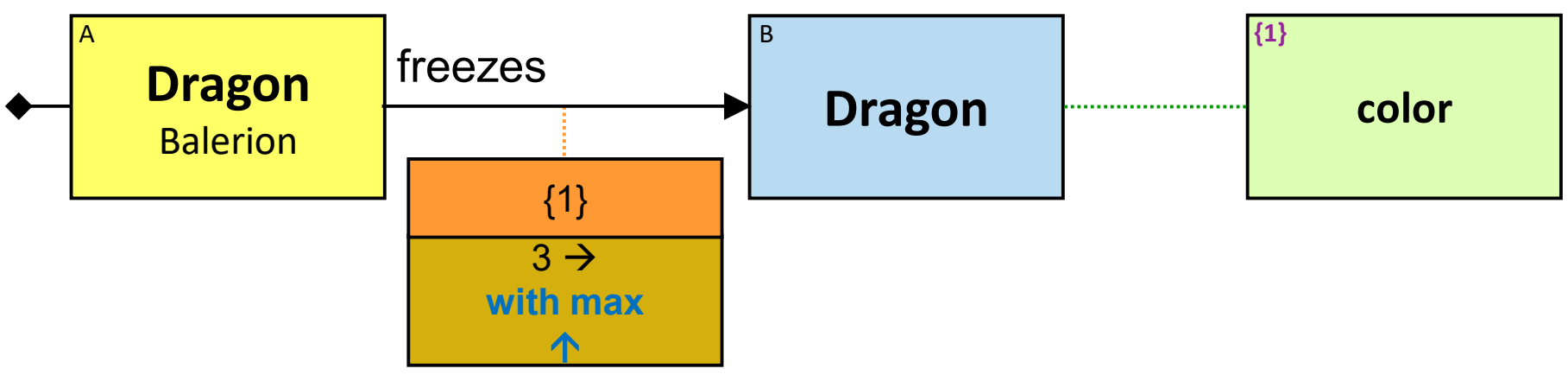

37. EXTENDED A1 AGGREGATOR

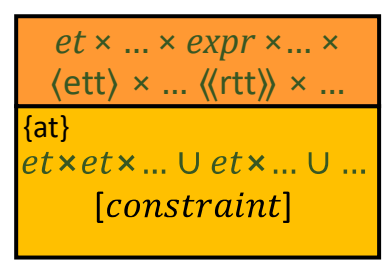

***Q261:*** *Any color of which there are at least ten horses*

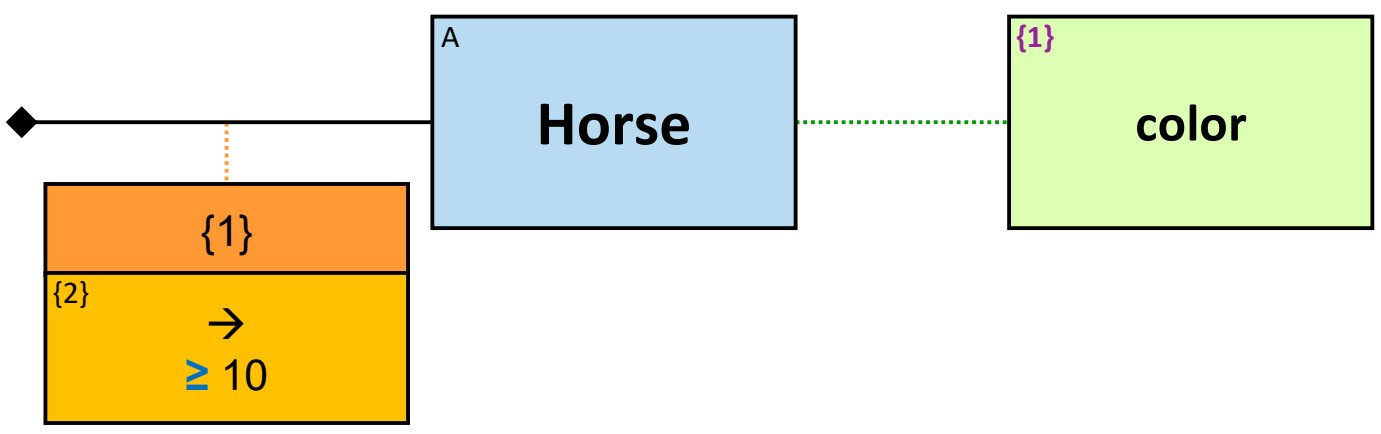

{2} is a property of each unique assignment to {1} in *S*.

***Q115:*** *Any person who at a certain date became an owner of more than five horses* (two versions)

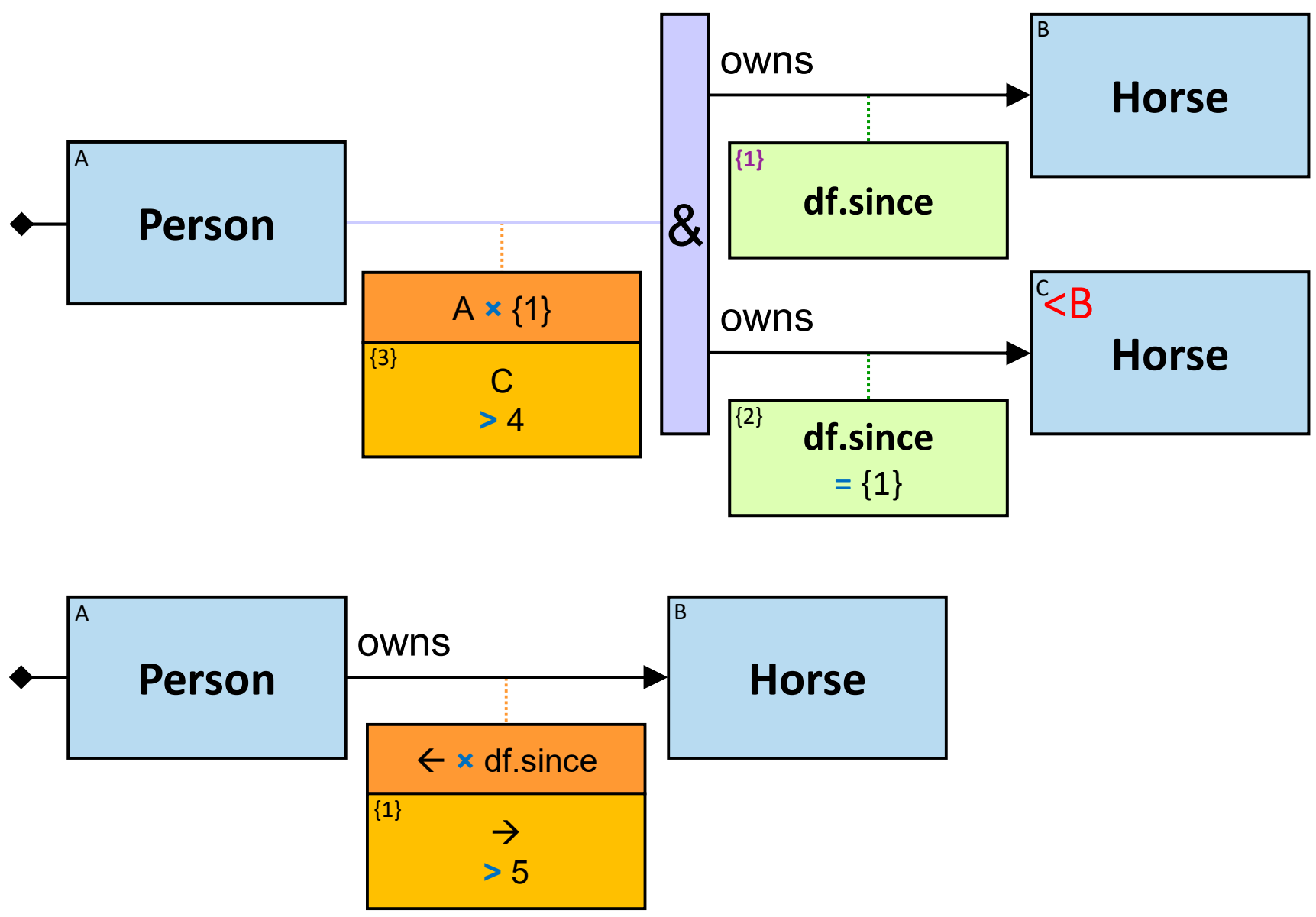

{1} is a property of each unique assignment to A × *df.since* in *S*.

***Q289:*** *Any person who at a certain three-day interval became an owner of more than five horses*

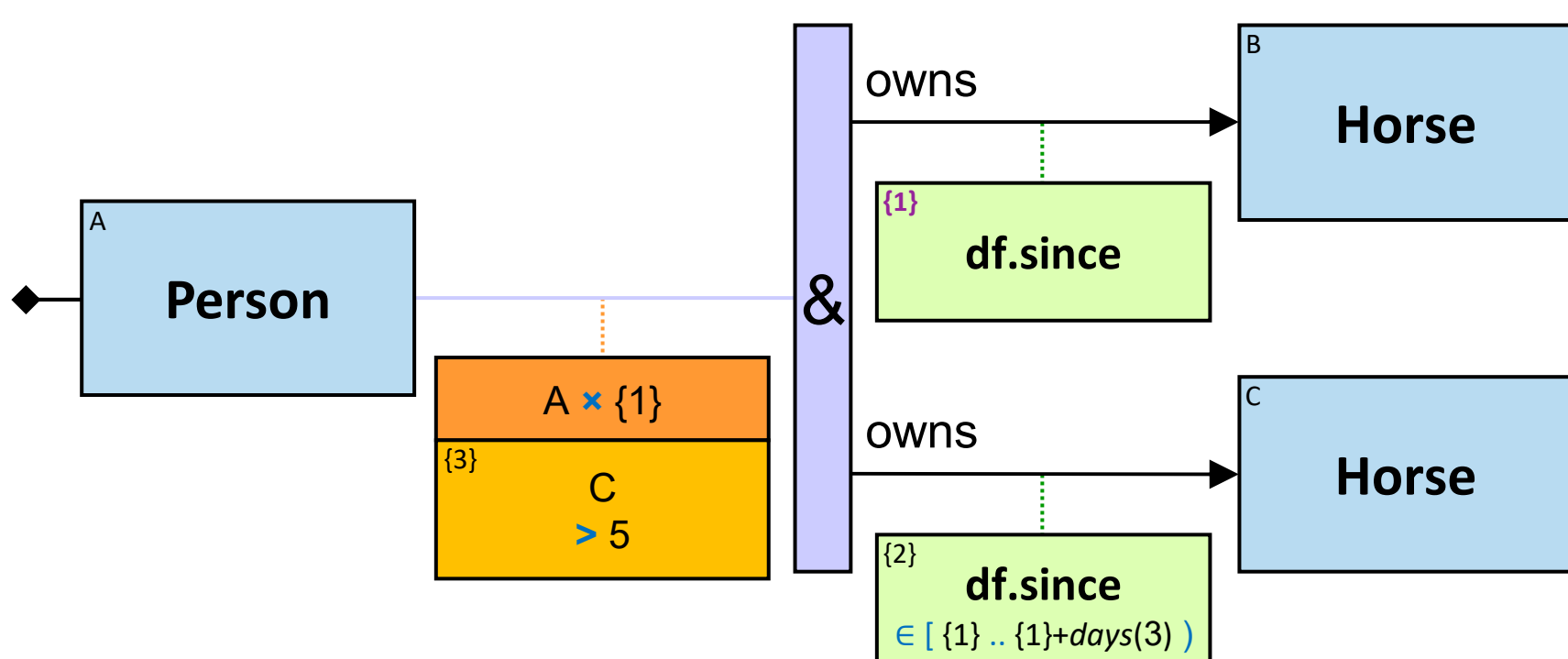

When $n$ intervals overlap, the intersection contains the start-time of at least one of the intervals (it also contains the end-time of at least one of the intervals).

***Q283:** Any person who at a certain day owned at least five horses*

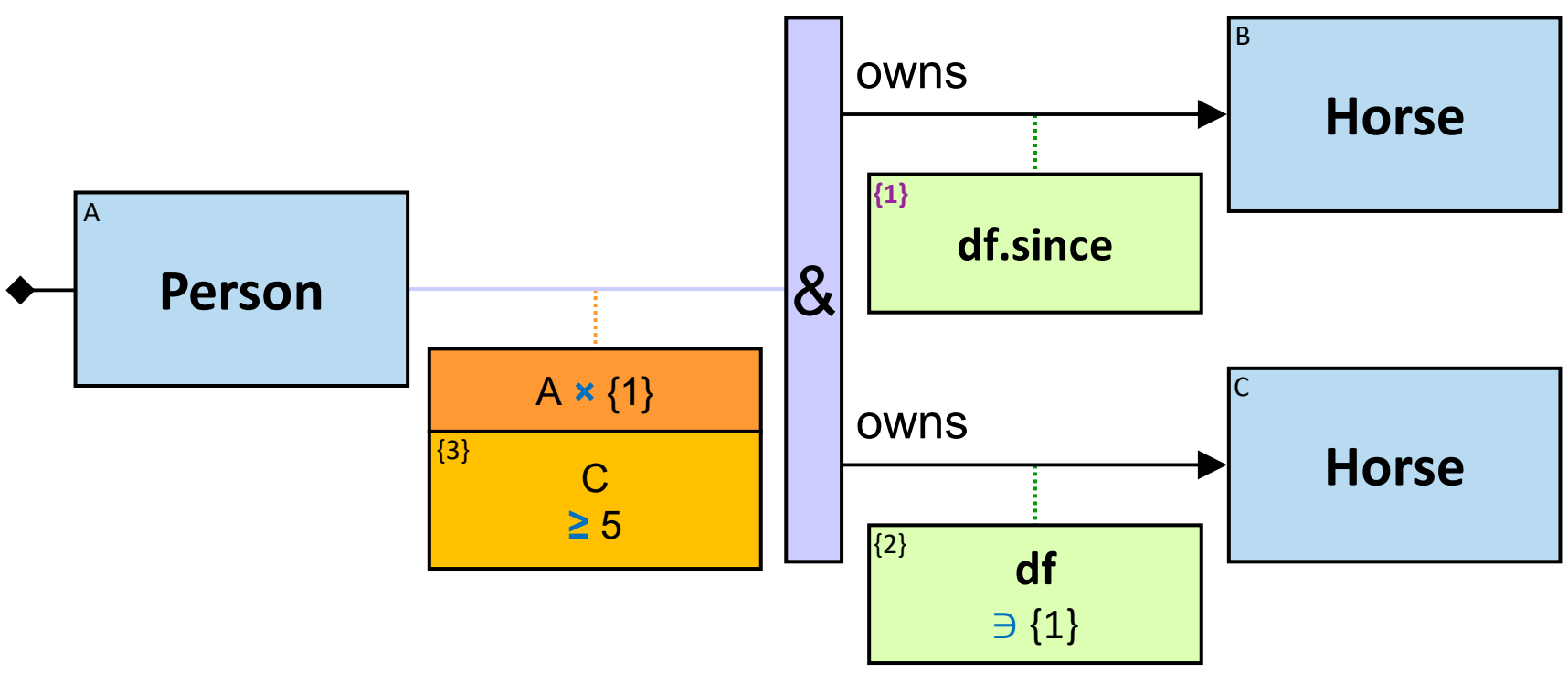

***Q284:** Any person who owned the same five horses – for at least ten consecutive days* (version 1)

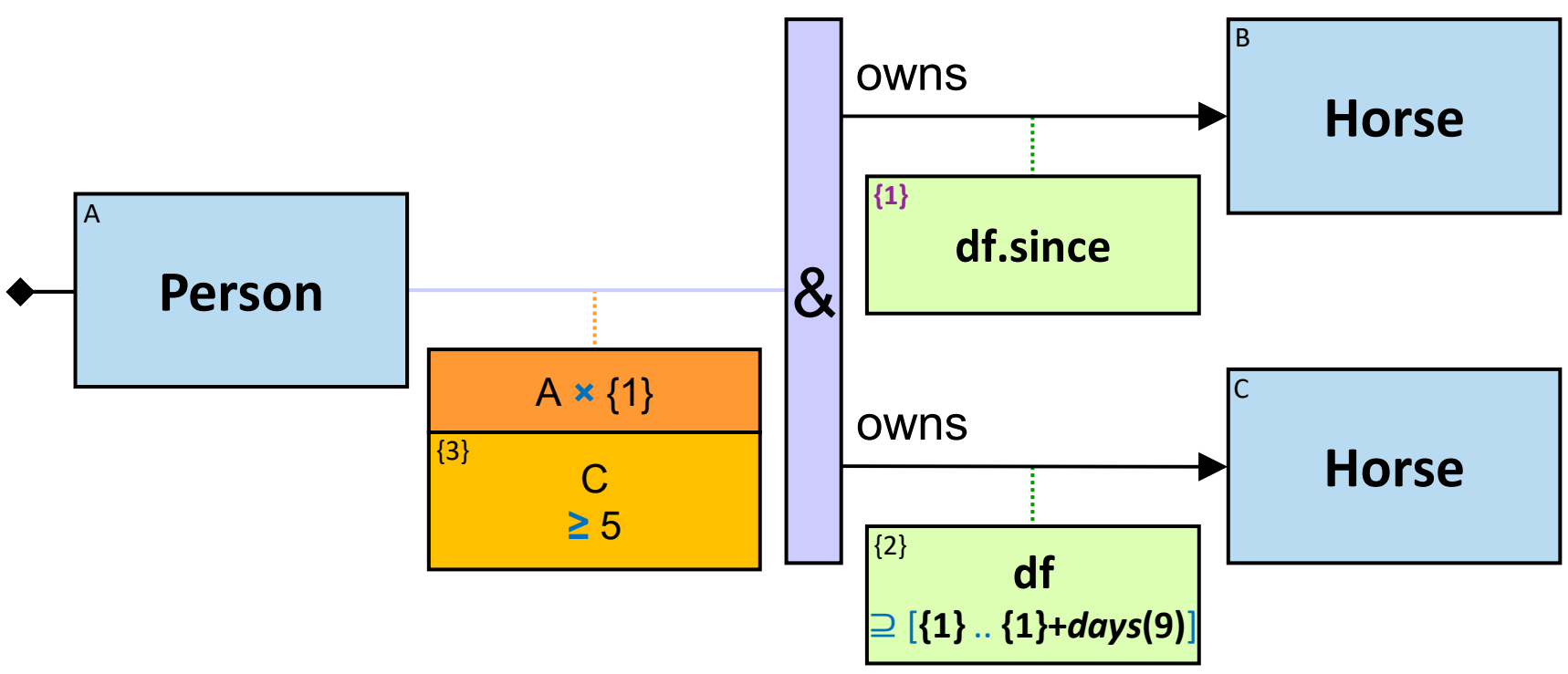

***Q217:** Any dragon and the dragons it froze – on days it froze between one and five dragons*

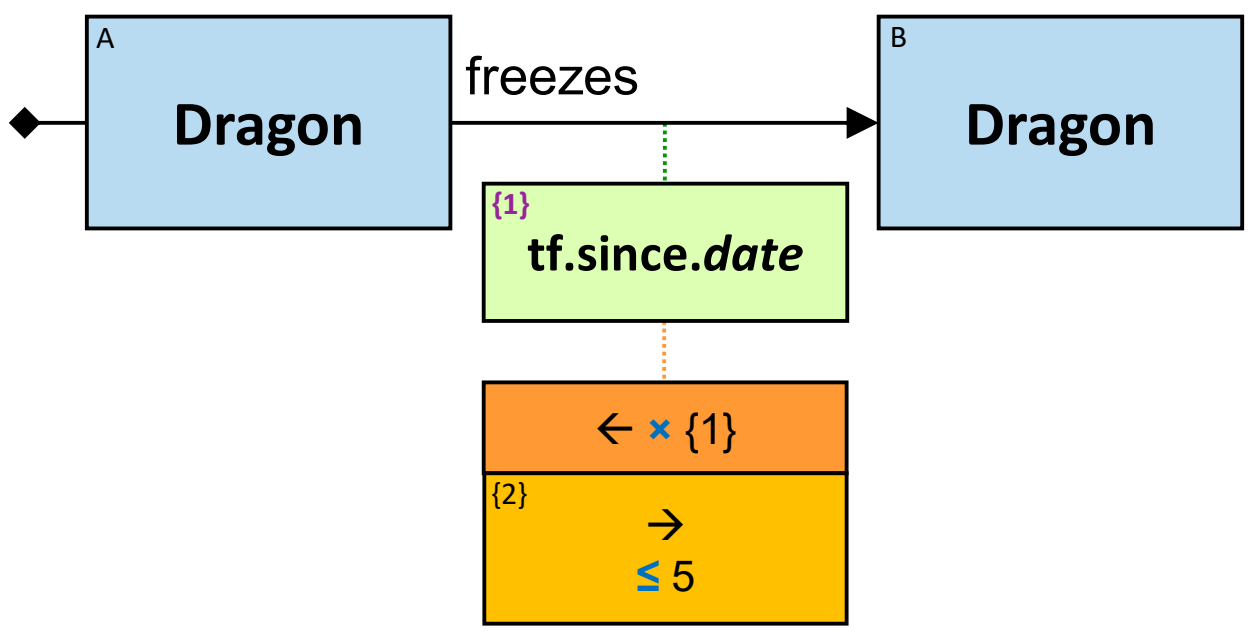

***Q114:** Any person who owns at least five horses of the same color* (version 2)

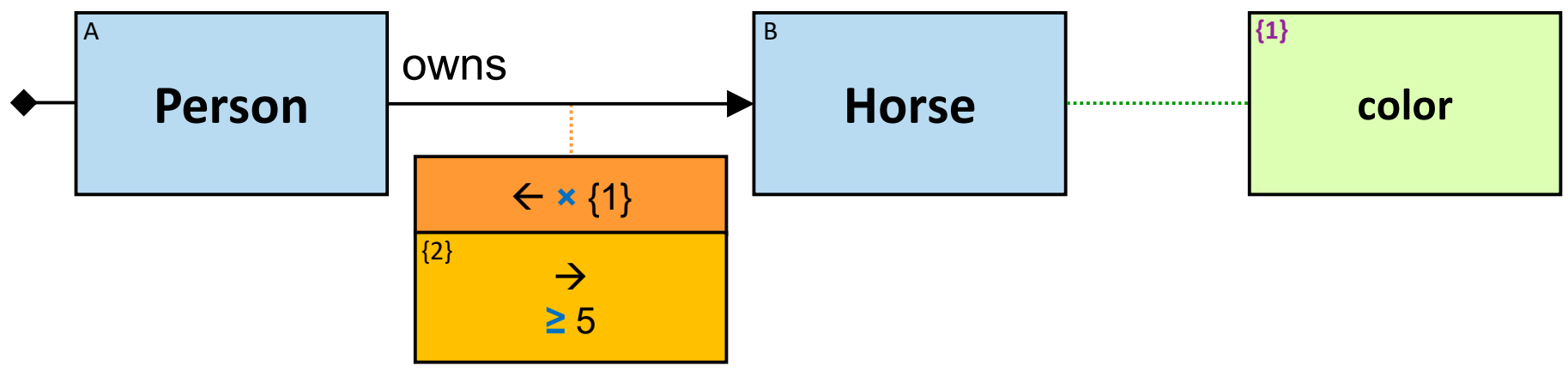

***Q218:*** *Any person who owns at least five entities of the same type* (version 2)

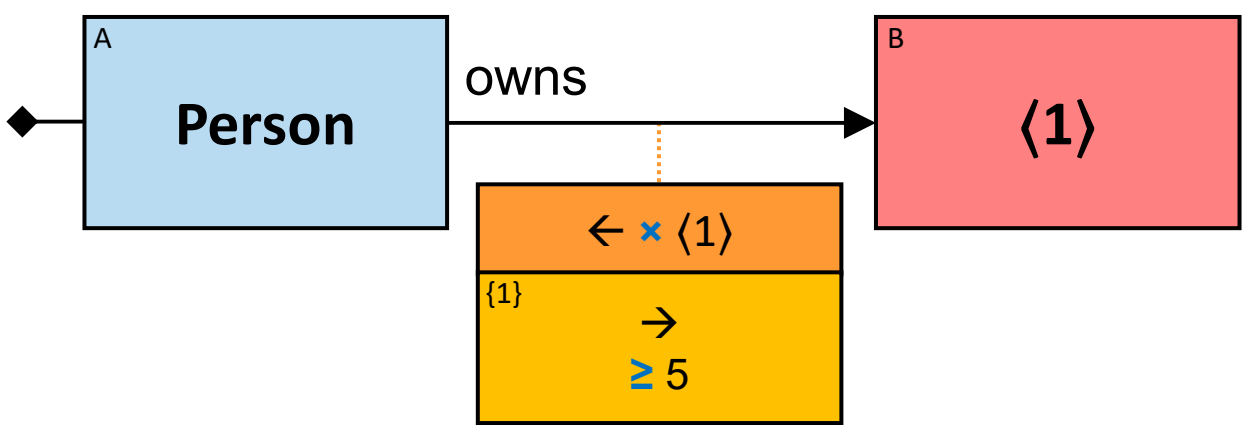

***Q157:*** *Any person who owns between one and three horses of the same color – for at least six colors*

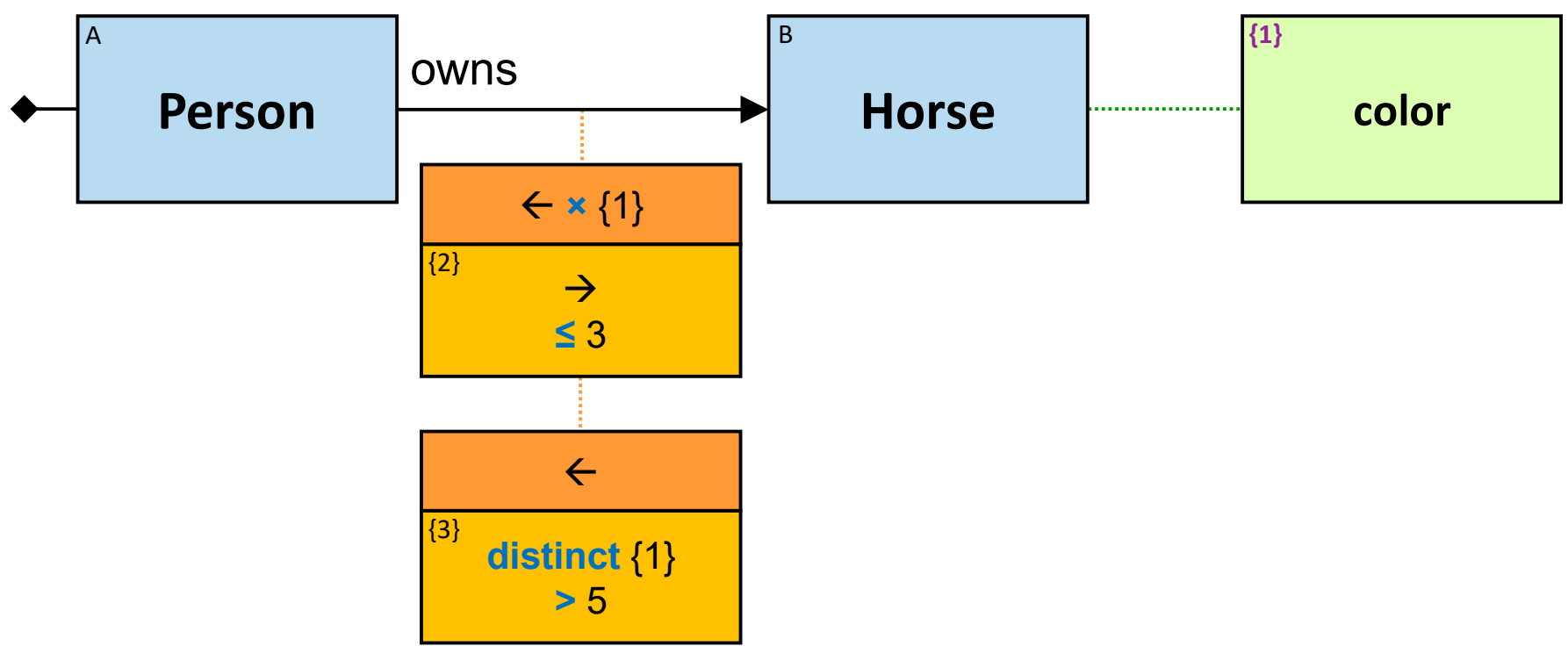

***Q214:*** *Any person who owns entities of at least four types. For each type – at least five entities*

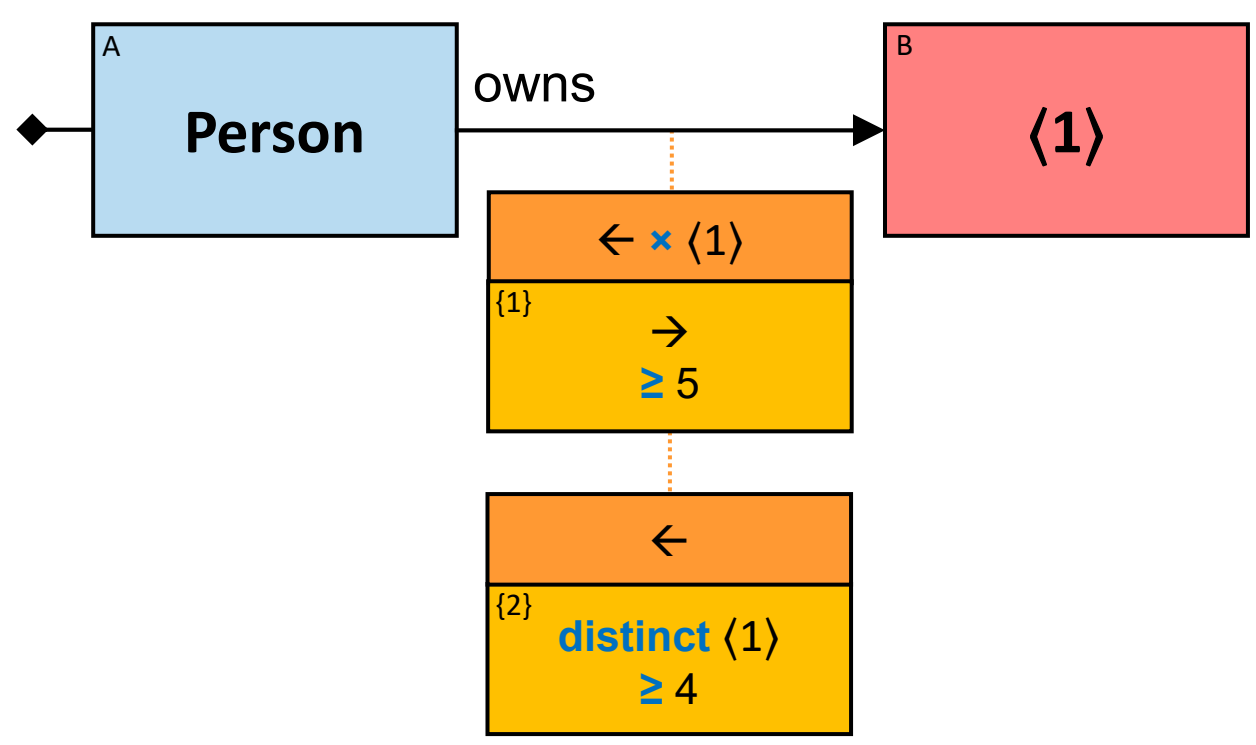

***Q235:*** *Any dragon that the number of dragons it froze on average – on months it froze dragons in – is at least ten*

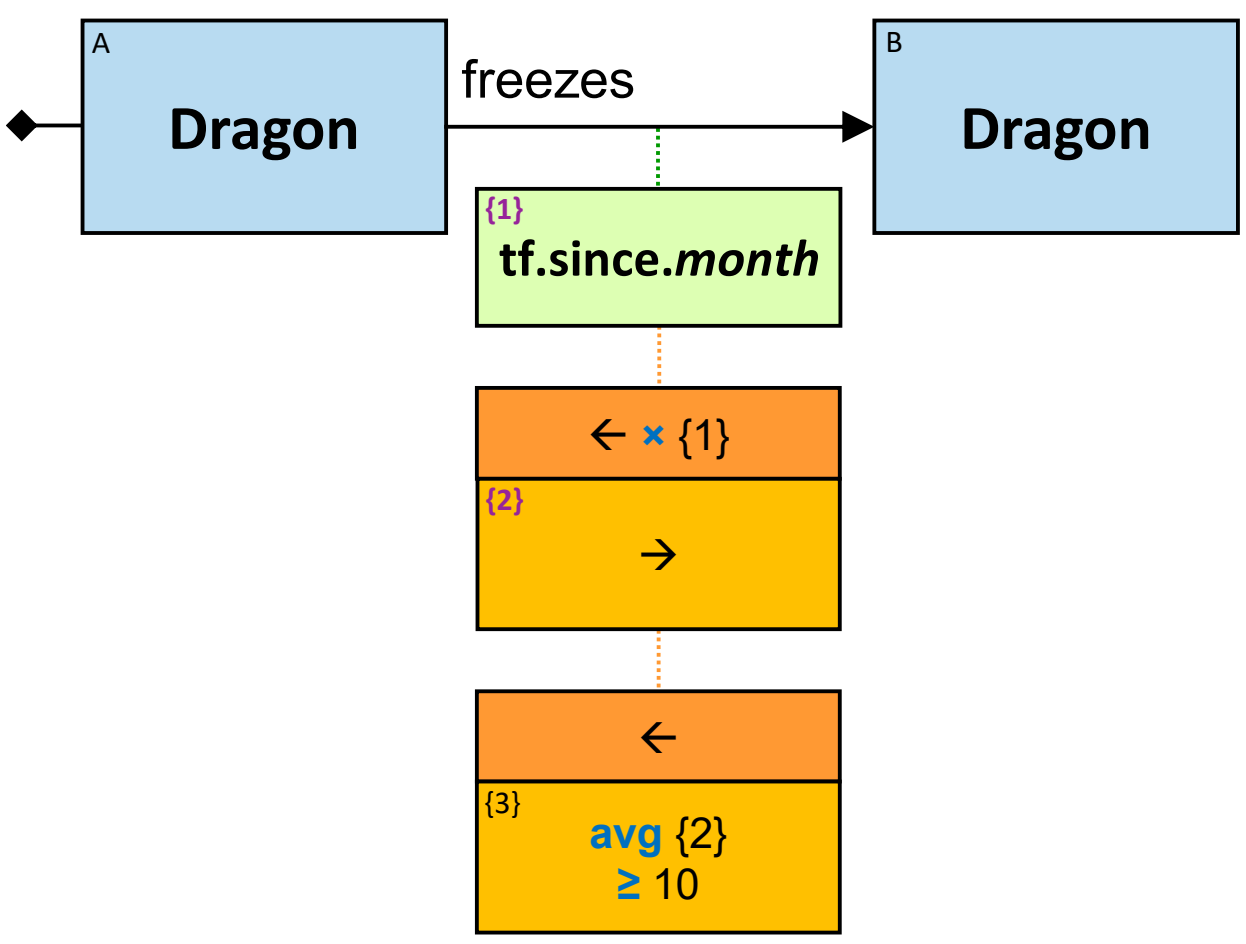

***Q153:*** *Any dragon A that in each of at least 11 days – froze between one and five dragons*

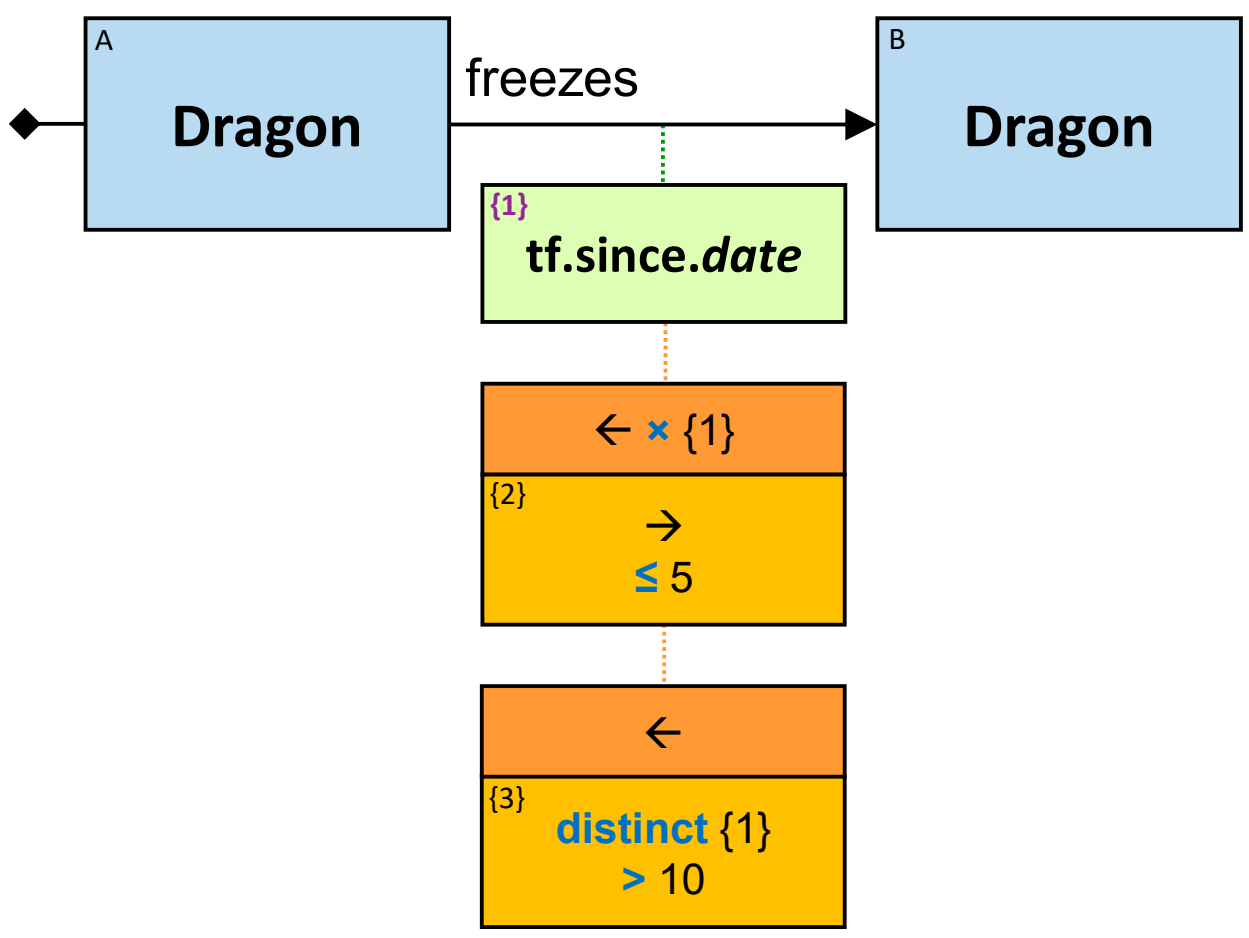

***Q252:*** *Any dragon B that in each of at least 11 days – was frozen by a dragon that on that day froze between one and five dragons*

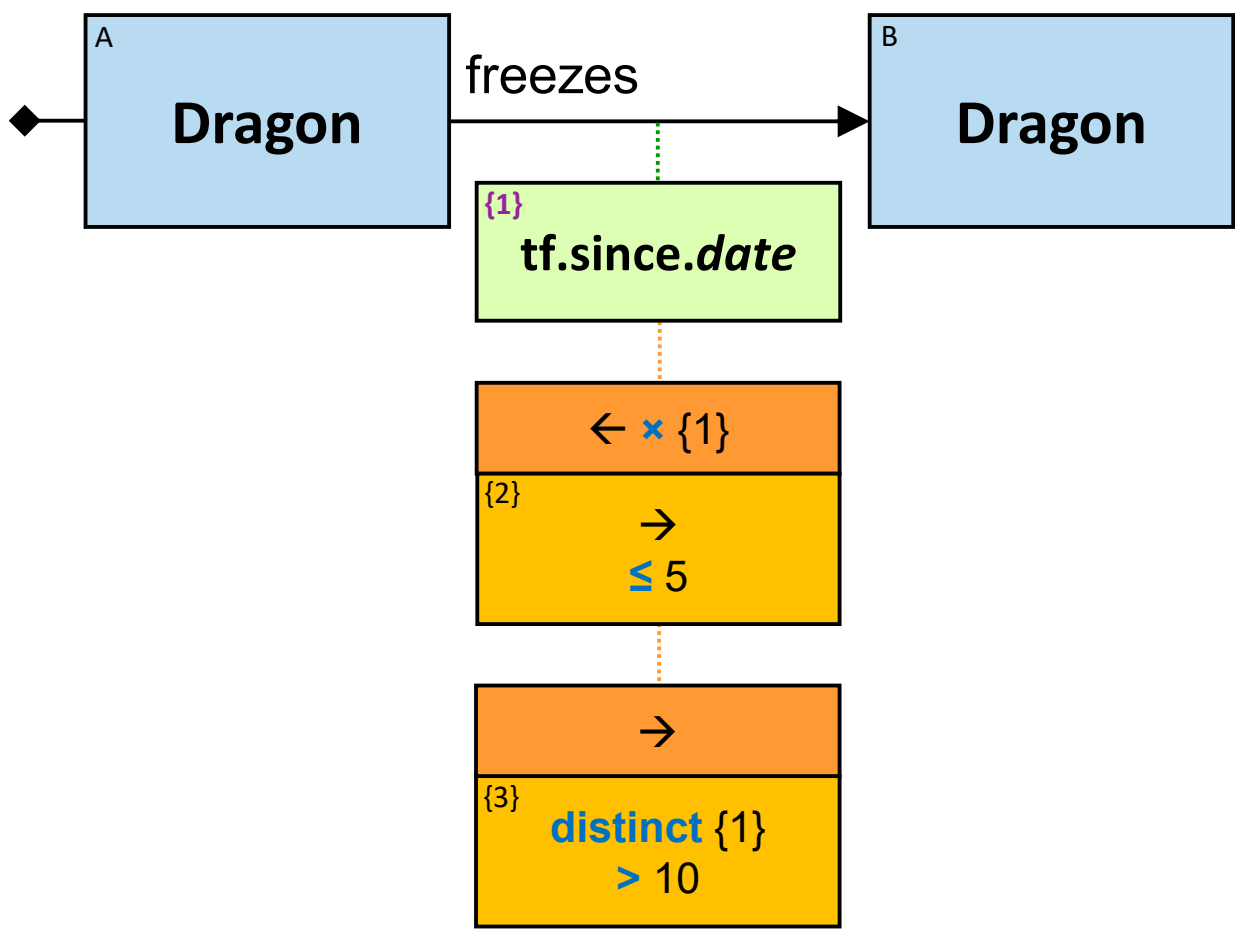

***Q255:*** *Any dragon whose name-length is equal to the number of days in each of which it froze ten dragons*

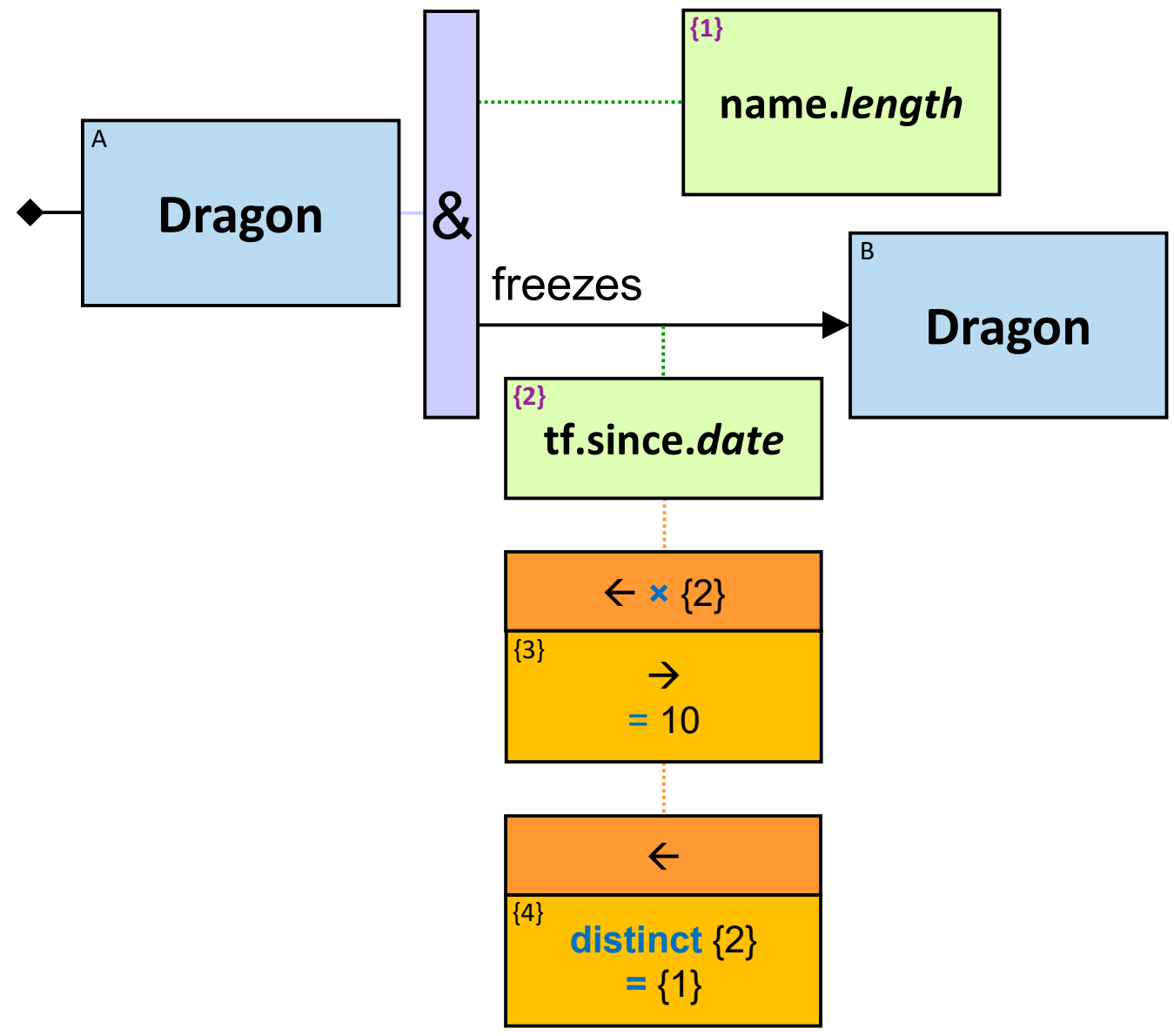

***Q278:*** *Any pair of dragons (A, B) that in each of at least 11 days – A froze between one and five dragons, one of them is B*

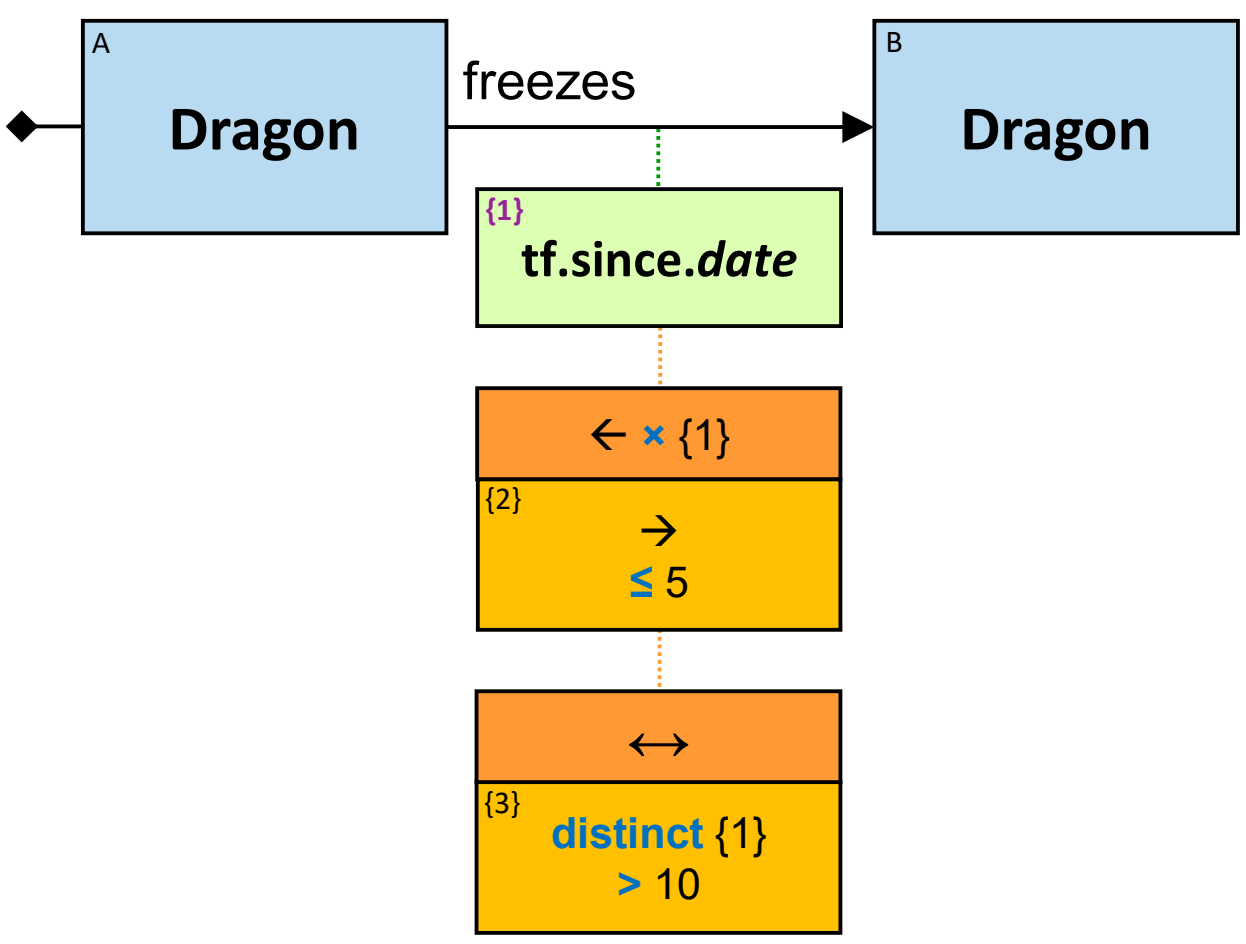

***Q254:*** *Any dragon that in each year it froze dragons – it froze more than three dragons*

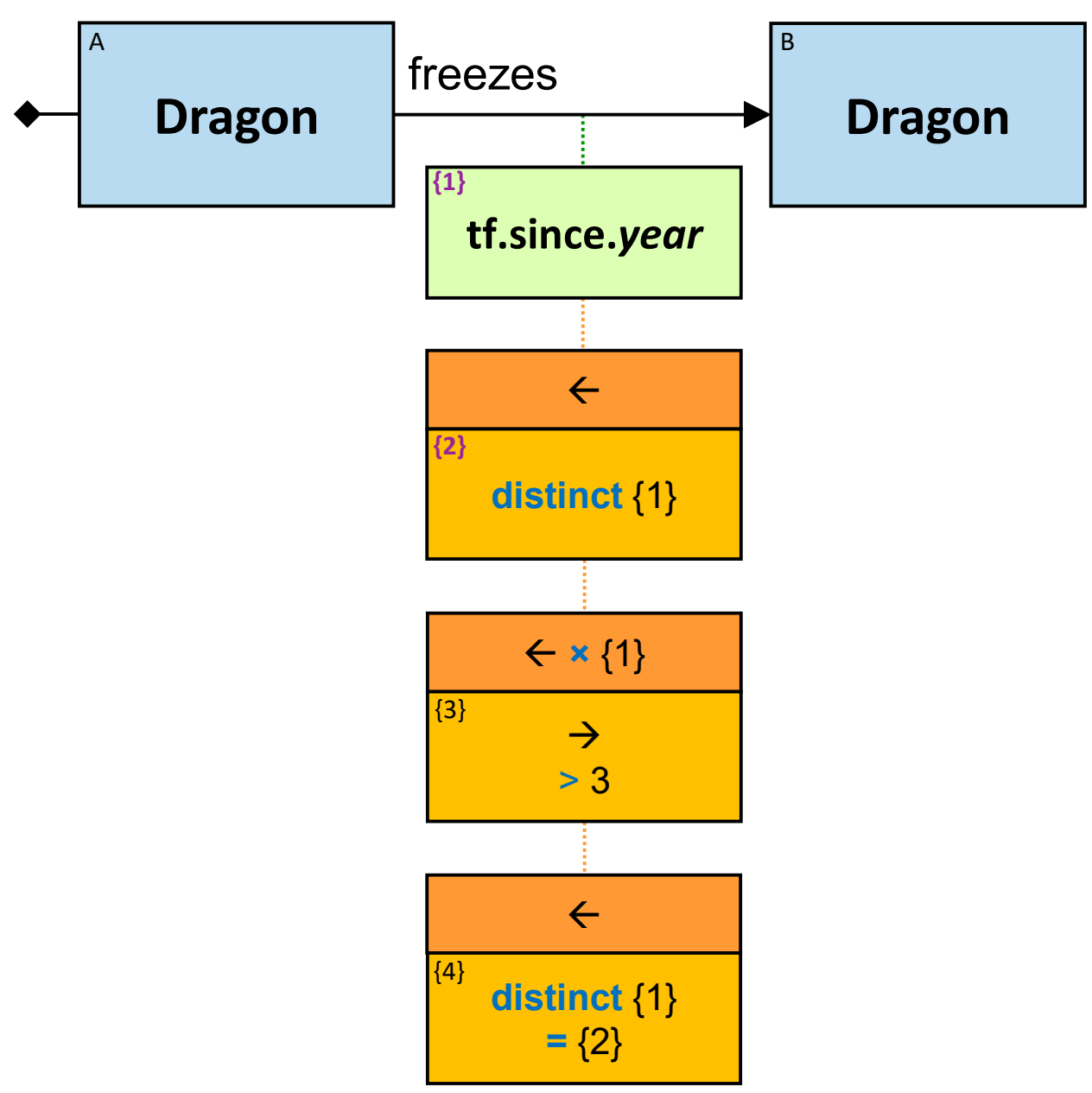

***Q306:*** *The three dragons that the number of days in each of which each froze at least five dragons – is maximal*

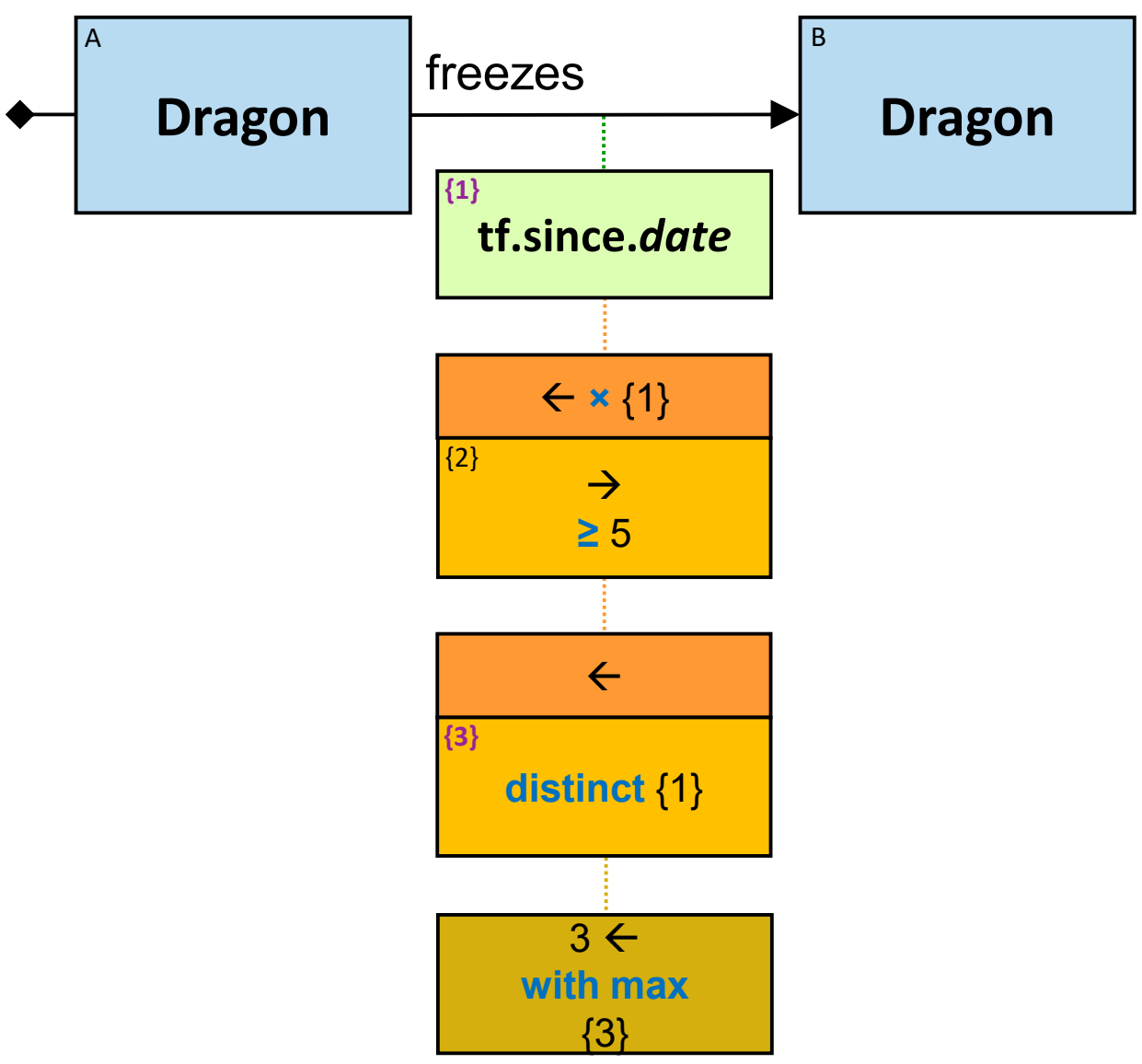

***Q285:*** Any person who at a certain day owned at least five horses and at least five dragons

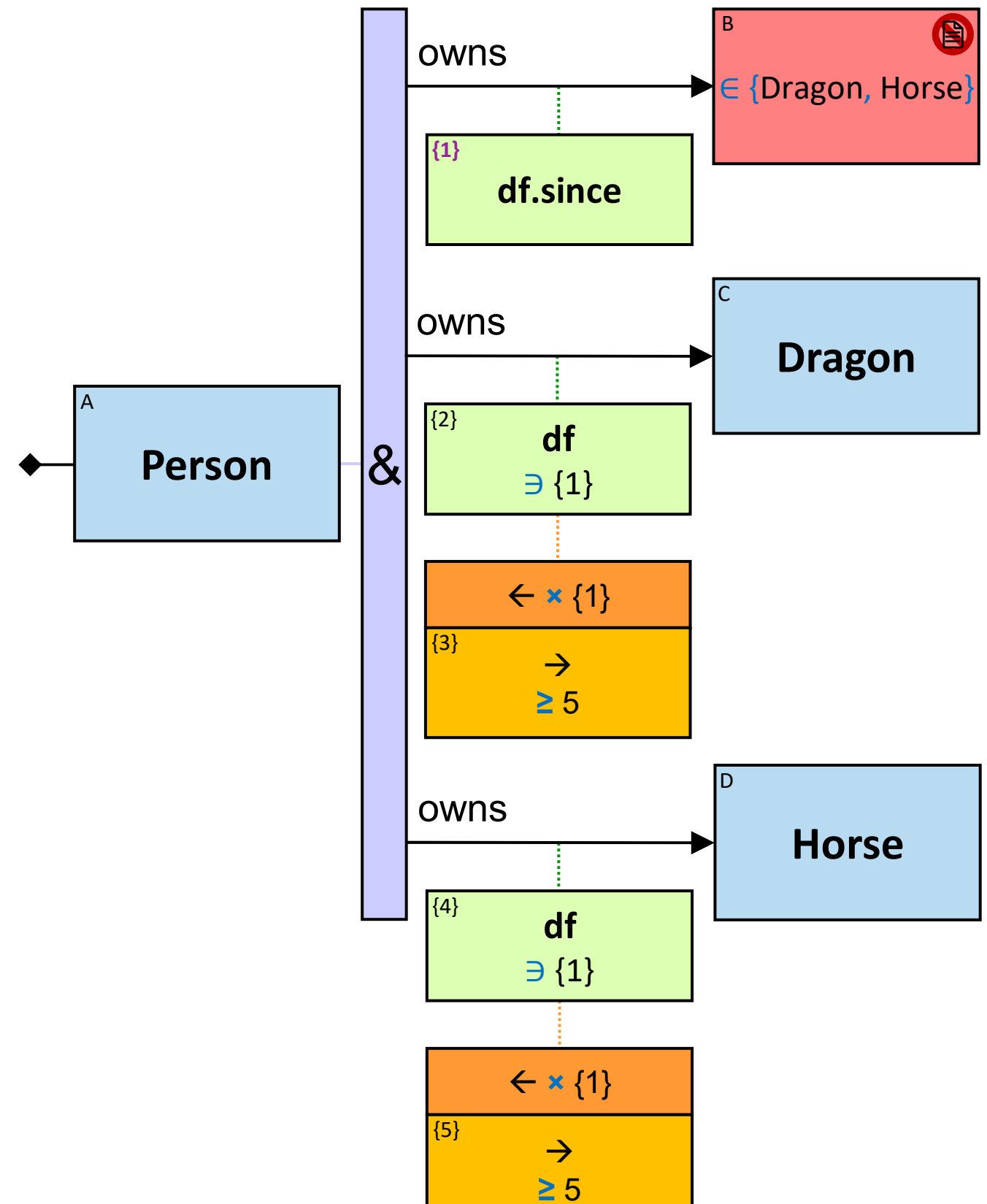

*Q286: Any person who at each day of at least ten (not necessarily consecutive) days – owned at least five horses (not necessarily the same horses each day)*

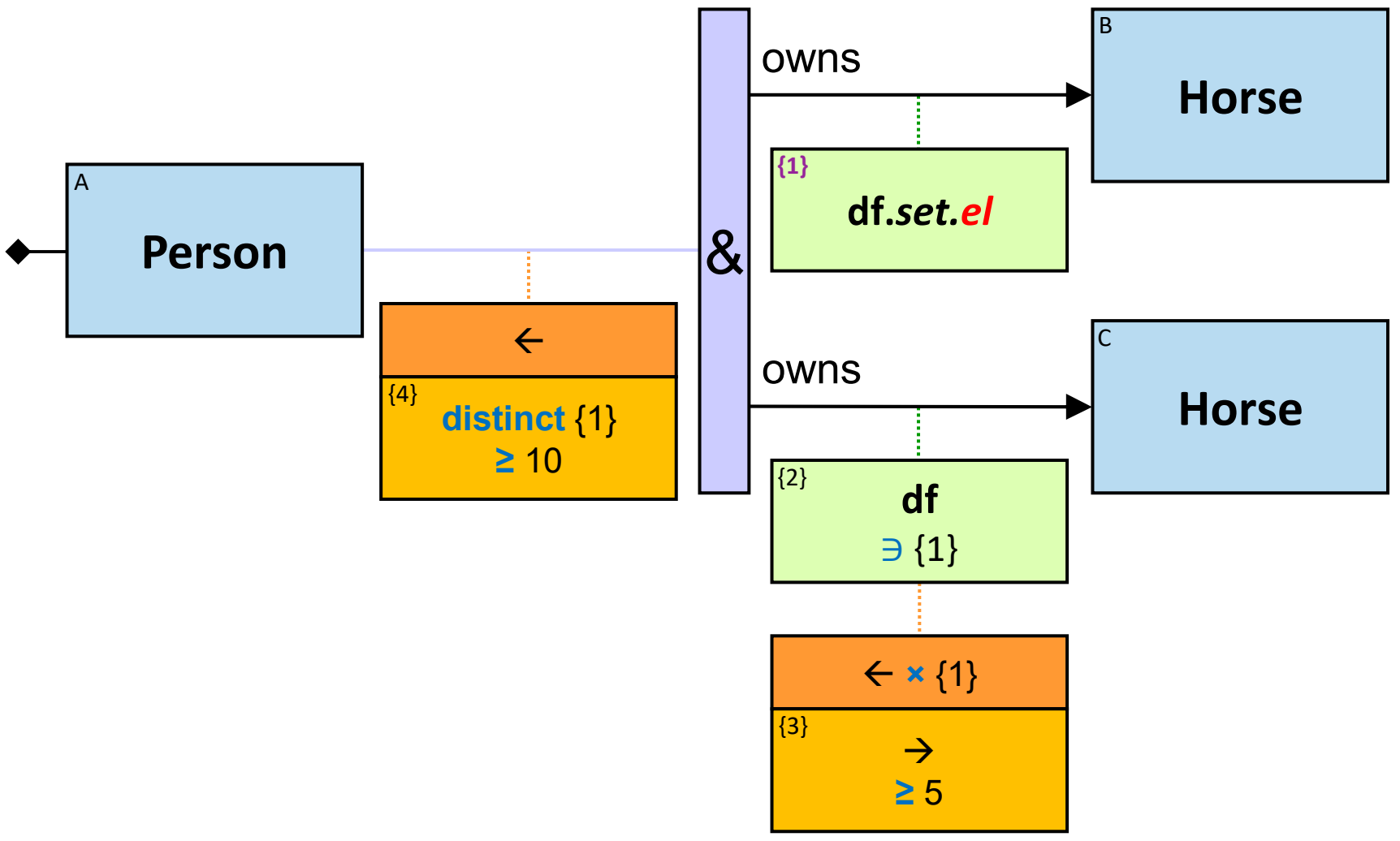

*Q158: Any dragon that in each of at least ten days – the number of dragons it froze is greater than the number of dragons that froze it*

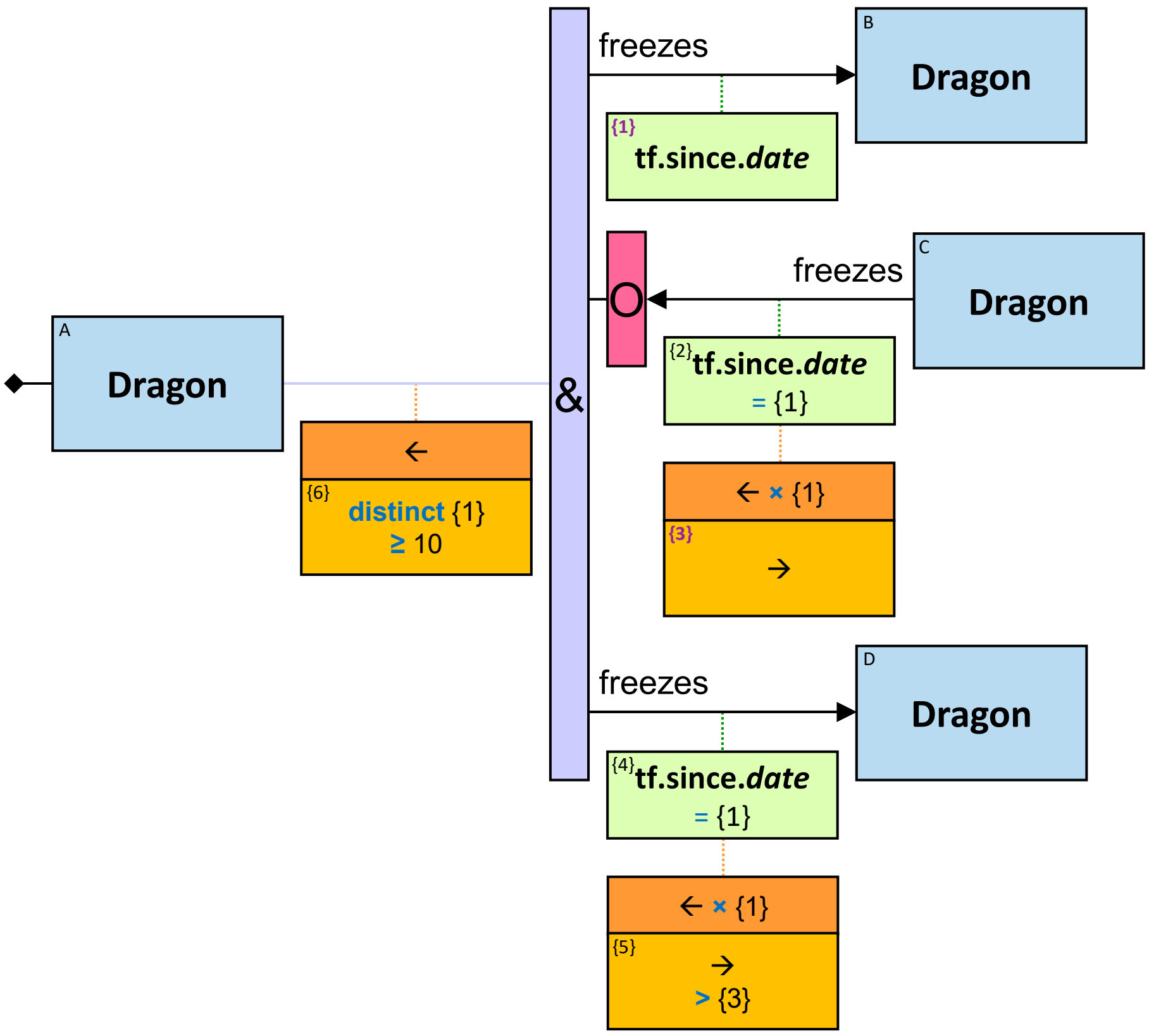

Note the location of {6}. If it was below {1}, the evaluation order was different. As it is, {6} counts distinct assignments to {1} for which the *All* quantifier was satisfied.

***Q260:*** *Any dragon that in each of at least ten days: (i) the number of dragons it froze is greater than the number of dragons that froze it, and (ii) it froze/was frozen at least five times*

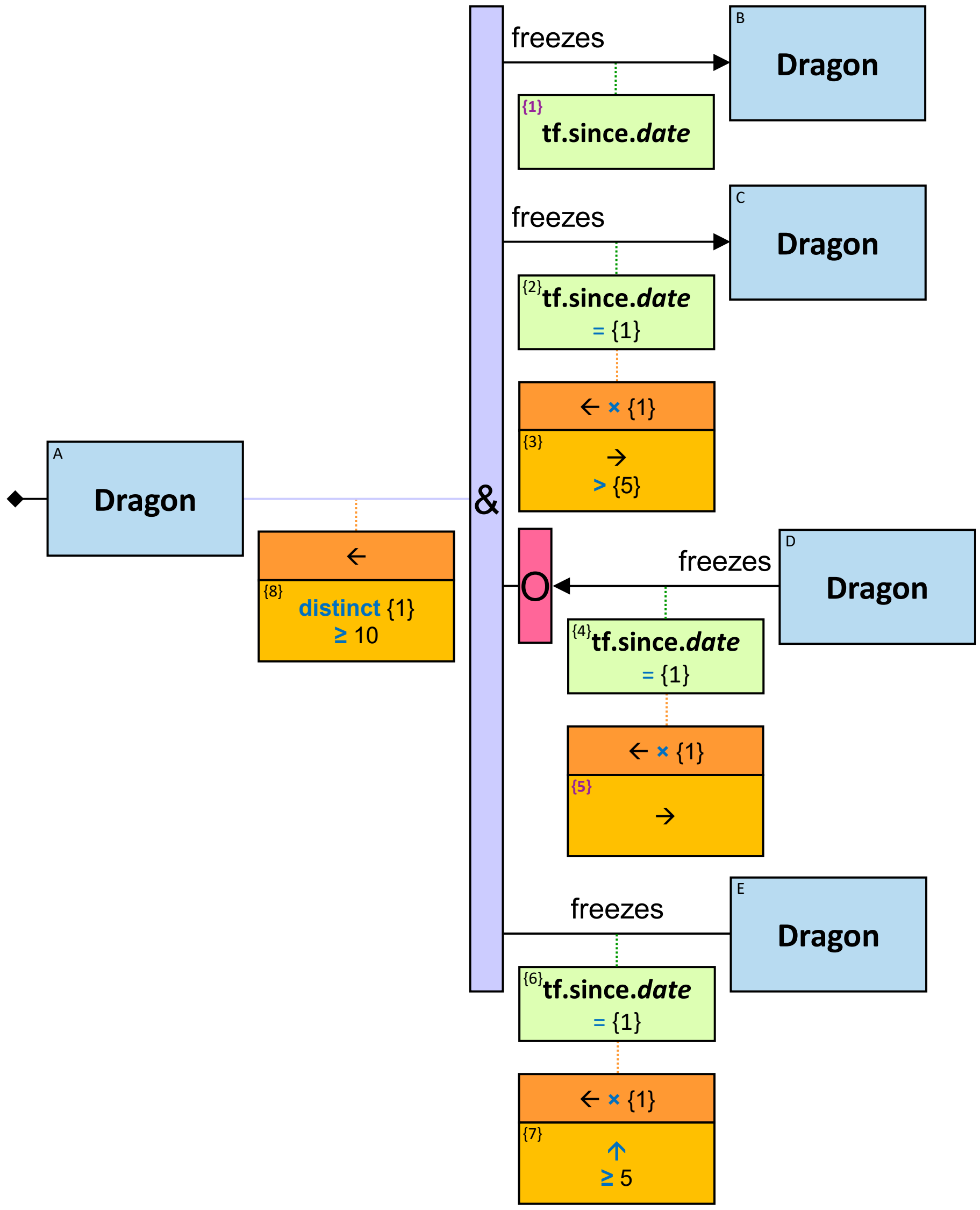

*Q350: Any pair of people who own the same number of entities of each type*

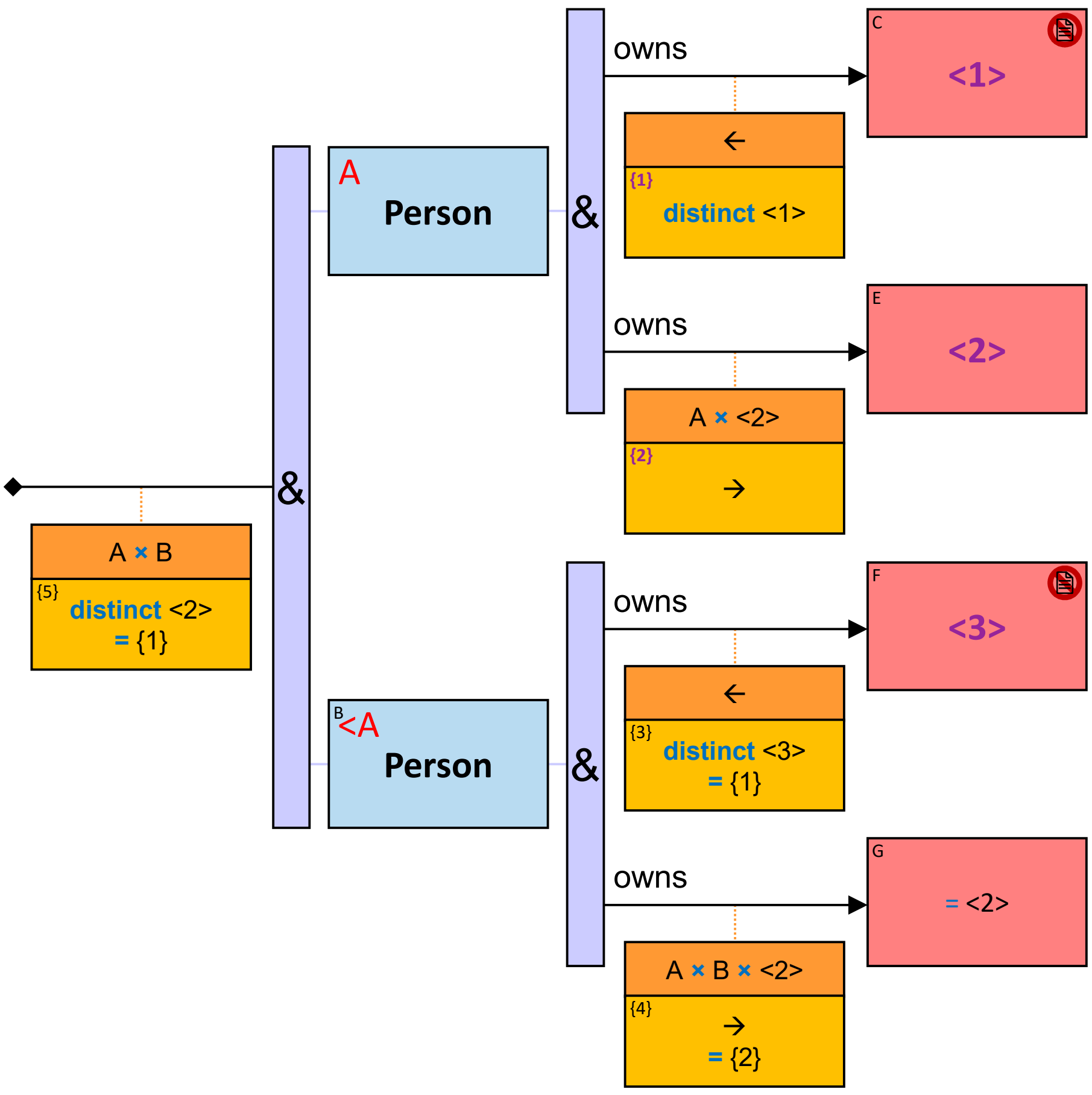

*Q332: Any person whose horses are all of rare colors. A rare color is a color of less than 1% of the horses* (two versions)

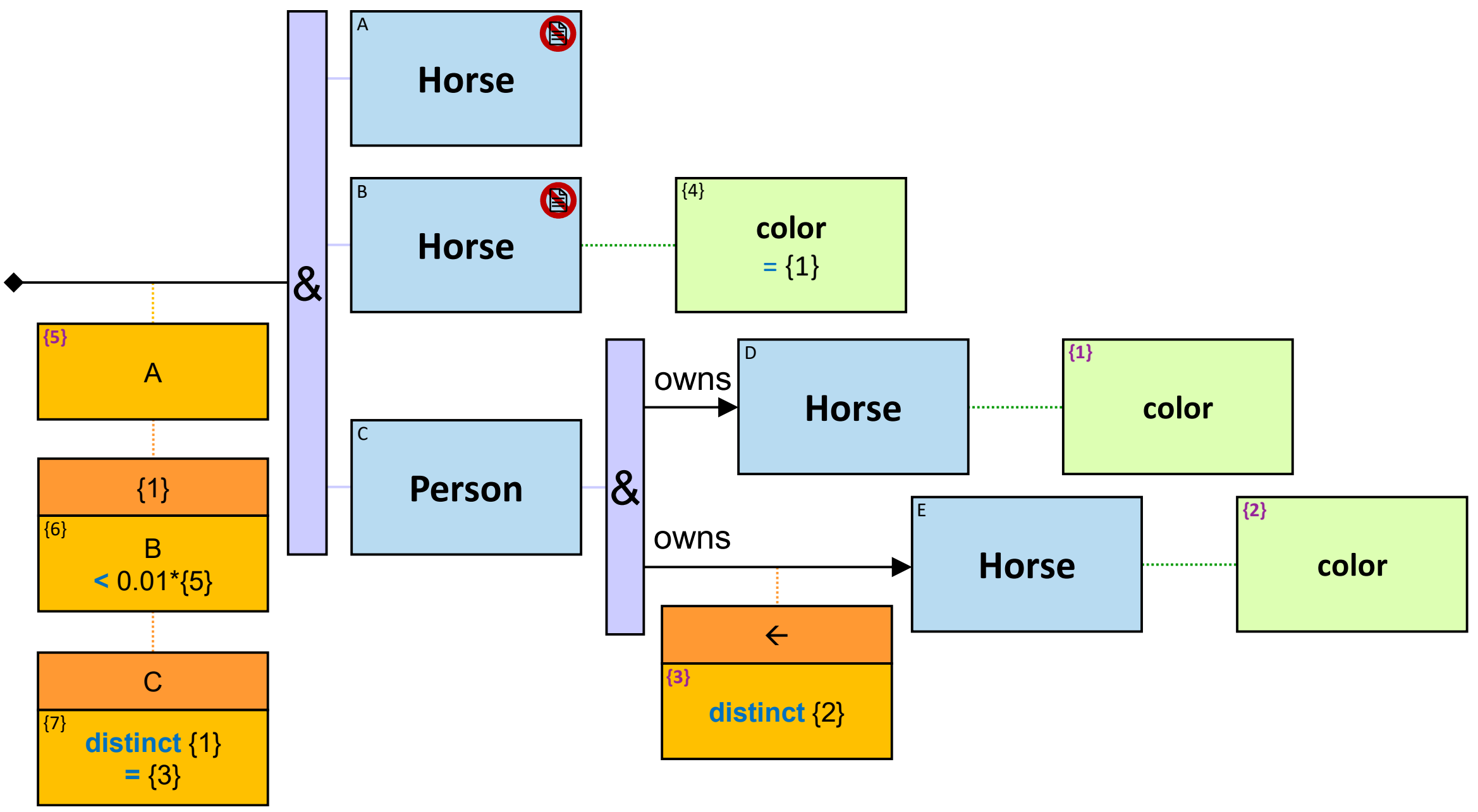

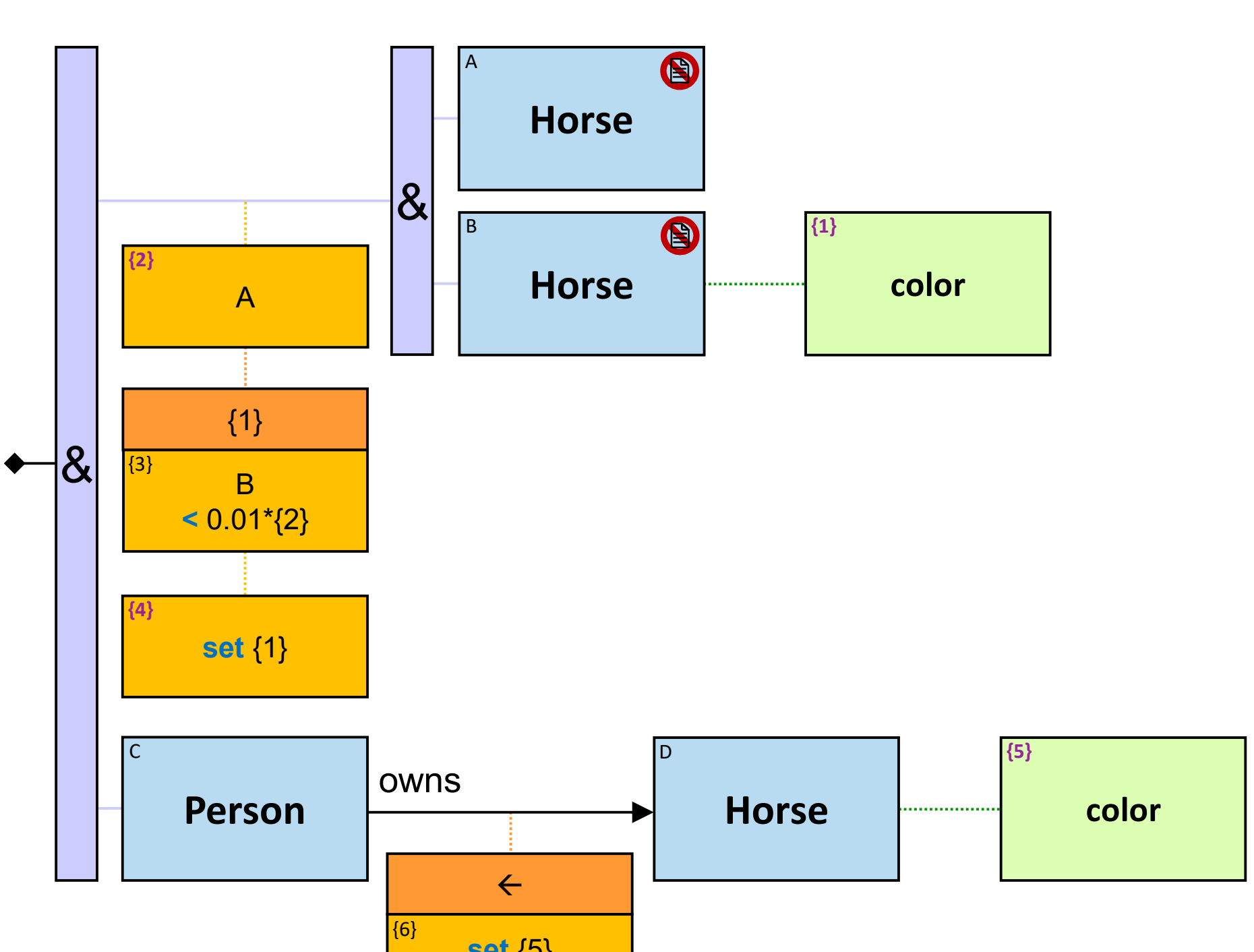

*Q338: Any person who owns at least ten horses, at least half of which are of rare colors. A rare color is a color of less than 1% of the horses* (two versions)

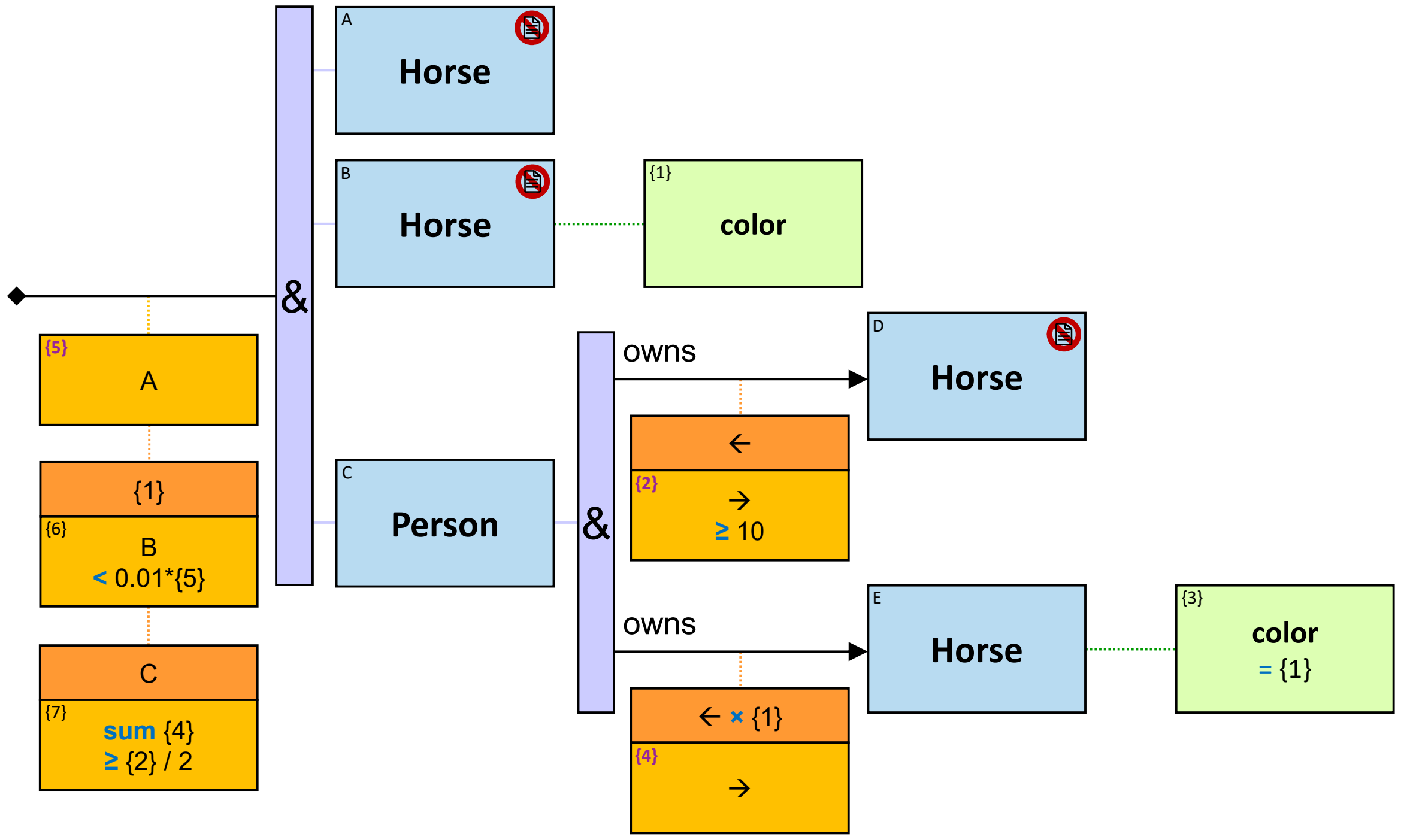

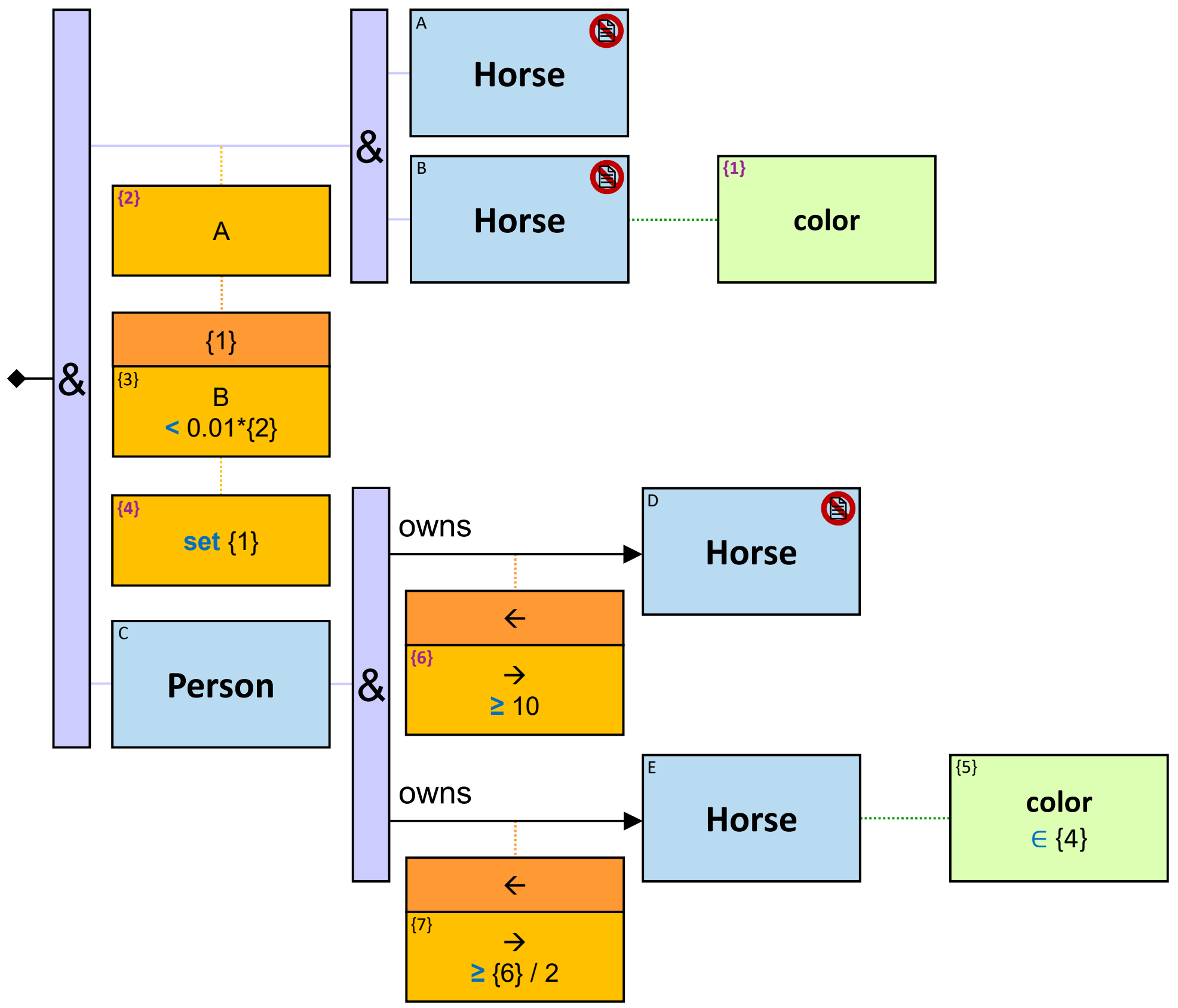

## 38. EXTENDED A2 AGGREGATOR

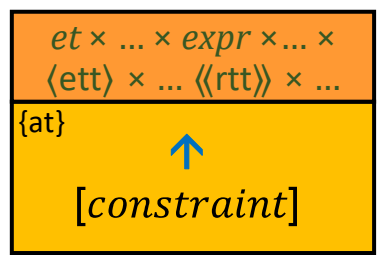

***Q330:*** *Any day in which more than five horse-ownerships started*

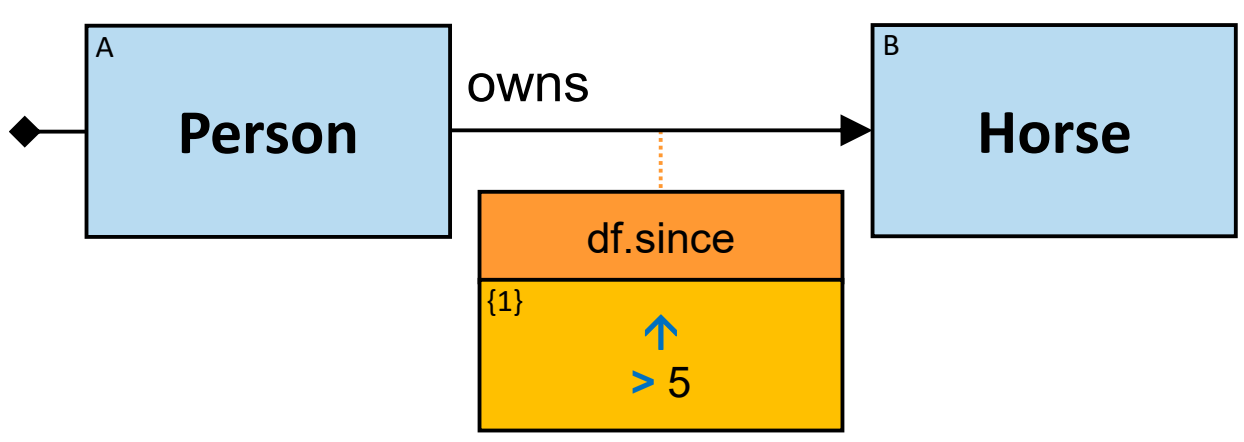

***Q270:*** *Any person A and A's horses of the colors for which there are more than five horse-ownerships by a person*

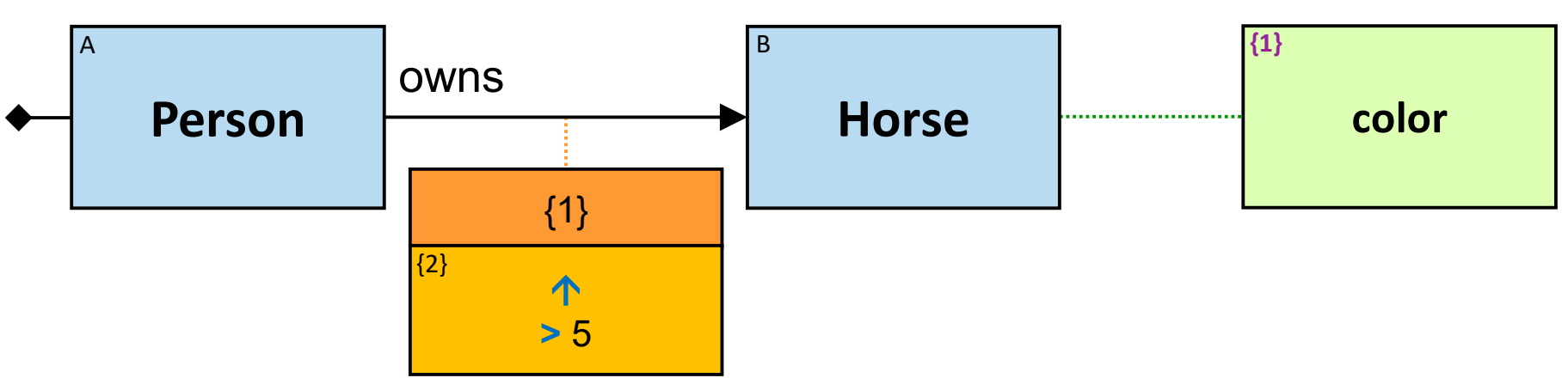

## 39. EXTENDED A3 AGGREGATOR

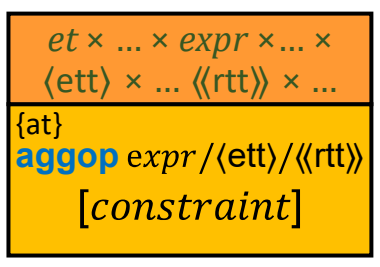

***Q331:*** *Any day in which the horse ownerships that started lasted on average for at least ten years*

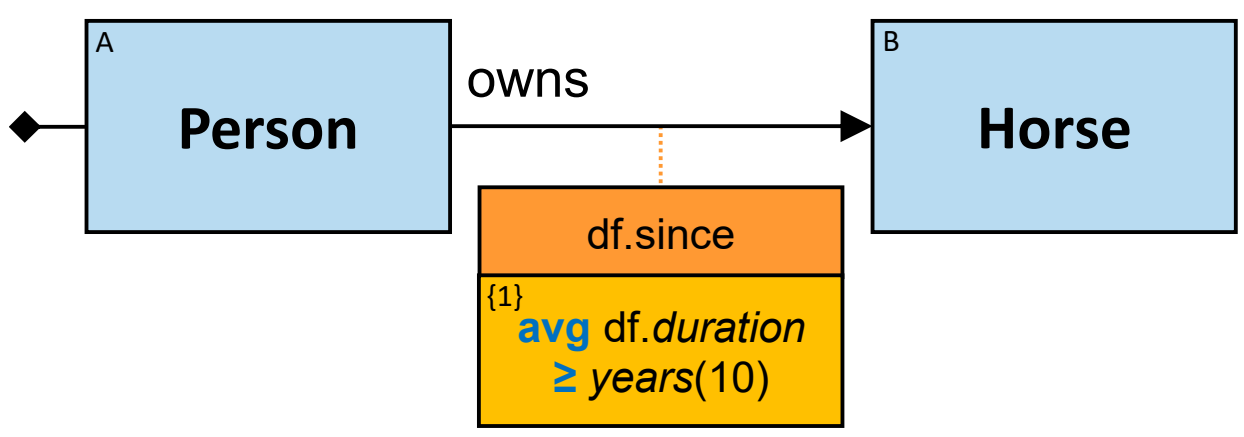

***Q271:*** *Any person A and A's horses – of the horse colors for which the average ownership start date is at least January 1, 1010*

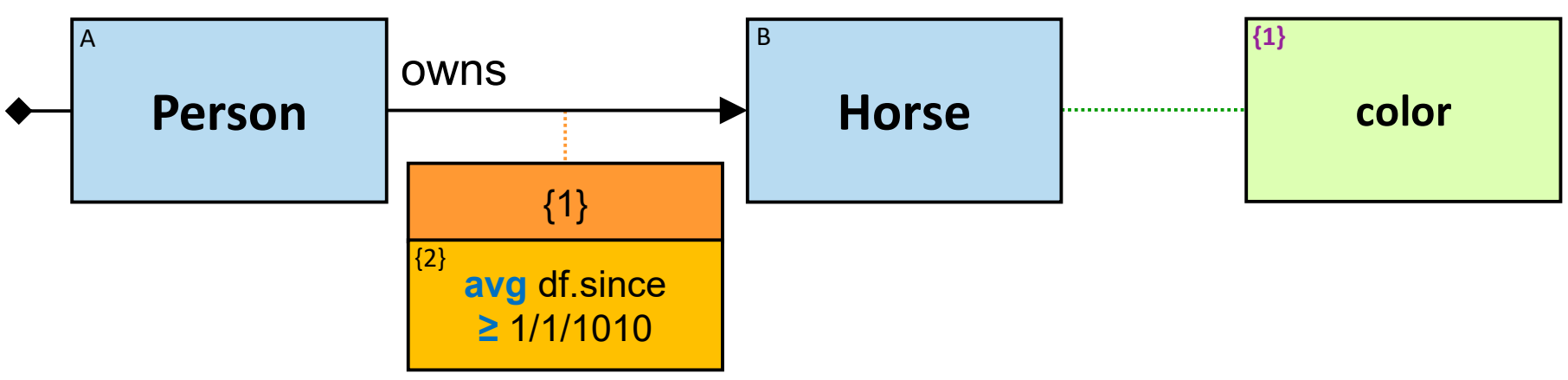

***Q263:*** *Any horse of each color of which the average horses' weight is greater than 450 Kg*

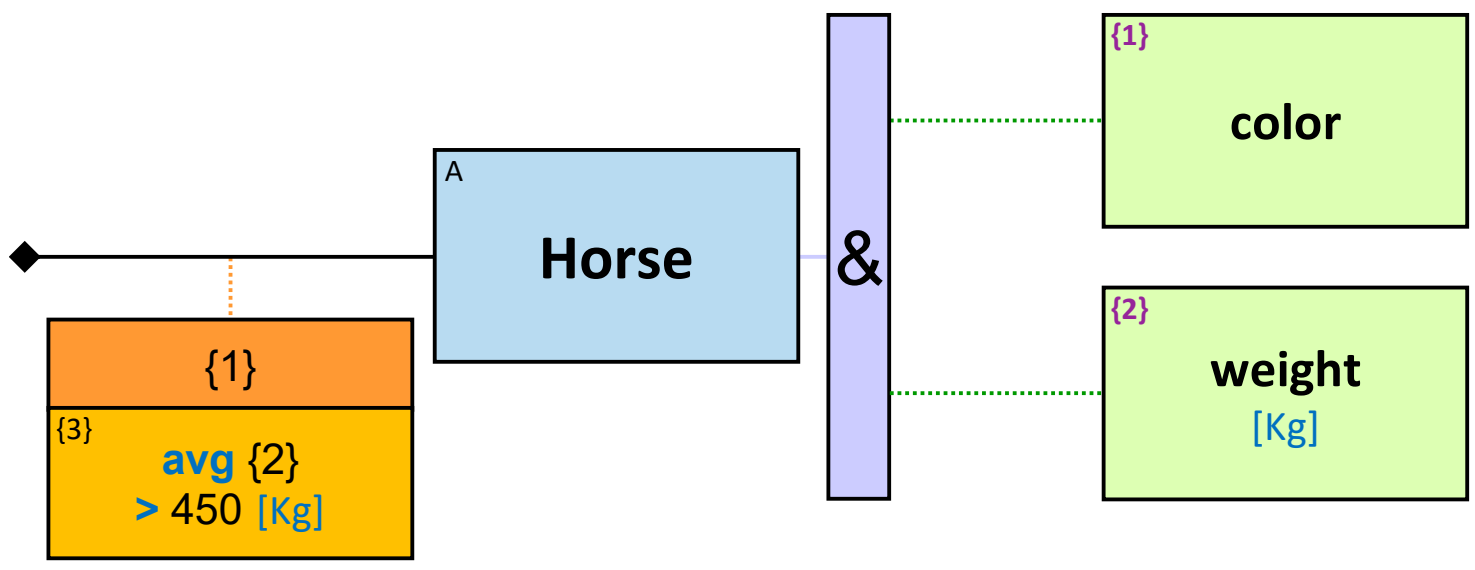

***Q265:*** *Any horse of each color of which the average horses' owners' height is at least 180 cm*

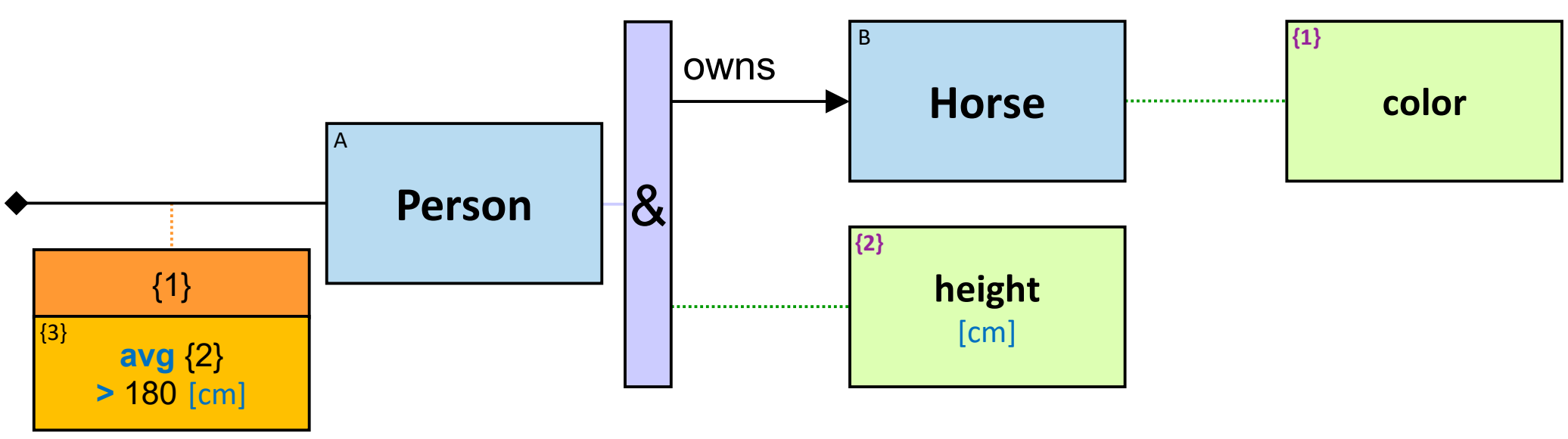

***Q155:*** *Any dragon that in each of at least 11 days – froze dragons for more than 100 minutes cumulatively*

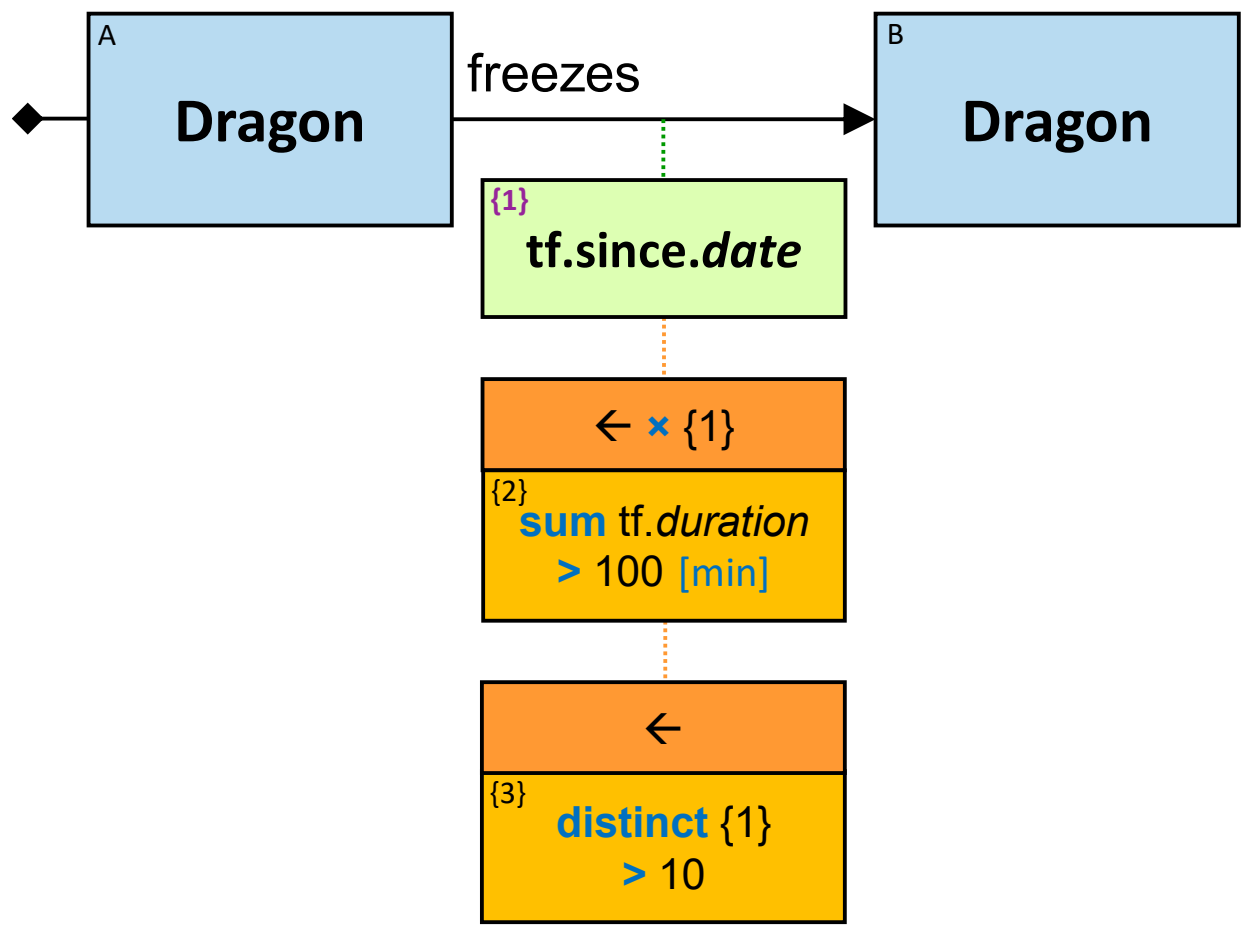

***Q156:*** *Any dragon that froze dragon for more than 1000 minutes cumulatively – in the days it froze dragons for more than 100 minutes*

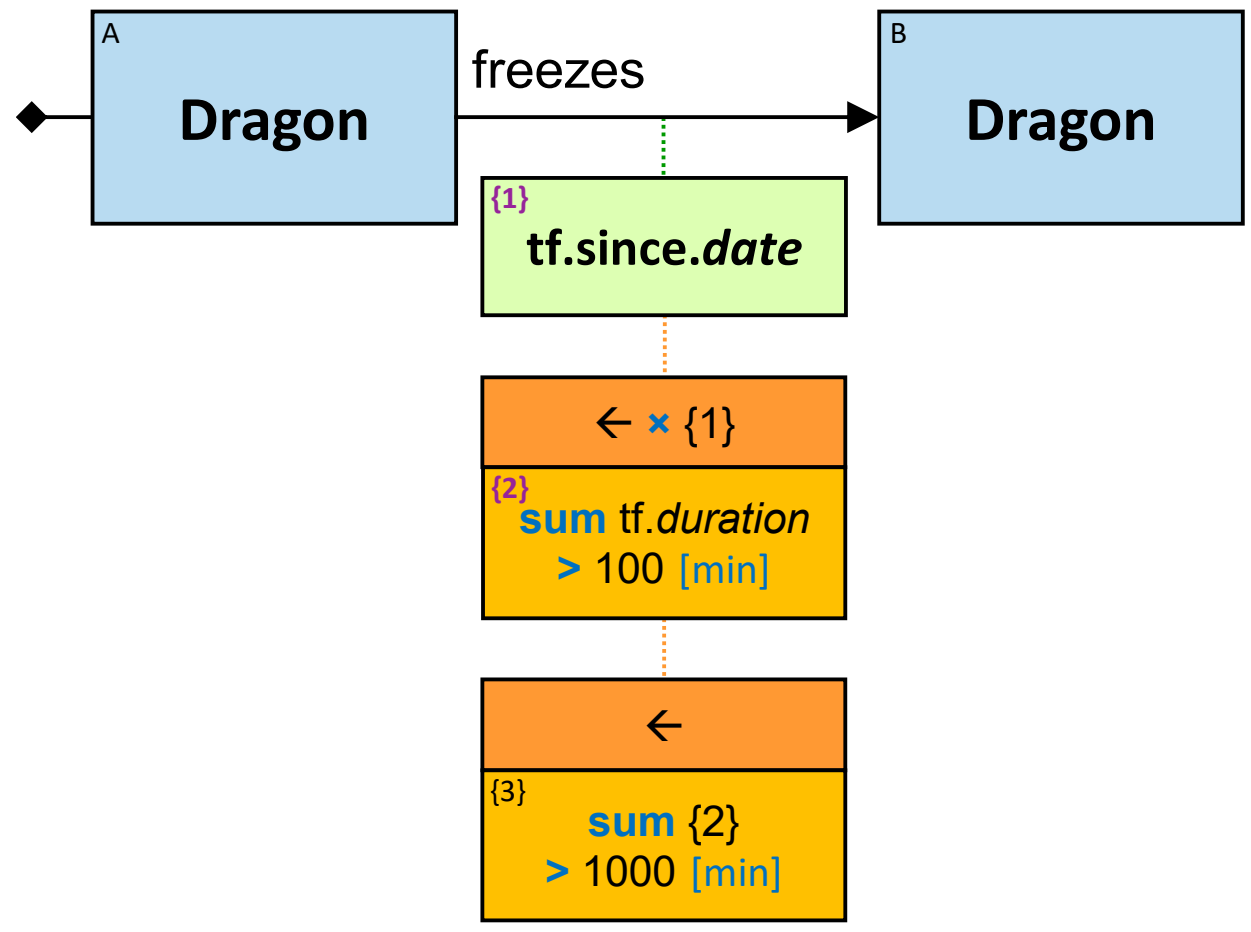

*Q154:* *Any dragon that in each of at least four years – in each of at least 11 days – froze between one and five dragons*

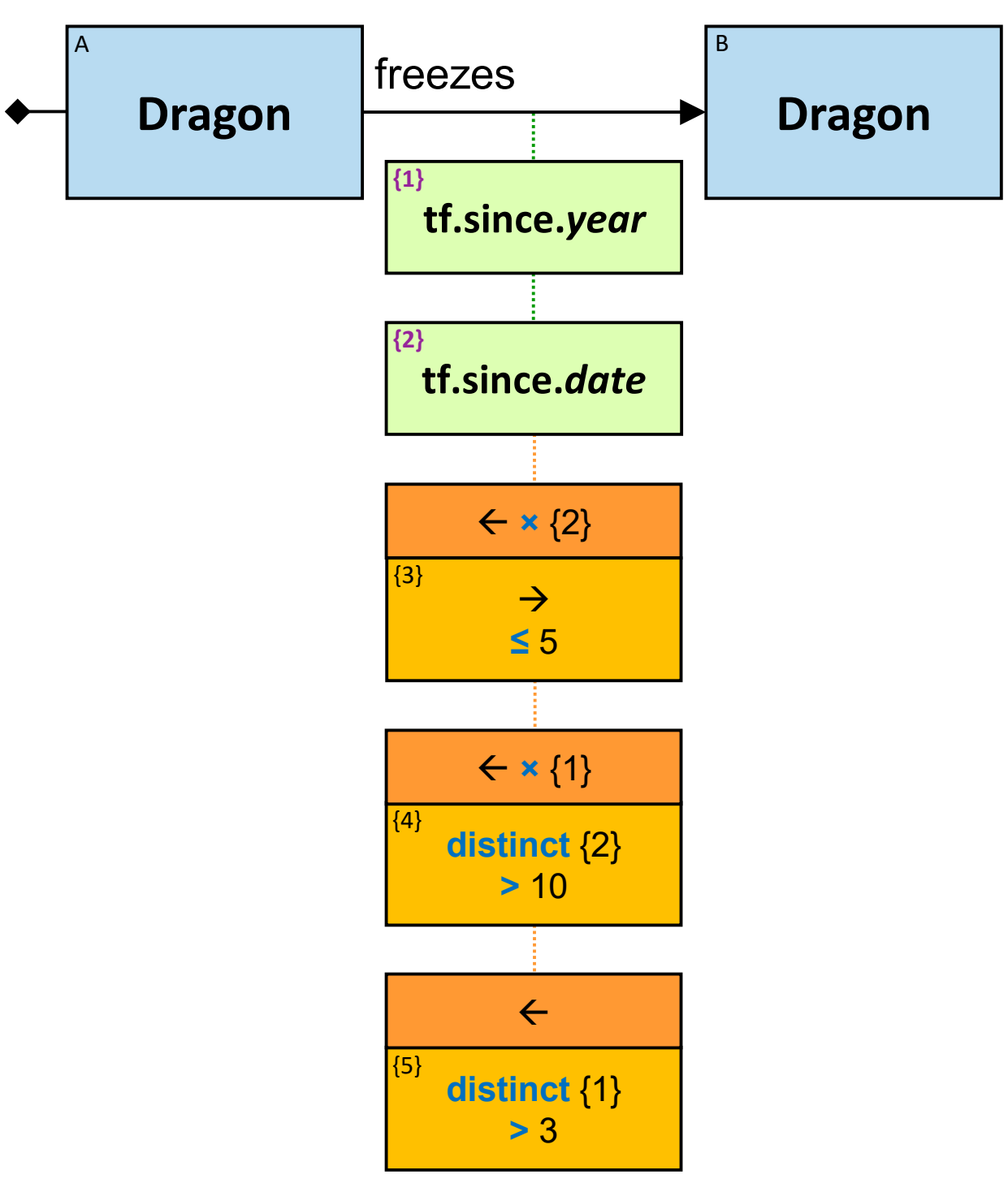

*Q159:* *Any* ***dragon*** *for which there are more days where (it froze/was frozen at least five times, and the number of dragons* ***it*** *froze is greater than the number of dragons that froze* ***it****) than days where (it froze/was frozen at least five times, and the number of dragons that froze* ***it*** *is greater than the number of dragons* ***it*** *froze)*

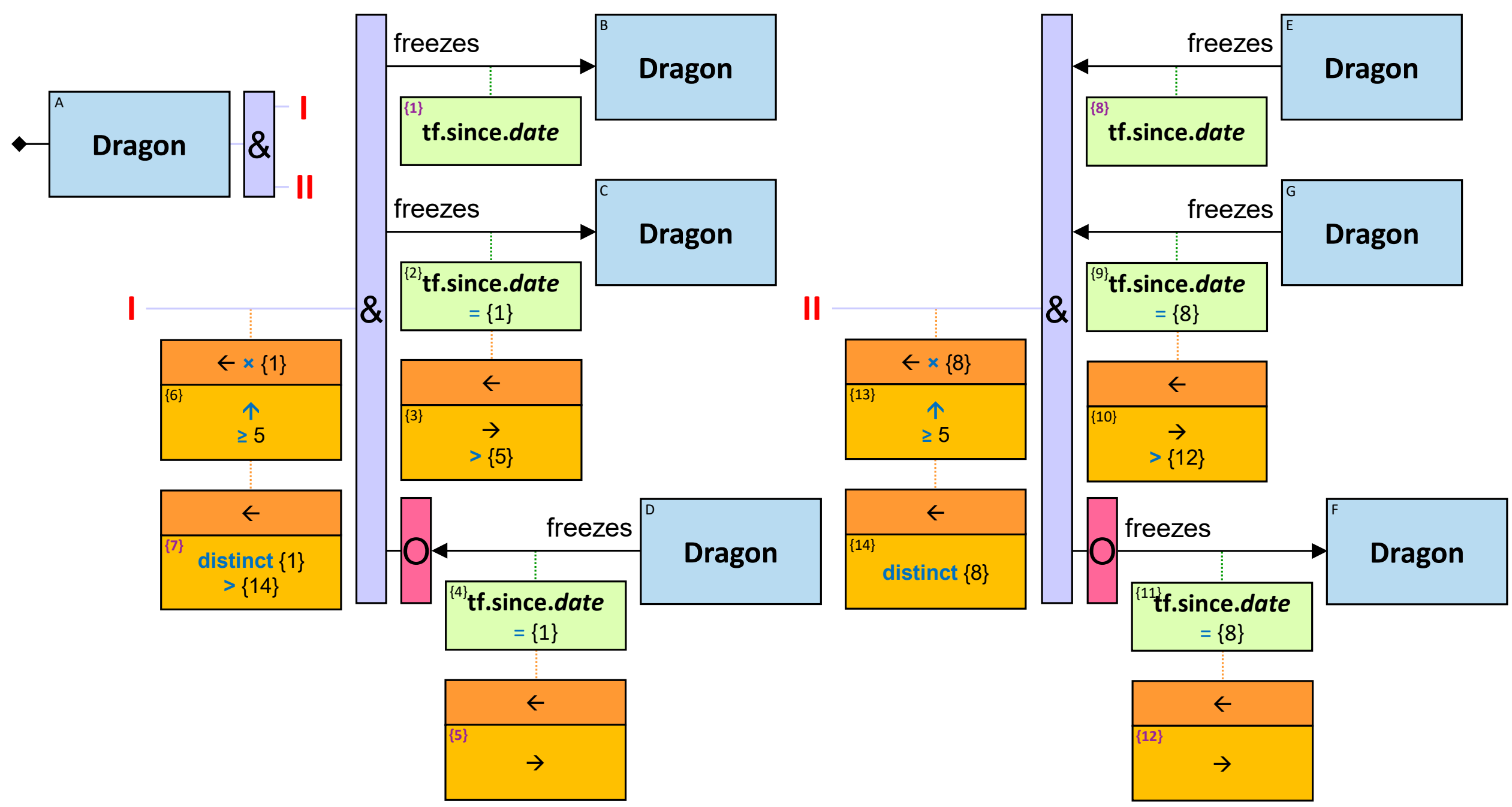

(Compare with Q260)

***Q361:*** *Any person A who owns horses of at least five colors, and the sets of weights of A's horses of each of these colors are identical*

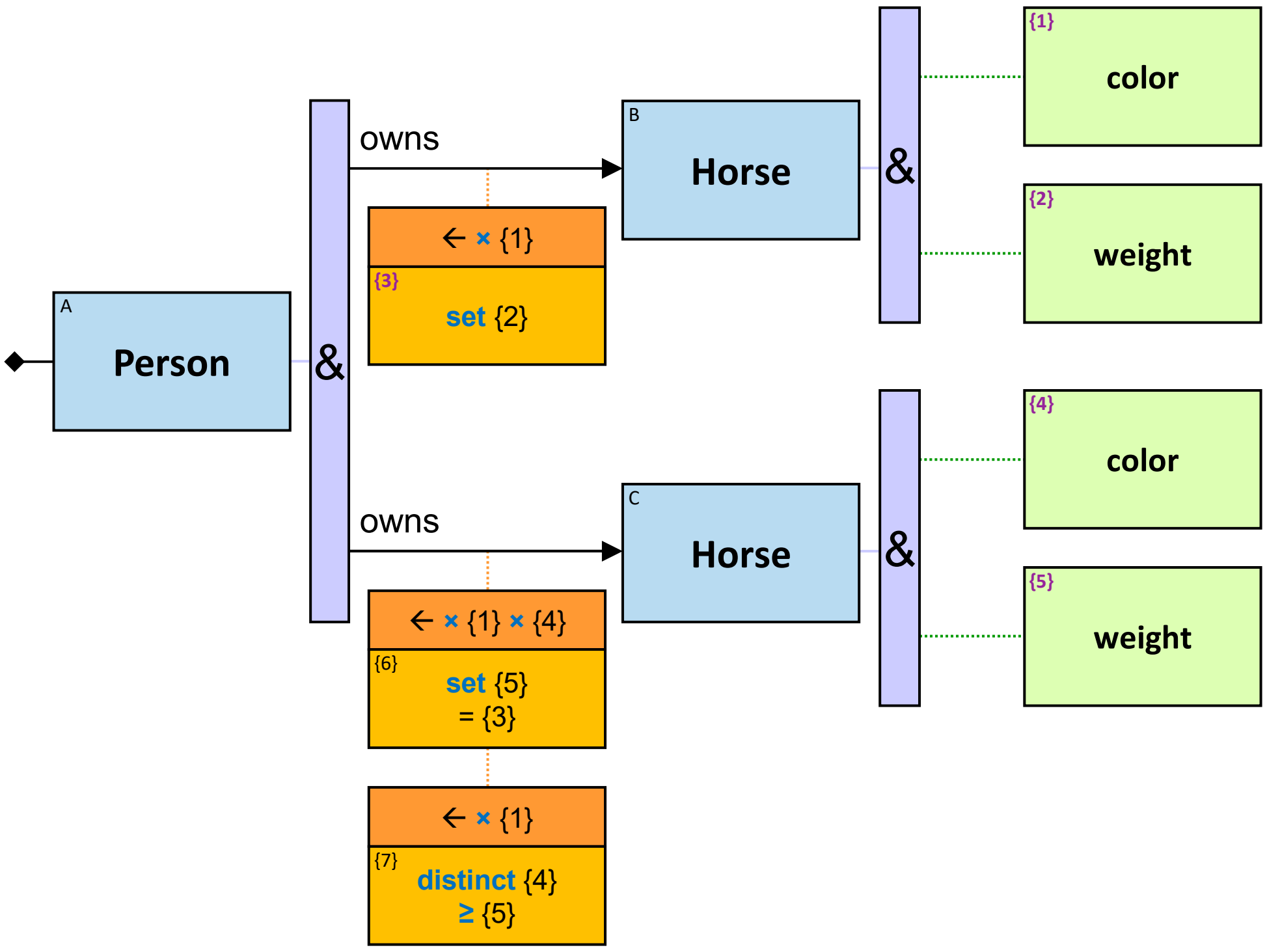

***Q320:*** *Any person A where there are more horses of the same colors as A's owned horses than dragons of the same colors as A's owned dragons* (version 2)

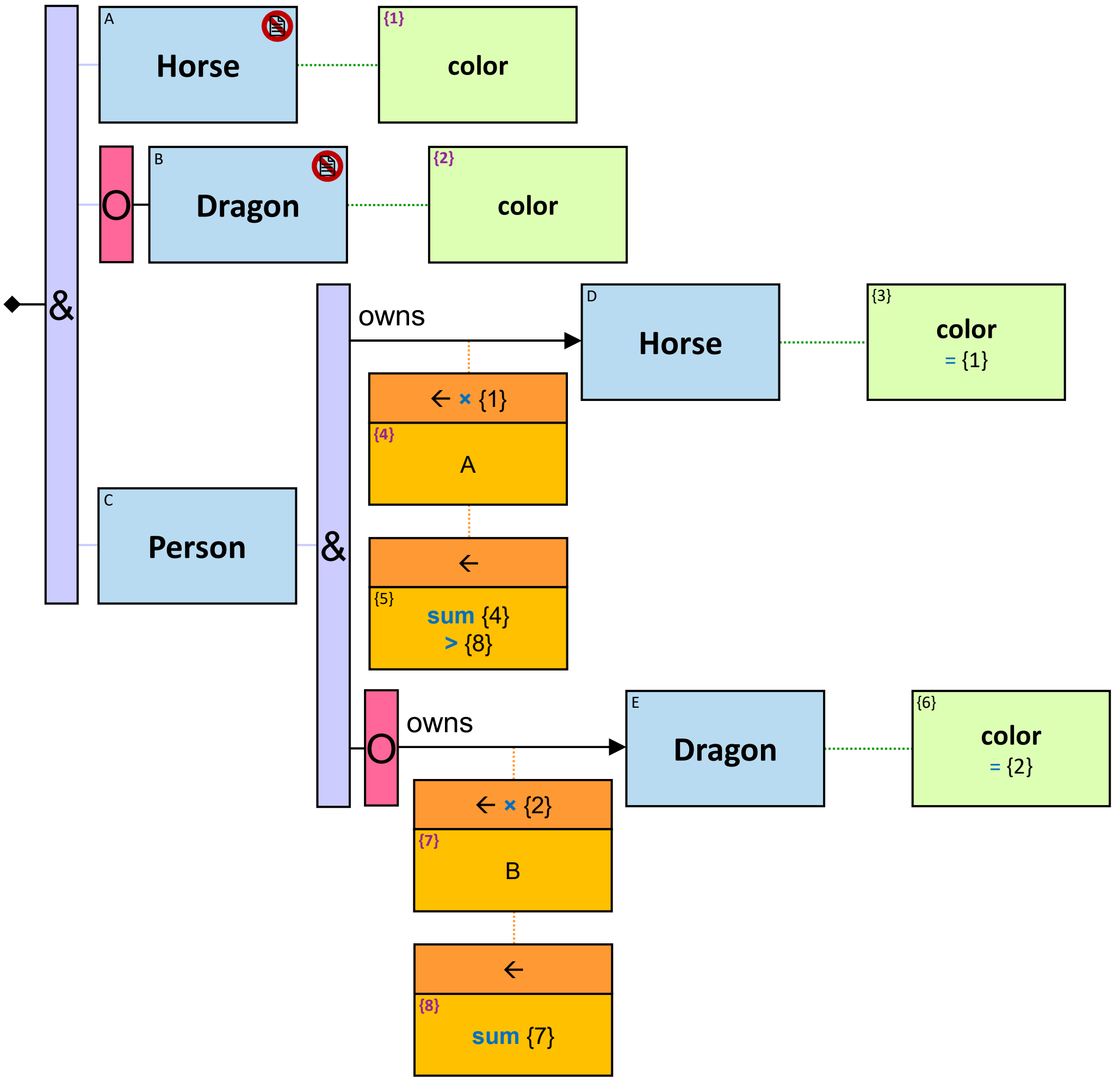

*Q321: Any person A where more horses than dragons have the same name-length as a horse/dragon A owns* (version 2)

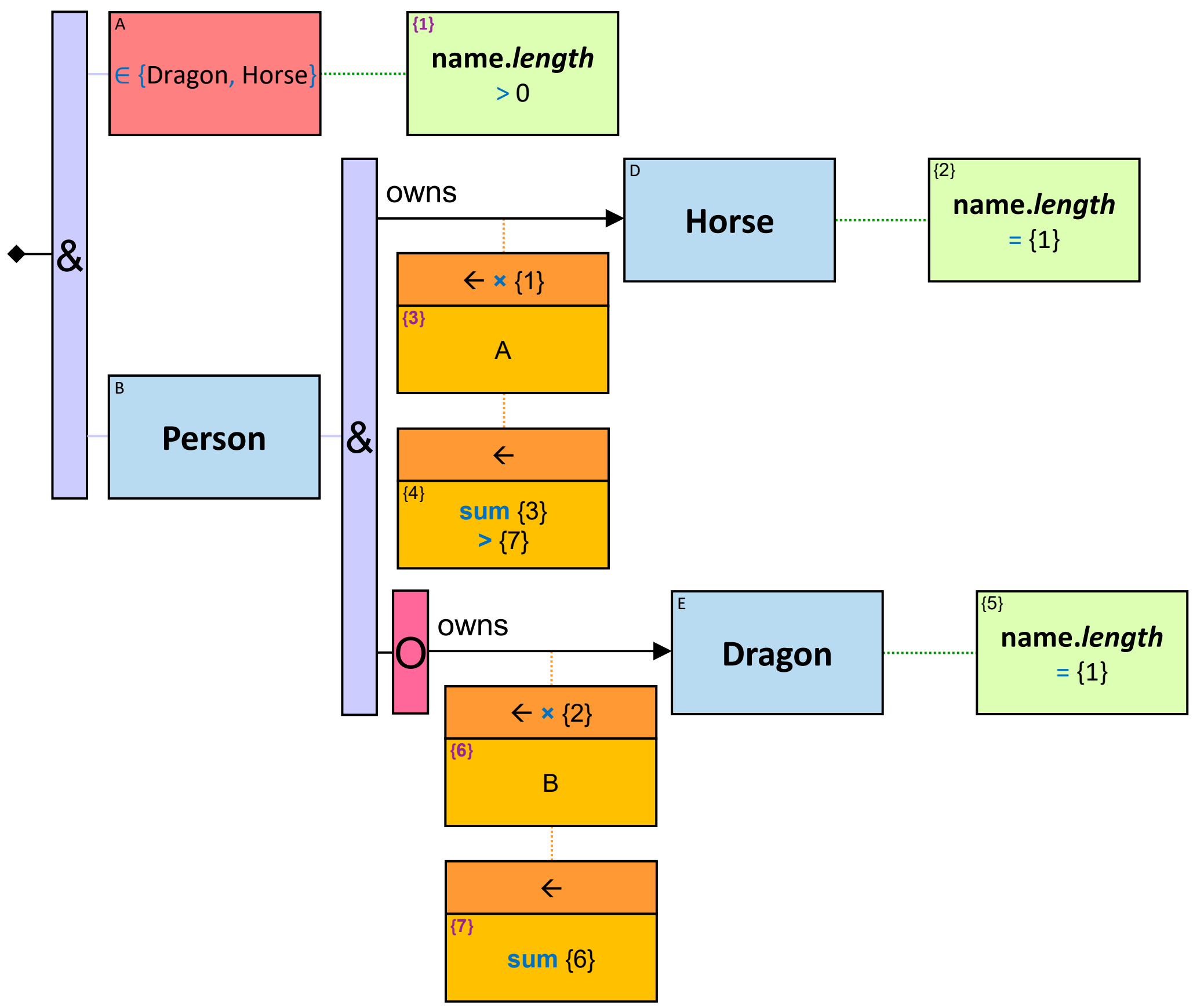

## 40. EXTENDED M1 AGGREGATOR

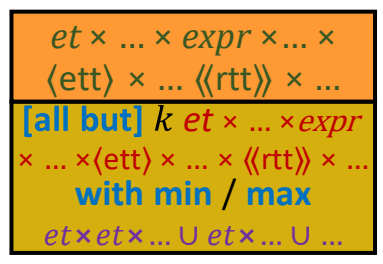

*Q220: Any person A and A's horses of the three colors of which A owns the largest number of horses*

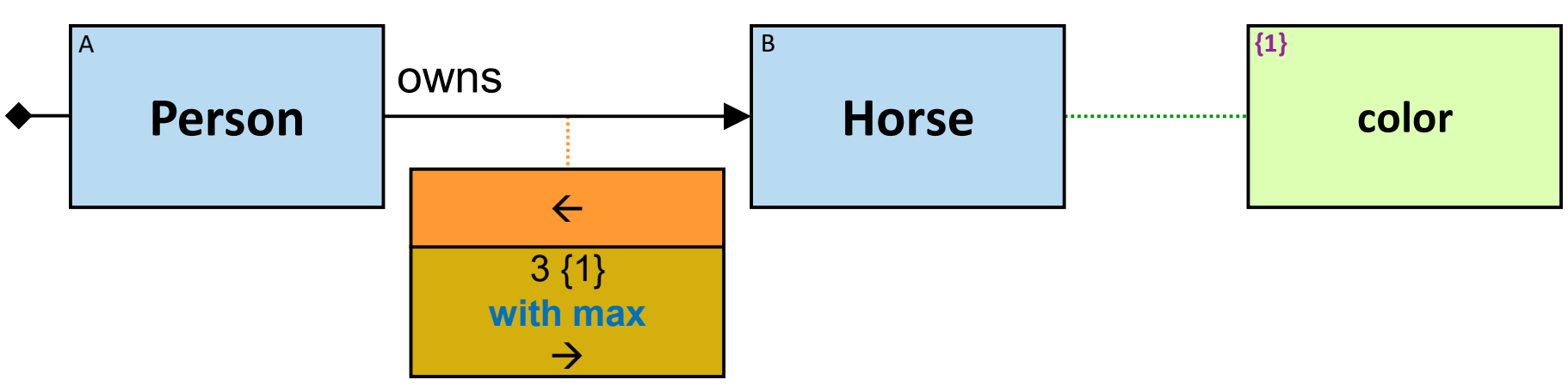

***Q262:*** *Any horse of the three most common horse colors*

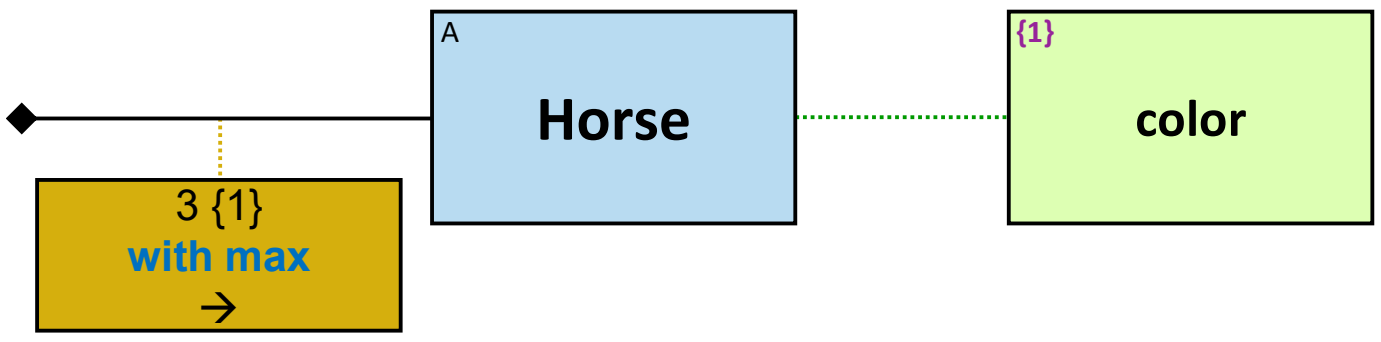

***Q224:*** *Any person A and A's horses of the three colors with the smallest positive number of horse owners* (two versions)

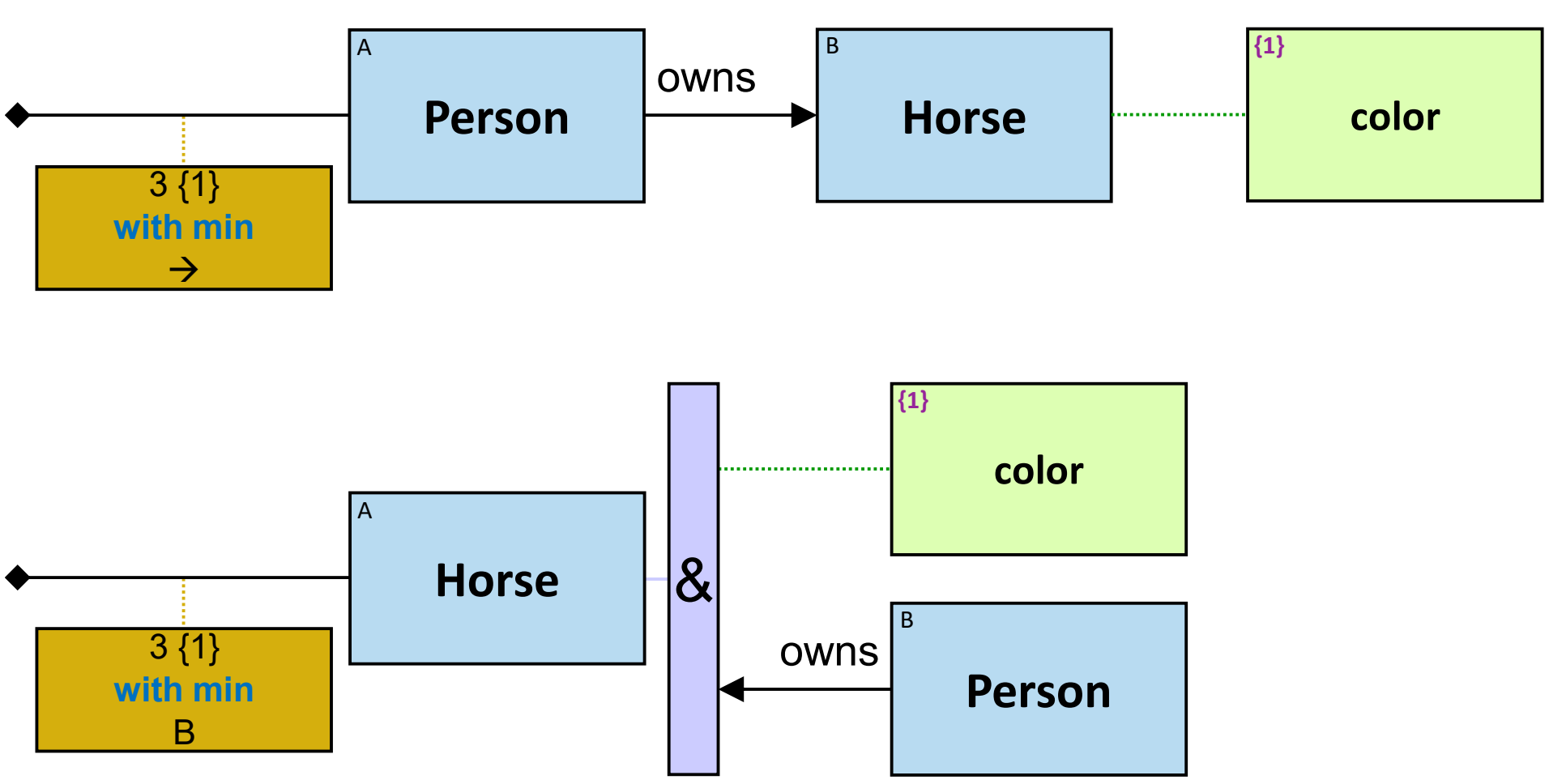

***Q342:*** *The three colors for which the number of pairs of dragons of that color where at least one of them froze the other – is maximal* (version 1)

Note that if a pair of dragons mutually froze each other – we need to count this pair only once.

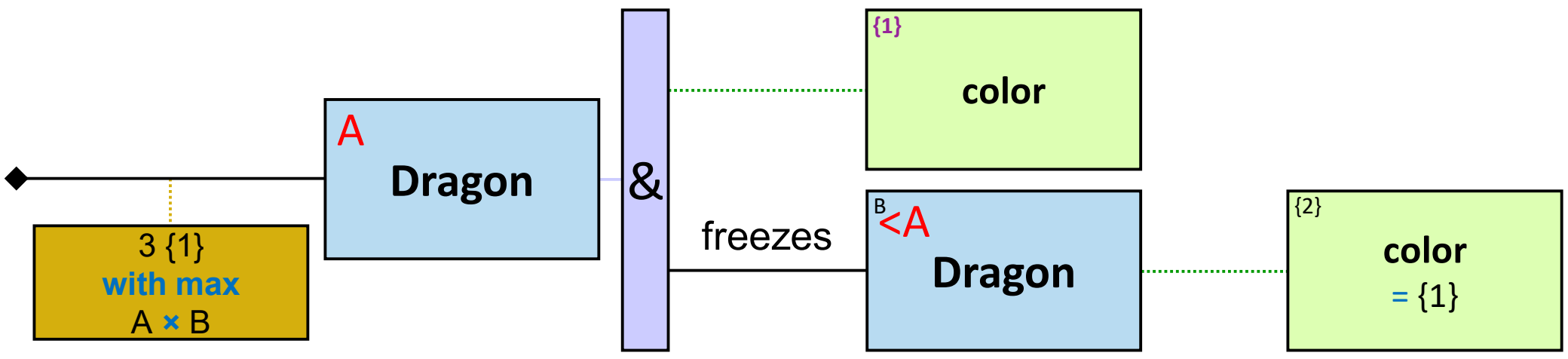

***Q343:*** *The three color-pairs (1, 2) for which the number of dragon pairs where a dragon of color 1 froze a dragon of color 2 is maximal*

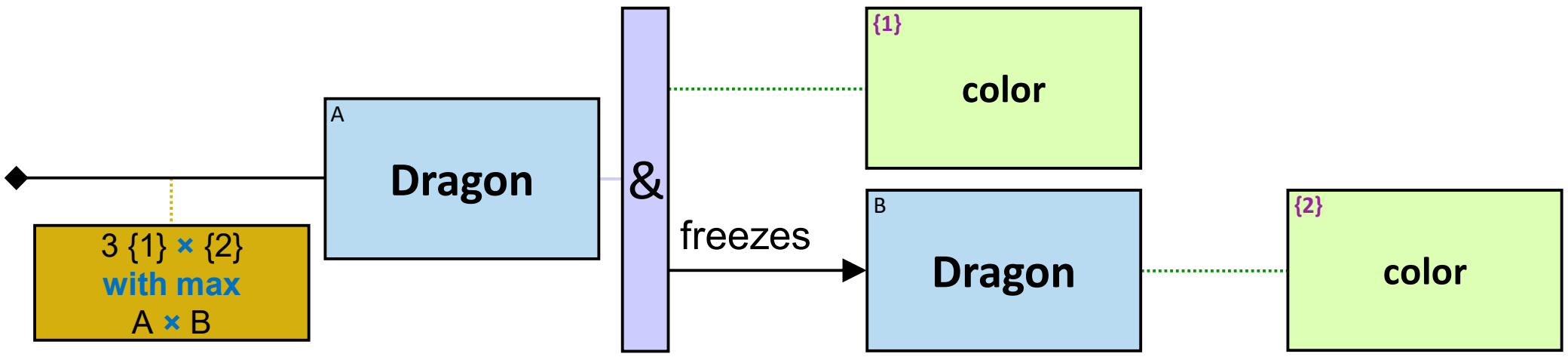

***Q251:*** *The three pairs (person, color) with most horses of this color owned by that person*

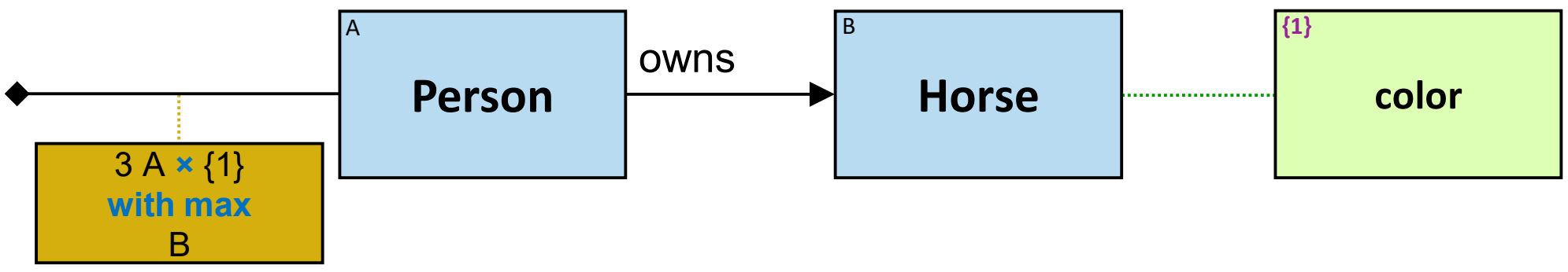

***Q211:*** *For each total number of freezes by a dragon: the three dragons that froze the largest number of dragons*

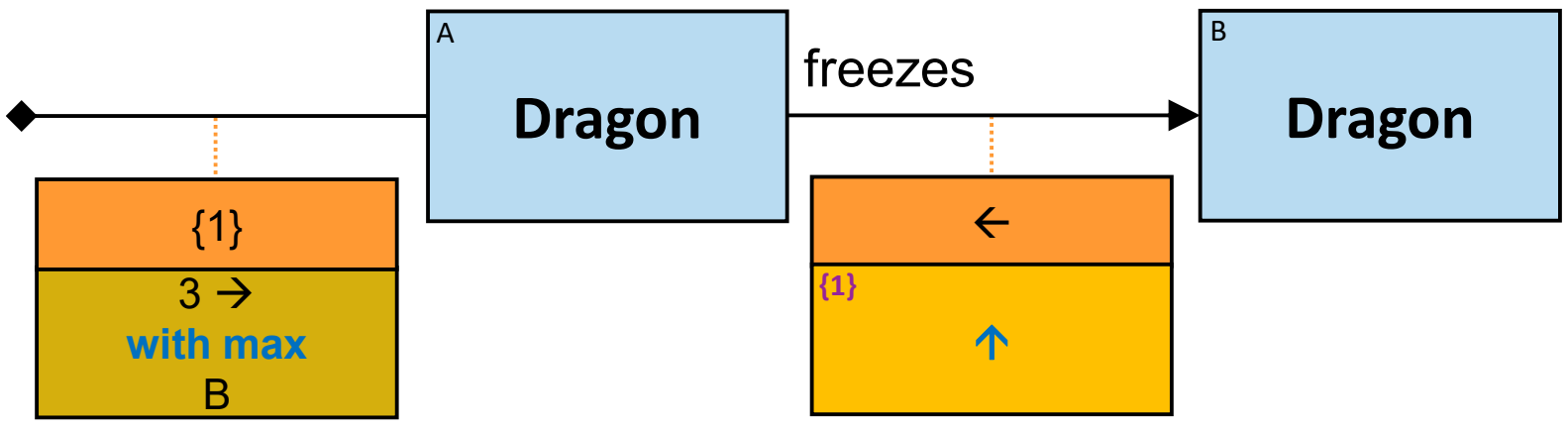

***Q212:*** *The three most common total number of freezes by a dragon*

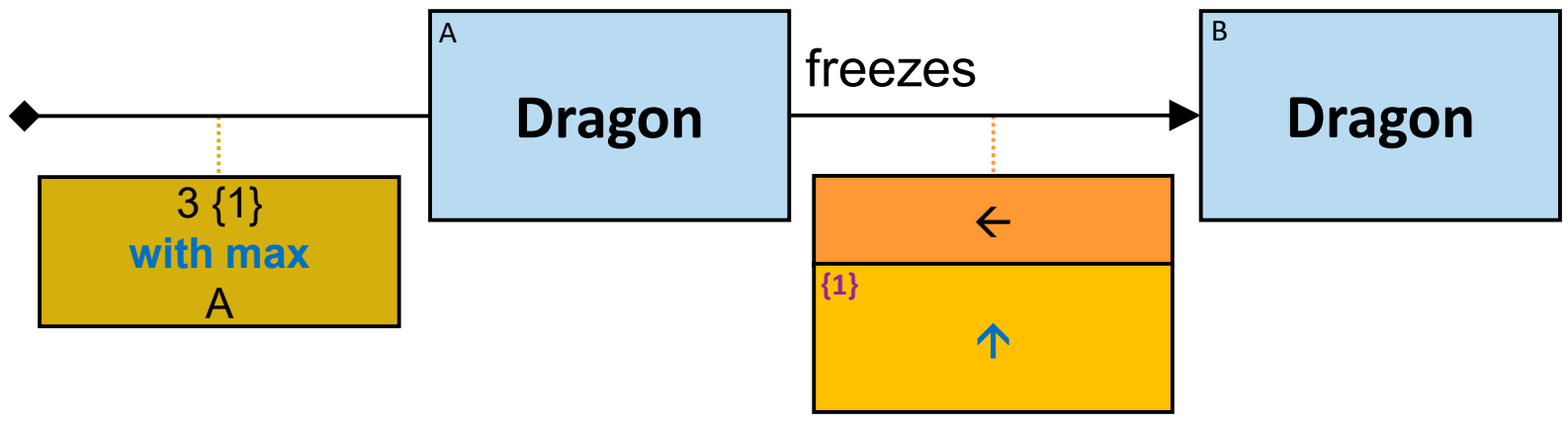

***Q325:*** *For each of the three most common horse colors: the three people who own the largest number of horses of this color*

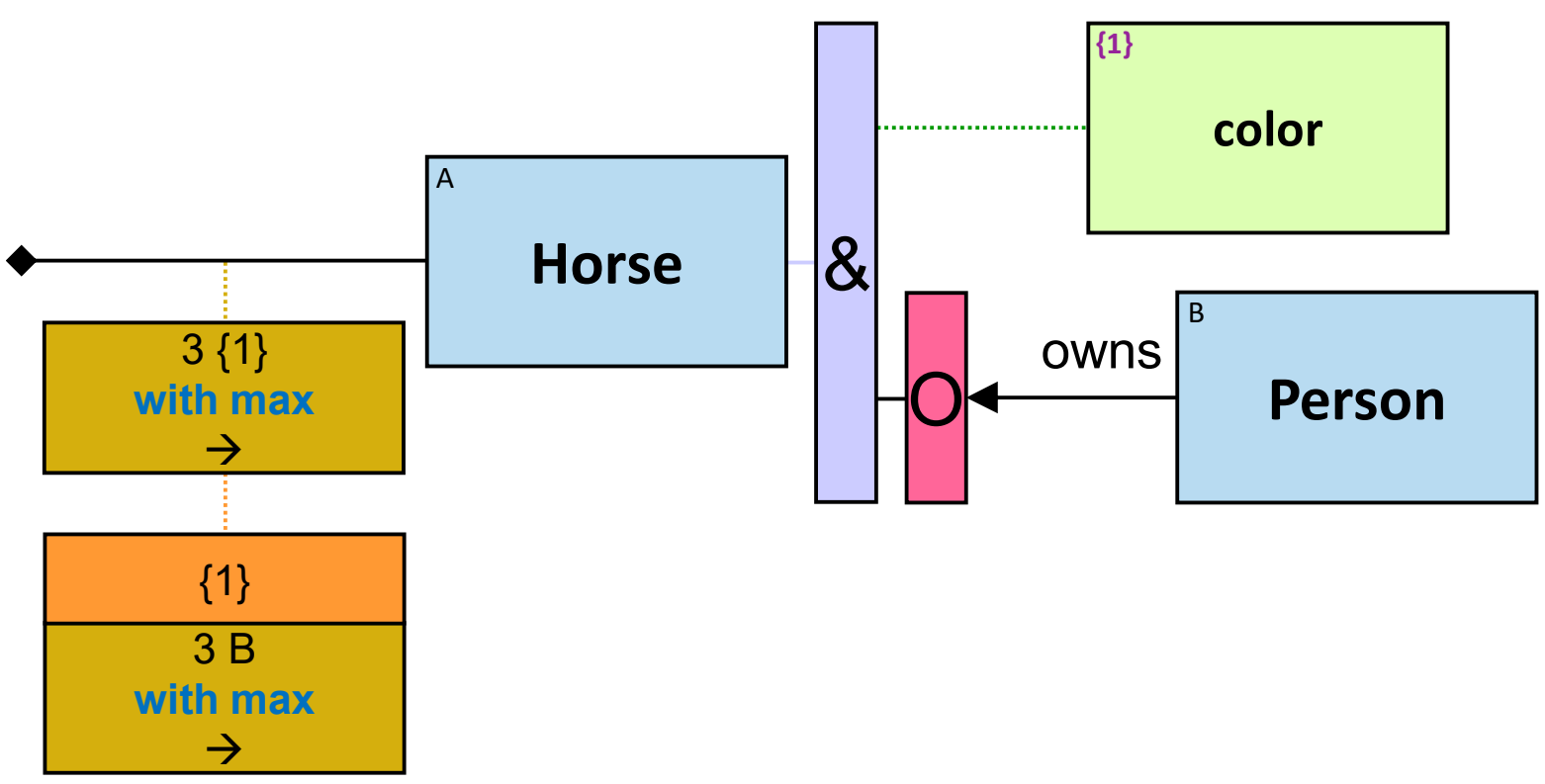

***Q326:*** *For the three most common horse colors (combined): the three people who own the largest number of horses of these colors*

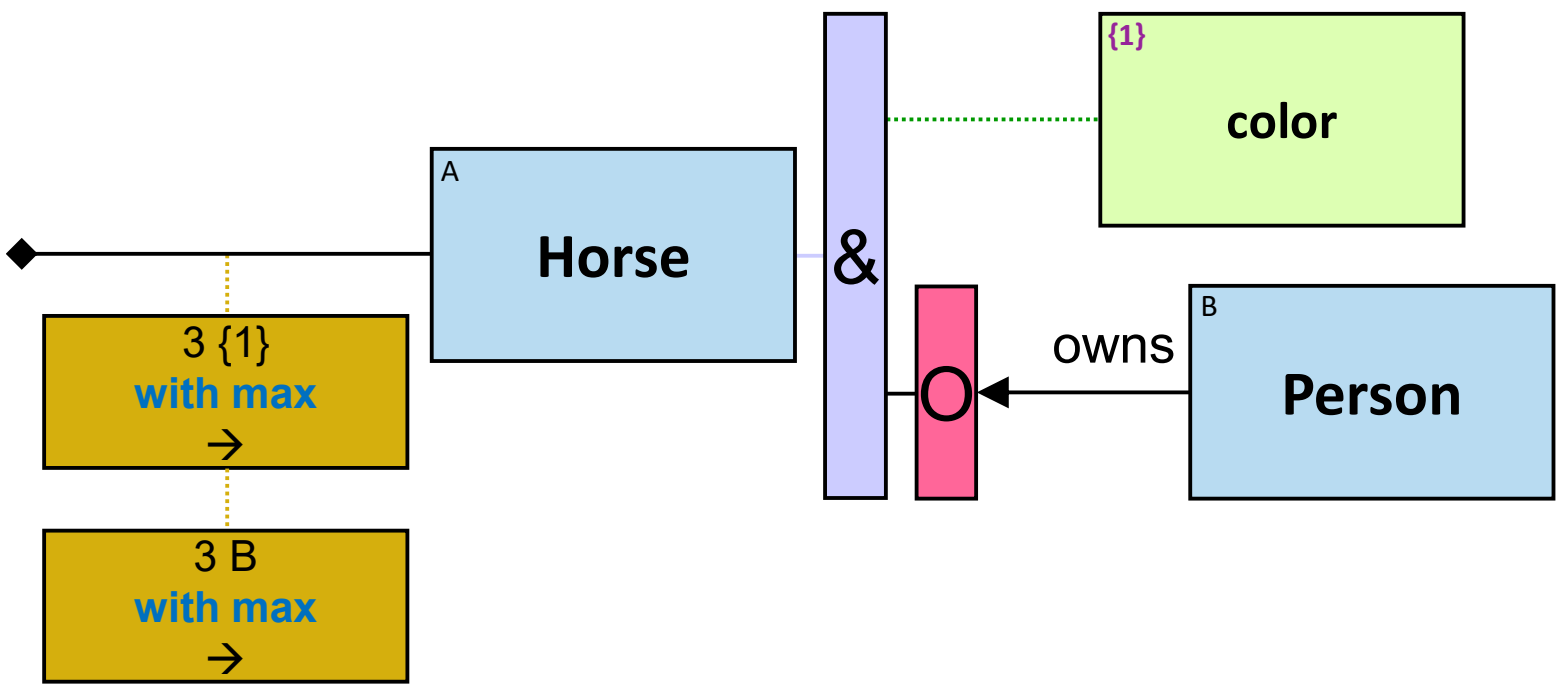

## 41. EXTENDED M2 AGGREGATOR

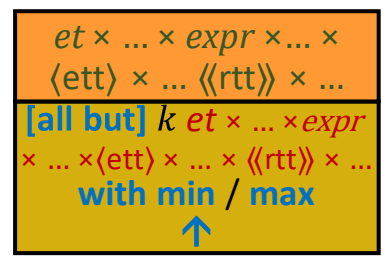

***Q215:*** *For each color of dragons that Balerion froze – the three dragons it froze the largest number of times*

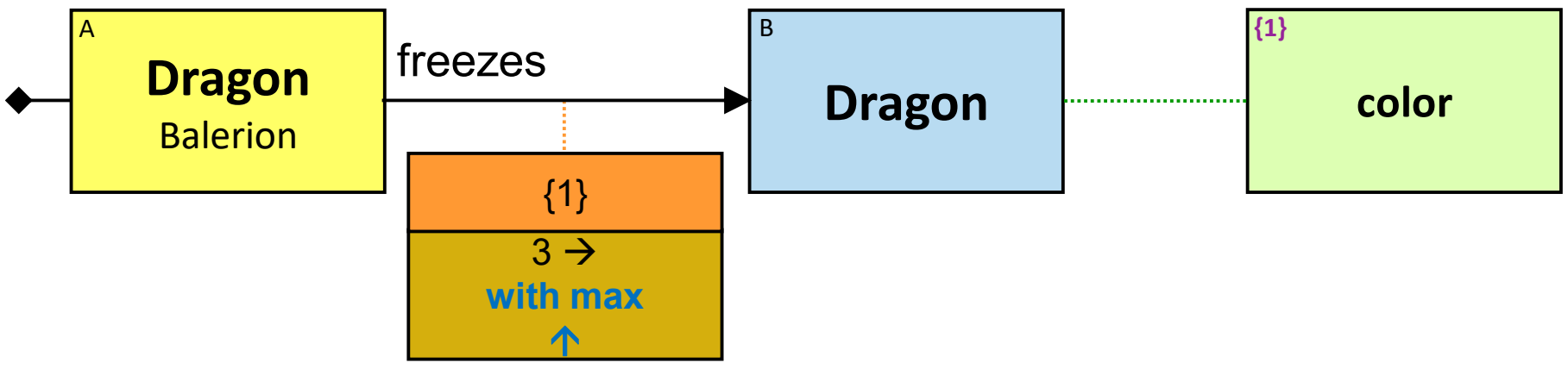

***Q221:*** *Balerion and the dragons it froze – of the three colors it froze dragons the largest number of times*

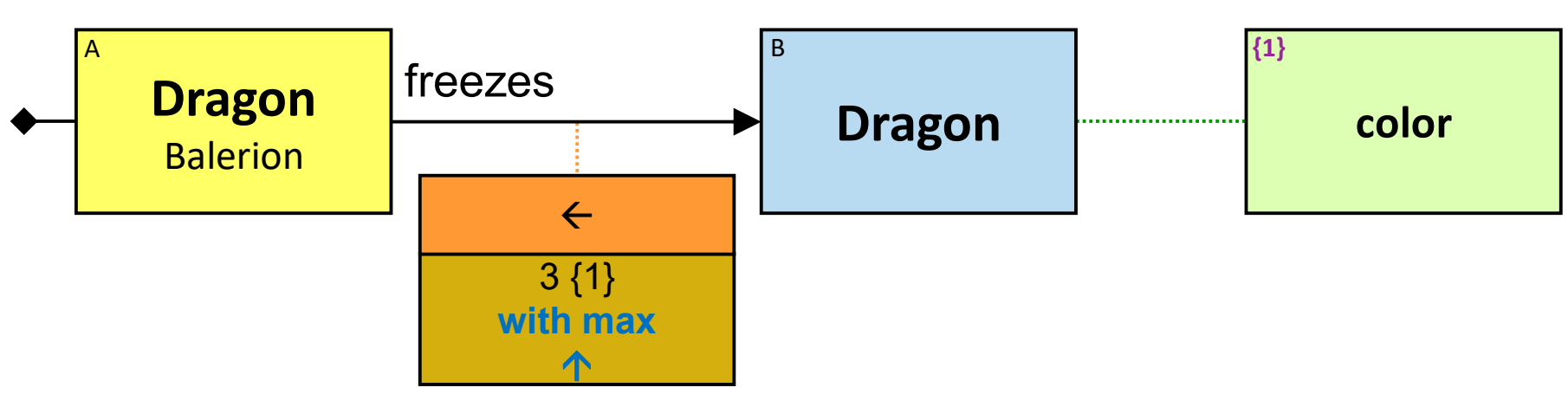

***Q280:*** *Any dragon and the dragons it froze – of the three colors it froze dragons the largest number of times*

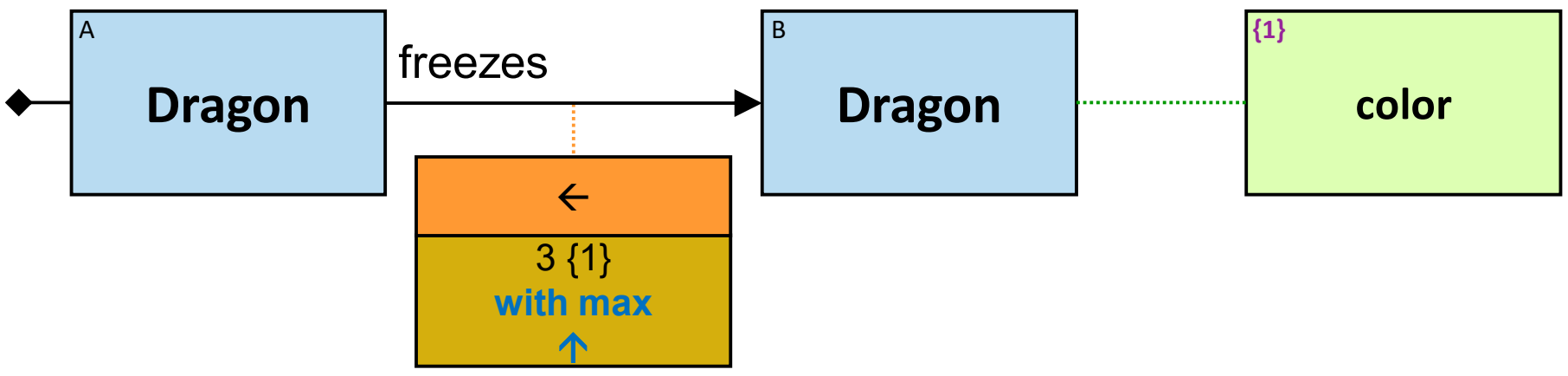

***Q253:** For each number of dragon's owners – the three dragons Balerion froze the largest number of times*

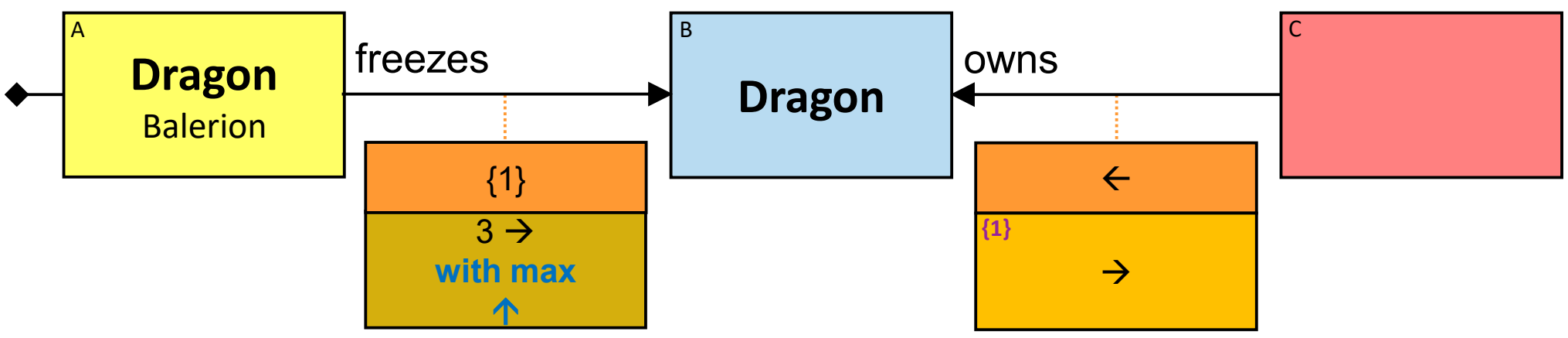

***Q225:** Any person A and A's horses of the three colors with the smallest positive number of horse ownerships by a person*

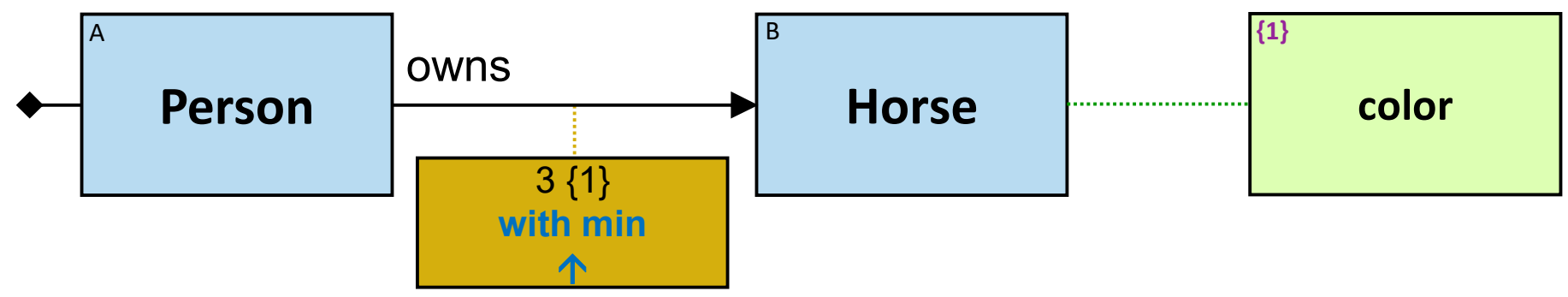

***Q281:** Any dragon and the dragons it froze – of the three colors of which dragons were frozen the largest number of times*

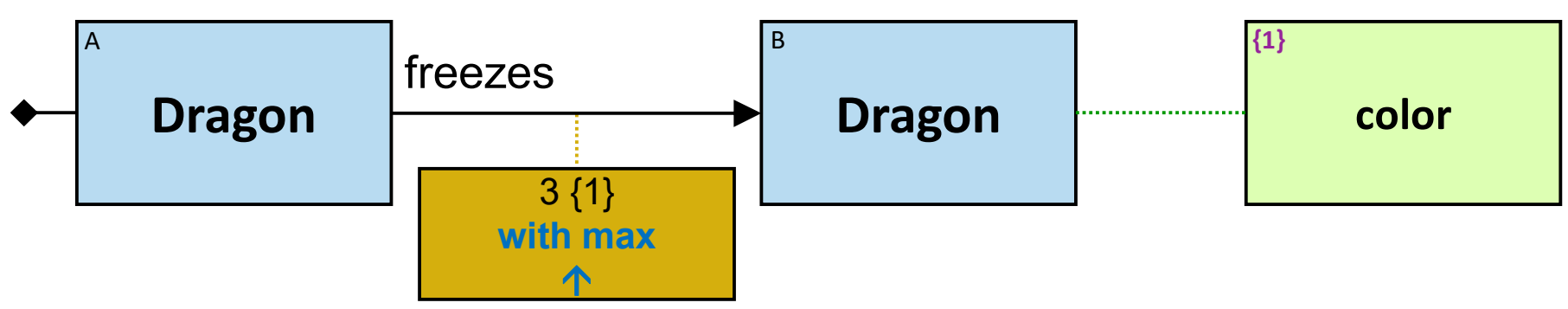

***Q319:** Any dragon and the dragons it froze – of the $4^{th}$-$6^{th}$ colors of which dragons were frozen the largest number of times* (two versions)

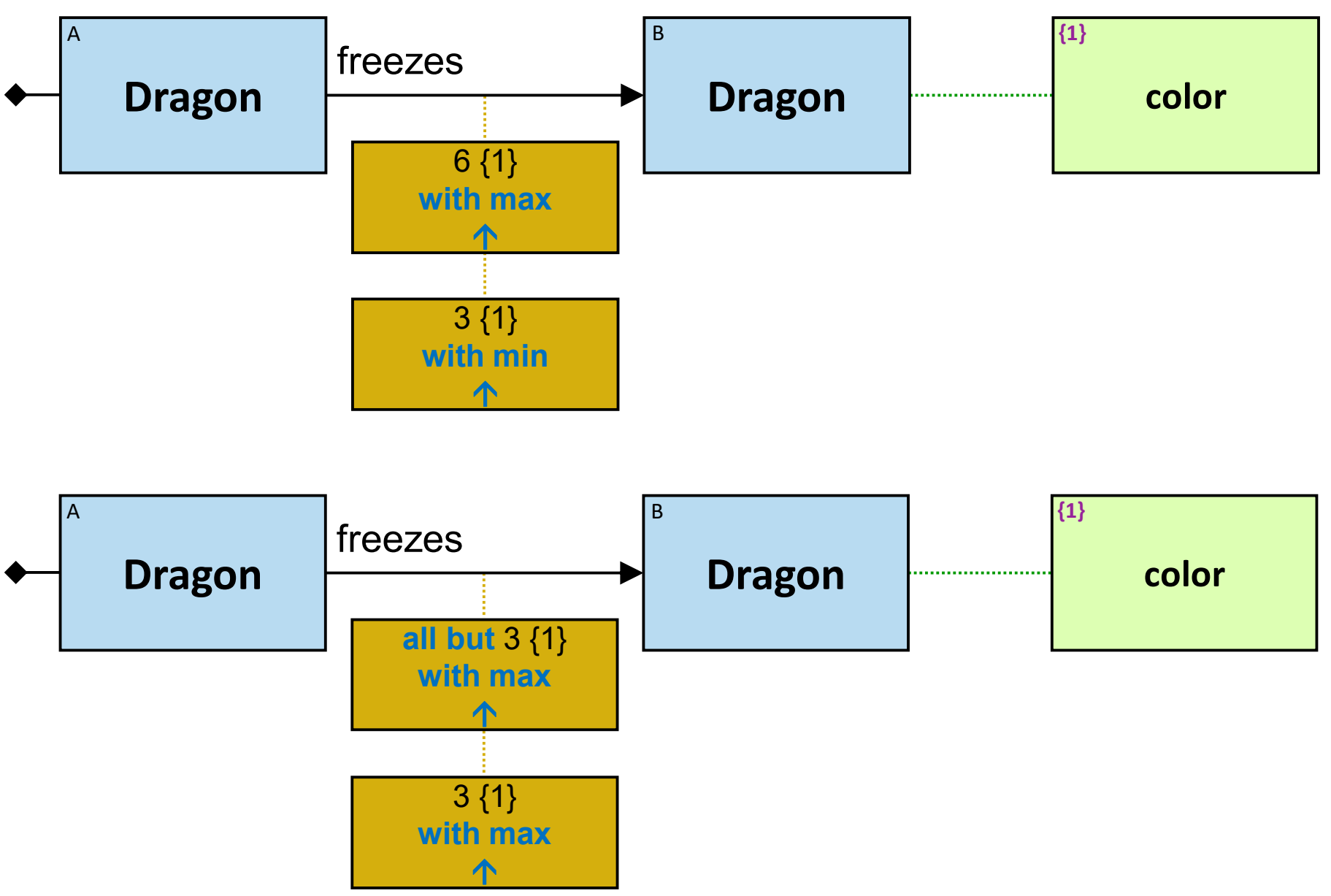

***Q282:** Balerion and the dragons it froze – of the three colors of which dragons were frozen the largest number of times*

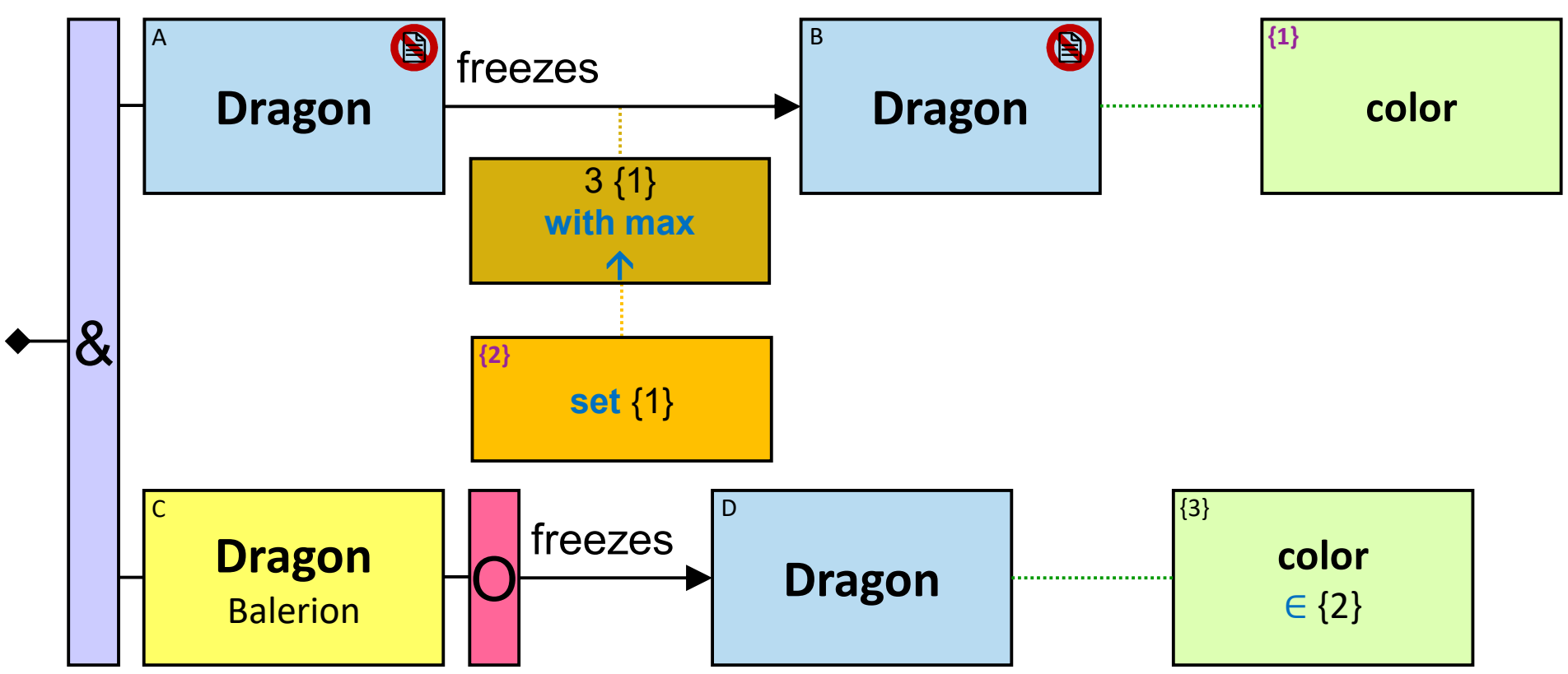

***Q209:** The three most owned entity-types*

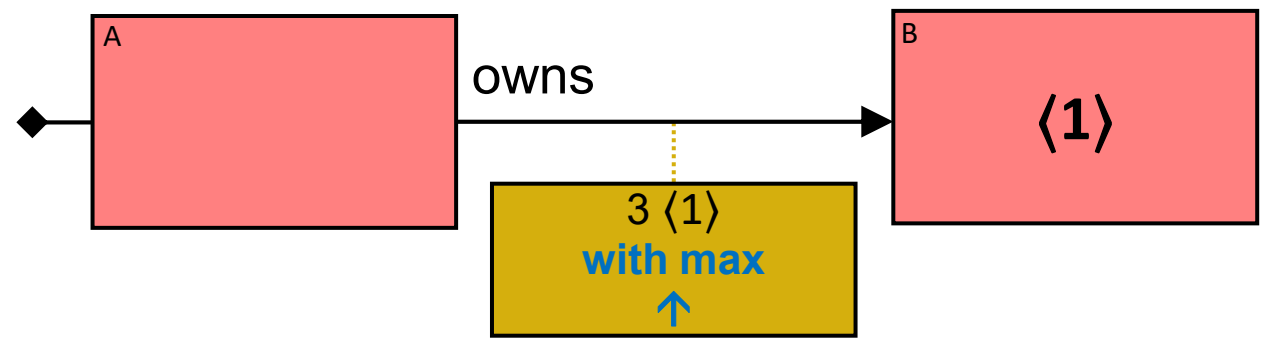

***Q210:** The three pairs (owner entity-type, owned entity-type) with most 'owns' relationships*

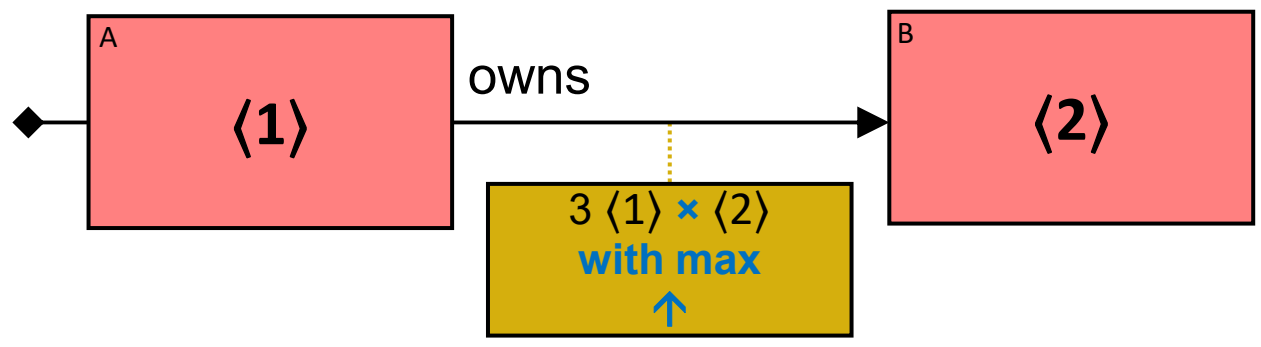

***Q369:** The three most common relationship-types and the number of relationships of each of these types*

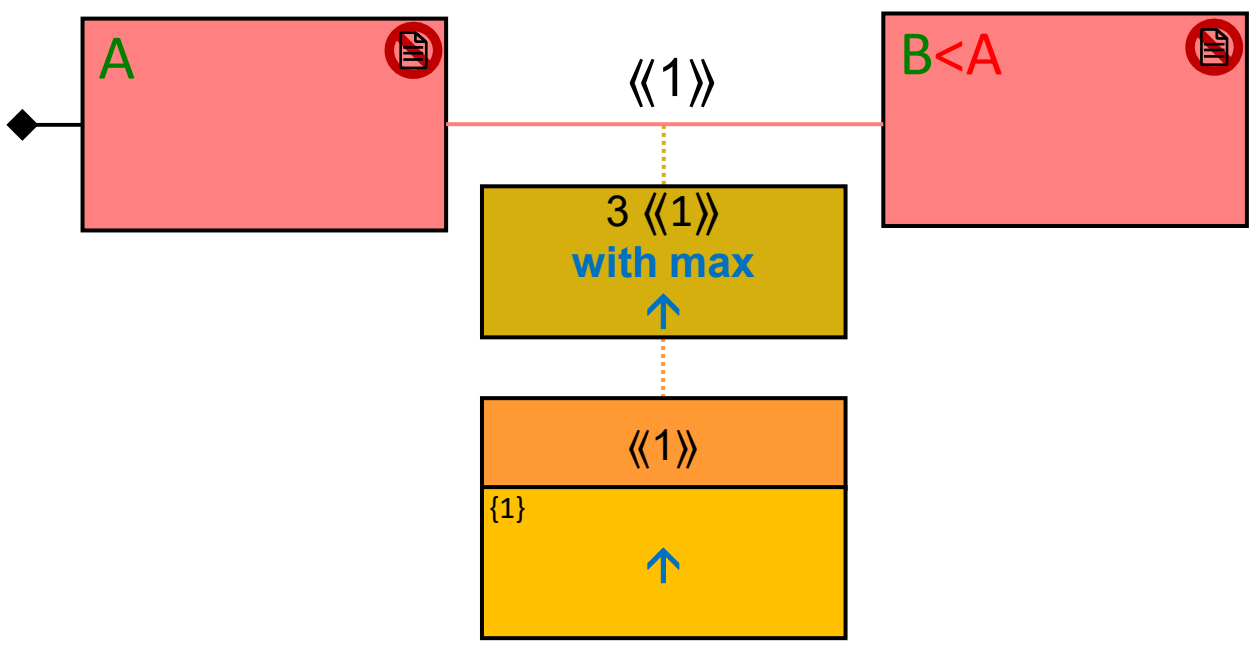

***Q370:** The most common relationship-type between entities of the same type*

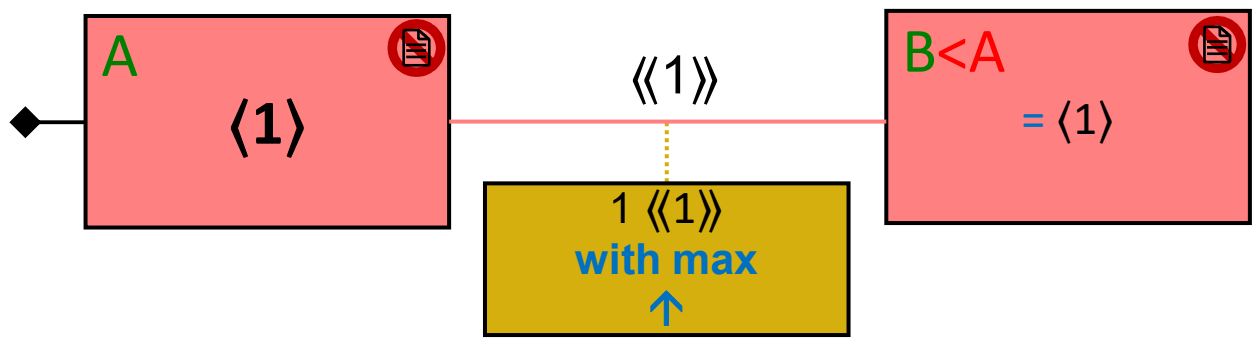

## 42. EXTENDED M3 AGGREGATOR

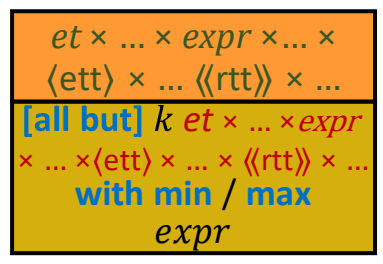

***Q275:*** *Any person A and A's horses of the three colors of which A owns the heaviest horse*

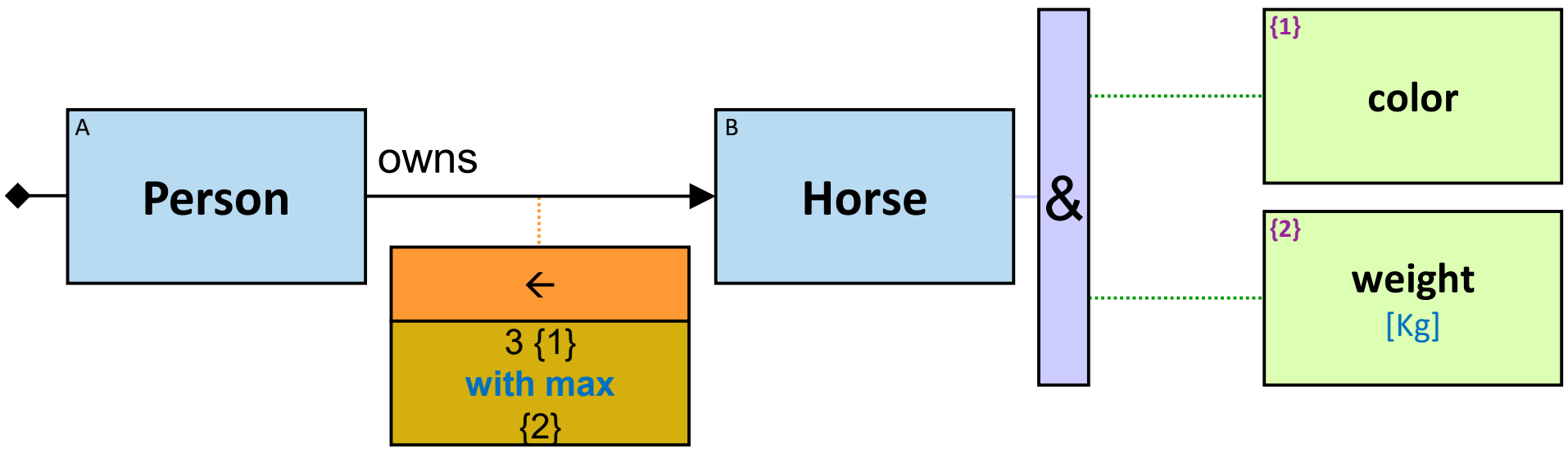

Even if A's three heaviest horses are of the same color, we would still get all A's horses for two more colors (those with the next heaviest horses)

***Q223:*** *Any person A and A's horses of the three colors for which the cumulative weight of A's horses is maximal*

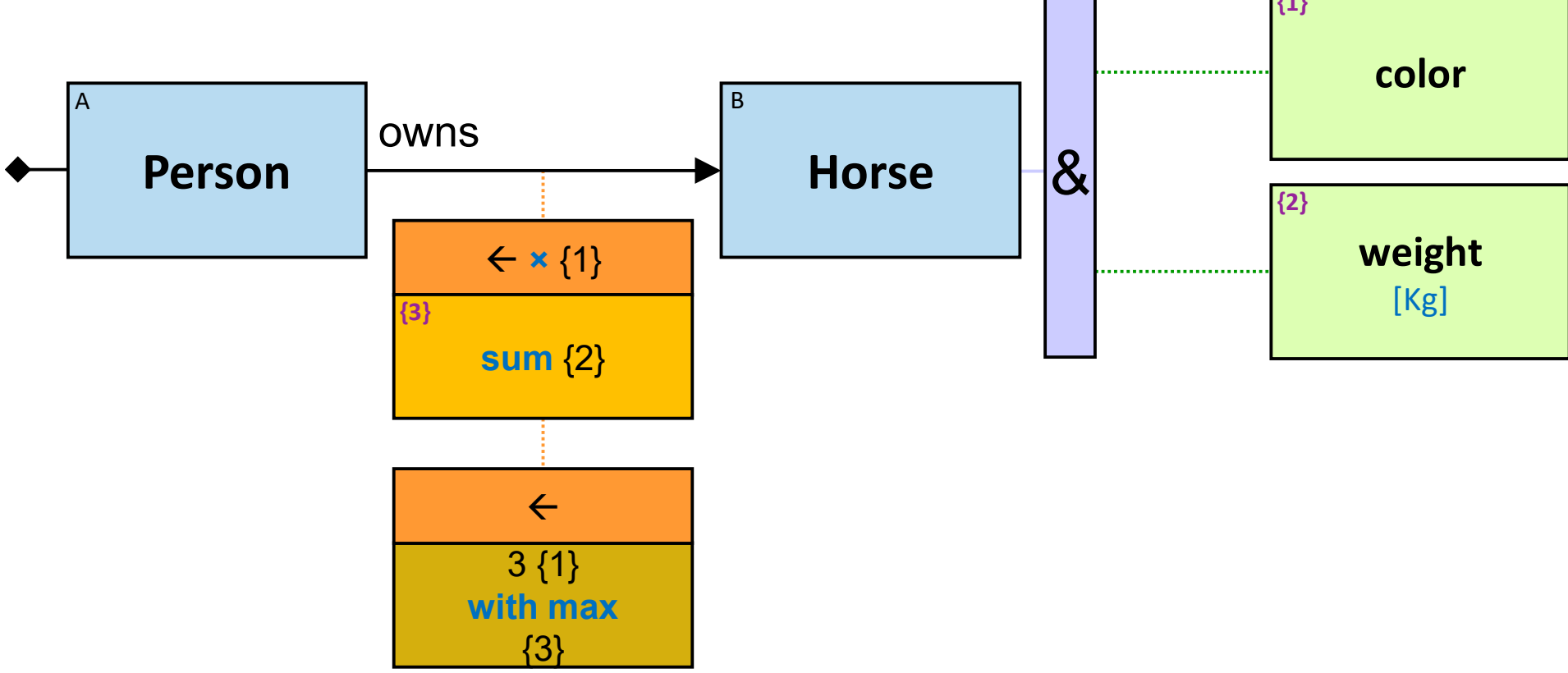

***Q269:*** *Any person A and A's horses of the three colors for which the average ownership start date is the latest*

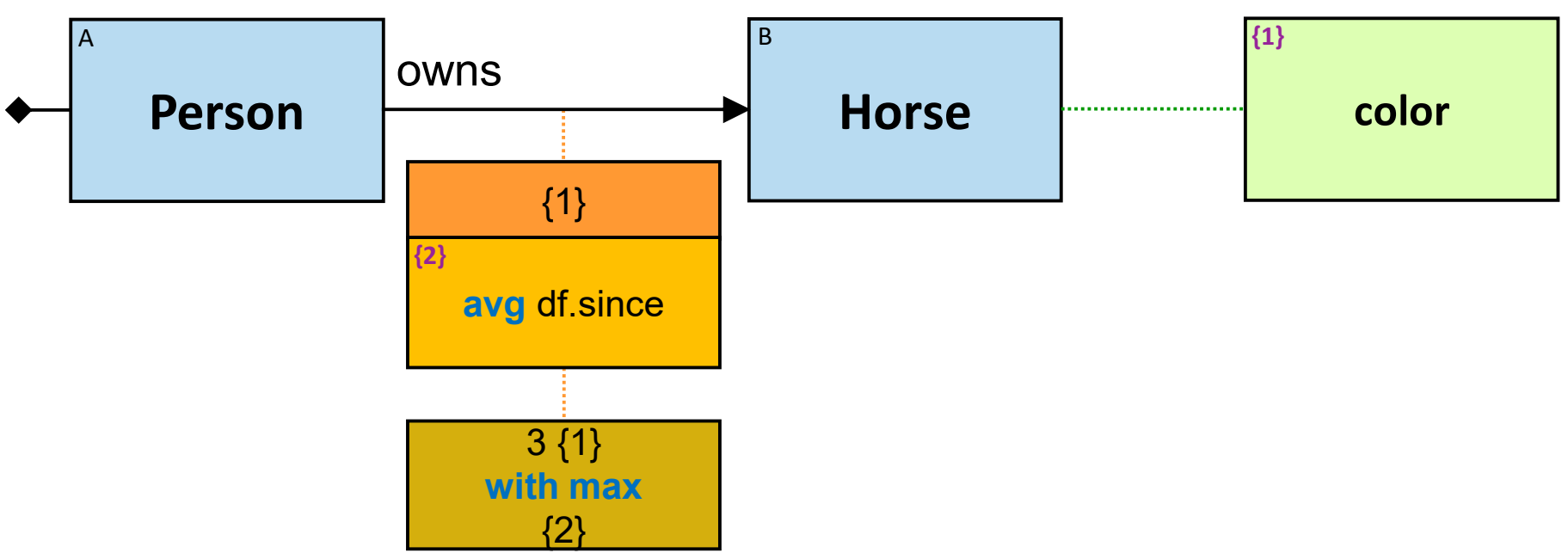

*Q276: Any person A and A's horses of the three colors for which the earliest ownership start date is the earliest*

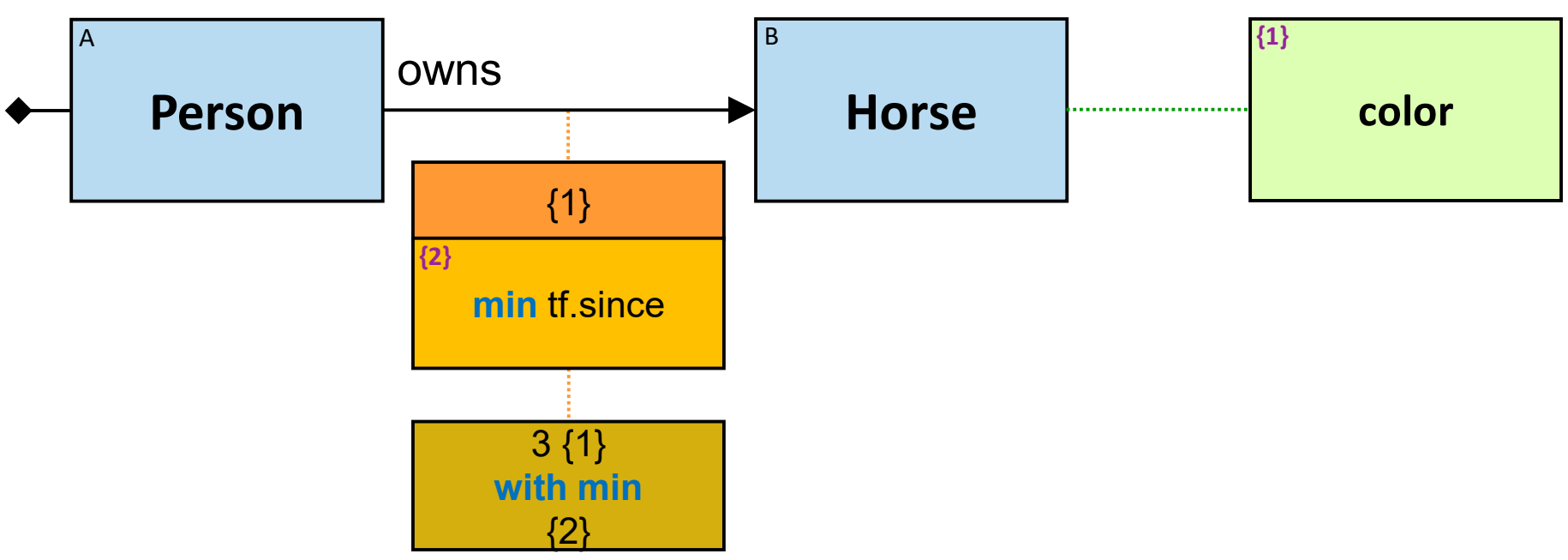

Even if the three horses with the earliest ownership start date are of the same color, we would still get all people and their horses of two more colors (of the next-earliest ownership start dates)

*Q222: Any person A and A's horses of the three colors for which A's average ownership start date is the latest*

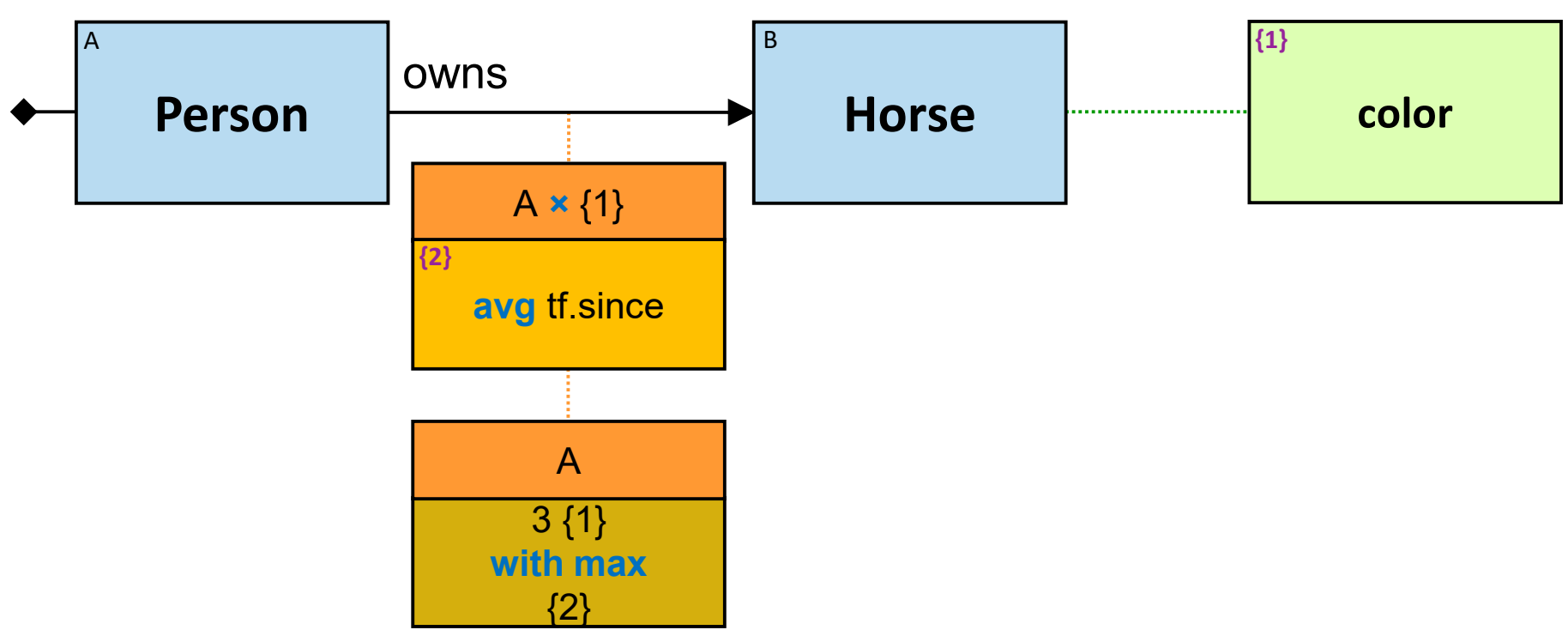

*Q216: Any dragon that Balerion froze in the three 30-day timeframes in which it froze dragons the largest number of times*

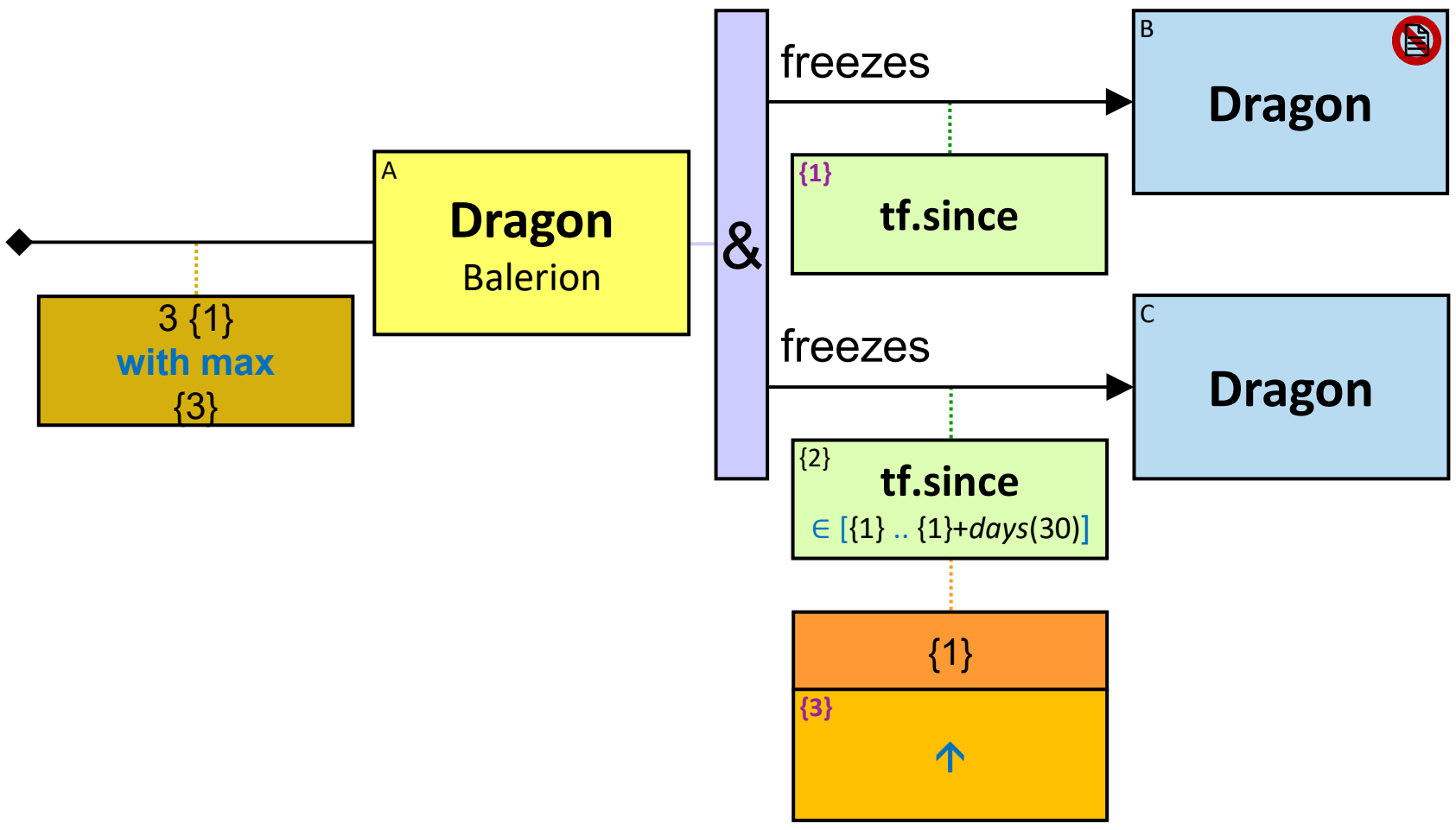

***Q373:** The three 30-day timeframes in which dragons were frozen the largest number of times*

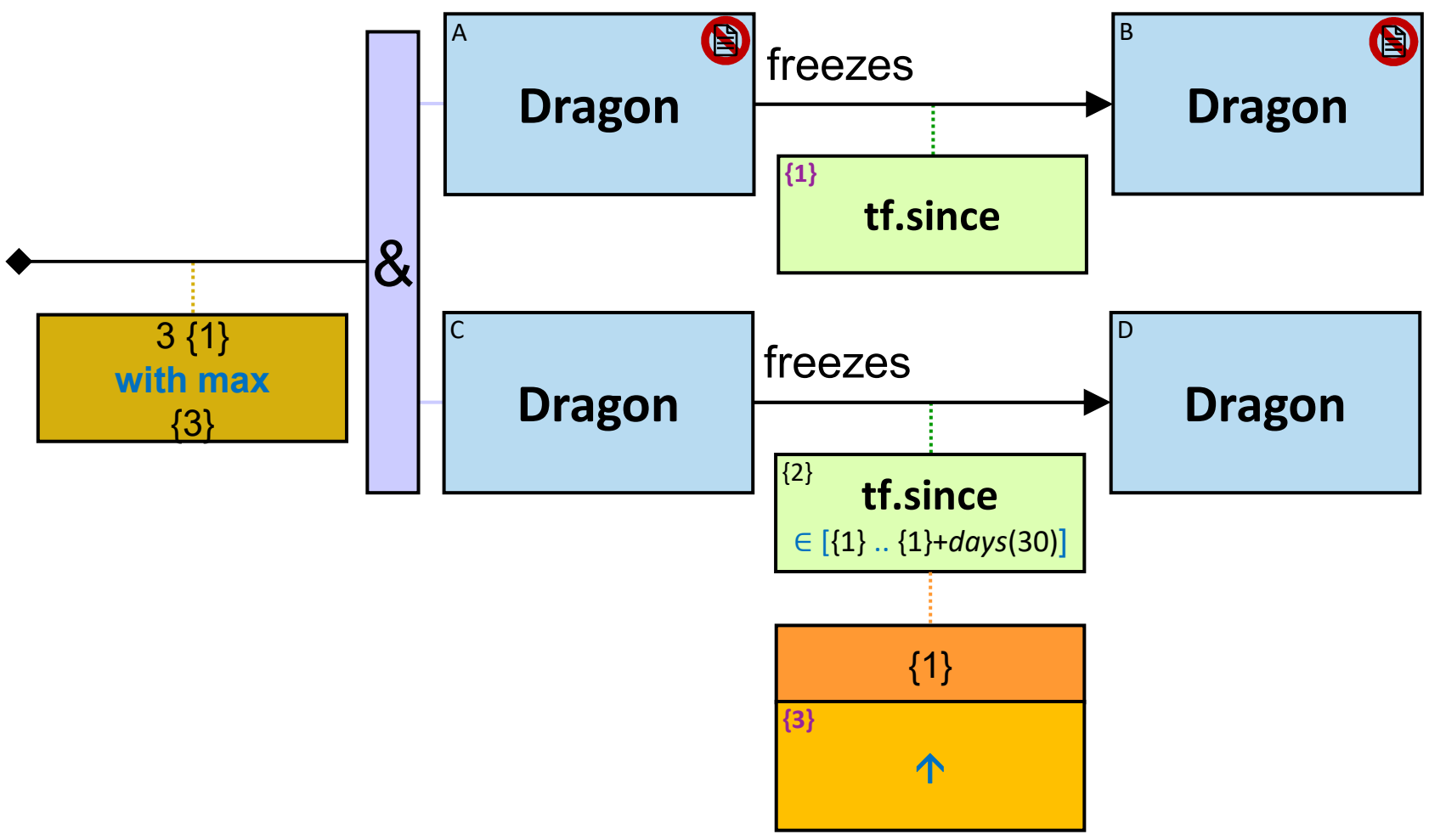

***Q374:** The three most prolonged timeframes during which no dragon was frozen*

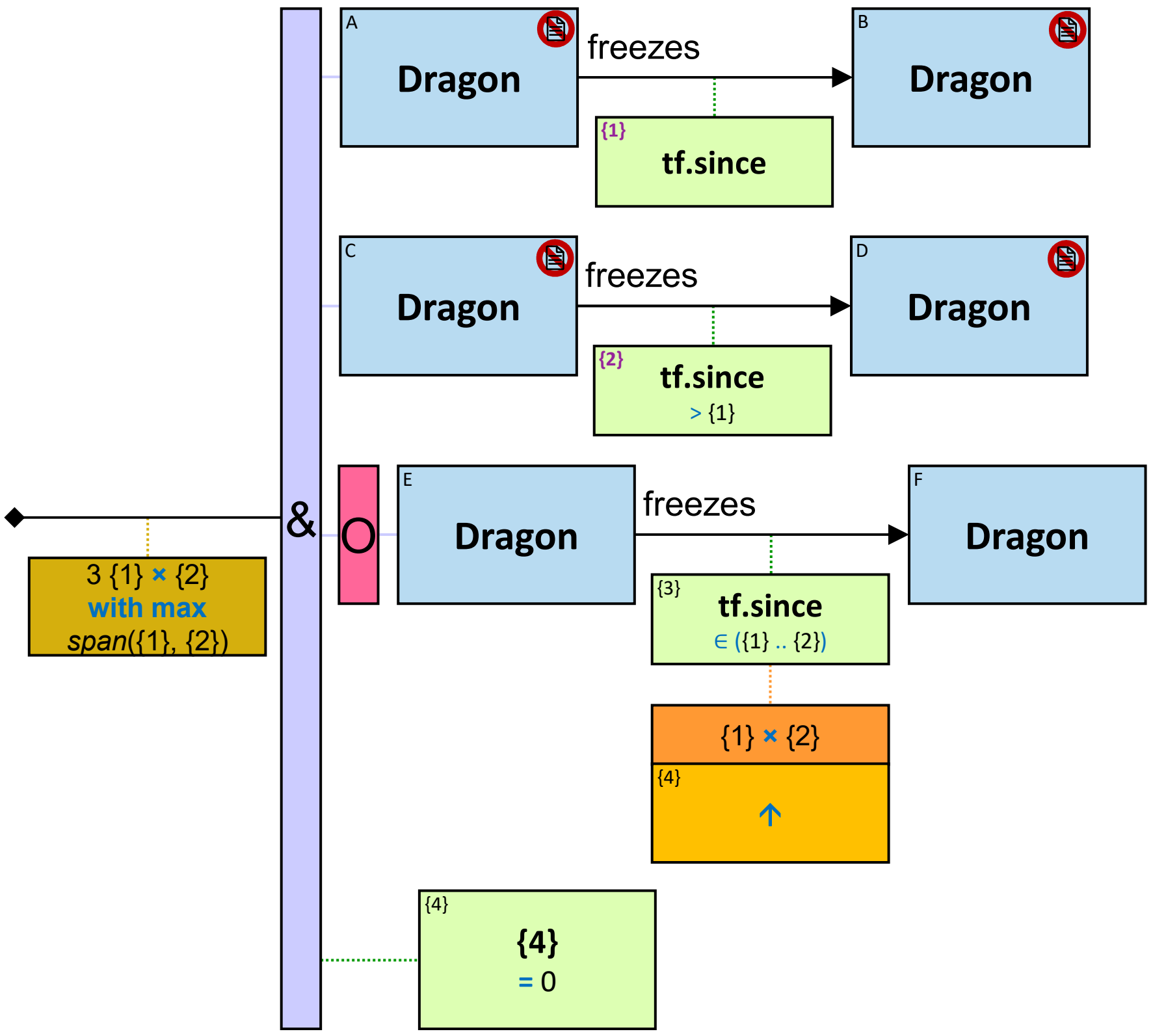

***Q264:** Any horse of the three colors of which the average horses' weight is maximal*

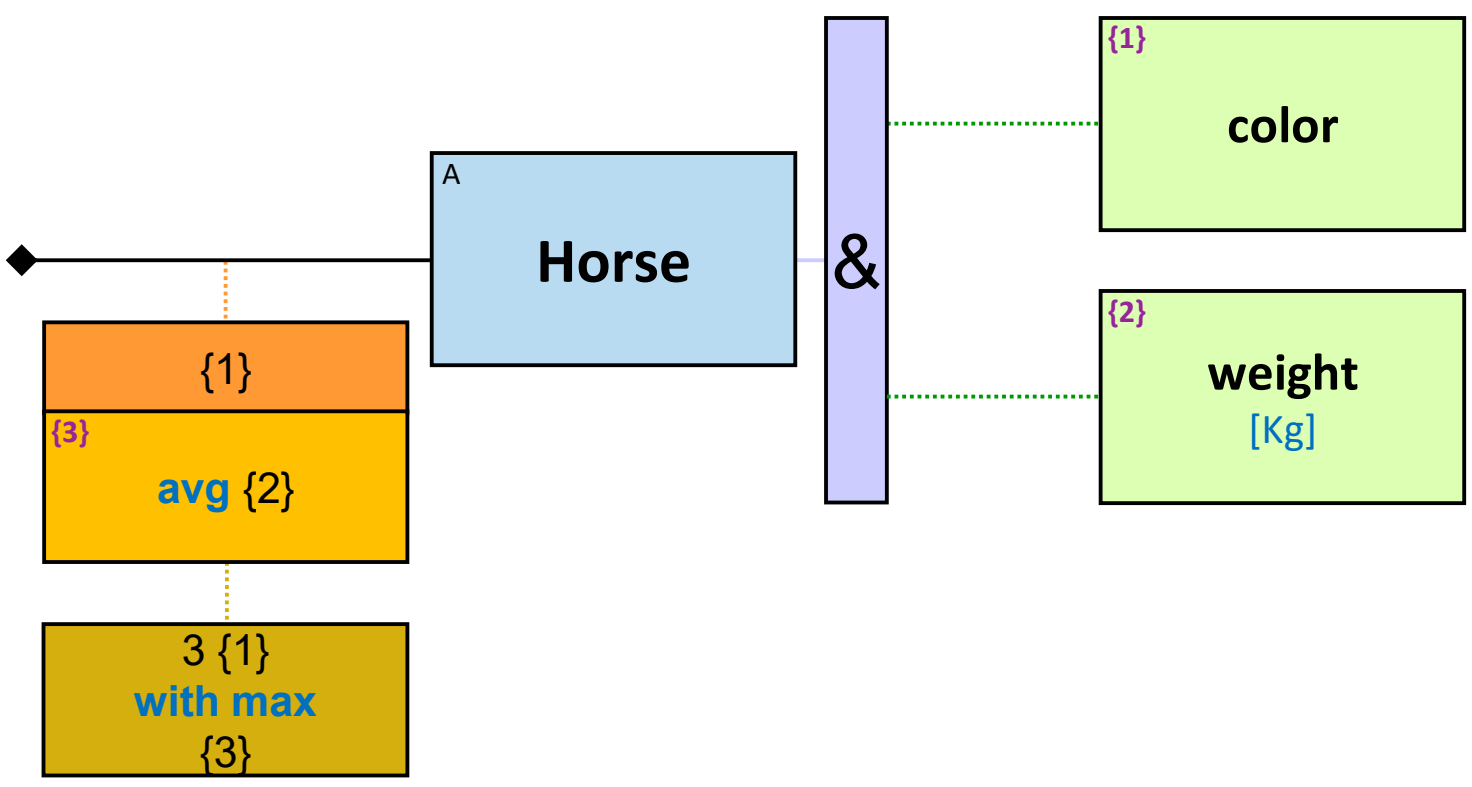

***Q226:** Any person A and A's horses – of the three horse-colors for which the average height of the horse owners is minimal*

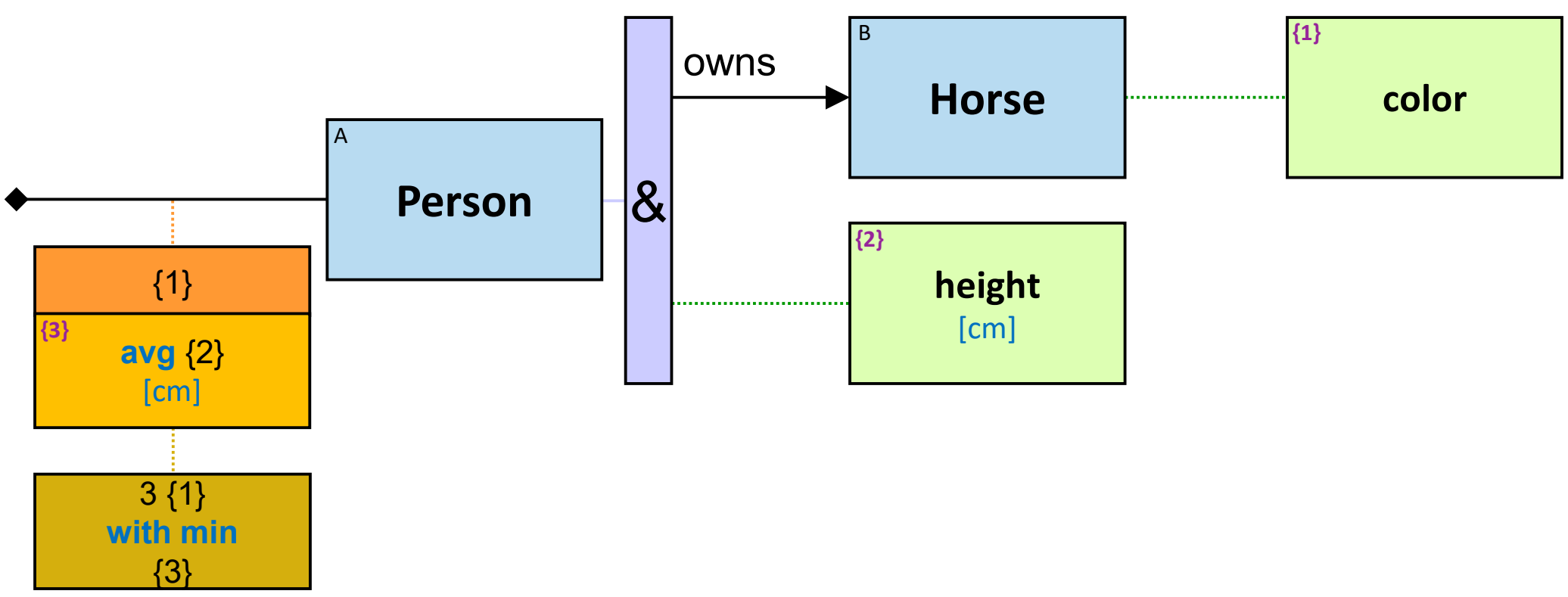

***Q342:** The three colors for which the number of pairs of dragons of that color where at least one of them froze the other – is maximal* (versions 2, 3)

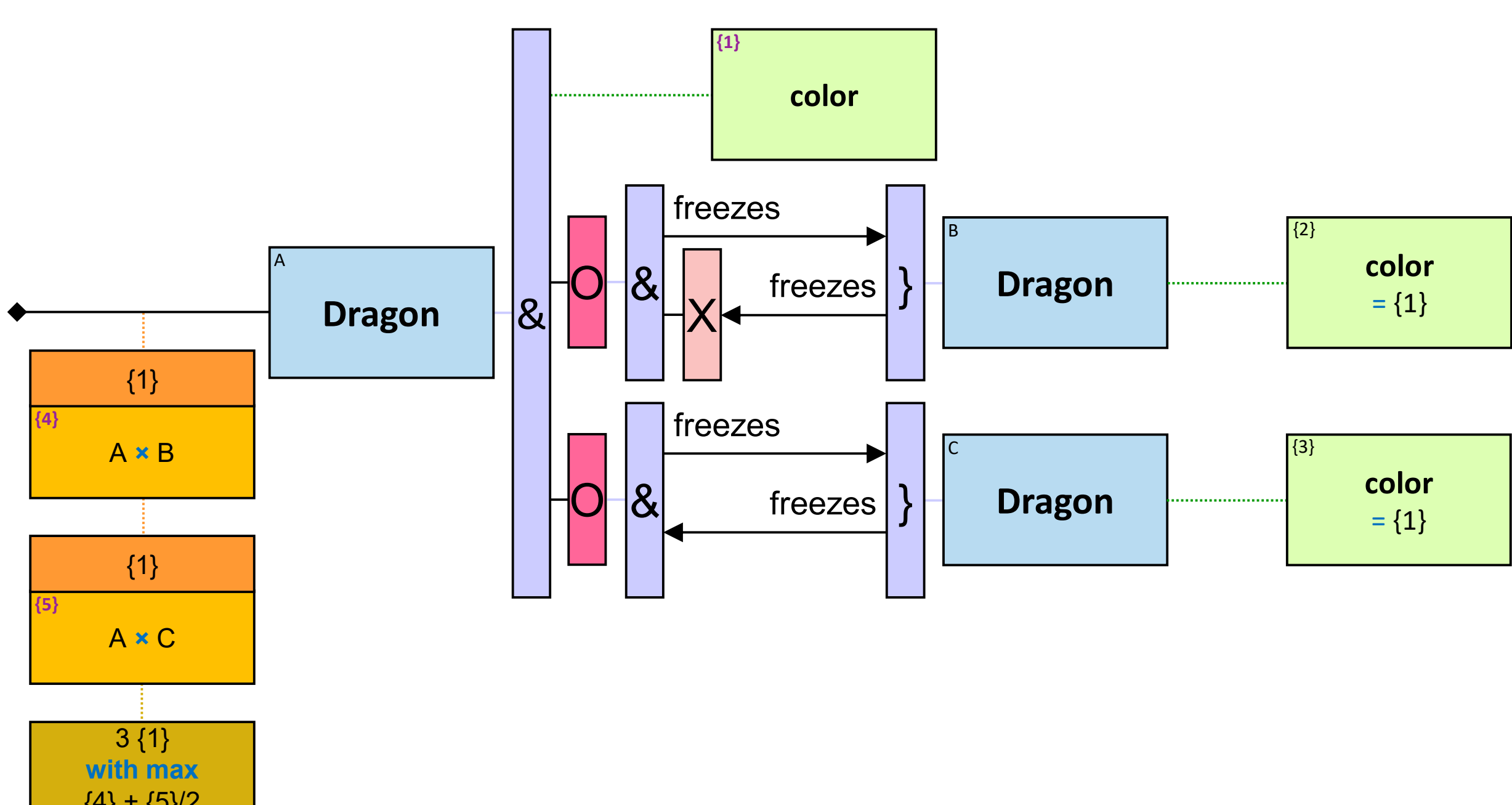

Note that if a pair of dragons mutually froze each other – we need to count this pair only once.

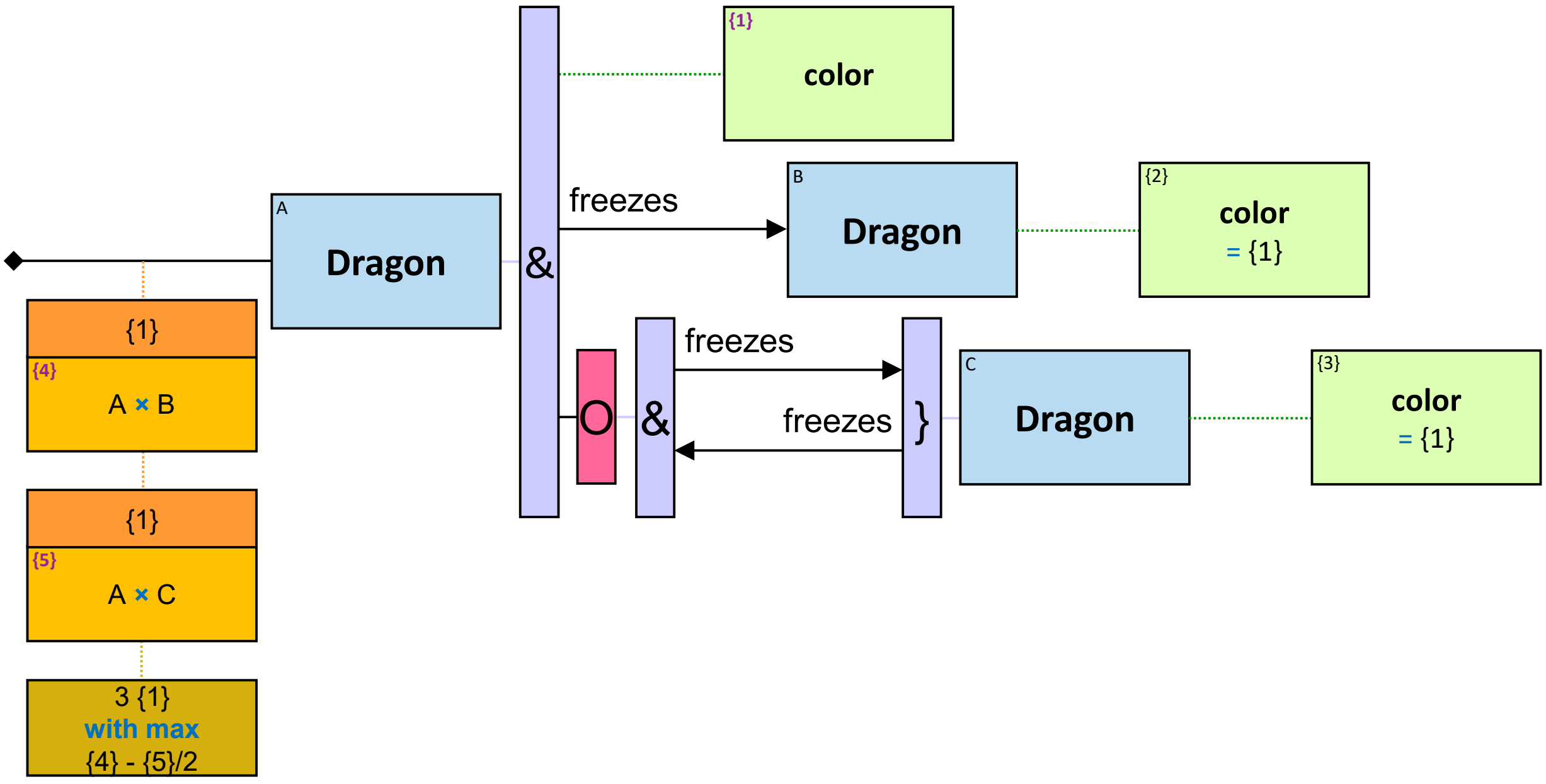

## 43. EXTENDED R1 AGGREGATOR

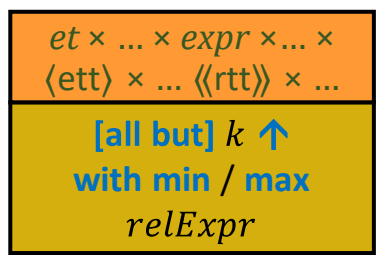

***Q273:** For each horse color – the three horse-ownerships with the latest ownership start date*

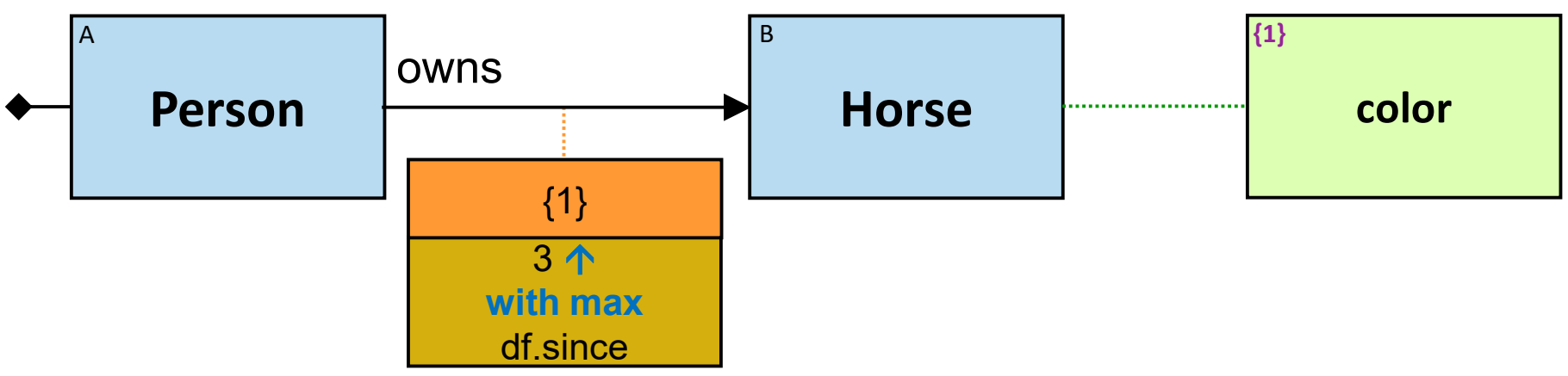

## 44. MULTIVALUED FUNCTIONS AND EXPRESSIONS

A *multivalued function* is a function that may have more than one value in its range for each value in its domain. In V1, expressions composed of multivalued functions are evaluated separately for each range value.

V1 supports the following multivalued functions:

el($S_t$) ⟶ $t$ – a single element of a set; has no evaluations when $S_t$ is empty
subset($S_t$) ⟶ $t$ – a single subset of a set; has no evaluations when $S_t$ is empty
el($B_t$) ⟶ $t$ – a single element of a bag; has no evaluations when $B_t$ is empty
subbag($B_t$) ⟶ $t$ – a single subbag of a bag; has no evaluations when $B_t$ is empty

In the following examples, *Person* has the following property: *nicknames*: *set(string)*.

***Q307:** Any person that has a nickname containing an 'a'*

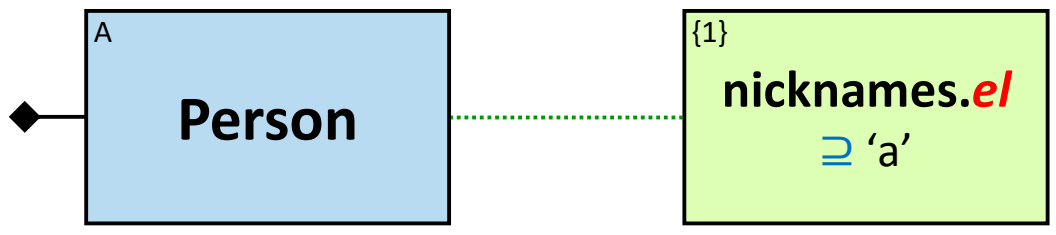

{1} is a *multivalued expression*. For each person, {1} is evaluated for each nickname. If at least one nickname satisfies the constraint – this person is a valid assignment.

***Q308:** Any person that has a nickname containing both (an 'a' or an 'A') and (a 'b' or a 'B')*

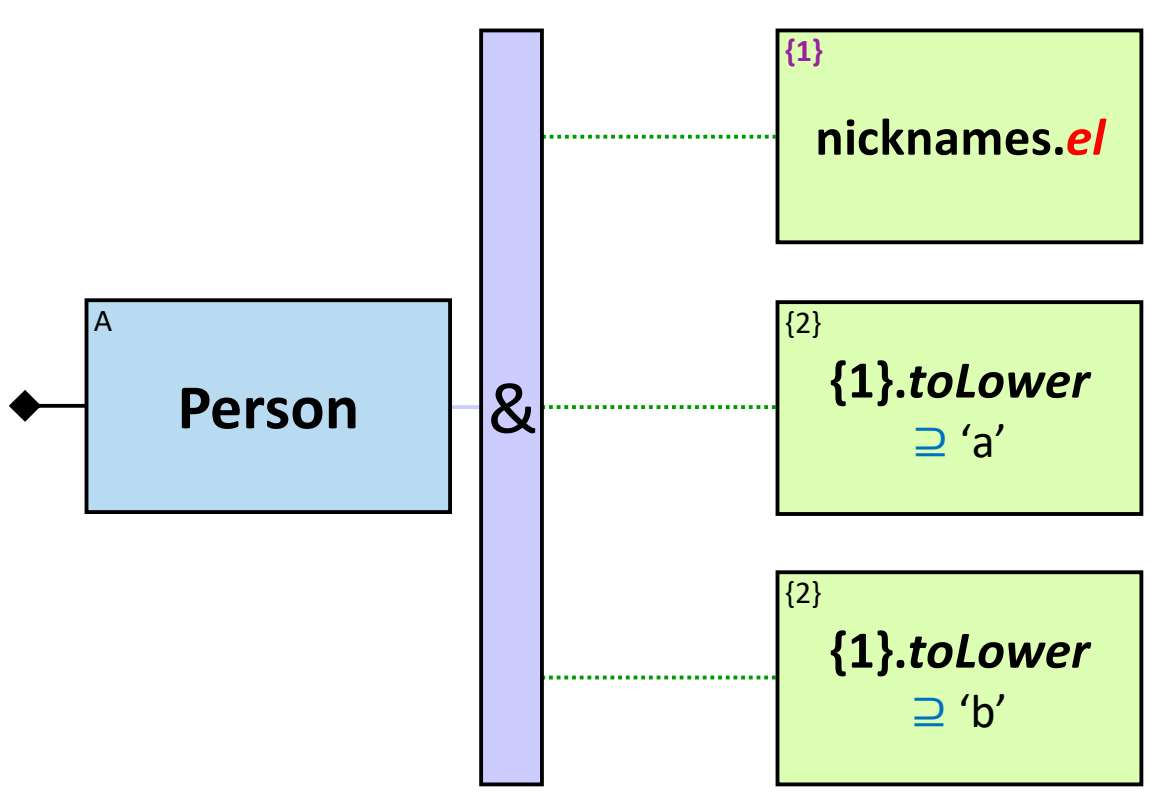

For each person, {1} is evaluated for each nickname. {2} is a multivalued expression as well and is evaluated for each value of {1}. If {1}, {2} and {3} satisfies the constraints for at least one nickname – the person is a valid assignment.

An A3 aggregator can be used for aggregating the evaluations of a multivalued expression:

- The aggregator appears below an *All* quantifier-input
- The per clause is '⟵'
- There are only entity's expressions right of the quantifier
- There is exactly one multivalued expression right of the quantifier

Consider a multivalued composite property, *names*, where each name has two sub-properties: (*first*: *string*, *last*: *string*):

***Q316:*** *Any person with a name (first: 'John', last: 'Doe')*

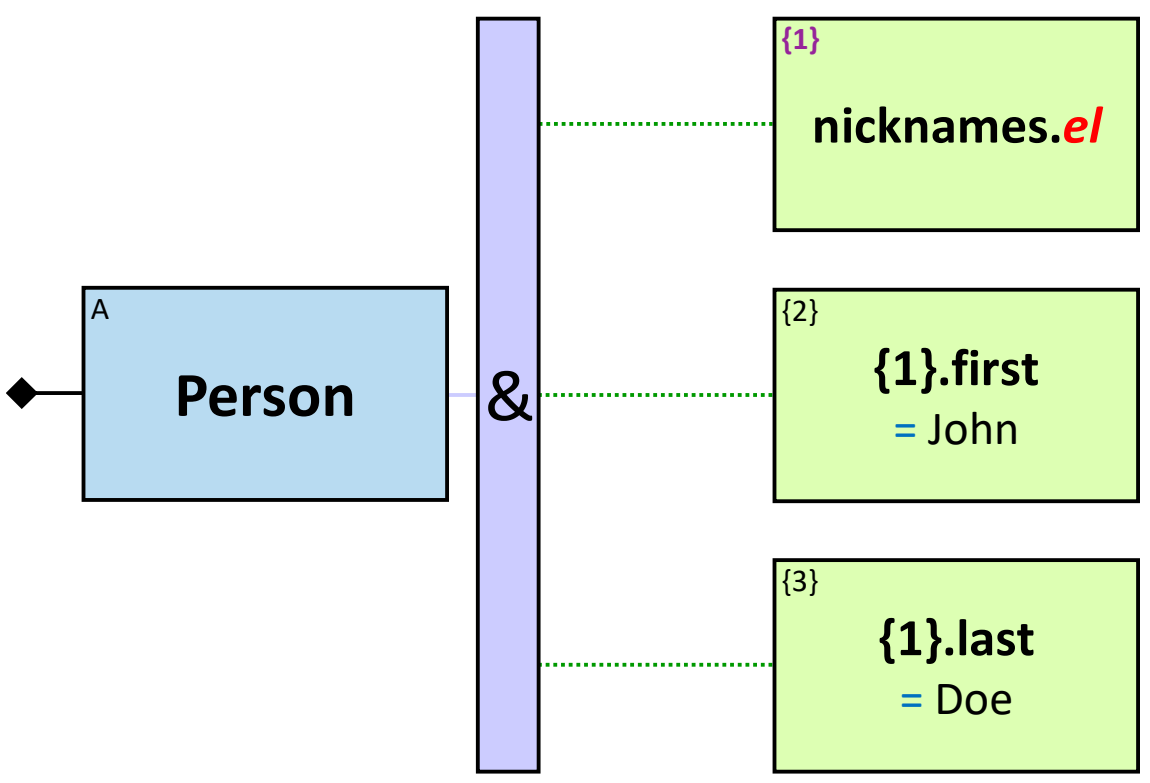

***Q309:*** *Any person that has at least two nicknames* (two versions)

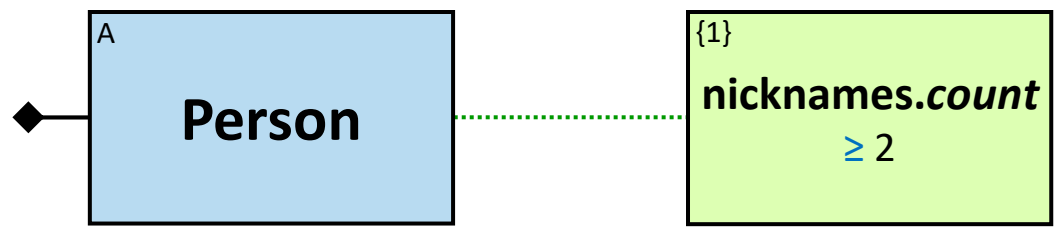

The *count* function returns the number of values a multivalued expression has. {1} is a single-valued expression.

Aggregate constraints can be defined over the number of distinct *non-null* values of a multivalued expression:

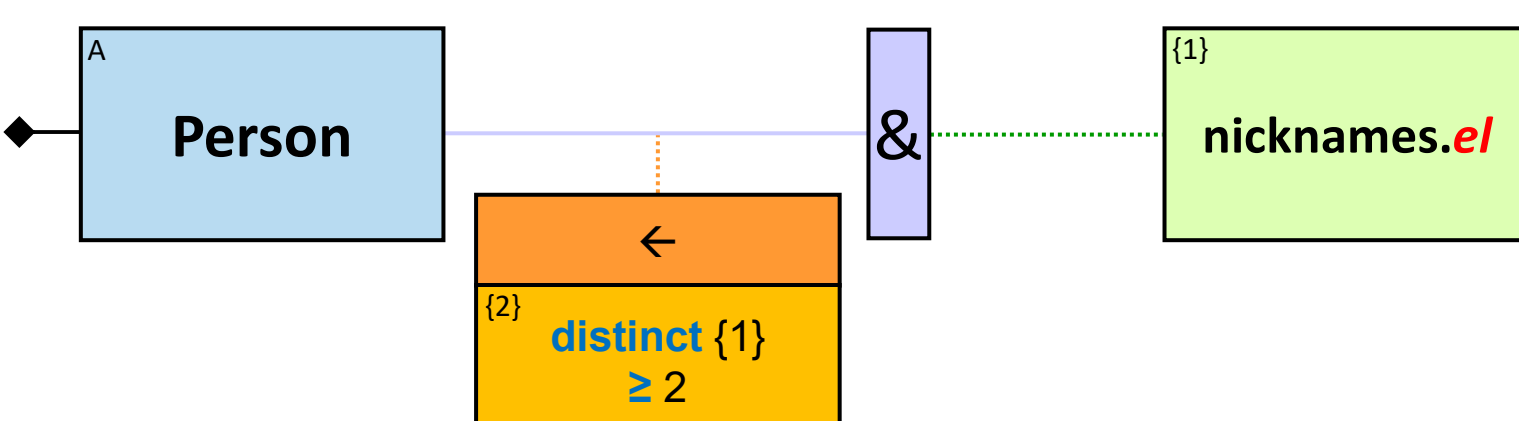

***Q310:*** *Any person that has at least two nicknames – each contains an 'a'* (two versions)

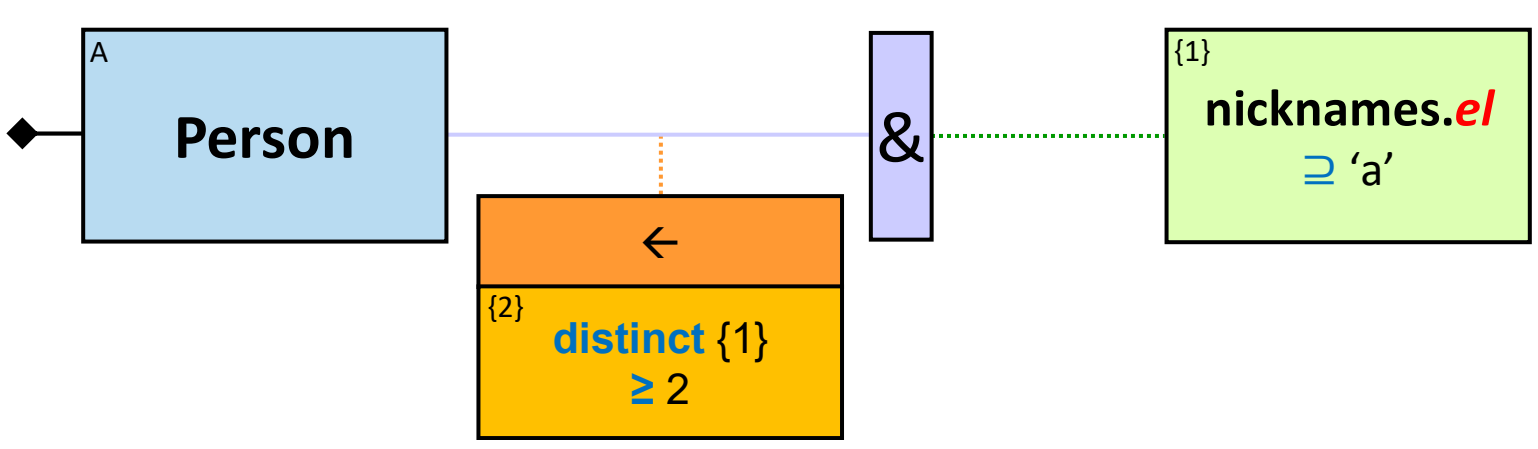

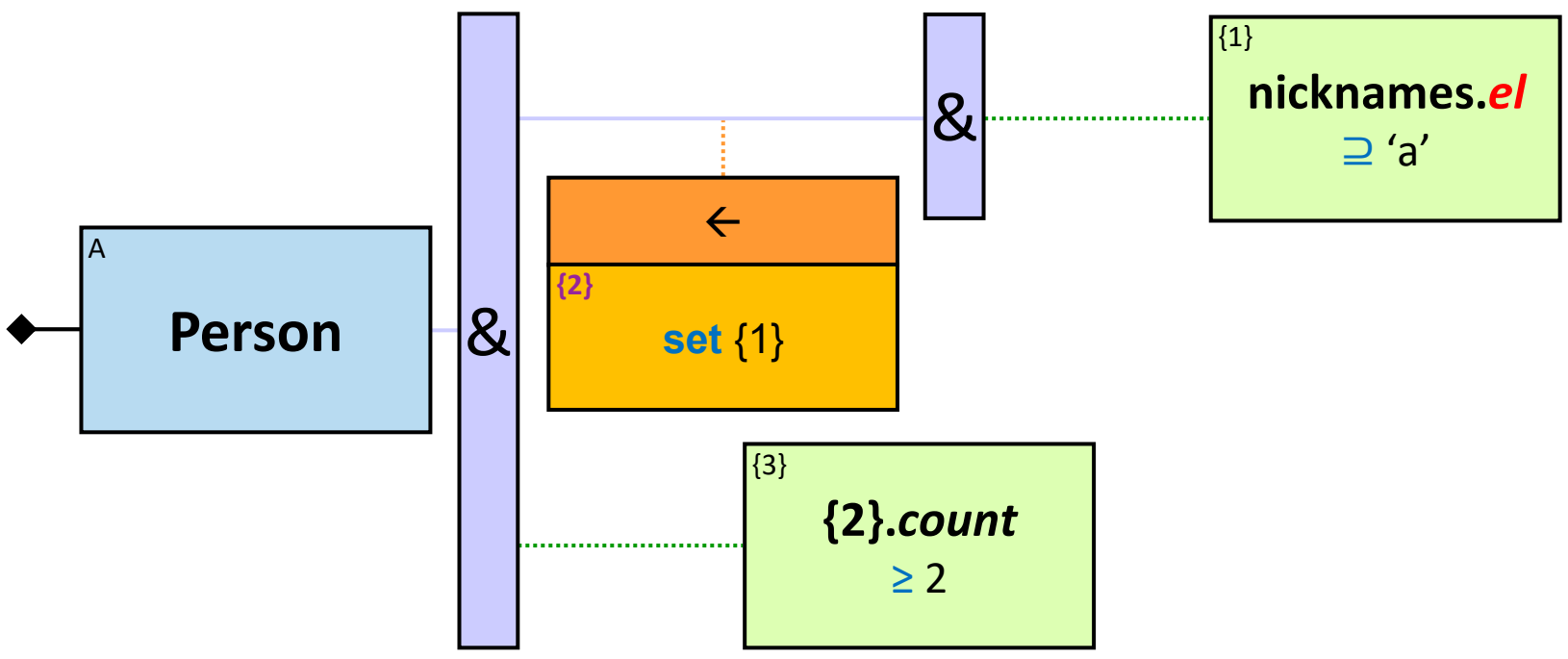

***Q311:*** *Any person that has at least five nicknames – each contains either an 'a' or a 'b'*

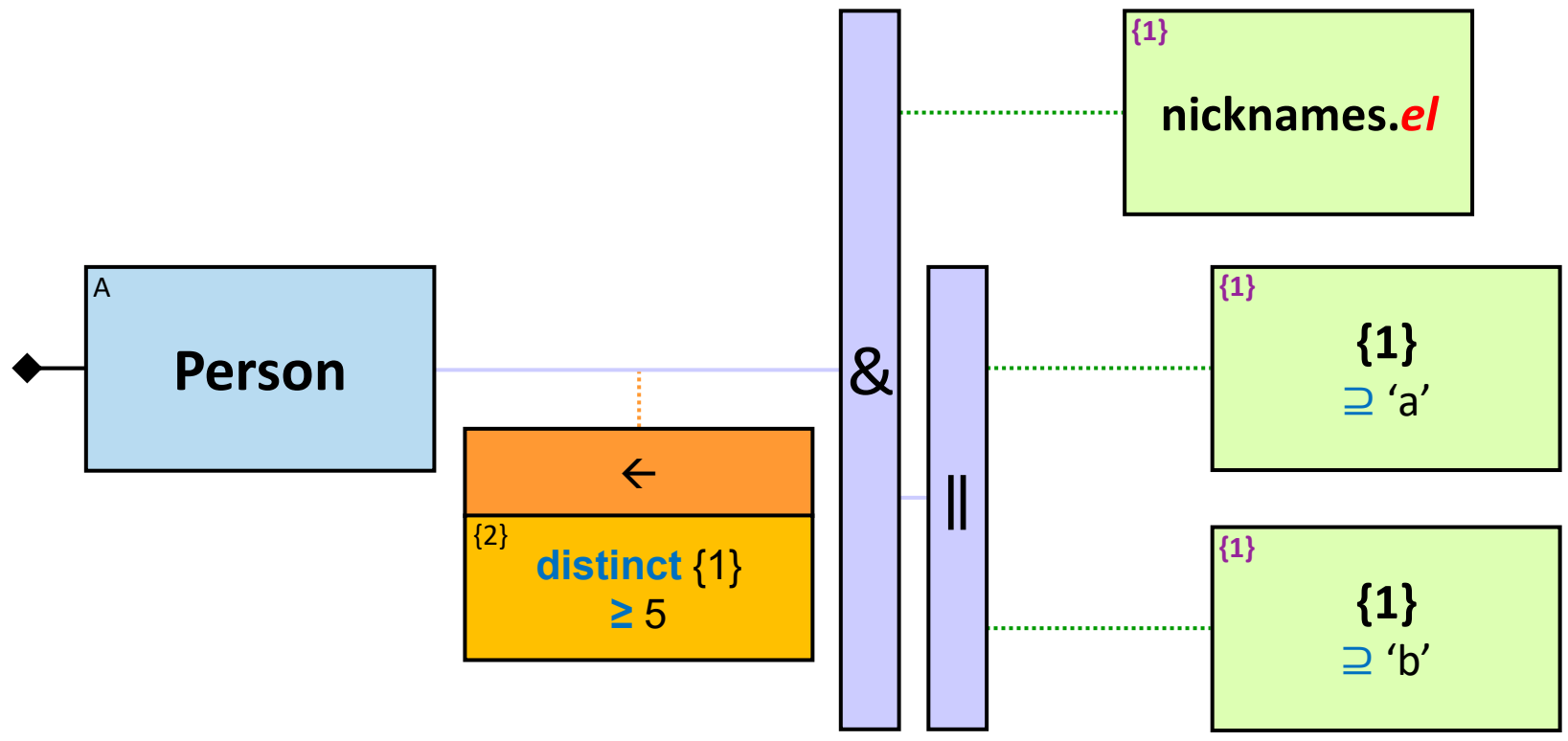

***Q312:*** *Any person with more nicknames containing a 'b' or a 'B' than nicknames containing an 'a' or an 'A'*

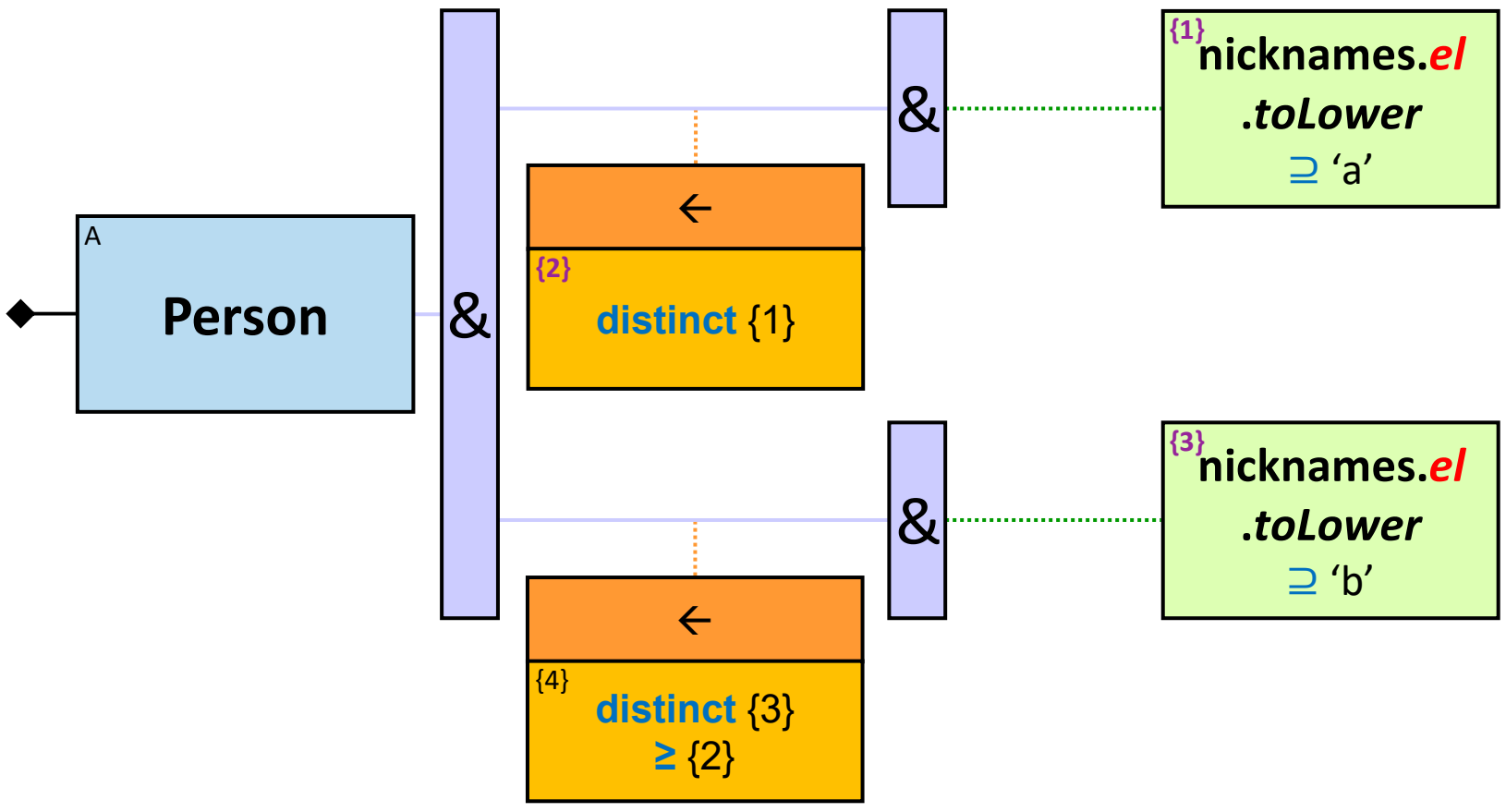

Note that if a nickname contains both an 'a' and a 'b' – it would be assigned to both {1} and {4}.

***Q313:*** *Any person A that has at least one nickname and all A's nicknames contain an 'a'*

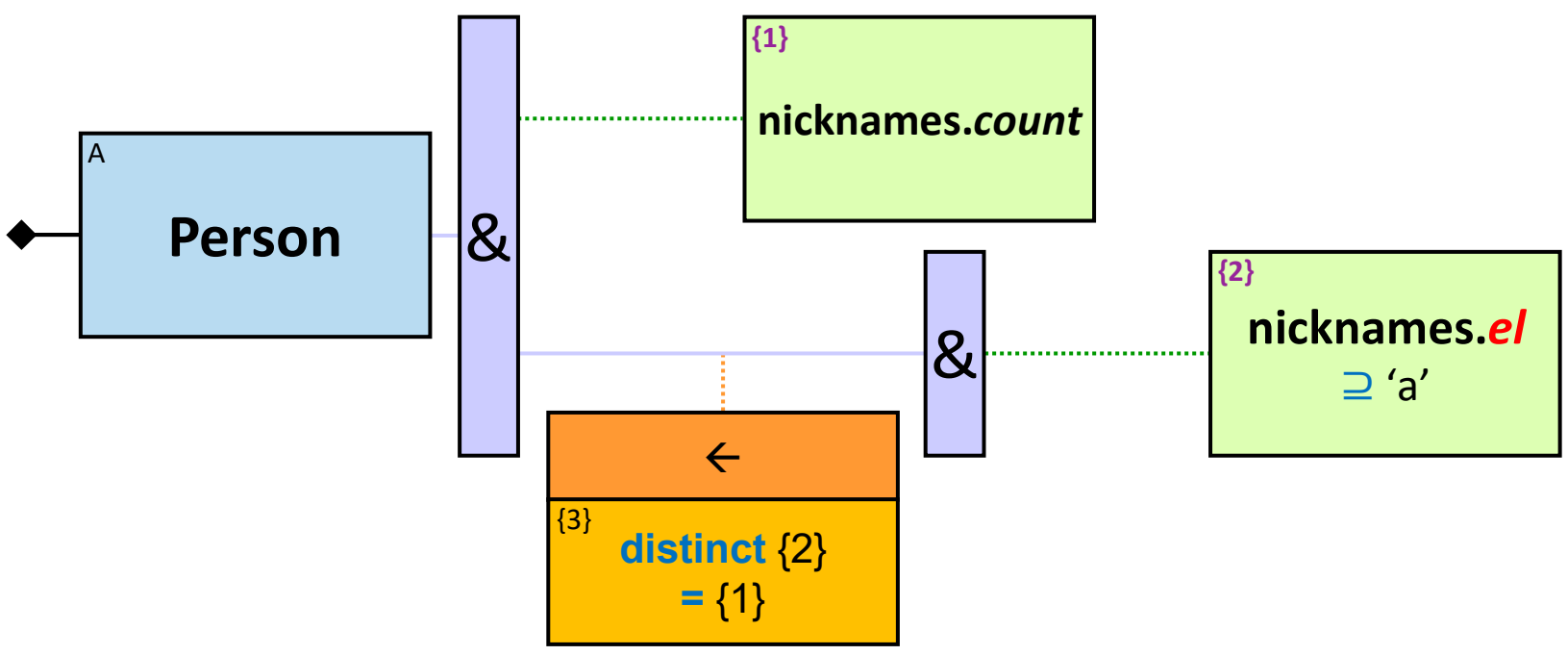

***Q314:*** *Any person that has at least one nickname, but none of its nicknames contain an 'a'*

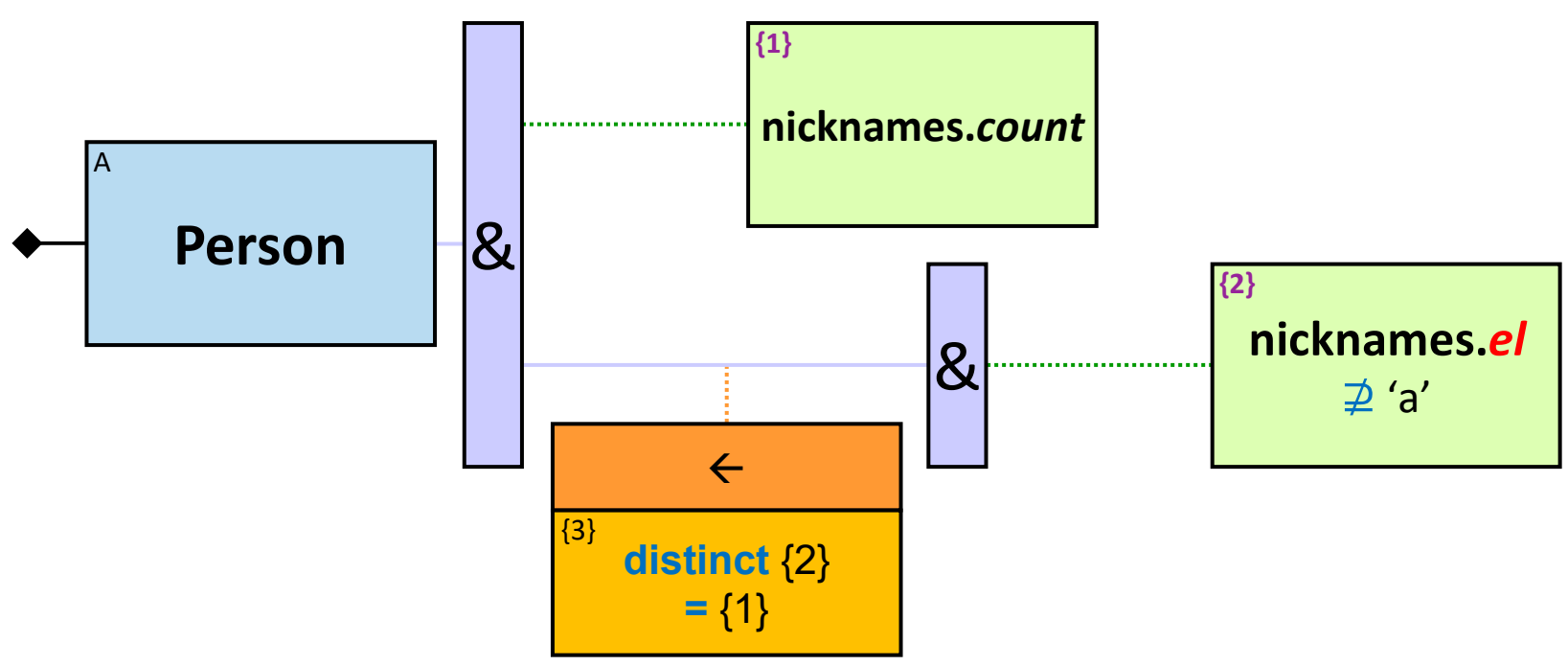

***Q315:*** *Any person A that the average length of A's three shortest nicknames is at least three*

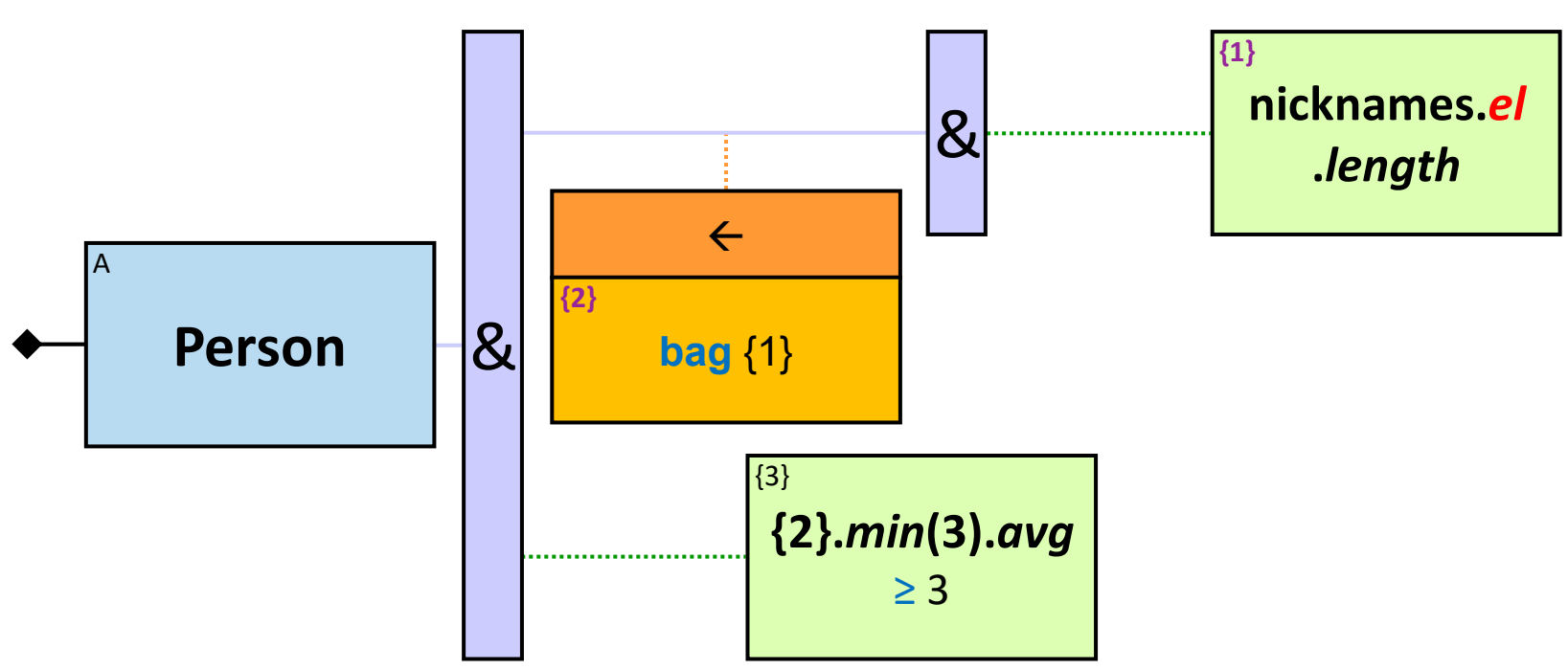

***Q348:*** *Any person A that the combined weight of three, four or five of A's horses – is exactly 1000 Kg*

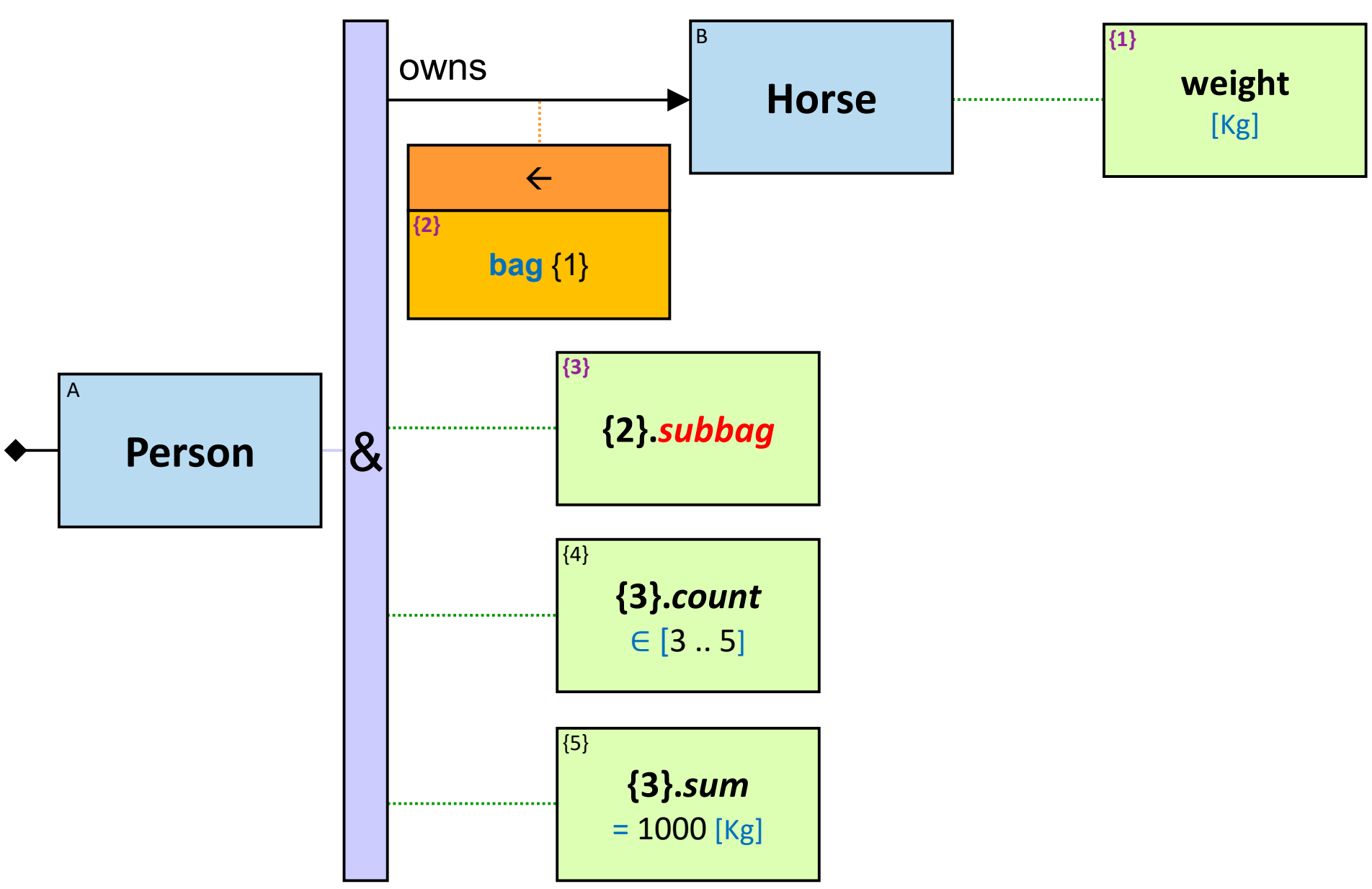

***Q349:*** *Any person A that A's horses can be split into three groups with an identical combined weight*

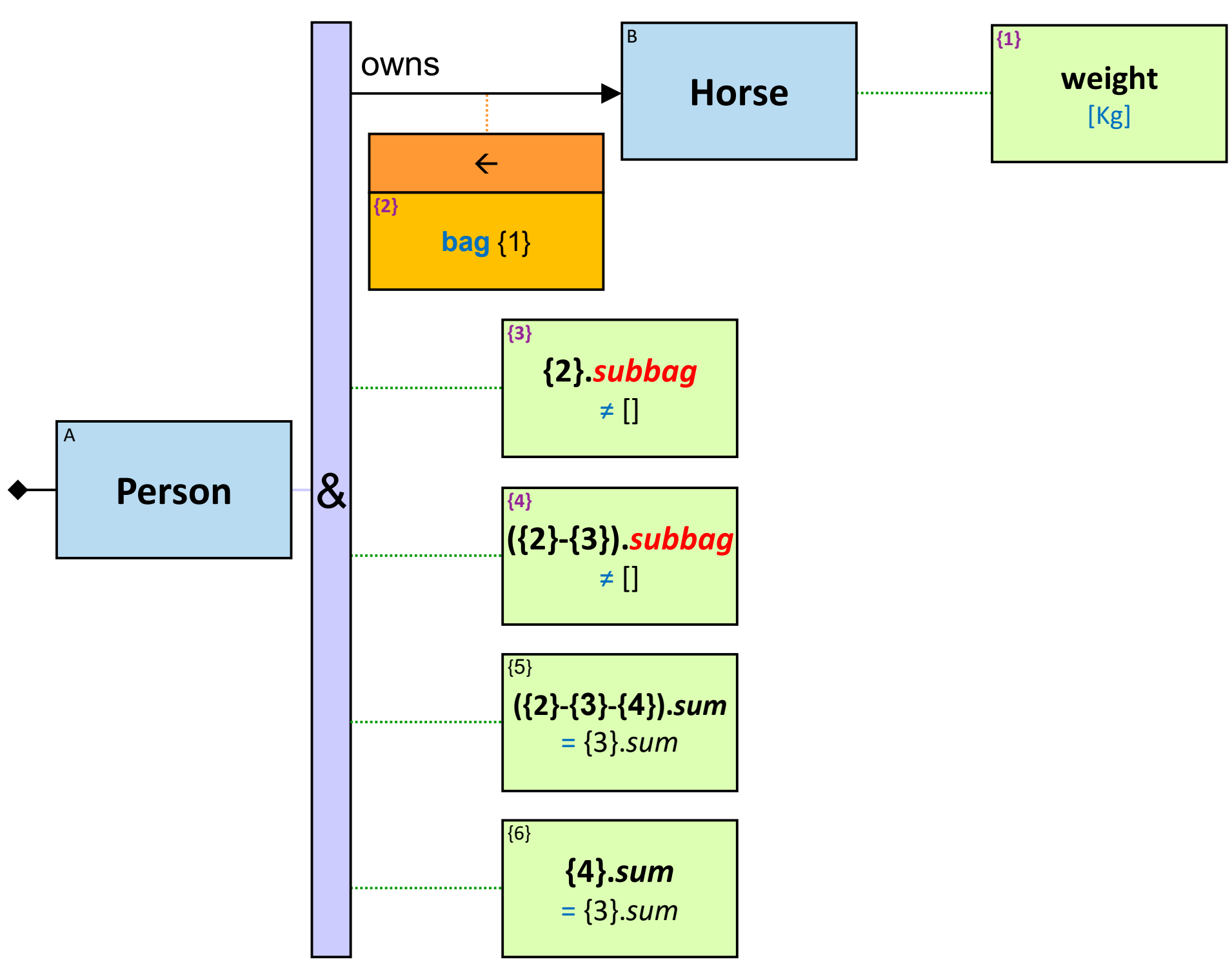

Relationships may have multivalued expressions as well.

***Q284:** Any person who owned the same five horses – for at least ten consecutive days* (version 2)

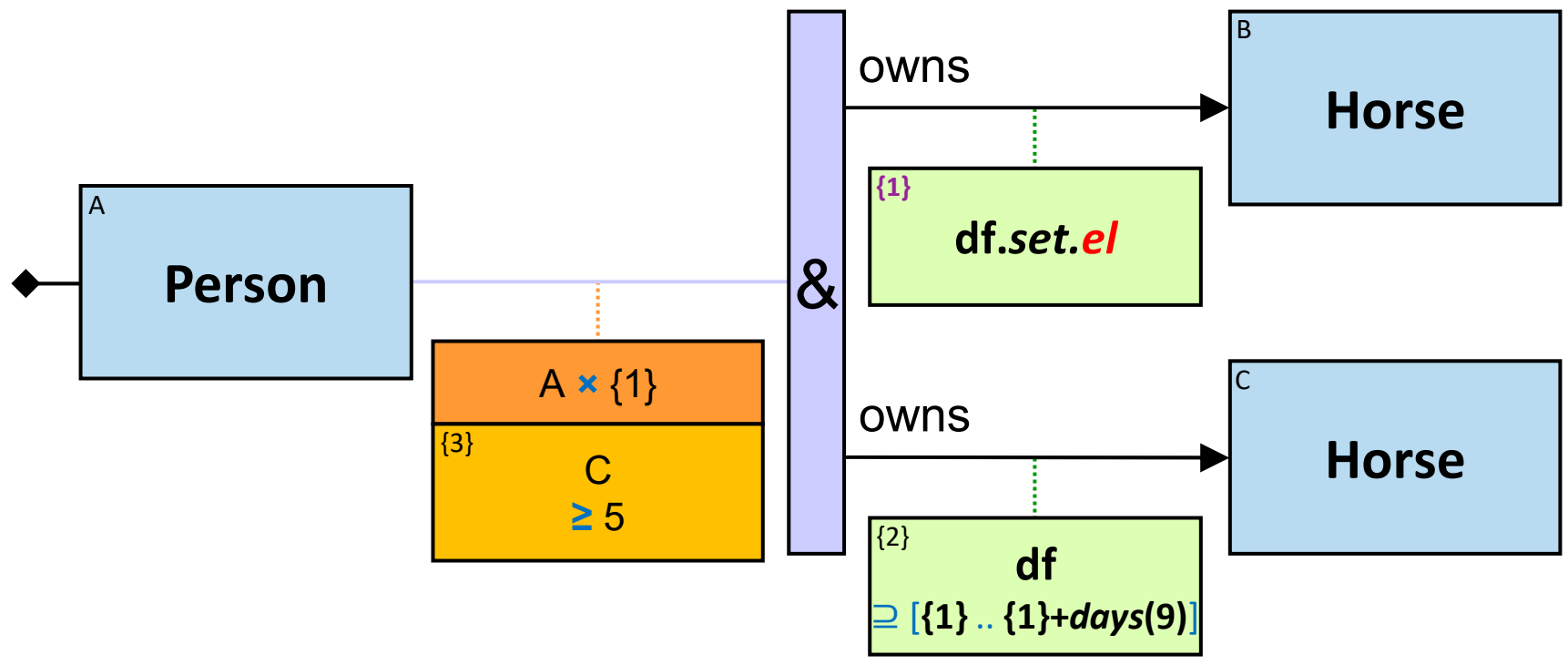

The *set* function returns a set of all individual dates within the *df* dateframe. *el* is a single element of this set (a single date). Each value is evaluated separately.

***Q340:** Any person who owned the same five horses – in at least ten (not necessarily consecutive) days*

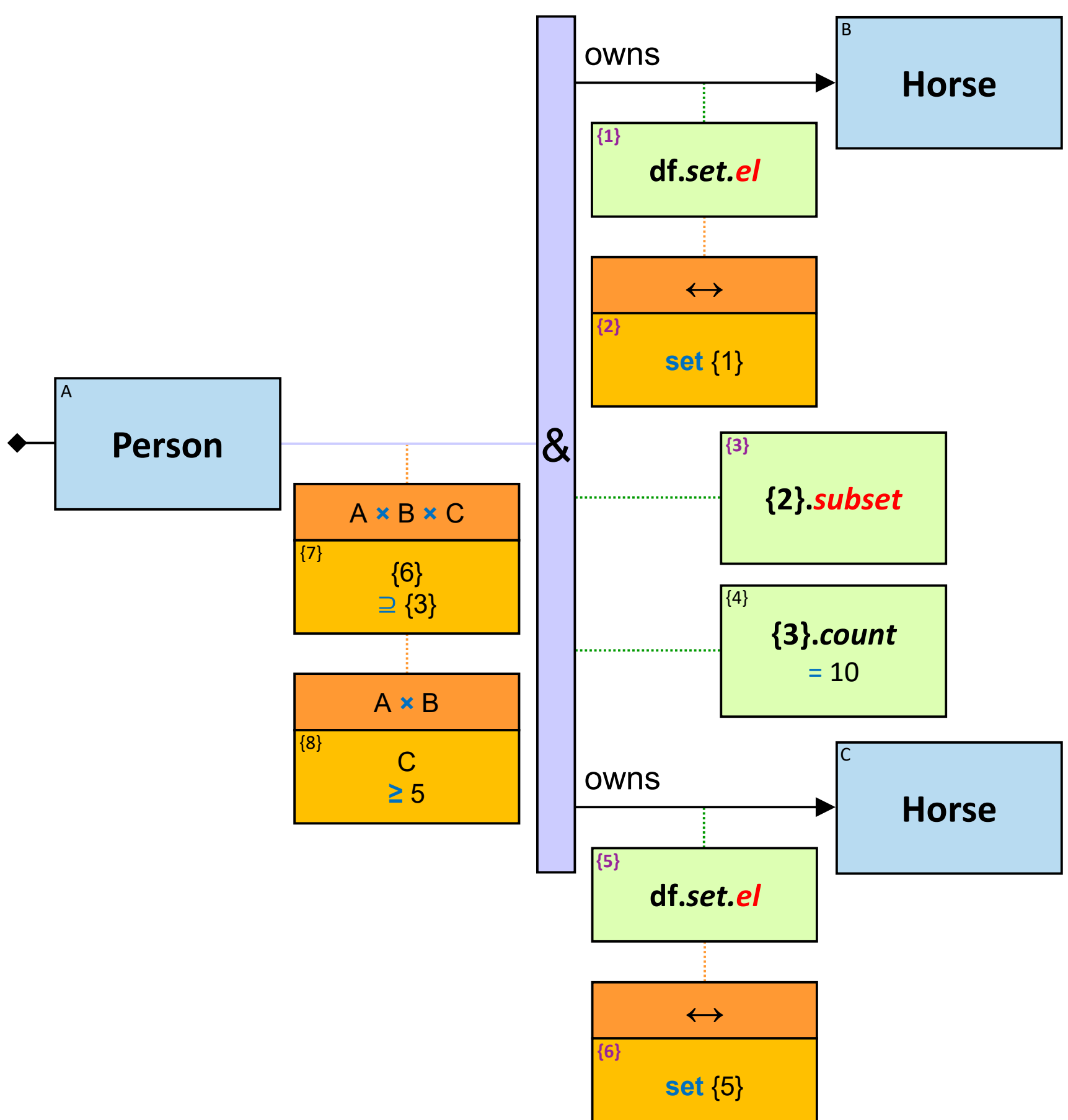

{2} and {5} are sets of all horse ownership dates, aggregated over all *owns* relationships.

Expression-tag {3} represents one subset of set expression {2}. The constraint in {4} limits assignments to {3} to subsets of {2} with exactly ten elements. {2}, {3} and {4} are properties of A × B.

*Q287: Any person who at each day of at least ten (not necessarily consecutive) days – owned the same two horses*

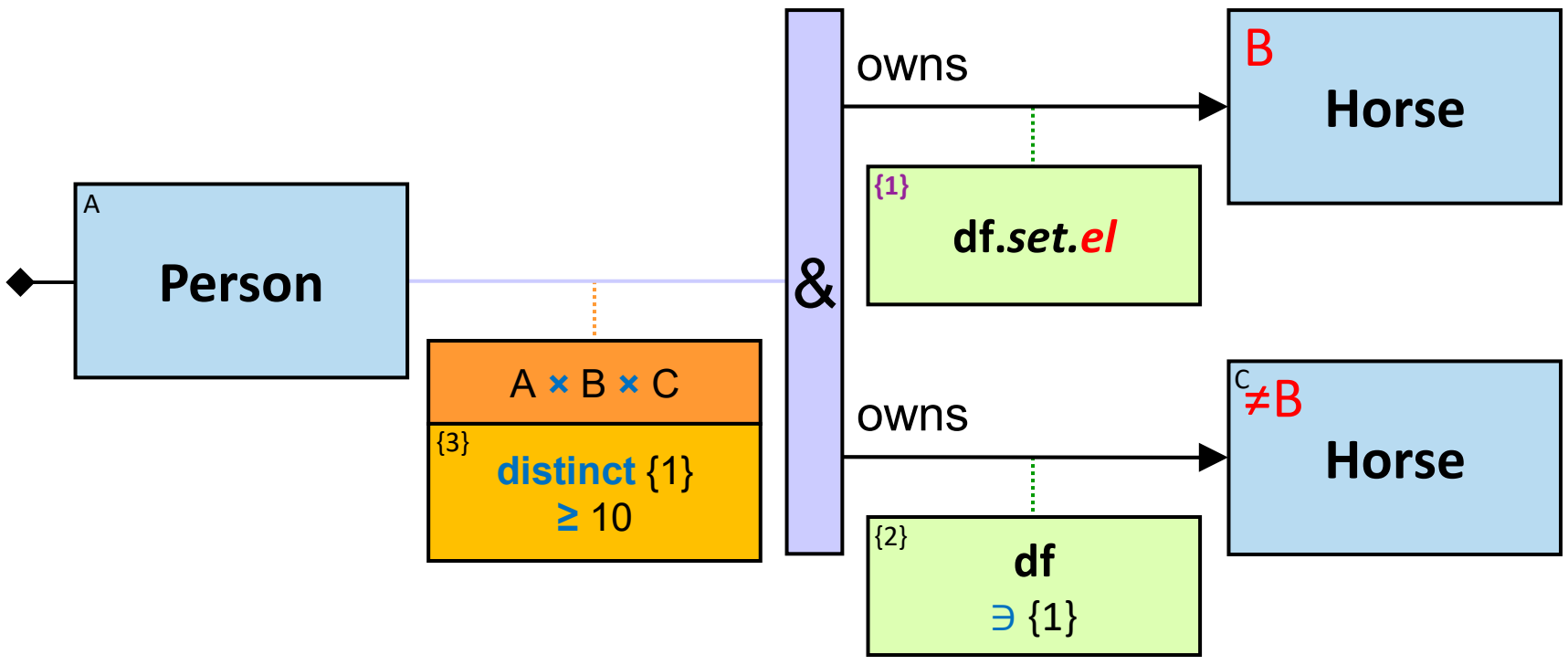

*Q327: Any person who at each day of at least ten consecutive days – owned at least five horses (not necessarily the same horses each day)*

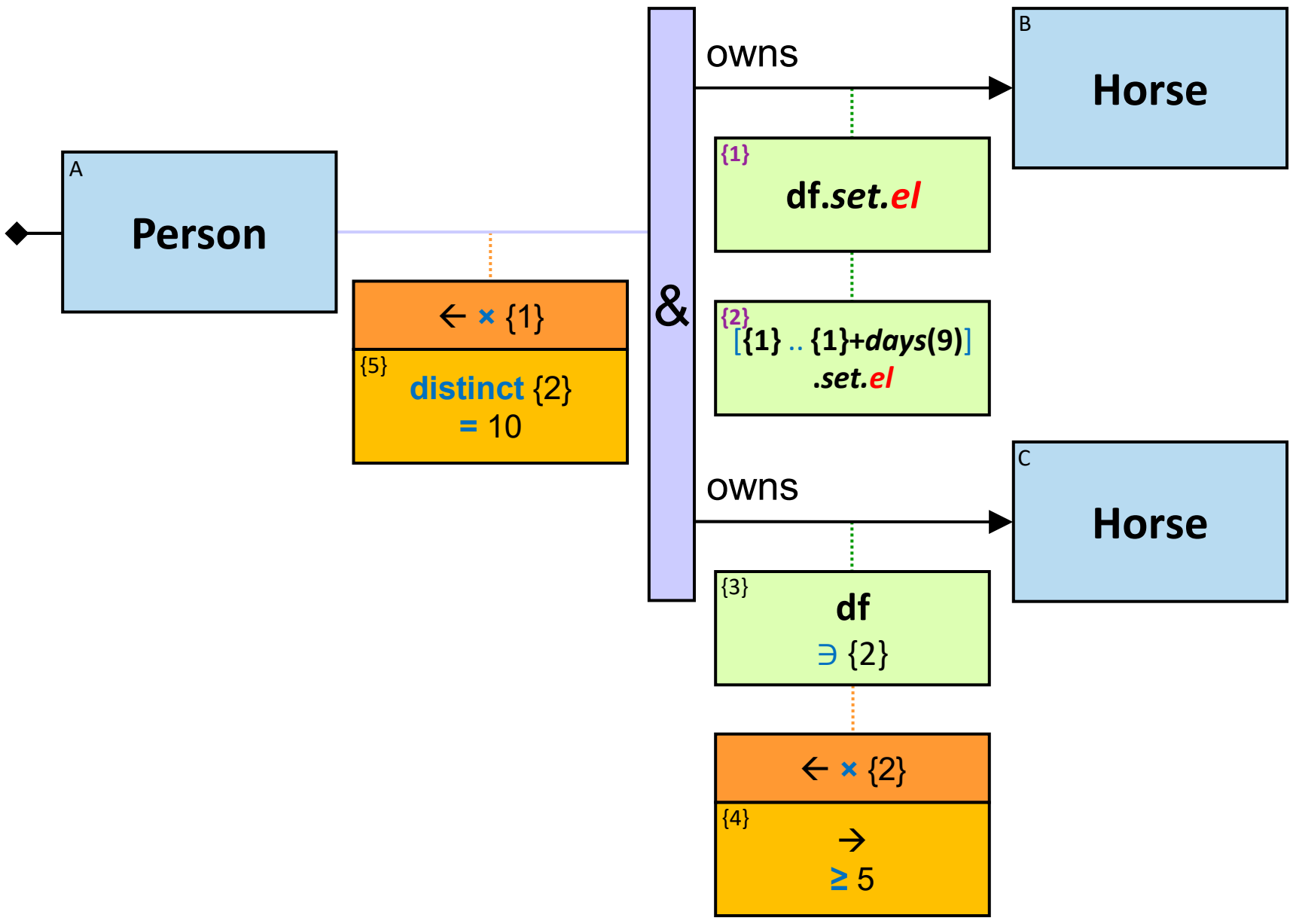

***Q318:*** *Any dragon that Balerion froze/fired at in the three 30-day timeframes in which it froze/fired at dragons the largest number of times*

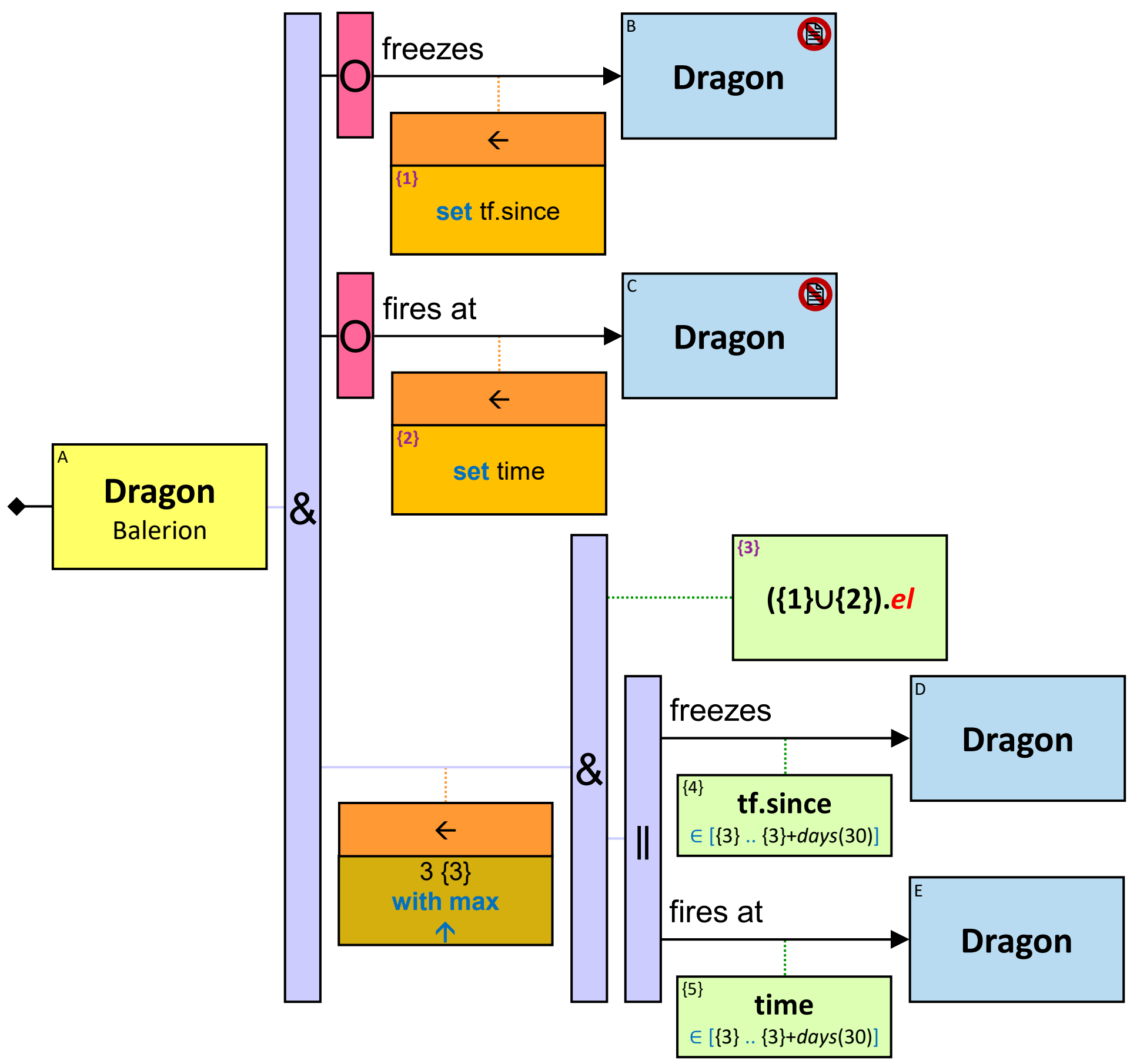

Note that {1} and {2} are defined right of an 'O', hence evaluated to an empty set when the optional part has no assignment. Since $s \cup \varnothing = s$, the pattern is valid even if Balerion froze no dragon or if it fired at no dragon.

{3} is a property of A, but, being an aggregated multivalued expression, must be defined right of the aggregator.

*Q371: Any dragon that Balerion froze/fired at in the three non-overlapping 30-day timeframes in which it froze/fired at dragons the largest number of times*

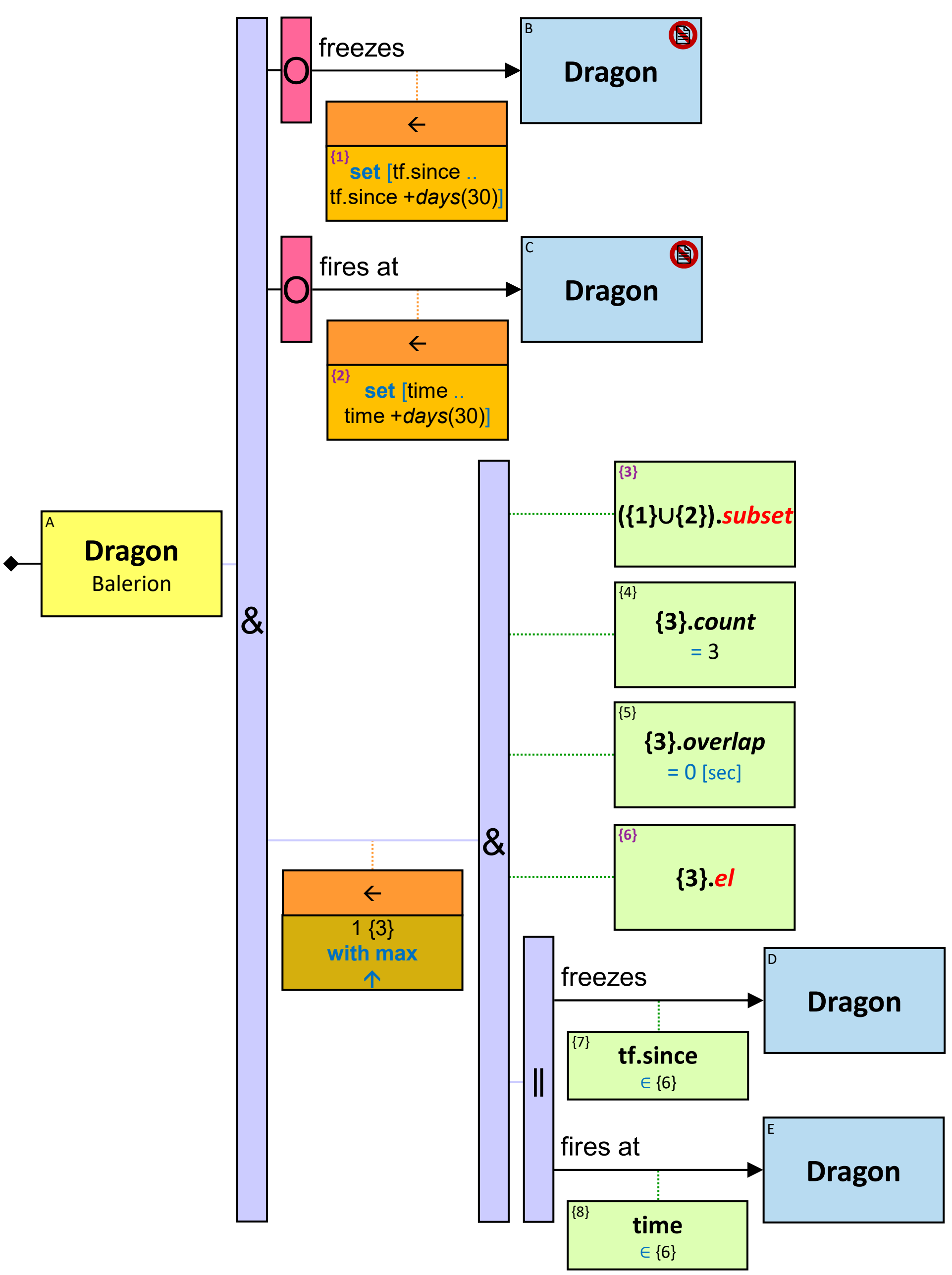

{1}, {2} and {3} are sets of intervals of *datetimeframe*. In {5}, {3} is coerced to set of *datetimeframes*. {4} and {5} limit assignment to {3} to sets with three non-overlapping members.

{6} is a multivalued expression evaluated per each value of the multivalued expression {3}. The M2 aggregator is evaluated for all assignments to {6} of each assignment to {3}.

## 45. APPLICATION: SPATIOTEMPORALITY

Dragon-spotting is a major hobby around the world. Dragon-spotters document the time, location, and of course – the identity of the dragon they spotted. The location may be a point (longitude and latitude) or a small area – when they are not sure about the exact coordinates. Most observations are attributed to the spotter, but in many ancient documents – the dragon-spotter is unknown.

We will extend our sample schema with the following relationship-type:

- *spotted*: {(*Person*, *Dragon*), (*Null*, *Dragon*)} – *time*: *datetime*, *loc*: *location*

*spotted* is a relationship between a dragon-spotter (which may be unknown) and the dragon spotted. The properties of the relationship are the time and the location where the dragon was spotted.

***location*** is an *opaque data type* that contains a geographic point (e.g., latitude and longitude) or a geographic shape (a circle, an ellipse, a polygon, etc.)

Constraint operators over *location* expressions:

| **Operator (*op*)** | **Operands type (result is false / true / unknown)** |
|---|---|
| ⊃, ⊇, ⊅, ⊉ ([not] [proper] superspace of) | both operands of the same type: location |
| ⊂, ⊆, ⊄, ⊈ ([not] [proper] subspace of) | both operands of the same type: location |

Functions over *location* expressions:

| **Function** | **Notes** |
|---|---|
| *dist*(*location*, *location*) → *float* | the distance [Km] between these two locations |

Two entity-types are defined as well:

- **Landmark** – loc: *location*
- **City** – loc: *location*

Note that we chose to hold the location of stationary entities (landmark, city) as entity properties and hold the location of mobile entities (dragon) as relationship properties.

***G1:*** Any dragon observed at Dragonmont Peak

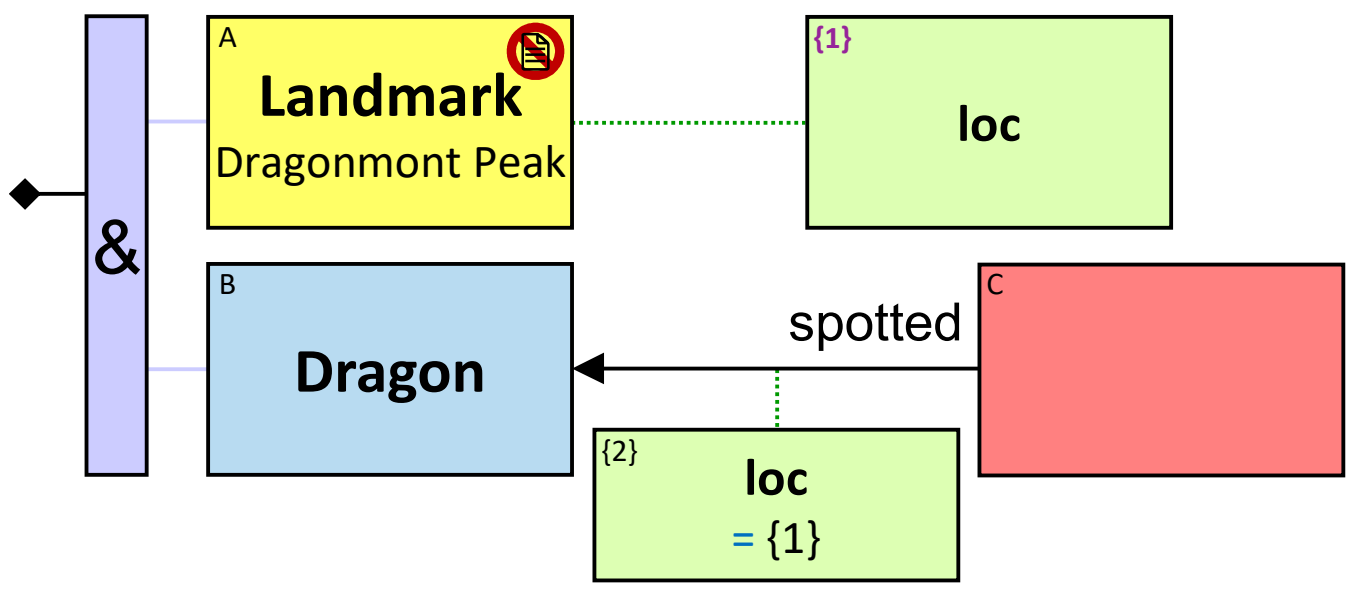

***G2:** Any dragon observed within 5 Km from Dragonmont Peak*

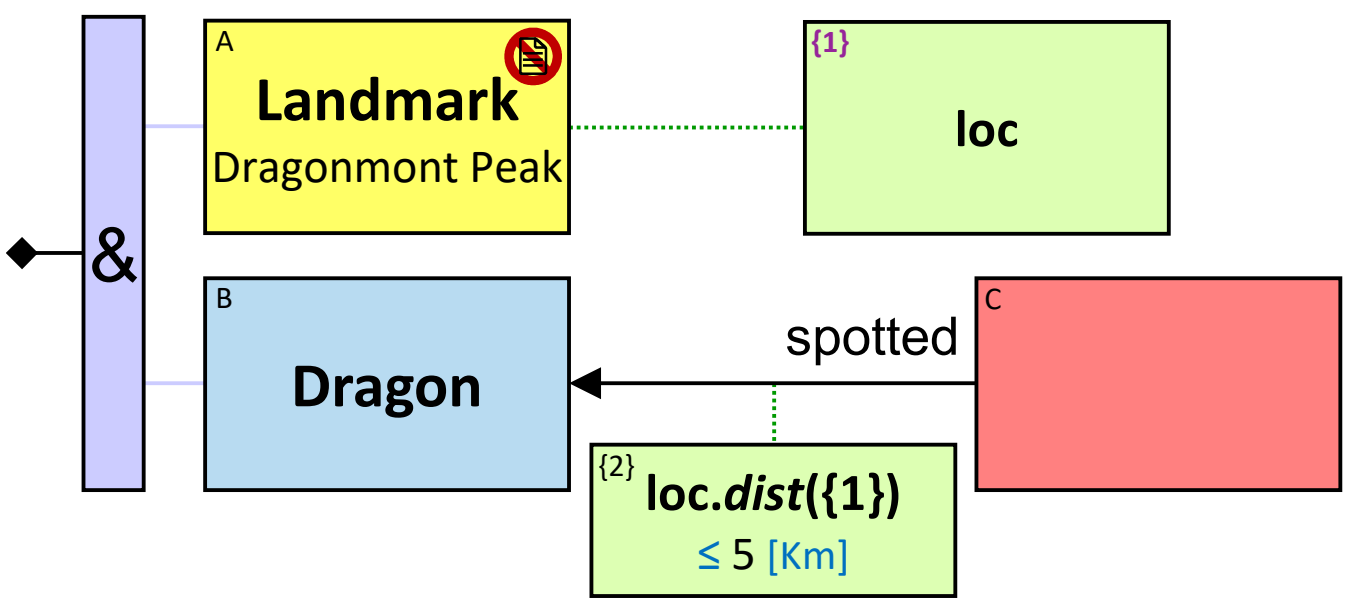

***G3:** Any dragon observed within 5 Km from Dragonmont Peak at least five times during a 7-day period*

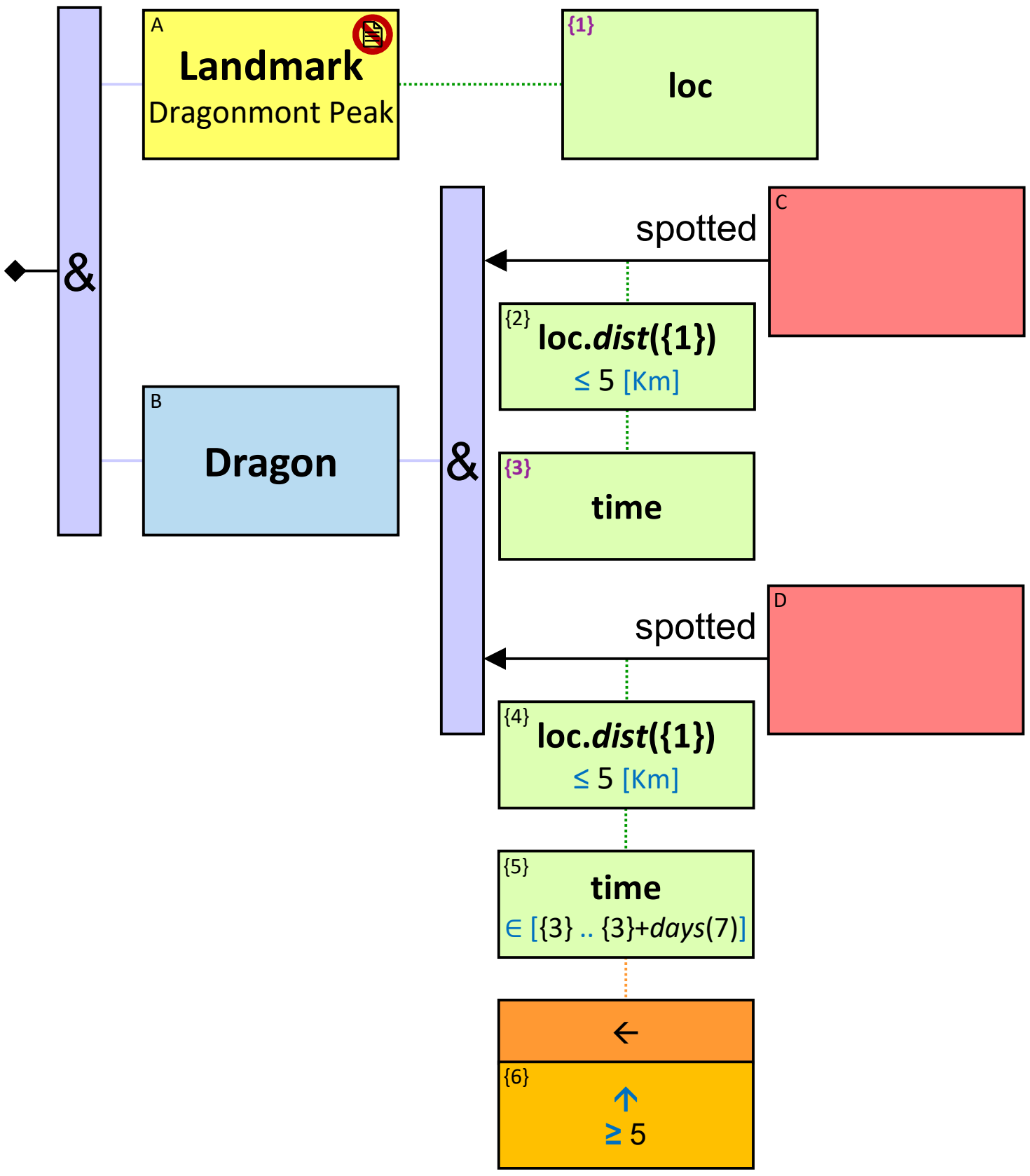

***G4:*** *Any dragon observed within 5 Km from Dragonmont Peak between 4 AM and 7 AM in at least five days during a 7-day period*

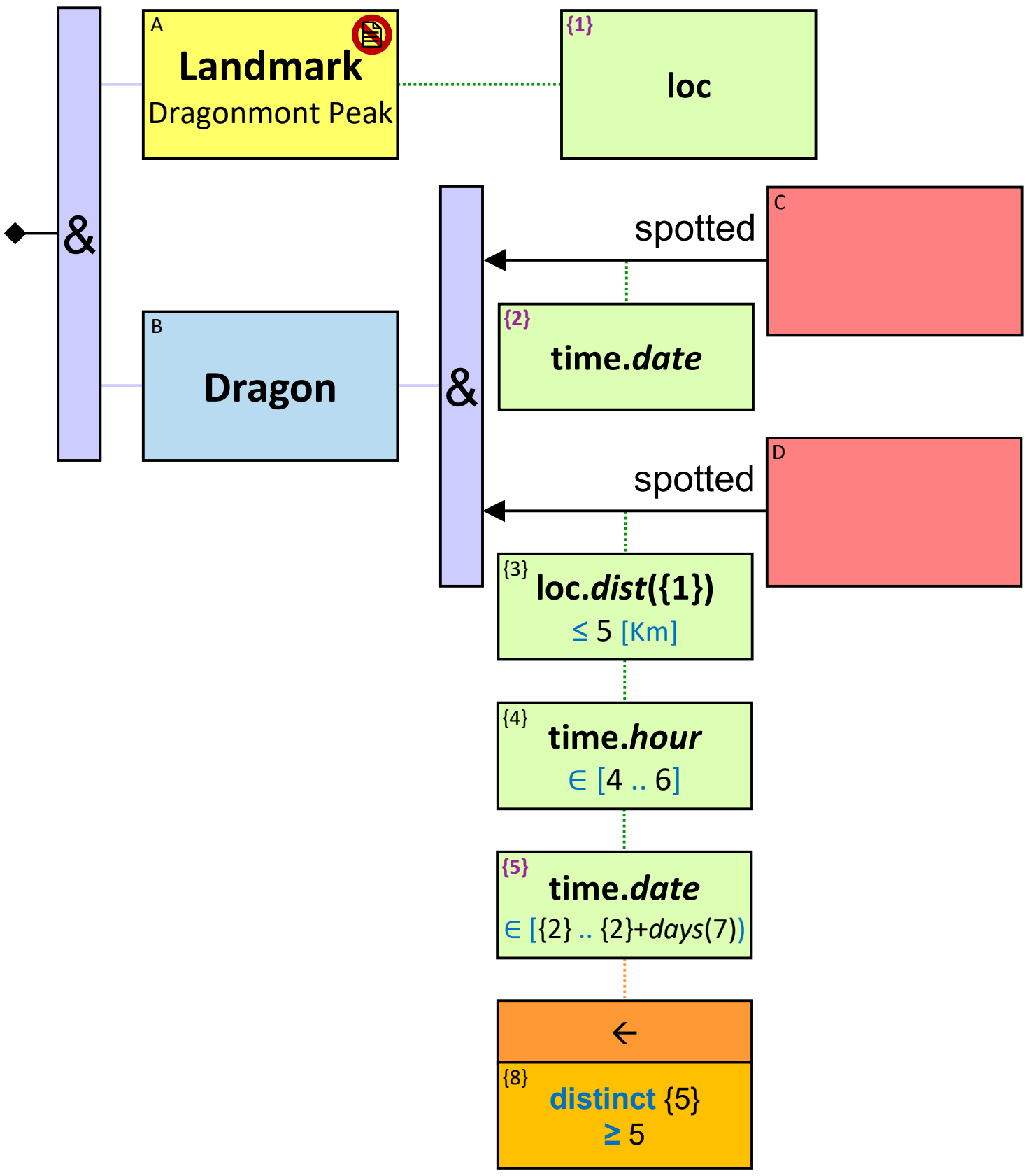

***G5:*** *Any dragon observed within 5 Km from Dragonmont Peak in at least five separate weeks*

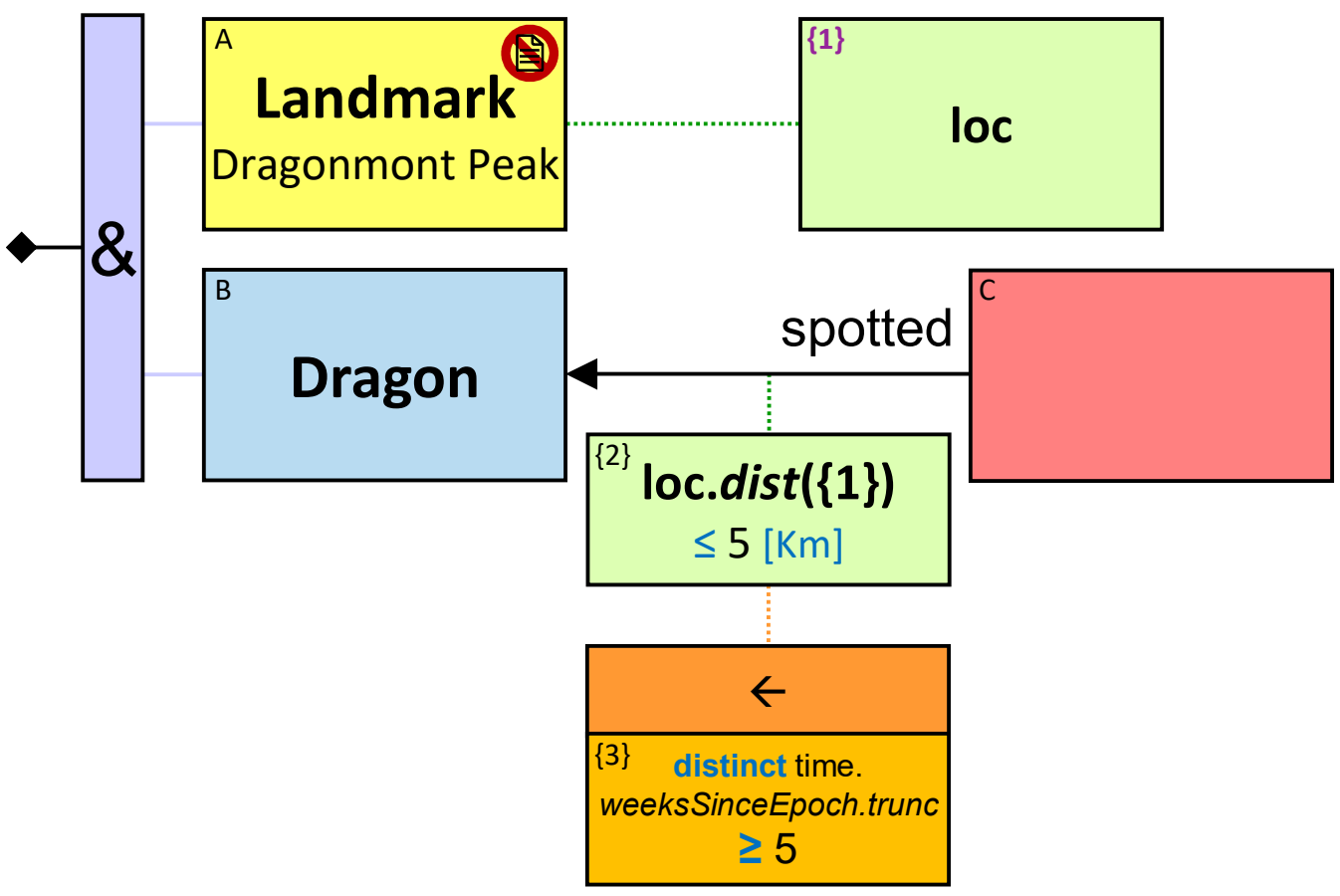

***G6:** Any week in which at least five dragons were observed within 5 Km from Dragonmont Peak*

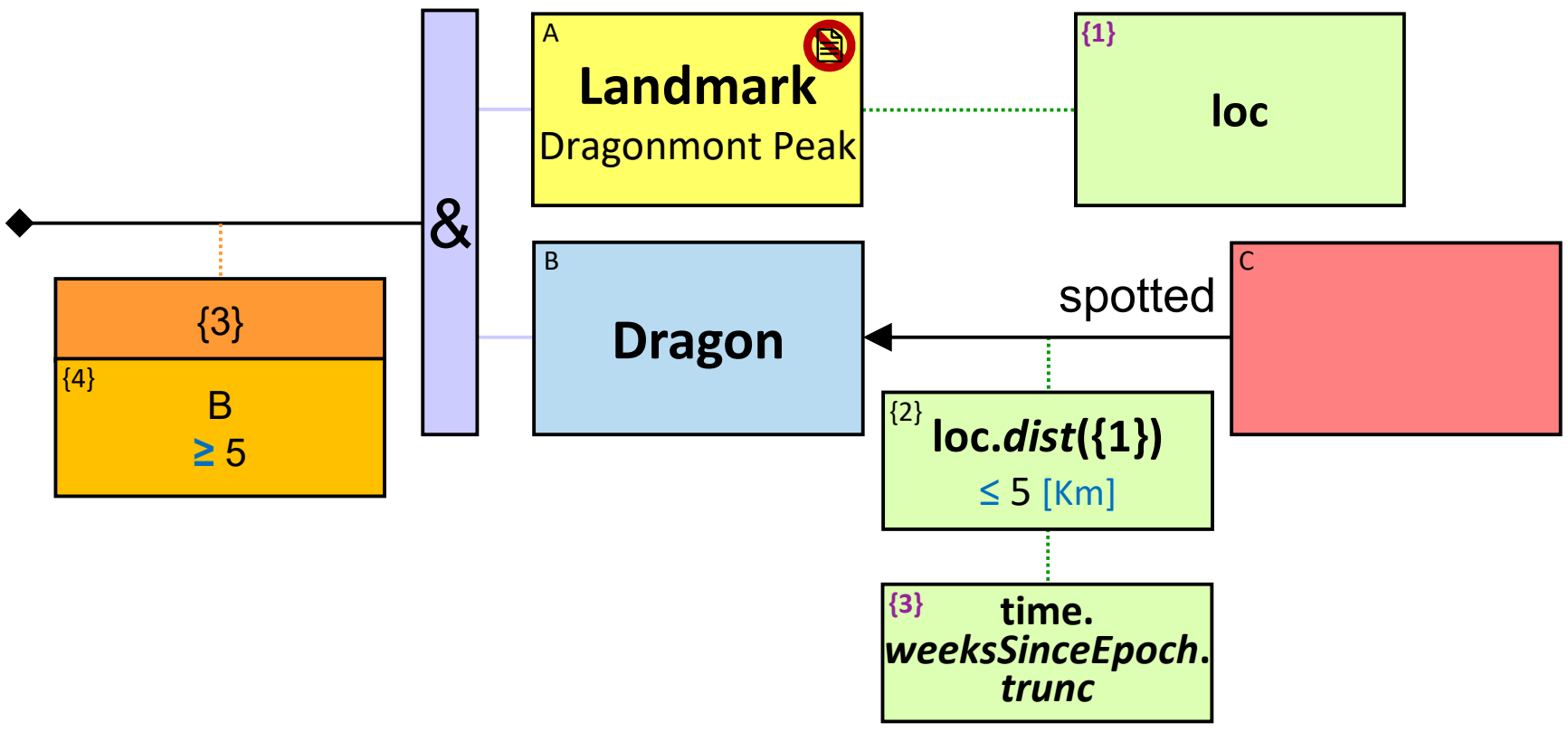

***G7:** Any dragon with conflicting observations (given that dragons cannot fly faster than 300 Km/h)*

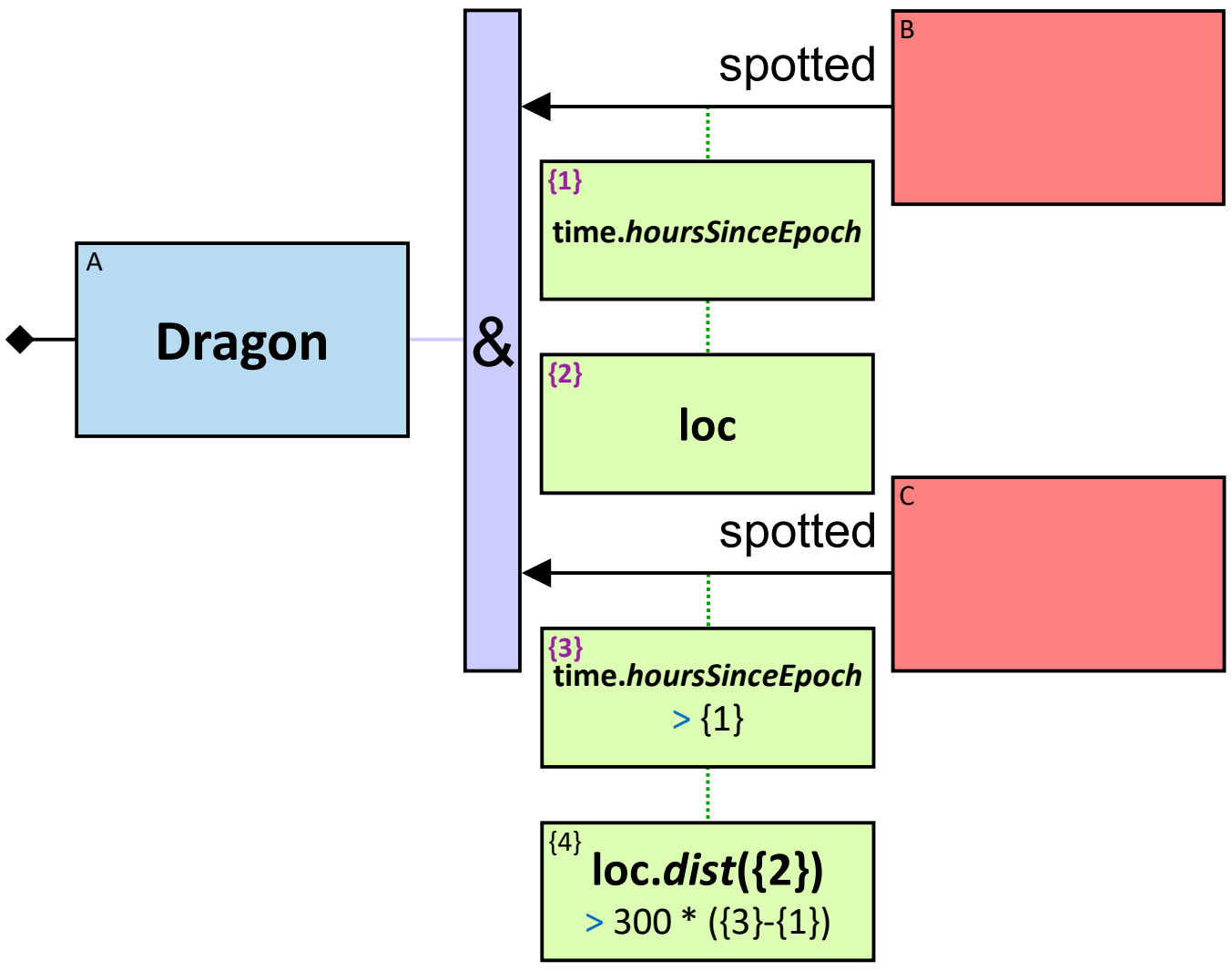

*G8: Any Sarnorian B for whom at least three of B's dragons were spotted within 5 Km from Dragonmont Peak – in a timeframe of 24 hours*

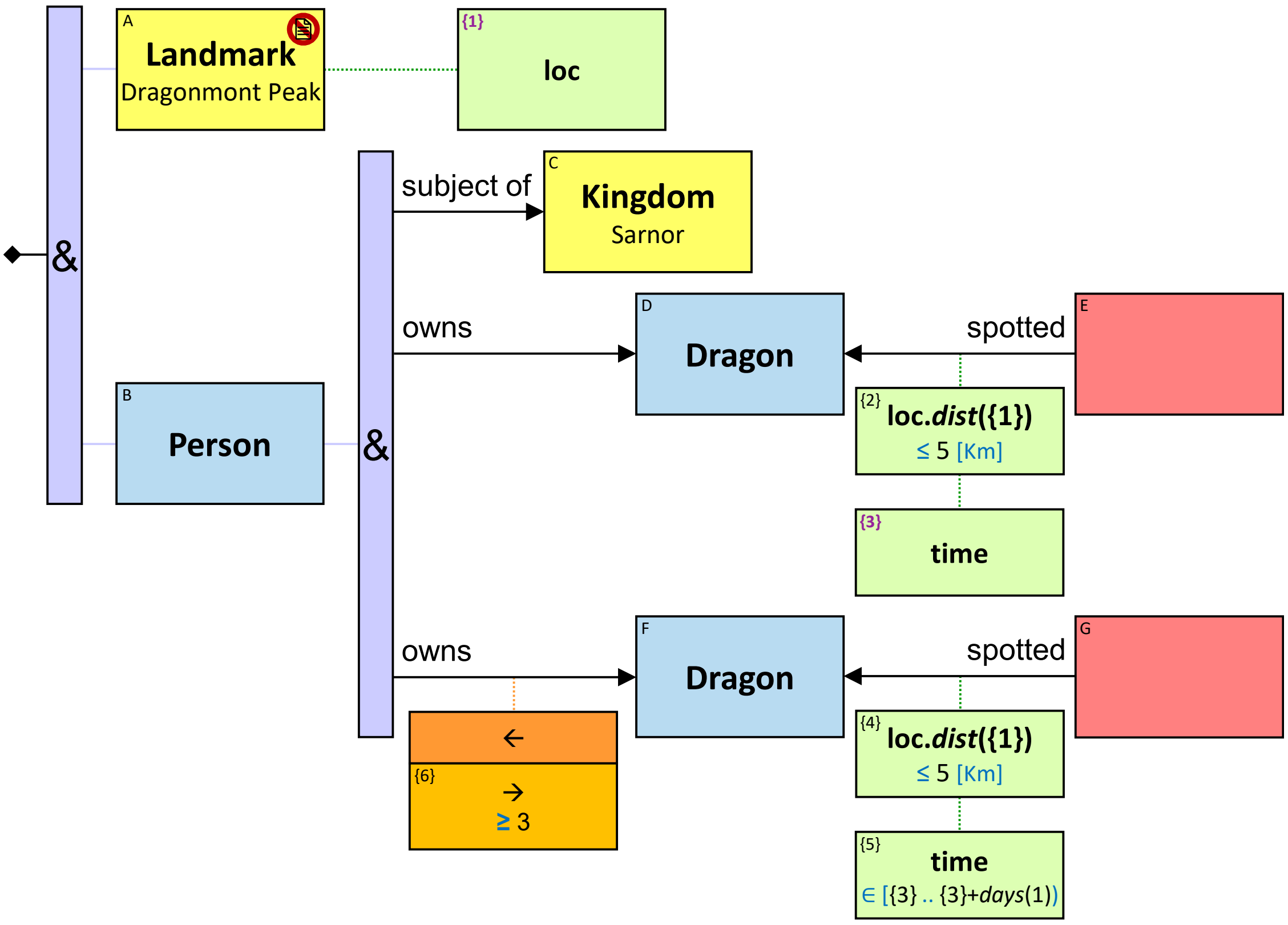

***G9:** Any dragon that stayed within 5 Km from Dragonmont Peak for at least one week (To ensure this – the dragon should have been spotted there at least once per 6 hours and not spotted at any other location throughout that period)*

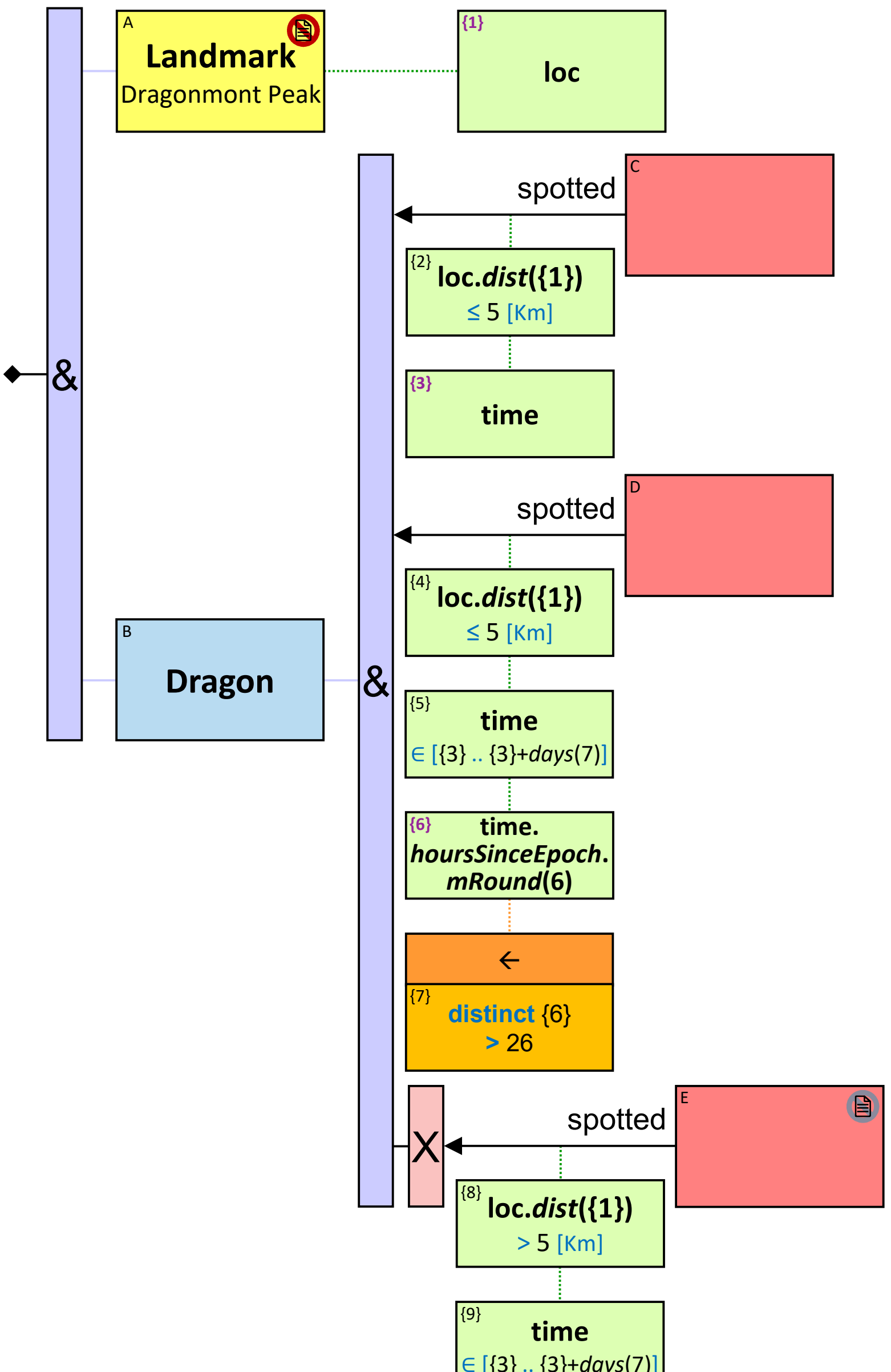

***G10:** Any pair of dragons that were spotted within 5 Km from each other at least five observation-pairs (To ensure this – paired observation should differ in up to 5 minutes; observation pairs should differ in at least 24 hours)*

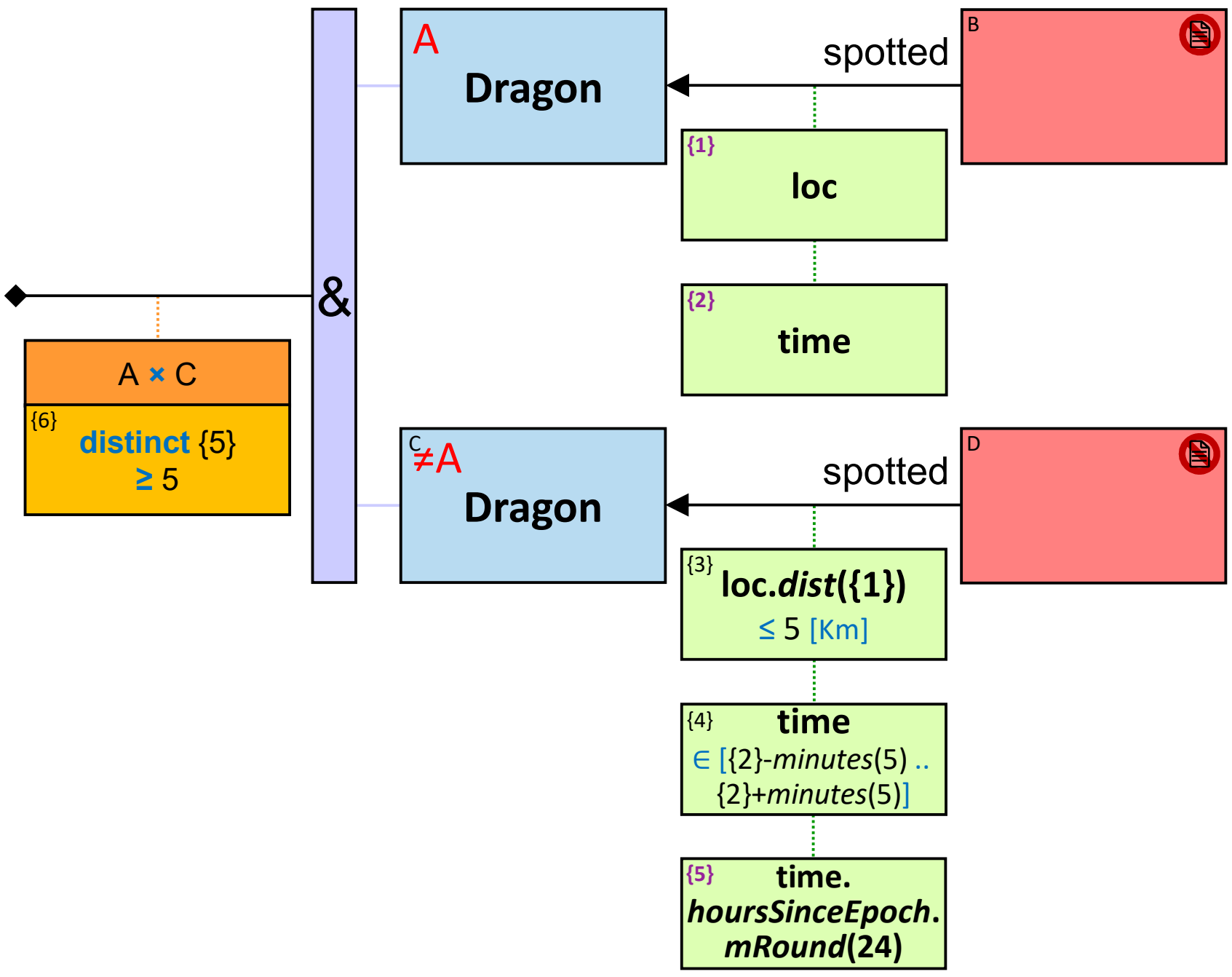

***G11:** Any pair of dragons A and B, where at least ten times in the last 300 days A was spotted at a distance of less than 1 Km from places where B was spotted between one day and five days before*

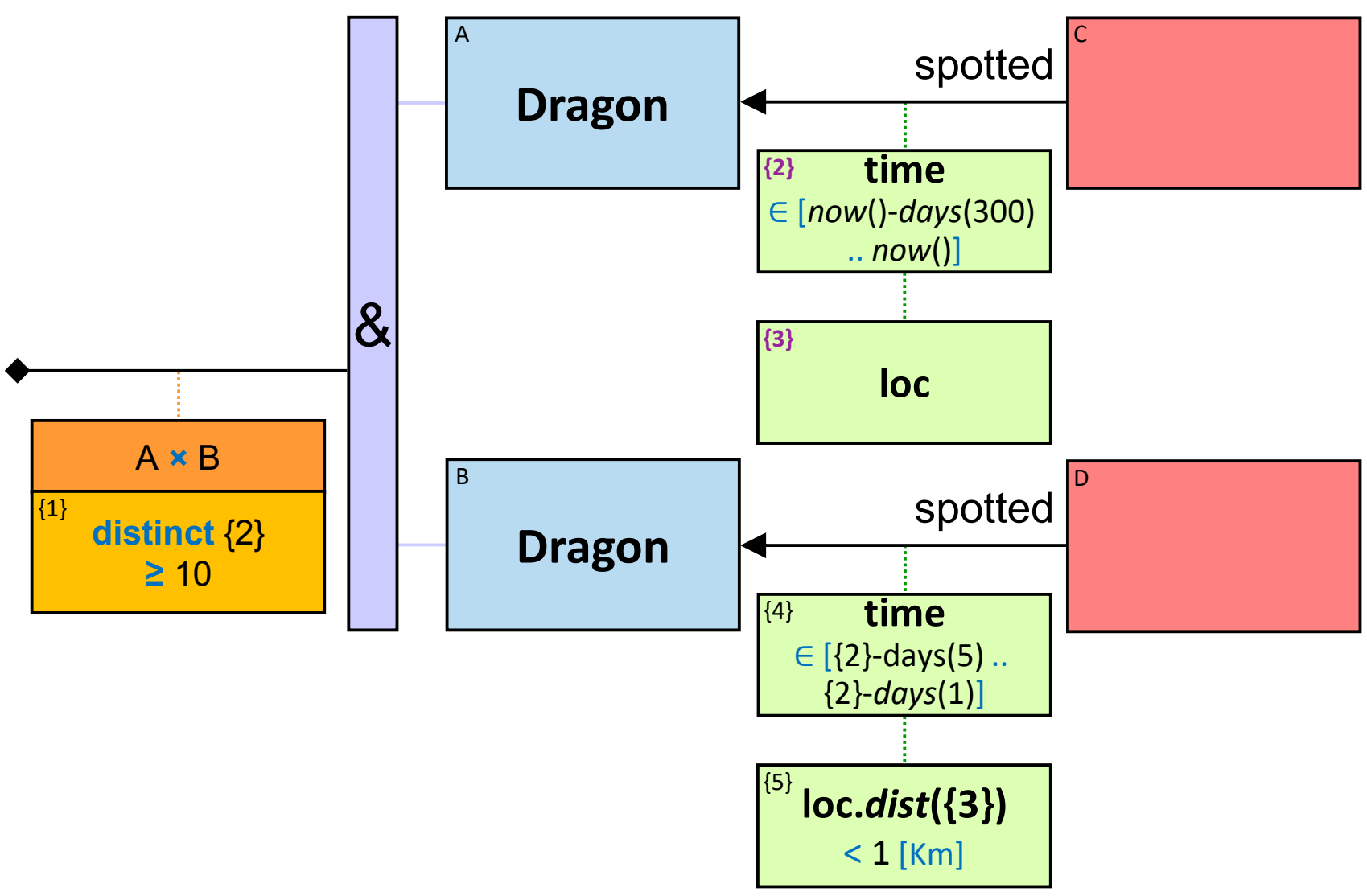

***G12:*** *Any pair of dragons that were spotted within 5 Km from each other at least five observation-pairs during a 30-day period (To ensure this – paired observation should differ in up to 5 minutes; observation pairs should differ in at least 24 hours)*

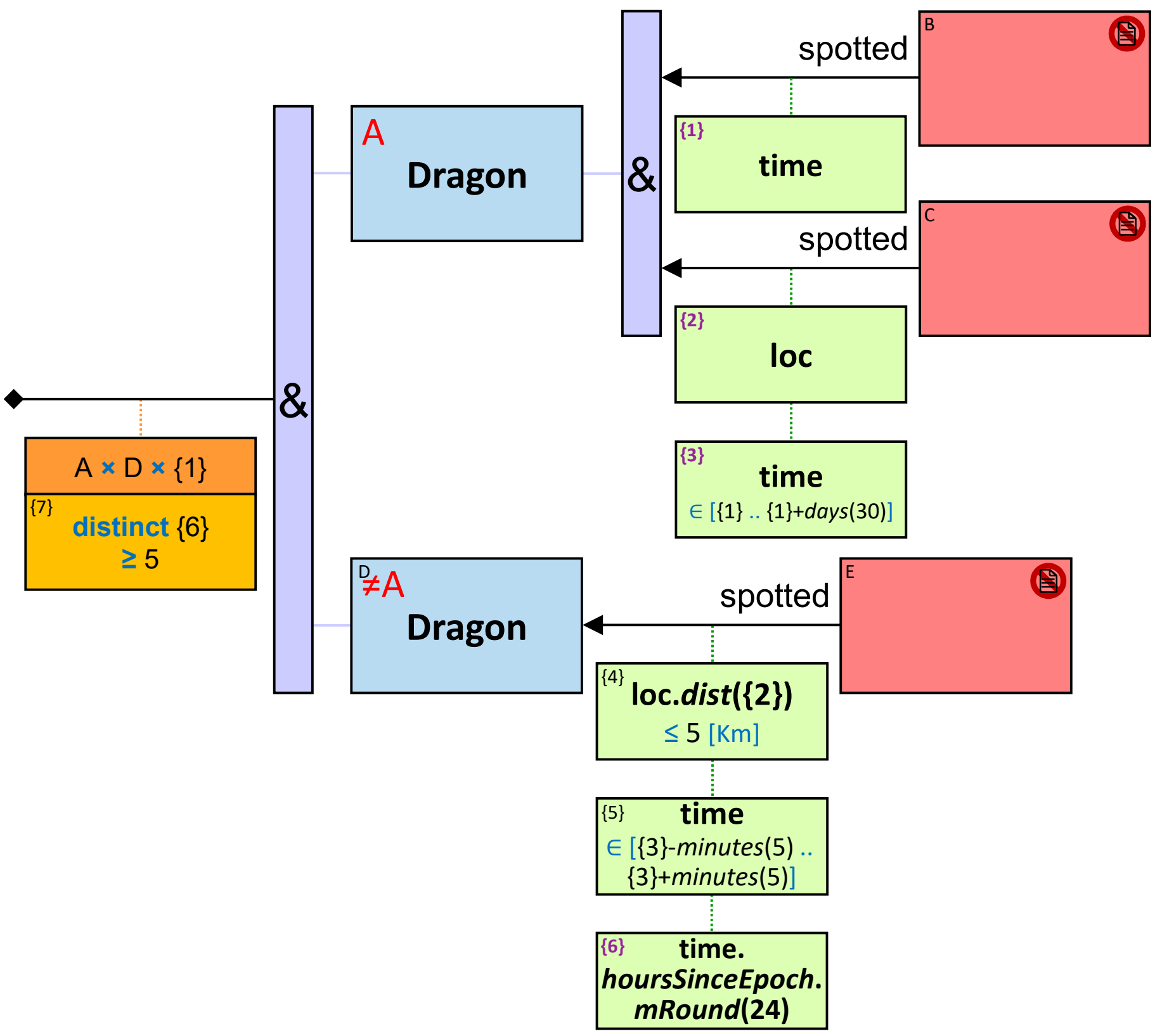

***G13:*** *Any Sarnorian E whose dragons seem to "spy" on Balerion (at least one of E's dragons was spotted at most 1 Km from Balerion throughout a ten-day period. Paired observation should differ in up to 5 minutes; observation pairs should differ in no more than 1 hour)*

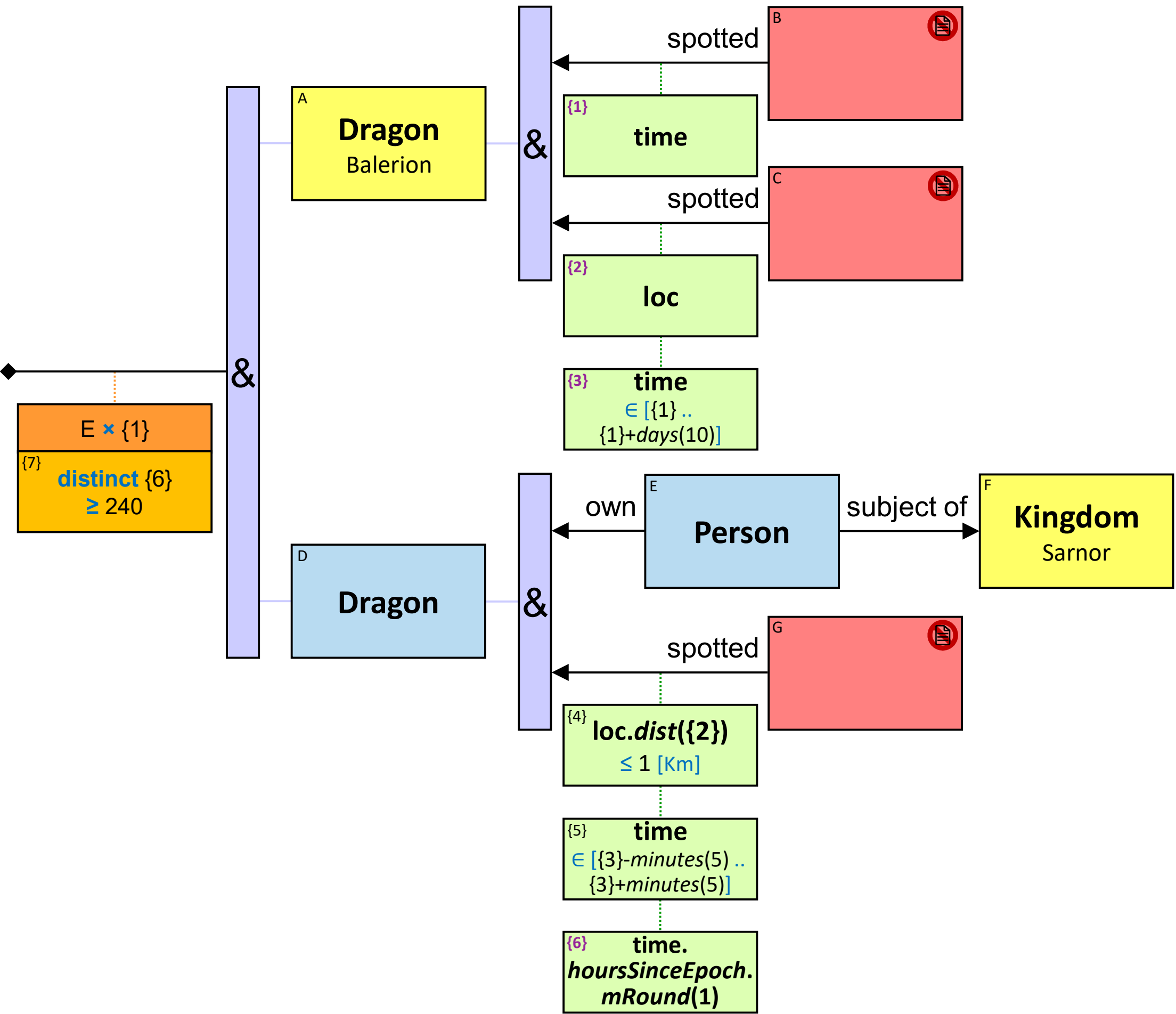

***G14:** Any pair of dragons that were spotted not less than 1000 Km from each other throughout a period of at least one year (To ensure this – paired observation should differ in up to 30 minutes; observation pairs should differ in no more than 24 hours)*

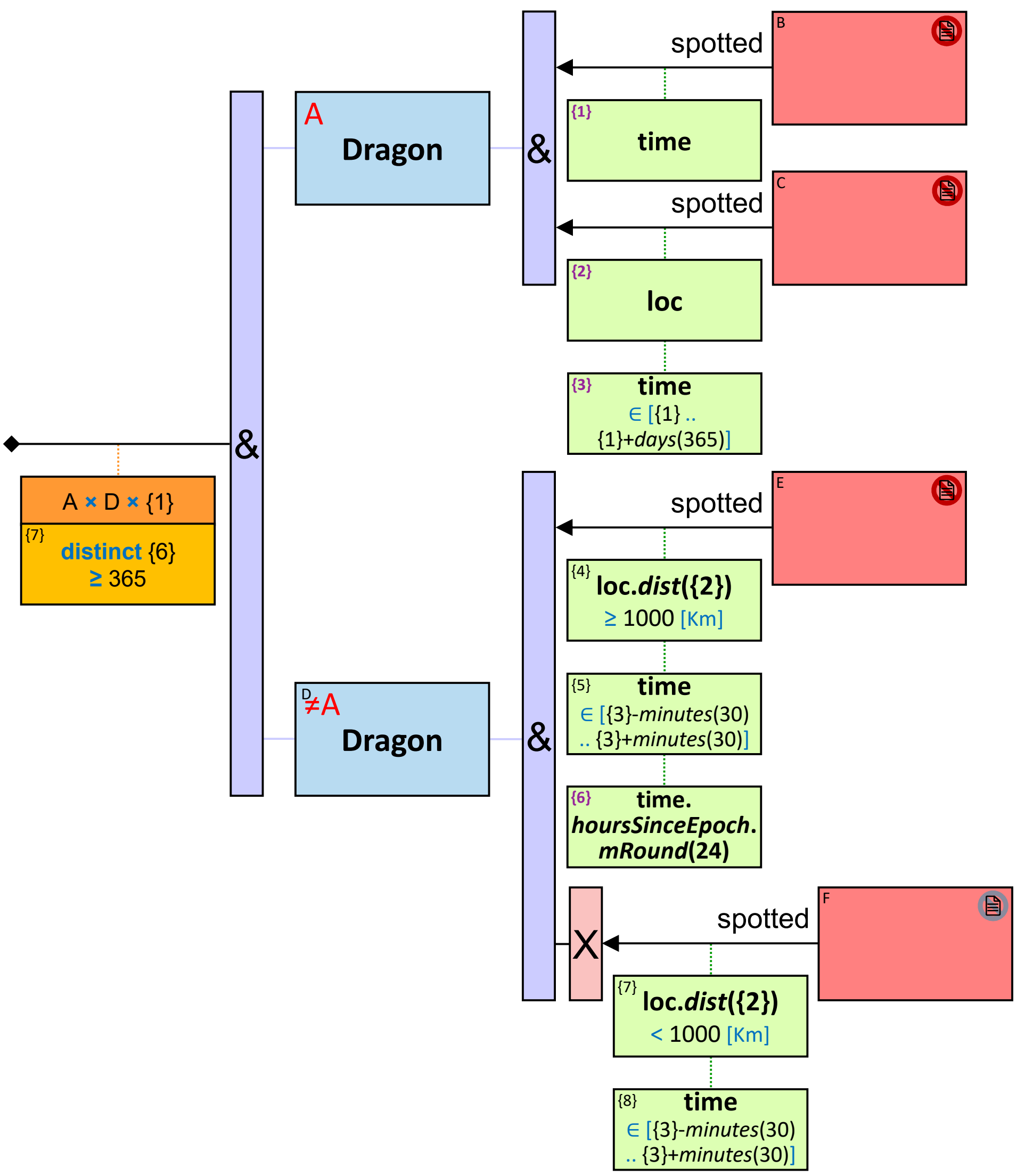

***G15:** Any landmark where at least five dragons were spotted within 5 Km from it*

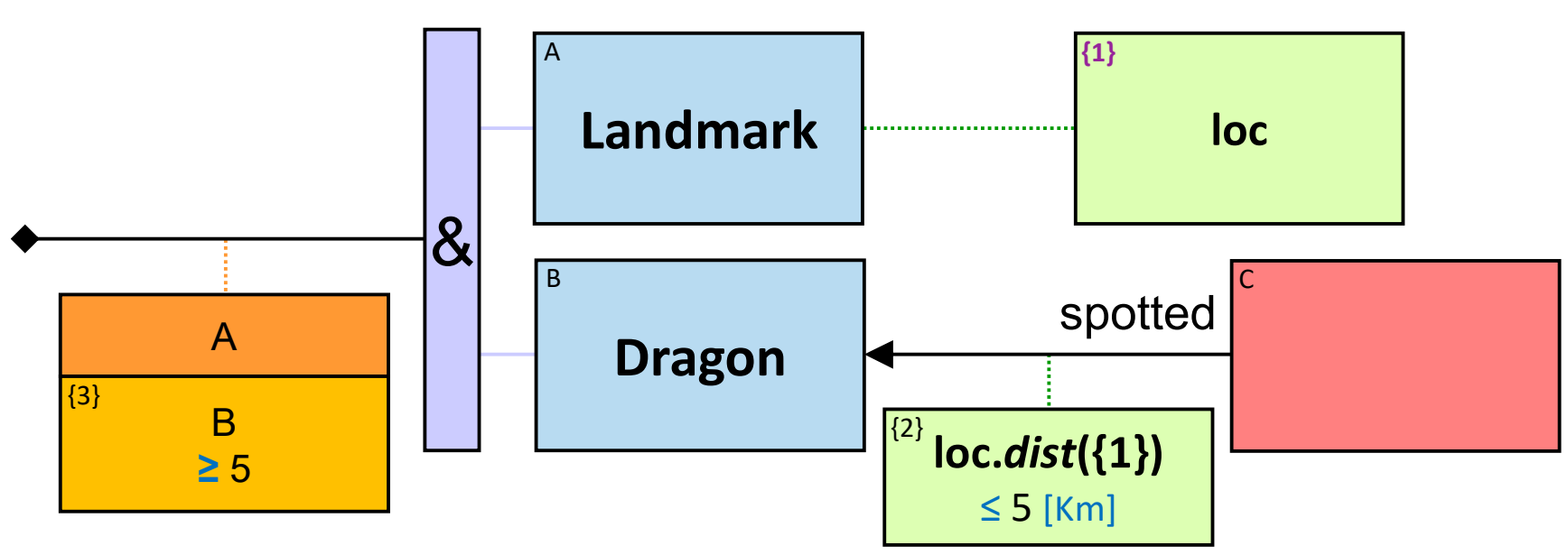

***G16:** Any city in which at least five dragons were spotted throughout a 2-minutes period*

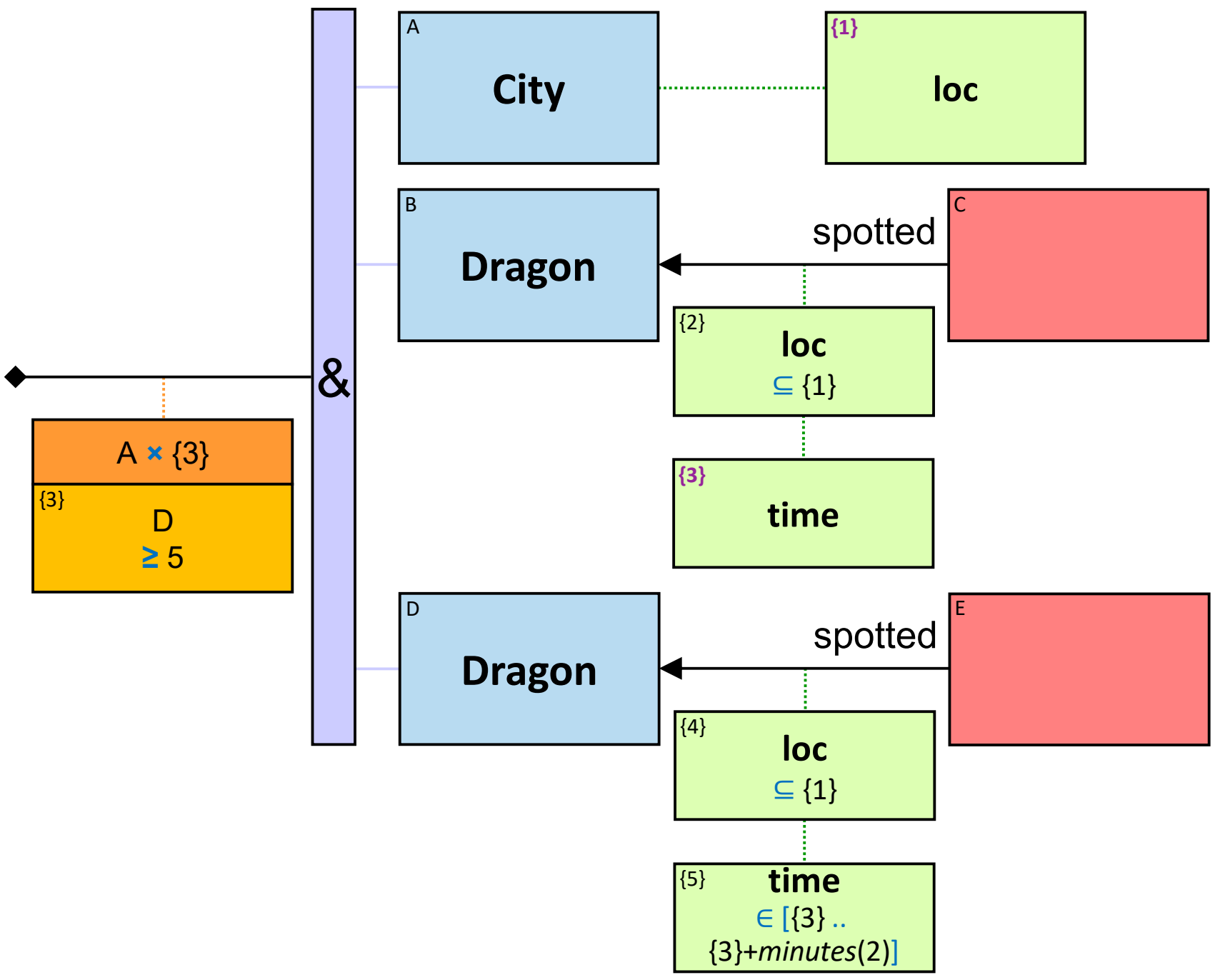

***G17:** Any landmark where at least five dragons were spotted within 5 Km from it throughout a 7-day period*

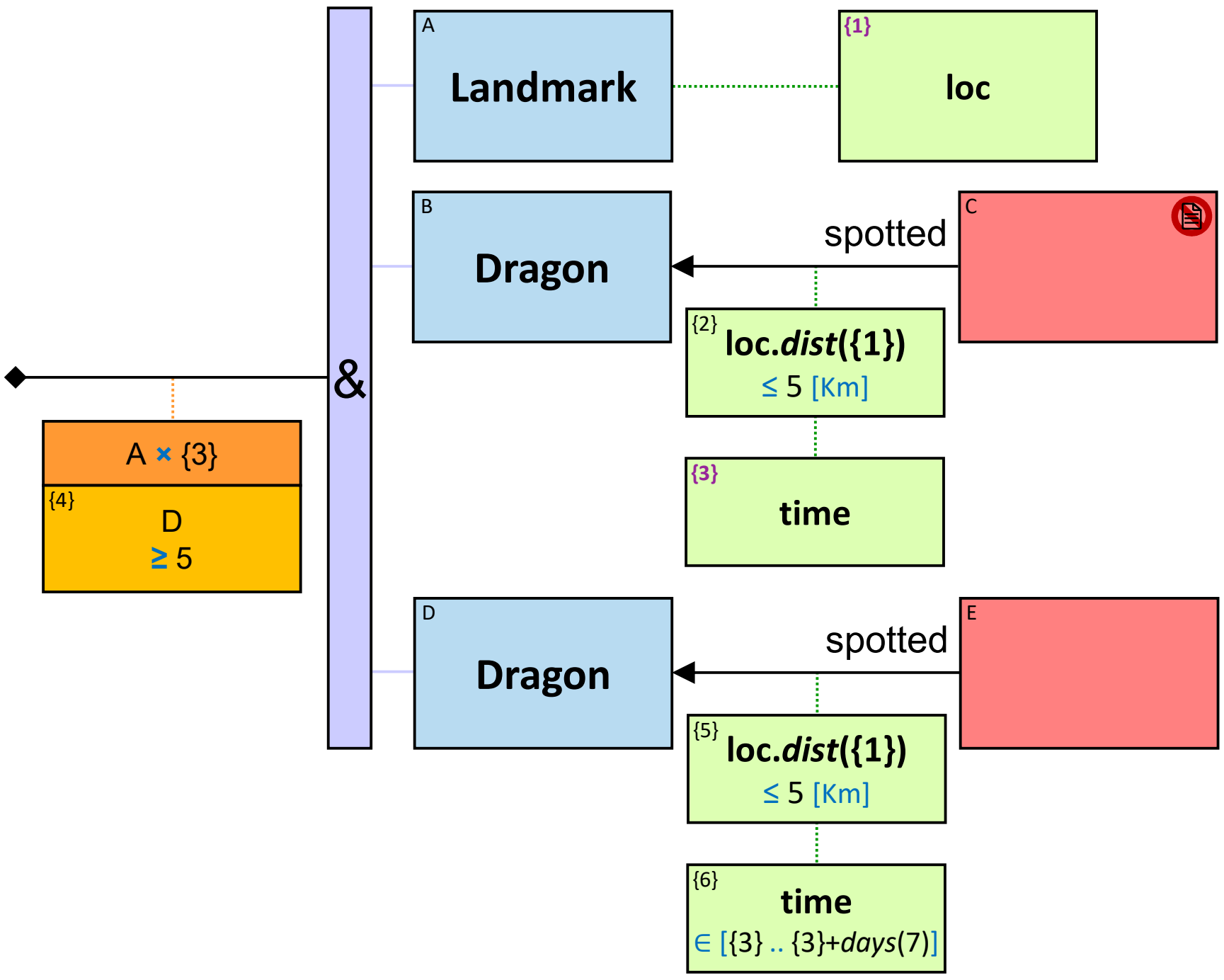

***G18:*** *Any dragon spotted in at least ten cities*

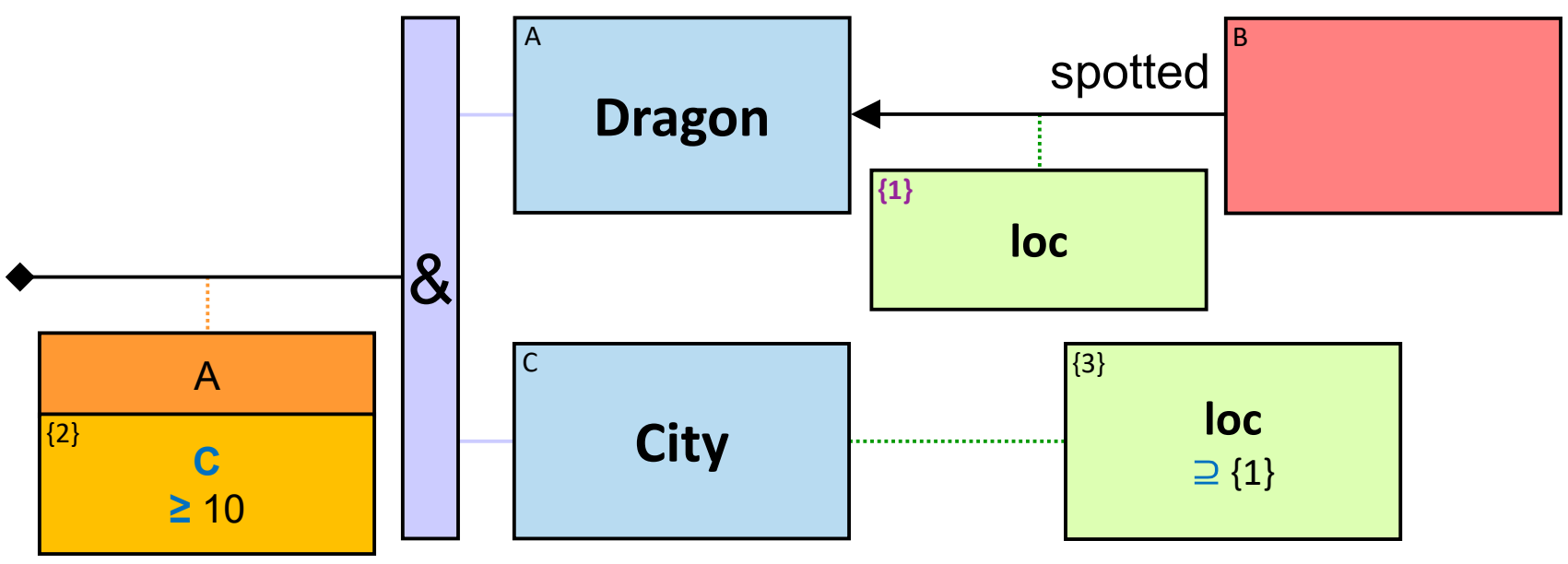

***G19:*** *Any dragon spotted in at least ten cities throughout a ten-day period*

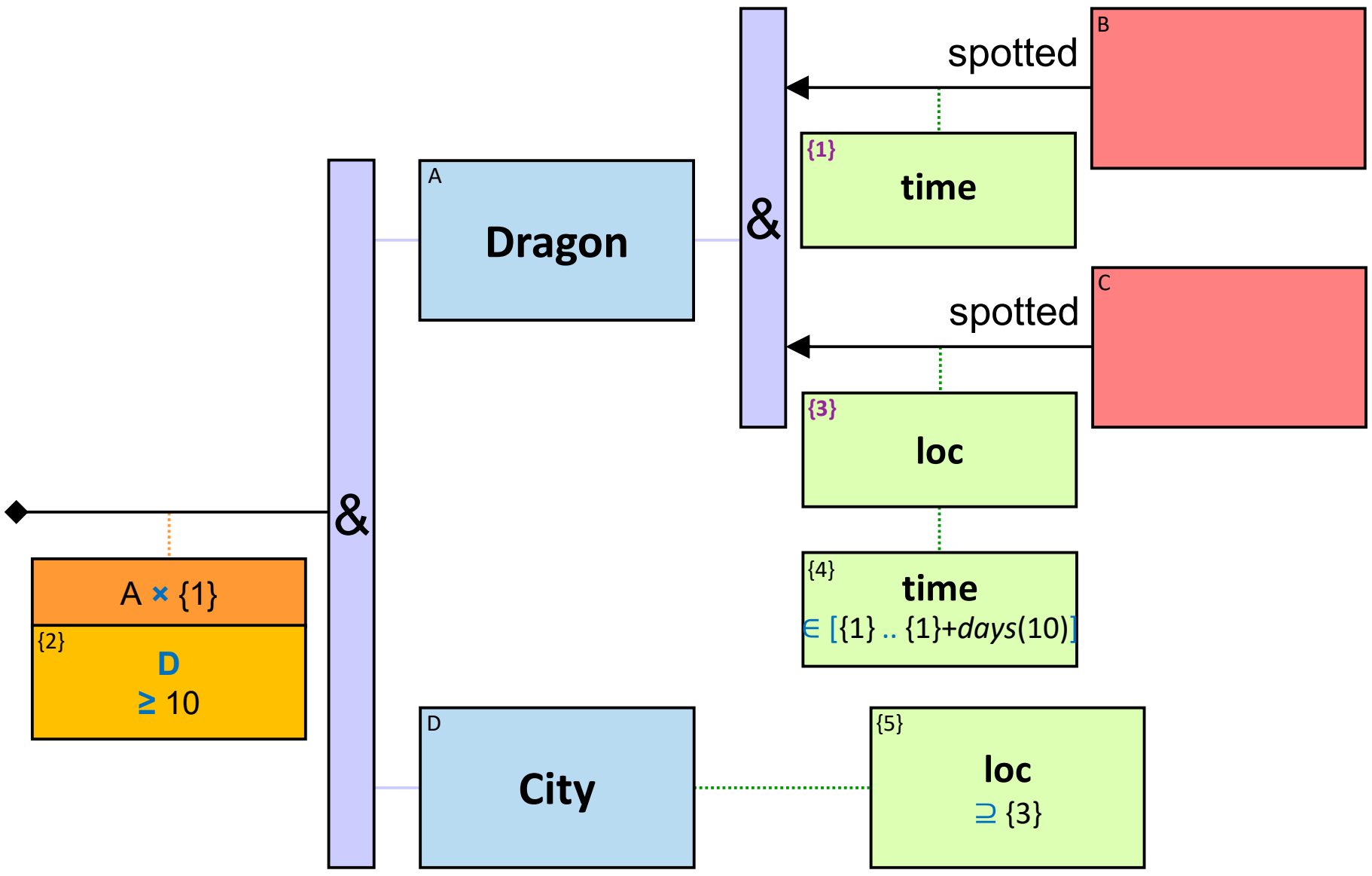

*G20: Any dragon that traveled at least 2000 Km in a 24-hours period*

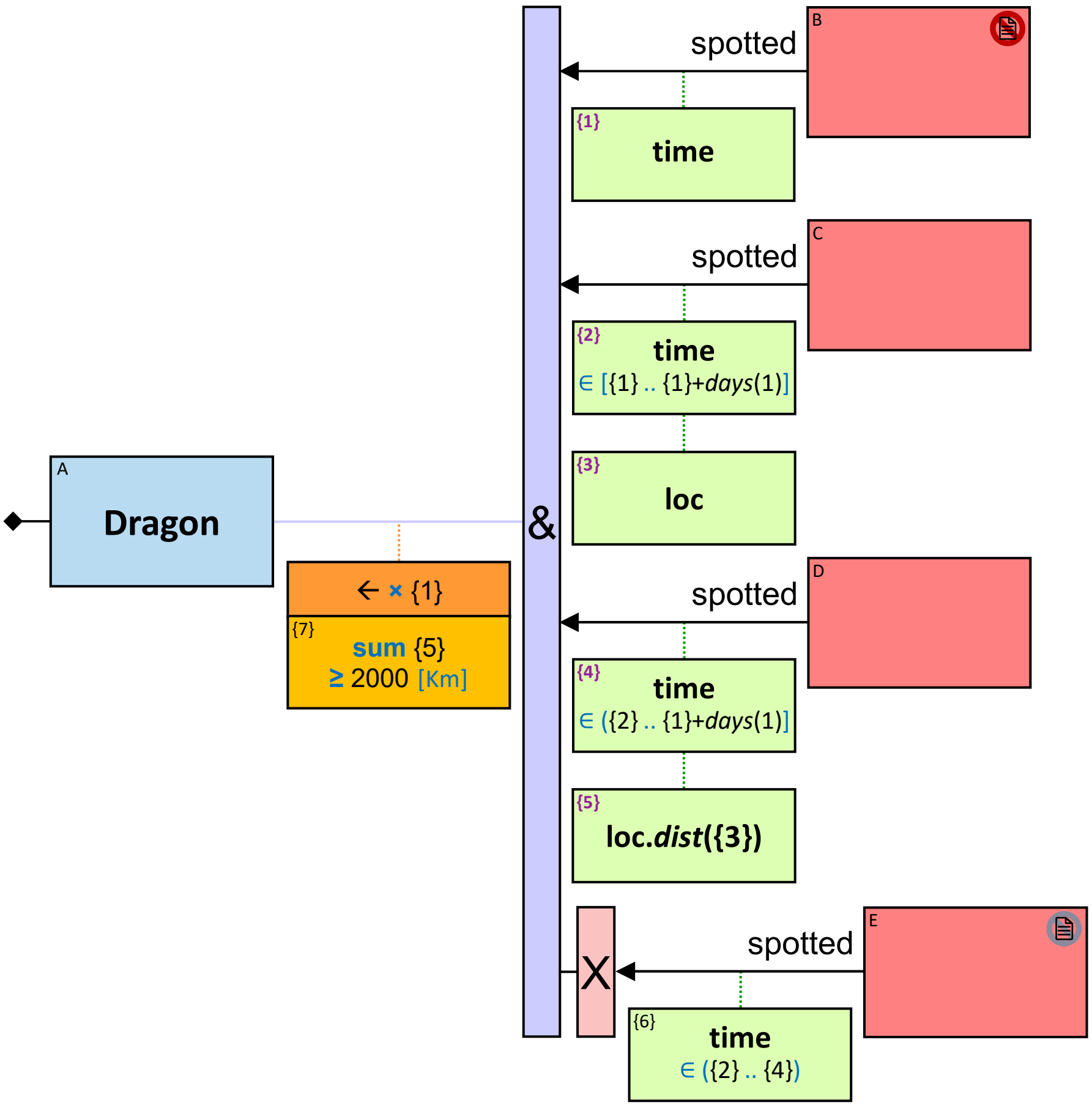

*G21: Any dragon-spotter that spotted the same dragon at two locations – at least 1000 Km apart*

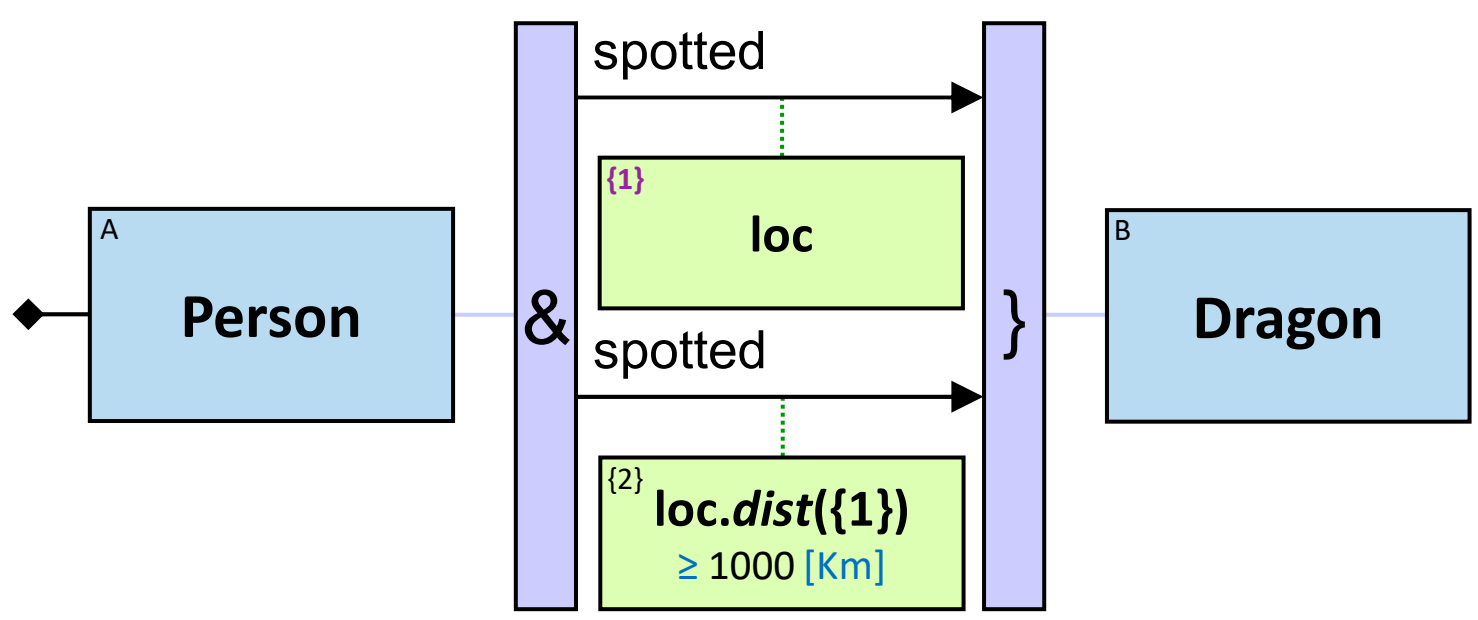

## 46. ACKNOWLEDGEMENTS

The author would like to thank Roman Margolis for fruitful discussions and helpful suggestions during this project's early stages and Thomas Frisendal for remarks on an early draft.